\documentstyle[11pt,german]{report}
\title{Ableitung von Feldgleichungen aus dem Prinzip des fermionischen Projektors}
\author{Felix Finster\\
Harvard University, Department of Mathematics}
\date{"uberarbeitete und leicht gek"urzte Fassung\\ Dezember 1996}

\makeatletter
\long\def\@makefntext#1{\parindent 1em\noindent
        \hbox to 1.8em{\hss$^{\@thefnmark}$}#1}
\makeatother

\newtheorem{Def}{Def.}[section]
\newtheorem{Thm}[Def]{Theorem}

\newtheorem{Lemma}[Def]{Lemma}

\newtheorem{Satz}[Def]{Satz}
\newcommand{\Beweis}{\em{Beweis:}}
\newcommand{\QED}{\ \hfill $\Box$ \\[1em]}

\newcommand{\Aslsh}{\mbox{ $\!\!A$ \hspace{-1.2 em} $/$}}

\newcommand{\Equ}[1]{\begin{equation} \label{eq:#1}}
\newcommand{\EndEqu}{\end{equation}}
\newcommand{\Ref}[1]{(\ref{eq:#1})}
\newcommand{\bra}{\mbox{$< \!\!$ \nolinebreak}}
\newcommand{\ket}{\mbox{\nolinebreak $>$}}
\newcommand{\C}{\:\mbox{\rm I \hspace{-1.25 em} {\bf C}}}
\newcommand{\R}{\mbox{\rm I \hspace{-.8 em} R}}
\newcommand{\1}{\mbox{\rm 1 \hspace{-1.05 em} 1}}
\newcommand{\Z}{\mbox{\rm \bf Z}}
\newcommand{\sZ}{\mbox{\rm \bf \scriptsize Z}}
\newcommand{\sR}{\mbox{\rm \scriptsize I \hspace{-.8 em} R}}
\newcommand{\N}{\mbox{\rm I \hspace{-.8 em} N}}

\newcommand{\Pdd}{\mbox{$\partial$ \hspace{-1.2 em} $/$}}
\newcommand{\slsh}{\mbox{ \hspace{-1.1 em} $/$}}

\newcommand{\Tr}{\mbox{Tr\/}}
\newcommand{\tr}{\mbox{tr\/}}
\newcommand{\Rg}{\mbox{Rg\/}}

\newcommand{\lep}{{\mbox{\tiny lep}}}
\newcommand{\qu}{{\mbox{\tiny qu}}}
\newcommand{\vektoriell}{{\mbox{\tiny vektoriell}}}
\newcommand{\axial}{{\mbox{\tiny axial}}}

\newcommand{\eff}{{\mbox{\scriptsize eff}}}

\newcommand{\el}{{\mbox{\scriptsize el.}}}
\newcommand{\spc}{\;\;\;\;\;\;\;\;\;\;}
\newcommand{\lint}{\: \Diamond \hspace{-0.98 em} \int}
\newcommand{\wint}{\: \bigtriangleup \hspace{-1.09 em} \int}
\newcommand{\vint}{\: \bigtriangledown \hspace{-0.98 em} \int}

\newcommand{\inti}{\int^\infty_{-\infty}}

\newcommand{\intL}{{\mbox{ \footnotesize{L}}} \hspace{-1.3 em} \int}
\newcommand{\fintL}{{\mbox{ \footnotesize{L}}} \hspace{-1.6 em} \int}
\newcommand{\sintL}{{\mbox{ \tiny{L}}} \hspace{-1.0 em} \int}
\newcommand{\intR}{{\mbox{ \footnotesize{R}}} \hspace{-1.3 em} \int}
\newcommand{\fintR}{{\mbox{ \footnotesize{R}}} \hspace{-1.6 em} \int}
\newcommand{\sintR}{{\mbox{ \tiny{R}}} \hspace{-1.0 em} \int}
\newcommand{\intLR}{{\mbox{ \footnotesize{L/R}}} \hspace{-2.3 em} \int }
\newcommand{\intRL}{{\mbox{ \footnotesize{R/L}}} \hspace{-2.3 em} \int }
\newcommand{\tintLR}{ \tilde{\mbox{\footnotesize{L}}} \!/
        \tilde{\mbox{\footnotesize{R}}} \hspace{-2.2 em} \int }

\newcommand{\slint}{{\scriptstyle \Diamond \hspace{-0.72 em}} \int}
\newcommand{\swint}{{\scriptstyle \bigtriangleup \hspace{-0.72 em}} \int}
\newcommand{\svint}{{\scriptstyle \bigtriangledown \hspace{-0.72 em}} \int}

\newcommand{\I}{\Im}
\newcommand{\Li}{{\cal L}}

\newcommand{\pe}{p^{(1)}}
\newcommand{\ke}{k^{(1)}}

\newcommand{\T}{{\mbox{T}}}
\newcommand{\Texp}{{\mbox{Texp}}}

\newcommand{\iso}{{\mbox{\scriptsize{iso}}}}

\begin{document}

\maketitle
\newpage
\tableofcontents

\chapter{Einleitung}
\label{kap1}
Die vorliegende Arbeit baut auf "Uberlegungen in \cite{Physdip} auf.
Dort wird vorgeschlagen, lokale Eichfreiheiten in der Physik durch
die Willk"ur in der Wahl der Basis eines Ska\-lar\-pro\-dukt\-rau\-mes
bei Vorgabe gewisser, als fundamental angesehener Operatoren zu erkl"aren.
Es wird gezeigt, da"s dieses Konzept im Rahmen der relativistischen
Quantenmechanik (also ohne zweite Quantisierung) zu einer einheitlichen
Beschreibung der Elektrodynamik und Gravitation als Eichtheorie f"uhrt.
Um selbstkonsistent zu sein, wollen wir zu Beginn einige Begriffe und
Konstruktionen aus \cite{Physdip} zusammenstellen.

Wir betrachten zun"achst die freie Diracgleichung $(i \Pdd -m) \: \Psi=0$.
Als Zustandsraum $H$ w"ahlen wir die vierkomponentigen Wellenfunktionen auf dem
Minkowski-Raum mit dem lorentzinvarianten, indefiniten Skalarprodukt
\Equ{1_1}
\bra \Psi, \Phi \ket \;=\; \int_{\sR^4} \overline{\Psi}(x) \:
	\Phi(x) \; d^4x \spc ,
\EndEqu
dabei bezeichnet $\overline{\Psi}=\Psi^* \gamma^0$ den
adjungierten Spinor.
Die Raumzeit beschreiben wir mit den hermiteschen, miteinander
kommutierenden Operatoren $(X^i)_{i=0,\ldots,3}$. Diese Operatoren sind
genau wie der Ortsoperator der nichtrelativistischen Quantenmechanik als
Multiplikationsoperatoren mit den Koordinatenfunktionen definiert
\[ (X^i \: \Psi)(x) \;=\; x^i \; \Psi(x) \spc . \]
Wir k"onnen die Raumzeit als das Spektrum der $X^i$ ansehen.

Im n"achsten Schritt fassen wir $H$ als abstrakten Skalarproduktraum
(also nicht mehr als speziellen Funktionenraum) auf und sehen die
Operatoren $X^i, i \Pdd$ als Ausgangspunkt der Diractheorie an.
Auf diese Weise erh"alt man unmittelbar lokale Eichfreiheiten:
Da $H$ ein abstrakter Skalarproduktraum ist, mu"s die Darstellung
der Vektoren aus $H$ als Wellenfunktionen mit Hilfe der Operatoren
$X^i$ konstruiert werden.
Dazu w"ahlt man eine ``Eigenvektorbasis''\footnote{Wir verwenden die
in der Physik gebr"auchliche bra/ket-Schreibweise, $\bra.|.\ket$ ist das
Skalarprodukt \Ref{1_1}. Die Gleichungen \Ref{1_2}, \Ref{1_3} sind aus
mathematischer Sicht unbefriedigend, weil die Operatoren $X^i$ ein
kontinuierliches Spektrum und damit keine
normierbaren Eigenvektoren besitzen. Mit etwas gr"o"serem
Aufwand (Spektralma"se, Radon-Nikodym Theorem) l"a"st sich die Konstruktion
jedoch mathematisch sauber durchf"uhren, siehe \cite{Mathdip}.}
$|x \alpha \ket$, $x \in \R^4$, $\alpha = 1,\ldots,4$ der $X^i$
\begin{eqnarray}
\label{eq:1_2}
X^i \: |x \alpha \ket &=& x^i \: |x \alpha \ket \\
\label{eq:1_3}
\bra x \alpha | y \beta \ket &=& \delta^4(x-y) \: \delta_{\alpha \beta} \:
	s_\alpha \;\;\;,\spc s_1=s_2=1, \; s_3=s_4=-1
\end{eqnarray}
und definiert zu $\Psi \in H$ die zugeh"orige Wellenfunktion $\Psi^\alpha(x)$
durch
\Equ{1_4}
\Psi^\alpha(x) \;=\; \bra x \alpha | \Psi \ket \spc .
\EndEqu
Entscheidend ist dabei, da"s die ``Basisvektoren'' $|x \alpha \ket$ nicht
eindeutig bestimmt sind, sondern gem"a"s
\Equ{1_5}
|x \alpha \ket \;\longrightarrow\; \sum_{\beta=1}^4 U_{\alpha \beta}(x)
	\: |x \beta \ket \;\;\;,\spc U(x) \in U(2,2)
\EndEqu
transformiert werden k"onnen. Bei dieser Transformation bleiben n"amlich
die Bedingungen \Ref{1_2}, \Ref{1_3} erhalten, wie man direkt
verifiziert. Gem"a"s der Definitionsgleichung \Ref{1_4} entspricht
\Ref{1_5} einer Transformation
\[ \Psi(x) \;\longrightarrow\; U^{-1}(x) \: \Psi(x) \]
der Wellenfunktionen, was als lokale $U(2,2)$-Eichfreiheit interpretiert
werden kann. Die Eichgruppe wirkt bei uns also direkt auf die
Spinorkomponenten, die $U(1)$-Pha\-sen\-trans\-for\-ma\-tio\-nen der
Elektrodynamik ergeben sich als ein Spezialfall.

Um auch die Dynamik mit dieser $U(2,2)$-Eichsymmetrie zu beschreiben, mu"s
man den freien Diracoperator $i \Pdd$ verallgemeinern: Zun"achst lassen wir
wie in der Allgemeinen Relativit"atstheorie krummlinige Koordinaten
als gleichberechtigte Bezugssysteme zu\footnote{Wir nehmen im folgenden zur
Einfacheit an, da"s die Raumzeit durch eine einzige Karte
beschrieben werden kann. Den allgemeineren Fall, da"s die zugeh"orige
Lorentzmannigfaltigkeit topologisch nicht trivial ist, erh"alt man
wie gewohnt durch Verkleben von Karten. Dies f"uhrt auf
den Begriff der Operatormannigfaltigkeit, siehe \cite{Mathdip}.}.
Zu einem allgemeinen Koordinatensystem $(x^i)$ definiert man die
Orts-/Zeit\-ope\-ra\-to\-ren $(X^i)$ wieder als Multiplikationsoperatoren
mit den Koordinatenfunktionen.
Der Diracoperator $G$ wird als ein hermitescher Differentialoperator erster
Ordnung auf $H$ definiert. In einem speziellen Bezugssystem und einer
speziellen Eichung kann man ihn also in der Form
\Equ{1_7}
G \;=\; i G^j(x) \: \frac{\partial}{\partial x^j} + B(x)
\EndEqu
mit geeigneten $(4 \times 4)$-Matrizen $G^j(x), B(x)$ darstellen, die
von den Koordinaten und der Eichung abh"angen.
"Ahnlich dem Einsteinschen "Aquivalenzprinzip fordern wir, da"s man durch
geeignete Wahl des Bezugssystems und der Eichung erreichen kann, da"s
$G$ lokal die Form des freien Diracoperators annimmt.
Zu jedem Punkt $p$ der Raumzeit soll es also ein Koordinatensystem und
eine Eichung geben, so da"s $G^j(p)=\gamma^j$, $B(p)=0$.

Dadurch, da"s in die Definition des Diracoperators nur eine lokale Bedingung
an die Matrixfelder $G^j, B$ eingeht, enth"alt $G$ im allgemeinen
Potentiale, die nicht global wegtransformiert werden k"onnen.
Es zeigt sich, da"s wir auf diese Weise genau das elektromagnetische Feld
und Gravitationsfeld eingef"uhrt haben. Durch die verallgemeinerte
Diracgleichung $(G-m) \: \Psi=0$ wird die Ankopplung dieser Felder
an die Fermionen auf physikalisch sinnvolle Weise beschrieben.

Um zu verstehen, wie die Gravitation durch die $U(2,2)$-Symmetrie zu einer
Eichtheorie wird, mu"s man die Beziehung zwischen Koordinaten- und
Eichtransformationen untersuchen.
"Ublicherweise werden bei einem Wechsel des Bezugssystems sowohl die
Raum/Zeit-Koordinaten als auch die Spinorkomponenten transformiert.
Bei unserer Beschreibung bleiben die Spinorkomponenten bei Koordinatenwechseln
unver"andert, man kann das "ubliche Transformationsverhalten der Spinoren
aber durch eine anschlie"sende Eich\-trans\-for\-ma\-tion realisieren.
Auf diese Weise sind Koordinatentransformationen mit Eichtransformationen
verkn"upft, und man kann die Freiheiten in der Koordinatenwahl mit
entsprechenden $U(2,2)$-Eichfreiheitsgraden identifizieren.
Dabei wird ausgenutzt, da"s die durch die bilinearen Kovarianten
$\sigma^{ij} \in su(2,2)$ erzeugte Untergruppe von $U(2,2)$ eine
"Uberlagerung der Lorentzgruppe ist.

Zur vollst"andigen Beschreibung der Wechselwirkung zwischen Eichfeldern
und Fermionen m"ussen wir noch die klassischen Feldgleichungen
(Maxwell- und Einsteingleichungen) aufstellen.
Wichtig ist, da"s wir dazu keine weiteren mathematischen Strukturen
einf"uhren m"ussen, weil alle ben"otigten Objekte aus dem Diracoperator
konstruiert werden k"onnen.
Insbesondere brauchen wir im Gegensatz zu den "ublichen Eichtheorien nicht
eine eichkovariante Ableitung $D_j=\partial_j - i e A_j$ mit Eichpotentialen
$A_j$ als zus"atzlichen physikalischen Gr"o"sen zu definieren.

Zur Konstruktion der Eichpotentiale und Feldst"arken aus dem Diracoperator
arbeitet man in der Darstellung \Ref{1_7} und nutzt das bekannte
Koordinaten- und Eichtransformationsverhalten der Matrixfelder $G^j, B$ aus.
Wir beschreiben das Vorgehen schematisch: "Uber die Definitionsgleichung
\Equ{1_6a}
	g^{jk}(x) \;=\; \frac{1}{2} \left\{ G^j(x), G^k(x) \right\}
\EndEqu
erh"alt man die Lorentzmetrik und daraus den Levi-Civita-Zusammenhang
$\nabla$. Mit geeigneten Kombinationen des matrixwertigen Tensors
$\nabla_j \: G^k$ kann man die Eichung global bis auf die
$U(1)$-Eichtransformationen der Elektrodynamik fixieren. In einer solchen
Eichung ist $B=\Aslsh$, wobei $A$ das elektromagnetische Potential
bezeichnet.
Auf kanonische Weise erh"alt man aus dieser Konstruktion die Spinableitung
$D$, welche der eichkovarianten Ableitung der "ublichen Eichtheorien
entspricht. Die Kr"ummung des Spinzusammenhanges setzt sich aus dem
elektromagnetischen Feldst"arketensor und dem Riemannschen Kr"ummungstensor
zusammen. Mit diesen Tensoren stellt man
die klassische Lagrangedichte auf und erh"alt durch Variation die
Maxwell- und Einsteingleichungen.
F"ur das Variationsprinzip ist zu beachten, da"s die zu variierenden
Potentiale und Felder aus dem Diracoperator abgeleitet sind. Dadurch
mu"s man den Diracoperator selbst als dynamische Gr"o"se auffassen: bei
Variationen wird der Diracoperator ver"andert, wodurch mittelbar
auch die Potentiale und Felder variiert werden.

Wir bemerken, da"s der Diracoperator bei unserer Beschreibung der
relativistischen Quantenmechanik eine zweifache Rolle spielt. Auf der einen
Seite hat man "uber die Eigenwertgleichung $G \Psi = m \Psi$ eine
Beziehung zwischen den Fermionen und dem Spektrum von $G$.
Auf der anderen Seite werden aus $G$ die Eichpotentiale konstruiert,
so da"s der Diracoperator die Felder der Eichbosonen bestimmt.
Allgemein sieht man, da"s wir lediglich die Operatoren $X^i, G$ auf $H$
als fundamentale physikalische Objekte auffassen m"ussen, alle
weiteren Gr"o"sen k"onnen daraus abgeleitet werden.
Dies ist begrifflich sehr einfach und bildet die Grundlage f"ur unsere
weiteren Konstruktionen.

Es ist klar, da"s das bisherige System f"ur ein realistisches
physikalisches Modell noch zu einfach ist. Um weitere Quantenzahlen wie
Isospin, Colour oder Leptonenzahl zu ber"ucksichtigen, m"ussen wir
die Anzahl der Komponenten der Wellenfunktionen erh"ohen.
Wir f"uhren an dieser Stelle noch nicht die genaue Konstruktion durch,
sondern wollen nur die Eichgruppe bei einer beliebigen Anzahl von
Komponenten untersuchen. Dazu betrachten wir $(p+q)$-komponentige
Wellenfunktionen und ersetzen das Skalarprodukt $\overline{\Psi} \Phi$
bei Diracspinoren durch ein Skalarprodukt der Signatur $(p,q)$.
Analog zu \Ref{1_1} definieren wir also durch
\begin{eqnarray*}
\bra \Psi, \Phi \ket &=& \int_{\sR^4} \: \sum_{\alpha=1}^{p+q} s_\alpha \;
	\overline{\Psi}^\alpha(x) \: \Phi^\alpha(x) \; d^4x
	\spc {\mbox{mit}} \\
s^1 \!\!\!&=&\!\!\! \cdots=s^p=1 \;\;,\spc s^{p+1}=\cdots=s^{p+q}=-1
\end{eqnarray*}
ein indefinites Skalarprodukt auf den Wellenfunktionen,
dabei ist $\overline{\Psi}=\Psi^*$ die komplex konjugierte Wellenfunktion.
Wir nennen $p+q$ die {\em{Spindimension}}\index{Spindimension}
des Systems.
Eine ``Eigenvektorbasis'' $|x \alpha\ket$, $x \in \R^4$, $\alpha=1,\ldots,
p+q$ der $X^i$ ist wieder durch die Gleichungen
\Equ{1_10}
X^i \: |x \alpha \ket = x^i \: |x \alpha \ket \;\;\;,\;\;\;\;
	\bra x \alpha | y \beta \ket = \delta^4(x-y) \: \delta_{\alpha \beta}
	\: s_\alpha
\EndEqu
gegeben. Die Willk"ur in der Wahl der $|x \alpha \ket$ f"uhrt jetzt
auf lokale Eichfreiheiten mit der Eichgruppe $U(p,q)$.
Wir sehen also, da"s eine Vergr"o"serung der Spindimension
eine gr"o"sere Eichgruppe zur Folge hat. Mit den zus"atzlichen
Eichfreiheitsgraden sollten sich zus"atzliche Wechselwirkungen
beschreiben lassen.

Man beachte, da"s die Eichgruppe bei uns bereits durch die Anzahl der
Komponenten der Wellenfunktionen festgelgt ist.
Auf diese Weise sind wir bei der Modellbildung gegen"uber den "ublichen
Eichtheorien stark eingeschr"ankt, bei denen die Eichgruppe und die
Ankopplung der Fermionen an die Eichfelder willk"urlich gew"ahlt werden
k"onnen.

Soweit die allgemeine Wiederholung der f"ur uns wichtigen Ergebnisse aus
\cite{Physdip}.
Es stellt sich die Frage, weshalb wir "uberhaupt versuchen wollen,
ausgehend von diesen "Uberlegungen ein realistisches physikalisches
Modell aufzubauen, obwohl wir doch bisher ohne zweite Quantisierung mit
klassischen Fermion- und Eichfeldern arbeiten.
Der Autor ist der Ansicht, da"s die relativistische Quantenfeldtheorie
in ihrer jetzigen Form (mit kanonischer Quantisierung oder in der
Formulierung mit Pfadintegralen
"uber Feldkonfigurationen) aus physikalischer und mathematischer Sicht
unbefriedigend ist, und da"s ein grundlegend anderer Ansatz ben"otigt wird,
um die Feldquantisierung wirklich zu verstehen.

Im n"achsten Abschnitt \ref{1_ab1} werden wir uns schrittweise von der
bisherigen klassischen Beschreibung l"osen und den mathematischen
Rahmen f"ur die Formulierung von Gleichungen schaffen, welche wir
``Gleichungen der diskreten Raumzeit'' nennen. Dabei wird kein
Zusammenhang zu einer Quantisierung der Felder erkennbar sein; es ist
zun"achst auch nicht klar, ob diese Konstruktionen physikalisch sinnvoll
sind. In Abschnitt \ref{1_ab2} wird dann qualitativ beschrieben, wie
man aus den Gleichungen der diskreten Raumzeit in einem bestimmten
Grenzfall, dem sogenannten Kontinuumslimes, wieder klassische
Gleichungen erh"alt. Erst "uber den Kontinuumslimes lassen sich die
Gleichungen der diskreten Raumzeit in eine f"ur uns gewohnte und damit
physikalisch interpretierbare Form bringen.
Die Diskussion des Kontinuumslimes f"uhrt in die eigentliche Thematik
dieser Arbeit ein, denn wir werden uns haupts"achlich damit
besch"aftigen, den Kontinuumslimes mathematisch zu fundieren
und f"ur verschiedene Modelle zu untersuchen.
In Abschnitt \ref{1_ab3} sind die Ergebnisse f"ur ein System
zusammengestellt, das der Fermionkonfiguration des Standardmodells
nachgebildet ist. In Abschnitt \ref{1_ab5} werden wir schlie"slich
auf die Feldquantisierung zur"uckkommen.

\section{Das Prinzip des fermionischen Projektors}
\label{1_ab1}
\subsection{Diskretisierung der Raumzeit}
\label{1_ab11}
Die Annahme, da"s die bekannten physikalischen Gleichungen auf
beliebigen L"angenskalen g"ultig sind, f"uhrt auf Schwierigkeiten,
wenn man zu Systemen in der Gr"o"senordnung der Planck-L"ange "ubergeht.
Bei einer recht naiven Betrachtungsweise treten Inkonsistenzen auf,
weil beispielsweise die gravitative Wechselwirkung der Energiefluktuationen
des Vakuums zu gro"s wird.
Ber"ucksichtigt man mit Renormierungsgruppenrechnungen das
``floating'' der Kopplungskonstanten, so stellt man fest, da"s bei den
zugeh"origen Ener\-gie\-ska\-len die Kopplungen der elektromagnetischen,
starken und schwachen Wechselwirkung etwa gleich gro"s werden, was
manchmal als eine ``Vereinigung'' dieser Kr"afte interpretiert wird.
Auch die UV-Divergenzen in der perturbativen Quantenfeldtheorie scheinen
darauf hinzudeuten, da"s die Physik f"ur sehr kleine Abst"ande modifiziert
werden mu"s. Man h"atte dann n"amlich einen nat"urlichen
Cutoff f"ur sehr gro"se Impulse, was das Renormierungsprogramm aus
theoretischer Sicht rechtfertigen w"urde.

Aus diesen Gr"unden ist die Meinung verbreitet, da"s bei Abst"anden
von etwa $10^{-40}$ Metern neue physikalische Effekte auftreten.
Wir wollen annehmen, da"s die Raumzeit auf der Skala der Planck-L"ange
in diskrete Punkte aufgel"ost wird.
Um diese Vorstellung mathematisch zu verwirklichen, ersetzen wir die
$X^i$ (zun"achst in einem festen Bezugssystem)
durch hermitesche Operatoren, die weiterhin miteinander
kommutieren, aber ein diskretes Spektrum besitzen. Das gemeinsame
Spektrum dieser ``diskretisierten Orts/Zeit-Operatoren'' $X^i$, also die
Menge
\[ M \;=\; \{ x \in \R^4 \;|\; \exists u \in H \;{\mbox{ mit }}\;
	X^i u = x^i u \} \spc , \]
ist als unsere ``diskretisierte Raumzeit'' anzusehen.
Wir wollen annehmen, da"s die gemeinsamen Eigenr"aume $e_x$
der $X^i$,
\[  e_x \;=\; \{ u \;|\; X^i u = x^i u \} \;\;\;\;,\spc x \in M \spc , \]
$(p+q)$-dimensionale Unterr"aume von $H$ sind, auf denen das
Skalarprodukt $\bra .|. \ket$ die Signatur $(p,q)$ besitzt.
Dann k"onnen wir eine Basis $|x \alpha \ket$,
$x \in M$, $\alpha=1,\ldots,p+q$ w"ahlen mit
\Equ{1_c3}
X^i \: |x \alpha \ket = x^i \: |x \alpha \ket \;\;\;\;,\spc
	\bra x \alpha | y \beta \ket = \delta_{xy} \: \delta_{\alpha \beta}
	\: s_\alpha \spc .
\EndEqu
Diese Gleichungen unterscheiden sich von \Ref{1_10} nur durch die
Ersetzung $\delta^4(x-y) \rightarrow \delta_{xy}$.
F"ur unsere weiteren Konstruktionen ist es g"unstig, die Projektoren
\begin{eqnarray}
\label{eq:1_8a}
E_x &=& \sum_{\alpha=1}^{p+q} s_\alpha \: |x \alpha \ket \bra x \alpha |
\end{eqnarray}
auf die Eigenr"aume $e_x$ einzuf"uhren.
Als gemeinsame Spektralprojektoren der $X^i$ sind die Operatoren $E_x$
eichinvariant (also unabh"angig von der Wahl der Basis $|x \alpha \ket$)
definiert.

In einem anderen Bezugssystem $\tilde{x}=\tilde{x}(x)$ erhalten wir die
diskretisierten Raumzeitpunkte $\tilde{M}$ und Spektralprojektoren
$\tilde{E}_{\tilde{x}}$ durch die Transformation
\Equ{1_c0}
\tilde{M}=\tilde{x}(M) \;\;\;\;,\spc \tilde{E}_{\tilde{x}(x)} = E_x
	\spc .
\EndEqu
Da nach dem allgemeinen "Aquivalenzprinzip alle Bezugssysteme gleichberechtigt
sind, darf keine der durch Koordinatentransformation aus $M$
hervorgehenden Mengen vor einer anderen ausgezeichnet sein.
Durch geeignete Koordinatentransformationen kann man aber den Abstand und
die relative Lage der Raumzeitpunkte beliebig ver"andern.
Um konsequent zu sein, ist das nur sinnvoll, wenn wir auf $M$ jede
Abstands- und Ordnungsrelation aufgeben. Darum fassen wir $M$
von nun an als Punktmenge ohne zus"atzliche mathematische Struktur
auf, sie dient lediglich als Indexmenge f"ur die Spektralprojektoren.
Nach dieser Verallgemeinerung geht $M$ bei Koordinatentransformationen
gem"a"s \Ref{1_c0} in eine "aquivalente Menge $\tilde{M}$ "uber, die wieder
mit $M$ identifiziert werden kann. Darum k"onnen wir $M$ und nach
\Ref{1_c0} auch $(E_x)_{x \in M}$ als vom Bezugssystem unabh"angige Gr"o"sen
auffassen. Sie sind durch die koordinateninvarianten Relationen
\begin{eqnarray}
\label{eq:1_a1}
&{\mbox{$E_x(H)$ ist f"ur alle $x \in M$ ein Unterraum der
	Signatur $(p,q)$}}& \\
\label{eq:1_a2}
&E_x \:E_y \;=\; \delta_{xy} \: E_x \;\;\;\;,\spc
	\sum_{x \in M} E_x \;=\; \1&
\end{eqnarray}
charakterisiert. Wir sehen die Projektoren $(E_x)_{x \in M}$ als die
fundamentalen Operatoren zur Beschreibung der Raumzeit an.
Die Orts-/Zeitoperatoren $X^i$ k"onnen daraus abgeleitet werden:
Jedes Bezugssystem entspricht einer Injektion
\Equ{1_c1}
\underline{x} \;:\; M \:\hookrightarrow\: \R^4 \spc ,
\EndEqu
die zugeh"origen Orts-/Zeitoperatoren $X^i$ sind durch die Gleichungen
\Equ{1_c2}
X^i \;=\; \sum_{p \in M} \underline{x}^i(p) \: E_p
\EndEqu
gegeben. Wir haben die Injektion \Ref{1_c1} zur besseren Unterscheidung
von den diskreten Raumzeitpunkten $x \in M$ mit dem Index
`$\underline{\:\:}$' gekennzeichnet.

Zus"atzlich wollen wir annehmen, da"s $M$ nur aus endlich vielen Punkten
besteht. Das
entspricht der Vorstellung, da"s das Volumen der Raumzeit beschr"ankt,
das Universum also r"aumlich geschlossen und zeitlich endlich ist.
Diese Voraussetzung ist f"ur unser weiteres Vorgehen nicht
entscheidend; wer will, kann die Annahme $\# M < \infty$ auch nur als eine
technische Vereinfachung ansehen.

Wir werden die Raumzeit also durch einen endlichdimensionalen,
indefiniten Skalarproduktraum $H$ und die Spektralprojektoren $E_x$,
\Ref{1_a1}, \Ref{1_a2}, mit $x$ aus einer endlichen Indexmenge $M$
beschreiben. Wir nennen $(H, M, E)$
{\em{diskrete Raumzeit}}\index{diskrete Raumzeit}.
Wir k"onnen eine Basis $|x \alpha \ket$, $x \in M$, $\alpha=1,\ldots,p+q$
von $H$ w"ahlen mit
\Equ{1_g0}
E_x \: |y \alpha \ket \;=\; \delta_{xy} \: |y \alpha \ket \;\;\;\;,\spc
	\bra x \alpha | y \beta \ket \;=\; \delta_{xy} \: \delta_{\alpha \beta}
	\: s_\alpha \spc .
\EndEqu
Eine solche Basis wird {\em{Eichung}}\index{Eichung} genannt.
In einem speziellen Bezugssystem \Ref{1_c1}, \Ref{1_c2} geht die
Eichung in eine Eigenvektorbasis \Ref{1_c3} der $X^i$ "uber.

\subsubsection*{kurze Diskussion des Begriffs der diskreten Raumzeit}
Wir wollen die Definition der diskreten Raumzeit etwas diskutieren.
Zun"achst sollte man beachten, da"s die diskrete Raumzeit durch die
Signatur $(p, q)$ und $\# M$ bereits (bis auf Isomorphismen) vollst"andig
bestimmt ist. Insbesondere gibt es in $M$ eine
Permutationssymmetrie\index{Permutationssymmetrie};
wir haben also die Freiheit, beliebige Punkte der Raumzeit miteinander
zu vertauschen.
Damit ist der Begriff der diskreten Raumzeit viel allgemeiner gefa"st
als der eines Gitters (als wesentlicher Unterschied kann man in der
diskreten Raumzeit nicht von ``benachbarten Gitterpunkten'' oder
``Gitterl"ange'' sprechen; die diskrete Raumzeit besteht,
anschaulich ausgedr"uckt, eher aus einer ``losen Ansammlung von Punkten'').
Umgekehrt kann man eine Gittertheorie auch als Theorie in der diskreten
Raumzeit beschreiben, indem man das Gitter nur noch als Punktmenge auffa"st.
Daf"ur ist allerdings notwendig, da"s die Theorie auch ohne
die zus"atzlichen Strukturen des Gitters formuliert werden kann.

Wir haben die diskrete Raumzeit mit der Intention definiert, den
Minkowski-Raum (oder allgemeiner eine Lorentzmannigfaltigkeit) auf der
Planck-Skala zu diskretisieren und auf diese Weise die
UV-Probleme der Kontinuumsbeschreibung zu beseitigen.
Im Gegensatz zu Regularisierungen in der Quantenfeldtheorie haben wir die
Diskretisierung nicht nur aus technischen Gr"unden eingef"uhrt
(etwa um UV-Divergenzen zu vermeiden), sondern haben die Vorstellung,
da"s die diskrete Raumzeit physikalische Realit"at ist.
Aus diesem Grund wollen wir in dieser Arbeit versuchen, die Physik
intrinisisch in der diskreten Raumzeit zu formulieren.
Das bedeutet konkreter, da"s alle physikalischen Objekte Operatoren
auf $H$ sein m"ussen; die physikalischen Gleichungen sind mit diesen
Operatoren und den Projektoren $(E_x)_{x \in M}$ aufzustellen.

Damit diese intrinsische Formulierung der Physik in der diskreten
Raumzeit nicht der "ublichen Kontinuumsbeschreibung widerspricht, darf
die diskrete Natur der Raumzeit bei Systemen, die sehr gro"s gegen"uber
der Planck-L"ange sind, nicht erkennbar sein.
In diesem Fall sollte die Kontinuumsbeschreibung also eine zul"assige
N"aherung sein. Anders ausgedr"uckt, mu"s es m"oglich sein, in einem
bestimmten Grenzfall von der diskreten Raumzeit ins Kontinuum
"uberzugehen. Diesen Grenz"ubergang nennen wir
{\em{Kontinuumslimes}}\index{Kontinuumslimes}.
Auf den ersten Blick scheint die Definition der diskreten Raumzeit zu
allgemein, um den Kontinuumslimes sinnvoll durchf"uhren zu k"onnen.
Insbesondere ist unklar, warum man in diesem Grenzfall trotz der
Permutationssymmetrie der Raumzeit-Punkte die topologische Struktur
des Kontinuums erhalten sollte.
Dazu mu"s man beachten, da"s die Permutationssymmetrie i.a. verloren
geht, sobald zus"atzliche Operatoren auf $H$ eingef"uhrt werden.
Wir haben die qualitative Vorstellung, da"s diese zus"atzlichen Operatoren
die Permutationssymmetrie in einer Weise brechen, die im Kontinuumslimes
auf die lokale und kausale Struktur einer Lorentzmannigfaltigkeit f"uhrt.
Um die Beschreibung in der diskreten Raumzeit deutlich von der
Kontinuumsbeschreibung zu trennen, werden wir den Kontinuumslimes
erst im n"achsten Abschnitt \ref{1_ab2} mathematisch pr"azisieren
und genauer besprechen.

Ein Einwand, der oft gegen eine Diskretisierung der Raumzeit vorgebracht
wird, ist die Tatsache, da"s dabei die kontinuierlichen Symmetrien des
Minkowski-Raumes verloren gehen.
Wir weisen darauf hin, da"s diese Symmetrien in der Natur durch
die vorhandenen Teilchen und Felder ohnehin zerst"ort sind. Man kann
deshalb alle "au"seren Symmetrien der Raumzeit (genau wie den Begriff
``Vakuum'') streng genommen nur als eine Idealisierung der Wirklichkeit ansehen.
Aus diesem Grund bereitet es keine prinzipiellen Probleme, auf diese
Symmetrien ganz zu verzichten.
Nat"urlich k"onnte es sein, da"s man sich durch die Aufgabe der
Lorentzsymmetrie technische Probleme einhandelt. Das wird bei unserem
weiteren Vorgehen aber nicht der Fall sein.

Im Gegensatz zu den "au"seren Symmetrien bleibt die Eichsymmetrie bei
der Diskretisierung erhalten. Sie entspricht in der diskreten Raumzeit
der Freiheit der Basiswahl in den $(p+q)$-dimensionalen Unterr"aumen
$E_x(H)$ von $H$. Bei allen in dieser Arbeit untersuchten Systemen
werden Koordinaten- und Eichtransformationen
miteinander verkn"upft sein, so wie dies weiter oben f"ur die Diracgleichung
erw"ahnt wurde und in \cite{Physdip} genauer beschrieben ist.
Die Lorentzgruppe tritt also auch in der diskreten Raumzeit als
Untergruppe der Eichgruppe auf, was die durch die Diskretisierung
aufgegebene Lorentzsymmetrie des Minkowski-Raumes f"ur manche
"Uberlegungen ersetzen kann.

\subsection{Projektion auf besetzte Fermionzust"ande}
\label{1_ab12}
Bevor in der diskreten Raumzeit sinnvolle Gleichungen aufgestellt werden
k"onnen, m"ussen wir weitere Operatoren auf $H$ einf"uhren.
In der klassischen Kontinuumsbeschreibung wird das Sytem durch die
fermionischen Wellenfunktionen $\Psi_a$ und den Diracoperator \Ref{1_7}
charakterisiert. Die Wellenfunktionen erf"ullen die Diracgleichung;
aus dem Diracoperator k"onnen die bosonischen Potentiale und Felder
konstruiert und damit klassischen Feldgleichungen aufgestellt werden.
Es ist an dieser Stelle nicht klar, ob und wie der Diracoperator und die
Konstruktion der klassischen Feldgleichungen in die diskrete Raumzeit
"ubertragen werden kann. Darum beginnen wir in einem abstrakten Ansatz
nur mit den Wellenfunktionen $\Psi_a$, die in der diskreten Raumzeit
Elemente des endlichdimensionalen Vektorraums $H$ sind.

Ein System mit einem Fermion beschreiben wir mit einem Vektor $\Psi \in H$.
In einer speziellen Eichung $|x \alpha \ket$
definieren wir die zugeh"orige Wellenfunktion $\Psi^\alpha(x)$ durch
\[ \Psi^\alpha(x) \;=\; \bra x \alpha | \Psi \ket \spc . \]
Wir bezeichnen den Projektor auf einen Unterraum $Y \subset X$ im
folgenden mit $P_Y$. "Aquivalent zur Wellenfunktion $\Psi^\alpha(x)$
l"a"st sich das System auch mit dem Projektor $P_{< \Psi >}$
auf den von $\Psi$ erzeugten Unterraum beschreiben. Bei einer Normierung
$\bra \Psi | \Psi \ket = \pm 1$ der Wellenfunktion haben wir
\Equ{1_a5}
P_{< \Psi >} \;=\; \pm \: |\Psi \ket \bra \Psi | \spc .
\EndEqu

F"ur ein System mit $m$ Fermionen $\Psi_1, \ldots, \Psi_m$
bilden wir in Verallgemeinerung von \Ref{1_a5} den Projektor $P$ auf den
von $(\Psi_a)_{a=1,\ldots,m}$ aufgespannten Unterraum von $H$, also
\[ P \;:=\; P_{<\Psi_1, \ldots, \Psi_m>} \spc . \]
Wir nennen $P$ den
{\em{fermionischen Projektor}}\index{fermionischer Projektor}
des Systems. In dieser
Arbeit werden wir ein Vielfermionsystem stets mit dem fermionischen
Projektor beschreiben.

\subsubsection*{Vergleich zum Fockraum-Formalismus}
Die Verwendung des fermionischen Projektors unterscheidet sich wesentlich
vom "ublichen Fockraum-Formalismus\index{Fockraum-Formalismus,
fermionischer} der Quantenfeldtheorie. Darum
m"ussen wir uns zun"achst davon "uberzeugen, da"s auch die Beschreibung
mit dem fermionischen Projektor physikalisch sinnvoll ist.

Im Formalismus der zweiten Quantisierung h"atten wir das System der
Fermionen $\Psi_1, \ldots, \Psi_m$ mit der antisymmetrischen
Produktwellenfunktion
\Equ{1_16}
\Psi^{\alpha_1 \cdots \alpha_m} (x_1,\ldots,x_m) \;=\;
	\left| \det \bra \Psi_i | \Psi_j \ket \right|^{-\frac{1}{2}} \;
	\sum_{\sigma \in S(m)} (-1)^{|\sigma|} \:
	\Psi^{\alpha_1}_{\sigma(1)}(x_1) \cdots
	\Psi^{\alpha_m}_{\sigma(m)}(x_m)
\EndEqu
beschrieben. Die Wellenfunktionen der Form \Ref{1_16} werden auch
($m$-Teilchen-)Hartree-Fock-Zust"ande genannt. Sie spannen den
$m$-Teilchen-Fockraum $F^m = \bigwedge^m H$ auf.
Ein allgemeiner Fermionzustand ist als Vektor des Fockraumes
$F=\bigoplus_{m=0}^\infty F^m$ eine
beliebige Linearkombination von Hartree-Fock-Zust"anden.
Wir verwenden f"ur das von $\bra .|. \ket$ induzierte Skalarprodukt
auf dem Fockraum zur Deutlichkeit die Schreibweise
$\bra .|. \ket_F$.

Um einen ersten Zusammenhang zwischen dem fermionischen Projektor und dem
Fockraum-Formalismus herzustellen, ordnen wir jedem Projektor $P_Y$
auf einen Unterraum $Y=\bra \Psi_1, \ldots, \Psi_m \ket$ von $H$ die
antisymmetrische Wellenfunktion \Ref{1_16} zu. Diese Abbildung ist
sinnvoll (also bis auf einen Phasenfaktor unabh"angig von der Wahl
der Basis in $Y$) definiert und bijektiv.
Also entspricht jeder Projektor genau einem
Hartree-Fock-Zustand des Fockraumes. Durch diese Konstruktion
wird die Beschreibung mit dem fermionischen Projektor zu einem
Spezialfall des Fockraum-Formalismus;
insbesondere "ubertr"agt sich die Ununterscheidbarkeit der
Teilchen\index{Ununterscheidbarkeit der Fermionen} und
das Pauli-Prinzip\index{Pauli-Prinzip}.
Die Beschreibungen sind aber mathematisch nicht "aquivalent, da ein
Vektor des Fockraumes i.a. eine nicht-triviale Linearkombination von
Hartree-Fock-Zust"anden ist.

Wir wollen untersuchen, wie sich dieser mathematische Unterschied
physikalisch auswirkt.
Bei einer Naturbeschreibung durch den fermionischen Projektor
$P_{<\Psi_1, \ldots, \Psi_m>}$ mu"s die gemeinsame Wellenfunktion aller
Fermionen des Universums ein Hartree-Fock-Zustand sein.
Diese Tatsache ist nur von bedingtem Interesse, da wir uns
bei physikalischen Beobachtungen immer auf ein kleines Teilsystem beschr"anken
m"ussen. Die effektive Wellenfunktion des Teilsystems braucht jedoch kein
Hartree-Fock-Zustand zu sein:
Wir nehmen an, da"s unser Teilsystem in einem Gebiet $N \subset M$
lokalisiert ist. Wir spalten den Zustandsraum in der Form
\[ H \;=\; H(N) \oplus H(M \setminus N) \spc{\mbox{mit}}\spc
	H(A) \;:=\; \bigoplus_{x \in A} E_x(H) \;\;,\;\;\; A \subset M \]
auf. Dann sind alle (Einteilchen-)Observablen ${\cal{O}}$ unseres
Teilsystems auf $H(M \setminus N)$ trivial,
\Equ{1_a6}
{\cal{O}}_{|H(M \setminus N)} \;=\; \1_{|H(M \setminus N)} \spc .
\EndEqu
Wir zerlegen die Zust"ande $\Psi_j$ in der Form
\[ \Psi_j \;=\; \Psi_j^N + \Psi_j^{M \setminus N} \spc{\mbox{mit}}\spc
	\Psi_j^N \in H(N) \;,\;\;\; \Psi_j^{M \setminus N} \in
	H(M \setminus N) \spc . \]
Wir setzen in \Ref{1_16} ein und erhalten f"ur die Vielteichen-Wellenfunktion
den Ausdruck
\Equ{1_21}
\Psi \;=\; \left| \det \bra \Psi_i, \Psi_j \ket \right|^{-\frac{1}{2}} \;
	\sum_{\pi \in {\cal{P}}(m)}
	(-1)^{|\pi|} \left(\bigwedge_{j \in \pi} \Psi_j^N \right) \wedge
	\left(\bigwedge_{j \not \in \pi} \Psi_j^{M \setminus N} \right) \spc ,
\EndEqu
wobei ${\cal{P}}(m)$ die Potenzmenge von $\{1, \ldots, m\}$ bezeichnet.
F"ur Messungen in unserem System ist der Erwartungswert $\bra
\Psi | {\cal{O}} | \Psi \ket_F$ zu berechnen\footnote{Wir bemerken zur
Deutlichkeit, da"s dieser Erwartungswert nicht mit dem Erwartungswert
einer Messung in der nichtrelativistischen Quantenmechanik "ubereinstimmt.
Im Kontinuum (also vor Diskretisierung der Raumzeit oder nach Bildung
des Kontinuumslimes) wird im Skalarprodukt $\bra .|. \ket$ n"amlich
gem"a"s \Ref{1_1} auch "uber die Zeit integriert. Man kann aber einen
Zusammenhang herstellen, indem man Operatoren ${\cal{O}}$ mit
spezieller Zeitabh"angigkeit betrachtet (beispielsweise solche, die auf
die Wellenfunktionen nur in einem kurzen Zeitintervall $[t, t +\Delta t]$
wirken).}, dabei wirken die Operatoren
${\cal{O}}$ auf
dem Fockraum gem"a"s
\[ {\cal{O}}(\Psi_1 \wedge \cdots \wedge \Psi_m) \;=\;
	({\cal{O}} \Psi_1) \wedge \cdots \wedge \Psi_m \:+\:
	\Psi_1 \wedge ({\cal{O}} \Psi_2) \cdots \wedge \Psi_m \:+\:
	\Psi_1 \wedge \cdots \wedge ({\cal{O}} \Psi_m) \;\;\;\; . \]
Es ist g"unstig, den Erwartungswert mit dem statistischen Operator
$S$ umzuschreiben,
\[ \bra \Psi | {\cal{O}} | \Psi \ket_F \;=\; \tr_F (S \: {\cal{O}})
	\spc {\mbox{mit}} \spc S \;=\; | \Psi \ket \bra \Psi |_F \spc , \]
wobei $\tr_F$ die Spur "uber den Fockraum bezeichnet.
Wegen \Ref{1_a6} k"onnen wir n"amlich die partielle Spur "uber
$H(M \setminus N)$ bilden und erhalten mit \Ref{1_21}
\begin{eqnarray}
\label{eq:1_a8}
\bra \Psi | {\cal{O}} | \Psi \ket &=& \tr_{F^N}(S^N \: {\cal{O}})
	\spc {\mbox{mit}} \\
\label{eq:1_a7}
S^N &=& \sum_{k=0}^\infty \sum_{ \scriptsize
	\begin{array}{cc} \scriptsize \pi, \pi^\prime \in {\cal{P}}(m) , \\
		\# \pi = \# \pi^\prime = k
	\end{array} }
	c_{\pi, \pi^\prime} \; | \wedge_{i \in \pi} \Psi_i^N \ket
		\bra \wedge_{j \in \pi^\prime} \Psi_j^N |_{F^N} \\
c_{\pi, \pi^\prime} &=& (-1)^{|\pi| + |\pi^\prime|} \; \bra \wedge_{i \not \in \pi}
	\Psi_i^{M \setminus N} | \wedge_{j \not \in \pi^\prime}
	\Psi_j^{M \setminus N} \ket_F \spc , \nonumber
\end{eqnarray}
wobei $\tr_{F^N}$ die Spur "uber den von $H(N)$ erzeugten Fockraum $F^N$
bezeichnet. Das in $N$ lokalisierte Teilsystem l"a"st sich also mit
einem statistischen Operator $S^N$ auf $F^N$ beschreiben, der aus
gemischten Zust"anden zu verschiedener Teilchenzahl aufgebaut ist.
Da die Konstanten $c_{\pi, \pi^\prime}$ von den Wellenfunktionen
$\Psi^{M \setminus N}$ au"serhalb unseres Teilsystems abh"angen,
k"onnen sie praktisch beliebig sein. Wenn wir die Anzahl $m$
der Teilchen des Gesamtsystems gegen Unendlich gehen lassen, kann mit
\Ref{1_a7} jeder statistische Operator dargestellt werden, der die
Teilchenzahl im Teilsystem nicht "andert.
Da wir uns f"ur Einteilchen-Observablen auf statistische Operatoren
beschr"anken k"onnen, die auf dem Teilchenzahloperator diagonal sind
(die au"serdiagonalen Beitr"age fallen bei der Berechnung der Spur
\Ref{1_a8} weg), l"a"st sich das Teilsystem folglich mit einem allgemeinen
statistischen Operator beschreiben, insbesondere mit dem statistischen
Operator eines reinen Fockraum-Zustandes
\[ S^N = |\Psi^N \ket \bra \Psi^N|_{F^N} \;\;\;,\spc \Psi^N \in F^N \spc . \]
Diese "Uberlegung l"a"st sich mit etwas mehr mathematischem Aufwand
auch auf Vielteilchen-Observablen "ubertragen.

Wir kommen zu dem Schlu"s, da"s die Beschreibung des Vielfermionsystems
mit dem fermionischen Projektor zum Fockraum-Formalismus physikalisch
"aquivalent ist.
F"ur theoretische "Uberlegung m"ussen wir ber"ucksichtigen, da"s
der fermionische Projektor lediglich einem Hartree-Fock-Zustand entspricht;
bei praktischen Problemstellungen kann man aber nach Belieben zur
Fockraum-Darstellung "ubergehen.

\subsection{Die Gleichungen der diskreten Raumzeit}
\label{1_ab13}
Wie bereits in Abschnitt \ref{1_ab11} angesprochen, wollen wir die
Physik intrinsisch in der diskreten Raumzeit beschreiben. Wir k"onnen
diese Vorstellung nun pr"azisieren und stellen dazu das {\em{Prinzip
des fermionischen Projektors}}\index{Prinzip des fermionischen Projektors}
auf:
\begin{quote}
Das physikalische System wird durch den fermionischen
Projektor $P$ in der diskreten Raumzeit vollst"andig beschrieben. Die
physikalischen Gleichungen sind allein mit dem fermionischen Projektor
in der diskreten Raumzeit aufzustellen, sie m"ussen also mit den
Operatoren $P$, $(E_x)_{x \in M}$ auf $H$ formuliert werden.
\end{quote}
Wir nennen die mit $P, E_x$ aufgestellten Gleichungen die
{\em{Gleichungen der diskreten Raumzeit}}\index{Gleichungen der
diskreten Raumzeit}.

Das Prinzip des fermionischen Projektors kann nicht aus bekannten
physikalischen Gleichungen oder Prinzipien abgeleitet werden.
An dieser Stelle ist nicht erkennbar, ob es auf mathematisch interessante
Gleichungen f"uhrt oder sogar physikalisch sinnvoll ist.
Es handelt sich also um ein `ad hoc' aufgestelltes Postulat, dessen
Konsequenzen in dieser Arbeit untersucht werden sollen.

\subsubsection*{kurze Diskussion des Prinzips des fermionischen Projektors}
Wir wollen das Prinzip des fermionischen Projektors kurz diskutieren.
Auf den ersten Blick mag es erstaunlich erscheinen, da"s das physikalische
System bereits durch den fermionischen Projektor vollst"andig beschrieben
sein soll. Wir haben die folgende qualitative Vorstellung:
Im Kontinuumslimes (der bisher noch nicht mathematisch eingef"uhrt wurde)
sollten die Wellenfunktionen $\Psi_a$ des fermionischen Projektors in
Eigenzust"ande des Diracoperators "ubergehen, also
\Equ{1_b0}
(G - m_a) \: \Psi_a \;=\; 0
\EndEqu
mit $G$ gem"a"s \Ref{1_7}. Wir haben die Hoffnung, da"s die Potentiale
$G^j, B$ in \Ref{1_7} bereits "uber die Diracgleichungen \Ref{1_b0}
eindeutig bestimmt sind, so da"s der Diracoperator aus dem
Kontinuumslimes des fermionischen Projektors konstruiert werden kann.
Nach den Konstruktionen in \cite{Physdip} sind dann auch die klassischen Felder
durch den fermionischen Projektor festgelegt.
Folglich braucht man nur den fermionischen Projektor als fundamentales
physikalisches Objekt anzusehen; alle weiteren physikalischen Gr"o"sen
(insbesondere die fermionischen Wellenfunktionen, Dirac-Str"ome,
Energie-Impuls-Tensoren, klassischen Eichfelder und die Metrik) k"onnen
daraus im Kontinuumslimes abgeleitet werden.
Diese Vorstellung werden wir noch wesentlich pr"azisieren.

Wir "uberlegen, welche physikalische Annahmen dem Prinzip des
fermionischen Projektors zugrunde liegen:
Wie wir in Abschnitt \ref{1_ab11} gesehen haben, f"uhrt die Formulierung
der Theorie in der diskreten Raumzeit "uber die Willk"ur der
Basiswahl in den Unterr"aumen $E_x(H) \subset H$ auf lokale Eichfreiheiten.
Da $M$ nur eine Punktmenge ist, sind gem"a"s unserer "Uberlegung an
\Ref{1_c1}, \Ref{1_c2} alle Bezugssysteme gleichberechtigt.
Daraus scheint das "Aquivalenzprinzip zu folgen, den genauen
Zusammenhang werden wir aber erst nach Pr"azisierung des Kontinuumslimes
im n"achsten Abschnitt \ref{1_ab2} herstellen k"onnen.
Die Verwendung eines fermionischen Projektors impliziert schlie"slich,
wie in Abschnitt \ref{1_ab12} beschrieben, die Ununterscheidbarkeit der
Fermionen und das Pauli-Prinzip.
Damit sind wichtige physikalische Prinzipien im Prinzip des fermionischen
Projektors implizit enthalten.
Allerdings fehlt bei unserem Ansatz die Lokalit"ats- und
Kausalit"atsforderung f"ur die physikalischen Gleichungen. Wir sehen
die Lokalit"at und Kausalit"at nicht als fundamentale physikalische
Prinzipien an. Damit unsere Beschreibung sinnvoll ist, m"ussen wir
die Lokalit"at und Kausalit"at aber im Kontinuumslimes erhalten.

\subsubsection*{ein Beispiel f"ur Gleichungen der diskreten Raumzeit}
Gem"a"s dem Prinzip des fermionischen Projektors m"ussen die Gleichungen der
diskreten Raumzeit aus den Operatoren $P, E_x$ aufgebaut werden.
Wegen der Orthogonalit"at der $E_x$, \Ref{1_a2}, und der Idempotenz
$P^2=P$ des fermionischen Projektors k"onnen wir bei Operatorprodukten
immer annehmen, da"s die Faktoren $E_x, P$ abwechselnd auftreten.
Die Gleichungen m"ussen also aus Termen der Form
\Equ{1_25}
E_{x_1} \:P\: E_{x_2} \:P\: \cdots \:P\: E_{x_{n-1}} \:P\: E_{x_n}
\EndEqu
aufgebaut werden.
Damit ist zwar die mathematische Struktur der Gleichungen der diskreten
Raumzeit grob festgelegt; es ist aber noch v"ollig unklar, wie die Gleichungen
konkret aussehen sollten.

Wir gehen dieses Problem an dieser Stelle noch nicht systematisch an,
sondern werden nur ein Beispiel f"ur Gleichungen der diskreten Raumzeit
angeben.
Dieses Beispiel ist zwar zu einfach und f"uhrt nicht auf sinnvolle
physikalische Gleichungen, aus mathematischer Sicht hat es mit den
eigentlich interessanten Gleichungen aber gro"se "Ahnlichkeit.
F"ur die qualitativen "Uberlegungen in der Einleitung wird dieses
Beispiel ausreichend sein.

In Analogie zur klassischen Feldtheorie wollen wir ein Variationsprinzip
aufstellen. Um aus den Operatorprodukten \Ref{1_25} Skalare zu bilden,
verwenden wir die Spur.
Um die Abh"angigkeit von den Parametern $x_j$ zu beseitigen, setzen
wir die $x_j$ in Gruppen gleich und summieren "uber $M$. Mit dieser Methode
erh"alt man z.B. den Ausdruck
\Equ{1_27}
	\sum_{x,y \in M} \; \tr (E_x \:P\: E_y \:P\: E_x \:P) \spc .
\EndEqu
Wir k"onnen annehmen, da"s die Raumzeitpunkte $x_j$ wenigstens in Zweiergruppen
zusammengefa"st sind, denn ansonsten kann man die Summe "uber $x_j$
mit Hilfe der Vollst"andigkeitsrelation in \Ref{1_a2} ausf"uhren.
Beispielsweise kann man in \Ref{1_27} "uber $y$ summieren und erh"alt
\[ =\; \sum_{x \in M} \; \tr (E_x \:P\: E_x \:P) \spc . \]
Dieser Ausdruck ist als Wirkung mathematisch zu einfach, weil die
Operatoren $E_x P E_y$ f"ur  $x \neq y$ gar nicht eingehen.
Wir w"ahlen als unser Beispiel die einfachste
{\em{Wirkung}}\index{Wirkung}, bei der
"uber zwei Parameter $x, y$ summiert wird,
\Equ{1_29}
S \;=\; \sum_{x,y \in M} \; \tr (E_x \:P\: E_y \:P\: E_x \:P\: E_y \:P)
	\spc .
\EndEqu
F"ur das Variationsprinzip suchen wir nach lokalen Extremalstellen
der Wirkung bei stetigen Variationen des fermionischen Projektors.

Wir leiten die zugeh"origen
{\em{Euler-Lagrange-Gleichungen}}\index{Euler-Lagrange-Gleichungen}
ab: Wir betrachten eine stetige Variation $P(\tau)$ des fermionischen
Projektors $P$ mit $P(0)=P$.
Da f"ur einen Projektor der Ausdruck $\tr (P)=\Rg (P)$ eine ganze Zahl
ist, bleibt der Rang von $P$ bei der stetigen Variation unver"andert.
Folglich k"onnen wir die Variation durch unit"are
Transformationen beschreiben. Es gibt also eine Schar unit"arer
Transformationen $U(\tau)$ mit $U(0)=\1$, so da"s
\[      P(\tau) \;=\; U(\tau) \:P\: U^{-1}(\tau) \spc . \]
\label{unit}
In erster Ordnung in $\tau$ haben wir $U = 1 +i \tau A, \;
U^{-1}=1-i \tau A$ mit einem hermiteschen Operator $A$ und folglich
$\delta P = i \: [A, \: P]$.
Um die Variation von \Ref{1_29} zu berechnen, nutzen
wir die Symmetrie in $x, y$ und die zyklische Invarianz der Spur aus
\begin{eqnarray*}
\delta S &=& 4 i \sum_{x,y \in M} \tr ( E_x \:[A,P]\:
	E_y \:P\: E_x \:P\: E_y \:P ) \\
&=& 4 i \sum_{x,y \in M} \tr \left( A \; [P,\: E_x \:P\: E_y \:P\: E_x
	\:P\: E_y] \right) \spc .
\end{eqnarray*}
Da dieser Ausdruck f"ur einen beliebigen hermiteschen Operator $A$
verschwinden soll, folgt die Bedingung
\begin{eqnarray}
\label{eq:1_30}
[P, \: Q]&=&0 \spc {\mbox{mit}} \\
\label{eq:1_30b}
Q &=& \sum_{x,y \in M} E_x \:P\: E_y \:P\:
	E_x \:P\: E_y \spc .
\end{eqnarray}
Als Euler-Lagrange-Gleichungen erh"alt man also die Kommutatorgleichung
\Ref{1_30}, dabei ist $Q$ ein zusammengesetzter Ausdruck in den Operatoren
$E_x, P$.

\section{Der Kontinuumslimes}
\label{1_ab2}
Im vorangehenden Abschitt \ref{1_ab1} haben wir mit dem Prinzip des
fermionischen Projektors festgelegt, da"s wir ein physikalisches System
mit dem fermionischen Projektor $P$ intrinsisch in der diskreten Raumzeit
$(H, M, E)$ beschreiben wollen. Mit der Wirkung \Ref{1_29} und den
Euler-Lagrange-Gleichungen \Ref{1_30}, \Ref{1_30b} wurde an einem
Beispiel erl"autert, wie die Gleichungen der diskreten Raumzeit im Prinzip
aussehen k"onnten.
Aus mathematischer Sicht besteht jetzt unsere Aufgabe darin, L"osungen
der Gleichungen der diskreten Raumzeit zu finden. Wir sollten also verschiedene
Variationsprinzipien genauer mathematisch
studieren und anschlie"send "uberlegen, ob das Prinzip des fermionischen
Projektors physikalisch sinnvoll ist.
Leider kann das Problem nicht so direkt angegangen werden:
F"ur eine kleine Zahl von Raumzeitpunkten (also z.B. f"ur
$\#M = 2,3,4$) lassen sich die Euler-Lagrange-Gleichungen direkt als
Matrixgleichungen analysieren.
Als Diskretisierung der Raumzeit sollte $M$ aber aus sehr vielen Punkten
bestehen. In diesem Fall werden die Matrixgleichungen beliebig kompliziert.
Wir kennen keine mathematische Methode, mit der die Euler-Lagrange-Gleichungen
f"ur gro"ses $\#M$ sinnvoll behandelt werden k"onnten.
Es scheint hoffnungslos, die Gleichungen der diskreten Raumzeit allgemein
exakt l"osen zu wollen.

Wegen dieser mathematischen Schwierigkeiten ist es wichtig, da"s wir
zun"achst eine anschauliche Vorstellung davon entwickeln, was die
Gleichungen der diskreten Raumzeit "uber den fermionischen Projektor
aussagen.
Dazu werden wir versuchen, einen Kontakt zur Kontinuumsbeschreibung
herzustellen.
Die Motivation f"ur dieses Vorgehen ist unsere physikalische Vorkenntnis:
Wenn das Prinzip des fermionischen Projektors physikalisch
sinnvoll sein soll, mu"s es die Kontinuumsbeschreibung der
relativistischen Quantenmechanik als Grenzfall liefern. Dieser Grenzfall
sollte sich aus den Gleichungen der diskreten Raumzeit direkt gewinnen
lassen. Wir haben also die Hoffnung, durch eine geeignete N"aherung
der Gleichungen der diskreten Raumzeit einen Zusammenhang zu den
bekannten Differentialgleichungen (Diracgleichung, klassische
Feldgleichungen) herstellen zu k"onnen.

\subsection{Beschreibung des Vakuums}
\label{1_ab21}
Als erste Ann"aherung an die physikalischen Begriffe des Kontinuums wollen
wir "uberlegen, was darunter zu verstehen ist, da"s der fermionische
Projektor ``das Vakuum\index{Vakuum} beschreibt''.
Dazu stellen wir den fermionischen Projektor in einer speziellen Eichung
$|x \alpha \ket$ als Matrix dar,
\Equ{1_c5}
(P \Psi)^\alpha(x) \;=\; \sum_{\beta=1}^{p+q} \sum_{y \in M}
	\bar{P}^\alpha_\beta(x,y) \: \Psi^\beta(y) \;\;\;\;\; {\mbox{mit}} \;\;\;\;\;
	\bar{P}^\alpha_\beta(x,y) \;=\; s_\alpha \: \bra x \alpha | P | y \beta \ket
	\spc .
\EndEqu
Um einen Zusammenhang zur Kontinuumsbeschreibung herstellen zu
k"onnen, m"ussen wir voraussetzen, da"s sich diese Gleichung sinnvoll ins
Kontinuum "ubertragen l"a"st. Dazu soll es ein Bezugssystem \Ref{1_c1},
\Ref{1_c2} und eine Eichung mit den folgenden Eigenschaften geben:
Die diskreten Raumzeitpunkte $\underline{x}(M) \subset \R^4$ sollen (bzgl. der
euklidischen Norm des $\R^4$) in einem mittleren
Abstand von der Gr"o"senordnung der Planck-L"ange angeordnet sein.
Wir betrachten Wellenfunktionen $\Psi^\beta(y)$, $y \in \underline{x}(M)$,
die sich nur auf
L"angenskalen ver"andern, welche sehr gro"s gegen"uber der Planck-L"ange
sind. Solche {\em{makroskopische Wellenfunktionen}}\index{makroskopische
Wellenfunktion} lassen sich sinnvoll
ins Kontinuum "ubertragen, indem man $\Psi(z)$ f"ur $z \in \R^4 \setminus
\underline{x}(M)$ mit dem Funktionswert $\Psi(y)$ an einem benachbarten
diskreten Raumzeitpunkt $y \approx z$, $y \in \underline{x}(M)$ gleichsetzt.
Wir fordern, da"s sich \Ref{1_c5} f"ur makroskopische Wellenfunktionen in
guter N"aherung als Integral
\Equ{1_c6}
(P \Psi)^\alpha(x) \;\approx\; \sum_{\beta=1}^{p+q} \int_{\sR^4} d^4y \;
	P^\alpha_\beta(x,y) \: \Psi^\beta(y)
\EndEqu
mit einer geeigneten Funktion (oder allgemeiner Distribution)
$P^\alpha_\beta(x,y)$ auf $\R^4 \times \R^4$ schreiben l"a"st.
Wir werden $P^\alpha_\beta(x,y)$ mit dem Integralkern eines bekannten
Operators des Kontinuums identifizieren.

Der gerade hergestellte Zusammenhang zum Kontinuum ist mathematisch
nicht ganz befriedigend. Wir haben offen gelassen, wie die diskreten
Raumzeitpunkte genau im Min\-kow\-ski-Raum angeordnet sind und haben den
Begriff der ``makroskopischen Wellenfunktion'' nicht sauber definiert.
Der "Ubergang zum Kontinuum lie"se sich mathematisch noch pr"azisieren
(beispielsweise als schwacher Limes einer Folge von fermionischen
Projektoren und diskreten Raumzeiten), wir werden darauf aber
bewu"st verzichten.
Dadurch soll hervorgehoben werden, da"s wir uns "uber Einzelheiten der
Einbettung der diskreten Raumzeit ins Kontinuum nicht festlegen k"onnen.
Wir m"ussen akzeptieren, da"s sich aus der Kontinuumsbeschreibung nur sehr
schwache Informationen "uber den fermionischen Projektor gewinnen lassen,
und m"ussen versuchen, mit dem etwas vagen Zusammenhang zwischen der
diskreten Raumzeit und dem Kontinuum auszukommen.

Wir f"uhren f"ur diesen "Ubergang zum Kontinuum eine geeignete Notation ein: Wir
schreiben
\Equ{1_d0}
P^\varepsilon(x,y) \;\equiv\; E_x \:P\: E_y \spc ,
\EndEqu
wobei der Parameter $\varepsilon$ die Diskretisierungsl"ange des
Bezugssystems angibt ($\varepsilon$ also von der Gr"o"senordnung
der Planck-L"ange).
Mit einer Matrixschreibweise in den Spinoren stimmt der Faktor
$\bar{P}^\alpha_\beta(x,y)$ in \Ref{1_c5} mit $P^\varepsilon(x,y)$ "uberein;
die Matrix $P^\alpha_\beta(x,y)$ in \Ref{1_c6} bezeichnen wir entsprechend mit
$P(x,y)$.
F"ur den "Ubergang von \Ref{1_c5} nach \Ref{1_c6} schreiben wir
symbolisch
\Equ{1_c7}
P^\varepsilon(x,y) \;\leadsto\; P(x,y)
\EndEqu
und bezeichnen $P(x,y)$ als den {\em{Kontinuumslimes von
$P^\varepsilon(x,y)$}}.

\subsubsection*{Aufbau von Diracseen}
Wir k"onnen nun im Kontinuum mathematisch sauber weiterarbeiten und
wollen festlegen, wie der Kontinuumslimes $P(x,y)$ des fermionischen
Projektors konkret aussieht.
Da $M$ nur aus endlich vielen Punkten besteht, k"onnen wir im
Kontinuumslimes nur ein Gebiet $\Omega \subset \R^4$ von endlichem Volumen
beschreiben. Da $\Omega$ beliebig gro"s gew"ahlt werden kann, spielt diese
Einschr"ankung im folgenden aber keine Rolle; zur Einfachheit lassen wir sie
ganz weg und tun so, als w"are $\Omega = \R^4$.

Wir beginnen bei Spindimension 4 mit dem System von nur einer Fermionsorte mit
Masse $m$. Der Kontinuumslimes des fermionischen Projektors sollte aus L"osungen
der freien Diracgleichung bestehen, es folgt
\[ (i \Pdd_x - m) \: P(x,y) \;=\; 0 \spc . \]
Genauer bauen wir $P(x,y)$ aus allen L"osungen negativer Energie auf, also
\Equ{1_c8}
P(x,y) \;=\; c \int \frac{d^4k}{(2 \pi)^4} \: (k \slsh + m) \:
	\delta(k^2 - m^2) \: \Theta(-k^0) \; e^{-i k(x-y)}
\EndEqu
mit einer Normierungskonstanten $c$, auf die wir in der
Einleitung nicht n"aher eingehen wollen. Wir nennen \Ref{1_c8} einen
{\em{Diracsee des Kontinuums}}\index{Diracsee des Kontinuums},
mathematisch ist $P(x,y)$ eine temperierte Distribution.
Durch das Festlegen des Kontinuumslimes haben wir "uber \Ref{1_c7} auch
im fermionischen Projektor der diskreten Raumzeit einzelne
Fermionzust"ande besetzt. Genauer ist $P^\varepsilon(x,y)$ aus allen makroskopischen
Wellenfunktionen $\Psi^\alpha(x)$ aufgebaut, die bei der "Ubertragung ins
Kontinuum in negative-Energie-L"osungen der freien Diracgleichung
"ubergehen. Zus"atzlich kann $P^\varepsilon(x,y)$ auf Wellenfunktionen
$\Psi$ projezieren, die nicht makroskopisch sind;
"uber diese Wellenfunktionen k"onnen wir aber keine Aussagen machen.
Wir nennen $P^\varepsilon(x,y)$ einen {\em{Diracsee der diskreten
Raumzeit}}\index{Diracsee der diskreten Raumzeit}.

Die Konstruktion l"a"st sich unmittelbar auf Systeme mit mehreren
Fermionsorten "ubertragen:
Wir unterscheiden die verschiedenen Teilchensorten durch einen Index
$^{(i)}$, $i=1,\ldots, K$. Die Wellenfunktionen $\Psi^{(i)}$ erf"ullen
die Diracgleichungen $(i \Pdd - m^{(i)}) \Psi^{(i)}=0$, dabei sind
$m^{(i)}$ die Massen der Fermionen. Wir k"onnen
jede Fermionsorte analog zu \Ref{1_c8} durch einen Diracsee
$P^{(i)}$ beschreiben. Um den fermionischen Projektor $P(x,y)$
aufzubauen, kombinieren wir zwei Konstruktionselemente:
Zun"achst kann man die Projektoren zu Teilchensorten verschiedener Massen
addieren. Wir w"ahlen also eine Zerlegung $(I_\alpha)_{\alpha=1,\ldots,B}$
von $\{1, \ldots, K \}$ und bilden neue Projektoren
\Equ{1_23}
P^{\{\alpha\}}(x,y) \;=\; \sum_{i \in I_\alpha} P^{(i)}(x,y) \spc .
\EndEqu
Im zweiten Schritt setzen wir als Wellenfunktionen die direkte Summe von
Diracspinoren an und lassen die $P^{\{\alpha\}}$ auf die einzelnen
direkten Summanden wirken.
Mit einer Matrixschreibweise in den Komponenten der Wellenfunktionen
haben wir dann also
\Equ{1_24}
P(x,y) \;=\; \left( \begin{array}{ccc}
		P^{\{1\}}(x,y) & & 0 \\
		& \ddots & \\
		0 & & P^{\{B\}}(x,y)
	\end{array} \right) \spc ,
\EndEqu
dabei sind die Matrixeintr"age selbst $(4 \times 4)$-Matrizen.
Die Spindimension ist $4B$, die Signatur $(2B, 2B)$.

Wir bezeichnen einen fermionischen Projektor $P^\varepsilon(x,y)$, der
\Ref{1_c7}, \Ref{1_24} erf"ullt, als {\em{fermionischen Projektor
des Vakuums}}\index{fermionischer Projektor des Vakuums}.

\subsubsection*{kurze Erl"auterung der Konstruktion}
Wir wollen unsere Beschreibung des Vakuums noch etwas erl"autern.
Mit dem fermionischen Projektor des Vakuums haben wir eine spezielle
Klasse von Projektoren konstruiert.
Zwar haben wir die Form des fermionischen Projektors mit \Ref{1_c7}
nicht im Detail bestimmt, wir haben aber trotzdem viele Informationen
"uber das physikalische System bei der Konstruktion verwendet:
Mit \Ref{1_24} haben wir festgelegt, aus welchen Fermionsorten das
System aufgebaut ist.
Die Beschreibung des Vakuums mit vollst"andig gef"ullten Diracseen
\Ref{1_c8} entspricht ganz der urspr"unglichen Vorstellung von Dirac, mit welcher
man das Problem der negativen Energiezust"ande der Diracgleichung beseitigen
und die Paarerzeugung auf einfache Weise verstehen kann.
Man sollte beachten, da"s der Diracsee bei uns nicht nur eine formale Konstruktion
ist, sondern da"s wir die Diracseen im fermionischen Projektor als
physikalische Realit"at ansehen. Die Konstruktion f"uhrt in der diskreten
Raumzeit auf keine prinzipiellen Schwierigkeiten, weil die Diracseen
nur aus endlich vielen Zust"anden aufgebaut sind.
Mit der Einbettung \Ref{1_c1} und der Verwendung des Differentialoperators
$i \Pdd$ ging in die Konstruktion der Diracseen die Topologie und die
differenzierbare Struktur des Kontinuums, also kurz gesagt die Lokalit"at, ein.
Mit den Diracmatrizen $\gamma^j$ wurde implizit die Minkowski-Metrik
$\eta^{jk}=\frac{1}{2} \{\gamma^j, \gamma^k\}$ und damit letztlich die
Lichtkegelstruktur, also die Kausalit"at, des Minkowski-Raumes verwendet.

Es widerspricht der von uns geforderten intrinsischen Formulierung der
Physik in der diskreten Raumzeit, da"s in die Konstruktion des Vakuums
die freie Diracgleichung und damit insbesondere die Lokalit"at und
Kausalit"at eingeflossen sind.
Darum kann die Beschreibung des Vakuums nur eine
erste Vorbereitung f"ur eine Kontinuumsbeschreibung sein.
Wir werden nun den Kontinuumslimes allgemein konstruieren und
dabei auf alle Strukturen des Minkowski-Raumes ganz explizit
verzichten.

\subsection{Allgemeine Definition des Kontinuumslimes von $P$}
\label{1_ab22a}
Wir gehen von einem allgemeinen fermionischen Projektor $P$ in der
diskreten Raumzeit $(M, H, E)$ aus und wollen einen Zusammenhang
zum Kontinuum herstellen.
Als Kontinuum fassen wir den Minkowski-Raum im folgenden lediglich als
differenzierbare Mannigfaltigkeit auf. Wir verzichten also auf die kausale
und metrische Struktur und lassen beliebige Diffeomorphismen
des $\R^4$ zu.

\subsubsection*{Anordnen der Raumzeitpunkte mit Diffeomorphismen}
Wir beginnen mit einer beliebigen Teilmenge $N \subset \R^4$ mit
$\#M=\#N$ und fassen $N$ als die diskreten Raumzeitpunkte auf.
Durch einen geigneten Diffeomorphismus des $\R^4$ k"onnen wir erreichen,
da"s die Raumzeitpunkte in einer Teilmenge $\Omega \supset N$ des
$\R^4$ gleichm"a"sig verteilt sind und einen mittleren Abstand in der
Gr"o"senordnung der Planck-L"ange haben. Etwas genauer bedeutet
diese {\em{Anordnungsvorschrift}}\index{Anordnungsvorschrift}
folgendes: Wir bezeichnen eine
Teilmenge $A \subset \Omega$ als makroskopisch, wenn ihre Ausdehnung
sehr gro"s gegen"uber der Planck-L"ange ist. F"ur jede makroskopische
Teilmenge $A$ sollen die Punkte $N \cap A$ einen mittleren Abstand von
der Gr"o"senordnung der Planck-L"ange haben.

Die Begriffe ``gleichm"a"sig verteilt'' und ``mittlerer Abstand'' sind wieder
nicht sauber definiert; wir verzichten analog wie im vorangehenden Abschnitt
\ref{1_ab21} auf eine mathematische Pr"azisierung.

\subsubsection*{Beschreibung der Permutationssymmetrie als innere
	Symmetrie}
Nachdem wir die diskreten Raumzeitpunkte $N$ in eine f"ur den
Kontinuumslimes sinnvolle relative Lage gebracht haben, m"ussen wir mit
einer Bijektion
\Equ{1_f0}
\underline{x} \;:\; M \:\rightarrow\: N \subset \R^4
\EndEqu
ein Bezugssystem festlegen. Leider ist $\underline{x}$
nur bis auf Permutationen\index{Permutationssymmetrie
als innere Symmetrie}
bestimmt; wir k"onnen also gem"a"s
\Equ{1_f1}
\underline{x} \:\rightarrow\: \underline{x} \circ \sigma
	\;\;\;\;,\spc \sigma \in S(M)
\EndEqu
zu einer anderen Bijektion "ubergehen, dabei bezeichnet
$S(M)$ die Gruppe der Permutationen in $M$.

Die Freiheit in der Wahl der Abbildung $\underline{x}$ l"a"st sich
als innere Symmetrie umschreiben:
Wir definieren in einer Eichung $|x \alpha \ket$ die unit"aren
Operatoren $U(\sigma)$, $\sigma \in S(M)$ durch
\[ U(\sigma) \;=\; \sum_{x \in M} \sum_{\alpha=1}^{p+q}
	\:s_\alpha \: |\sigma(x) \:\alpha \ket \bra x \alpha | \spc . \]
Sie bilden die Unterr"aume $E_x(H) \subset H$ isometrisch in
$E_{\sigma(x)}(H)$ ab. Aus der Pseudo-Orthonormalit"at \Ref{1_g0}
folgt
\begin{eqnarray*}
U(\bar{\sigma}) \: U(\sigma) &=& \sum_{x,y \in M} \:s_\alpha \:
	|\bar{\sigma}(y) \: \alpha \ket \:\delta_{y, \sigma(x)} \:
	\bra x \alpha | \\
&=& \sum_{x \in M} \:s_\alpha\: |(\bar{\sigma} \circ \sigma)(x) \:
	\alpha \ket \bra x \alpha | \;=\; U(\bar{\sigma} \sigma) \spc ,
\end{eqnarray*}
so da"s $U$ eine unit"are Darstellung von $S(M)$ auf $H$ ist.
Au"serdem haben wir
\[ E_{\sigma(x)} \;=\; U(\sigma) \:E_x\: U(\sigma)^{-1} \spc . \]
Folglich k"onnen wir das Verhalten von $E_x$ unter Permutationen
\Ref{1_f1} auch beschreiben, indem wir alle Operatoren auf $H$
mit $U(\sigma)$ unit"ar transformieren.
Der fermionische Projektor verh"alt sich dabei nach der Formel
\[ P \;\rightarrow\; U(\sigma)^{-1} \:P\: U(\sigma) \spc . \]

Gem"a"s dieser Konstruktion sind alle Bijektionen \Ref{1_f0} unit"ar
"aquivalent, so da"s wir uns ohne Einschr"ankung willk"urlich f"ur
ein $\underline{x}$ entscheiden k"onnen.

\subsubsection{Einschr"ankung auf makroskopische Wellenfunktionen}
Nach diesen Vorbereitungen l"a"st sich der Kontinuumslimes ganz
analog wie f"ur das Vakuum durchf"uhren:
Wir definieren makroskopische Wellenfunktionen $\Psi$ durch die
Bedingung, da"s $\Psi^\beta(y)$ im Bezugssystem \Ref{1_f0}
nur auf L"angenskalen variiert, die sehr gro"s gegen die Planck-L"ange
sind. Die Matrix
\Equ{1_g1}
P^\varepsilon(x,y) \;\equiv\; E_x \:P\: E_y
\EndEqu
kann bei Einschr"ankung auf makroskopische Wellenfunktionen sinnvoll
ins Kontinuum "ubertragen werden und geht in den Integralkern eines
geeigneten Operators "uber. Wir schreiben symbolisch
\Equ{1_z0}
P^\varepsilon(x,y) \;\leadsto\; P(x,y)
\EndEqu
und bezeichnen $P(x,y)$ als {\em{Kontinuumslimes des fermionischen
Projektors}}\index{Kontinuumslimes des fermionischen Projektors}.

\subsubsection*{kurze Diskussion, Ableitung des "Aquivalenzprinzips}
Wir wollen die Konstruktion des Kontinuumslimes kurz diskutieren.
Die Menge $N$ wurde zu Beginn willk"urlich vorgegeben.
Dies ist keine Einschr"ankung, weil die relative
Lage der diskreten Raumzeitpunkte mit Diffeomorphismen beliebig
ver"andert werden kann.
Au"serdem h"angt die Definition des Kontinuumslimes von der Wahl
des Koordinatensystems und der Bijektion \Ref{1_f0} ab, was durch den Index
`$^\varepsilon$' symbolisch angezeigt wird.
Das willk"urliche Herausgreifen einer Bijektion $\underline{x}$ ist
nach der Beschreibung der Permutationssymmetrie als innere Symmetrie
ebenfalls keine Einschr"ankung.

Folglich h"angt der Kontinuumslimes letztlich nur von der Wahl des
Ko\-or\-di\-na\-ten\-sys\-tems ab. Als einzige Bedingung haben wir dabei
die Anordnungsvorschrift zu erf"ullen. Wir haben also die Freiheit, im
$\R^4$ Diffeomorphismen durchzuf"uhren, falls die diskreten Raumzeitpunkte
auch in den neuen Koordinaten gleichm"a"sig mit mittlerem Abstand
in der Gr"o"senordnung der Planck-L"ange angeordnet sind.
Insbesondere k"onnen wir {\em{makroskopische
Koordinatentransformationen}}\index{Koordinatentransformation,
makroskopisch}, also Transformationen $x \rightarrow
y(x)$ mit makroskopischen Funktionen $y^i$, durchf"uhren.
Da man sich in der Allgemeinen Relativit"atstheorie bei einem Wechsel
des Bezugssystems auf makroskopische Transformationen
beschr"anken kann, folgt die Invarianz des Kontinuumslimes unter
allgemeinen Koordinatentransformationen, also das
"Aquivalenzprinzip\index{"Aquivalenzprinzip}.

Es ist wichtig zu beachten, da"s die Freiheiten in der Koordinatenwahl
mit den makroskopischen Koordinatentransformationen noch nicht
ersch"opft sind. Zus"atzlich sind viele nicht-makroskopische
Koordinatentransformationen zul"assig, beispielsweise solche, welche
die diskreten Raumzeitpunkte permutieren. Wenn unsere Beschreibung
physikalisch sinnvoll sein soll, m"ussen wir solche
Koordinatentransformationen ausschlie"sen k"onnen. Dazu mu"s $P$
die Permutationssymmetrie vollst"andig brechen, und es mu"s
(bei gegebenem $P$) eine kanonische Wahl der Bijektion \Ref{1_f0}
geben. Darauf werden wir nach expliziterer Untersuchung des fermionischen
Projektors auf Seite \pageref{1_kont} zur"uckkommen.

Insgesamt kommen wir zu dem Schlu"s, da"s das Prinzip des
fermionischen Projektors im Kontinuumslimes auf jeden Fall das
"Aquivalenzprinzip liefert. Die lokale Struktur des Kontinuums erh"alt
man aber nur unter Annahme zus"atzlicher Bedingungen an $P$.
Um die Lokalit"at konsistent zu begr"unden, werden wir diese zus"atzlichen
Bedingungen aus den Gleichungen der diskreten Raumzeit ableiten m"ussen.

\subsection{Die Methode der St"orung des Vakuums}
\label{1_ab22}
F"ur eine zweckm"a"sige physikalische Beschreibung ist es oft g"unstig,
ein System (z.B. wenige, schwach gekoppelte Fermionen) als eine
St"orung des Vakuums aufzufassen.
Aus diesem Grund wollen wir den Kontinuumslimes eines allgemeinen
fermionischen Projektors $\tilde{P}$ auf unsere Beschreibung des
Vakuums zur"uckf"uhren.
Zur Einfachheit beschr"anken wir uns auf ein System \Ref{1_c8}
mit einem Diracsee, die Konstruktion l"a"st sich aber unmittelbar auf
zusammengesetzte Systeme \Ref{1_24} "ubertragen.

Wir gehen aus von einem Bezugssystem \Ref{1_c1}, \Ref{1_c2}.
Es gibt einen fermionischen Projektor $P$, der in diesem
Bezugssystem das Vakuum beschreibt ($P$ erh"alt man beispielsweise
durch Regularisierung der Distribution $P(x,y)$).
Wir w"ahlen makroskopische Wellenfunktionen
$\Psi_1,\ldots, \Psi_f$ und $\Phi_1,\ldots, \Phi_a$, die im Kontinuum
in positive- bzw. negative-Energie-L"osungen der freien Diracgleichung
\[ (i \Pdd - m) \: \Psi_j \;=\; (i \Pdd - m) \: \Phi_j \;=\; 0 \]
"ubergehen. Wir setzen $Y={\mbox{Im }} P$, also $P=P_Y$.
Nach den "Uberlegungen in Abschnitt \ref{1_ab21} sind die
negativen-Energie-L"osungen $\Phi_j$ im Diracsee $P^\varepsilon(x,y)$
enthalten, also $\Phi_1, \ldots, \Phi_a \in Y$.
Folglich k"onnen wir mit
\Equ{1_d1}
\bar{P} \;:=\; P_Y \:+\: P_{\bra \Psi_1, \ldots, \Psi_f \ket}
	\:-\: P_{\bra \Phi_1, \ldots, \Phi_a \ket}
\EndEqu
im Diracsee $a$ L"ocher erzeugen und $f$ Fermionen hinzuf"ugen,
$\bar{P}$ ist ebenfalls ein Projektor.
Im n"achsten Schritt f"uhren wir eine beliebige unit"are Transformation
des Projektors durch, also
\Equ{1_d2}
\tilde{P} \;:=\; U \:\bar{P}\:U^* \spc {\mbox{mit einem unit"aren
	Operator $U$ auf $H$}} \spc .
\EndEqu
Wir bezeichnen $\tilde{P}$ als {\em{gest"orten fermionischen
Projektor}}\index{fermionischer Projektor, gest"orter}
und nennen die Konstruktion \Ref{1_d1}, \Ref{1_d2} die {\em{Methode der
St"orung des Vakuums}}\index{Methode der St"orung des Vakuums}.

Wir m"ussen uns davon "uberzeugen, da"s mit der Methode der St"orung
des Vakuums tats"achlich jeder fermionische Projektor
$\tilde{P}$ gebildet werden kann:
Es sei ein fermionischer Projektor $\tilde{P}$ gegeben.
Wir w"ahlen als Bezugssystem die Bijektion \Ref{1_f0} aus der
Definition des Kontinuumslimes und bilden einen zugeh"origen
fermionischen Projektor $P$ des Vakuums.
Der Rang der Projektoren $P$, $\tilde{P}$ wird i.a. verschieden sein,
${\mbox{Rg }} P \neq {\mbox{Rg }} \tilde{P}$.
Durch geeignetes Erzeugen von L"ochern oder zus"atzlichen Zust"anden
k"onnen wir aber erreichen, da"s
${\mbox{Rg }} \bar{P} = {\mbox{Rg }} \tilde{P}$ ist. Dann sind
$\bar{P}$, $\tilde{P}$ unit"ar "aquivalent.

\subsubsection*{nichtlokale St"orungen des Diracoperators}
Wir untersuchen nun, wie sich die Methode der St"orung des Vakuums im
Kontinuum beschreiben l"a"st:
Die Definitionsgleichung \Ref{1_d1} geht im Kontinuumslimes bei geeigneter
Normierung der Wellenfunktionen $\Psi_j, \Phi_j$ (auf die wir in
der Einleitung wieder nicht eingehen) in die Distributionsgleichung
\Equ{1_d3}
\bar{P}(x,y) \;=\; P(x,y) \:+\: \sum_{j=1}^f \Psi_j(x) \: \overline{\Psi_j}(y)
	\:-\: \sum_{j=1}^a \Phi_j(x) \: \overline{\Phi_j}(y)
\EndEqu
"uber. Die unit"are Transformation \Ref{1_d2} "ubersetzt sich in der Form
\Equ{1_d4}
\tilde{P} \;=\; U \:\bar{P} \: U^*
\EndEqu
mit einem geeigneten unit"aren Operator $U$ des Kontinuums, dabei
bezeichnen $\bar{P}$ und $\tilde{P}$ die Operatoren mit Integralkernen
$\bar{P}(x,y)$ bzw. $\tilde{P}(x,y)$.

Die Transformation \Ref{1_d4} l"a"st sich alternativ auch als St"orung des
Diracoperators umschreiben: Die Distribution $\bar{P}(x,y)$ erf"ullt
die Diracgleichung
\[ (i \Pdd_x - m) \: \bar{P}(x,y) \;=\; 0 \spc . \]
Mit \Ref{1_d4} folgt f"ur den Kontinuumslimes des
gest"orten fermionischen Projektors
\[ U (i \Pdd - m) U^{-1} \: \tilde{P} \;=\; 0 \spc . \]
Diese Gleichung kann mit der Abk"urzung ${\cal{B}}:=U(i \Pdd)U^{-1} - i \Pdd$
auch in der Form
\Equ{1_36a}
(i \Pdd + {\cal{B}} - m) \: \tilde{P} \;=\; 0
\EndEqu
geschrieben werden. Der Operator ${\cal{B}}$ ist i.a.
nichtlokal\index{nichtlokale St"orung des Diracoperators}, er l"a"st
sich als Integraloperator
\Equ{1_37}
({\cal{B}} \Psi)(x) \;=\; \int d^4y \; B(x,y) \: \Psi(y)
\EndEqu
mit einer geeigneten matrixwertigen Distribution $B(x,y)$ ausdr"ucken.

\subsubsection{ein erster Kontakt zu klassischen Feldgleichungen}
\index{klassische Feldgleichungen, erster Kontakt}
Wir haben die Methode der St"orung des Vakuums f"ur einen allgemeinen
fermionischen Projektor $\tilde{P}$ durchgef"uhrt.
Diese Allgemeinheit ist allerdings nur von theoretischem Interesse:
Die Methode der St"orung des Vakuums ist n"utzlich, weil sich damit
der fermionische Projektor um das Vakuum entwickeln l"a"st.
Eine solche Entwicklung ist aber nur sinnvoll, wenn $\tilde{P}$
wirklich als ``St"orung'' des Vakuums angesehen werden kann, also
wenn die Anzahl $f$, $a$ der Fermionen nicht zu gro"s ist und
wenn sich $U$ nur wenig von der Identit"at unterscheidet.
Wir werden sp"ater sehen, da"s wir uns in realistischen Situationen
tats"achlich auf den Fall beschr"anken k"onnen, da"s der Operator
$U-\1$ (in einer dann n"aher spezifizierten Weise) klein ist.

Die Vorstellung einer kleinen St"orung in \Ref{1_d2} vereinfacht
auch die physikalische Begriffsbildung.
Man kann dann n"amlich auch f"ur den gest"orten fermionischen
Projektor von ``Diracseen'' sprechen und \Ref{1_d1} als Einf"uhrung
von $f$ Fermionen und $a$ Antifermionen in das System interpretieren.
Wir wollen mit dieser Vorstellung versuchen, einen ersten
Zusammenhang zwischen den Gleichungen der diskreten Raumzeit
und klassischen Feldgleichungen herzustellen.
Im ersten Schritt betrachten wir die Situation in der diskreten
Raumzeit: Wir nehmen an, da"s ein fermionischer Projektor $P$ des Vakuums
die Euler-Lagrange-Gleichungen eines geeigneten Variationsprinzips
erf"ullt. Nach Einf"uhrung von Fermionen und Antifermionen gem"a"s
\Ref{1_d1} werden diese Gleichungen i.a. verletzt sein. Damit auch
das erhaltene Vielfermionsystem die Euler-Lagrange-Gleichungen
erf"ullt, mu"s $\bar{P}$ geeignet modifiziert werden.
Um nicht zus"atzliche Fermionen in das System einzuf"uhren,
mu"s diese St"orung die Form einer unit"aren Transformation \Ref{1_d2} haben.
Wir erwarten, da"s $\tilde{P}$ f"ur geeignetes $U$ eine L"osung
der Euler-Lagrange-Gleichungen ist; die genaue Form der Transformation
\Ref{1_d2} in Abh"angigkeit von den Fermionen in \Ref{1_d1} wird durch die
spezielle Form des Variationsprinzips festgelegt.
Durch die unit"are Transformation werden auch die Wellenfunktionen $\Psi_j,
\Phi_j$ beeinflu"st, was schlie"slich als eine Wechselwirkung der Fermionen
interpretiert werden kann.

Im n"achsten Schritt untersuchen wir, wie sich diese Wechselwirkung
im Kontinuum beschreiben l"a"st: Die Einf"uhrung der Fermionen und
Antifermionen zeigt sich in \Ref{1_d3} im Auftreten klassischer Wellenfunktionen.
Die unit"are Transformation kann gem"a"s \Ref{1_36a} als St"orung des
Diracoperators geschrieben werden.
F"ur sinnvolle Gleichungen der diskreten Raumzeit
mu"s diese St"orung lokal sein. Genauer mu"s aus den
Euler-Lagrange-Gleichungen die Bedingung
\Equ{1_38a}
	{\cal{B}} \;=\; i \left(G^j(x) -\gamma^j\right) \:
	\frac{\partial}{\partial x^j} + B(x)
\EndEqu
folgen, so da"s sich der Diracoperator in \Ref{1_36a} auf \Ref{1_7} reduziert.
Wenn wir dies im Moment einfach annehmen, liefern die Gleichungen der
diskreten Raumzeit einen Zusammenhang zwischen den Wellenfunktionen
in \Ref{1_d3} und den bosonischen Potentialen im Diracoperator \Ref{1_7}.
Genau dieser Zusammenhang mu"s auch durch die klassischen Feldgleichungen
gegeben sein. Man beachte, da"s wir durch die unit"are Transformation
\Ref{1_d4} in einem Schritt sowohl die klassischen Bosefelder einf"uhren als
auch die Ankopplung dieser Felder an die Fermionen beschreiben.

Der Zusammenhang zu den klassischen Feldgleichungen ist im Moment sehr
qualitativ. Bevor wir ihn in Abschnitt \ref{1_ab25} pr"azisieren k"onnen, m"ussen
wir im n"achsten Abschnitt untersuchen, wie sich die Gleichungen der
diskreten Raumzeit ins Kontinuum "ubertragen lassen.

\subsection{Asymptotische Entwicklung}
Nachdem wir f"ur den fermionischen Projektor eine Kontinuumsbeschreibung
eingef"uhrt haben, kommen wir zu der Frage, wie f"ur zusammengesetzte
Ausdr"ucke in $P, E_x$ ein sinnvoller Kontinuumslimes gebildet werden kann.
Wir werden dieses Problem in der Einleitung nur am Beispiel der
Euler-Lagrange-Gleichungen \Ref{1_30}, \Ref{1_30b} diskutieren;
die meisten Konstruktionen lassen sich aber f"ur andere
Gleichungen "ahnlichen Typs ganz analog durchf"uhren.

Wir betrachten einen allgemeinen fermionischen Projektor $P$. Der
Operator $Q$, \Ref{1_30b}, hat in einem speziellen Bezugssystem die
Form
\Equ{1_g2}
Q^\varepsilon(x,y) \;\equiv\; E_x \:Q\: E_y \;=\;
	P^\varepsilon(x,y) \: P^\varepsilon(y,x) \: P^\varepsilon(x,y)
	\spc .
\EndEqu
Als ersten Ansatz zur Kontinuumsbeschreibung k"onnte man versuchen, einfach
die Faktoren in \Ref{1_g2} durch ihren Kontinuumslimes ersetzen. Der sich
ergebende Ausdruck
\Equ{1_g3}
P(x,y) \: P(y,x) \: P(x,y)
\EndEqu
ist aber mathematisch nicht sinnvoll, wie man schon f"ur
den fermionischen Projektor des Vakuums sieht:
Im Vakuum ist $P(x,y)$ gem"a"s \Ref{1_c8}, \Ref{1_24} eine temperierte
Distribution. Bei expliziter Ausf"uhrung des Fourierintegrals in \Ref{1_c8}
erh"alt man Beitr"age der Form
\Equ{1_39}
	\delta^\prime((y-x)^2) \;\;,\;\;\;\; \delta((y-x)^2) \;\;,\;\;\;\;
	\frac{1}{(y-x)^4} \;\;,\;\;\;\; \frac{1}{(y-x)^2} \spc ,
\EndEqu
dabei ist $(y-x)^2 \equiv (y-x)_j \: (y-x)^j$; die Distributionen
$(y-x)^{-2}$ und $(y-x)^{-4}$ sind als Hauptwert bzw. Ableitung des
Hauptwertes definiert.
Folglich besitzen die Distributionen $P(x,y)$, $P(y,x)$ auf dem Lichtkegel,
also f"ur $(y-x)^2=0$, Singularit"aten und Pole; sie k"onnen nicht wie
in \Ref{1_g3} miteinander multipliziert werden.

Wir k"onnen erwarten, da"s sich dieses mathematische Problem bei
sorgf"altigerer Bildung des Kontinuumslimes beheben l"a"st.
Trotzdem sollten die Singularit"aten auf dem Lichtkegel auch bei einer
genaueren Analyse eine entscheidende Rolle spielen:
Der fermionische Projektor $P^\varepsilon(x,y)$ ist eine Diskretisierung
der Distribution $P(x,y)$ auf der Skala der Planck-L"ange.
Die Singularit"aten von $P(x,y)$ zeigen sich in der diskreten Raumzeit
darin, da"s $P^\varepsilon(x,y)$ auf dem Lichtkegel Werte von
der Gr"o"senordnung $\sim \varepsilon^{-p}$ annimmt. Die Exponenten
$p$ kann man f"ur die verschiedenen Beitr"age in \Ref{1_39} mit einem
Skalierungsargument bestimmen,
\begin{eqnarray*}
P^\varepsilon(x,y) \;\sim\; \varepsilon^{-2} &\spc{\mbox{f"ur}}\spc&
	\delta^\prime((y-x)^2) , \; (y-x)^{-4} \\
P^\varepsilon(x,y) \;\sim\; \varepsilon^{-1} &\spc{\mbox{f"ur}}\spc&
	\delta((y-x)^2) , \; (y-x)^{-2} \spc .
\end{eqnarray*}
Bei der Bildung von $Q^\varepsilon(x,y)$ k"onnen wir die
Exponenten $p_j$ der einzelnen Faktoren $P^\varepsilon$ in
\Ref{1_g2} addieren, also symbolisch
\[ Q^\varepsilon(x,y) \;\sim\; \varepsilon^{-q}
	\spc{\mbox{mit}}\spc q=p_1+p_2+p_3 \spc . \]
F"ur "ubliche physikalische Systeme ist die Planck-L"ange um viele
Gr"o"senordnungen kleiner als alle anderen L"angenskalen des Systems.
Folglich erwarten wir, da"s die Beitr"age $\sim \varepsilon^{-p}$
zu $P^\varepsilon, Q^\varepsilon$ mit hohen Exponenten $p$ wesentlich
gr"o"ser als diejenigen mit niedrigen Exponenten sind.
Genauer sollte jeder zus"atzliche Faktor $\varepsilon$ die Beitr"age
um einen dimensionslosen Faktor
\Equ{1_38b}
{\mbox{Planck-L"ange}} \:\times\: {\mbox{Energie}}
\spc{\mbox{oder}}\spc \frac{{\mbox{Planck-L"ange}}}{{\mbox{Fermi-L"ange}}}
\EndEqu
abschw"achen. Da $\Ref{1_38b}$ bei typischen Energieskalen von der
Gr"o"senordnung $< 10^{-20}$  ist, scheint es eine sehr gute N"aherung
zu sein, die Beitr"age zu niedrigeren Exponenten ganz zu vernachl"assigen.

Um diese "Uberlegung mathematisch zu pr"azisieren, m"ussen
wir $\varepsilon$ als variablen Parameter auffassen und den Grenzfall
$\varepsilon \rightarrow 0$ untersuchen:
Wir bilden f"ur beliebiges $\varepsilon$ eine Regularisierung
$P^\varepsilon(x,y)$ auf der L"angenskala $\varepsilon$.
F"ur einen bestimmten Wert $\varepsilon$ in der Gr"o"senordnung der
Planck-L"ange soll $P^\varepsilon(x,y)$ der physikalische fermionische
Projektor sein, f"ur alle anderen Werte von $\varepsilon$ sehen wir
$P^\varepsilon$ nur als mathematische Hilfs\-kons\-truk\-tion an.
Wir bilden die Operatoren $Q^\varepsilon$ und die Kommutatoren
$[P^\varepsilon, Q^\varepsilon]$. Im Limes $\varepsilon \rightarrow 0$
treten in $Q^\varepsilon$, $[P^\varepsilon, Q^\varepsilon]$
Divergenzen auf. Genauer k"onnen wir eine Entwicklung nach der
Polordnung durchf"uhren
\begin{eqnarray}
\label{eq:1_g4}
Q^\varepsilon(x,y) &=& \frac{1}{\varepsilon^n} \: Q^{(0)}(x,y) \:+\:
	\frac{1}{\varepsilon^{n-1}} \: Q^{(1)}(x,y) \:+\: \cdots \\
\label{eq:1_g5}
[P^\varepsilon, Q^\varepsilon](x,y) &=&
	\frac{1}{\varepsilon^n} \: E^{(0)}(x,y) \:+\:
	\frac{1}{\varepsilon^{n-1}} \: E^{(1)}(x,y) \:+\: \cdots
	\spc ,
\end{eqnarray}
dabei sind $Q^{(j)}(x,y), E^{(j)}(x,y)$ temperierte Distributionen.
Wir nennen diese Entwicklung {\em{asymptotische
Entwicklung}}\index{asymptotische Entwicklung}.

\subsubsection*{die Planckn"aherung}
Da $\varepsilon$ f"ur den physikalischen fermionischen Projektor ein
sehr kleiner Parameter ist, sind die Reihen \Ref{1_g4}, \Ref{1_g5}
parit"atisch geordnet, d.h.
\[ \frac{1}{\varepsilon^n} \: Q^{(0)} \:\gg\: \frac{1}{\varepsilon^{n-1}} \: Q^{(1)}
	\:\gg\: \cdots \;\;\;\;,\spc
	\frac{1}{\varepsilon^n} \: E^{(0)} \:\gg\:
	\frac{1}{\varepsilon^{n-1}} \: E^{(1)} \:\gg\: \cdots \spc . \]
Wir nutzen dies f"ur eine N"aherung der Euler-Lagrange-Gleichungen aus:
Die Operatoren $E^{(j)}$ h"angen von den Fermionen und dem St"oroperator
${\cal{B}}$ ab, also symbolisch $E^{(j)}=E^{(j)}[\Psi, {\cal{B}}]$.
In der Euler-Lagrange-Gleichung $[P^\varepsilon, Q^\varepsilon]=0$ mu"s der
f"uhrende Beitrag von \Ref{1_g5} fast verschwinden, wir setzen n"aherungsweise
\Equ{1_h0}
E^{(0)}[\Psi, {\cal{B}}] \;=\; 0 \spc .
\EndEqu
F"ur den n"achsten Summanden in \Ref{1_g5} mu"s man ber"ucksichtigen,
da"s man durch sehr kleine St"orungen von $E^{(0)}$ Beitr"age in $E^{(1)}$
kompensieren kann. Man erh"alt so die Bedingung
\Equ{1_h1}
E^{(1)}[\Psi, {\cal{B}}] \;=\; \frac{d}{d\lambda} E^{(0)}[\Psi, {\cal{B}}+
	\lambda {\cal{B}}_1] _{|\lambda=0}
\EndEqu
mit einem frei w"ahlbaren Operator ${\cal{B}}_1$.
F"ur alle weiteren Summanden in \Ref{1_g5} geht man analog vor; allgemein
k"onnen Beitr"age in $E^{(k)}$ durch sehr kleine St"orungen in 
$E^{(0)}, \ldots, E^{(k-1)}$ kompensiert werden\footnote{Um diese Gleichungen
formal abzuleiten, setzt man den St"oroperator ${\cal{B}}$ als Potenzreihe
in $\varepsilon$ an,
\[ {\cal{B}}_\varepsilon \;=\; {\cal{B}} + \varepsilon {\cal{B}}_1 +
	\varepsilon^2 {\cal{B}}_2 + \cdots \spc . \]
F"ur die Operatoren $E^{(j)}$ erh"alt man mit einer Taylorentwicklung
\begin{eqnarray*}
E^{(j)}[\Psi, {\cal{B}}_\varepsilon] &=& E^{(j)}[\Psi, {\cal{B}}]
	\:+\: \varepsilon \: \frac{d}{d\lambda} E^{(j)}[\Psi,
	{\cal{B}}+\lambda {\cal{B}}_1]_{|\lambda=0} \\
&&+\: \frac{\varepsilon^2}{2} \:
	\frac{d^2}{d\lambda^2} E^{(j)}[\Psi, {\cal{B}}+ \lambda
	{\cal{B}}_1]_{|\lambda=0} \:+\:
	\varepsilon^2 \: \frac{d}{d\lambda} E^{(j)}[\Psi, {\cal{B}}+\lambda
	{\cal{B}}_2]_{|\lambda=0} \:+\: \cdots \spc .
\end{eqnarray*}
Man setzt diese Entwicklungsformeln in die Reihe \Ref{1_g5} ein und fordert,
da"s die Beitr"age jeder Ordnung in $\varepsilon^{-p}$ verschwinden.
Die St"oroperatoren ${\cal{B}}_1, {\cal{B}}_2, \ldots$ treten wegen
$\varepsilon \ll 1$ nicht als physikalische St"orungen in Erscheinung.
Sie k"onnen "ahnlich wie Lagrangesche Multiplikatoren beliebig sein
und schw"achen die Bedingungen an die Operatoren $E^{(j)}[\Psi, {\cal{B}}]$ ab.}.
Wir nennen die verwendete N"aherung
{\em{Planckn"aherung}}\index{Planckn"aherung} und die
Gleichungen \Ref{1_h0}, \Ref{1_h1}, u.s.w. die {\em{Gleichungen der
Planck\-n"a\-he\-rung}}\index{Gleichungen der Planckn"aherung}.
Die Gleichungen der Planckn"aherung sind als Distributionsgleichungen
im Kontinuum wohldefiniert; sie lassen sich wesentlich einfacher als die
Euler-Lagrange-Gleichungen der diskreten Raumzeit analysieren.
Die relativen Fehler der Planckn"aherung sind von der Gr"o"senordnung
\Ref{1_38b} und folglich kleiner als die Me"sgenauigkeit in "ublichen
Experimenten.

\subsubsection*{Vergleich zur Renormierung}
\index{Renormierung, Vergleich zur asymptotischen Entwicklung}
Wir wollen die Konstruktion der asymptotischen Entwicklung kurz diskutieren.
Nach dem Prinzip des fermionischen Projektors sollten wir von den Operatoren
$P, E_x$ ausgehen und daraus den Kontinuumslimes ableiten.
Bei der asymptotischen Entwicklung sind wir aber genau umgekehrt vorgegangen:
wir haben mit dem Kontinuumslimes $P(x,y)$ begonnen und daraus durch
Regularisierung den fermionischen Projektor $P^\varepsilon(x,y)$ der
diskreten Raumzeit gebildet.
Der Grund f"ur dieses Vorgehen ist rein technischer Art:
Die asymptotische Entwicklung macht nur Sinn, wenn $\varepsilon$ ein variabler
Parameter ist. Wir m"ussen also eine ganze Familie $(P^\varepsilon)_{\varepsilon
\in (0, \varepsilon_0)}$ von fermionischen Projektoren betrachten, was nur
durch Regularisierung von $P(x,y)$ realisiert werden kann.

Es ist wichtig zu beachten, da"s wir trotz dieser Konstruktionsmethode
am Prinzip des fermionischen Projektors festhalten.
Wir sehen eine spezielle Diskretisierung $P^\varepsilon$ von $P(x,y)$
als das fundamentale physikalische Objekt an, allerdings k"onnen wir "uber
Einzelheiten der Regularisierung keine Aussagen machen.
Dieses indirekte Vorgehen bei der asymptotischen Entwicklung
ist nur dann sinnvoll, wenn es auf das Regularisierungsverfahren
letztlich nicht ankommt. Die eigentliche Schwierigkeit wird
darin bestehen zu zeigen, da"s die mit der asymptotischen Entwicklung
abgeleiteten Ergebnisse von Einzelheiten der Regularisierung unabh"angig sind.
Wir bemerken, da"s wir f"ur die Konstruktion von $P^\varepsilon$ zur
technischen Einfachheit nicht mit Diskretisierungen der Raumzeit, sondern mit
Regularisierungen im Kontinuum arbeiten werden. Genauer werden wir die
Distributionen durch Faltung regularisieren, also
\[ P^\varepsilon(x,y) \;:=\; (P * \eta_\varepsilon)(x,y) \spc
	{\mbox{mit einer glatten Funktion $\eta_\varepsilon$}} \spc . \]

Das technische Vorgehen bei der asymptotischen Entwicklung hat "Ahnlichkeit
mit der Renormierung der Quantenfeldtheorie. Dort f"uhrt man auch eine
Regularisierung ein, beispielsweise durch Diskretisierung der Theorie auf
einem Gitter mit Gitterl"ange $\varepsilon$. Anschlie"send zeigt man, da"s die
Regularisierung (bei gleichzeitiger Umskalierung der nackten Massen und
Kopplungskonstanten) entfernt werden kann, was im Beispiel des Gitters
dem Limes $\varepsilon \rightarrow 0$ entspricht.
Im Gegensatz zum Renormierungsprogramm f"uhren wir aber den Grenz"ubergang
$\varepsilon \rightarrow 0$ nicht durch, sondern sehen eine auf der
L"angenskala der Planck\-l"an\-ge regularisierte Theorie als die physikalische
Theorie an.
Dieser Unterschied hat zur Folge, da"s wir auch physikalisch me"sbare
Gr"o"sen mit der Diskretisierung in Verbindung bringen k"onnen.
Insbesondere werden wir sehen, da"s die Gravitationskonstante mit der
L"angenskala $\varepsilon$ der Diskretisierung verkn"upft ist, und k"onnen
$\varepsilon$ mit der Planck-L"ange ausdr"ucken.

\subsection{Qualitative Beschreibung einiger Ergebnisse}
\label{1_ab25}
In den vorangehenden Abschnitten haben wir die Methoden bereitgestellt,
mit denen die Euler-Lagrange-Gleichungen im Kontinuumslimes
untersucht werden k"onnen.
Schematisch m"ussen wir nun folgenderma"sen vorgehen:
Zun"achst mu"s der Kontinuumslimes $\tilde{P}(x,y)$ des gest"orten
fermionischen Projektors f"ur m"oglichst allgemeine St"oroperatoren
${\cal{B}}$ explizit berechnet werden. Dazu konstruiert man L"osungen
der nichtlokalen Diracgleichung \Ref{1_36a}. Nach Regularisierung
der Distributionen $\tilde{P}(x,y)$ kann die asymptotische Entwicklung
\Ref{1_g4}, \Ref{1_g5} durchgef"uhrt werden. Anschlie"send untersucht
man die Gleichungen der Planckn"aherung.
Auf diese Weise hat man die Gleichungen der diskreten Raumzeit
letztlich in Kontinuumsgleichungen in den Parametern $[\Psi, {\cal{B}}]$
umgeschrieben.

Dieses Programm ist allgemein genug gefa"st, um neben einer expliziten
Ableitung klassischer Feldgleichungen auch die noch offen gebliebenen
theoretischen Fragen zu beantworten. Genauer m"ussen wir noch
die Lokalit"at und Kausalit"at des Kontinuums konsistent aus dem Prinzip
des fermionischen Projektors begr"unden. Au"serdem stellt sich die
allgemeine Frage, warum die Gleichungen der diskreten Raumzeit im
Kontinuumslimes in lokale Differentialgleichungen "ubergehen.

Die Berechnung von $\tilde{P}(x,y)$ und die asymptotische Entwicklung
sind zu umfangreich, um in der Einleitung im Detail dargestellt zu werden.
Darum m"ussen wir in der folgenden Diskussion auf
sp"atere Ergebnisse Bezug nehmen.

\subsubsection*{Lokalit"at der St"orungen des Diracoperators}
\index{Diracoperator, Lokalit"at}
In der Diracgleichung \Ref{1_36a} tritt ein allgemeiner nichtlokaler
St"oroperator ${\cal{B}}$ auf. In Anhang G wird (f"ur die eigentlich
interessante Wirkung \Ref{wirk}) gezeigt, da"s der Operator
${\cal{B}}$ f"ur alle L"osungen $[\Psi, {\cal{B}}]$ der Gleichungen
der Planckn"aherung die Form einer lokalen St"orung \Ref{1_38a} hat.
Dieses Ergebnis l"a"st sich auch in unserem Beispiel \Ref{1_29}
einsehen: Die Singularit"aten des formalen Produkts
\Ref{1_g3} auf dem Lichtkegel bedeuten in der diskreten Raumzeit,
da"s der Operator $Q^\varepsilon(x,y)$ seinen Hauptbeitrag
auf wenige Punkte $(x,y)$ (n"amlich die Punkte in unmittelbarer
N"ahe des Lichtkegels) konzentriert.
Eine solche Situation ist f"ur das Variationsprinzip in dem Sinne
stabil, da"s sie bei durch die Euler-Lagrange-Gleichungen
zugelassenen St"orungen von $P$ erhalten bleibt.
Ausgedr"uckt im Kontinuumslimes d"urfen die St"orungen des Diracoperators
also die Singularit"aten von $\tilde{P}(x,y)$ auf dem Lichtkegel nicht zerst"oren.
Bei nichtlokalen St"orungen des Diracoperators werden diese Singularit"aten
aber ``ausgeschmiert'' und verschwinden schlie"slich, wie man an expliziten
Rechnungen sieht.

Wir m"ussen noch pr"azisieren, unter welchen Voraussetzungen
das Ergebnis von Anhang G anwendbar ist und beschreiben dazu
gleich allgemein, wie der St"oroperator ${\cal{B}}$ in dieser Arbeit
behandelt wird:
Wir f"uhren zun"achst eine St"orungsentwicklung nach ${\cal{B}}$
durch. Im Rahmen der asymptotischen Entwicklung k"onnen wir
die Beitr"age zu $Q^{(j)}$, $E^{(j)}$ in beliebiger Ordnung in ${\cal{B}}$
berechnen und alle Beitr"age explizit aufsummieren.
Auf diese Weise kommen wir schlie"slich zu nicht-perturbativen
Ergebnissen. Die einzige Einschr"ankung f"ur diese Methode besteht
darin, da"s die asymptotische Entwicklung sinnvoll sein mu"s. Der
St"oroperator ${\cal{B}}$ mu"s also so gew"ahlt werden, da"s in
\Ref{1_g4}, \Ref{1_g5} die Beitr"age h"oherer Ordnung in $\varepsilon$
stark abfallen.
F"ur lokale St"orungen \Ref{1_38a} ist dies keine Einschr"ankung,
da dann $\tilde{P}(x,y)$, wie wir gerade beschrieben haben,
Singularit"aten auf dem Lichtkegel besitzt.
Der nichtlokale Anteil der St"orung des Diracoperators mu"s aber
(in einer nicht genau spezifizierten Weise) klein sein.

\subsubsection*{die Lokalit"at und Kausalit"at des Kontinuums}
\label{1_kont}
Mit der Lokalit"at des St"oroperators ${\cal{B}}$ k"onnen wir die
lokale und kausale Struktur des Kontinuums intrinsisch aus dem
Prinzip des fermionischen Projektors begr"unden:
Als Folge der Lokalit"at von ${\cal{B}}$ besitzt $\tilde{P}(x,y)$
Singularit"aten auf einem Lichtkegel (im Fall mit Gravitation ist
dies der Lichtkegel der Lorentzmetrik).
In der diskreten Raumzeit ist der Hauptbeitrag des Operators
$Q^\varepsilon(x,y)$ folglich auf die Punkte $(x,y)$ in unmittelbarer
N"ahe des Lichtkegels konzentriert. Wir k"onnen also zu gegebenem
$x \in M$ intrinsisch diejenigen Punkte $y \in M$ auszeichnen, die
in unserem Bezugssystem in unmittelbarer N"ahe des Lichtkegels
um $x \in \R^4$ zu liegen kommen.
In diesem Sinne kann man aus L"osungen der
Euler-Lagrange-Gleichungen die Lichtkegelstruktur und damit die
Topologie, also die Kausalit"at und Lokalit"at, konstruieren.
\index{Lokalit"at des Kontinuums}
\index{Kausalit"at des Kontinuums}

Zur Deutlichkeit erl"autern wir diese intrinsische Konstruktion der
Kausalit"at und Lokalit"at mit den Begriffen von Abschnitt \ref{1_ab22a}:
Wir geben eine L"osung $P$ der Euler-Lagrange-Glei\-chun\-gen und eine
Menge $N \subset \R^4$ von diskreten Raumzeitpunkten vor, welche
die Anordnungsvorschrift erf"ullen. Dann k"onnen wir die Wahl der
Bijektion \Ref{1_f0} durch die Bedingung weitgehend festlegen,
da"s der Operator $Q^\varepsilon(x,y)$ seinen Hauptbeitrag
bez"uglich einer beliebigen kausalen Struktur in unmittelbarer Umgebung
des Lichtkegels konzentriert.
Wenn wir die Bijektion \Ref{1_f0} festhalten (was wir nach dem
Umschreiben der Permutationssymmetrie als innere Symmetrie ohne
Beschr"ankung tun k"onnen),
l"a"st sich diese zus"atzliche Bedingung auch als Einschr"ankung
f"ur die Wahl des Koordinatensystems auffassen.
Insbesondere k"onnen wir dann i.a.
keine Koordinatentransformationen durchf"uhren, bei welchen die
Raumzeitpunkte $N$ permutiert werden. Neben makroskopischen
Koordinatentransformationen sind nur noch {\em{mikroskopische
Koordinatentransformationen}}\index{Koordinatentransformation,
mikroskopisch} $x \rightarrow y(x)$ m"oglich, bei
denen die Funktionswerte $y^i(x)-x^i$ von der Gr"o"senordnung der
Planck-L"ange sind.
Da mikroskopische Koordinatentransformationen f"ur die
Kontinuumsbeschreibung irrelevant sind, k"onnen wir uns also
tats"achlich auf die makroskopischen Diffeomorphismen der
Allgemeinen Relativit"atstheorie beschr"anken.

Da diese Argumentation nur unter der Voraussetzung einer ``kleinen''
Nichtlokalit"at von ${\cal{B}}$ zul"assig ist, erhalten wir f"ur die
intrinsische Konstruktion der Lokalit"at und Kausalit"at die
folgende Einschr"ankung:
Wir gehen von einem fermionischen Projektor $P$ des Vakuums aus.
Sein Kontinuumslimes besitzt wegen \Ref{1_24} Singularit"aten auf
dem Lichtkegel. Bei Einf"uhrung von Fermionen gem"a"s \Ref{1_d1}
bleiben die Singularit"aten auf dem Lichtkegel nach \Ref{1_d3}
erhalten, so da"s die asymptotische Entwicklung sinnvoll ist.
Nun betrachten wir eine stetige Schar unit"arer Transformationen
$U(\tau)$ mit $U(0)=\1$ und bilden die Variation
\[ \tilde{P}(\tau) \;=\; U(\tau) \:\bar{P}\: U(\tau)^{-1} \]
des fermionischen Projektors.
Im Kontinuum geht $\tilde{P}$ in eine Variation der Form
\Ref{1_d4} oder, "aquivalent, in eine Variation ${\cal{B}}(\tau)$
der St"orung des Diracoperators "uber.
Wir nehmen an, da"s die Projektoren $\tilde{P}(\tau)$ (bis auf Beitr"age
von der Gr"o"senordnung der Wellenfunktionen $\Psi$) die
Euler-Lagrange-Gleichungen erf"ullen. Dann mu"s ${\cal{B}}(\tau)$
eine Schar lokaler Operatoren sein
(man beachte, da"s wir f"ur dieses ``Stetigkeitsargument'' mit beliebig kleinen
nichtlokalen Beitr"agen zu ${\cal{B}}(\tau)$ auskommen).
Folglich bleiben die Singularit"aten von $\tilde{P}(x,y)$ auf dem
Lichtkegel erhalten, so da"s
die asymptotische Entwicklung f"ur beliebiges $\tau$ g"ultig bleibt.
Wir k"onnen die intrinsische Konstruktion der Lokalit"at und Kausalit"at
also f"ur alle fermionischen Projektoren $\tilde{P}$ anwenden, die man
auf die gerade beschriebene Weise als stetige Deformation $\tilde{P}(\tau)$,
$\tilde{P}(0)=\bar{P}$ von L"osungen $\tilde{P}(\tau)$ der Gleichungen
der diskreten Raumzeit bilden kann.
Damit scheinen alle physikalisch interessanten F"alle abgedeckt zu sein.
Wir k"onnen aber nicht ausschlie"sen, da"s es weitere L"osungen der
Euler-Lagrange-Gleichungen gibt, die m"oglicherweise eine ganz andere
Struktur als die von uns durch Variation konstruierten fermionischen
Projektoren besitzen.

\subsubsection*{die klassischen Gleichungen sind Differentialgleichungen}
Aus der gerade begr"undeten lokalen und kausalen Struktur des
Kontinuums folgt noch nicht unmittelbar, da"s die Gleichungen
der diskreten Raumzeit im Kontinuumslimes in kausale Differentialgleichungen
"ubergehen. Wir wollen den Zusammenhang nun etwas genauer beschreiben.

Wir erkl"aren zun"achst, warum man Differentialgleichungen erh"alt:
\index{Lokalit"at der klassischen Gleichungen}
Aus \Ref{1_36a} und der Lokalit"at der St"orung ${\cal{B}}$ folgt
unmittelbar, da"s die Wellenfunktionen der Fermionen L"osungen einer
Diracgleichung mit Diracoperator \Ref{1_7} sind.
Die klassischen Feldgleichungen (z.B. die Maxwell- und Einsteingleichungen)
m"ussen aus der asymptotischen Entwicklung der
Euler-Lagrange-Gleichungen abgeleitet werden.
Die Euler-Lagrange-Gleichungen sind in dem Sinne nichtlokale
Gleichungen, da"s darin die Operatoren $P^\varepsilon(x,y),
Q^\varepsilon(x,y)$ auch f"ur makroskopisch entfernte Raumzeitpunkte
$x, y$ eingehen. Darum ist die Lokalit"at des Kontinuumslimes nicht
offensichtlich. Wir geben ein allgemeines Argument f"ur
Differentialgleichungen:
In der Distribution $\tilde{P}(x,y)$ treten Beitr"age in den
St"orpotentialen $(G^j - \gamma^j)$, $B$ und deren partiellen Ableitungen
auf. Genauer hat die Abh"angigkeit von den Potentialen die Form
konvexer Linienintegrale, also als typisches Beispiel
\[ \tilde{P}(x,y) \;=\; \int_0^1 d\alpha \; (\Box B)(\alpha y
	+ (1-\alpha)x) \;\cdots \;+\; \cdots \spc . \]
In dem regularisierten Produkt $Q^\varepsilon(x,y)$ treten ebenfalls
solche Linienintegrale auf. Die f"uhrenden Divergenzen der
Euler-Lagrange-Gleichungen f"ur $\varepsilon \rightarrow 0$ treten in
$Q^\varepsilon(x,y)$ am Ursprung, also f"ur $x=y$ auf.
Damit reduzieren sich die konvexen Linienintegrale auf lokale Beitr"age
in den Potentialen und deren Ableitungen.
Die Beitr"age der Fermionen (also die Dirac-Str"ome und fermionischen
Energie-Impuls-Tensoren) zu $Q^\varepsilon(x,y)$ sind zu f"uhrender
Divergenz ebenfalls am Ursprung lokalisiert.
Insgesamt erh"alt man lineare Relationen zwischen diesen Tensorfeldern,
also klassische Differentialgleichungen.

F"ur die Kausalit"at des Kontinuumslimes\index{Kausalit"at
der klassischen Gleichungen} der Euler-Lagrange-Gleichungen
haben wir keine allgemeine Begr"undung. Man kann sich aber f"ur die
Diracgleichung und die klassischen Feldgleichungen wie "ublich durch
Konstruktion einer retardierten Greensfunktion von der Kausalit"at
"uberzeugen.

\subsubsection{die klassischen Feldgleichungen f"ur dynamische Eichfelder}
Wir schlie"sen die allgemeinen "Uberlegungen zur Lokalit"at und Kausalit"at
des Kontinuumslimes ab und wollen nun etwas konkreter auf die Ableitung
der klassischen Feldgleichungen eingehen.
Dazu werden wir an verschiedenen St"oroperatoren ${\cal{B}}$
die Ergebnisse sp"aterer Rechnungen anschaulich beschreiben.

Wir beginnen mit dem Fall $G^j \equiv \gamma^j$ ohne Gravitationsfeld,
also einem allgemeinen lokalen Potential ${\cal{B}}=B(x)$.
Wie in Kapitel \ref{kap2} genau erkl"art wird, kann die
Distribution $\tilde{P}(x,y)$ als L"osung der Diracgleichung \Ref{1_36a}
explizit berechnet werden.
Bei Durchf"uhrung der asymptotischen Entwicklung treten
in $E^{(j)}$ Ausdr"ucke in den Wellenfunktionen $\Psi$, dem Potential
$B$ und dessen partiellen Ableitungen $\partial^\gamma B$ auf,
wobei $\gamma$ einen Multiindex bezeichnet.
Wir k"onnen diese Ausdr"ucke jeweiligen Operatoren $E^{(j)}$
zuordnen:
Die Dirac-Str"ome und fermionischen Energie-Impuls-Tensoren treten
erstmals in $E^{(2)}$ bzw. $E^{(3)}$ auf. Die Terme der Form
\Equ{1_42}
	\partial^{\gamma_1} B \: \cdots \: \partial^{\gamma_p} B
\EndEqu
sind im Operator $E^{(j)}$ mit
\Equ{1_41}
	j \;=\; p - 1 + \sum_{k=1}^p |\gamma_k|
\EndEqu
zu finden. Folglich kommen die (Noether-)Str"ome
und Energie-Impuls-Tensoren der klassischen Bosefelder in den
gleichen Operatoren wie die entsprechenden fermionischen Ausdr"ucke
vor, was sinnvoll erscheint.
Au"serdem sehen wir an \Ref{1_41}, da"s in der asymptotischen
Entwicklung Terme h"oherer Ordnung in $B$ im Vergleich zu Termen
niedrigerer Ordnung  um Potenzen der Planck-L"ange kleiner sind.
Diese Tatsache wird eine Begr"undung daf"ur liefern, da"s in die
Feldgleichungen die klassischen Potentiale nur bis zu einer
bestimmten Potenz eingehen (also da"s beispielsweise die
Maxwell-Gleichungen lineare Gleichungen sind).

Es zeigt sich, da"s die meisten Freiheitsgrade der Matrix $B$
zu einem gro"sen Beitrag in $E^{(0)}$ f"uhren und deswegen nicht
auftreten d"urfen. Genauer brauchen wir nur {\em{vektorielle}} und
{\em{axiale Potentiale}}\index{Potential, vektoriell}
\index{Potential, axial}
zu betrachten, also
\begin{eqnarray}
\label{eq:1_45}
B(x) &=& (V \!\! \slsh_{ij}(x))_{i,j = 1,\ldots,B} \;+\; \rho \:
	(A \! \slsh_{ij}(x))_{i,j = 1,\ldots,B} \spc ,
\end{eqnarray}
dabei bezeichnet $\rho \equiv \gamma^5$ die pseudoskalare
Diracmatrix.
Die bosonischen Potentiale haben nun (trotz der lokalen $U(2B,2B)$-Eichsymmetrie)
die Form wie bei einer $U(B) \otimes U(B)$-Eichtheorie.
Um die Unterschiede zwischen den vektoriellen und axialen
Potentialen herauszuarbeiten, betrachten
wir das Verhalten der Potentiale bei Eichtransformationen:
Das vektorielle Potential l"a"st sich mit einer lokalen Eichtransformation
\[ \Psi(x) \;\longrightarrow\; e^{i \Lambda(x)} \: \Psi(x) \spc {\mbox{mit}}
	\spc \Lambda(p)=1 {\mbox{ und }}
	(\partial_j \Lambda)(p) = V_j(p) \]
in jedem Raumzeitpunkt $p$ lokal zum Verschwinden bringen.
Folglich k"onnen wir aus $V$ (ohne Bildung von Ableitungen) keine
eichinvarianten Gr"o"sen konstruieren. In unseren Rechnungen zeigt sich
das darin, da"s $V$ im Operator $E^{(0)}$
nicht beitr"agt. In den berechneten Formeln f"ur $P^\varepsilon, Q^\varepsilon$
treten zwar die sogenannten {\em{Eichterme}}\index{Eichterme} auf, die das Potential
enthalten und das Verhalten unter Eichtransformationen beschreiben, diese
Terme fallen aber bei Einsetzen in \Ref{1_g5} weg.
Das axiale Potential $A$ l"a"st sich dagegen nicht lokal
wegeichen\footnote{\label{2_foot5}
Das sieht man am einfachsten so: Der Diracoperator
\Equ{1_47}
i \Pdd + \rho \: (\Pdd \Lambda)
\EndEqu
wirkt auf die links- und rechtsh"andige Komponente $\Psi_{L\!/\!R} :=
\frac{1}{2}(1 \mp \rho) \Psi$ der Wellenfunktion wie
$(i \Pdd \mp (\Pdd \Lambda)$.
Um das Potential in \Ref{1_47} zum Verschwinden zu bringen,
mu"s man folglich die Transformation
\Equ{1_58}
\Psi_L \:\rightarrow\: e^{i \Lambda} \: \Psi_L \;\;\;\;,\spc
	\Psi_R \:\rightarrow\: \Psi_R \: e^{-i \Lambda}
\EndEqu
durchf"uhren, also insgesamt
\[ \Psi \;\rightarrow\; \left( \frac{1}{2} (1-\rho) \: e^{i \Lambda}
	\;+\; \frac{1}{2} (1+\rho) \: e^{-i \Lambda} \right)  \Psi \spc . \]
Das ist aber keine unit"are Transformation und damit auch keine
Eichtransformation.}, was sich in unseren Rechnungen an zwei Stellen
auswirkt: Zun"achst f"uhrt $A$ zu einem Beitrag in $E^{(0)}$.
In den berechneten Formeln f"ur $P^\varepsilon, Q^\varepsilon$ treten
n"amlich die sogenannten {\em{Pseudoeichterme}}\index{Pseudoeichterme}
auf, die zwar eine "ahnliche Form wie die Eichterme haben, aber bei
Einsetzen in \Ref{1_g5} nicht verschwinden.
Au"serdem hat man in $E^{(2)}$ zus"atzlich einen Term der Form
\Equ{1_49}
	m^2 \; \rho A \:\cdots \spc ,
\EndEqu
den sogenannten {\em{Massenterm}}\index{Massenterm}, dabei setzt sich
$m^2$ aus den Massen der Fermionen zusammen.

Wir untersuchen nun die Gleichungen der Planckn"aherung \Ref{1_h0},
\Ref{1_h1}. Damit die Pseudoeichterme in $E^{(0)}$ verschwinden, mu"s
das axiale Potential $A$ bestimmte Bedingungen erf"ullen.
Wenn wir annehmen, da"s die axialen Potentiale in der dadurch
zugelassenen Weise tats"achlich auftreten, erhalten wir in $E^{(1)}$
Kreuzterme zwischen den Pseudoeichtermen und den Eichtermen
und damit einschr"ankende Bedingungen f"ur die vektoriellen Potentiale.
Wir sehen also, da"s die Gleichungen \Ref{1_h0} und \Ref{1_h1} die
M"oglichkeiten in der Wahl der Eichpotentiale stark einschr"anken.
Diesen Effekt nennen wir {\em{Reduktion der dynamischen
Eichfreiheitsgrade}}\index{Reduktion der dynamischen Eichfreiheitsgrade}.
Wir k"onnen die Bedingungen an die Potentiale auch durch Einf"uhrung
{\em{effektiver Eichgruppen}}\index{Eichgruppe, effektive}
f"ur gewisse Linearkombinationen der
vektoriellen und axialen Potentiale schreiben. Es zeigt sich genauer,
da"s wir zu vektoriellen und rechtsh"andigen Potentialen $V, L$,
\Equ{1_y0}
B \;=\; V \!\!\slsh \;+\; \frac{1}{2}(1 + \rho) \: L\!\slsh \spc .
\EndEqu
"ubergehen m"ussen, falls das Vakuum Fermionsorten einer
ausgezeichneten H"andigkeit (also z.B. linksh"andige Neutrinos)
enth"alt. Das rechtsh"andige Potential koppelt nur an die linksh"andige
Komponente der Fermionen an (also wie beispielsweise das
$W$-Potential im Standardmodell).
Im Hinblick auf die Physik sollten sich als effektive Eichgruppen
f"ur die vektoriellen und rechtsh"andigen Potentiale die Gruppen
$U(1) \times SU(3)$ bzw. $SU(2)$ des Standardmodells ergeben.

Wir kommen zum Operator $E^{(2)}$: Er enth"alt sowohl die
(Dirac-)Str"ome der Fermionen
als auch zweite Ableitungen des Potentials $B$.
Die Gleichungen der Planckn"aherung liefern eine lineare Beziehung zwischen
diesen Gr"o"sen, also klassische Feldgleichungen\index{klassische
Feldgleichungen f"ur effektive Eichfelder}.
Die Ankopplung der Fermionen an die Felder ist bereits durch die Form der
Potentiale in \Ref{1_y0} festgelegt.
Aus den Proportionalit"atsfaktoren k"onnen die (nackten)
Kopplungskonstanten berechnet werden.

\subsubsection*{massive Eichbosonen, kurzer Vergleich zum
	Higgs-Mechanismus}
\label{spontan} \index{massive Eichbosonen}
\index{Higgs-Mechanismus}
Es zeigt sich, da"s die Bosonen des rechtsh"andigen Potentials $L$
als Folge des Massenterms \Ref{1_49} automatisch eine Ruhemasse
besitzen.
Es scheint auf den ersten Blick erstaunlich, da"s wir im Gegensatz
zu den "ublichen Eichtheorien zur Massenerzeugung der Eichbosonen
ohne den Higgs-Mechanismus der spontanen Symmetriebrechung
auskommen. Eine erste Erkl"arung besteht darin, da"s die rechtsh"andigen
Potentiale gar keiner Symmetrie des Systems entsprechen.
Rechtsh"andige Potentiale f"uhren n"amlich (analog wie das axiale Potential
in Fu"snote \ref{2_foot5} auf Seite \pageref{2_foot5})
zu einer relativen verallgemeinerten Phasenverschiebung der links- und
rechtsh"andigen Komponente von $P$. Dadurch wird die chirale Symmetrie im
fermionischen Projektor zerst"ort, was sich letztlich in den Pseudoeichtermen
und Massentermen zeigt.

Diese Argumentation ist aber zu stark vereinfacht.
Um die Situation genauer zu analysieren, fassen wir die Diracmatrizen
wie im Diracoperator \Ref{1_7} als dynamische Matrixfelder auf:
Nach der Umformung
\begin{eqnarray}
\label{eq:1_60}
i \Pdd + \frac{1}{2} (1 + \rho) \: L \! \slsh &=& i \gamma^j \:
	\left( \partial_j - i C_j \right) \\
{\mbox{mit }} \;\; C_j &:=& \frac{1}{2} \: L_j - \frac{1}{12} \:
	\varepsilon_{jklm} \: L^k \: \sigma^{lm} \nonumber
\end{eqnarray}
hat der Ausdruck $\partial_j - i C_j$ die Form der eichkovarianten Ableitung
in den Yang-Mills-Theorien (man beachte, da"s die Matrix $C_j$ bzgl. des
Spinskalarproduktes hermitesch ist).
Daher k"onnen wir $C_j$ durch eine Eichtransformation in jedem
Raumzeitpunkt $p$ zum Verschwinden bringen, z.B. durch
\[ \Psi \;\longrightarrow\; U \: \Psi \spc{\mbox{mit}} \spc
	U(x) \;=\; e^{-i \: C_j(p) \: (x^j-p^j)} \spc . \]
Der Diracoperator hat dann die Form
\begin{eqnarray}
\label{eq:1_61}
U (i \Pdd + B) U^{-1} &=& i G^j \: \frac{\partial}{\partial x^j}
	+ \tilde{B}(x) \\
{\mbox{mit}} \spc \tilde{B}(p) &=& 0 \nonumber \\
G^j(p) &=& \gamma^j \;\;\;,\spc \partial_k G^j \;=\; \frac{1}{3} \:
	\varepsilon^j_{\;klm} \: L^l \: \gamma^m \;\;\; . \nonumber
\end{eqnarray}
Durch die lokale Eichtransformation haben wir also das rechtsh"andige
Potential in $p$ zum Verschwinden gebracht, daf"ur h"angen jetzt die
Matrixfelder $G^j$ von $L$ ab.

Nun hat die Situation gro"se "Ahnlichkeit mit dem Higgs-Mechanismus.
Nach spontaner Symmetriebrechung mit einem Higgs-Feld kann man
n"amlich die Potentiale der spontan gebrochenen Eichfreiheitsgrade
ebenfalls lokal wegtransformieren, wenn man eine allgemeine
Form des Higgs-Feldes zul"a"st, also das Higgs-Feld nicht mehr
in den ``flachen Richtungen'' des Higgs-Potentials fixiert.
In diesem Sinne wird die Rolle des Higgs-Feldes bei uns von den
Matrixfeldern $G^j$ "ubernommen.
Wenn man die Analogie genauer untersuchen m"ochte, tritt das Problem
auf, da"s wir nicht auf einfache Weise den Kontinuumslimes der Wirkung
bilden k"onnen, und dadurch beispielsweise nicht wissen, was dem
``Sektflaschenpotential'' beim Higgs-Mechanismus entspricht.
Wir k"onnen nur ganz allgemein sagen, da"s die Nebenbedingung $P^2=P$
bei der Variation verhindert, da"s die Matrixfelder $G^j$ im Vakuum
verschwinden.

\subsubsection*{das Gravitationsfeld}
\label{gravi} \index{Gravitationsfeld}
Wir gehen nun zum allgemeineren Diracoperator \Ref{1_7} "uber
und untersuchen Variationen der Matrixfelder $G^j$.
Von diesen Variationen k"onnen einige durch Eichtransformationen in
St"orungen durch lokale Potentiale ${\cal{B}}=B(x)$ umgewandelt werden
(so wie beispielsweise beim "Ubergang von \Ref{1_61} zu \Ref{1_60}),
andere f"uhren bei asymptotischer Entwicklung auf stark singul"are
Beitr"age und d"urfen deswegen nicht auftreten.
Letztlich k"onnen wir uns auf diejenigen St"orungen beschr"anken,
bei denen man in der Blockmatrixdarstellung \Ref{1_24} f"ur jedes
$P^{\{\alpha\}}$ ein Gravitationsfeld einf"uhrt.

Wir "uberlegen zun"achst, warum das Graviationsfeld in allen Bl"ocken
$P^{\{\alpha\}}$ gleich sein mu"s:
Bei der Berechnung von $\tilde{P}(x,y)$ im Gravitationsfeld stellt
man fest, da"s die f"uhrenden Beitr"age, die sogenannten
{\em{Diffeomorphismenterme}}\index{Diffeomorphismenterm},
eine Koordinatentransformation beschreiben.
Dies ist nach dem "Aquivalenzprinzip auch einsichtig.
Als Folge der Diffeomorphismenterme verschieben sich
die Punkte $(x, y)$, an denen $Q^\varepsilon(x,y)$
f"ur $\varepsilon \rightarrow 0$ singul"ar wird.
Wenn der Operator $Q^\varepsilon$ Spuren enth"alt, werden die
Beitr"age in den verschiedenen Bl"ocken miteinander verkn"upft.
Als Folge m"ussen dann die Singularit"aten in allen Bl"ocken an
den gleichen Punkten $(x,y)$ auftreten, was wiederum ein
einheitliches Gravitationsfeld impliziert.
Etwas genauer sieht man das Prinzip am Beispiel zweier Diracseen
$P^{(1)}, P^{(2)}$ gleicher Masse und $P = P^{(1)} \oplus P^{(2)}$:
in dem Ausdruck
\Equ{1_62}
P^\varepsilon(x,y) \: P^\varepsilon(y,x) - \frac{1}{8} \:
	\Tr \left( P^\varepsilon(x,y) \: P^\varepsilon(y,x) \right)
\EndEqu
heben sich die f"uhrenden Divergenzen auf dem Lichtkegel nur dann weg,
wenn das Gravitationsfeld in beiden Bl"ocken "ubereinstimmt.

Wir werden sehen, da"s physikalisch interessante Wirkungen auch aus anderen
Gr"unden mit Kombinationen "ahnlich zu \Ref{1_62} gebildet werden m"ussen.
Dadurch wird das Gravitationsfeld in einer physikalisch
sinnvollen Weise auftreten.

Wir kommen zur Ableitung der zugeh"origen Feldgleichungen:
Bei der Berechnung von $P^\varepsilon, Q^\varepsilon$ zeigt sich, da"s im
Operator $E^{(1)}$ der Einstein-Tensor auftritt. Da die Energie-Impuls-Tensoren
der Fermionen und Eichfelder dagegen in $E^{(3)}$ zu finden sind,
k"onnen wir die Planckn"aherung nicht anwenden.
Da"s der Beitrag des Einstein-Tensors in der asymptotischen Entwicklung
viel gr"o"ser als derjenige des Energie-Impuls-Tensors ist, kann als
Begr"undung daf"ur angesehen werden, da"s das Gravitationsfeld so
schwach an Materie ankoppelt. Man erh"alt schlie"slich eine Gleichung der Form
\[  G_{ij} \;=\; c \varepsilon^2 \; T_{ij} \spc , \]
wobei die Konstante $c$ im konkreten Modell explizit berechnet werden kann.
Das sind die Einstein-Gleichungen, der Faktor $c \varepsilon^2$ kann
mit der Gravitationskonstanten identifiziert werden.
Auf diese Weise k"onnen wir $\varepsilon$ direkt durch die Planck-L"ange
ausdr"ucken.

\section{Das Modell}
\label{1_ab3}
Die bisherige Beschreibung war so allgemein wie m"oglich gehalten und
sollte unser Konzept und das grobe Vorgehen skizzieren. Wir haben
qualitativ gesehen, da"s man im Kontinuumslimes einige Ergebnisse erh"alt,
die physikalisch sinnvoll erscheinen.
In Kapitel 4 und Kapitel 5 (das noch nicht vollst"andig ausgearbeitet und noch
nicht getippt ist) werden diese Resultate pr"azisiert, und es wird versucht, ein
realistisches Modell aufzubauen. Wir wollen an dieser Stelle das Modell
definieren und die wichtigsten Ergebnisse auflisten.

Zun"achst m"ussen wir den fermionischen Projektor des Vakuums definieren.
Dazu bauen wir drei Diracseen aus Fermionen der Masse $m^{(i)}$ und
drei Diracseen aus linksh"andigen masselosen Fermionen auf, also
\begin{eqnarray*}
P^{(i)}(x,y) &:=& \int \frac{d^4k}{(2 \pi)^4} \; (k \slsh + m^{(i)}) \:
	\delta(k^2 - (m^{(i)})^2) \: \Theta(-k^0) \; e^{-ik(x-y)} \;\;\;\;\;
	i=1,2,3 \\
P^{(i)}(x,y) &:=& \int \frac{d^4k}{(2 \pi)^4} \; \frac{1}{2} (1 - \rho) \:
	k \slsh \: \delta(k^2) \: \Theta(-k^0) \; e^{-ik(x-y)} \;\;\;\;\;\spc\;\;\;\;\:
	i=4,5,6 \spc .
\end{eqnarray*}
Wir addieren die Diracseen gem"a"s \Ref{1_23} und bilden einen massiven und
einen chiralen Fermionblock
\[ P^{\{1\}} \;=\; P^{(1)} + P^{(2)} + P^{(3)} \;\;\;\;,\spc
	P^{\{2\}} \;=\; P^{(4)} + P^{(5)} + P^{(6)} \spc . \]
Als Kontinuumslimes $P(x,y)$ des fermionischen Projektors des Vakuums setzen wir
als Spezialfall von \Ref{1_24} die direkte Summe von 7 massiven Bl"ocken und einem
chiralen Block an,
\Equ{1_70}
P(x,y) \;:=\; \left( P^{\{1\}}(x,y) \right)^7 \oplus P^{\{2\}}(x,y) \spc .
\EndEqu
Die Spindimension ist $32$, die Eichgruppe $U(16,16)$.
Mit den massiven Bl"ocken wollen wir sowohl die Quarks $u, s, t$ bzw.
$d, c, b$ als auch die massiven Leptonen $e, \mu, \tau$ beschreiben;
der chirale Block soll die Neutrinos modellieren.
Die Potenz $7$ begr"undet sich damit, da"s wir einen massiven Leptonblock und,
wegen der Colour- und Isospinfreiheitsgrade, $3 \times 2 = 6$ Quarkbl"ocke
ben"otigen. Man beachte, da"s die Massen $m^{(i)}$ der massiven Fermionen
in jedem Block gleich sind.

Als Wirkung\index{Wirkung} in der diskreten Raumzeit w"ahlen wir
\Equ{wirk}
S \;=\; \sum_{x,y \in M} \; \sum_{r=1}^{12} \; \sum_{\{p\}_r {\mbox{\scriptsize{ mit }}}
	|\{p\}_r|=12} c_{\{p\}_r} \: \prod_{j=1}^r \; \tr \left(
	(E_x \:P\: E_y \: P)^{p_j} \right) \spc ,
\EndEqu
dabei durchl"auft die Summe $\sum_{\{p\}_r}$ alle Konfigurationen der
ganzzahligen Parameter $p_1, \ldots, p_r$ mit $1 \leq p_1 \leq \cdots
\leq p_r$ und $p_1 + \cdots + p_r = 12$; $c_{\{p\}_r}$ sind beliebige reelle Parameter.
Die Form dieser Wirkung kann man folgenderma"sen einsehen:
Man beachte zun"achst, da"s $S$ im Operator $E_x \:P\: E_y \:P$ homogen vom
Grade 12 ist. Zur Bildung eines solchen Polynoms kann man die einzelnen
Faktoren $E_x \:P\: E_y \:P$ in Gruppen zusammenfassen und in jeder Gruppe getrennt
die Spur bilden. Die Wirkung ist als eine Linearkombination dieser Terme aufgebaut.

Als Euler-Lagrange-Gleichungen\index{Euler-Lagrange-Gleichungen} erh"alt man
die Kommutatorgleichung
\begin{eqnarray}
\label{eq:1_71}
[P, \: Q] &=& 0 \spc {\mbox{mit}} \\
\label{eq:1_72}
Q &=& \sum_{x,y \in M} \: \sum_{q=1}^{12} \beta^{(q)}_{xy} \:
	(E_x \:P\: E_y \:P)^{q-1} \; E_x \:P\: E_y \spc ,
\end{eqnarray}
dabei sind die reellen Funktionen $\beta^{(q)}_{xy}$ homogene Polynome
vom Grade $12-q$ in $E_x \:P\: E_y \:P$.

In Kapitel 5 werden die Koeffizienten $c_{\{p\}_r}$ explizit bestimmt.
F"ur den Kontinuumslimes erh"alt man die folgenden Ergebnisse:
\begin{enumerate}
\item Die Gleichungen der Planckn"aherung \Ref{1_h0}, \Ref{1_h1}
legen die Struktur der Gleichungen des Kontinuums fest:
Durch Reduktion der dynamischen Freiheitsgrade erh"alt man
f"ur die vektoriellen bzw. rechtsh"andigen Potentiale in \Ref{1_y0}
die effektiven Eichgruppen $U(1) \times SU(3)$ bzw. $SU(2)$.
Der fermionische Projektor zerf"allt auf dem Spinorraum
in vier $(8 \times 8)$-Bl"ocke, wodurch die urspr"ungliche Symmetrie
zwischen den massiven Fermionbl"ocken zerst"ort wird.
Wir schreiben symbolisch
\Equ{1_74}
P \;=\; \left[ \begin{array}{ccc} u & u & u \\ d & d & d \end{array} \right.
	\left| \begin{array}{c} \nu \\ e \end{array} \right] \spc ,
\EndEqu
dabei entspricht jede Spalte einem $(8 \times 8)$-Block in \Ref{1_70}.
Die effektiven Eichpotentiale koppeln genau wie im Standardmodell
an die Fermionen an. Insbesondere k"onnen wir die relativen elektrischen
Ladungen der Quarks und Leptonen zu $\frac{2}{3}, -\frac{1}{3}$ bzw $-1$
berechnen; die Neutrinos koppeln nicht an das elektromagnetische
Feld an.
\item Die $SU(2)$-Eichgruppe ist in der auf Seite \pageref{spontan}
beschriebenen Weise spontan gebrochen. Die zugeh"origen Eichbosonen
haben eine Ruhemasse, die mit den Fermionmassen $m^{(i)}$ ausgedr"uckt werden
kann; das in \Ref{1_74} diagonale Eichboson
wird mit dem elektromagnetischen gemischt, wobei sich auch seine Masse
"andert (das mu"s ich noch genau ausrechnen \ldots).
\item Die Kopplungskonstanten k"onnen berechnet werden, man erh"alt \ldots
	(das mu"s ich auch noch genau ausrechnen).
\item Das Gravitationsfeld tritt auf sinnvolle Weise auf, so wie das auf
Seite \pageref{gravi} beschrieben ist.
\item Als Unterschied zum Standardmodell erh"alt man eine
sogenannte Eichbedingung zwischen den Potentialen der $W$- und $Z$-Bosonen.
\end{enumerate}
Es ist nicht klar, ob und, wenn ja, wie die Wirkung \Ref{wirk} unmittelbar
physikalisch interpretiert werden kann.
Die Wirkung folgt in dem Sinne zwangsl"aufig, da"s andere Wirkungen
"ahnlicher Form im Kontinuumslimes nicht auf sinnvolle Gleichungen f"uhren.
Die im Moment eher spekulative Frage, ob unser Variationsprinzip als
``fundamental'' anzusehen ist oder es sich beispielsweise durch
Entwicklung aus einer anderen, einfacheren Wirkung ergibt, wollen wir hier
nicht diskutieren.

In die Euler-Lagrange-Gleichungen geht besonders f"ur die
Reduktion der dynamischen Eichfreiheitsgrade entscheidend ein, da"s die
Neutrinos nur in einer H"andigkeit vorkommen und masselos sind.
Es ist f"ur das Variationsprinzip auch wichtig, da"s die Massen der
massiven Fermionen in jedem Block gleich sind.

Diese Massenbedingung f"ur die Fermionen scheint auf den ersten Blick
unphysikalisch zu sein, auch stimmen die berechneten Werte f"ur die
Kopplungskonstanten nicht mit den physikalischen Werten "uberein. Dazu mu"s man
allgemein beachten,
da"s wir hier mit den nackten Massen und Kopplungskonstanten arbeiten,
die aufgrund der Selbstwechselwirkung nicht mit den effektiven Konstanten
"ubereinstimmen. Darauf werden wir im n"achsten Abschnitt etwas genauer
zur"uckkommen.

\section{\ldots und die Feldquantisierung?}
\label{1_ab5}
Wir wollen uns noch einmal das Ergebnis der bisherigen Konstruktionen
klarmachen: Nach dem Prinzip des fermionischen Projektors mu"s das
physikalische System mit dem fermionischen Projektor $P$ in der diskreten
Raumzeit formuliert werden. Die physikalische Wechselwirkung wird durch
die Gleichungen der diskreten Raumzeit beschrieben, f"ur die wir ein
Variationsprinzip ansetzen.
Leider haben wir f"ur die Gleichungen der diskreten Raumzeit kein
anschauliches Verst"andnis, was vor allem daran liegt, da"s wir darin
nicht mit den gewohnten physikalischen Begriffen und mathematischen
Objekten arbeiten k"onnen.
Aus diesem Grund haben wir als speziellen Grenzfall den Kontinuumslimes
untersucht: Wir erhalten die lokale und kausale
Struktur einer Lorentzmannigfaltigkeit. Analog zu \Ref{1_d3} k"onnen
wir die fermionischen Wellenfunktionen vom Diracsee abspalten, welcher bei
der Kontinuumsbeschreibung nicht mehr auftritt.
Mit der asymptotischen Entwicklung und der Planckn"aherung
k"onnen wir die durch die Euler-Lagrange-Gleichungen beschriebene
Wechselwirkung in einer f"ur uns vertrauten Form als Wechselwirkung
zwischen Fermionen und Eichfeldern umschreiben.
Wir haben au"serdem gesehen, da"s unsere Behandlung der Fermionen zum
Fockraum-Formalismus physikalisch "aquivalent ist. Die Fermionen werden also
in zweiter Quantisierung beschrieben, die Eichfelder dagegen als
klassische Felder.

Eine M"oglichkeit f"ur unser weiteres Vorgehen w"urde darin bestehen,
die im Kontinuumslimes erhaltenen klassischen Felder auf
die gewohnte Weise zu quantisieren (z.B. mit Pfadintegralen). Das halten
wir aber nicht f"ur besonders sinnvoll, weil dann die Feldquantisierung als
nachtr"aglich eingef"uhrter Effekt nur unbefriedigend begr"undet w"are.
Wir wollen anstatt dessen "uberlegen, ob sich mit dem Prinzip des
fermionischen Projektors auch die Feldquantisierung verstehen l"a"st.

Als Vorbereitung wollen wir zun"achst herausarbeiten, wozu quantisierte
Bosefelder eigentlich ben"otigt werden, also was die ``Quantisierung'' dieser
Felder genau physikalisch ausmacht.
Dazu werden wir untersuchen, inwieweit man bereits mit
den klassischen Eichfeldern einen Bezug zur Quantenfeldtheorie herstellen
kann. Wir beschr"anken uns zur Einfachheit auf eine Teilchensorte
und die elektromagnetische Wechselwirkung, die "Uberlegungen lassen sich
aber unmittelbar auf den allgemeinen Fall (auch mit Gravitation)
"ubertragen. Bei der Beschreibung eines wechselwirkenden Fermions
erh"alt man im Kontinuumslimes das gekoppelte System
von Differentialgleichungen
\Equ{1_73}
(i \Pdd + e \Aslsh - m) \: \Psi \;=\; 0 \;\;\;,\spc F^{ij}_{\;\;,j} \;=\;
	e \: \overline{\Psi} \gamma^i \Psi \spc .
\EndEqu
Diese Gleichungen verlieren ihre G"ultigkeit, wenn
man zu Energien in der Gr"o"senordnung der Planck-Energie "ubergeht,
weil dann die Planckn"aherung nicht mehr g"ultig ist.
Die Euler-Lagrange-Gleichungen sollten unser System dann zwar immer
noch beschreiben, wir k"onnen "uber die Form der Wechselwirkung (zur Zeit)
aber keine Aussagen machen. Zur Einfachheit nehmen wir im folgenden an, da"s
die Fermionen bei so hohen Energien nicht mehr wechselwirken.
Auf diese Weise erhalten wir in den klassischen Maxwellgleichungen einen
nat"urlichen Cutoff f"ur sehr hohe Impulse.

Bei der perturbativen Beschreibung der Wechselwirkung \Ref{1_73} erh"alt
man Feynman-Graphen. \label{feynman} \index{Feynman-Graphen}
Dazu gehen wir genau vor wie in \cite{bjorken}:
Man entwickelt $\Psi, A$ nach Potenzen von $e$
\[ \Psi \;=\; \sum_{j=0}^\infty e^j \: \Psi^{(j)} \;\;\;,\spc
	A \;=\; \sum_{j=0}^\infty e^j \: A^{(j)} \]
und setzt in die Differentialgleichungen \Ref{1_73} ein.
In diesen Gleichungen m"ussen die Terme jeder Ordnung in $e$ verschwinden,
man l"ost jeweils nach dem h"ochsten auftretenden Index $^{(j)}$ auf.
In Lo\-rentz\-ei\-chung erh"alt man so die formalen Relationen
\Equ{1_65a}
\Psi^{(j)} = - \sum_{k+l=j-1} (i \Pdd -m)^{-1} \: \left(
	\Aslsh^{(k)} \: \Psi^{(l)} \right) \;,\;\;\;\;
	A^{(j)}_i = - \sum_{k+l=j-1}  \Box^{-1} \: \left( 
	\overline{\Psi}^{(k)} \gamma_i \Psi^{(l)} \right) \;\;\; ,
\EndEqu
die man duch iteratives Einsetzen in eine explizite Form bringen kann.
Unter Ber"ucksichtigung der Paarerzeugung erhalten wir weitere Graphen,
die geschlossene Fermionlinien enthalten, wegen des Pauli-Prinzips
mit den richtigen relativen Vorzeichen.
Auf diese Weise erh"alt man alle Feynman-Graphen.

Wir kommen zur Renormierung\index{Renormierung}.
Da wir alle Feynman-Graphen der Quantenfeldtheorie
erhalten, besteht der einzige Unterschied bei unserer Betrachtungsweise
in dem nat"urlichen Cutoff f"ur sehr gro"se Impulse. Damit verschwinden
alle UV-Divergenzen, die Abweichungen zwischen den nackten und effektiven
Massen und Kopplungskonstanten wird endlich.
Man kann (zumindest im Prinzip) die effektiven Konstanten durch die
nackten Konstanten ausdr"ucken, indem man alle Beitr"age der
Selbstwechselwirkung aufsummiert.
Wir k"onnen die Situation auch mit der Renormierungsgruppe beschreiben:
An Renormierungsgruppenrechnungen sieht man, da"s die effektiven Massen
und Kopplungskonstanten von der Energie abh"angen. Die effektiven
Konstanten etwa bei der Planck-Energie sind als unsere nackten Konstanten
anzusehen\footnote{Da wir die nackten Kopplungskonstanten bestimmen
und verschiedene Relationen zwischen den nackten Massen ableiten k"onnen,
sollten sich durch Berechnung der zugeh"origen effektiven Gr"o"sen
unsere Vorhersagen gut testen lassen.
Da diese Rechnung sehr aufwendig sind, konnten wir
uns damit noch nicht n"aher befassen. Unsere Situation ist
"ahnlich wie bei den GUTs, wo alle Kopplungskonstanten in der Lagrangedichte
"ubereinstimmen und erst durch
die Selbstwechselwirkung ihre physikalischen Werte annehmen.
Durch Vergleich mit diesen Rechnungen k"onnen wir qualitativ sagen, da"s
die Abweichung zwischen nackten und effektiven Konstanten gro"s
sein sollte.}.

Die Renormierbarkeit der effektiven Kontinuumstheorie ist f"ur uns
wichtig, damit die Selbstwechselwirkung nur durch eine "Anderung der
Massen und Kopplungskonstanten ausgedr"uckt werden kann. Sie ist
f"ur eine sinnvolle Theorie aber
nicht unbedingt notwendig; beispielsweise
ist die Renormierbarkeit von Graphen irrelevant, die (mit unserem Cutoff)
so klein sind, da"s wir sie ganz vernachl"assigen k"onnen.
Au"serdem m"ussen wir uns dar"uber
im Klaren sein, da"s die Einf"uhrung des Cutoffs eine N"aherung ist,
von der wir nicht wissen, ob sie tats"achlich sinnvoll ist.
Um die Selbstwechselwirkung bei hohen Impulsen zu verstehen, m"u"ste
man die Euler-Lagrange-Gleichungen ohne die Planckn"aherung studieren.

Wir gehen hier genauer auf die Feynman-Regeln und die Renormierung
ein, um darauf hinzuweisen, da"s man die gesamte perturbative
Quantenfeldtheorie bereits mit klassischen
Bosefeldern erh"alt, wenn man die gekoppelte Wechselwirkung zwischen dem
klassischen Feld und den Fermionen untersucht.
Mit zweiter Quantisierung der Eichfelder kann man die Feynman-Graphen
zwar mit dem Wick-Theorem auf "ubersichtlichere Weise ableiten; es ist aber
an dieser Stelle weder aus mathematischer noch aus physikalischer Sicht
notwendig, von klassischen zu quantisierten Bosefeldern
"uberzugehen.
Insbesondere sollte man sich klarmachen, da"s alle Pr"azisionstests
der QFT (z.B. Lamb-Shift, anomaler g-Faktor) in Wirklichkeit gar kein
Test f"ur die Feldquantisierung sind. Man braucht sich eine Photonlinie
im Feynman-Graphen nicht als ``Austausch eines virtuellen Photons''
vorzustellen; der Photonpropagator kann auch einfach als der Operator
$-\Box^{-1}$ in \Ref{1_65a} angesehen werden, der bei der St"orungsentwicklung
der gekoppelten Differentialgleichungen \Ref{1_73} auftritt.
Auch die Gleichung $E=\hbar \omega$, die in anschaulicher Vorstellung
die Energie ``eines'' Photons angibt, macht "uber die Quantisierung
des elektromagnetischen Feldes keine Aussage. Das sieht man folgenderma"sen:
In der Physik tritt der Begriff der Energie an zwei verschiedenen Stellen auf.
In der klassischen Feldtheorie erh"alt man die Energie als Erhaltungsgr"o"se
aus der Translationsinvarianz der Lagrangedichte. In der Quantentheorie
ist die Summe der Frequenzen der Wellenfunktionen und Potentiale bei
Wechselwirkungen
erhalten, weil in der St"orungsrechnung ebene Wellen unterschiedlicher
Wellenzahl aufeinander orthogonal stehen. Diese ``klassische'' und
``quantenmechanische'' Energie sind "uber die Gleichung $E=\hbar \omega$
miteinander verkn"upft. Die Planck-Konstante kann dabei ohne
Bezug auf das elektromagnetische Feld bestimmt werden (beispielsweise "uber
die Compton-Wellenl"ange des Elektrons).
Da die klassische und quantenmechanische Energie bei
Wechselwirkungen getrennt erhalten
bleiben, mu"s die Gleichung $E=\hbar \omega$ ganz allgemein gelten.
(Die klassische Energie, die von einer Photonlinie der Frequenz
$\omega$ "ubertragen wird, ist also wirklich $\hbar \omega$.)

Nach diesen "Uberlegungen gibt es nur wenige Effekte, welche die Quantisierung
des elektromagnetischen Feldes "uberpr"ufen. Genau gesagt sind das die
folgenden Beobachtungen:
\begin{enumerate}
\item die Planck-Strahlung
\item der Casimir-Effekt
\item der Welle-Teilchen-Dualismus beim elektromagnetischen Feld, also
beispielsweise das Doppelspalt-Experiment
\end{enumerate}
F"ur die Ableitung des Planckschen Strahlungsgesetzes verwendet man, da"s
die Energie einer elektromagnetischen Wellenmode nicht
kontinuierliche, sondern nur in Stufen von $\hbar \omega$ ``quantisierte''
Werte annehmen kann.
Beim Casimir-Effekt mi"st man die Nullpunktsenergie der elektromagnetischen
Wellenmoden.

Um die Feldquantisierung zu verstehen, mu"s eine befriedigende
Erkl"arung f"ur die Beobachtungen 1.-3. gefunden werden.
Der Formalismus der Quantenfeldtheorie folgt aus diesen Beobachtungen
noch nicht. Bei der kanonischen Quantisierung nimmt man beispielsweise
an, da"s jede Wellenmode als quantenmechanischer harmonischer Oszillator
beschrieben werden kann. Das ist zwar plausibel, aber keine
zwingende Konsequenz aus der Diskretheit der Energiezust"ande.

Der Autor ist der Meinung, da"s diese Beobachtungen alle mit den
Euler-Lagrange-Gleichungen erkl"art werden k"onnen, wenn man Effekte
ber"ucksichtigt, die "uber die Planck\-n"a\-he\-rung hinausgehen.
Leider ist diese Vorstellung noch nicht mathematisch
ausgearbeitet. Trotzdem soll die Idee ausf"uhrlich
beschrieben werden, weil dieser Punkt die urspr"ungliche Motivation f"ur die
vorliegende Arbeit war.
Wir verlassen also an dieser Stelle den durch Rechnungen gut abgesicherten
Bereich und wollen in einem ersten Versuch vorschlagen, wie man die
Feldquantisierung und den Welle-Teilchen-Dualismus mit dem Prinzip des
fermionischen Projektors m"oglicherweise verstehen kann:
\\[1em]
Wir werden die Unterschiede zwischen der physikalischen Beschreibung in
der diskreten Raumzeit und der Kontinuumsn"aherung an verschiedenen
Beispielen herausarbeiten.
Es wird dabei gen"ugen, in der diskreten Raumzeit mit den klassischen
Begriffen zu arbeiten: eine elektromagnetische Welle in der diskreten
Raumzeit ist beispielsweise eine
Variation des fermionischen Projektors, die sich im Kontinuumslimes
mit einer St"orung des Diracoperators durch eine elektromagnetische
Welle ausdr"ucken l"a"st.

Wir beginnen mit einem einfachen Modell in der diskreten Raumzeit,
n"amlich einem vollst"andig gef"ullten Diracsee und einem elektromagnetischen Feld
in Form einer angeregten Wellenmode.
Wir wollen untersuchen, wie sich eine "Anderung der Amplitude der
elektromagnetischen Welle auswirkt.
Im Kontinuumslimes k"onnen wir die Amplitude beliebig
w"ahlen, denn die Maxwell-Gleichungen sind in jedem Fall erf"ullt.
Betrachtet man die Gleichungen der diskreten Raumzeit jedoch exakt, so
ist die Situation schwieriger: Die "Anderung der Amplitude wird auch jetzt
durch eine Variation von $P$ beschrieben.
Bei der St"orungsrechnung m"ussen wir aber in der diskreten Raumzeit
verschiedene Beitr"age mit ber"ucksichtigen, die wir im
Kontinuumslimes weglassen konnten. Diese zus"atzlichen Beitr"age
fallen in den Euler-Lagrange-Gleichungen nicht weg.
Wenn die Gleichungen f"ur einen Projektor $P$ erf"ullt sind,
k"onnen wir also
nicht erwarten, da"s sie auch dann noch gelten, wenn die Amplitude
der elektromagnetischen Welle ver"andert wird.
Allgemeiner ausgedr"uckt scheint es in der diskreten Raumzeit keine
stetige Schar $P(\tau)$ von L"osungen der Euler-Lagrange-Gleichungen zu
geben. Damit kann insbesondere die Amplitude der elektromagnetischen
Welle nur diskrete Werte annehmen.

Etwas anschaulicher kann man sich den Unterschied zwischen dem Kontinuumslimes
und der Beschreibung in der diskreten Raumzeit mit dem Rang des Projektors $P$
klarmachen: In der diskreten Raumzeit ist $\Rg(P)$ eine nat"urliche
Zahl. Wenn wir zu verschiedenen Werten von $\Rg(P)$ einen Projektor
unserer Form als L"osung der Euler-Lagrange-Gleichungen konstruieren,
wird die Amplitude der zugeh"origen elektromagnetischen Welle
i.a. verschiedenen sein.
Wir wollen zur Einfachheit annehmen, da"s es zu jedem
$m=\Rg(P)$ (in einem gewissen, sinnvollen Bereich von $m \in \N$) genau
einen solchen Projektor $P_m$ mit Amplitude $A_m$ gibt
(die St"orungsrechnung scheint anzudeuten, da"s das tats"achlich der Fall ist,
siehe Seite \pageref{2_quant}).
Da wir $\Rg(P)$ f"ur unser System nicht kennen, k"onnen wir $m$ beliebig
w"ahlen. Dadurch kann die Amplitude alle Werte in der diskreten Menge
$\{A_m\}$ annehmen.
In der Kontinuumsn"aherung ist $P$ dagegen ein Operator von unendlichem
Rang.
Deswegen ist einsichtig, da"s wir nun keine Einschr"ankung f"ur die
Amplitude der elektromagnetischen Welle erhalten; die Amplitude kann
kontinuierlich variiert werden.

Wir sehen also, da"s in der diskreten Raumzeit auf nat"urliche
Weise eine ``Quantisierung'' der Amplitude einer
elektromagnetischen Wellenmode auftreten sollte.
Bevor wir einen Zusammenhang zur Planck-Strahlung und dem Casimir-Effekt
herstellen k"onnen, m"ussen wir die "Uberlegung noch verfeinern:
Es scheint unrealistisch zu sein, eine elektromagnetische Welle zu
betrachten, die "uber die ganze Raumzeit ausgedehnt ist.
Darum untersuchen wir nun eine Welle, die in einem vierdimensionalen
Kasten lokalisiert ist (z.B. mit festen Randbedingungen). Der Kasten habe
Kangenl"ange $L$ in raumartiger und $T$ in zeitartiger Richtung.
Die Amplitude der Welle sollte in diesem Fall auch nur diskrete Werte
$\{A_j\}$
annehmen k"onnen. Die Quantisierungsstufen h"angen aber jetzt von
der Gr"o"se des Kastens, insbesondere von $T$ ab.
Qualitativ kann man sich "uberlegen, da"s bei kleinerem $T$ die Amplitude
der elektromagnetischen Welle gr"o"ser sein mu"s, damit der Projektor $P$
in vergleichbarer Weise gest"ort wird. Das bedeutet, da"s die Quantisierungsstufen
immer feiner werden, je gr"o"ser wir $T$ w"ahlen.
"Uber die klassische Energiedichte des elektromagnetischen Feldes k"onnen
wir die Amplituden $\{A_j\}$ in Quantisierungsstufen f"ur die Feldenergie
der Welle umrechnen.
Physikalisch ausgedr"uckt wird in unserem System zu einem Zeitpunkt $t$
eine elektromagnetische Welle erzeugt und zu einem sp"ateren Zeitpunkt
$t+T$ wieder vernichtet.
Da nach unserer obigen "Uberlegung bei allgemeinen Wechselwirkungen
und damit insbesondere bei der Erzeugung der elektromagnetischen Welle
zur Zeit $t$ die
Gleichung $E=\hbar \omega$ gilt, mu"s die Feldenergie in Stufen
von $\hbar \omega$ ``quantisiert'' sein\footnote{Wir lassen zur Einfachheit alle Arten
von Energiefluktuationen weg. Die Annahme, da"s sich die Feldenergie
bei einer Wechselwirkung zur Zeit $t$ um ein Vielfaches von $\hbar \omega$
"andert, ist nur eine N"aherung, weil bei der Beschreibung der
Wechselwirkung durch Feynman-Graphen Energieerhaltung erst nach beliebig
langer Zeit gilt.}.
Auf der anderen Seite hatten wir gerade gesehen, da"s die Quantisierungsstufen
von $T$ abh"angen.
Damit unser Vorgehen nicht auf Widerspr"uche f"uhrt, m"ussen wir $T$
so w"ahlen, da"s die Quantisierungsstufen f"ur die Feldenergie gerade
$\hbar \omega$ betragen.

Damit erhalten wir eine auf den ersten Blick eigenartige Bedingung: Wenn
wir zu einem Zeitpunkt eine elektromagnetische Welle erzeugen, so
mu"s diese zu einem bestimmten sp"ateren Zeitpunkt wieder vernichtet werden.
Eine solche zus"atzliche Bedingung, die keine Entsprechung im
Kontinuumslimes hat, nennen wir {\em{nichtlokale
Quantenbedingung}}\index{nichtlokale Quantenbedingung}.
Wir haben sie unter der Annahme unserer ``Quantisierung''
der Amplitude aus den Gleichungen der Planckn"aherung
(klassische Feldgleichungen, Beschreibung der Wechselwirkung durch
Feynman-Graphen) abgeleitet. Da die Euler-Lagrange-Gleichungen im
Kontinuumslimes in die klassischen Gleichungen "ubergehen, sollte eine
L"osung der Euler-Lagrange-Gleichungen die nichtlokalen Quantenbedingungen
automatisch erf"ullen.

Nat"urlich ist die gerade abgeleitete Bedingung physikalisch nicht sinnvoll.
Unser System ist mit nur einer Wellenmode aber auch noch sehr stark
idealisiert. Bevor wir weitere Schl"usse ziehen, wollen wir daher die
Situation in realistischeren Modellen betrachten:
Bei einem System mit mehreren Wellenmoden k"onnen wir im Gegensatz zur
kanonischen Quantisierung die verschiedenen Moden nicht als voneinander
unabh"angig ansehen; die Variation der Amplitude einer Welle ver"andert
die Quantisierungsstufen aller anderen Wellenmoden.
Diese gegenseitige Beeinflussung der elektromagnetischen Wellen ist
nichtlokal. Eine elektromagnetische Welle ver"andert also auch die
Energieniveaus von Wellen, die sich in gro"ser r"aumlicher Entfernung
befinden\footnote{Das sieht man in der St"orungsrechnung daran, da"s
sich eine St"orung durch ein elektromagnetisches Potential in $P(x,y)$
auch f"ur gro"se (raumartige) Abst"ande auswirkt.
Die Nichtlokalit"at kann man, genau wie im Austauschpotential
beim Hartree-Fock-Ansatz, mit dem Pauli-Prinzip verstehen: W"ahlt man
eine Orthonormalbasis $\Psi_j$ von $P(H)$ und ver"andert die
Wellenfunktionen $\Psi(x)$ lokal ab, so sind die Vektoren $\Psi_j$
i.a. nicht mehr orthogonal. Um den gest"orten Projektor zu bilden,
m"ussen die $\Psi_j$ erneut orthonormiert werden, wodurch sich der
neue Projektor auch global von dem urspr"unglichen Projektor unterscheidet.}.
Noch komplizierter wird die Lage, wenn man zus"atzlich
Fermionen in das System einbringt, weil die elektromagnetischen
Str"ome ebenfalls die Lage der Energieniveaus beeinflussen.

Die Komplexit"at dieser Situation hat zwei Konsequenzen: Zun"achst einmal
k"onnen wir "uber die genaue Lage der Energieniveaus praktisch keine
Aussage mehr machen, wir wissen nur noch, da"s die Quantisierungsstufen
$\hbar \omega$ betragen. Als Folge k"onnen wir die Energie des niedrigsten
Niveaus nur noch statistisch beschreiben. Wir nehmen zur
Einfachheit an, da"s sie in dem Intervall $[0, \hbar \omega )$
gleichm"a"sig verteilt ist. Dann erh"alt man f"ur die m"oglichen
Energiezust"ande jeder Wellenmode im Mittel die Werte
$(\frac{1}{2}+n)\: \hbar \omega$.
Als weitere Konsequenz sind die nichtlokalen Quantenbedingungen
jetzt so kompliziert, da"s wir sie nicht mehr n"aher spezifizieren k"onnen.
Es scheint aber durchaus m"oglich, da"s sie nun auch in einer physikalisch
realistischen Situation erf"ullt werden k"onnen. Wir haben die Vorstellung,
da"s durch die nichtlokalen Quantenbedingungen all das festgelegt
wird, was bei der statistischen Interpretation der Quantenmechanik als
``nicht determiniert'' oder ``zuf"allig'' gilt. Darauf werden wir bei der Diskussion
des Welle-Teilchen-Dualismus gleich genauer zur"uckkommen.

Nach diesen "Uberlegungen k"onnen wir die Beobachtungen 1. und 2.
erkl"aren: Da die Energie jeder Wellenmode in Stufen von $\hbar \omega$
``quantisiert'' ist, folgt das Plancksche Strahlungsgesetz;
aus der mittleren Energie $\frac{1}{2} \hbar \omega$ des ``Grundzustandes''
erh"alt man den Casimir-Effekt.
Wir sehen also, da"s wir unter unserer Annahme der ``Quantisierung'' der
Amplitude der elektromagnetischen Welle zu den gleichen Ergebnissen wie
mit kanonischer Quantisierung kommen. Der Grund liegt
darin, da"s wir mit den Feynman-Graphen und der Gleichung $E=\hbar \omega$
schon alle Formeln f"ur die quantitative Beschreibung zur Verf"ugung
haben und deswegen mit einer sehr allgemeinen Diskretheit der
Energiezust"ande auskommen.

Damit kommen wir zum Welle-Teilchen-Dualismus. Weil es sich dabei um
einen grundlegenden Effekt in der Quantenmechanik handelt, der bei Bosonen
und Fermionen in gleicher Weise auftritt, soll dieser Punkt etwas genauer
disktutiert werden. Zun"achst wollen wir unsere
Vorstellung der Quantisierung von Bose- und Fermifeldern vergleichen.
Es f"allt auf, da"s wir Bosonen und Fermionen ganz verschieden
beschreiben: die Wellenfunktionen der Fermionen spannen das Bild
des Projektors $P$ auf; die Bosonen entsprechen dagegen diskreten
Anregungsniveaus der klassischen Bosefelder, so wie das gerade
beschrieben wurde.
Der Fockraum oder ein "aquivalenter Formalismus tritt bei dieser
Beschreibung nicht auf.
Es mag unbefriedigend erscheinen, da"s dadurch die Analogie der
Quantenfeldtheorie in der Beschreibung von Bosonen und Fermionen, n"amlich
die blo"se Ersetzung von Kommutatoren durch Antikommutatoren, verloren geht.
Wir weisen darauf hin, da"s sich die \underline{elementaren}
Fermionen und Bosonen au"ser in ihrer Statistik noch in einem weitereren
wesentlichen Punkt voneinander unterscheiden. F"ur die Fermionen (Leptonen,
Quarks) hat man n"amlich einen Erhaltungssatz (Leptonenzahl, Baryonenzahl),
f"ur die Eichbosonen dagegen nicht.
Dieser Unterschied wird bei unserer Beschreibung ber"ucksichtigt: Jedes
Fermion entspricht einem Vektor in $P(H)$. Wir k"onnen Fermionen ineinander
umwandeln und "uber Wechselwirkung mit dem Diracsee in Paaren erzeugen
oder vernichten. Wir k"onnen aber die Gr"o"se $\Rg(P)$ bei Wechselwirkungen
nicht ver"andern, also beispielsweise nicht ein einzelnes Fermion vernichten.
Da die Eichbosonen lediglich diskreten Werten der Bosefelder entsprechen,
k"onnen sie durch Wechselwirkungen beliebig erzeugt und vernichtet werden,
sofern der Energie- und Impulssatz dabei erf"ullt sind.

Um den Zusammenhang zum Fockraum\index{Fockraum-Formalismus,
allgemeiner Zusammenhang} zu verdeutlichen, wollen wir kurz
untersuchen, wie wir zusammengesetzte Teilchen (z.B. Mesonen, Baryonen)
beschreiben. Sie sind alle aus den elementaren Fermionen
aufgebaut. Damit entspricht ein aus $p$ Komponenten zusammengesetztes
Teilchen einem Vektor aus $(P(H))^p$. Diese Darstellung ist f"ur
praktische Anwendungen aber ungeeignet. Es ist g"unstiger, f"ur die
elementaren Fermionen den Fockraum-Formalismus zu verwenden.
Dann erh"alt man als Erzeugungs-/Vernichtungsoperator f"ur das
zusammengesetzte Teilchen ein Produkt von $p$ fermionischen
Erzeugungs-/Vernichtungsoperatoren.
Falls $p$ gerade (ungerade) ist, k"onnen wir mit diesem Erzeugungsoperator
aus dem Vakuum einen bosonischen (fermionischen) Fockraum aufbauen.
Auf diese Weise erh"alt man bei zusammengesetzten Teilchen den
gewohnten Formalismus.
Man beachte jedoch, da"s dieser Formalismus bei uns keine grundlegende
Bedeutung besitzt.

Wegen unserer unterschiedlichen Behandlung der elementaren Fermionen
und Bosonen m"ussen wir f"ur den Welle-Teilchen-Dualismus
eine Erkl"arung finden, die von der speziellen
Beschreibungsweise dieser Teilchen unabh"angig ist.
Nach dem Prinzip des fermionischen Projektors m"ussen
alle physikalischen Objekte aus $P$ ableitbar sein.
F"ur ein Fermion ist das ein Vektor $\Psi \in H$, f"ur die Bosonen
die Eichfelder.
Damit ist das physikalische Objekt bei uns nicht das
punktf"ormige Teilchen, sondern die Welle selbst. Das scheint auf den
ersten Blick nicht sinnvoll zu sein, weil wir den Teilchencharakter
gar nicht ber"ucksichtigt haben.
Nach unserer Vorstellung kommt der Teilchencharakter lediglich durch eine
``Diskretheit'' der durch die Euler-Lagrange-Gleichungen beschriebenen
Wechselwirkung zustande.

Um zu pr"azisieren, was mit ``Diskretheit'' der Wechselwirkung gemeint ist,
wollen wir das Doppelspaltexperiment diskutieren.
Wir arbeiten mit einem Elektron, die "Uberlegung l"a"st sich aber
f"ur ein Photon direkt "ubertragen, wenn man die Wellenfunktion des Elektrons
durch das elektrische Feld ersetzt.
Wir lenken also ein Elektron "uber einen Doppelspalt auf einen
fotographischen Schirm. Beim Auftreffen auf dem Schirm tritt das Elektron
mit den Silberatomen des Films in Wechselwirkung, wodurch der Film
belichtet wird.
Im Kontinuumslimes erhalten wir die gleiche Situation wie in der
Wellenmechanik: die von beiden Spalten ausgehenden Zylinderwellen
"uberlagern sich und erzeugen auf dem Schirm ein Interferenzmuster.

"Ahnlich wie bei unserer Diskussion der elektromagnetischen Wellenmode
sollte die Kontinuumsn"aherung die physikalische Situation auch hier nur
grob beschreiben, bei exakter Betrachtung der Euler-Lagrange-Gleichungen in
der diskreten Raumzeit wird die Situation wesentlich komplizierter.
Wir wollen annehmen, da"s die durch die Euler-Lagrange-Gleichungen
beschriebene Wechselwirkung in dem Sinne ``diskret'' ist, da"s das
Elektron bevorzugt nur mit einem Silberatom des Schirms wechselwirkt.
Diese Annahme l"a"st sich schon im Kontinuumslimes plausibel machen:
Bei der Wechselwirkung des Elektrons mit dem Silberatom mu"s ein Elektron
des Atoms angeregt werden. Weil dazu eine gewisse Mindestenergie
ben"otigt wird, kann das auftreffende Elektron mit seiner kinetischen
Energie nur eine bestimmte (kleine) Anzahl von Atomen anregen.
Damit kann die Wechselwirkung zwischen Elektron und Schirm nur an
einzelnen Silberatomen stattfinden; es ist nicht m"oglich, die
kinetische Energie durch elektrische Anregung kontinuierlich auf den
Schirm zu "ubertragen.

Unter dieser Annahme erhalten wir auf dem Schirm einen belichteten Punkt,
so da"s der Eindruck eines punktf"ormigen Teilchens entsteht.
An welcher Stelle des Schirms das Elektron wechselwirkt, wird durch
die genaue Form des Projektors $P$ in der diskreten Raumzeit oder,
mit der oben eingef"uhrten Sprechweise, durch nichtlokale Quantenbedingungen
festgelegt.
Dabei wirkt sich die Nichtlokalit"at und Nichtkausalit"at der
Euler-Lagrange-Gleichungen aus.
Weil die nichtlokalen Quantenbedingungen so kompliziert sind,
k"onnen wir nicht vorhersagen, an welcher Stelle des Schirms das
Elektron wechselwirkt.
Selbst wenn wir das Experiment unter scheinbar gleichen "au"seren
Bedingungen wiederholen, wird die globale physikalische Situation
unterschiedlich sein. Damit k"onnen die nichtlokalen Quantenbedingungen
ganz verschieden sein, so da"s auch das Experiment ein anderes
Ergebnis liefert.
Aus diesem Grund k"onnen wir "uber den Ausgang des Experiments nur
statistische Aussagen machen.
Aus dem bekannten Kontinuumslimes der Euler-Lagrange-Gleichungen folgt, da"s
die Wahrscheinlichkeitsdichte durch $|\Psi|^2$ gegeben ist.

Wir vergleichen die erhaltene Situation mit der statistischen Deutung
der Quantenmechanik.
Wir kommen letztlich zum gleichen Ergebnis: auf dem
Schirm trifft ein punktf"ormiges Teilchen auf, f"ur den genauen Ort
k"onnen wir nur die Wahrscheinlichkeit angeben.
Wir begr"unden diese Beobachtungen aber ganz anders: das punktf"ormige
Teilchen mit der gerade beschriebenen ``Diskretheit'' der Wechselwirkung,
den fehlenden Determinismus mit der Nichtlokalit"at der
Euler-Lagrange-Gleichungen.
Als Konsequenz spielen der Me"sproze"s und der Beobachter bei uns
keine zentrale Rolle. Die Wellenfunktion gibt nicht nur den aktuellen
Wissensstand eines Beobachters an, sondern ist als das eigentliche
physikalische Objekt anzusehen. Bei einer Messung mu"s man nicht
zu einer anderen Wellenfunktion "ubergehen, weil der Beobachter neue
Informationen "uber das System erh"alt, sondern weil das System durch die
beim Me"sproze"s stattfindende Wechselwirkung ver"andert wird.

Damit wollen wir die "Uberlegungen zur Feldquantisierung und der
Interpretation der Quantenmechanik abschlie"sen.
Nat"urlich ist unsere Beschreibung der
Feldquantisierung und des Welle-Teilchen-Dualismus im Moment nicht mehr
als ein Deutungsversuch, der dem Leser einleuchten oder bei ihm auf
Ablehnung sto"sen kann.
Wir weisen aber darauf hin, da"s wir mit den Euler-Lagrange-Gleichungen
die mathematischen Mittel zur Verf"ugung haben, um unsere Annahmen
(diskrete Energieniveaus bei Wellenmoden, ``diskrete'' Wechselwirkung)
zu verifizieren und unsere Vorstellung gegebenenfalls zu pr"azisieren.
Wir haben uns damit bisher noch nicht besch"aftigt und uns
ganz auf den Kontinuumslimes konzentriert, weil wir vor Rechnungen
zur Feldquantisierung einen sauberen
Kontakt zur klassischen Theorie herstellen wollten.
Au"serdem konnten wir aus dem Kontinuumslimes konkretere Ergebnisse und damit
bessere Hinweise darauf erwarten, ob das Prinzip des fermionischen
Projektors physikalisch sinnvoll ist.

\chapter{Der fermionische Projektor im Kontinuum}
\label{kap2}
Gem"a"s den "Uberlegungen in der Einleitung wollen wir ein physikalisches
System in der diskreten Raumzeit durch den Projektor $P$ auf die besetzten
Fermionzust"ande beschreiben. Alle physikalischen Gleichungen sollen
unmittelbar mit $P$ und den Spektralprojektoren $E_x$ der diskreten
Raumzeit-Punkte $x \in M$ formuliert werden. Bevor wir dieses Programm
durchf"uhren k"onnen, m"ussen wir den fermionischen Projektor als
mathematisches Objekt einf"uhren und m"oglichst allgemein und explizit
untersuchen.

Es ist technisch einfacher, im Kontinuum $M=\R^4$ zu arbeiten: Wie in
der Einleitung beschrieben, besitzt der Operator
\[ P(x,y) \;\equiv\; E_x \:P\: E_y \]
als Distribution einen sinnvollen Kontinuumslimes. In diesem
Kapitel werden wir diese Distribution $P(x,y)$ studieren. Im
n"achsten Kapitel \ref{kap3} werden wir dann die Vorstellung einer diskreten
Raumzeit durch Regularisierung von $P(x,y)$ auf der L"angenskala der
Planck-L"ange umsetzen. Die genaue Regularisierungsvorschrift darf in
unsere Ergebnisse letztlich nicht eingehen.

\section{Der freie fermionische Projektor}
In diesem Abschnitt wollen wir den fermionischen Projektor $P(x,y)$ des
Vakuums konstruieren, den wir auch den {\em{freien fermionischen
Projektor}}\index{fermionischer Projektor, freier}
nennen. Die verschiedenen Fermionsorten sollen jeweils durch Diracseen
der Form
\Equ{2_a1}
\int \frac{d^4k}{(2 \pi)^4} \; (k\slsh + m) \: \delta(k^2 - m^2) \:
	\Theta(-k^0) \; e^{-i k (x-y)}
\EndEqu
beschrieben werden; der freie fermionische Projektor mu"s auf geeignete
Weise aus solchen Diracseen zusammengesetzt werden.

\subsection{Spektralzerlegung des freien Diracoperators}
Da \Ref{2_a1} aus Eigenzust"anden des freien Diracoperators $i \Pdd$
besteht, werden wir zun"achst dessen Spektralzerlegung etwas allgemeiner
untersuchen.
Aus mathematischer Sicht w"are ein Spektralsatz der Form
\Equ{2_b0}
i \Pdd \;=\; \int_\sigma m \: dp_m
\EndEqu
mit Spektralma"s $dp$ w"unschenswert, dabei bezeichnet $\sigma=\R \cup i \R$
das Spektrum des Diracoperators. Gleichung \Ref{2_b0} kann leicht
hergeleitet werden, indem man das Spektralma"s explizit aus den
Ebenen-Wellen-L"osungen der freien Diracgleichung konstruiert.
Bei einer "Ubertragung des Spektralsatzes auf den Fall mit Wechselwirkung
(also beispielsweise f"ur den Diracoperator $i \Pdd + e \Aslsh$ im
"au"seren elektromagnetischen Feld) treten aber Probleme auf.
Das liegt daran, da"s das Skalarprodukt von $H$ indefinit ist.
Die grundlegende Schwierigkeit sieht man schon bei endlicher Dimension:
ein hermitescher Operator von endlichem Rang
ist i.a. nicht diagonalisierbar, wenn Null-Eigenvektoren
(also Eigenvektoren $u$ mit $\bra u,u \ket=0$) auftreten.

Wegen dieser mathematischen Probleme behandeln wir auch den freien
Diracoperator vereinfacht: wir beschreiben die ``Eigenr"aume'' von
$i \Pdd$ durch Distributionen $p_m$, $k_m$ und leiten f"ur diese
Distributionen formale Rechenregeln ab.  Der Formalismus hat "Ahnlichkeit
mit der Diracnotation \Ref{1_2}, \Ref{1_3}, wenn man die Ortskoordinaten
durch die Variable $m$ ersetzt. Auf dieser Ebene wird sich sp"ater
auch der Fall mit Wechselwirkung befriedigend beschreiben lassen.

Es ist g"unstig, sowohl im Orts- als auch im Impulsraum zu arbeiten, dabei
bezeichnen wir wie "ublich die Impuls- und die Ortskoordinaten mit
$k$ bzw. $x$.
\begin{Def}
\label{2_def1}
Wir definieren f"ur $a \in \R$, $m \in \R\cup i \R$ die temperierten
Distributionen
\begin{eqnarray}
\label{eq:2_1}
P_a (k) &=& \delta(k^2-a) \\
\label{eq:2_3}
p_m (k) &=& \frac{|m|}{m} \: (k \slsh + m) \; \delta(k^2-m^2)
\end{eqnarray}
und f"ur $a \in \R^+$, $m \in \R$
\begin{eqnarray}
\label{eq:2_1a}
K_a (k) &=& \delta(k^2-a) \: \epsilon(k^0) \\
\label{eq:2_3a}
k_m (k) &=& \frac{|m|}{m} \: (k \slsh + m) \; \delta(k^2-m^2)
	\: \epsilon(k^0) \spc .
\end{eqnarray}
Wir fassen diese Distributionen auch als Multiplikationsoperatoren im
Impulsraum auf.
\end{Def}
F"ur $m=0$ ist die Definitionsgleichung \Ref{2_3}, \Ref{2_3a} nicht
eindeutig. In solchen F"allen bilden wir stets den Grenzwert
$0<m \rightarrow 0$, also
\[ p_0(k) \;:=\; k \slsh \: \delta(k^2) \;\;\;\;,\spc
	k_0(k) \;:=\; k\slsh \: \delta(k^2) \: \epsilon(k^0) \spc . \]

Da die Multiplikation bei Fouriertransformation in die Faltung "ubergeht,
sind im Ortsraum die Distributionen die Integralkerne
der zugeh"origen Operatoren, also beispielsweise
\begin{eqnarray*}
(p_m \: \Psi)(x) &=& \int d^4y \; p_m(x,y) \; \Psi(y)
	\;=\; \int d^4y \; p_m(x-y) \; \Psi(y) \spc .
\end{eqnarray*}
Die Distributionen $P_a, K_a$ und $p_m, k_m$ erf"ullen die
Klein-Gordon- bzw. Dirac-Gleichung
\begin{eqnarray}
(-\Box - a) \: P_a &=& (-\Box - a) \: K_a \;=\; 0  \\
\label{eq:2_a99}
(i \Pdd - m) \: p_m &=& \: (i \Pdd - m) \: k_m\: \;=\; 0 \spc .
\end{eqnarray}
Die L"osungen f"ur $a<0$, $m \in i \R$ sind
unphysikalisch; wir m"ussen sie aber trotzdem ber"ucksichtigen, um das
ganze Spektrum der Differentialoperatoren zu erfassen.

Wir wollen jetzt formale Rechenregeln f"ur Produkte der Operatoren
$P_a, K_a, p_m, k_m$ ableiten. Wenn ein Faktor der Form
$\delta(\alpha - \beta)$ auftritt, setzen wir dazu
in allen anderen Faktoren die Variablen $\alpha$ und $\beta$ gleich.
Auf diese Weise erhalten die formalen Distributionsprodukte
einen Sinn, denn die einzelnen Faktoren h"angen dann
von verschiedenen Variablen ab. Wir erhalten die Relationen
\begin{eqnarray}
\label{eq:2_5}
P_a \: P_b &=& \delta(k^2-a) \; \delta(k^2-b) \;=\; \delta(a-b) \; P_a \\
\label{eq:2_6}
P_a \: K_b &=& K_b \: P_a \;=\; \delta(k^2-a) \; \delta(k^2-b) \:
	\epsilon(k^0) \;=\; \delta(a-b) \; K_a \\
\label{eq:2_7}
K_a \: K_b &=& \delta(k^2-a) \: \epsilon(k^0) \; \delta(k^2-b) \:
	\epsilon(k^0) \;=\; \delta(a-b) \; P_a \\
p_m \: p_n &=& \frac{|mn|}{mn} \; (k\slsh+m)(k\slsh+n) \;
	\delta(k^2-m^2) \: \delta(k^2-n^2) \nonumber \\
&=& \delta(m^2-n^2) \; \frac{|mn|}{mn}
	\; (k^2 + (m+n) \: k\slsh + m n) \; \delta(k^2-n^2) \nonumber \\
&=& \frac{1}{2 m} \left( \delta(m-n) + \delta(m+n) \right) \; (m+n) \:
	\frac{|n|}{n} \; (k\slsh+n) \; \delta(k^2-n^2) \nonumber \\
\label{eq:2_11}
&=& \delta(m-n) \: p_m \\
\label{eq:2_12}
k_m \: p_n &=& p_n \: k_m \;=\; \delta(m-n) \: k_m \\
\label{eq:2_13}
k_m \: k_n &=& \delta(m-n) \: p_m \spc .
\end{eqnarray}
Au"serdem gelten die Vollst"andigkeitsrelationen
\begin{eqnarray}
\label{eq:2_30}
\inti P_a \: da &=& \inti \delta(k^2-a) \: da \;=\; \1  \\
\label{eq:2_31}
\int_{\sR \cup i \sR} p_m \: dm &=& \int_{\sR^+ \cup i \sR^+}
	\frac{|m|}{m} \;
	2m \: \delta(k^2-m^2) \: dm \nonumber \\
&=& \inti \delta(k^2-m^2) \: d(m^2) \;=\; \1
\end{eqnarray}
und die Spektrals"atze
\begin{eqnarray}
\label{eq:2_32}
\inti a \; P_a \: da &=& \inti a \; \delta(k^2-a) \: da \;=\;
	k^2 \;=\; -\Box_x  \\
\label{eq:2_33}
\int_{\sR \cup i \sR} m \; p_m \: dm &=& \int_{\sR^+ \cup i \sR^+} 2|m|
	\: k\slsh \; \delta(k^2-m^2) \: dm \;=\; k\slsh
	\;=\; i \Pdd_x \spc .
\end{eqnarray}
Wegen Gleichung \Ref{2_11} und \Ref{2_31}, \Ref{2_33} k"onnen wir
$p_m$ als die Spektralprojektoren des freien
Diracoperators auffassen. Die Distributionen $k_m$ unterscheiden sich
von $p_m$ durch ein relatives Minuszeichen f"ur die Zust"ande auf der
oberen bzw. unteren Massenschale. Wir bezeichnen die Ausdr"ucke
\Equ{2_88a}
\frac{1}{2}\:(p_m + k_m) \;\;\;,\spc \frac{1}{2}\:(p_m - k_m)
\EndEqu
manchmal als ``Projektoren'' auf die Eigenzust"ande positiver bzw. negativer
Energie, obwohl es sich dabei wegen der $\delta$-Normierung nat"urlich
mathematisch nicht um Projektoren handelt.

\subsection{Ansatz f"ur $P(x,y)$}
Wir wollen nun schrittweise den freien fermionischen Projektor aufbauen.
Eine massive Fermionsorte beschreiben wir durch einen Diracsee, also mit
der Notation \Ref{2_3}, \Ref{2_3a}
\Equ{2_a9}
P(x,y) \;=\; \frac{1}{2} \: (p_m - k_m)(x,y) \spc .
\EndEqu
In den einfachsten Systemen mit mehreren Fermionsorten zeigen alle
Fermionen die gleichen Wechselwirkungen. In diesem Fall addieren wir
die Diracseen, also bei $f \in \N$ Fermionsorten mit Massen
$m_a$, $a=1,\ldots, f$,
\Equ{2_a2}
P(x,y) \;=\; \sum_{a=1}^f \: \frac{1}{2} \: (p_{m_a} -
	k_{m_a})(x,y) \;\;\;\;,\spc m_a \neq m_b \;\;
	{\mbox{f"ur alle $a \neq b$}} \spc .
\EndEqu
Mit diesem fermionischen Projektor k"onnten beispielsweise die Leptonen
$e, \mu, \tau$ beschrieben werden. In Analogie zum Standardmodell nennen
wir die Fermionsorten in \Ref{2_a2} {\em{Familien}}\index{Familie} und den Index
$a$ {\em{Flavour-Index}}\index{Flavour-Index}.

In realistischen physikalischen Systemen gibt es Fermionsorten, die
auf unterschiedliche Weise wechselwirken (z.B. Quarks und Leptonen).
Darum scheint der Ansatz \Ref{2_a2} zu speziell und mu"s
verallgemeinert werden: Wir gehen zu Spindimension $4B$, $B \in \N$
"uber und w"ahlen f"ur $P(x,y)$ die direkte Summe von Projektoren der Form
\Ref{2_a2}
\Equ{2_a3}
P(x,y) \;=\; \bigoplus_{j=1}^B \left( \sum_{a=1}^f \: \frac{1}{2}
	\: (p_{m_{ja}} - k_{m_{ja}})(x,y) \right) \spc ,
\EndEqu
dabei ist $(m_{ja})$ eine Matrix
\[ (m_{ja})_{j=1,\ldots, B; \; a=1,\ldots, f} \spc
	{\mbox{mit}} \spc m_{ja} \neq m_{jb} \;\;
	{\mbox{f"ur alle $j$ und $a \neq b$}} \spc . \]
Wir nennen die einzelnen direkten Summanden in \Ref{2_a3} auch
{\em{Bl"ocke}}\index{Block} und den Index $j$
{\em{Block-Index}}\index{Block-Index}. F"ur $B=2$, $f=3$ erh"alt man ein
Modell f"ur die Isospinpartner $u, c, t \leftrightarrow d, s, b$.
F"ur $B=f=3$ und $m_{ia} = m_{ja} \; \forall i, j$ k"onnte
man die Quarks unter Ber"ucksichtigung der Colour-Freiheitsgrade
beschreiben.

F"ur ein realistisches physikalisches Modell fehlen noch Neutrinos, also
Femionen mit einer ausgezeichneten H"andigkeit. Damit die Lorentzkovarianz
gewahrt ist, m"ussen diese chiralen Fermionen masselos sein\footnote{Dieser Schlu"s h"angt damit
zusammen, da"s wir mit vierkomponentigen Diracspinoren arbeiten:
In der Diracgleichung $(i \Pdd - m) \Psi = 0$ sind die links- und
rechtsh"andige Komponente $\Psi_{L\!/\!R}:=\chi_{L\!/\!R} \Psi$
der Wellenfunktion miteinander gekoppelt
\[ 0 \;=\; \chi_{L\!/\!R} \: (i \Pdd - m ) \: \Psi \;=\;
	i \Pdd \:\chi_{R\!/\!L} \: \Psi \:-\: m \:\chi_{L\!/\!R} \: \Psi
	\;=\; i \Pdd \: \Psi_{R\!/\!L} - m \Psi_{L\!/\!R} \spc . \]
Nur f"ur $m=0$ sind $\Psi_{L\!/\!R}$ voneinander unabh"angig, so da"s
es Sinn macht, von chiralen Wellenfunktionen zu sprechen.

Verwendet man dagegen zweikomponentige Weyl-Spinoren, so l"a"st
sich einfach ein Massenparameter in die Weyl-Gleichung einf"ugen
\[ (i \sigma^j \partial_j - m) \: \Psi \;=\; 0 \spc . \]
Bei der Diskussion um eine m"ogliche Ruhemasse des $\mu$-Neutrinos
wird stets in der Weyl-Darstellung gearbeitet.
Wir bemerken, da"s die Weyl-Gleichung bei unserer Verkn"upfung
von Koordinaten- und Eichtransformationen nicht sinnvoll ist
(siehe auch \cite{Physdip}).}.
In Analogie zu \Ref{2_a9} beschreiben wir einen Diracsee chiraler Fermionen
durch den Ausdruck
\[ \chi_{L\!/\!R} \: \frac{1}{2} \: (p_0 - k_0)(x,y) \spc . \]
Wir verallgemeinern \Ref{2_a3} auf den Fall mit Neutrinos:
F"ur jeden Block $j$ definieren wir eine $(4 \times 4)$-Matrix $X_j$ mit
\[ X_j = \1 \;\;\;\;\;{\mbox{oder}}\;\;\;\;\; X_j = \chi_L
	\;\;\;\;\;{\mbox{oder}}\;\;\;\;\; X_j = \chi_R \spc . \]
Falls $X_j \neq \1$ ist, soll die Matrix $(m_{ja})$ verschwinden,
\Equ{2_a4}
X_j \neq \1 \spc {\mbox{impliziert}} \spc m_{ja} = 0 \;\;\;
	\forall a \spc .
\EndEqu
Wir f"ugen die chiralen Projektoren $X_j$ als Faktoren in \Ref{2_a3}
ein und erhalten
\Equ{2_a5}
P(x,y) \;=\; \bigoplus_{j=1}^B \;X_j\: \sum_{a=1}^f \: \frac{1}{2}
	\: (p_{m_{ja}} - k_{m_{ja}})(x,y) \spc .
\EndEqu
Dies ist unser allgemeiner Ansatz f"ur den freien fermionischen Projektor.
Wir nennen die direkten Summanden mit $X_j \neq \1$ auch
{\em{Neutrinobl"ocke}}\index{Neutrinoblock}.

\subsubsection*{kurze Diskussion des Ansatzes}
Unser Ansatz \Ref{2_a5} enth"alt einige spezielle Annahmen:
Zun"achst einmal tritt in allen Bl"ocken die gleiche Zahl von Familien
auf. Au"serdem haben wir ausgeschlossen, da"s ein Block sowohl
aus chiralen als auch aus massiven Fermionen aufgebaut ist.
Schlie"slich ist auch die Blockdiagonalit"at von \Ref{2_a5} eine
starke Bedingung f"ur den freien fermionischen Projektor.
Unser Ansatz sollte aber hinreichend allgemein sein, um die
Fermionen des Standardmodells beschreiben zu k"onnen.
Nat"urlich lie"se sich sp"ater untersuchen, inwieweit der Ansatz in dem
Sinne zwingend ist, da"s er bereits aus den Gleichungen der diskreten
Raumzeit folgt; wir werden solch allgemeine Fragen aber hier ausklammern.

Als andere m"ogliche Erweiterung von \Ref{2_a5} k"onnte man f"ur die
einzelnen Diracseen zus"atzliche Normierungskonstanten einf"uhren, also
\Equ{2_b2}
P(x,y) \;=\; \bigoplus_{j=1}^B \;X_j\: \sum_{a=1}^f \: c_{ja}
	\; \frac{1}{2}  \: (p_{m_{ja}} - k_{m_{ja}})(x,y)
	\spc {\mbox{mit $c_{ja} \in \R$}} \;\;\;\; .
\EndEqu
Diese Verallgemeinerung ist aber nicht sinnvoll:
In der diskreten Raumzeit soll $P$ ein Projektor sein. Im Kontinuum
$M=\R^4$ konnten wir dies nicht erreichen, weil die Eigenfunktionen
des Diracoperators
im $\R^4$ nicht normierbar sind; wir haben gem"a"s
\Ref{2_11} bis \Ref{2_13} eine $\delta$-Normierung verwendet.
Man kann aber auch im Minkowski-Raum mit Projektoren arbeiten,
indem man die Masse etwas ausschmiert. Genauer ersetzen wir die
Diracseen in \Ref{2_b2} gem"a"s
\Equ{2_b3}
\frac{1}{2} \: (p_m - k_m)(x,y) \;\longrightarrow\;
	\int_m^{m+\varepsilon} dm^\prime \; \frac{1}{2}
	\: (p_{m^\prime} - k_{m^\prime})(x,y)
\EndEqu
durch Integrale "uber den Massenparameter. Mit den Rechenregeln \Ref{2_11}
bis \Ref{2_13} kann man direkt "uberpr"ufen, da"s die rechte Seite von
\Ref{2_b3} ein Projektor ist.
Wir haben die Vorstellung, da"s \Ref{2_b3} n"aherungsweise die Situation
in der diskreten Raumzeit beschreibt. Nat"urlich ist diese Beschreibung
stark vereinfacht, sie ist aber f"ur unser Argument
ausreichend\footnote{\label{2_foot2}Im
endlichen Volumen l"a"st sich der Zusammenhang etwas genauer beschreiben:
Zur Einfachheit ersetzen wir den Minkowski-Raum durch den vierdimensionalen
Kasten $M \;=\; [0,L]^4$ mit periodischen Randbedingungen.
Dann d"urfen lediglich Impulse auf einem Gitter mit Gitterl"ange
$2 \pi / L$ auftreten. Die rechte Seite von \Ref{2_b3} geht in den Projektor
\Equ{2_b4}
P(x,y) \;=\; \left( \frac{2 \pi}{L} \right)^4 \;
	\sum_{k \in \frac{2 \pi}{L}\: \sZ^4} \: (k\slsh + |k|) \; \Theta(|k|-m) \:
	\Theta(m+\varepsilon-|k|) \; e^{-ik(x-y)}
\EndEqu
"uber ($|k|\equiv\sqrt{|k^2|}$), der aus diskreten Zust"anden aufgebaut ist.
Eine zus"atzliche Diskretisierung der Raumzeit auf einem Gitter liefert
einen Cutoff im Impulsraum, so da"s man einen Projektor von endlichem Rang
erh"alt.

Der Parameter $\varepsilon$ beschreibt die ``Breite'' des Diracsees;
f"ur eine sinnvolle Regularisierung sollte man
$\varepsilon \approx 2 \pi / L$ w"ahlen.}.
Da der Parameter $\varepsilon$ von der Geometrie der diskreten Raumzeit
abh"angt, mu"s er f"ur alle Diracseen gleich sein. Mit der N"aherung
\Equ{2_s3}
\int_m^{m+\varepsilon} dm^\prime \; \frac{1}{2}
	\: (p_{m^\prime} - k_{m^\prime})(x,y) \;\approx\;
	\varepsilon \; \frac{1}{2} \: (p_m - k_m)(x,y)
\EndEqu
folgt, da"s \Ref{2_b2} ein sinnvoller Grenzfall im unendlichen Volumen ist,
falls wir
\[ c_{ja} \;=\;  \varepsilon \spc {\mbox{f"ur alle $a, j$}} \]
w"ahlen. Dann stimmt \Ref{2_b2} bis auf einen f"ur uns unwichtigen
Vorfaktor $2 \varepsilon$ mit \Ref{2_a5} "uberein.

Die Beschreibung der Neutrinos scheint in \Ref{2_a5} auf den ersten
Blick problematisch zu sein. Die links- bzw. rechtsh"andigen
Neutrinobl"ocke haben die Form
\Equ{2_b5}
P(x,y) \;=\; \chi_{L\!/\!R} \: \sum_{a=1}^f \: \frac{1}{2} \:
	(p_0 - k_0)(x,y) \spc .
\EndEqu
Diese Distribution ist nilpotent,
\begin{eqnarray*}
P(x,z) \: P(z,y) &=& \chi_{L\!/\!R} \: \sum_{a=1}^f \: \frac{1}{2} \:
	(p_0 - k_0)(x,z) \; \chi_{L\!/\!R} \: \sum_{a=1}^f \:
	\frac{1}{2} \: (p_0 - k_0)(z,y) \\
&=& (\chi_{L\!/\!R} \: \chi_{R\!/\!L}) \: \sum_{a=1}^f \: \frac{1}{2} \:
	(p_0 - k_0)(x,z) \;
	\sum_{a=1}^f \: \frac{1}{2} \: (p_0 - k_0)(z,y) \;=\; 0
	\spc ,
\end{eqnarray*}
und geht folglich bei naiver Regularisierung (also z.B. im endlichen
Volumen mit Impuls-Cutoff) nicht in einen Projektor "uber.
Als weitere Schwierigkeit stimmen alle Summanden in \Ref{2_b5} "uberein,
so da"s die Neutrinofamilien im fermionischen Projektor nicht voneinander
unterschieden werden k"onnen.
Um einzusehen, da"s es sich dabei nur im unendlichen Volumen um
Probleme handelt, schmieren wir wie in \Ref{2_b3} die Massen auf der Skala
$\varepsilon$ aus und f"uhren au"serdem kleine Neutrinomassen $m_a$ mit
$|m_a - m_b|>\varepsilon \;\;\; \forall a \neq b$ ein,
also
\Equ{2_b6}
\Ref{2_b5} \;\longrightarrow\; \chi_{L\!/\!R} \: \sum_{a=1}^f \;
	\int_{m_a}^{m_a+\varepsilon} dm^\prime \;
	\frac{1}{2} \: (p_{m^\prime} - k_{m^\prime})(x,y) \spc .
\EndEqu
Nach dieser Ersetzung ist $P(x,y)$ nicht mehr nilpotent; die einzelnen
Diracseen sind im Impulsraum voneinander getrennt. Wegen des
Vorfaktors $\chi_{L\!/\!R}$ ist
\Ref{2_b6} nicht aus Eigenzust"anden des freien Diracoperators
aufgebaut. Das sollte aber keine Rolle spielen, falls $m_a, \varepsilon$
klein genug gew"ahlt werden, also insbesondere in der Gr"o"senordnung
\[ m_a, \varepsilon \;\approx\; ({\mbox{Ausdehnung der diskreten
	Raumzeit}})^{-1} \spc . \]
Wir bemerken, da"s die Unterscheidbarkeit der Neutrino-Flavours
auch aus experimenteller Sicht eine offene Frage ist.

\subsection{Die Asymmetriematrizen $X$, $Y$}
Bei der Konstruktion des freien fermionischen Projektors traten
drei Arten von Indizes auf, n"amlich der Dirac-Index
$\alpha, \beta=1,\ldots,4$, der Block-Index $j, k=1,\ldots,B$ und der
Flavour-Index $a, b=1,\ldots,f$. F"ur eine "ubersichtliche Notation ist
es g"unstig, diese Indizes zusammenzufassen. Dazu bilden wir das
Tensorprodukt $\C^4 \otimes \C^B \otimes \C^f$ der zugeh"origen
Vektorr"aume und verwenden auf dem Tensorprodukt eine Matrixschreibweise.
Insbesondere definieren wir die sogenannten
{\em{Asymmetriematrizen}}\index{Asymmetriematrix}
$X$, $Y$ durch
\Equ{2_b7}
X_{\alpha j a \: \beta k b} \;=\; (X_j)_{\alpha \beta} \: \delta_{jk} \:
	\delta_{ab} \;\;\;,\spc Y_{\alpha j a \: \beta k b} \;=\;
	\frac{1}{m} \: m_{ja} \: \delta_{\alpha \beta} \: \delta_{jk}
	\: \delta_{ab} \spc .
\EndEqu
Der Massenparameter $m$ in der Definitionsgleichung f"ur $Y$ kann
beliebig gew"ahlt werden. Er wurde eingef"uhrt, damit $Y$ eine
dimensionslose Gr"o"se ist; bei einer Entwicklung nach der Masse ist er
au"serdem hilfreich, um die Beitr"age verschiedener Ordnung leichter
auseinanderzuhalten.
Falls $X \neq \1$ ist, sagen wir, da"s $P$ eine
{\em{chirale Asymmetrie}}\index{chirale Asymmetrie}
besitzt. Im Fall $Y \neq m_a \: \delta_{\alpha \beta} \: \delta_{jk} \:
\delta_{ab}$ haben wir entsprechend eine
{\em{Massenasymmetrie}}\index{Massenasymmetrie}.
Die Bedingung \Ref{2_a4} l"a"st sich in der Form
\Equ{2_asymm}
XY \;=\; YX \;=\; Y
\EndEqu
umschreiben.

Mit dieser Matrixschreibweise mu"s man etwas aufpassen. Es ist n"amlich
zu beachten, da"s die Dirac-/Block-Indizes und der Flavour-Index
eine grundlegend verschiedene Rolle spielen:
Der Raum $\C^{4B}$ der Dirac-/Block-Indizes ist der Spinorraum;
die Wellenfunktionen $\Psi_{\alpha j}(x)$ sind darin Schnitte.
Auf dem Spinorraum ist das Spinskalarprodukt mit Signatur
$(2B, 2B)$ gegeben. Die lokalen Isometrietransformationen dieses
Skalarproduktes k"onnen, wie in der Einleitung beschrieben, als
$U(2B, 2B)$-Eichtransformationen interpretiert werden.
Den Flavour-Raum haben wir dagegen nur eingef"uhrt, um die Fermionfamilien
zu indizieren. Der Flavour-Index tritt im fermionischen Projektor
gem"a"s \Ref{2_a5}
lediglich als innerer Summationsindex auf. Eine Transformation in
$\C^{4B} \otimes \C^f$, bei der Spinor- mit Flavour-Indizes
gemischt werden, ist nicht sinnvoll.
Das Zusammenfassen des Spinorraumes und des Flavour-Raumes ist also
wirklich nur eine Vereinfachung der Notation und hat keine physikalische
Bedeutung.

Wir entwickeln den freien fermionischen Projektor nach der Masse:
Zun"achst stellen wir die Distributionen $p_m, k_m$ in einer
formalen Potenzreihe dar,
\Equ{2_b9}
p_m \;=\; \sum_{l=0}^\infty m^l \: p^{(l)} \spc,\spc
	k_m \;=\; \sum_{l=0}^\infty m^l \: k^{(l)} \spc .
\EndEqu
Bei Einsetzen in \Ref{2_a5} erh"alt man
\begin{eqnarray}
P(x,y) &=& \sum_{l=0}^\infty \: \bigoplus_{j=1}^B \: X_j \:
	\sum_{a=1}^f \: (m_{ja})^l \; \frac{1}{2} \:
	(p^{(l)} - k^{(l)})(x,y) \nonumber \\
\label{eq:2_b8}
&=& \sum_{l=0}^\infty \: m^l \: \Tr_{\cal{F}} (X Y^l) \; \frac{1}{2} \:
	(p^{(l)} - k^{(l)})(x,y) \spc ,
\end{eqnarray}
dabei bezeichnet $\Tr_{\cal{F}}$ die partielle Spur "uber den
Flavour-Raum,
\[ \Tr_{\cal{F}} (A)_{\alpha j \: \beta k} \;=\; \sum_{a=1}^f
	A_{\alpha j a \: \beta k a} \;\;\;,\spc
	A \in L(\C^4 \otimes \C^B \otimes \C^f) \spc . \]
Mit \Ref{2_asymm}, \Ref{2_b8} haben wir handliche Formeln zur Beschreibung
des freien fermionischen Projektors abgeleitet.

\subsection{Explizite Betrachtung im Ortsraum}
Wir wollen nun den freien fermionischen Projektor $P(x,y)$ als Funktion
von $x, y$ untersuchen.
Bei expliziter Berechnung der Fouriertransformierten von \Ref{2_1},
\Ref{2_1a} erh"alt man die Gleichungen
\begin{eqnarray}
\label{eq:2_20}
P_{m^2}(x) &=& \left\{ \displaystyle \begin{array}{ll}
	\displaystyle \frac{m^2}{8 \pi^2} \frac{
	Y_1(\sqrt{m^2 x^2})}{\sqrt{m^2 x^2}} & {\mbox{f"ur $x^2>0$}} \\[.3cm]
	\displaystyle \frac{m^2}{4 \pi^3} \frac{K_1(\sqrt{-m^2 x^2})}
	{\sqrt{-m^2 x^2}} & {\mbox{f"ur $x^2<0$}} \end{array} \right. \\
\label{eq:2_21}
K_{m^2}(x) &=& -\frac{i}{4 \pi^2} \: \delta(x^2) \: \epsilon(x^0)
	\;+\; \frac{i m^2}{8 \pi^2} \frac{J_1(\sqrt{m^2 x^2})}{\sqrt{m^2 x^2}}
	\: \Theta(x^2) \: \epsilon(x^0)
\end{eqnarray}
mit Besselfunktionen $J_1, Y_1, K_1$. In \Ref{2_20} ist der Pol auf dem
Lichtkegel als Hauptwert zu behandeln.
Die Distributionen $k_m, p_m$ erh"alt man durch Differentiation
\Equ{2_c9}
p_m \;=\; \frac{|m|}{m} \; (i \Pdd_x+m) P_{m^2} \;\;\;,\spc k_m \;=\;
	\frac{|m|}{m} \; (i \Pdd_x+m) K_{m^2} \nonumber \spc .
\EndEqu
Die Besselfunktionen besitzen die Reihendarstellungen
\begin{eqnarray}
\label{eq:2_d1}
J_1(x) &=& \sum_{j=0}^\infty \: \frac{(-1)^j}{j ! \: (j+1)!} \:
	\left( \frac{x}{2} \right)^{2j+1} \\
Y_1(x) &=& \frac{2}{\pi} \: \left( \log \left( \frac{x}{2} \right)
	\:+\: C_e \right) \:
	\sum_{j=0}^\infty \: \frac{(-1)^j}{j! \: (j+1)!} \:
	\left( \frac{x}{2} \right)^{2j+1} \nonumber \\
\label{eq:2_d2}
&&-\frac{2}{\pi} \: \frac{1}{x} \:-\: \frac{1}{\pi} \: \sum_{j=0}^\infty
	\: \frac{(-1)^j}{j! \: (j+1)!} \: \left( \frac{x}{2} \right)^{2j+1}
	\: (\Phi(j+1) + \Phi(j)) \\
K_1(x) &=& \left( \log \left( \frac{x}{2} \right) \:+\: C_e \right) \:
	\sum_{j=0}^\infty \: \frac{1}{j! \: (j+1)!} \:
	\left( \frac{x}{2} \right)^{2j+1} \nonumber \\
\label{eq:2_d3}
&&+\frac{1}{x} \:-\: \frac{1}{2} \: \sum_{j=0}^\infty
	\: \frac{1}{j! \: (j+1)!} \: \left( \frac{x}{2} \right)^{2j+1}
	\: (\Phi(j+1) + \Phi(j)) \spc ,
\end{eqnarray}
dabei ist $C_e$ die Eulersche Konstante und $\Phi$ die Funktion
\[ \Phi(n) \;=\; \sum_{k=1}^n \: \frac{1}{k} \spc (,\;\;\Phi(0)=0) \spc . \]
Durch Einsetzen dieser Reihendarstellungen in \Ref{2_20}, \Ref{2_21}
und \Ref{2_c9} erh"alt man Entwicklungsformeln f"ur $p_m$, $k_m$.

Wir diskutieren kurz die erhaltenen Ausdr"ucke: Die Distribution $k_m(x)$
verschwindet f"ur raumartiges $x$, w"ahrend $p_m(x)$ im ganzen
Minkowski-Raum beitr"agt. Auf dem Lichtkegel $x^2=0$ sind die
Distributionen singul"ar;\label{2_abn} die Ordnung der Singularit"at nimmt
dabei mit steigender Potenz in der Masse ab. In $p_m$ treten au"serdem
logarithmische Singularit"aten $\sim \ln(|x^2|) \; x^{2p}$ auf.
Au"serhalb des Lichtkegels sind die Distributionen glatte Funktionen,
die f"ur $x^2 \rightarrow \pm \infty$ asymptotisch abfallen.

Die Logarithmen in \Ref{2_d2}, \Ref{2_d3} f"uhren auf eine
Problem: die Distribution $p_m$ l"a"st sich entgegen unserem
Ansatz \Ref{2_b9} nicht in einer Potenzreihe in $m$ darstellen,
sondern besitzt nur eine Entwicklung der Form
\Equ{2_zz}
p_m \;=\; \sum_{l=0}^\infty m^l \: p^{(l)} \;+\; \log(m) \:
	\sum_{l=2}^\infty m^l \: q^{(l)} \spc .
\EndEqu
mit geeigneten Distributionen $q^{(l)}$.
Als m"oglichen Ausweg k"onnten wir \Ref{2_b9} durch die Gleichung
\begin{eqnarray}
\label{eq:2_d5}
P(x,y) &=& \sum_{l=0}^\infty \: m^l \: \Tr_{\cal{F}} (X Y^l) \:
	\frac{1}{2} \: (p^{(l)} - k^{(l)})(x,y) \\
\label{eq:2_d6}
&&+\sum_{l=2}^\infty \: m^l \: \Tr_{\cal{F}}(Y^l \: \log(m Y)) \:
	\frac{1}{2} \: q^{(l)}(x,y)
\end{eqnarray}
ersetzen. F"ur unsere Zwecke ist aber eine vereinfachte Behandlung
ausreichend: Nach Einsetzen von \Ref{2_d2}, \Ref{2_d3} in \Ref{2_20},
\Ref{2_c9} haben alle logarithmischen Faktoren die Form
\Equ{2_d4}
\log \left( \frac{1}{2} \: \sqrt{|m^2 \: x^2|} \right) + C_e \;=\;
	\frac{1}{2} \: \log |x^2| \:+\: \log m \:-\: \log 2 \:+\: C_e
	\spc .
\EndEqu
Die problematischen $\log m$-Terme treten also immer in Kombination
mit logarithmischen Singularit"aten auf dem Lichtkegel auf und
verschieben diese Singularit"at um eine Konstante.
Bei der Untersuchung der Gleichungen der diskreten Raumzeit werden
wir in Abschnitt \ref{4_ab5} die Bedingung ableiten, da"s bestimmte
logarithmische Singularit"aten $\sim x^{-2p} \log |x^2|$ des
fermionischen Projektors in den Gleichungen der diskreten Raumzeit
verschwinden
m"ussen. Als Folge werden dann automatisch auch die zugeh"origen
konstanten Beitr"age $\sim x^{-2p}$ wegfallen. Im Hinblick auf diese
Rechnungen kommt es uns auf die von $x^2$ unabh"angigen Summanden in
\Ref{2_d4} nicht an, so da"s wir alle Logarithmen modulo einer
reellen Konstanten behandeln k"onnen. Dazu f"uhren wir die Funktionen
\Equ{2_ln}
\ln(|x^2|) \;:=\; \log |x^2| \:+\: C
\EndEqu
ein. Wir k"onnten die Konstante $C$ explizit angeben,
\[ C(m) \;=\; 2 \log m \:-\: 2 \log 2 \:+\: 2 C_2 \spc , \]
die genaue Massenabh"angigkeit ist f"ur uns aber unwichtig. Mit dieser
Schreibweise k"onnen wir den zweiten Summanden \Ref{2_d6} mit den
logarithmischen Termen des ersten Summanden \Ref{2_d5} zusammenfassen
und so trotz der Beitr"age $\sim m^2 \: \log m$ mit dem Potenzreihenansatz
\Ref{2_b8} arbeiten. Die Reihe \Ref{2_b8} konvergiert dann
im Distributionssinn.

F"ur die ersten Entwicklungskoeffizienten hat man mit der
Abk"urzung $\xi \equiv y-x$
\begin{eqnarray*}
\begin{array}{rclcrcl}
p^{(0)}(x,y)  &=& \displaystyle -\frac{i}{2 \pi^3} \frac{\xi \slsh}{\xi^4}
	&,&
k^{(0)}(x,y) &=& \displaystyle \frac{1}{2 \pi^2} \; \xi\slsh \:
	\delta^\prime(\xi^2) \: \epsilon(\xi^0) \\[.3cm]
p^{(1)}(x,y)  &=& \displaystyle -\frac{1}{4 \pi^3} \frac{1}{\xi^2} &,&
k^{(1)}(x,y) &=& \displaystyle \frac{i}{4 \pi^2} \: \delta(\xi^2) \:
	\epsilon(\xi^0) \\[.3cm]
p^{(2)}(x,y) &=& \displaystyle -\frac{i}{8 \pi^3} \:
	\frac{\xi \slsh}{\xi^2} &,&
k^{(2)}(x,y) &=& \displaystyle -\frac{1}{8 \pi^2} \; \xi\slsh \:
	\delta(\xi^2) \: \epsilon(\xi^0) \\[.3cm]
p^{(3)}(x,y) &=& \displaystyle \frac{1}{16 \pi^3} \: \ln(|\xi^2|)
	&,&
k^{(3)}(x,y) &=& \displaystyle -\frac{i}{16 \pi^2} \; \Theta(\xi^2) \:
	\epsilon(\xi^0) \\[.3cm]
p^{(4)}(x,y) &=& \displaystyle \frac{i}{64 \pi^3} \:\xi\slsh\;
	\ln(|\xi^2|) &,&
k^{(4)}(x,y) &=& \displaystyle \frac{1}{64 \pi^2} \:\xi\slsh \;
	\Theta(\xi^2) \: \epsilon(\xi^0) \spc ,
\end{array}
\end{eqnarray*}
dabei bezeichnet $\xi_j \: \xi^{-4}$ die partielle Distributionsableitung
des Hauptwertes $\xi^{-2}$. Durch Einsetzen in \Ref{2_b8} erhalten wir
schlie"slich explizite Formeln f"ur den freien fermionischen Projektor.

Wie in der Einleitung beschrieben, spielen die Singularit"aten von $p_m,
k_m$ auf dem Lichtkegel f"ur uns eine entscheidende Rolle. Deshalb
wollen wir anschaulich "uberlegen, wie es bei der Fouriertransformation
zu diesen Singularit"aten kommt. \label{2_ansch}
Um die logarithmischen Singularit"aten zu vermeiden, betrachten
wir die Distribution $K_{m^2}$ und f"uhren einen Impuls-Cutoff $\lambda$
ein
\[ K_{m^2}^{(\lambda)}(x) \;:=\; \int \frac{d^4k}{(2 \pi)^4} \;
	\Theta(\lambda-|k^0|) \: \delta(k^2-m^2) \: \epsilon(k^0) \;
	e^{-i k x} \spc . \]
Durch den Cutoff wird die Distribution auf der L"angenskala $2 \pi/\lambda$
regularisiert; wir untersuchen den Grenzfall $\lambda \rightarrow \infty$.
Nach einer Umskalierung und Entwicklung nach der Masse\footnote{Zur
Vollst"andigkeit erw"ahnen wir, wie man mit dieser
Rechnung auch die logarithmischen Singularit"aten von $P_{m^2}$
verstehen kann: Bei Regularisierung und formaler Potenzreihenentwicklung
von $P_{m^2}$ erhalten wir analog zu \Ref{2_c9a} den Ausdruck
\Equ{2_c6}
P_{m^2}^{(\lambda)}(x) \;=\; \sum_{j=0}^\infty (-1)^j \: \lambda^{2-2j}
	\: m^{2j} \; \int \frac{d^4k}{(2 \pi)^4} \; \Theta(1-|k^0|)
	\: \delta^{(j)}(k^2) \; e^{-i k \: (\lambda x)} \spc .
\EndEqu
Um \Ref{2_c9a}, \Ref{2_c6} einen mathematischen Sinn zu geben, m"ussen
wir die Faktoren $\delta^{(j)}(k^2) \: \epsilon(k^0)$ bzw.
$\delta^{(j)}(k^2)$ als Distributionen definieren; dazu untersuchen wir
f"ur eine Schwartzfunktion $f$ die Gleichungen
\begin{eqnarray}
\label{eq:2_c7}
\int d^4k \; \delta^{(j)}(k^2) \: \epsilon(k^0) \; f(k) &=&
\left( \frac{d}{da} \right)^j_{|a=0} \: \int d^4k \;
	\delta(k^2 - a) \: \epsilon(k^0) \; f(k) \\
\label{eq:2_c8}
\int d^4k \; \delta^{(j)}(k^2) \: \epsilon(k^0) \; f(k) &=&
\left( \frac{d}{da} \right)^j_{|a=0} \: \int d^4k \;
	\delta(k^2 - a) \; f(k) \spc .
\end{eqnarray}
Die Ableitungen nach dem Parameter $a$ sind in der Umgebung
des Ursprungs $k=0$ problematisch.
Unter Ausnutzung des umgekehrten Vorzeichens von $\delta(k^2-a) \:
\epsilon(k^0)$ auf der oberen und unteren Massenschale kann man \Ref{2_c7}
("ahnlich wie der Distributionsableitung eines Hauptwertintegrals)
einen Sinn geben und so die Potenzreihe \Ref{2_c9a} mathematisch rechtfertigen.
In \Ref{2_c8} treten dagegen nicht-hebbare Divergenzen auf, so da"s
auch \Ref{2_c6} nicht existiert. Bei Regularisierung im Impulsraum
stellt man fest, da"s diese Divergenzen logarithmisch sind und folglich
gerade den $\log$-Terme in \Ref{2_d2}, \Ref{2_d3} entsprechen.}
\begin{eqnarray}
K_{m^2}^{(\lambda)}(x) &=& \lambda^2 \; \int \frac{d^4k}{(2 \pi)^4} \;
	\Theta(1-|k^0|) \: \delta \left(k^2 - \frac{m^2}{\lambda^2} \right)
	\: \epsilon(k^0) \; e^{-i k \: (\lambda x)} \nonumber \\
\label{eq:2_c9a}
&=& \sum_{j=0}^\infty (-1)^j \: \lambda^{2-2j} \: m^{2j} \;
	\int \frac{d^4k}{(2 \pi)^4} \; \Theta(1-|k^0|)
	\: \delta^{(j)}(k^2) \: \epsilon(k^0) \; e^{-i k \: (\lambda x)}
\end{eqnarray}
entspricht der Grenzwert $\lambda \rightarrow \infty$ in den Integralen
dem Limes, da"s die Ortskoordinate $\lambda x$ ins Unendliche
l"auft.
Die Wellenzahl $-i\lambda x$ in \Ref{2_c9a} wird in diesem Grenzfall immer
gr"o"ser. Die meisten Beitr"age des Integrals oszillieren sich immer besser
weg und fallen folglich f"ur gro"ses $\lambda$ ab.
Eine besondere Rolle spielt die Hyperebene $E(x)=\{k \;|\; \bra k,x \ket=0 \}$.
Alle Beitr"age in einer kleinen Umgebung dieser Ebene sind in \Ref{2_c9a}
in Phase. Falls $x$ ein Punkt auf dem Lichtkegel ist, liegt $E$
tangential zum Massenkegel. Da der Tr"ager des Integranden
auf dem Massenkegel liegt, haben wir dann in der Umgebung von $E$ einen
gro"sen Beitrag zu dem Fourierintegral \Ref{2_c9a}.
Damit f"allt das Integral in \Ref{2_c9a} f"ur $x^2=0$ im Grenzfall
$\lambda \rightarrow \infty$ weniger stark ab oder steigt sogar an, was
schlie"slich in \Ref{2_c9a} zu Divergenzen f"uhrt.

Diese "Uberlegung "ubertr"agt sich direkt auf beliebige Distributionen
mit Tr"ager im Innern des Massenkegels (also in der Menge $\{k^2 \geq 0\}$).
Verantwortlich f"ur die Singularit"aten auf dem Lichtkegel ist,
anschaulich gesagt, die Flanke des Integranden in der N"ahe des
Massenkegels\footnote{Mit diesem Argument l"a"st sich sogar die Ordnung
der Singularit"at beschreiben, wir betrachten als Beispiel den Fall $m=0$.
Als Fourierintegral untersuchen wir gem"a"s \Ref{2_c9a}
\Equ{2_25}
\int \frac{d^4k}{(2 \pi)^4} \; \Theta(1-|k^0|) \; \delta(k^2) \:
	\epsilon(-k^0) \; e^{-ik (Kx)} \spc .
\EndEqu
Wir hatten "uberlegt, da"s f"ur die Singularit"at auf dem Lichtkegel
das Integral "uber das Gebiet $\{-c \leq k \lambda x \leq
c \}$ mit festem $c \approx \pi$ entscheidend ist.
Darum ersetzen wir den oszillierenden Faktor $\exp(-ik \lambda x)$ in
\Ref{2_25} n"aherungsweise durch $\Theta(\pi - |k \lambda x|)$.
Wir f"uhren das Integral "uber $k^0$ aus
und w"ahlen Polarkoordinaten $(k=|\vec{k}|, \theta,\varphi)$.
Das Integral "uber $k$ skaliert sich nicht in $\lambda$, das
Integral "uber die Winkelkoordinaten verh"alt sich in niedrigster
Ordnung $\sim \lambda^{-1}$. Damit ist \Ref{2_25} proportional zu
$\lambda^{-1}$. Folglich hat \Ref{2_c9a} eine Divergenz $\sim \lambda$,
zeigt also tats"achlich das richtige Skalierungsverhalten, wie man durch
Vergleich mit $K^{(0)}(x)$ bei Regularisierung auf der
L"angenskala $\lambda^{-1}$ sieht.}.

\subsection{Der freie fermionische Projektor des Standardmodells}
Zur Erl"auterung wollen wir abschlie"send einen freien fermionischen
Projektor aufbauen, der die Fermionkonfiguration des Standardmodells
nachbildet. Wir w"ahlen $f=3$ und definieren auf dem
Flavour-Raum die Massenmatrizen der Lepton- und Quarkfamilien
\begin{eqnarray*}
M^\lep &=& \left( \begin{array}{ccc} m_e & 0 & 0 \\ 0 & m_\mu & 0 \\
	0 & 0 & m_\tau \end{array} \right) \\
M^u &=& \left( \begin{array}{ccc} m_u & 0 & 0 \\ 0 & m_c & 0 \\
	0 & 0 & m_t \end{array} \right) \;\;\;\;, \spc
M^d \;=\; \left( \begin{array}{ccc} m_d & 0 & 0 \\ 0 & m_s & 0 \\
	0 & 0 & m_b \end{array} \right) \spc .
\end{eqnarray*}
Die Neutrinos m"ussen als linksh"andige Fermionen masselos sein.

Wir betrachten zun"achst die Isospinpartner $\nu_e, \nu_\mu, \nu_\tau
\leftrightarrow e, \mu, \tau$ und $u, c, t \leftrightarrow d, s, b$
getrennt:
Zur Beschreibung der Leptonen und Quarks werden jeweils achtkomponentige
Wellenfunktionen ben"otigt. Bei der Zerlegung $\C^8 = \C^4 \oplus \C^4$
des Spinorraumes haben die Asymmetriematrizen die Form
\begin{eqnarray*}
X^\lep &=& \chi_L \oplus \1 \spc,\spc Y^\lep \;=\; \frac{1}{m} \:
	(0 \oplus M^\lep) \\
X^\qu &=& \1 \oplus \1 \spc \:\:,\spc Y^\qu \;=\; \frac{1}{m} \:
	(M^u \oplus M^d) \spc .
\end{eqnarray*}
Mit Gleichung \Ref{2_b8} erh"alt man die zugeh"origen freien
fermionischen Projektoren, die wir {\em{Lepton-}}\index{Leptonsektor} bzw.
{\em{Quark-Sektor}}\index{Quarksektor} nennen.

Zur Beschreibung der Fermionen des Standardmodells m"ussen wir
eine direkte Summe des Lepton- und Quarksektors bilden.
Um die Colour-Freiheitsgrade zu ber"ucksichtigen, bauen wir den Quarksektor
dreifach ein. Wir setzen also bei Spindimension 32
mit der Zerlegung $\C^{32}=(\C^4 \oplus \C^4)^4$
\begin{eqnarray}
\label{eq:2_c1}
X &=& X^\lep \oplus (X^\qu)^3 \;=\; (\chi_L \oplus \1) \:\oplus\: (\1 \oplus \1)^3 \\
\label{eq:2_c2}
Y &=& Y^\lep \oplus (Y^\qu)^3 \;=\; \frac{1}{m} \: \left( (0 \oplus M^\lep) \:\oplus\:
(M^u \oplus M^d)^3 \right) \spc .
\end{eqnarray}
Die Matrizen $X, Y$ wirken auf $\C^{96}=\C^4 \otimes \C^8 \otimes \C^3$. Der
freie fermionische Projektor ist wieder durch die Potenzreihe \Ref{2_b8}
gegeben.

Man beachte, da"s f"ur die Fermionmassen nicht die physikalischen Massen, sondern,
wie in Abschnitt \ref{1_ab5} der Einleitung beschrieben, die nackten
Massen bei Regularisierung der Theorie auf der Planck-Skala einzusetzen sind.
Es ist nicht klar, welche genauen Werte diese nackten Massen haben; auf jeden Fall
sollten sie sich f"ur die schweren Fermionen deutlich von den physikalischen
Massen unterscheiden. Da Proton und Neutron ann"ahernd die gleiche Masse besitzen,
ist es naheliegend, die Isospinabh"angigkeit der Quarkmassen allein auf die
Selbstwechselwirkung zur"uckzuf"uhren, also
\Equ{2_c3}
m_u = m_d \;\;\;,\spc m_c=m_s \;\;\;,\spc m_t=m_b
\EndEqu
anzunehmen. Aus physikalischer Sicht ist nicht ausgeschlossen, da"s es
auch Relationen zwischen den Lepton- und Quarkmassen gibt,
im einfachsten Fall
\Equ{2_c4}
m_u = m_d = m_e \;\;\;,\spc m_c=m_s = m_\mu\;\;\;,\spc m_t = m_b = m_\tau \spc .
\EndEqu
Die Relationen \Ref{2_c3}, \Ref{2_c4} sind im Moment eher spekulativ.
Wir werden sie zun"achst weglassen und die nackten Fermionmassen als
9 voneinander unabh"angige Parameter ansehen.

\section{St"orungen erster Ordnung}
Wie in der Einleitung beschrieben, m"ussen wir den fermionischen Projektor
unter allgemeinen St"orungen \Ref{1_36a} des Diracoperators
untersuchen.
Bevor wir dieses Problem im n"achsten Abschnitt \ref{2_ab3}
systematisch angehen, wollen wir die lineare N"aherung studieren.
Dazu beginnen wir mit der St"orungsrechnung f"ur die
Distributionen $p_m, k_m$. Die detaillierten Rechnungen wurden in
die Anh"ange A-D ausgelagert; wir werden
hier die formale Entwicklung durchf"uhren und die wichtigsten
Ergebnisse aus den Anh"angen diskutieren.
Anschlie"send werden wir erkl"aren, wie die St"orungsrechnung f"ur
den fermionischen Projektor $P$ auf diejenige f"ur $p_m, k_m$
zur"uckgef"uhrt werden kann.

\subsection{Formale St"orungsentwicklung f"ur $p_m, k_m$}
\label{2_ab21}
In diesem Abschnitt wollen wir bei Spindimension $4$ die Spektralprojektoren
des Di\-rac\-ope\-ra\-tors $i \Pdd + {\cal{B}}$ in erster Ordnung in ${\cal{B}}$
bestimmen. Zur Unterscheidung von den
Spektralprojektoren $p_m, k_m$ des freien Diracoperators bezeichnen
wir die gest"orten Gr"o"sen mit einer zus"atzlichen Tilde.
Gesucht sind also hermitesche Operatoren $\tilde{p}_m, \tilde{k}_m$ mit
\Equ{2_26}
(i \Pdd + {\cal{B}} - m) \: \tilde{p}_m = {\cal{O}}({\cal{B}}^2) \;\;\;,\spc
(i \Pdd + {\cal{B}} - m) \: \tilde{k}_m = {\cal{O}}({\cal{B}}^2) \spc ,
\EndEqu
au"serdem sollen in erster Ordnung die zu \Ref{2_11} bis \Ref{2_13} analogen
Relationen
\Equ{2_27}
\tilde{k}_m \: \tilde{k}_n \;=\; \tilde{p}_m \: \tilde{p}_n \;=\;
\delta(m-n) \: \tilde{p}_m \;\;\;,\spc
\tilde{k}_m \: \tilde{p}_n \;=\; \tilde{p}_m \: \tilde{k}_n \;=\;
\delta(m-n) \: \tilde{k}_m
\EndEqu
gelten.

Wir leiten zun"achst auf anschauliche, aber mathematisch nicht strenge
Weise einen Ansatz f"ur $\tilde{p}_m, \tilde{k}_m$ ab:
Die Distributionen $p_m, k_m$ sind aus Eigenzust"anden des freien
Diracoperators aufgebaut, also formal
\Equ{2_f1}
p_m(x,y), k_m(x,y) \;=\; \sum_a \: \Psi_a(x) \: \overline{\Psi}(y)
	\;\;\;\; {\mbox{mit}} \;\;\;\; (i \Pdd - m) \: \Psi_a \;=\; 0 \spc .
\EndEqu
Die gest"orten Zust"ande $\tilde{\Psi}_a$ erf"ullen die Gleichung
$(i \Pdd - m + {\cal{B}}) \: \tilde{\Psi}_a = {\cal{O}}({\cal{B}}^2)$
und k"onnen mit der in der relativistischen Quantenmechanik "ublichen
St"orungsrechnung behandelt werden. Man erh"alt
\[ \tilde{\Psi}_a(x) \;=\; \Psi_a(x) \;-\; \int d^4y \; s_m(x-y) \:
	({\cal{B}} \: \Psi_a)(y) \]
mit der Dirac-Greensfunktion $s_m$, welche durch die folgende Definition
gegeben ist.
\begin{Def}
Wir definieren f"ur $m \in \R$ die temperierte Distribution $s_m$
als Hauptwert
\Equ{2_4}
s_m (k) \;=\; \frac{1}{2} \: \lim_{\varepsilon \rightarrow 0} \left(
	\frac{1}{k\slsh-m+i\varepsilon} + \frac{1}{k\slsh-m-i\varepsilon}
	\right)
\EndEqu
und fassen $s_m$ im Impulsraum auch als Multiplikationsoperator auf.
\end{Def}
In Operatorschreibweise haben wir also
\Equ{2_f4}
\tilde{\Psi}_a \;=\; \Psi_a \:-\: s_m \:{\cal{B}}\: \Psi_a \;\;\;\;,\spc
	\overline{\tilde{\Psi}_a} \;=\; \overline{\Psi_a} \:-\:
	\overline{\Psi_a} \:{\cal{B}}\: s_m \spc .
\EndEqu
Wir setzen diese gest"orten Eigenzust"ande in \Ref{2_f1} ein und
erhalten in erster Ordnung
\begin{eqnarray}
\label{eq:2_28}
\tilde{p}_m &=& p_m - s_m \: {\cal{B}} \: p_m -  p_m \: {\cal{B}} \: s_m \\
\label{eq:2_29}
\tilde{k}_m &=& k_m - s_m \: {\cal{B}} \: k_m -  k_m \: {\cal{B}} \: s_m
	\spc .
\end{eqnarray}

Wir m"ussen verifizieren, da"s der Ansatz \Ref{2_28}, \Ref{2_29}
tats"achlich die Bedingungen \Ref{2_26}, \Ref{2_27} erf"ullt:
Aus der Operatorgleichung
\Equ{2_e0}
(i \Pdd - m) \: s_m \;=\; \1
\EndEqu
folgt unmittelbar \Ref{2_26}. F"ur die Greensfunktionen gelten
analog zu \Ref{2_5} bis \Ref{2_13} die formalen Rechenregeln
\begin{eqnarray}
p_m \: s_n &=& s_n \: p_m \;=\; \frac{(k\slsh+m)(k\slsh+n)}{k^2-n^2} \;
	\frac{|m|}{m} \: \delta(k^2-m^2) \nonumber \\
\label{eq:2_14}
&=& \frac{(m+n) \: (k\slsh+m)}{m^2-n^2} \; \frac{|m|}{m} \:
	\delta(k^2-m^2) \;=\; \frac{1}{m-n} \; p_m \\
\label{eq:2_15}
k_m \: s_n &=& s_n \: k_m \;=\; \frac{1}{m-n} \; k_m \\
\label{eq:2_16}
s_m \: s_n &=& \frac{1}{(k\slsh-m) \: (k\slsh-n)} \;=\;
	\frac{1}{m-n} \: (s_m-s_n) \spc ,
\end{eqnarray}
wobei wir alle Pole als Hauptwert behandeln. Die Relation
$\tilde{p}_m \: \tilde{p}_n \:=\: \delta(m-n) \: \tilde{p}_m$ 
erh"alt man man unter Verwendung von \Ref{2_11}, \Ref{2_14} durch die
Rechnung
\begin{eqnarray}
\tilde{p}_m \: \tilde{p}_n &=& p_m \: p_n - s_m \: {\cal{B}} \: p_m \; p_n
	- p_m \: {\cal{B}} \: s_m \; p_n
	- p_m \; p_n \: {\cal{B}} \: s_n - p_m \; s_n \: {\cal{B}} \: p_n
	+ {\cal{O}}({\cal{B}}^2) \nonumber \\
&=& \delta(m-n) \; (p_m \:-\: s_m \: {\cal{B}} \: p_m \:-\:  p_m \:
	{\cal{B}} \: s_m) \nonumber \\
&& \hspace*{1cm} - \frac{1}{n-m} \; p_m \: {\cal{B}} \: p_n - \frac{1}{m-n} \;
	p_m \: {\cal{B}} \: p_n + {\cal{O}}({\cal{B}}^2) \nonumber \\
\label{eq:2_g0}
&=& \delta(m-n) \: \tilde{p}_m \;+\; {\cal{O}}({\cal{B}}^2) \spc ,
\end{eqnarray}
die anderen Bedingungen in \Ref{2_27} folgen analog.

Etwas eleganter l"a"st sich die St"orungsrechnung erster Ordnung auch
mit einer unit"aren Transformation beschreiben:
\begin{Satz}
\label{a2_satz1}
Der Operator
\Equ{2_39}
U[{\cal{B}}] \;=\; 1 - \int_{\sR \cup i \sR} \!\!\! dm \;
	s_m \: {\cal{B}} \: p_m
\EndEqu
ist (in erster Ordnung in ${\cal{B}}$) unit"ar und
\Equ{2_40}
\tilde{p}_m \;=\; U \: p_m \: U^* \;\;\;,\spc \tilde{k}_m
	\;=\; U \: k_m \: U^* \spc .
\EndEqu
Zu jeder infinitesimalen unit"aren Transformation $V=1+iA$ (mit einem
hermiteschen Operator $A$) gibt es einen St"oroperator ${\cal{B}}$
mit $U[{\cal{B}}]=V$.
\end{Satz}
{\Beweis}
Mit Hilfe von \Ref{2_14} und der Vollst"andigkeitsrelation \Ref{2_31}
k"onnen wir den Propagator $s_m$ in der Form
\Equ{2_41}
s_m \;=\; \int_{\sR \cup i \sR} \!\!\! dm \; \frac{1}{m-m^\prime} \; p_m
\EndEqu
umschreiben. Damit haben wir
\Equ{2_yz}
\int_{\sR \cup i \sR} \!\!\! dm \;
	s_m \: {\cal{B}} \: p_m \;=\; \int_{\sR \cup i \sR} \!\!\! dm
	\int_{\sR \cup i \sR} \!\!\! dm^\prime \; \frac{1}{m-m^\prime}
	\; p_m \: {\cal{B}} \: p_{m^\prime} \spc .
\EndEqu
Da dies ein antihermitescher Operator ist, ist $U$ unit"ar.
Unter Verwendung von \Ref{2_41} erh"alt man weiterhin
\begin{eqnarray*}
U \: p_m \: U^* &=& p_m + \int_{\sR \cup i \sR} \!\!\! dm^\prime
	\int_{\sR \cup i \sR} \!\!\! dm^{\prime \prime} \;
	\frac{1}{m^\prime-m^{\prime \prime}} \;
	\left( p_m \: p_{m^\prime} \: {\cal{B}} \: p_{m^{\prime \prime}} -
	p_{m^\prime} \: {\cal{B}} \: p_{m^{\prime \prime}} \: p_m \right) \\
&=& p_m - p_m \: {\cal{B}} \: s_m - s_m \: {\cal{B}} \: p_m \;=\;
	\tilde{p}_m \spc ,
\end{eqnarray*}
die zweite Gleichung in \Ref{2_40} folgt analog.
Zu gegebenem $V=1+iA$ setzen wir
\[ {\cal{B}} \;=\; -i \int_{\sR \cup i \sR} \!\!\! dm \int_{\sR \cup i \sR}
	\!\!\! dm^\prime \; (m-m^\prime) \; p_m \:A\: p_{m^\prime} \]
und erhalten mit \Ref{2_31}, \Ref{2_41} schlie"slich
\begin{eqnarray*}
U[{\cal{B}}] &=& 1 + i \int_{\sR \cup i \sR} \!\!\! dm \int_{\sR \cup i
	\sR} \!\!\! dm^\prime \; \frac{1}{m-m^\prime} \; (m-m^\prime) \;
	p_m \: A \: p_{m^\prime} \\
&=& 1 + i A \;=\; V \spc .
\end{eqnarray*}
\QED

\subsubsection*{Uneindeutigkeit der St"orungsrechnung f"ur $k_m$}
\label{2_uneins} \index{Uneindeutigkeit der St"orungsrechnung}
Mit den Gleichungen \Ref{2_28}, \Ref{2_29} wird die Auswirkung der
St"orung ${\cal{B}}$ auf die Spektralprojektoren
$p_m, k_m$ in erster Ordnung vollst"andig beschrieben.
Es ist etwas unbefriedigend, da"s dieser Ansatz nicht zwingend
erscheint. Insbesondere h"atten wir bei der St"orungsrechnung f"ur die
Eigenzust"ande \Ref{2_f4} anstelle von \Ref{2_4} auch die retardierte
oder avancierte Greensfunktion verwenden k"onnen.
Wir wollen abschlie"send untersuchen, wie sich diese Uneindeutigkeit
der St"orungsrechnung f"ur $\tilde{\Psi}_a$ in den Gleichungen
f"ur $\tilde{p}_m, \tilde{k}_m$ auswirkt.

Zun"achst m"ussen wir die retardierten und avancierten Greensfunktionen
einf"uhren: Die Distribution $s_m$ ist als Ableitung der Greensfunktion
der Klein-Gordon-Gleichung
\Equ{2_f2}
S_{m^2}(k) \;:=\; \frac{1}{2} \: \lim_{\varepsilon \rightarrow 0} \left(
	\frac{1}{k^2 - m^2 + i \varepsilon} +
	\frac{1}{k^2 - m^2 - i \varepsilon} \right)
\EndEqu
darstellbar, genauer
\Equ{2_f3}
s_m \;=\; (i \Pdd_x + m) \: S_{m^2} \spc .
\EndEqu
Bei der Berechnung der Fouriertransformierten von \Ref{2_f2} erh"alt
man die explizite Formel
\Equ{2_22}
S_{m^2}(x) \;=\; -\frac{1}{4 \pi} \: \delta(x^2) \;+\; \frac{m^2}{8 \pi}
	\frac{J_1(\sqrt{m^2 x^2})}{\sqrt{m^2 x^2}} \: \Theta(x^2)
	\spc .
\EndEqu
Durch Vergleich mit \Ref{2_21} stellt man fest, da"s sich $S_{m^2}(x)$
und $K_{m^2}(x)$ nur um einen relativen Faktor $-i \pi \: \epsilon(x^0)$
voneinander unterscheiden. Wegen \Ref{2_f3}, \Ref{2_c9} gilt dasselbe
auch f"ur $s_m(x)$ und $k_m(x)$. Folglich k"onnen wir durch geeignete
Linearkombination von $s_m, k_m$ Greensfunktionen konstruieren, deren
Tr"ager nur innerhalb des oberen oder unteren Lichtkegels liegt.
\begin{Def}
\label{2_def_av}
Wir definieren die avancierte und retardierte Greensfunktion durch
\begin{eqnarray}
\label{eq:2_f7}
s_m^\vee &=& s_m \:+\: i \pi \: k_m \\
\label{eq:2_f8}
s_m^\wedge &=& s_m \:-\: i \pi \: k_m \spc .
\end{eqnarray}
\end{Def}

Diese Definition stimmt mit der "ublichen Festlegung
der Pole in der komplexen Ebene "uberein, also
\[ s^\vee(k) \;=\; \lim_{\varepsilon \rightarrow 0} \: \frac{k \slsh + m}
	{k^2 - m^2 - i \varepsilon k^0} \;\;\;\;,\spc
   s^\wedge(k) \;=\; \lim_{\varepsilon \rightarrow 0} \: \frac{k \slsh + m}
	{k^2 - m^2 + i \varepsilon k^0} \spc . \]
Mit der Schreibweise \Ref{2_f7}, \Ref{2_f8} wird aber deutlicher, da"s
sich die verschiedenen Greensfunktionen um ein Vielfaches von $k_m$
unterscheiden.

Wenn wir \Ref{2_e0} als Bestimmungsgleichung f"ur
die Greensfunktionen ansehen, k"onnen wir zu $s_m$ sogar  eine
beliebige L"osung der freien Diracgleichung hinzuaddieren. Damit unser
Ansatz nicht zu kompliziert wird, w"ahlen wir als Greensfunktion in
Verallgemeinerung von \Ref{2_f7}, \Ref{2_f8}
\begin{eqnarray}
\label{eq:2_g1}
s_m^< &:=& s_m \:+\: \alpha(m) \: p_m \:+\: \beta(m) \: k_m \\
\label{eq:2_g2}
s_m^> &:=& (s_m^<)^* \;=\; s_m \:+\: \overline{\alpha(m)} \: p_m
	\:+\: \overline{\beta(m)} \: k_m
\end{eqnarray}
mit komplexwertigen Funktionen $\alpha, \beta$. F"ur die gest"orten
Zust"ande $\tilde{\Psi}_a$ folgt gegen"uber \Ref{2_f4}
\[ \tilde{\Psi}_a \;=\; \Psi_a \:-\: s_m^< \: {\cal{B}} \: \Psi_a \;\;\;\;,\spc
	\overline{\tilde{\Psi}_a} \;=\; \overline{\Psi_a} \:-\:
	\overline{\Psi_a} \:{\cal{B}}\: s_m^> \]
und damit
\[ \tilde{p}_m \;=\; p_m \:-\: s_m^< \:{\cal{B}}\: p_m \:-\:
	p_m \:{\cal{B}}\: s_m^> \;\;\;\;,\spc
   \tilde{k}_m \;=\; k_m \:-\: s_m^< \:{\cal{B}}\: k_m \:-\:
	k_m \:{\cal{B}}\: s_m^> \spc . \]
Wir wollen untersuchen, f"ur welche Funktionen $\alpha, \beta$ die
Bedingungen \Ref{2_27} erf"ullt sind. In Analogie zu \Ref{2_g0} haben
wir nun
\begin{eqnarray*}
\tilde{p}_m \: \tilde{p}_n &=& \delta(m-n) \: \tilde{p}_m \:-\:
	p_m \: s_n^< \:{\cal{B}}\: p_n \:-\: p_m \:{\cal{B}}\: s_m^> \:
	p_n \\
&=& \delta(m-n) \: \tilde{p}_m \:-\: \delta(m-n) \:
	(\alpha(m) + \overline{\alpha(m)}) \: p_m \:{\cal{B}}\: p_m \\
&&+ \delta(m-n) \: \left[ \beta(m) \: k_m \:{\cal{B}}\: p_m \:+\:
	\overline{\beta(m)} \: p_m \:{\cal{B}}\: k_m \right] \spc .
\end{eqnarray*}
Folglich m"ussen die Bedingungen $\alpha(m)+\overline{\alpha(m)}=0$
und $\beta(m)=0$ gelten. Der Ansatz \Ref{2_g1}, \Ref{2_g2} vereinfacht
sich zu
\Equ{2_g3}
s_m^< \;=\; s_m \:+\: i \gamma(m) \: p_m \;\;\;\;,\spc
	s_m^> \;=\; s_m \:-\: i \gamma(m) \: p_m
\EndEqu
mit einer reellen Funktion $\gamma$. F"ur $\tilde{p}_m, \tilde{k}_m$
folgt
\begin{eqnarray}
\tilde{p}_m &=& p_m \:-\: s_m \:{\cal{B}}\: p_m \:-\: i \gamma(m)
	\: p_m \:{\cal{B}}\: p_m \:-\: p_m \:{\cal{B}}\: s_m \:+\:
	i \gamma(m) \: p_m \:{\cal{B}}\: p_m \nonumber \\
\label{eq:2_g4}
&=& p_m \:-\: s_m \:{\cal{B}}\: p_m \:-\: p_m \:{\cal{B}}\: s_m \\
\tilde{k}_m &=& k_m \:-\: s_m \:{\cal{B}}\: k_m \:-\: i \gamma(m)
	\: p_m \:{\cal{B}}\: k_m \:-\: k_m \:{\cal{B}}\: s_m \:+\:
	i \gamma(m) \: k_m \:{\cal{B}}\: p_m \nonumber \\
\label{eq:2_g5}
&=& k_m \:-\: s_m \:{\cal{B}}\: k_m \:-\: k_m \:{\cal{B}}\: s_m
	\:-\: i \gamma(m) \: (p_m \:{\cal{B}}\: k_m \:-\:
	k_m \:{\cal{B}}\: p_m) \spc .
\end{eqnarray}
Man kann direkt verifizieren, da"s \Ref{2_g4}, \Ref{2_g5} alle
Bedingungen \Ref{2_27} und \Ref{2_26} erf"ullt.
Nach \Ref{2_g4} scheint $\tilde{p}_m$ von der speziellen Greensfunktion
unabh"angig zu sein, in die St"orungsrechnung f"ur $k_m$ geht
gem"a"s \Ref{2_g5} dagegen die Wahl der Greensfunktion explizit ein.

Dieses Ergebnis l"a"st sich auch direkt einsehen: Die Distributionen
$\tilde{p}_m$ sind als Spektralprojektoren des gest"orten Diracoperators
$i \Pdd + {\cal{B}}$, also durch die Gleichungen
\Equ{2_q4}
\tilde{p}_m \: \tilde{p}_n \;=\; \delta(m-n) \: \tilde{p}_m \;\;\;,\;\;\;\;\;
	\int_\sigma \tilde{p}_m \: dm \;=\; \1 \;\;\;,\;\;\;\;\;
	\int_\sigma m \: \tilde{p}_m \: dm \;=\; i \Pdd + {\cal{B}} \spc ,
\EndEqu
unabh"angig von einer St"orungsrechnung definiert. Daher ist klar,
da"s die Freiheit in der Wahl der Greensfunktion nicht in die Formeln
f"ur $\tilde{p}_m$ eingeht.
Bei der Distribution $k_m$ haben wir f"ur die Zust"ande auf der oberen
und unteren Massenschale gem"a"s \Ref{2_3a} ein relatives Minuszeichen
eingef"uhrt. F"ur den freien Diracoperator ist die Aufspaltung der
Eigenr"aume in Zust"ande positiver und negativer Energie eindeutig.
Durch die St"orung des Diracoperators werden aber die Zust"ande der
beiden Massenschalen miteinander gemischt, so da"s wir nicht mehr
auf kanonische Weise von positiven und negativen Energiezust"anden
sprechen k"onnen. Da in $\tilde{k}_m$ zwischen diesen Zust"anden
dennoch durch ein relatives Vorzeichen unterschieden werden mu"s,
enth"alt die Definition von $\tilde{k}_m$ eine gewisse Willk"ur.
Diese Willk"ur entspricht der Uneindeutigkeit der St"orungsrechnung
f"ur $\tilde{k}_m$.

Wegen der Uneindeutigkeit der St"orungsrechnung mag es auf den
ersten Blick nicht sinnvoll erscheinen, "uberhaupt mit der Distribution
$\tilde{k}_m$ zu arbeiten. Wir erkl"aren, warum und in welchem Sinn
$\tilde{k}_m$ f"ur uns trotzdem n"utzlich ist:
Nach den "Uberlegungen in der Einleitung beschreiben wir die St"orung
des fermionischen Projektors durch eine allgemeine unit"are Transformation
\Ref{1_d4}. Um einen Kontakt zur "ublichen Formulierung
physikalischer Wechselwirkungen herzustellen, schreiben wir diese
unit"are Transformation gem"a"s \Ref{1_36a} in eine St"orung des
Diracoperators um. Nach Satz \ref{a2_satz1} l"a"st sich jede
unit"are Transformation durch eine geeignete St"orung des Diracoperators
beschreiben. Damit erf"ullt die St"orungsrechnung \Ref{2_28}, \Ref{2_29}
genau den gew"unschten Zweck.
Man kann sich "uberlegen, da"s sich mit der alternativen St"orungsrechnung
\Ref{2_g5} ebenfalls jede unit"are Transformation realisieren l"a"st.
In diesem Sinne sind die verschiedenen Varianten der St"orungsrechnung
also gleichwertig.
Die Uneindeutigkeit der St"orungsrechnung betrifft somit nur die Frage,
welcher St"oroperator ${\cal{B}}$ zur Beschreibung einer bestimmten
unit"aren Transformation verwendet werden soll. Dabei handelt es sich
nicht um eine grundlegende Frage; es geht lediglich darum,
mit welcher Methode der St"orungsrechnung der wechselwirkende fermionische
Projektor am besten mit Potentialen und klassischen Feldern beschrieben
werden kann.

In Abschnitt \ref{2_ab31} werden wir die Uneindeutigkeit der St"orungsrechnung
in allgemeinerem Rahmen untersuchen. Es wird sich zeigen,
da"s \Ref{2_29} die g"unstigste Definition f"ur $\tilde{k}_m$ ist.

\subsection{St"orungsrechnung im Ortsraum}
\label{2_ab22}
Mit \Ref{2_28}, \Ref{2_29} haben wir die St"orungsrechnung f"ur die
Spektralprojektoren zwar formal
durchgef"uhrt; wir haben aber noch keine Vorstellung davon, wie die
Distributionen $\tilde{p}_m(x,y), \tilde{k}_m(x,y)$ konkret aussehen.
Um den Zusammenhang zwischen der St"orung des Diracoperators und dessen
Spektralzerlegung besser zu verstehen, wurden die Gleichungen
\Ref{2_28}, \Ref{2_29} in den Anh"angen A-D f"ur
verschiedene St"oroperatoren ${\cal{B}}$ im Ortsraum ausgewertet.
Als Ergebnis erh"alt man Formeln f"ur $\tilde{p}_m(x,y), \tilde{k}_m(x,y)$,
an denen sich das singul"are Verhalten dieser Distributionen auf dem
Lichtkegel explizit ablesen l"a"st.
In diesem Abschnitt wollen wir die Technik dieser Rechnungen schematisch
beschreiben und einige wichtige Ergebnisse aus den Anh"angen
A-D zusammenstellen.

\subsubsection*{grundlegende Methode der Rechnung}
Um das Prinzip der Rechnungen zu erkl"aren, gen"ugt es, die St"orung
${\cal{B}}=e \Aslsh$ durch ein elektromagnetisches Potential zu betrachten.
Au"serdem beschr"anken wir uns zun"achst auf die St"orungsrechnung f"ur
$k_m$ und den Fall $m=0$.
Wir wollen also gem"a"s \Ref{2_29} die Gleichung
\[ \tilde{k}_0 \;=\; k_0 \:-\: e (s_0 \:\Aslsh\: k_0 \:+\:
	k_0 \:\Aslsh\: s_0) \]
in eine explizitere Form bringen. Zun"achst ziehen wir mit Hilfe von
\Ref{2_f3}, \Ref{2_c9} zwei partielle Ableitungen nach au"sen
\Equ{2_g1a}
=\; k_0 \:-\: e \:(i \Pdd)(S_0 \:\Aslsh\: K_0 \:+\:
	K_0 \:\Aslsh\: S_0)(i \Pdd) \spc .
\EndEqu
Die Distributionen $S_0, K_0$ haben nach \Ref{2_22}, \Ref{2_21} im
Ortsraum den Tr"ager auf dem Lichtkegel
\[ S_0(x) \;=\; -\frac{1}{4 \pi} \: \delta(x^2) \;\;\;\;,\spc
	K_0(x) \;=\; -\frac{i}{4 \pi^2} \: \delta(x^2) \: \epsilon(x^0)
	\spc . \]
Damit k"onnen wir die Terme $S_0 \:\Aslsh\: P_0$, $P_0 \:\Aslsh\: S_0$
in \Ref{2_g1a} als Integrale "uber den Schnitt zweier Lichtkegel umschreiben,
also formal
\Equ{2_g8}
(S_0 \:\Aslsh\: K_0)(x,y), \; (K_0 \:\Aslsh\: S_0)(x,y) \;=\;
	\int d^4z \; \delta((x-z)^2) \: \delta((y-z)^2) \; \Aslsh(z) \;
	\cdots \;\;\;\; .
\EndEqu
Wir nennen Integrale dieses Typs
{\em{Lichtkegelintegrale}}\index{Lichtkegelintegral}.
Lichtkegelintegrale lassen sich technisch recht gut handhaben.
Insbesondere kann man die Randwerte auf dem Lichtkegel explizit
berechnen; man erh"alt dabei Linienintegrale "uber das Argument,
also z.B.
\Equ{2_g2a}
\lim_{z \rightarrow y {\mbox{\scriptsize{ mit }}} (x-y)^2=0}
	(S_0 \:\Aslsh\: K_0)(x,z), \; (K_0 \:\Aslsh\: S_0)(x,z)
	\;=\; \inti d\lambda \; \Aslsh(\lambda y - (1-\lambda) x) \;
	\cdots \;\;\;\; .
\EndEqu

Das verbleibende Problem besteht darin, die beiden partiellen
Ableitungen $i \Pdd$ in \Ref{2_g1a} zu berechnen.
Wir beschreiben die Methode zur Einfachheit nur f"ur
$(S_0 \:\Aslsh\: K_0)(x,y)$ mit $x=0$ und
partielle Ableitungen nach der Variablen $y$:
Die Funktion $f(y):=(S_0 \:\Aslsh\: K_0)(0, y)$ ist harmonisch,
\Equ{2_g3b}
\Box f(y) \;=\; (S_0 \:\Aslsh\: (\Box K_0))(0, y) \;=\; 0 \spc ,
\EndEqu
au"serdem sind die Randwerte von $f$ auf dem Lichtkegel gem"a"s
\Ref{2_g2a} explizit bekannt. Falls $f$ eine glatte Funktion ist, haben
wir also
\Equ{2_g7}
f_{|\{y \;|\; y^2=0\}} \;=\; f_0 \spc {\mbox{mit einer gegebenen
	Funktion $f_0$}} \spc .
\EndEqu
Wir m"ussen die partiellen Ableitungen auf dem Lichtkegel
$\partial_j f_{|\{y \;|\; y^2=0\}}$ mit Hilfe von $f_0$ ausdr"ucken.
F"ur die Richtungsableitungen tangential
zum Lichtkegel k"onnen wir einfach $f_0$ ableiten
\Equ{2_g4a}
u^j \partial_j f(y) \;=\; u^j \partial_j f_0(y) \spc {\mbox{falls
	$y^2=0$ und $uy=0$}} \spc .
\EndEqu
Damit gen"ugt es, die Ableitung noch in einer beliebigen transversalen
Richtung zu bestimmen. Wir schreiben die Wellengleichung \Ref{2_g3b} mit
Lichtkegelkoordinaten $u=\frac{1}{2}(t+r)$, $v=\frac{1}{2}(t-r)$,
$\vartheta$, $\varphi$ um
\Equ{2_g5a}
\left( \frac{\partial^2}{\partial u \: \partial v} \:+\: \frac{1}{r}
	\frac{\partial}{\partial v} \:-\: \frac{1}{r} \frac{\partial}
	{\partial u} \:-\: \frac{1}{r^2} \: \Delta_s \right)
	f(u,v,\vartheta,\varphi) \;=\; 0 \spc ,
\EndEqu
dabei bezeichnet $\Delta_s$ den sph"arischen Laplace-Operator.
Auf dem oberen Lichtkegel k"onnen $\partial_u$, $\partial_\vartheta$,
$\partial_\varphi$ als tangentiale Ableitungen gem"a"s \Ref{2_g4a}
ausgef"uhrt werden. Da \Ref{2_g5a} nur erste Ableitungen nach $v$
enth"alt, k"onnen wir aus dieser Gleichung die transversale Ableitung
$\partial_v f$ bestimmen. Dazu schreiben wir \Ref{2_g5a} in der Form
\[ \frac{\partial}{\partial u} \left( u \: \frac{\partial}{\partial v}
	f(u,0,\vartheta,\varphi) \right) \;=\; \left(\frac{\partial}
	{\partial u} + \frac{1}{u} \: \Delta_s \right)
	f_0(u,0,\vartheta,\varphi) \;\;\;,\spc u>0, \: v=0 \]
um und integrieren "uber $u$
\begin{eqnarray}
\frac{\partial}{\partial v} f(u_0, 0, \vartheta, \varphi) &=&
	\frac{1}{u_0} \int_0^{u_0} \: \frac{\partial}{\partial u}
	\left( u \: \frac{\partial}{\partial v} f(u,0,\vartheta,\varphi)
	\right) \; du \nonumber \\
\label{eq:2_g6}
&=& \frac{1}{u_0} \int_0^{u_0} \left(\frac{\partial}
	{\partial u} + \frac{1}{u} \: \Delta_s \right)
	f_0(u,0,\vartheta,\varphi) \; du \spc .
\end{eqnarray}
Diese Rechnung in speziellen Koordinaten ist zwar nicht besonders
elegant, man sieht daran aber am schnellsten, da"s die partiellen
Ableitungen $\partial_j f$ auf dem Lichtkegel als Linienintegrale "uber
Ableitungen der Randwerte $f_0$ darstellbar sind. Da $\partial_j f$
wieder eine harmonische Funktion ist, kann das Verfahren iteriert
werden und liefert so auch Formeln f"ur die h"oheren partiellen
Ableitungen von $f$.

Nach leichter Verallgemeinerung dieser Methode k"onnen die Ableitungen
in \Ref{2_g1a} ausgef"uhrt werden. Gemeinsam mit \Ref{2_g2a} kann man das
Verhalten von $\tilde{k}_0(x,y)$ auf dem Lichtkegel mit geschachtelten
Linienintegralen "uber das Potential $\Aslsh$ und dessen partielle Ableitungen
beschreiben. Diese geschachtelten Linienintegrale k"onnen schlie"slich
in einfache Linienintegrale umgeschrieben werden.

Die Methode \Ref{2_g6} der Integration partieller Ableitungen l"angs
des Lichtkegels ist nicht neu. In \cite{Fl} beispielsweise wird damit
die Wellenausbreitung von Singularit"aten untersucht. Wir haben diese
Technik erweitert und zur expliziten Berechnung von $\tilde{k}_m(x,y)$,
$\tilde{p}_m(x,y)$ ausgenutzt.

Die bisherige Beschreibung war stark vereinfacht. Wir erw"ahnen kurz die
auftretenden Komplikationen und notwendigen Verallgemeinerungen:
Zun"achst einmal ist die harmonische Funktion $f$ i.a. nicht glatt,
sondern besitzt auf dem Lichtkegel Unstetigkeiten und
Singularit"aten. Man kann also nicht mit \Ref{2_g7} arbeiten, sondern
mu"s bestimmte Grenzwerte von $f(z)$ f"ur $z \rightarrow y$ und
$z^2 \neq 0$, $y^2=0$ betrachten. Dies wird in Anhang A
genau beschrieben. Im Fall $m \neq 0$ treten keine harmonischen
Funktionen auf, so da"s unsere Methode nicht mehr anwendbar ist.
Bei einer Entwicklung von $\tilde{k}_m$ nach $m$ lassen sich aber die
Beitr"age jeder Ordnung mit den
{\em{inneren Lichtkegelintegralen}}\index{Lichtkegelintegral, inneres}
bestimmen. In Anhang B werden diese Entwicklungsbeitr"age
bis zur Ordnung ${\cal{O}}(m^5)$ bestimmt.
Bei der St"orungsrechnung f"ur $p_0(x,y)$ tritt die Schwierigkeit auf,
da"s die zu \Ref{2_g8} analogen Terme
\[ (S_0 \:\Aslsh\: P_0)(x,y),\; (P_0 \:\Aslsh\: S_0)(x,y) \]
nicht mehr Lichtkegelintegrale sind, sondern wegen \Ref{2_20}
eine kompliziertere Form haben. Insbesondere treten nun
logarithmische Singularit"aten auf dem Lichtkegel auf, die in
Anhang C mit den {\em{verallgemeinerten
Lichtkegelintegralen}}\index{Lichtkegelintegral, verallgemeinertes}
behandelt werden.
In Anhang D haben wir schlie"slich Methoden der inneren
und verallgemeinerten Lichtkegelintegrale kombiniert, um $\tilde{p}_m$
bei Entwicklung nach $m$ bis zur Ordnung ${\cal{O}}(m^3)$ zu berechnen.

Die Methoden, die wir gerade f"ur das elektromagnetische Feld beschrieben
haben, lassen sich auf viele andere St"orungen des Diracoperators
"ubertragen. In den Anh"angen A-D werden allgemeine
Matrixpotentiale $(A^\alpha_\beta(x))_{\alpha, \beta = 1,\ldots, 4}$ und
verschiedene St"orungen durch Differentialoperatoren behandelt.

\subsubsection*{Beschreibung einzelner Ergebnisse aus Anhang
	A-D}
Wir wollen nun die Ergebnisse der St"orungsrechnung im Ortsraum
an verschiedenen Beispielen diskutieren.
Dazu beginnen wir wieder mit der St"orung ${\cal{B}}=e \Aslsh$ durch
ein elektromagnetisches Potential. F"ur die Distribution $k_0(x,y)$
ist der St"orungsbeitrag
\Equ{2_i5}
\Delta k_0(x,y) \;:=\; \tilde{k}_0(x,y) - k_0(x,y) \;=\; -e
	(s_0 \:\Aslsh\: k_0 \:+\: k_0 \:\Aslsh\: s_0)(x,y)
\EndEqu
in Theorem \ref{a1_theorem1} auf Seite \pageref{a1_theorem1}
explizit angegeben.
Wir m"ussen zun"achst die verwendete Notation erkl"aren:
Die Integrale $\int_x^y$, $\int_x^z$ bezeichnen Linienintegrale
l"angs der Verbindungsstrecken $\overline{xy}$ bzw. $\overline{xz}$
mit Integrationsvariable $\alpha$, also
\[ \int_x^y f \;\equiv\; \int_0^1 f(\alpha y + (1-\alpha) x) \: d\alpha
	\;\;\;,\spc
   \int_x^z f \;\equiv\; \int_0^1 f(\alpha z + (1-\alpha) x) \: d\alpha
	\spc . \]
F"ur die $\delta$-Distribution auf dem oberen und unteren Lichtkegel
wird die Schreibweise $l^\vee, l^\wedge$ verwendet,
\[ l^\vee(\xi) \;=\; \delta(\xi^2) \: \Theta(\xi^0) \;\;\;\;,\spc
	l^\wedge(\xi) \;=\; \delta(\xi^2) \: \Theta(-\xi^0) \spc . \]
Wie bereits im Theorem angegeben, ist $\xi \equiv y-x$,
$\zeta \equiv z-x$.
Das Symbol $\slint$ bezeichnet schlie"slich ein spezielles Lichtkegelintegral
vom Typ \Ref{2_g8}. Die genaue Definition ist an dieser Stelle nicht
entscheidend; wichtig ist, da"s die Funktion $\slint_x^y f$ nur dann
beitr"agt, wenn $y-x$ im oberen Lichtkegel liegt, also
\[ \lint_x^y f \;=\; 0 \;\;\;{\mbox{falls}}\;\;\;
	y \;\not \in \; \I^\vee_x := \left\{ y \;|\; (y-x)^2>0
	{\mbox{ und }} y^0-x^0>0 \right\} \spc . \]
Die Randwerte auf dem Lichtkegel sind wie in \Ref{2_g2a} Linienintegrale
"uber $f$, genauer
\[ \lim_{\I^\vee_x \ni u \rightarrow y \in \partial \I^\vee_x} \lint_x^u f \;=\;
	\frac{\pi}{2} \int_x^y f \spc . \]
Die Randwerte der Lichtkegelintegrale \Ref{a1_115} bis \Ref{a1_118} sind
in Satz \ref{a1_randwert} auf Seite \pageref{a1_randwert} zur besseren
"Ubersicht separat aufgef"uhrt.

Da die Formel von Theorem \ref{a1_theorem1} auf den ersten Blick
etwas un"ubersichtlich ist, wollen wir sie genauer diskutieren.
Zun"achst f"allt auf, da"s die einzelnen Beitr"age \Ref{a1_111} bis
\Ref{a1_118} nach der St"arke der Singularit"at auf dem Lichtkegel
geordnet sind: \Ref{a1_111} verh"alt sich auf dem Lichtkegel wie
$\delta^\prime(\xi^2)$, die Summanden \Ref{a1_112} bis \Ref{a1_114}
besitzen eine $\delta(\xi^2)$-Singularit"at, bei den Lichtkegelintegralen
\Ref{a1_115} bis \Ref{a1_118} tritt schlie"slich nur noch eine
Unstetigkeit $\sim \Theta(\xi^2)$ auf.
Wir haben also $\Delta k_0(x,y)$ um den Lichtkegel entwickelt und alle
Beitr"age bis zur Ordnung ${\cal{O}}(\xi^2)$ explizit berechnet, dabei
bezeichnet ${\cal{O}}(\xi^2)$ alle Distributionen $f(x,y)$ mit der
Eigenschaft, da"s $|(x-y)^{-2} \: f(x,y)|$ regul"ar ist.
Eine solche {\em{Lichtkegelentwicklung}}\index{Lichtkegelentwicklung}
ist sinnvoll, weil es
uns auf das Verhalten von $\Delta k_0(x,y)$ auf dem Lichtkegel ankommt.
Die schw"acher singul"aren Beitr"age werden in der Planckn"aherung stets
gegen"uber den st"arker singul"aren Beitr"agen vernachl"assigbar sein.
Alle nicht berechneten Beitr"age der Ordnung ${\cal{O}}(\xi^2)$ sind
f"ur uns tats"achlich irrelevant.

Wir wollen die einzelnen Beitr"age etwas detaillierter betrachten: Das
elektromagnetische Potential $A$ tritt lediglich im f"uhrenden Term
\Ref{a1_111} auf; alle anderen Summanden sind eichinvariant aus dem
Feldst"arketensor $F_{ij}$ und dem Stromtensor $j_k$ aufgebaut.
Um die Bedeutung von \Ref{a1_111} zu verstehen, betrachten wir den
Spezialfall $A_j=\partial_j \Lambda$. In diesem Fall kann das
elektromagnetische Potential durch die $U(1)$-Eichtransformation
\[ \Psi(x) \;\longrightarrow\; \exp (-ie \Lambda(x)) \: \Psi(x) \]
global zum Verschwinden gebracht werden; dabei geht
$\tilde{k}_m(x,y)$ in die freie Distribution $k_m(x,y)$ "uber.
Folglich k"onnen wir $\tilde{k}_m$ exakt angeben,
\[ \tilde{k}_m(x,y) \;=\; \exp \left( ie \Lambda(x) - ie \Lambda(y) \right)
	\; k_m(x,y) \spc , \]
und erhalten in erster Ordnung in den Potentialen
\Equ{2_i7}
\Delta k_m(x,y) \;=\; ie \: (\Lambda(x) - \Lambda(y)) \: k_m(x,y)
	\;=\; -ie \left( \int_x^y A_j \xi^j \right) \; k_m(x,y) \spc .
\EndEqu
Theorem \ref{a1_theorem1} liefert f"ur $m=0$ das gleiche Ergebnis, denn
wegen $F_{ij}=j_k=0$ verschwinden \Ref{a1_112} bis \Ref{a1_118}.
Wir sehen auf diese Weise, da"s der Beitrag \Ref{a1_111} f"ur das
richtige Eichtransformationsverhalten verantwortlich ist. Er wird
{\em{Eichterm}}\index{Eichterm} genannt. Wir bezeichnen die anderen Beitr"age
\Ref{a1_113}, \Ref{a1_114} als {\em{Feldst"arketerme}}\index{Feldst"arketerm}
und die Summanden \Ref{a1_112}, \Ref{a1_118} als
{\em{Stromterme}}\index{Stromterm}.
Die Lichtkegelintegrale \Ref{a1_115} bis \Ref{a1_117}
enthalten schlie"slich h"ohere Ableitungen $\Box F_{jk}$, $\Box j_k$
von Feldst"arke und Maxwellstrom.

Wir haben f"ur die Spektralprojektoren $\tilde{k}_m, \tilde{p}_m$
weitere Formeln "ahnlicher Form abgeleitet:
In Theorem \ref{a2_theorem2} auf Seite \pageref{a2_theorem2}
sind die Beitr"age von $\Delta k_m(x,y)$ bei Entwicklung nach $m$ bis
zur Ordnung ${\cal{O}}(m^5)$ aufgelistet, dabei ist
\[ \Theta^\vee(\xi) \;:=\; \Theta(\xi^2) \: \Theta(\xi^0) \;\;\;\;,\spc
	\Theta^\wedge(\xi) \;:=\; \Theta(\xi^2) \: \Theta(-\xi^0) \spc . \]
Der Eichterm \Ref{a2_c2}
beschreibt eine lokale Phasentransformation; zus"atzlich treten die
Feldst"arketerme \Ref{a2_c3}, \Ref{a2_c5}, \Ref{a2_c6}, \Ref{a2_c8},
\Ref{a2_c10}, \Ref{a2_c11} und die Stromterme \Ref{a2_c4}, \Ref{a2_c7},
\Ref{a2_c9}, \Ref{a2_c12} auf. In Satz \ref{a2_satzb9} auf Seite
\pageref{a2_satzb9} sind die Randwerte der Lichtkegelintegrale
\Ref{a2_c3} bis \Ref{a2_c7} angegeben.
F"ur die Distribution $p_m$ wurde die St"orungsrechnung in Theorem
\ref{a3_theorem1} auf Seite \pageref{a3_theorem1} und Theorem \ref{a4_thm1}
auf Seite \pageref{a4_thm1} durchgef"uhrt. Es treten ganz analoge
Beitr"age wie bei der St"orungsrechnung f"ur $k_m$ auf, nur haben die
Singularit"aten auf dem Lichtkegel eine andere Form.
Man beachte insbesondere das $\log(|\xi^2|)$-Verhalten der Stromterme
\Ref{a3_118a}, \Ref{a4_113f}, \Ref{a4_116f}; in Verallgemeinerung
der Schreibweise \Ref{2_ln} h"angt dabei die Konstante $C$ auch
vom Maxwellstrom ab.

An diesen Formeln l"a"st sich allgemein ablesen, da"s die Singularit"aten
von $\tilde{p}_m, \tilde{k}_m$ auf dem Lichtkegel mit steigender Ordnung
in $m$ schw"acher werden. F"ur die freien Distributionen haben wir das
schon auf Seite \pageref{2_abn} festgestellt; es gilt aber auch f"ur alle
St"orbeitr"age. Dazu vergleiche man z.B. die Stromterme \Ref{a1_112},
\Ref{a2_c7}, \Ref{a2_c12} oder die Feldst"arketerme \Ref{a2_c3},
\Ref{a2_c8}. Erst aufgrund dieser Tatsache ist bei einer
Lichtkegelentwicklung von $\tilde{p}_m, \tilde{k}_m$ eine Taylorentwicklung
nach $m$ sinnvoll. Beispielsweise sind die Stromterme $\sim m^4$,
\Ref{a2_c12}, in Planckn"aherung gegen"uber \Ref{a2_c7} vernachl"assigbar.
Tats"achlich werden alle nicht berechneten St"orungsbeitr"age der Ordnung
${\cal{O}}(m^5)$ f"ur uns keine Rolle spielen.

Wir werden zu Beginn des n"achsten Kapitels \ref{kap3} "uberlegen, warum
auch die Analogie der Entwicklungsformeln f"ur $\Delta k_m$
und $\Delta p_m$ allgemeinen Charakter hat. Im Moment gen"ugt es, wenn
wir dies empirisch festhalten. Wegen der Analogie werden wir f"ur den Rest
des Abschnitts oft nur die Distribution $\Delta k_m$ diskutieren.

Bei den gerade besprochenen Formeln f"ur $\Delta p_m(x,y), \Delta k_m(x,y)$
handelt es sich einfach um mathematische Ergebnisse l"angerer Rechnungen.
Trotzdem wollen wir versuchen zu beschreiben, wie die verschiedenen
Beitr"age bei unserer Vorstellung des fermionischen Projektors anschaulich
zu verstehen sind: Gem"a"s \Ref{2_a9} l"a"st sich mit den Distributionen
$p_m, k_m$ ein Diracsee im Vakuum beschreiben. Mit der St"orung ${\cal{B}}=e \Aslsh$
des Diracoperators f"uhren wir in das System ein "au"seres
elektromagnetisches Feld ein. Den Diracsee mit elektromagnetischem
Feld beschreiben wir analog zu \Ref{2_a9} durch $\tilde{p}_m, \tilde{k}_m$;
er ist formal aus Fermionen mit Wellenfunktionen $\tilde{\Psi}_a$ aufgebaut,
\Equ{2_h1}
\frac{1}{2} \: (\tilde{p}_m - \tilde{k}_m)(x,y) \;=\; \sum_a
	\tilde{\Psi}_a(x) \: \overline{\tilde{\Psi}_a(y)} \spc .
\EndEqu
Im Fall $A_j=\partial_j \Lambda$ wird lediglich die Phase der
Wellenfunktionen transformiert, was in \Ref{2_h1} zu einer
Phasenverschiebung f"uhrt. Diese Phasenverschiebung wird durch die
Eichterme beschrieben. Im allgemeinen Fall ist die Situation komplizierter,
weil auf die Fermionen zus"atzlich elektromagnetische Kr"afte wirken.
Die Wellenfunktionen $\tilde{\Psi}_a$ geben die quantenmechanische
Bewegung der Fermionen im "au"seren Feld an. Gem"a"s \Ref{2_h1}
wird diese kollektive Bewegung im Kraftfeld auch durch $\tilde{p}_m,
\tilde{k}_m$ beschrieben und f"uhrt insbesondere auf die Feldst"arke-
und Stromterme.

Damit gehen wir zur Besprechnung anderer St"orungen des Diracoperators
"uber. Die St"orung ${\cal{B}}=e \rho \Aslsh$ durch ein {\em{axiales
Potential}}\index{Potential, axiales} hat "Ahnlichkeit mit derjenigen durch
ein elektromagnetisches Feld. F"ur $m=0$ k"onnen wir die St"orungsrechnung
f"ur ${\cal{B}}=e \Aslsh$ "ubernehmen, denn nach den Umformungen
\begin{eqnarray*}
\Delta k_0 &=& -e \:(s_0 \:\rho \Aslsh\: k_0 \:+\: k_0 \:\rho \Aslsh\: s_0)
	\;=\; e \rho \: (s_0 \:\Aslsh\: k_0 \:+\: k_0 \:\Aslsh\: s_0) \\
\Delta p_0 &=& -e \:(s_0 \:\rho \Aslsh\: p_0 \:+\: p_0 \:\rho \Aslsh\: s_0)
	\;=\; e \rho \: (s_0 \:\Aslsh\: p_0 \:+\: p_0 \:\Aslsh\: s_0)
\end{eqnarray*}
brauchen wir nur die Ergebnisse von Theorem \ref{a1_theorem1} und
Theorem \ref{a3_theorem1} mit dem Faktor $-\rho$ zu multiplizieren.
Bei den St"orungsbeitr"agen h"oherer Ordnung in der Masse ist der
Zusammenhang nicht ganz so einfach; die Ergebnisse sind in
Theorem \ref{a2_thm13}, Satz \ref{a2_satz13} auf Seite \pageref{a2_thm13},
\pageref{a2_satz13} und in Theorem \ref{a4_thm13} auf Seite
\pageref{a4_thm13} aufgelistet.
In \Ref{a2_x8}, \Ref{a2_x9} sind $\svint$, $\swint$ innere Lichtkegelintegrale;
bei Lichtkegelentwicklung f"uhren sie auf Beitr"age $\sim \xi^2 \:
\Theta^\vee(\xi)$ bzw. $\sim \xi^2 \: \Theta^\wedge(\xi)$.
Anstelle der Eichterme treten nun die
{\em{Pseudoeichterme}}\index{Pseudoeichterme}
\Ref{a2_x1}, \Ref{a2_x2}, \Ref{a2_x7} auf.
Sie zeigen auf dem Lichtkegel das gleiche singul"are Verhalten, haben
aber mit
\begin{eqnarray*}
ie \int_x^y \rho \: A_j \xi^j \; \left[ k_m(x,y),
	p_m(x,y) \right] && \spc {\mbox{in gerader Ordnung in $m$}} \\
ie \int_x^y \rho \: \frac{1}{2}  \left[ \xi\slsh, \Aslsh \right] \;
	\left[ k_m(x,y), p_m(x,y) \right] &&
	\spc {\mbox{in ungerader Ordnung in $m$}}
\end{eqnarray*}
eine etwas andere Form. Die Summanden \Ref{a2_x3}, \Ref{a2_x4},
\Ref{a2_x5}, \Ref{a2_x8} modifizieren die Feldst"arke- und Stromterme
in \Ref{a2_x1}.
Als wesentlicher Unterschied zur St"orung durch ein elektromagnetisches
Potential kommen mit \Ref{a2_x6}, \Ref{a2_x9} zus"atzliche Beitr"age im
axialen Potential vor, die wir {\em{Massenterme}}\index{Massenterm}
nennen. Wie in der Einleitung beschrieben, h"angt das Auftreten der
Pseudoeichterme und Massenterme damit zusammen, da"s dem axialen
Potential keine lokale Eichsymmetrie entspricht.

Wir kommen zur St"orung durch Gravitationsfelder. Wie in \cite{Physdip}
erkl"art und in der Einleitung kurz wiederholt wurde, beschreiben
wir die Gravitation mit dem allgemeinen Diracoperator \Ref{1_7},
aus dem die Lorentzmetrik gem"a"s \Ref{1_6a} konstruiert wird.
Koordinaten- und Eichtransformationen sind "uber eine Untergruppe der
Eichgruppe miteinander verkn"upft.
Mit unserer St"orungsrechnung erster Ordnung k"onnen wir selbstverst"andlich
nur eine linearisierte Gravitationstheorie beschreiben. In den Anh"angen
A-D haben wir die verallgemeinerten Diracmatrizen in der Form
\Equ{2_k3}
G^j(x) \;=\; \gamma^j \:+\: \sum_{k=0}^3 \:h^{jk}(x) \: \gamma_k
\EndEqu
mit einem (ohne Einschr"ankung) symmetrischen Tensorfeld $h^{jk}$
angesetzt und $h^{jk}$ in linearer N"aherung behandelt. F"ur die
Lorentzmetrik \Ref{1_6a} erh"alt man
\Equ{2_k4}
g_{ij}(x) \;=\; \eta_{ij} \:-\: 2 h_{ij}(x)
\EndEqu
mit der Minkowski-Metrik $\eta_{ij}$. Der Ansatz \Ref{2_k3} f"uhrt gem"a"s
\Ref{1_7} auf eine St"orung des Diracoperators durch einen
Differentialoperator erster Ordnung (im Grad der Ableitung).

Die Ergebnisse der St"orungsrechnung sind f"ur $k_m(x,y)$
in Theorem \ref{a1_theorem2}, Theorem \ref{thm_kpm} auf den Seiten
\pageref{a1_theorem2}, \pageref{thm_kpm} und f"ur $p_m(x,y)$
in Theorem \ref{thm_gp0}, Theorem \ref{thm_kpm} auf den Seiten
\pageref{thm_gp0}, \pageref{thm_kpm} zusammengestellt.
F"ur die Diskussion dieser Formeln betrachten wir zun"achst den
Spezialfall, da"s die Metrik \Ref{2_k4} durch eine infinitesimale
Koordinatentransformation
\[ x^i \;\longrightarrow\; x^i \:+\: \kappa^i(x) \]
aus der Minkowski-Metrik hervorgeht, das bedeutet
\Equ{2_k5}
h_{ij} \;=\; \frac{1}{2} \: (\partial_i \kappa_j - \partial_j \kappa_i)
	\spc .
\EndEqu
Da wir das Verhalten von $k_m(x,y)$ bei Koordinatentransformationen
kennen, k"onnen wir $\tilde{k}_m(x,y)$ direkt angeben
\begin{eqnarray}
\tilde{k}_m(x,y) &=& k_m(x-\kappa(x), y-\kappa(y)) \nonumber \\
&=& k_m(x,y) \:-\: \kappa^k(x) \: \frac{\partial}{\partial x^k} k_m(x,y)
	\:-\: \kappa^k(y) \: \frac{\partial}{\partial y^k} k_m(x,y)
	\nonumber \\
\label{eq:2_k6}
&=& k_m(x,y) \:-\: (\kappa^k(y) - \kappa^k(x)) \:
	\frac{\partial}{\partial y^k} k_m(x,y) \spc .
\end{eqnarray}
Wir wollen untersuchen, auf welche Weise Theorem \ref{a1_theorem2},
Theorem \ref{thm_kpm} auf das gleiche Ergebnis f"uhrt.
Dazu setzen wir in \Ref{2_k5} den f"uhrenden Term \Ref{a1_212},
\Ref{a2_g11} ein
\begin{eqnarray*}
\lefteqn{ -\left(\int_x^y h^k_j \right) \: \xi^j \: \frac{\partial}
	{\partial y^k} k_m(x,y) \;=\; -\frac{1}{2} \int_x^y
	(\partial_j \kappa^k + \partial^k \kappa_j) \:\xi^j \:
	\frac{\partial}{\partial y^k} k_m(x,y) } \\
&=&-\int_x^y (\partial_j \kappa^k) \: \xi^j \: \frac{\partial}
	{\partial y^k} k_m(x,y) \:+\: \frac{1}{2}
	\int_x^y (\partial_j \kappa^k - \partial^k \kappa_j) \:
	\xi^j \: \frac{\partial}{\partial y^k} k_m(x,y) \spc .
\end{eqnarray*}
Das erste Integral kann partiell integriert werden,
\Equ{2_k7}
\;=\; -(\kappa^k(y) - \kappa^k(x)) \: \frac{\partial}{\partial y^k}
	k_m(x,y) \:+\: \frac{1}{2}
	\int_x^y (\partial_j \kappa^k - \partial^k \kappa_j) \:
	\xi^j \: \frac{\partial}{\partial y^k} k_m(x,y) \spc ,
\EndEqu
und liefert den gesuchten Ausdruck \Ref{2_k6}. Das Integral in \Ref{2_k7}
ist auf dem Lichtkegel schw"acher singul"ar als der erste Summand,
denn dort f"allt die st"arkste Singularit"at $\sim K_{m^2}^{\prime\prime}(\xi^2)$
der Distribution
\begin{eqnarray*}
\frac{\partial}{\partial y^k} k_m(x,y) &=& \frac{\partial}{\partial y^k} \:
	(i \Pdd_x + m) \: K_{m^2}(\xi^2) \\
&=& \frac{\partial}{\partial y^k} \left(-2i \xi\slsh
	K_{m^2}^\prime(\xi^2) \:+\: m \: K_{m^2}(\xi^2) \right) \\
&=& -4i \:\xi_k\:\xi\slsh\: K_{m^2}^{\prime \prime}(\xi^2) \:-\:
	2(i \gamma_k - m \xi_k) \:K_{m^2}^\prime(\xi^2)
\end{eqnarray*}
wegen der Antisymmetrie des Vektorfeldes
$\partial_j \kappa^k - \partial^k \kappa_j$ weg.
Wir kommen zu dem Schlu"s, da"s die Summanden\Ref{a1_212},
\Ref{a2_g11} die
f"uhrende Singularit"at auf dem Lichtkegel richtig beschreiben.
Die Beitr"age \Ref{a1_213}, \Ref{a1_214}, \Ref{a2_g12}, \Ref{a2_g17},
\Ref{a2_g18} werden ben"otigt, um den zweiten Summanden in \Ref{2_k7}
zu kompensieren. Weil die zugeh"orige Rechnung etwas aufwendiger
ist, wollen wir darauf hier nicht n"aher eingehen.
Da der Kr"ummungstensor verschwindet, fallen alle anderen Summanden
der Lichtkegelentwicklung weg.

Die Beitr"age \Ref{a1_212}, \Ref{a2_g11}, \Ref{a1_213}, \Ref{a1_214},
\Ref{a2_g12}, \Ref{a2_g17} sind allgemein f"ur das richtige
Verhalten bei Koordinatentransformationen verantwortlich.
Wir nennen sie {\em{Diffeomorphismenterme}}\index{Diffeomorphismenterm}.
Alle anderen Beitr"age sind kovariant aus dem Riemannschen
Kr"ummungstensor und dessen Ableitungen aufgebaut. Die Summanden
\Ref{a1_215}, \Ref{a1_220}, \Ref{a2_g13}, \Ref{a2_g19}, \Ref{a2_g20}
enthalten den Ricci- oder Einsteintensor und werden
{\em{Kr"ummungsterme}}\index{Kr"ummungsterm} genannt.

Mit den elektromagnetischen und axialen Potentialen sowie dem
Gravitationsfeld haben wir nun alle klassischen Felder untersucht, die
"ublicherweise im Zusammenhang mit der Diracgleichung betrachtet werden.
Um einen besseren "Uberblick zu bekommen, haben wir in
den Anh"angen weitere St"orungen des Diracoperators behandelt.
Wir erw"ahnen abschlie"send die St"orung durch skalare und pseudoskalare
Potentiale: Die {\em{skalare St"orung}}\index{St"orung, skalare}
${\cal{B}} = \Xi(x)$ mit einer reellen Funktion $\Xi$ wird f"ur $k_m$ in
Theorem \ref{theorem_sk0}, Theorem \ref{a2_theorem_sm} auf den Seiten
\pageref{theorem_sk0}, \pageref{a2_theorem_sm} und f"ur $p_m$ in
Theorem \ref{theorem_sp0}, Theorem \ref{a4_theorem_spm} auf den Seiten
\pageref{theorem_sp0}, \pageref{a4_theorem_spm} untersucht. Falls $\Xi$
nicht von $x$ abh"angt, k"onnen wir die skalare St"orung in der
Diracgleichung einfach mit der Masse zusammenfassen
\Equ{2_k4a}
(i \Pdd \:-\: (m - \Xi)) \: \tilde{\Psi}_a \;=\; 0 \spc .
\EndEqu
Daher ist einleuchtend, da"s die Beitr"age \Ref{a1_s2}, \Ref{a21_s2},
\Ref{a2_sa}, \Ref{a2_sb} eine lokale Massenverschiebung von $k_m$
beschreiben. Die restlichen Summanden enthalten Ableitungen von $\Xi$
und sind auf dem Lichtkegel schw"acher singul"ar.
Bei der {\em{pseudoskalaren St"orung}}\index{St"orung, pseudoskalare}
${\cal{B}}=i \rho \Xi(x)$ mit einer reellen Funktion $\Xi$ l"a"st sich der Fall
$m=0$ nach den Umformungen
\begin{eqnarray*}
\Delta k_0 &=& -i \: (s_0 \:\rho \Xi\: k_0 \:+\: k_0 \:\rho \Xi\: s_0)
	\;=\; i \rho \: (s_0 \:\Xi\: k_0 \:+\: k_0 \:\Xi\: s_0) \\
\Delta p_0 &=& -i \: (s_0 \:\rho \Xi\: p_0 \:+\: p_0 \:\rho \Xi\: s_0)
	\;=\; i \rho \: (s_0 \:\Xi\: p_0 \:+\: p_0 \:\Xi\: s_0)
\end{eqnarray*}
auf die St"orungsrechnung f"ur skalare St"orungen zur"uckf"uhren; wir
m"ussen nur die Ergebnisse von Theorem \ref{theorem_sk0}, Theorem
\ref{a2_theorem_sm} mit einem Faktor $-i \rho$ multiplizieren.
F"ur $m \neq 0$ sind die Ergebnisse in Theorem \ref{a2_theorem_psm}
und Theorem \ref{a4_theorem_pspm} auf Seite \pageref{a2_theorem_psm},
\pageref{a4_theorem_pspm} zusammengestellt.
Analog zu \Ref{2_k4a} haben wir f"ur konstantes $\Xi$ die Diracgleichung
\Equ{2_x4}
(i \Pdd \:-\: (m - i \rho \Xi)) \: \tilde{\Psi}_a \;=\; 0 \spc ,
\EndEqu
so da"s mit \Ref{a1_s2}, \Ref{a2_pz} eine dynamische axiale
Fermionmasse eingef"uhrt wird.

\subsection{St"orungsrechnung f"ur $P(x,y)$ mit Massenasymmetrie}
\label{2_ab23}
Nachdem die St"orungsrechnung f"ur $p_m, k_m$
durchgef"uhrt ist, k"onnen wir uns nun der St"orungsrechnung f"ur
den fermionischen Projektor zuwenden. Wir bezeichnen den
gest"orten fermionischen Projektor $\tilde{P}(x,y)$ wie in der
Einleitung mit einer zus"atzlichen Tilde.
Im Fall eines Diracsees \Ref{2_a9} brauchen wir nur $p_m, k_m$
durch die gest"orten Distributionen zu ersetzen
\[ \tilde{P}(x,y) \;=\; \frac{1}{2} \: (\tilde{p}_m - \tilde{k}_m)(x,y)
	\spc . \]
Im allgemeinen Fall \Ref{2_b8} ist die Situation komplizierter,
sobald der St"oroperator ${\cal{B}}$ nicht mit den Asymmetriematrizen
$X, Y$ kommutiert. In diesem Abschnitt beschr"anken wir uns mit der
Annahme $X=\1$ auf die St"orungsrechnung mit Massenasymmetrie,
im n"achsten Abschnitt wird der Fall mit zus"atzlicher chiraler
Asymmetrie behandelt.

\subsubsection*{ein einfaches Beispiel}
\label{2_beispiel}
Um das eigentliche Problem bei der St"orungsrechnung mit Massenasymmetrie
herauszuarbeiten, beginnen wir mit einem Beispiel und betrachten
bei Spindimension 8 und bei einer Teilchenfamilie den freien fermionischen Projektor
\Equ{2_i0}
P(x,y) \;=\; \bigoplus_{j=1}^2 \: \frac{1}{2} \: (p_{m_j} - k_{m_j})(x,y)
	\spc .
\EndEqu
In Analogie zum Standardmodell nennen wir den Blockindex $j$ auch
Isospinindex und den zugeh"origen Vektorraum $\C^2 = \C^2_\iso$
Isospinraum.

Im Fall $m_1=m_2=m$ ohne Massendrehung, also
\[ P(x,y) \;=\; \frac{1}{2} \: (p_m - k_m)(x,y) \:\otimes\: \1_\iso \spc , \]
l"a"st sich die St"orungsrechnung auf ganz naheliegende Weise
durchf"uhren: Der gest"orte Diracoperator hat bei der Zerlegung
$\C^8=\C^4 \otimes \C^2_\iso$ des Spinorraumes die Form
\[ i \Pdd \otimes \1 \:+\: {\cal{B}} \;=\; i \Pdd \otimes \1 \:+\:
	\sum_{i=0}^3 {\cal{B}}^i \otimes \sigma^i \]
mit Pauli-Matrizen $\sigma^0=\1, \vec{\sigma}$.
Wir setzen in Analogie zu \Ref{2_28}, \Ref{2_29}
\begin{eqnarray}
\Delta p_m[{\cal{B}}^i] &=& -s_m \:{\cal{B}}^i\: p_m \:-\:
	p_m \:{\cal{B}}\: s_m \nonumber \\
\Delta k_m[{\cal{B}}^i] &=& -s_m \:{\cal{B}}^i\: k_m \:-\:
	k_m \:{\cal{B}}\: s_m \spc {\mbox{und}} \nonumber \\
\label{eq:2_nn}
\tilde{P}(x,y) &=& \frac{1}{2} \: (p_m - k_m)(x,y) \:\otimes\: \1
	\:+\: \sum_{i=0}^3 \: \frac{1}{2} \left( \Delta p_m[{\cal{B}}^i]
	- \Delta k_m[{\cal{B}}^i] \right)(x,y) \:\otimes\: \sigma^i
	\; . \spc
\end{eqnarray}
"Aquivalent l"a"st sich die St"orungsrechnung in Verallgemeinerung
von Satz \ref{a2_satz1} auch mit der unit"aren Transformation
\Equ{2_ia}
U[{\cal{B}}] \;=\; 1 \:-\: \sum_{i=0}^3 \int_{\sR \cup i \sR} dm \;
	s_m \:{\cal{B}}^i\: p_m \:\otimes\: \sigma^i
\EndEqu
beschreiben, also
\Equ{2_i1}
\tilde{P}(x,y) \;=\; (U \:P\: U^*)(x,y) \spc .
\EndEqu

Im Fall $m_1 \neq m_2$ mit Massendrehung k"onnte man versuchen,
einfach die unit"are Transformation \Ref{2_i1} auf den freien fermionischen
Projektor \Ref{2_i0} anzuwenden.
Im Operatorkalk"ul erh"alt man dabei unter Verwendung von \Ref{2_yz}
\begin{eqnarray*}
\tilde{P} - P &=& \sum_{i=0}^3 \int_{\sR \cup i \sR} dm \;
	\left( -(s_m \:{\cal{B}}^i\: p_m \:\otimes\: \sigma^1) \:P \:+\:
	P \: (s_m \:{\cal{B}}^i\: p_m \:\otimes\: \sigma^1) \right) \\
&=& \sum_{i=0}^3 \int_{\sR \cup i \sR} dm \;
	\left( -(s_m \:{\cal{B}}^i\: p_m \:\otimes\: \sigma^1) \:P \:-\:
	P \: (p_m \:{\cal{B}}^i\: s_m \:\otimes\: \sigma^1) \right) \spc .
\end{eqnarray*}
Wir betrachten die diagonalen und au"serdiagonalen Isospinbeitr"age
getrennt. Nach Einsetzen von \Ref{2_i0} k"onnen wir die Integration
"uber $m$ mit Hilfe von \Ref{2_11}, \Ref{2_12} ausf"uhren und erhalten
bei einer Blockmatrixdarstellung im Isospin und mit der Abk"urzung
$l_m=\frac{1}{2}(p_m-k_m)$
\begin{eqnarray}
\label{eq:2_ba}
&=&-\sum_{i=0,3} \left( \begin{array}{cc} s_{m_1} & 0 \\ 0 & s_{m_2}
	\end{array} \right) {\cal{B}}^i \sigma^i
	\left( \begin{array}{cc} l_{m_1} & 0 \\ 0 & l_{m_2}
	\end{array} \right) +
	\left( \begin{array}{cc} l_{m_1} & 0 \\ 0 & l_{m_2}
	\end{array} \right) {\cal{B}}^i \sigma^i
	\left( \begin{array}{cc} s_{m_1} & 0 \\ 0 & s_{m_2}
	\end{array} \right) \;\;\;\;\;\;\;\;\;\; \\
\label{eq:2_bb}
&&-\sum_{i=1,2} \left( \begin{array}{cc} s_{m_2} & 0 \\ 0 & s_{m_1}
	\end{array} \right) {\cal{B}}^i \sigma^i
	\left( \begin{array}{cc} l_{m_1} & 0 \\ 0 & l_{m_2}
	\end{array} \right) +
	\left( \begin{array}{cc} l_{m_1} & 0 \\ 0 & l_{m_2}
	\end{array} \right) {\cal{B}}^i \sigma^i
	\left( \begin{array}{cc} s_{m_2} & 0 \\ 0 & s_{m_1}
	\end{array} \right) \\
\label{eq:2_i2}
&=& -\sum_{i=0,3} \sigma^i \; \left( \begin{array}{cc}
	s_{m_1} {\cal{B}}^i l_{m_1} +
	l_{m_1} {\cal{B}}^i s_{m_1} & 0 \\
	0 & s_{m_2} {\cal{B}}^i l_{m_2} +
	l_{m_2} {\cal{B}}^i s_{m_2} \end{array} \right) \\
\label{eq:2_i3}
&&- \sum_{i=1,2} \sigma^i \; \left( \begin{array}{cc}
	s_{m_1} {\cal{B}}^i l_{m_1} +
	l_{m_2} {\cal{B}}^i s_{m_2} & 0 \\
	0 & s_{m_2} {\cal{B}}^i l_{m_2} +
	l_{m_1} {\cal{B}}^i s_{m_1} \end{array} \right) \spc .
\end{eqnarray}
Die Matrixeintr"age in \Ref{2_i2} sind genau von der Form wie bei
der St"orungsrechnung f"ur $p_m, k_m$. In den au"serdiagonalen Beitr"agen
\Ref{2_i3} kommen dagegen Kombinationen der Form
\Equ{2_i4}
s_{m_1} \:{\cal{B}}\: p_{m_1} \:+\: p_{m_2} \:{\cal{B}}\: s_{m_2}
	\;\;\;\;,\spc
s_{m_1} \:{\cal{B}}\: k_{m_1} \:+\: k_{m_2} \:{\cal{B}}\: s_{m_2}
\EndEqu
vor, bei welchen in beiden Summanden verschiedene Massenparameter auftreten.

\subsubsection*{nichtlokale Linienintegrale}
Wir wollen die Terme in \Ref{2_i4} exemplarisch an dem Ausdruck
\Equ{2_i6}
\Delta k_{m_1, m_2} \;:=\; -e \: (s_{m_1} \:\Aslsh\: k_{m_1} \:+\:
	k_{m_2} \:\Aslsh\: s_{m_2})
\EndEqu
mit einem Potential $A$ studieren. F"ur $m_1=m_2$ stimmt $\Delta k_{m_1, m_2}$
mit $\Delta k_m$, \Ref{2_i5}, "uberein, so da"s dieser Grenzfall in den Anh"angen
und im vorigen Abschnitt ausf"uhrlich behandelt wurde. Wir konzentrieren
uns im folgenden auf die Eichterme, die wir mit dem Symbol `$\asymp$'
kennzeichnen, also
\Equ{2_i8}
\Delta k_{m, m}(x,y) \;\asymp\; -ie \left(\int_x^y A_j \xi^j \right)
	\; k_m(x,y) \spc .
\EndEqu
Die Eichterme beschreiben gem"a"s \Ref{2_i7} das Verhalten von $k_m$
bei Eichtransformationen.
Man kann sich die Eichsymmetrie der St"orungsrechnung f"ur $k_m$ auch
im Operatorkalk"ul klarmachen: Im Spezialfall $A_j=\partial_j \Lambda$
hat man
\begin{eqnarray*}
\Delta k_{m,m} &=& -e \: (s_m \:(\Pdd \Lambda)\: k_m \:+\:
	k_m \:(\Pdd \Lambda)\: s_m) \\
&=& ie \left( s_m \:[i \Pdd - m, \: \Lambda]\: k_m \:+\:
	k_m \:[i \Pdd - m, \: \Lambda]\: s_m \right) \spc ,
\end{eqnarray*}
wobei die Funktion $\Lambda$ im Kommutator als Multiplikationsoperator aufgefa"st wird.
Wir nutzen aus, da"s der Operator $(i \Pdd - m)$ mit $s_m, k_m$ kommutiert
und wenden \Ref{2_a99}, \Ref{2_e0} an
\begin{eqnarray}
&=& ie \left( ((i \Pdd - m) s_m) \:\Lambda\: k_m \:-\:
	s_m \:\Lambda\: ((i \Pdd - m) k_m) \right. \nonumber \\
&& \; +\left. ((i \Pdd - m) k_m) \:\Lambda\: s_m \:-\:
	k_m \:\Lambda\: ((i \Pdd - m) s_m) \right) \nonumber \\
\label{eq:2_j6}
&=& ie \: (\Lambda \: k_m \:-\: k_m \:\Lambda) \spc .
\end{eqnarray}
Im Ortsraum stimmt diese Formel mit \Ref{2_i8} "uberein.

Wir wissen im Moment nicht, wie das Analogon zu den Eichtermen \Ref{2_i8}
f"ur $\Delta k_{m_1, m_2}$ und $m_1 \neq m_2$ aussieht.
Im Grenzfall $A_j=\partial_j \Lambda$ k"onnen wir aber die Operatorrechnung
\Ref{2_j6} "ubertragen und erhalten
\begin{eqnarray*}
\Delta k_{m_1, m_2} &=& ie \left(
	s_{m_1} \:[i \Pdd-m_1, \:\Lambda]\: k_{m_1} \:+\:
	k_{m_2} \:[i \Pdd-m_2, \:\Lambda]\: s_{m_2} \right) \\
&=& ie \left( \Lambda \: k_{m_1} \:-\: k_{m_2} \: \Lambda \right) \spc ,
\end{eqnarray*}
also im Ortsraum
\Equ{2_i9}
\Delta k_{m_1, m_2}(x,y) \;=\; ie \left( \Lambda(x) \: k_{m_1}(x,y) \:-\:
	\Lambda(y) \: k_{m_2}(x,y) \right) \spc .
\EndEqu
Nach dieser Rechnung ist plausibel (und wird in Anhang C
bewiesen), da"s die Eichterme die Form
\begin{eqnarray}
\Delta k_{m_1, m_2}(x,y) &\asymp& -\frac{ie}{2} \: \int_{-\infty}^\infty
	d\lambda \; \epsilon(\lambda) \: A_j(\lambda y + (1-\lambda) x)
	\: \xi^j \; k_{m_1}(x,y) \nonumber \\
\label{eq:2_q1}
&&+\frac{ie}{2} \: \int_{-\infty}^\infty d\lambda \; \epsilon(\lambda-1)
	\: A_j(\lambda y + (1-\lambda) x) \: \xi^j \; k_{m_2}(x,y) \spc
\end{eqnarray}
haben. Als wesentlicher Unterschied zu \Ref{2_i8} reichen die
Linienintegrale "uber die Potentiale nun bis ins Unendliche.
Wir nennen in der St"orungsrechnung auftretende unbeschr"ankte
Linienintegrale allgemein
{\em{nichtlokale Linienintegrale}}\index{nichtlokales Linienintegral}.

Das Auftreten nichtlokaler Linienintegrale l"a"st sich auch
mit einer Eichtransformation im Isospinraum einsehen:
Wir gehen zur"uck zur St"orungsrechnung f"ur $P(x,y)$,
\Ref{2_i2}, \Ref{2_i3}, und betrachten als St"oroperator die
$U(2)$-Potentiale ${\cal{B}}^i=\Pdd \Lambda^i$ mit reellen Funktionen
$\Lambda^i$. Analog zur Rechnung \Ref{2_j6} folgt
\Equ{2_j0}
\tilde{P}(x,y) \;=\; (\1 \otimes V(x)) \:P(x,y)\: (\1 \otimes V(y)^*)
	\;\;\;\; {\mbox{mit}} \;\;\;\; V(x) \;=\; 1 + i \sum_{k=0}^3
	\Lambda^k(x) \: \sigma^k \;\;\; .
\EndEqu
Wir betrachten f"ur festes $x, y$ die spezielle Situation, da"s
das Matrixfeld $V$ in einer Umgebung ${\cal{U}}$ von
$\overline{xy}$ konstant ist, also $V_{|{\cal{U}}}=V_0$.
Da der freie fermionische Projektor f"ur $m_1 \neq m_2$ auf dem Isospinraum
nicht trivial ist, h"angt der gest"orte Projektor gem"a"s \Ref{2_j0}
explizit von $V_0$ ab. Auf der anderen Seite verschwinden die St"orpotentiale
${\cal{B}}^i=\Pdd \Lambda^i$ als partielle Ableitungen von $V$ in der Menge
${\cal{U}}$ und damit l"angs $\overline{xy}$.
Insgesamt folgt, da"s in $\tilde{P}(x,y)$ auch das Potential au"serhalb der
Verbindungsstrecke $\overline{xy}$ eingehen mu"s.
In der St"orungsrechnung zeigt sich dies daran, da"s in den
au"serdiagonalen Matrixelementen \Ref{2_i3} bei Termen der Form
\Ref{2_i9} nichtlokale Linienintegrale vorkommen.

\subsubsection*{das Problem bei nichtlokalen Linienintegralen}
Wir wollen nun die Schwierigkeit der nichtlokalen Linienintegrale allgemein
(also ohne Bezug auf die St"orungsrechnung mit Massenasymmetrie)
diskutieren.

Die nichtlokalen Linienintegrale f"uhren auf ein prinzipielles Problem,
wenn wir einen Zusammenhang zu klassischen Feldgleichungen herstellen
wollen: In Abschnitt \ref{4_ab5} werden wir f"ur den
klassischen Grenzfall der Gleichungen der diskreten Raumzeit die
Lichtkegelentwicklung eines zusammengesetzten Ausdrucks in
$\tilde{P}(x,y)$ untersuchen. Wir brauchen an dieser Stelle noch keine
Einzelheiten dieser Rechnungen vorwegzunehmen; es gen"ugt zu wissen,
da"s es dabei letztlich nur auf das Verhalten von $\tilde{P}(x,y)$ am
Ursprung, also f"ur $x \approx y$ ankommt.
Unter dieser Annahme kann man den klassischen Grenzfall bereits
schematisch verstehen; wir betrachten als Beispiel das
elektromagnetische Feld. Die im vorigen Abschnitt angesprochenen Stromterme
\Ref{a1_118}, \Ref{a3_118a} liefern zu $\tilde{p}_m(x,y),
\tilde{k}_m(x,y)$ und damit auch zu $\tilde{P}(x,y)$
einen Beitrag der Form
\Equ{2_l1}
\int_x^y j_k \gamma^k \; ({\mbox{Distribution in $(y-x)$}}) \spc .
\EndEqu
Ein einzelner Fermionzustand f"uhrt zu einer St"orung $\Psi(x) \:
\overline{\Psi(y)}$ des fermionischen Projektors $\tilde{P}(x,y)$.
Im Grenzfall $x=y$ sind diese Beitr"age proportional zum Maxwell-
und Diracstrom $j^k(x)$, $\overline{\Psi(x)} \gamma^k \Psi(x)$.
Wir k"onnen hoffen, bei geeigneter Wahl der Gleichungen der diskreten
Raumzeit eine Relation zwischen diesen Vektorfeldern, genauer gesagt die
Maxwellgleichungen
\[ j^k(x) \;=\; e \: \overline{\Psi(x)} \:\gamma^k\: \Psi(x) \spc , \]
zu erhalten.
Im Fall mit nichtlokalen Linienintegralen treten zus"atzlich zu
\Ref{2_l1} unbeschr"ankte Integrale "uber die Potentiale auf, beispielsweise
\Equ{2_l2}
\inti d\lambda \: \epsilon(\lambda) \: \gamma^k \:
	j_k(\lambda y + (1-\lambda) x) \;
	({\mbox{Distribution in $(y-x)$}}) \spc .
\EndEqu
In dieses Integral geht auch f"ur $x \approx y$ der Maxwellstrom l"angs
einer Geraden ein, die bis ins Unendliche l"auft. Folglich k"onnen wir
im Limes $y \rightarrow x$ keine lokalen Gleichungen mehr erwarten.
Durch nichtlokale Linienintegrale scheint also die Lokalit"at der
klassischen Feldgleichungen gef"ahrdet.

Bei genauerer Untersuchung der St"orbeitr"age wird dieses Problem
noch deutlicher: Im Linienintegral in \Ref{2_l1} existiert der Limes
$y \rightarrow x$. Um zu sehen, ob das im nichtlokalen Linienintegral
\Ref{2_l2} auch der Fall ist, betrachten wir f"ur festes $x$ den Punkt
$y$ l"angs der Geraden $y=\alpha z + (1-\alpha) x$.
Mit einer Variablensubstitution erh"alt man
\[ \inti d\lambda \: \epsilon(\lambda) \: \gamma^k \:
	j_k(\lambda y+(1-\lambda)x)
	\;=\; \frac{1}{\alpha} \: \inti d\lambda
	\: \epsilon(\lambda) \: \gamma^k \: j_k(\lambda z + (1-\lambda x))
	\spc . \]
Man sieht an dieser Formel, da"s das nichtlokale Linienintegral f"ur
$y \rightarrow x$ einen
Pol besitzt. Die beiden Linienintegrale in \Ref{2_l1}, \Ref{2_l2} zeigen
also ein unterschiedliches Verhalten am Ursprung. Dies hat zur Folge, da"s
Beitr"age der Form \Ref{2_l1}, \Ref{2_l2} in den Gleichungen der diskreten
Raumzeit auseinandergehalten werden k"onnen. Anders ausgedr"uckt, liefern
die Gleichungen der diskreten Raumzeit unabh"angige Bedingungen f"ur
diese Beitr"age.
Wie gerade beschrieben wurde, k"onnen wir hoffen, da"s die Bedingungen
an die Beitr"age der Form \Ref{2_l1} klassische Feldgleichungen liefern.
Da die Terme \Ref{2_l2} nicht durch andere Beitr"age (etwa Diracstr"ome)
kompensiert werden k"onnen, implizieren die Bedingungen an diese Terme,
da"s keine nichtlokalen Linienintegrale "uber den Maxwellstrom
auftreten d"urfen.
Mit dieser sch"arferen Bedingung ist die Lokalit"at der klassischen
Gleichungen wieder sichergestellt: wenn alle nichtlokalen Linienintegrale
\Ref{2_l2} als Folge der Gleichungen der diskreten Raumzeit verschwinden,
d"urfen lediglich Beitr"age der Form \Ref{2_l1} auftreten, was zwangsl"aufig
auf lokale Feldgleichungen f"uhrt.

Diese Argumentation ist nat"urlich nicht v"ollig befriedigend, weil wir
Ergebnisse sp"aterer Rechnungen qualitativ vorwegnehmen mu"sten.
Au"serdem haben wir uns zur Einfachheit auf die Stromterme beschr"ankt
(die "Uberlegung gilt analog f"ur viele andere St"orungsbeitr"age,
z.B. Massenterme, Kr"ummungsterme oder Pseudoeichterme).
Unsere Diskussion der Stromterme \Ref{2_l1}, \Ref{2_l2} dient auch nur
als Motivation f"ur die allgemeine mathematische\index{Lokalit"atsforderung}
\Equ{2_fl}
{\mbox{ {\bf{Forderung:}} $\;\;$
	\parbox[t]{10cm}{In der St"orungsrechnung d"urfen
	keine nichtlokalen Linienintegrale auftreten.}}}
\EndEqu
An dieser Forderung werden wir w"ahrend der gesamten Arbeit festhalten;
wir werden sie aber an verschiedenen Stellen hinterfragen und mit weiteren
Argumenten st"utzen. Sie wird sowohl Bedingungen an die Methode der
St"orungsrechnung als auch an den St"oroperator ${\cal{B}}$ liefern.

\subsubsection*{Durchf"uhrung der St"orungsrechnung}
Gem"a"s unserer Forderung \Ref{2_fl} mu"s die St"orungsrechnung
\Ref{2_ba}, \Ref{2_bb} so modifiziert werden, da"s keine nichtlokalen
Linienintegrale mehr auftreten. Wir f"uhren die Konstruktion gleich
allgemein durch: Wir f"ugen in die Reihenentwicklungen von
$p_m, k_m, s_m$
\begin{eqnarray*}
p_m &=& \sum_{l=0}^\infty m^l \: p^{(l)} \:+\: \log(m) \sum_{l=2}^\infty
	m^l \: q^{(l)} \\
k_m &=& \sum_{l=0}^\infty m^l \: k^{(l)} \;\;\;,\spc
	s_m \;=\; \sum_{l=0}^\infty m^l \: s^{(l)}
\end{eqnarray*}
(wir haben zur Deutlichkeit die $\log m$-Terme gem"a"s \Ref{2_zz}
mitber"ucksichtigt)\\
die Massenmatrix $Y$ ein und definieren
\begin{eqnarray}
\label{eq:2_o0}
p_{[m]} &:=& \sum_{l=0}^\infty m^l \:Y^l p^{(l)} \:+\: \log(m Y)
	\sum_{l=2}^\infty m^l \:Y^l\: q^{(l)} \\
\label{eq:2_o1}
k_{[m]} &:=& \sum_{l=0}^\infty m^l \:Y^l\: k^{(l)} \;\;\;,\spc
	s_{[m]} \;:=\; \sum_{l=0}^\infty m^l \:Y^l\: s^{(l)} \spc .
\end{eqnarray}
Den Index `$[m]$' lassen wir auch oft weg.
Der freie fermionische Projektor \Ref{2_d5}, \Ref{2_d6} kann
in der Form
\Equ{2_o2}
P(x,y) \;=\; \frac{1}{2} \: \Tr_{\cal{F}} \left( (p - k)(x,y) \right)
\EndEqu
geschrieben werden; nach Absorbieren der $\log m$-Terme mit der
Notation \Ref{2_ln} erh"alt man \Ref{2_b8}.
Die Distributionen $p, k$ und $s$ sind L"osungen bzw. die Greensfunktion
der Diracgleichung mit Massenmatrix
\begin{eqnarray*}
(i \Pdd - m Y) \:p &=& (i \Pdd - m Y) \:k \;=\; 0 \\
(i \Pdd - m Y) \:s &=& \1 \spc .
\end{eqnarray*}
Wir setzen in Analogie zu \Ref{2_28}, \Ref{2_29}
\Equ{2_m0}
\tilde{p} \;=\; p - s \:{\cal{B}}\: p - p \:{\cal{B}}\: s \;\;\;,\spc
	\tilde{k} \;=\; k - s \:{\cal{B}}\: k - k \:{\cal{B}}\: s
\EndEqu
und schlie"slich
\Equ{2_m1}
\tilde{P}(x,y) \;=\; \frac{1}{2} \: \Tr_{\cal{F}} \left(
	(\tilde{p} - \tilde{k})(x,y) \right) \spc .
\EndEqu
Diese St"orungsrechnung l"a"st sich auch mit einer unit"aren Transformation
beschreiben:
\begin{Satz}
\label{a2_satz5}
Der Operator
\Equ{2_n7}
U[{\cal{B}}] \;=\; 1 - \int_{\sR \cup i \sR} \!\!\! dm \;
	s_{[m]} \: {\cal{B}} \: p_{[m]}
\EndEqu
ist als Operator auf $H \otimes \C^f$ (in erster Ordnung in ${\cal{B}}$)
unit"ar und
\begin{eqnarray}
\tilde{p} &=& U \: p \: U^* \;\;\;,\spc \tilde{k}
	\;=\; U \: k \: U^* \\
\label{eq:2_n5}
\tilde{P} &=& \frac{1}{2} \: \Tr_{\cal{F}} \left( U \: (p-k) \: U^* \right)
\end{eqnarray}
Zu jeder infinitesimalen unit"aren Transformation $V=1+iA$ (mit einem
hermiteschen Operator $A$) gibt es einen St"oroperator ${\cal{B}}$
mit $U[{\cal{B}}]=V$.
\end{Satz}
{\Beweis}
Da $Y$ gem"a"s \Ref{2_b7} eine Diagonalmatrix ist, kann der Beweis von
Satz \ref{a2_satz1} w"ortlich "ubernommen werden.
\QED
Man beachte, da"s die Operatoren $s, k, p, U$ nicht auf dem
Zustandsraum $H$, sondern auf $H \otimes \C^f$ wirken. Damit "ubernimmt
der Flavour-Raum, den wir zun"achst nur zur Indizierung der Familien
im freien fermionischen Projektor eingef"uhrt haben, in der St"orungsrechnung
mit Massenasymmetrie eine wichtigere Rolle.

\subsubsection*{physikalische Interpretation}
Zur Diskussion der St"orungsrechnung \Ref{2_m0}, \Ref{2_m1} gehen wir
zur"uck zu Beispiel \Ref{2_i0}. Im Gegensatz zu \Ref{2_ba}, \Ref{2_bb}
hat man nun f"ur den gest"orten freien Projektor
\begin{eqnarray}
\lefteqn{ \tilde{P} \;=\; P } \nonumber \\
\label{eq:2_m2}
&& -\sum_{i=0}^3
	\left( \begin{array}{cc} s_{m_1} & 0 \\ 0 & s_{m_2}
		\end{array} \right) {\cal{B}}^i \sigma^i
	\left( \begin{array}{cc} l_{m_1} & 0 \\ 0 & l_{m_2}
		\end{array} \right) +
	\left( \begin{array}{cc} l_{m_1} & 0 \\ 0 & l_{m_2}
		\end{array} \right) {\cal{B}}^i \sigma^i
	\left( \begin{array}{cc} s_{m_1} & 0 \\ 0 & s_{m_2}
		\end{array} \right) \; . \spc
\end{eqnarray}
Als wesentlicher Unterschied sind nun auch die im Isospin au"serdiagonalen
Beitr"age zu $\tilde{P}$ in $s, l$ symmetrisch, anstelle von \Ref{2_i4} treten
Kombinationen der Form
\Equ{2_m3}
s_{m_1} \:{\cal{B}}\: p_{m_2} \:+\: p_{m_1} \:{\cal{B}}\: s_{m_2}
	\;\;\;\;,\spc
s_{m_1} \:{\cal{B}}\: k_{m_2} \:+\: k_{m_1} \:{\cal{B}}\: s_{m_2}
\EndEqu
auf. Durch Lichtkegelentwicklung kann man explizit verifizieren,
da"s in \Ref{2_m3} keine nichtlokalen Linienintegrale auftreten, worauf
wir aber hier nicht n"aher eingehen.

Wir wollen versuchen, den Unterschied zwischen \Ref{2_ba}, \Ref{2_bb} und
der St"orungsrechnung \Ref{2_m2} anschaulich zu interpretieren. Dazu
beginnen wir mit dem Grenzfall \Ref{2_nn} ohne Massenasymmetrie, in welchem
die verschiedenen Varianten der St"orungsrechnung "ubereinstimmen.
Die Eigenzust"ande des gest"orten Diracoperators sind Linearkombinationen
der freien Eigenzust"ande, also formal
\Equ{2_n8}
\tilde{\Psi}_a \;=\; \sum_b c_{ab} \: \Psi_b \spc .
\EndEqu
mit komplexen Koeffizienten $c_{ab}$.
Die Mischung der freien Zust"ande findet sowohl zwischen Zust"anden der
gleichen Masse als auch zwischen Zust"anden verschiedener Masse statt.
F"ur die St"orungsrechnung eines Zustandes $\Psi_a$ mit Masse $m$ spalten
wir die Summe \Ref{2_n8} in der Form
\Equ{2_n9a}
\tilde{\Psi}_a \;=\; \sum_{b \:|\: (i \partial \!\!\!/ - m) \Psi_b=0} c_{ab} \:
	\Psi_b \;+\;
	\sum_{b \:|\: (i \partial \!\!\!/ - m) \Psi_b \neq 0} c_{ab} \: \Psi_b
\EndEqu
auf. Diese Gleichung ist wegen des kontinuierlichen Spektrums des Diracoperators
nat"urlich mathematisch nicht sinnvoll, sie ist f"ur ein anschauliches Verst"andnis
der St"orungsrechnung aber dennoch n"utzlich.
Die zweite Summe in \Ref{2_n9a} beschreibt die Ver"anderung der
Wellenfunktion $\Psi(x)$ durch die "au"sere St"orung ${\cal{B}}$; durch die
erste Summe werden die Zust"ande des entarteteten Unterraumes miteinander
gemischt.
Wir f"uhren jetzt in Gedanken ein Streuexperiment durch. In diesem Fall geht
$\tilde{\Psi}_a$ f"ur $t \rightarrow \pm \infty$ in einen Eigenzustand
des freien Diracoperators "uber, also
\Equ{2_n9}
(i \Pdd - m) \: \tilde{\Psi}_a(\vec{x}, t) \;=\; 0 \spc {\mbox{f"ur $t < t_0$
	und $t > t_1$}}
\EndEqu
(die asymptotischen Zust"ande f"ur $t \rightarrow \mp \infty$ werden in der
Streutheorie oft ``in-'' und ``out-Zust"ande'' genannt; die Streuung findet im
Zeitraum $t_0 \leq t \leq t_1$ statt).\\
Zur Beschreibung des Streuexperimentes kommt es nur auf die erste Summe
in \Ref{2_n9a} an. Die zweite Summe ist wichtig, um die genauen Vorg"ange
w"ahrend des Streuprozesses zu studieren; da wir in der Asymptotik
$t \rightarrow \pm \infty$ aber L"osungen der freien Diracgleichung erhalten,
f"allt die zweite Summe in diesem Grenzfall weg.
In diesem Sinne k"onnen wir die St"orungsrechnung \Ref{2_nn} als
Wechselwirkung zwischen den Fermionen (und Antifermionen) der Masse $m$
interpretieren.

Wir "ubertragen dieses Bild auf den Fall \Ref{2_i0} mit Massenasymmetrie
und die St"orungsrechnung \Ref{2_ba}, \Ref{2_bb}:
Der freie fermionische Projektor ist im ersten und zweiten Isospinblock
aus Fermionen der Masse $m_1$ bzw. $m_2$ aufgebaut.
Die Diagonalbeitr"age \Ref{2_ba} beschreiben eine Wechselwirkung
der Fermionen innerhalb jedes Blocks. Die Au"serdiagonalbeitr"age
\Ref{2_bb} f"uhren zu einer Wechselwirkung der Zust"ande mit Masse
$m_1$ im ersten mit Zust"anden derselben Masse $m_1$ im
zweiten Isospinblock; entsprechend findet eine Wechselwirkung
zwischen den Zust"anden mit Masse $m_2$ in beiden Isospinbl"ocken
statt.
Diese Vorstellung entspricht nicht dem "ublichen Bild einer physikalischen
Wechselwirkung. Es scheint nicht sinnvoll zu sein, da"s die Fermionen
des oberen Diracsees mit den Zust"anden zur Masse $m_1$ des zweiten
Isospinblocks wechselwirken. Denn im zweiten Isospinblock sind keine
Zust"ande zur Masse $m_1$ besetzt; diese Zust"ande sollten in der
Theorie gar nicht in Erscheinung treten.
Ganz analog scheint eine Wechselwirkung des unteren Diracsees mit Zust"anden
zur Masse $m_2$ im oberen Isospinblock der physikalischen Beobachtung
zu widersprechen.

Die St"orungsrechnung \Ref{2_m2} scheint die Physik besser zu beschreiben,
denn dort findet eine Wechselwirkung zwischen den Fermionen im ersten
Isospinblock mit Masse $m_1$ und den Fermionen im zweiten Isospinblock
mit Masse $m_2$ statt. Wir k"onnen die St"orung des Diracoperators also als
Wechselwirkung derjenigen Zust"ande auffassen, aus denen der freie
fermionische Projektor aufgebaut ist.

Allgemeiner ausgedr"uckt legt die St"orungsrechnung \Ref{2_n5} im
Gegensatz zu \Ref{2_i1} fest, welche Fermionfamilien miteinander
in Wechselwirkung treten. Genauer k"onnen wir die St"orungsrechnung
\Ref{2_m1} im Fall mehrerer Teilchenfamilien folgenderma"sen interpretieren:
Da der Flavour-Index in \Ref{2_m0} als freier Index auftritt,
braucht der St"oroperator ${\cal{B}}$ auf dem Flavour-Raum nicht
notwendigerweise trivial zu sein. Falls ${\cal{B}}$ auf dem Flavour-Raum
diagonal ist, k"onnen wir die St"orung \Ref{2_m1} als eine Wechselwirkung
der Fermionen innerhalb jeder Familie auffassen.
Man kann auch eine Wechselwirkung zwischen Fermionen aus
verschiedenen Familien beschreiben; dazu mu"s der St"oroperator
au"serdiagonale Flavour-Anteile enthalten.
Solche Flavour-mischenden St"orungen sind f"ur ein realistisches
physikalisches Modell tats"achlich notwendig, insbesondere bei der
CKM-Matrix in der schwachen Wechselwirkung.

Mit \Ref{2_ba}, \Ref{2_bb} und \Ref{2_m1} sind wir f"ur das Beispiel
\Ref{2_i0} verschiedenen
M"oglichkeiten begegnet, wie die St"orungsrechnung durchgef"uhrt werden
k"onnte. Zur Deutlichkeit beschreiben wir abschlie"send, wie mit dieser
Uneindeutigkeit umzugehen ist:
Genau wie f"ur $\tilde{k}_m$ ab Seite \pageref{2_uneins}
beschrieben, sind auch hier die verschiedenen Varianten der St"orungsrechnung
in dem Sinne gleichwertig, da"s damit jede unit"are Transformation
\Ref{1_d4} als geeignete St"orung des Diracoperators darstellbar ist.
Aus diesem Grund k"onnen wir uns willk"urlich f"ur eine der Varianten
entscheiden. Die St"orungsrechnung \Ref{2_m1} hat den Vorteil, da"s
sie f"ur lokale Potentiale die Forderung \Ref{2_fl} erf"ullt.
Alternativ k"onnten wir auch mit \Ref{2_ba}, \Ref{2_bb} arbeiten.
Damit in der St"orungsrechnung keine nichtlokalen Linienintegrale auftreten,
m"u"sten wir dann aber zur Beschreibung der klassischen Wechselwirkungen
mit nichtlokalen St"oroperatoren ${\cal{B}}$ anstelle der lokalen Potentiale
arbeiten (und zwar so, da"s sich die Nichtlokalit"at der
Linienintegrale und des St"oroperators gerade kompensieren).
Dies w"are zwar mathematisch machbar, erscheint aber unpraktikabel.

Um zu illustrieren, da"s eine unit"are Transformation \Ref{1_d4}
bei den verschiedenen Varianten der St"orungsrechnung durch
unterschiedliche St"orungen des Diracoperators beschrieben wird,
betrachten wir das Beispiel einer Eichtransformation
\[ \tilde{P}(x,y) \;=\; U(x) \: P(x,y) \: U(y)^{-1} \spc {\mbox{mit
	$U(x) \in U(8)$}} \spc . \]
Bei der St"orungsrechnung \Ref{2_ba}, \Ref{2_bb} m"ussen wir gem"a"s
\Ref{2_ia}, \Ref{2_i1}
\[ {\cal{B}} \;=\; [i \Pdd, \: U] \]
w"ahlen; bei der Methode \Ref{2_m1} ist nach \Ref{2_m0} dagegen
\[ {\cal{B}} \;=\; [i \Pdd - m Y, \: U] \]
zu setzen. Im Fall $[Y, U] \neq 0$ stimmen diese beiden St"oroperatoren
nicht "uberein.

\subsection{St"orungsrechnung f"ur $P(x,y)$ mit zus"atzlicher
chiraler Asymmetrie}
\label{2_ab24}
Wir kommen zur allgemeinen St"orungsrechnung mit Massenasymmetrie
und chiraler Asymmetrie. Der freie fermionische Projektor \Ref{2_b8}
l"a"st sich mit den Operatoren \Ref{2_o0}, \Ref{2_o1} in der Form
\Equ{2_o4}
P(x,y) \;=\; \frac{1}{2} \: \Tr_{\cal{F}} \left( X \: (p-k)(x,y) \right)
\EndEqu
umschreiben. Durch den zus"atzlichen Faktor $X$ werden gegen"uber
\Ref{2_o2} in den Sektoren mit $X_j \neq \1$ chirale Fermionzust"ande
herausprojeziert. Folglich kommen alle Zust"ande,
aus denen \Ref{2_o4} aufgebaut ist, auch im zugeh"origen fermionischen
Projektor ohne chirale Asymmetrie \Ref{2_o2} vor. Es scheint daher
sinnvoll, die St"orungsrechnung f"ur die einzelnen Fermionzust"ande
genau wie im vorigen Abschnitt durchzuf"uhren und aus diesen gest"orten
Zust"anden anschlie"send den gest"orten fermionischen Projektor
$\tilde{P}$ aufzubauen.
Diese Methode f"uhrt in direkter Verallgemeinerung von \Ref{2_n5}
auf die Gleichungen
\begin{eqnarray}
\label{eq:2_o5}
\tilde{P} &=& \frac{1}{2} \: \Tr_{\cal{F}} \left( U \: X (p-k) \: U^*
	\right) \\
\label{eq:2_o6}
&=& P \:-\: \frac{1}{2} \: \Tr_{\cal{F}} \left( s \:{\cal{B}}\: X(p-k)
	\:+\: X(p-k) \:{\cal{B}}\: s \right) \spc .
\end{eqnarray}
Man sieht, da"s bei der "Ubertragung der Gleichungen \Ref{2_m0}
auf den Fall mit chiraler Asymmetrie die Matrix $X$ jeweils bei den
Faktoren $p, k$ einzuf"ugen ist.

\subsubsection*{Bedingungen an den St"oroperator ${\cal{B}}$}
Mit \Ref{2_o5}, \Ref{2_o6} haben wir die St"orungsentwicklung
vollst"andig durchgef"uhrt\footnote{Wir bemerken zur Vollst"andigkeit,
da"s der alternative Ansatz zur St"orungsrechnung
\begin{eqnarray*}
\Delta(X p) &=& -(Xs) \:{\cal{B}}\: (Xp) \:-\: (Xp) \:{\cal{B}}\: (Xs) \\
\Delta(X k) &=& -(Xs) \:{\cal{B}}\: (Xk) \:-\: (Xk) \:{\cal{B}}\: (Xs) \\
\tilde{P} &=& P \:+\: \frac{1}{2} \: \Tr_{\cal{F}} \left( \Delta (Xp) -
	\Delta (Xk) \right)
\end{eqnarray*}
nicht sinnvoll ist. Die zu \Ref{2_n7} analoge Transformation
\[ U[{\cal{B}}] \;=\; 1 - \int_{\sR \cup i \sR} \!\!\! dm \;
	(X s_{[m]}) \: {\cal{B}} \: (p_{[m]} X) \]
ist n"amlich nicht unit"ar; au"serdem l"a"st sich damit (wegen der
Singularit"at von $X$) nicht jede infinitesimale Transformation
$V=1+iA$ realisieren.}. Es bleibt zu untersuchen, ob, und wenn ja,
f"ur welche Operatoren ${\cal{B}}$ in der St"orungsrechnung nichtlokale
Linienintegrale auftreten.

Dazu beginnen wir mit dem Beispiel eines $U(B)$-Potentials
$(A_{ij})_{i,j=1,\ldots,B}$ und betrachten als einen der in
\Ref{2_o6} auftretenden Beitr"age den Ausdruck
\[ - s \:\Aslsh\: (Xk) \:-\: (Xk) \:\Aslsh\: s \spc . \]
Zur Einfachheit werden wir nur den Grenzfall $m=0$ untersuchen und
setzen dazu
\[ \Delta k_X \;:=\; -s_0 \:\Aslsh\: (X k_0) - (X k_0) \:\Aslsh\: s_0
	\spc . \]
Wir k"onnen "ahnlich wie im Beispiel ab Seite \pageref{2_beispiel}
vorgehen: F"ur eine "ubersichtliche Notation spalten wir die chirale
Asymmetriematrix in der Form
\Equ{2_x3}
X \;=\; \chi_L \:X_L \:+\: \chi_R \:X_R
\EndEqu
mit Matrizen $X_{L\!/\!R}$ auf, die auf dem Raum der Diracspinoren
trivial sind. Formal sind die Matrizen $X_{L\!/\!R}$ analog zu
\Ref{2_b7} durch
\[ X^j_{L\!/\!R} \;=\; \left\{ \begin{array}{ll} 1 & {\mbox{falls
	$X_j=\1$ oder $X_j=\chi_{L\!/\!R}$}} \\
	0 & {\mbox{falls $X_j=\chi_{R\!/\!L}$}} \end{array} \right.
	\;\;\; {\mbox{und}} \;\;\; (X_{L\!/\!R})_{\alpha j a \:
	\beta k b} \;=\; X^j_{L\!/\!R} \:\delta_{\alpha \beta} \:
	\delta_{jk} \: \delta_{ab} \]
gegeben. Im Spezialfall $A_j=\partial_j \Lambda$ kann man
$\Delta k_X$ im Operatorkalk"ul berechnen,
\begin{eqnarray}
\chi_{L\!/\!R} \: \Delta k_X &=& i \chi_{L\!/\!R} \left(
	s_0 \:[i \Pdd, \Lambda]\: X_{L\!/\!R} \: k_0 \:+\:
	X_{L\!/\!R} \: k_0 \:[i \Pdd, \Lambda]\: s_0 \right) \nonumber \\
&=& i \chi_{L\!/\!R} \left( (i \Pdd s_0) \:\Lambda\: X_{L\!/\!R} \: k_0
	\:-\: s_0 \:\Lambda\: X_{L\!/\!R} \: (i \Pdd k_0) \right.
	\nonumber \\
&& \left. \spc + X_{L\!/\!R} \:(i \Pdd k_0) \:\Lambda\: s_0 \:-\:
	X_{L\!/\!R} \:k_0\: \Lambda\: (i \Pdd s_0) \right) \nonumber \\
\label{eq:2_o8}
&=& i \chi_{L\!/\!R} \left( \Lambda \: X_{L\!/\!R} \: k_0 \:-\:
	X_{L\!/\!R} \:k_0\: \Lambda \right) \spc .
\end{eqnarray}
An dieser Formel l"a"st sich genau wie bei \Ref{2_q1} die Form der
Eichterme f"ur ein allgemeines Potential $A$ ablesen, n"amlich
\begin{eqnarray}
\Delta k_X(x,y) &\asymp& -\frac{i}{2} \: \inti d\lambda \; \epsilon(\lambda)
	\: A_j(\lambda y + (1-\lambda) x) \: \xi^j \;  X k_0(x,y)
	\nonumber \\
\label{eq:2_o9}
&&+\frac{i}{2} \:X k_0(x,y) \; \inti d\lambda \; \epsilon(\lambda-1) \:
	A_j(\lambda y + (1-\lambda) x) \: \xi^j \spc . \spc
\end{eqnarray}
Wir sehen also, da"s auch als Folge der chiralen Asymmetrie nichtlokale
Linienintegrale auftreten k"onnen. In unserem Beispiel verschwinden
sie, falls die chirale Asymmetriematrix mit dem St"orpotential kommutiert,
\Equ{2_o7}
	[X,\:A] \;=\; 0 \spc .
\EndEqu

Diese "Uberlegung war mathematisch nicht streng; wir m"u"sten \Ref{2_o9}
noch durch asymp\-to\-ti\-sche Entwicklung von $\Delta k_X$ beweisen.
Durch Verfeinerung und mathematische Pr"azisierung dieser Methode lassen
sich aber alle lokalen Potentiale bestimmen, welche die Forderung \Ref{2_fl}
erf"ullen. Um uns nicht in Einzelheiten zu verlieren, begn"ugen wir uns
hier mit einer Veranschaulichung der Ergebnisse. Die zugeh"origen
Rechnungen sind "uber die Anh"ange A-E verteilt.

Zun"achst einmal l"a"st sich die Bedingung \Ref{2_o7} an das Potential $A$
abschw"achen: Bei dem Diracoperator
\Equ{2_p0}
	i \Pdd + \Aslsh \;\;\;\;,\spc [X, A]=0
\EndEqu
mit einem $U(B)$-Potential $A$ treten bei der St"orungsrechnung in
Verallgemeinerung von \Ref{2_o9} auch in h"oherer Ordnung in $m$ keine
nichtlokalen Linienintegrale auf.
Wir f"uhren nun in den Diracoperator gem"a"s
\Equ{2_z9}
i \Pdd \:+\: \Aslsh \:-\: m (U^{-1} \:Y\: U - Y)
\EndEqu
eine zus"atzliche skalare St"orung ein, dabei ist $U$ ein unit"ares
$U(B)$-Matrixfeld. Die Zust"ande $\Psi$ des fermionischen Projektors
erf"ullen dann die Diracgleichung
\[ (i \Pdd \:+\: \Aslsh \:-\: m \: U^{-1} \:Y\: U) \: \Psi \;=\; 0 \spc . \]
Wir k"onnen die zus"atzliche skalare St"orung auch so interpretieren,
da"s wir von der festen Matrix $Y$ zu einer dynamischen Massenmatrix
$U^{-1}(x) \:Y\: U(x)$ "ubergegangen sind. Darum ist einsichtig,
da"s die Bedingung \Ref{2_asymm} nun durch
\[ X \: U^{-1} \:Y\: U \;=\; U^{-1} \:Y\: U \: X \;=\; U^{-1} \:Y\: U \]
oder, etwas einfacher, durch
\Equ{2_y0}
UXU^{-1} \:Y \;=\; Y \: UXU^{-1} \;=\; Y
\EndEqu
ersetzt werden mu"s. Es zeigt sich, da"s der Diracoperator
\Ref{2_z9} mit \Ref{2_y0} ebenfalls der Lokalit"atsforderung
\Ref{2_fl} gen"ugt, was als Verallgemeinerung unseres
Ergebnisses bei konstanter Massendrehung auch plausibel ist.
Wir f"uhren nun die Eichtransformation $\Psi(x) \rightarrow
\tilde{\Psi}(x) = U(x) \: \Psi(x)$ durch. Die Wellenfunktionen
$\tilde{\Psi}$ erf"ullen die Diracgleichung
\[ ( U (i \Pdd + \Aslsh) U^{-1} \:-\: m Y) \:\tilde{\Psi} \;=\; 0
	\spc . \]
Da bei Eichtransformationen alle Integralkerne nur lokal
transformiert werden, treten schlie"slich auch beim Diracoperator
\Equ{2_p1}
U(i \Pdd + \Aslsh) U^{-1} \;=\; i \Pdd \:+\: U \Aslsh U^{-1} \:+\:
	i U (\Pdd U^{-1})
\EndEqu
in Verbindung mit \Ref{2_y0} keine nichtlokalen Linienintegrale auf.

Um explizit in der St"orungsrechnung zu verifizieren, da"s der
Diracoperator \Ref{2_p1} die Lokalit"atsforderung \Ref{2_fl} erf"ullt,
betrachten wir den zu \Ref{2_o9} analogen Beitrag
\begin{eqnarray}
\lefteqn{ \Delta k_X(x,y) \;=\; -\frac{i}{2} \: \inti d\lambda \; \epsilon(\lambda)
	\: (U A_j U^{-1} + i U (\partial_j U^{-1}))_{|\lambda y + (1-\lambda) x}
	\: \xi^j \;  X k_0(x,y) } \nonumber \\
\label{eq:2_r0}
&&+\frac{i}{2} \:X k_0(x,y) \; \inti d\lambda \; \epsilon(\lambda-1) \:
	(U A_j U^{-1} + i U (\partial_j U^{-1}))_{|\lambda y + (1-\lambda) x}
	\: \xi^j \spc . \spc
\end{eqnarray}
Wir entwickeln das Matrixfeld $U$ in der Form
$U(x)=1 + i \Lambda(x) + {\cal{O}}(\Lambda^2)$ und erhalten in erster
Ordnung in $\Lambda, A$
\begin{eqnarray*}
&=& -\frac{i}{2} \: \inti d\lambda \: \left( \epsilon(\lambda)
	\: A_j \xi^j \: X - \epsilon(\lambda-1) \: X \: A_j \xi^j \right) \;
	k_0(x,y) \\
&&-\frac{i}{2} \: \inti d\lambda \; \epsilon(\lambda)
	\: \partial_j \Lambda(\lambda y + (1-\lambda) x)
	\: \xi^j \;  X k_0(x,y) \\
&&+\frac{i}{2} \:X k_0(x,y) \; \inti d\lambda \; \epsilon(\lambda-1) \:
	\partial_j \Lambda(\lambda y + (1-\lambda) x)
	\: \xi^j \spc .
\end{eqnarray*}
Wir sehen, da"s $\Delta k_X(x,y)$ in einen Beitrag des Potentials $A$ und
einen Beitrag der Eichtransformation $\Lambda$ zerf"allt.
Im ersten Integral k"onnen wir die Kommutatorgleichung \Ref{2_o7} anwenden,
im zweiten Integral kann man partiell integrieren
\begin{eqnarray}
\Delta k_X(x,y) &=& -\frac{i}{2} \left( \int_x^y A_j \xi^j \right) \: X k_0(x,y)
	\nonumber \\
\label{eq:2_r1}
&&+i \Lambda(x) \: X k_0(x,y) \:-\: i X k_0(x,y) \: \Lambda(y) \spc .
\end{eqnarray}
Nach diesen Umformungen sind alle nichtlokalen Linienintegrale
verschwunden.

Das bei \Ref{2_p1} verwendete Verfahren l"a"st sich auf chirale
Transformationen erweitern: Wir f"uhren in den Diracoperator
\Ref{2_p0} eine zus"atzliche St"orung durch ein axiales $U(B)$-Potential
ein. Es ist g"unstig, die vektoriellen und axialen Potentiale als
chirale Potentiale umzuschreiben. Der Diracoperator hat dann die Form
\Equ{2_p2}
i \Pdd \:+\: \chi_L \: \Aslsh_R \:+\: \chi_R \: \Aslsh_L \spc
	{\mbox{mit $U(B)$-Potentialen $A_{L\!/\!R}$}} \spc .
\EndEqu
In der St"orungsrechnung treten keine nichtlokalen Linienintegrale
auf, falls
\[ [X_L,\: A_L] \;=\; [X_R,\:A_R] \;=\; 0 \spc . \]
Wir f"uhren jetzt in Verallgemeinerung von \Ref{2_p1} f"ur die links-
und rechtsh"andige Komponente getrennt eine lokale Phasentransformation
durch. Wir gehen also von \Ref{2_p2} zum Diracoperator
\begin{eqnarray}
\lefteqn{ \chi_L \: U_R \:(i \Pdd + \Aslsh_R) \:U_R^{-1} \:+\:
	\chi_R \: U_L \:(i \Pdd + \Aslsh_L) \:U_L^{-1} } \nonumber \\
\label{eq:2_p3}
&=& i \Pdd \:+\: \chi_L \:(U_R \Aslsh_R U_R^{-1} + i U_R (\Pdd U_R^{-1}))
	\:+\: \chi_R \:(U_L \Aslsh_L U_L^{-1} + i U_L (\Pdd U_L^{-1}))
\end{eqnarray}
"uber, dabei sind $U_{L\!/\!R}$ zwei unit"are $U(B)$-Matrixfelder.
Die Bedingung \Ref{2_asymm} mu"s nun mit der Notation \Ref{2_x3}
durch
\Equ{2_y1}
Y \: U_{L\!/\!R} X_{L\!/\!R} U_{L\!/\!R}^{-1} \;=\; U_{L\!/\!R} X_{L\!/\!R}
	U_{L\!/\!R}^{-1} \: Y \;=\; Y
\EndEqu
ersetzt werden.
Um zu sehen, da"s in der St"orungsrechnung tats"achlich keine
Nichtlokalit"aten auftreten, k"onnen wir genau wie f"ur den
Diracoperator \Ref{2_p1} argumentieren. In den nichtlokalen
Linienintegralen treten n"amlich (auch in h"oherer Ordnung
in $m$) immer entweder die links- oder die rechtsh"andigen Potentiale
auf. Die Beitr"age haben also die Form \Ref{2_r0}, wenn wir bei $U, A$
einen Index $L$ oder $R$ hinzuf"ugen, und lassen sich analog wie
\Ref{2_r1} in eine lokale Form bringen.
Man beachte, da"s der "Ubergang von \Ref{2_p2} zu \Ref{2_p3}
f"ur $U_L \neq U_R$ keine Eichtransformation ist.

Nach diesen Vorbereitungen k"onnen wir den allgemeinen Fall besprechen:
Gem"a"s Rechnung \Ref{2_o8} ist einsichtig, da"s alle nichtlokalen
Linienintegrale verschwinden, falls wir die chirale Asymmetriematrix in der
St"orungsrechnung ausklammern k"onnen
\Equ{2_p4}
s_0 \:{\cal{B}}\: (X k_0) \:+\: (X k_0) \:{\cal{B}}\: s_0 \;=\;
	X \left( s_0 \:{\cal{B}}\: k_0 \:+\: k_0 \:{\cal{B}}\: s_0
	\right) \spc .
\EndEqu
Zur besseren "Ubersicht spalten wir ${\cal{B}}$ in den geraden und
ungeraden Anteil auf,
\[ {\cal{B}} \;=\; {\cal{B}}^g + {\cal{B}}^u \spc {\mbox{mit}} \spc
	[ {\cal{B}}^g, \rho] \;=\; \{ {\cal{B}}^u, \rho \} \;=\; 0 \spc . \]
Da $s_0$ ungerade ist, l"a"st sich \Ref{2_p4} nach Multiplikation mit
$\chi_{L\!/\!R}$ in die Bedingungen
\Equ{2_p5}
\chi_{R\!/\!L} \: [X_{L\!/\!R},\: {\cal{B}}^u] \;=\; \chi_{R\!/\!L}
	\left( X_{L\!/\!R} \: {\cal{B}}^g \:-\:
	{\cal{B}}^g \: X_{R\!/\!L} \right) \;=\; 0
\EndEqu
an den St"oroperator umschreiben. Wir k"onnen nun analog zu \Ref{2_p3}
eine lokale chirale $U(B)$-Transformation durchf"uhren und erhalten
schlie"slich f"ur den Diracoperator
\Equ{2_p6}
(\chi_L \: U_R + \chi_R \: U_L) (i \Pdd + {\cal{B}}) (\chi_R \: U_R^{-1}
	+ \chi_L \: U_L^{-1}) \spc .
\EndEqu
Mit \Ref{2_p6} und den Bedingungen \Ref{2_y1}, \Ref{2_p5} haben wir
(abgesehen von trivialen Erweiterungen) die allgemeinste lokale St"orung
des Diracoperators gefunden, die der Lokalit"atsforderung \Ref{2_fl} gen"ugt.

Leider k"onnen wir die Form der chiralen Transformationen in \Ref{2_y1},
\Ref{2_p6} (und streng genommen auch schon Bedingung \Ref{2_y0})
an dieser Stelle nicht sauber begr"unden. Dazu m"u"ste man n"amlich
die St"orungsbeitr"age h"oherer Ordnung in der Masse betrachten, die
gegen"uber \Ref{2_r0} eine kompliziertere Form haben.
Bei der Diskussion endlicher St"orungen in Abschnitt \ref{2_ab33}
werden wir aber an expliziten Formeln genauer sehen, warum gerade
f"ur den Diracoperator \Ref{2_p6} in Verbindung mit \Ref{2_y1},
\Ref{2_p5} alle nichtlokalen Linienintegrale verschwinden.

\section{Endliche St"orungen}
\label{2_ab3}
Die St"orungsrechnung erster Ordnung des vorigen Abschnittes kann
selbstverst"andlich nur einen ersten Eindruck des wechselwirkenden
fermionischen Projektors vermitteln. Wir m"ussen die Ergebnisse mit
nicht-perturbativen Methoden absichern und erg"anzen.
Dazu wurden in Anhang E einzelne St"orbeitr"age in
beliebiger Ordnung berechnet und explizit aufsummiert.
In diesem Abschnitt werden wir die St"orungsrechnung h"oherer Ordnung
allgemein beschreiben und die wichtigsten Ergebnisse aus Anhang
E zusammenstellen.

Bevor wir mit der formalen St"orungsentwicklung beginnen, wollen wir
kurz beschreiben, in welchem Sinn unsere Behandlung mathematisch zu
verstehen ist. Zur Einfachheit betrachten wir dazu die Spektralprojektoren
$\tilde{p}, \tilde{k}$. Wir stehen vor dem Problem, exakte L"osungen
der gest"orten Diracgleichung
\Equ{2_q0}
(i \Pdd - m + {\cal{B}}) \:\tilde{p}_m \;=\;
	(i \Pdd - m + {\cal{B}}) \:\tilde{k}_m \;=\; 0
\EndEqu
zu finden. Eine solche St"orung einer linearen partiellen
Differentialgleichung ist aus theoretischer Sicht i.a. unproblematisch
(im Gegensatz zu den St"orungsentwicklungen der Quantenfeldtheorie);
wir erwarten daher, da"s die Distributionen $\tilde{p}_m(x,y),
\tilde{k}_m(x,y)$ f"ur gen"ugend kleine (aber endliche) St"orungen
${\cal{B}}$ wohldefiniert sind. Dies werden wir aber nicht beweisen
und auch nicht untersuchen.

Wir f"uhren eine St"orungsentwicklung nach ${\cal{B}}$ durch, die
Summe "uber die Ordnung der St"orungstheorie ist dabei als formale
Summe anzusehen. Die einzelnen St"orungsbeitr"age lassen sich als
Operatorprodukte der Form
\Equ{2_t6}
C_1 \: {\cal{B}} \: C_2 \:{\cal{B}}\: \cdots \:{\cal{B}}\: C_l
	\:{\cal{B}}\: C_{l+1}
\EndEqu
schreiben, wobei die Faktoren $C_j$ f"ur die Operatoren $p_m$, $k_m$ oder
$s_m$ stehen.
In Anhang E wird gezeigt, da"s diese Operatorprodukte und damit
auch die St"orungsbeitr"age jeder Ordnung wohldefiniert und endlich sind.
In der Sprache der perturbativen Quantenfeldtheorie liegt das daran,
da"s bei der Entwicklung nach der "au"seren St"orung ${\cal{B}}$ nur
Tree-Graphen auftreten.
Wir entwickeln die St"orungsbeitr"age jeder Ordnung um den Lichtkegel
und stellen fest, da"s sich die erhaltenen Formeln
explizit aufsummieren lassen. Die Exis\-tenz dieser Summe "uber die
Ordnung der St"orungstheorie ist zwar ein deutlicher Hinweis f"ur die
Konvergenz der St"orungsentwicklung, sie liefert aber keinen
strengen Konvergenzbeweis (denn wir arbeiten ja nur
mit den Formeln der Lichtkegelentwicklung und nicht mit den
St"orungsbeitr"agen selbst).
Um diese mathematische Unsauberkeit k"ummern wir uns jedoch nicht und
fassen die abgeleiteten Formeln als nicht-perturbative
Lichtkegelentwicklung f"ur $\tilde{p}_m$, $\tilde{k}_m$ auf.

\subsection{Formale St"orungsentwicklung f"ur $p_m, k_m$}
\label{2_ab31}
In diesem Abschnitt wollen wir die St"orungsentwicklung f"ur
$p_m, k_m$ von Abschnitt \ref{2_ab21} auf endliche St"orungen
verallgemeinern.
Zun"achst einmal k"onnen wir Satz \ref{a2_satz1} direkt auf h"ohere
Ordnung St"orungstheorie "ubertragen:
\begin{Satz}
\label{a2_satz2}
Der Operator
\Equ{2_69}
U \;=\; \int_{\sR \cup i \sR} dm \; \sum_{l=0}^\infty \:
	(-s_m \: {\cal{B}})^l \: p_m
\EndEqu
ist unit"ar und
\Equ{2_70}
(i \Pdd + {\cal{B}} - m) \: U p_m U^* \;=\; (i \Pdd + {\cal{B}} - m)
	\: U k_m U^* \;=\; 0 \spc .
\EndEqu
\end{Satz}
{\Beweis}
Wir ordnen die Reihen im formalen Produkt $U^* U$ mit der
Cauchy'schen Produktformel um
\begin{eqnarray}
U^* \: U &=& \int_{\sR \cup i \sR} dm \int_{\sR \cup i \sR} dm^\prime \;
	\sum_{l_1, l_2=0}^\infty \;
	p_m \: (- {\cal{B}} \: s_m)^{l_1} \: (- s_{m^\prime} \:
	{\cal{B}})^{l_2} \: p_{m^\prime} \nonumber \\
\label{eq:2_104a}
&=& \int_{\sR \cup i \sR} dm \int_{\sR \cup i \sR} dm^\prime \;
	\sum_{l=0}^\infty \; (-1)^l \; \sum_{p=0}^l \;
	p_m \: ({\cal{B}} \: s_m)^p \: (s_{m^\prime} \: {\cal{B}})^{l-p}
	\: p_{m^\prime} \;\;\; . \spc
\end{eqnarray}
Bei den Summanden $l>0$ k"onnen wir die Rechenregeln \Ref{2_14}, \Ref{2_16}
anwenden und erhalten
\begin{eqnarray*}
\lefteqn{ \sum_{p=0}^l \: p_m \: ({\cal{B}} \: s_m)^p \: (s_{m^\prime} \:
	{\cal{B}})^{l-p} \: p_{m^\prime} } \\
&=& \frac{1}{m-m^\prime} \; \sum_{p=1}^{l-1} p_m \: ({\cal{B}} \: s_m)^{p-1}
	\: {\cal{B}} \: (s_m-s_{m^\prime}) \: {\cal{B}}
	(s_{m^\prime} \: {\cal{B}})^{l-p-1} \: p_{m^\prime} \\
&&+ \frac{1}{m-m^\prime} \; \left( p_m \: {\cal{B}} \: (s_{m^\prime} \:
	{\cal{B}})^{l-1} \: p_{m^\prime} \;-\;
	p_m \: ({\cal{B}} \: s_m)^{l-1} \: {\cal{B}} \: p_{m^\prime} \right)
	\;=\; 0 \spc ,
\end{eqnarray*}
denn die Summe "uber $p$ ist teleskopisch.
Folglich bleibt in \Ref{2_104a} nur der Summand f"ur $l=0$ "ubrig, und es
folgt
\[ U^* \: U \;=\; \int_{\sR \cup i \sR} dm \int_{\sR \cup i \sR} dm^\prime
	\; p_m \: p_{m^\prime} \;=\; \1 \spc . \]
Zum Beweis von \Ref{2_70} wendet man auf die Gleichungen
\begin{eqnarray*}
U \: p_m &=& \sum_{l=0}^\infty (- s_m \: {\cal{B}})^l \: p_m \spc,\spc
U \: k_m \;=\; \sum_{l=0}^\infty (- s_m \: {\cal{B}})^l \: k_m
\end{eqnarray*}
den Operator $(i \Pdd - m)$ an und berechnet das Ergebnis explizit.
\QED

\subsubsection*{das Lokalit"atsproblem der Spektralzerlegung}
Mit Hilfe dieses Satzes scheinen wir die St"orungsentwicklung unmittelbar
durchf"uhren zu k"onnen. Wenn wir n"amlich $\tilde{p}_m, \tilde{k}_m$ durch
\Equ{2_q11}
\tilde{p}_m \;:=\; U \:p_m\: U^* \spc,\spc
	\tilde{k}_m \;:=\; U \:k_m\: U^*
\EndEqu
definieren, sind die Diracgleichung \Ref{2_q0} und, wegen der Unitarit"at
von $U$, auch \Ref{2_27} exakt erf"ullt.

Leider ist die Situation nicht ganz so einfach. Um das Problem der
St"orungsrechnung \Ref{2_q11} zu erkennen, betrachten wir ein
elektromagnetisches Potential ${\cal{B}}=\Aslsh$ und untersuchen die
Lichtkegelentwicklung von $\tilde{p}_m$:
Im Spezialfall $A_j=\partial_j \Lambda$ haben wir
\[ s_m \:(\Pdd \Lambda)\: p_m \;=\; -i s_m \:[i \Pdd - m, \: \Lambda]
	\:p_m \;=\; -i \Lambda\:p_m \spc . \]
Genau wie bei \Ref{2_q1} k"onnen wir aus dieser Gleichung den
f"uhrenden Beitrag der Lichtkegelentwicklung von $s_m \:\Aslsh\: p_m$
ablesen, n"amlich
\Equ{2_t0}
(s_m \:\Aslsh\: p_m)(x,y) \;\asymp\; \frac{i}{2} \inti d\lambda\;
	\epsilon(\lambda) \: A_j(\lambda y + (1-\lambda) x) \: \xi^j
	\; p_m(x,y) \spc .
\EndEqu
Nach Multiplikation mit $s_m$ kann diese Formel iteriert werden, so da"s
man auch h"ohere Operatorprodukte um den Lichtkegel entwickeln kann.
Man erh"alt so beispielsweise
\begin{eqnarray}
\left( (s_m \: \Aslsh)^n \: p_m \right)(x,y) &\asymp&
	\left(\frac{i}{2} \right)^n
	\inti \!\! d\lambda_1 \; \epsilon(\lambda_1) \inti \!\! d\lambda_2
	\: \epsilon(\lambda_2-\lambda_1) \cdots
	\inti \!\! d\lambda_n \; \epsilon(\lambda_n-\lambda_{n-1}) \nonumber \\
\label{eq:2_q3}
&& \hspace*{2cm} \times \; A_{j_1}(z_1) \: \xi^{j_1} \cdots A_{j_n}(z_n) \:
	\xi^{j_n} \; p_m(x,y)
\end{eqnarray}
mit $z_j=\lambda_j y + (1-\lambda_j) x$.
Im unit"aren Operator $U$, \Ref{2_69}, treten also geschachtelte
nichtlokale Linienintegrale auf. Dies ist noch kein Problem, denn
gem"a"s der Lokalit"atsforderung \Ref{2_fl} m"ussen die nichtlokalen
Linienintegrale lediglich in den zusammengesetzten Ausdr"ucken \Ref{2_q11}
f"ur $\tilde{p}_m, \tilde{k}_m$ verschwinden.
In erster Ordnung haben wir bei der Diskussion der Formeln
\Ref{2_28}, \Ref{2_29} gesehen, da"s
tats"achlich keine nichtlokalen Linienintegrale auftreten.
Der Beitrag zu $\tilde{p}_m$ zweiter Ordnung hat die Form
\Equ{2_t9}
\Delta p_m^{[2]} \;=\; s_m \:\Aslsh\: s_m \:\Aslsh\: p_m +
	p_m \:\Aslsh\: s_m \:\Aslsh\: s_m +
	s_m \:\Aslsh\: p_m \:\Aslsh\: s_m \spc .
\EndEqu
Durch Iteration von \Ref{2_t0} erh"alt man analog zu \Ref{2_q3} f"ur
die einzelnen Summanden die Entwicklungsformeln
\begin{eqnarray*}
(s_m \Aslsh s_m \Aslsh p_m)(x,y) &\!\!\!\asymp\!\!\!&
	-\frac{1}{4} \inti \!\!d\lambda_1 \: \epsilon(\lambda_1)
	\inti \!\!d\lambda_2 \: \epsilon(\lambda_2 - \lambda_1) \:
	A_j(z_1) \: \xi^j \: A_k(z_2) \: \xi^k \; p_m(x,y) \\
(p_m \Aslsh s_m \Aslsh s_m)(x,y) &\!\!\!\asymp\!\!\!&
	-\frac{1}{4} \inti \!\!d\lambda_1 \: \epsilon(\lambda_1-1)
	\inti \!\!d\lambda_2 \: \epsilon(\lambda_2-1) \:
	A_j(z_1) \: \xi^j \: A_k(z_2) \: \xi^k \; p_m(x,y) \\
(s_m \Aslsh p_m \Aslsh s_m)(x,y) &\!\!\!\asymp\!\!\!&
	\frac{1}{4} \inti \!\!d\lambda_1 \: \epsilon(\lambda_1)
	\inti \!\!d\lambda_2 \: \epsilon(\lambda_2-1) \:
	A_j(z_1) \: \xi^j \: A_k(z_2) \: \xi^k \; p_m(x,y) \;\;\; .
\end{eqnarray*}
An diesen Formeln sieht man, da"s in $\Delta p_m^{[2]}$ die nichtlokalen
Linienintegrale nicht wegfallen.
Folglich ist die Forderung \Ref{2_fl} f"ur die St"orungsentwicklung
\Ref{2_q11} in h"oherer Ordnung verletzt.

Nichtlokale Linienintegrale waren auch schon bei der St"orungsrechnung
erster Ordnung mit Massenasymmetrie oder chiraler Asymmetrie aufgetreten.
An dieser Stelle ist das Problem aber prinzipieller Art:
Wir hatten "uberlegt, da"s die Spektralprojektoren $\tilde{p}_m$ durch
die Gleichungen \Ref{2_q4} unabh"angig von einer St"orungsrechnung definiert
sind. Durch Einsetzen kann man verifizieren, da"s diese Spektralprojektoren
mit dem Ausdruck in \Ref{2_q11} "ubereinstimmen.
Folglich k"onnen wir die nichtlokalen Linienintegrale nicht einfach durch
Modifikation der St"orungsrechnung beseitigen, es handelt
sich um ein allgemeines Problem bei der Spektralzerlegung des
gest"orten Diracoperators.
Wir nennen dieses Problem das {\em{Lokalit"atsproblem der
Spektralzerlegung}}\index{Lokalit"atsproblem der Spektralzerlegung}.

\subsubsection*{der Ausweg: nichtunit"are St"ortransformationen}
Um das Lokalit"atsproblem der Spektralzerlegung zu umgehen, m"ussen wir
den Ansatz \Ref{2_q11} erweitern und gehen zu den Definitionsgleichungen
\Equ{2_s1}
\tilde{p}_m \;:=\; V \:p_m\: V^* \spc,\spc
	\tilde{k}_m \;:=\; V \:k_m\: V^*
\EndEqu
mit einem Operator $V$ "uber, der nicht notwendigerweise unit"ar ist.
Wir nennen eine solche Transformation der freien in die gest"orten
Gr"o"sen {\em{nichtunit"are St"ortransformation}}\index{nichtunit"are
St"ortransformation}.

Mit nichtunit"aren St"ortransformationen geben wir die Untersuchung
der Spektralprojektoren des gest"orten Diracoperators auf.
Die Operatoren \Ref{2_s1} werden zwar L"osungen der gest"orten
Diracgleichung \Ref{2_q0} sein und sind folglich auch orthogonal,
\begin{eqnarray*}
\tilde{p}_m \: \tilde{p}_n &=& \tilde{k}_m \: \tilde{k}_n \;=\;
	\tilde{p}_m \: \tilde{k}_n \;=\; \tilde{k}_m \: \tilde{p}_n
	\;=\; 0 \spc {\mbox{f"ur $m \neq n$}} \spc ,
\end{eqnarray*}
sie erf"ullen aber nicht die $\delta$-Normierungsbedingung und
Vollst"andigkeitsrelation, also i.a.
\begin{eqnarray}
\label{eq:2_s4}
\tilde{p}_m \: \tilde{p}_n &=& V\:p_m\:V^*\: V\:p_n\:V^*
	\;\neq\; V \:p_m\: p_n \: V^*
	\;=\; \delta(m-n) \: \tilde{p}_m \\
\label{eq:2_s5}
\tilde{k}_m \: \tilde{k}_n &\neq& \delta(m-n) \: \tilde{p}_m \;\;\;,\;\;\;
\tilde{k}_m \: \tilde{p}_n,\: \tilde{p}_m \: \tilde{k}_n
	\;\neq\; \delta(m-n) \: \tilde{k}_m \spc \\
\label{eq:2_s6}
\lefteqn{ \int_{\sR \cup i \sR} \tilde{p}_m \: dm \;=\; V \left(
	\int_{\sR \cup i \sR}
	p_m \: dm \right) V^* \;=\; V V^* \;\neq\; \1 \spc . }
\end{eqnarray}
Aus funktionalanalytischer Sicht machen die Gleichungen \Ref{2_s1} also
keinen Sinn.

Auf den ersten Blick scheinen nichtunit"are St"ortransformationen auch
Gleichung \Ref{1_d2} zu widersprechen, in welcher die Unitarit"at von
$U$ f"ur die Idempotenz $\tilde{P}^2=\tilde{P}$ des fermionischen
Projektors notwendig ist.
Aus diesem Grund wollen wir zun"achst allgemein begr"unden, warum es
f"ur die Kontinuumsbeschreibung des wechselwirkenden fermionischen
Projektors trotz allem sinnvoll ist, mit nichtunit"aren
St"or\-trans\-for\-ma\-tio\-nen zu arbeiten.
Wir werden au"serdem "uberlegen, was die Nichtunitarit"at
bei unserer Vorstellung der diskreten Raumzeit bedeutet.
Im n"achsten Unterabschnitt wird dann der Operator $V$ explizit
konstruiert. Diese etwas technische Konstruktion wird auch
die ab Seite \pageref{2_uneins} angesprochene Uneindeutigkeit der
St"orungsrechnung f"ur $k_m$ beseitigen.

Zur besseren physikalischen Anschauung betrachten wir im folgenden einen
Diracsee
\[ l_m \;:=\; \frac{1}{2} \: (p_m - k_m) \;\;\;\;\;,\spc
	\tilde{l}_m \;:=\; \frac{1}{2} \: (\tilde{p}_m - \tilde{k}_m)
	\spc , \]
die "Uberlegung gilt aber analog f"ur beide Operatoren $p_m, k_m$ getrennt.
Zun"achst schreiben wir die Definitionsgleichung \Ref{2_s1} in der Form
\Equ{2_s0}
\tilde{l}_m \;=\; U \: C l_m \: U^* \;\;\;\;\;,\spc
	[ C, \: l_m] \;=\; 0
\EndEqu
mit einem unit"aren Operator $U$ und einem hermiteschen Operator $C$
um\footnote{Um diese Transformation einzusehen, kann man den endlichdimensionalen
Fall betrachten: Falls ${\cal{B}}$ klein genug gew"ahlt wird, ist $V$
invertierbar. Dann sind die Operatoren $l_m$ und $V l_m V^*$ von
gleichem Rang $r$. Wir zerlegen die Operatoren spektral,
\[ l_m \;=\; \sum_{j=1}^r \lambda_j \: P_j \;\;\;\;,\spc
	V \:l_m\: V^* \;=\; \sum_{j=1}^r \nu_j \: Q_j \spc
	{\mbox{mit $\lambda_j, \nu_j \neq 0$}} \spc , \]
dabei sind $P_j, Q_j$ die Projektoren auf eindimensionale Eigenr"aume
(im Fall mit Entartung spalten wir die Spektralprojektoren willk"urlich
in die Summe von Projektoren auf eindimensionale Unterr"aume auf).
Wenn wir
\[ C \;=\; \sum_{j=1}^r \frac{\nu_j}{\lambda_j} \: P_j \]
setzen, besitzen die Operatoren $V l_m V^*$ und $C l_m$ die gleichen
Eigenwerte (mit gleicher Vielfachheit) und sind folglich unit"ar
"aquivalent.}.
Der Operator $C$ kann die Eigenwerte von $l_m$ modifizieren, die
Eigenvektoren bleiben aber unver"andert. Konkreter k"onnen wir
erreichen, da"s $C$ f"ur jeden Wellenvektor $k$ die
Projektoren auf die beiden Spinzust"ande mit einem Faktor
multipliziert, also
\begin{eqnarray}
(C l_m)(x,y) &=& \int \frac{d^4k}{(2 \pi)^4} \: (k \slsh + m) \:
	\Theta(-k^0) \: \delta(k^2-m^2) \; e^{-ik(x-y)} \nonumber \\
\label{eq:2_s2}
&&\hspace*{2cm} \times \left[ \frac{f_1(k)}{2} \: (1 + \rho v \slsh(k)) \:+\:
	\frac{f_2(k)}{2} \: (1 - \rho v \slsh(k)) \right]
\end{eqnarray}
mit skalaren Funktionen $f_1, f_2$ und einem Vektorfeld $v$ mit
$k_j v^j(k)=0$ und $v^2 \equiv -1$.
F"ur kleine St"orungen ist $f_1 \approx f_2 \approx 1$.
Die Distributionen $C l_m$ sind L"osungen der freien Diracgleichung, sie
k"onnen aber selbstverst"andlich nicht als Spektralprojektoren von
$i \Pdd$ aufgefa"st werden.

Es ist g"unstig, wenn wir die Transformation \Ref{2_s0} in Gedanken in
zwei Schritte zerlegen:
Zun"achst einmal beeinflu"st der Operator $C$ die Methode, wie der
Diracsee aus den Ebenen-Wellen-L"osungen der Diracgleichung aufgebaut
wird. Genauer wird der freie Diracsee nicht mehr durch den Operator  $l_m$,
sondern gem"a"s \Ref{2_s2} durch $C l_m$ beschrieben.
Anschlie"send werden durch die unit"are Transformation $U$ genau wie
in \Ref{2_q11} alle Fermionzust"ande des Diracsees gest"ort.
Bei dieser Sichtweise unterscheiden sich die St"orungsentwicklungen
\Ref{2_q11}, \Ref{2_s1} nur um den Faktor $C$, und wir k"onnen uns
auf die Diskussion des freien Diracsees beschr"anken.

Nach dieser "Uberlegung m"ussen wir zeigen, da"s auch die Distribution
$C l_m$ (und nicht nur $l_m$) als sinnvoller Kontinuumslimes eines
fermionischen Projektors in der diskreten Raumzeit aufgefa"st werden
kann. Im ersten Schritt argumentieren wir dazu in Verallgemeinerung
von \Ref{2_b3} im Minkowski-Raum durch Ausschmieren des
Impulsintegrals: Die Faktoren $f_1, f_2$ in \Ref{2_s2} lassen sich bei
Regularisierung im Impulsraum dadurch ber"ucksichtigen, da"s man
die ``Breite'' des Diracsees f"ur beide Spineinstellungen variabel
gestaltet. Genauer betrachten wir f"ur einen kleinen Parameter
$\varepsilon$ das Integral
\begin{eqnarray}
\lefteqn{ P_\varepsilon(x,y) \;:=\; \int \frac{d^4k}{(2 \pi)^4} \:
	(k \slsh + m) \: \Theta(|k|-m) \: \Theta(-k^0) \;
	e^{-ik (x-y)} } \nonumber \\
\label{eq:2_s7}
&&\times \!\left[
	\frac{1}{2} \left( 1 + \rho v \slsh(k) \right)
	\Theta(m + \varepsilon f_1(k) - |k|) \:+\:
	\frac{1}{2} \left( 1 - \rho v \slsh(k) \right)
	\Theta(m + \varepsilon f_2(k) - |k|) \right] \;\;. \spc
\end{eqnarray}
Der Operator $P_\varepsilon$ ist ein Projektor, wie man explizit verifiziert.
Entsprechend zu \Ref{2_s3} haben wir die N"aherung
\[ P_\varepsilon(x,y) \;\approx\; \varepsilon \: (C l_m)(x,y) \spc . \]
Wir sehen also, da"s sich der Kontinuumslimes eines
Diracsees "aquivalent zu
$l_m$ auch mit dem Operator $C l_m$ beschreiben l"a"st.
Diese Freiheit in der Kontinuumsbeschreibung h"angt letztlich damit
zusammen, da"s wir im fermionischen Projektor nur mit wenigen diskreten
und nicht mit einer kontinuierlichen Schar von Diracseen arbeiten.
Dadurch sind Probleme der $\delta$-Normierung \Ref{2_s4}, \Ref{2_s5}
und der Vollst"andigkeit \Ref{2_s6} f"ur uns irrelevant.
Allgemeiner kommen wir zu dem Schlu"s, da"s die Beschreibung
des fermionischen Projektors mit nichtunit"aren St"ortransformationen
auf keine prinzipiellen oder begrifflichen Schwierigkeiten f"uhrt.

Im n"achsten Schritt wollen wir die erwartete Situation in der diskreten
Raumzeit genauer betrachten. Da wir "uber die Form des fermionischen
Projektors in der diskreten Raumzeit keine Einzelheiten kennen, mu"s
die Argumentation zwangsl"aufig etwas qualitativ bleiben; wir werden
darauf in dieser Arbeit auch nicht wieder zur"uckkommen.
Die "Uberlegung ist aber trotzdem interessant, weil sie eine Best"atigung
f"ur unseren Deutungsversuch der Feldquantisierung in Abschnitt
\ref{1_ab5} liefert:\label{2_quant}
Wir f"uhren in den St"oroperator zur Deutlichkeit einen Parameter
$\lambda$ ein, betrachten also den Diracoperator $i \Pdd + \lambda {\cal{B}}$.
Als physikalisches Beispiel kann man an eine St"orung durch eine
(klassische) elektromagnetische Welle denken, der Parameter $\lambda$ gibt
ihre Amplitude an.
Die Operatoren $U, C$ in \Ref{2_s0} h"angen von $\lambda$ ab. An
der St"orungsrechnung erster Ordnung haben wir gesehen, da"s sich $U$
in f"uhrender Ordnung linear in $\lambda$ verh"alt. Da das
Lokalit"atsproblem wie beschrieben ein Effekt zweiter Ordnung ist,
h"angt $C$ in f"uhrender Ordnung quadratisch von $\lambda$ ab,
also insgesamt
\[ U(\lambda) \;=\; \1 + \lambda \: U^{[1]} + {\cal{O}}(\lambda^2)
	\;\;\;\; , \spc
	C(\lambda) \;=\; \1 + \lambda^2 \: C^{[2]} + {\cal{O}}(\lambda^3)
	\spc . \]
Die Funktionen $f_1, f_2$ in \Ref{2_s2} verhalten sich folglich ebenfalls
quadratisch in $\lambda$
\Equ{2_s8}
f_{1\!/\!2}(\lambda) \;=\; 1 + \lambda^2 \: f^{[2]}_{1\!/\!2}
	+ {\cal{O}}(\lambda^3) \spc .
\EndEqu
In der diskreten Raumzeit m"ussen wir das Integral in \Ref{2_s7}
durch eine diskrete Summe "uber die Fermionzust"ande ersetzen.
Die stetige Ver"anderung des Integrationsgebietes in \Ref{2_s7}
bei Variation von $\lambda$ l"a"st sich mit einer endlichen Summe
aber nicht beschreiben.
Anders ausgedr"uckt, l"a"st sich das Integral in \Ref{2_s7}
nur f"ur diskrete Werte des Parameters $\lambda$ gut durch eine
endliche Summe "uber Fermionzust"ande approximieren.
Dies k"onnen wir als ``Quantisierungsbedingung'' f"ur $\lambda$
auffasen\footnote{Diese ``Quantisierung'' l"a"st sich etwas genauer
unter den Voraussetzungen von Fu"snote \ref{2_foot2} auf Seite
\pageref{2_foot2} beschreiben:
Im endlichen Volumen geht $P_\varepsilon(x,y)$ in die Summe
\begin{eqnarray*}
P_\varepsilon(x,y) &=& \left( \frac{2 \pi}{L} \right)^4
	\sum_{k \in \frac{2 \pi}{L}\: \sZ^4} \:
	(k \slsh + m) \: \Theta(|k|-m) \: \Theta(-k^0) \;
	e^{-ik (x-y)} \\
&&\hspace*{1cm} \times \!\left[
	\frac{1}{2} \left( 1 + \rho v \slsh(k) \right)
	\Theta(m + \varepsilon f_1(k) - |k|) \:+\:
	\frac{1}{2} \left( 1 - \rho v \!\slsh(k) \right)
	\Theta(m + \varepsilon f_2(k) - |k|) \right]
\end{eqnarray*}
"uber, bei zus"atzlicher Diskretisierung der Raumzeit auf einem Gitter wird die
Summe endlich. Bei kontinuierlicher Variation der Funktionen $f_1, f_2$
"andert sich die Anzahl der Summanden bei diskreten Parameterwerten sprunghaft.}.
Da $f_1, f_2$ gem"a"s \Ref{2_s8} quadratisch von $\lambda$
abh"angen, erhalten wir quantitativ die Bedingung
\[ \lambda^2 \;=\; c \: (n + d) \spc,\spc n \in \N \]
mit einer unbestimmten Konstanten $d \in [0,1)$ und einem Parameter $c$,
der von der Geometrie der St"orung abh"angt.
Im Beispiel der elektromagnetischen Welle haben wir auf diese Weise
genau die in Abschnitt \ref{1_ab5} verwendete Amplitudenbedingung
hergeleitet.

\subsubsection*{Durchf"uhrung der kausalen St"orungsentwicklung}
Wir wollen nun die Gleichungen \Ref{2_s1} mathematisch ableiten und
den Operator $V$ konstruieren. Dabei werden wir stets mit einem
lokalen St"oroperator ${\cal{B}}$ (also einem lokalen Potential oder einem
Differentialoperator) arbeiten, die Endformeln k"onnen aber
auch f"ur allgemeine St"oroperatoren verwendet werden.

Die Grundidee der Konstruktion besteht darin, die St"orungsentwicklung
f"ur $p_m, k_m$ auf diejenige f"ur die avancierte und retardierte
Greensfunktion zur"uckzuf"uhren. F"ur diese Greensfunktionen
\Ref{2_f7}, \Ref{2_f8} l"a"st sich n"amlich die St"orungsrechnung auf
kanonische Weise durchf"uhren: Der Tr"ager der Distribution $s^\vee_m$
liegt im oberen Lichtkegel. Folglich geht in das Operatorprodukt
\Equ{2_t1}
(s^\vee_m \:{\cal{B}}\: s^\vee_m)(x,y) \;=\;
	\int d^4z \; s^\vee_m(x,z) \: {\cal{B}}_z \: s^\vee_m(z,y)
\EndEqu
der St"oroperator ${\cal{B}}_z$ nur f"ur solche $z$ ein, die im Schnitt
des oberen Lichtkegels um $x$ mit dem unteren Lichtkegel um $y$ liegen.
In diesem Sinn ist der Ausdruck \Ref{2_t1} {\em{kausal}}.
Insbesondere liegt der Tr"ager von \Ref{2_t1} wieder im oberen Lichtkegel.
Durch Iteration folgt, da"s auch die h"oheren Produkte
\[ s^\vee_m \:{\cal{B}}\: s^\vee_m \:{\cal{B}}\: \cdots {\cal{B}}\:
	s^\vee_m \:{\cal{B}}\: s^\vee_m \]
kausal sind und den Tr"ager im oberen Lichtkegel besitzen.
Wir definieren die gest"orte avancierte Greensfunktion als Summe "uber
diese Operatorprodukte
\Equ{2_t2}
\tilde{s}_m^\vee \;=\; \sum_{k=0}^\infty \left(- s_m^\vee \: {\cal{B}}
	\right)^k \: s_m^\vee \spc .
\EndEqu
F"ur die retardierte Greensfunktion setzen wir analog
\Equ{2_t3}
\tilde{s}_m^\wedge \;=\; \sum_{k=0}^\infty \left(- s_m^\wedge \: {\cal{B}}
	\right)^k \: s_m^\wedge \spc .
\EndEqu
Wie man direkt nachrechnet, erf"ullen $\tilde{s}^\vee_m$,
$\tilde{s}^\wedge_m$ tats"achlich die Bestimmungsgleichung der Greensfunktion
\Equ{2_t4}
(i \Pdd - m + {\cal{B}}) \: \tilde{s}^\vee_m \;=\;
(i \Pdd - m + {\cal{B}}) \: \tilde{s}^\wedge_m \;=\; \1 \spc .
\EndEqu
Man beachte, da"s die St"orungsrechnung f"ur die Greensfunktionen durch
die Forderung eindeutig wird, da"s der Tr"ager von $\tilde{s}^\vee_m,
\tilde{s}^\wedge_m$ im oberen bzw. unteren Lichtkegel liegt.
Als Folge der Kausalit"at treten bei dieser St"orungsrechnung keine
nichtlokalen Linienintegrale auf.

Wir kommen zur St"orungsrechnung f"ur $k_m$: Nach Definition \ref{2_def_av}
k"onnen wir $k_m$ durch die avancierte und retardierte Greensfunktion
ausdr"ucken
\[ k_m \;=\; \frac{1}{2 \pi i} \: (s^\vee_m - s^\wedge_m) \spc . \]
Wir "ubertragen diese Relation auf den wechselwirkenden Fall und verwenden
sie als Definitionsgleichung f"ur $\tilde{k}_m$.
\begin{Def}
\label{2_def7}
Wir setzen
\begin{eqnarray}
\label{eq:2_tm}
\tilde{k}_m &=& \frac{1}{2 \pi i} \: \left(\tilde{s}_m^\vee -
	\tilde{s}_m^\wedge \right)
\end{eqnarray}
mit $\tilde{s}^\vee_m$, $\tilde{s}^\wedge_m$ gem"a"s \Ref{2_t2}, \Ref{2_t3}.
\end{Def}
Wegen \Ref{2_t4} erf"ullt $\tilde{k}_m$ die Diracgleichung \Ref{2_q0}.
Da $\tilde{k}_m$ au"serdem im Grenzfall ${\cal{B}} \rightarrow 0$ in die
freie Distribution $k_m$ "ubergeht und die die Lokalit"atsbedingung
\Ref{2_fl} erf"ullt, scheint Definition \ref{2_def7} sinnvoll zu sein.

Wir m"ussen $\tilde{k}_m$ in die Form \Ref{2_s1} bringen. Dazu
schreiben wir die Summen \Ref{2_t2}, \Ref{2_t3} in \Ref{2_tm} aus
\Equ{2_t7}
\tilde{k}_m \;=\; \frac{1}{2 \pi i} \sum_{l=0}^\infty \left(
	(- s^\vee_m \:{\cal{B}})^l \: s^\vee_m \:-\:
	(- s^\wedge_m \:{\cal{B}})^l \: s^\wedge_m \right) \spc ,
\EndEqu
setzen \Ref{2_f7}, \Ref{2_f8} ein und multiplizieren aus.
Man erh"alt Operatorprodukte der Form \Ref{2_t6}, wobei die Faktoren $C_j$
f"ur $p_m$ oder $k_m$ stehen. Die Beitr"age mit einer geraden Zahl
von Faktoren $k_m$ haben f"ur die avancierte und retardierte
Greensfunktion das gleiche Vorzeichen und heben sich in \Ref{2_t7}
weg. Die Beitr"age mit einer ungeraden Anzahl von $k_m$'s treten in
beiden Greensfunktionen jeweils genau einmal auf und haben
umgekehrtes relatives Vorzeichen. Mit der Notation
\[ C_m(Q, n) \;=\; \left\{ \begin{array}{ll}
	k_m & {\mbox{falls $n \in Q$}} \\
	s_m & {\mbox{falls $n \not \in Q$}} \end{array} \right.
	\;\;\;, \spc Q \subset \N \]
k"onnen wir also \Ref{2_t7} in der Form
\begin{eqnarray}
\tilde{k}_m &=& \sum_{l=0}^\infty \: (-1)^l \!\!\! \sum_{ \scriptsize
	\begin{array}{cc} \scriptsize Q \in {\cal{P}}(l+1) , \\
		\scriptsize \# Q \; {\mbox{ungerade}}
	\end{array} }  (i \pi)^{\#Q-1} \nonumber \\
\label{eq:2_61}
&& \hspace*{1cm} \times \;
	C_m(Q,1) \: {\cal{B}} \: C_m(Q,2) \: {\cal{B}} \cdots
	{\cal{B}} \: C_m(Q,l) \: {\cal{B}} \: C_m(Q,l+1) \spc
\end{eqnarray}
umschreiben, wobei ${\cal{P}}(n)$ die Potenzmenge von
$\{1,\cdots,n\}$ bezeichnet.
Diese Summe "uber Operatorprodukte l"a"st sich in ein Produkt
zusammengesetzter Ausdr"ucke zerlegen, aus denen sich die Form des
Operators $V$ in \Ref{2_s1} ablesen l"a"st. Da die Kombinatorik etwas
un"ubersichtlich ist, geben wir gleich die Formel f"ur $V$ an.
\begin{Satz}
\label{2_satz3}
Der Operator
\begin{eqnarray}
V &=& \int_{\sR \cup i \sR} dm \;
	\sum_{l=0}^\infty \: (-1)^l \!\!\! \sum_{ \scriptsize
	\begin{array}{cc} \scriptsize Q \in {\cal{P}}(l) , \\
		\scriptsize \# Q \; {\mbox{gerade}}
	\end{array} }  \frac{(\#Q-1)!!}{(\#Q/2)! \cdot 2^{\#Q/2} }
	\; (i \pi)^{\#Q} \nonumber \\
\label{eq:2_62}
&& \hspace*{1cm} \times \;
	C_m(Q,1) \: {\cal{B}} \: C_m(Q,2) \cdots C_m(Q,l-1) \: {\cal{B}}
	\: C_m(Q,l) \: {\cal{B}} \: p_m \spc
\end{eqnarray}
erf"ullt die Gleichung
\[ \tilde{k}_m \;=\; V \: k_m \: V^* \spc . \]
\end{Satz}
{\Beweis}
Wir setzen zur Abk"urzung
\Equ{2_63}
c(n) \;=\; \frac{(2n-1)!!}{n! \cdot 2^n} \spc .
\EndEqu
Dann gilt f"ur alle $n$ die Gleichung\footnote{Das sieht man am einfachsten
mit der ``erzeugenden Funktion'' $f(x) =\sum_{n=0}^\infty c(n) \: x^n$. Aus
\Ref{2_64} folgt mit der Cauchy'schen Produktformel
\[ f^2(x) \;=\; \sum_{n=0}^\infty \left(\sum_{q=0}^n c(q) \: c(n-q) \right)
	x^n \;=\; \sum_{n=0}^\infty x^n \;=\; \frac{1}{1-x} \]
und somit $f(x)=(1-x)^{-1/2}$. Durch Taylorentwicklung dieser Funktion
erh"alt man die Koeffizienten \Ref{2_63}. }
\Equ{2_64}
\sum_{q=0}^n \; c(q) \; c(n-q) \;=\; 1 \spc .
\EndEqu
Bei der Berechnung von $V \: k_m \: V^*$ k"onnen wir mit Hilfe der Relation
\Ref{2_12} beide $m$-Integrale ausf"uhren und erhalten
\begin{eqnarray}
V \: k_m \: V^* &=& \sum_{l_1, l_2 =0}^\infty \; (-1)^{l_1+l_2}
\sum_{ \scriptsize
	\begin{array}{cc} \scriptsize Q_1 \in {\cal{P}}(l_1) , \\
		\scriptsize \# Q_1 \; {\mbox{gerade}}
	\end{array} }
\sum_{ \scriptsize
	\begin{array}{cc} \scriptsize Q_2 \in {\cal{P}}(l_2) , \\
		\scriptsize \# Q_2 \; {\mbox{gerade}}
	\end{array} } \!\!\!
	(i \pi)^{\#Q_1 + \#Q_2} \; c\left( \frac{\#Q_1}{2} \right)
	\: c \left( \frac{\#Q_2}{2} \right) \nonumber \\
\label{eq:2_65}
&& \hspace*{-2cm} \times \; C_m(Q_1, 1) \: {\cal{B}} \cdots {\cal{B}} \:
	C_m(Q_1, l_1) \:
	{\cal{B}} \: k_m \: {\cal{B}} \: C_m(Q_2, 1) \: {\cal{B}} \cdots
	{\cal{B}} \: C_m(Q_2, l_2) \;\;\; .
\end{eqnarray}
Jede Kombination der Operatorprodukte tritt genau $\frac{1}{2}(\#Q_1
+ \#Q_2)+1$ mal auf, denn der in \Ref{2_65} ausgeschriebene Faktor $k_m$
kann der insgesamt $1., 3., 5., \ldots$ Faktor $k_m$ des Produktes sein.
Wir fassen diese Summanden jeweils zusammen und erhalten
\begin{eqnarray*}
U \: k_m \: U^* &=& \sum_{l=0}^\infty \; (-1)^{l}
\sum_{ \scriptsize
	\begin{array}{c} \scriptsize Q \in {\cal{P}}(l+1) , \\
		\scriptsize \# Q \; {\mbox{ungerade}}
	\end{array} } \!\!\!
	(i \pi)^{\#Q-1} \; \sum_{\scriptsize
	\begin{array}{c} \scriptsize q=0, \\
		\scriptsize q\; {\mbox{gerade}}
	\end{array} }^{\# Q}
	c\left(\frac{q}{2}\right) \; c\left(\frac{\#Q-1-q}{2}\right) \\
&&\times \; C_m(Q, 1) \: {\cal{B}} \: C_m(Q,2) \cdots C_m(Q,l) \:
	{\cal{B}} \: C_m(Q, l+1) \;\;\; .
\end{eqnarray*}
Nach Einsetzen von \Ref{2_64} und Vergleich mit \Ref{2_61} folgt die
Behauptung.
\QED

Nachdem wir den Operator $V$ kennen, k"onnen wir nun auch die
St"orungsrechnung f"ur $p_m$ durchf"uhren. Dazu verwenden wir
\Ref{2_s1} als Definitionsgleichung.
\begin{Def}
\label{2_def8}
Wir setzen
\Equ{2_123a}
\tilde{p}_m \;=\; V \: p_m \: V^*
\EndEqu
mit $V$ gem"a"s \Ref{2_62}.
\end{Def}
Wir m"ussen verifizieren, da"s $\tilde{p}_m$ die Diracgleichung erf"ullt.
\begin{Lemma}
\label{2_lem0}
Es gilt
\[ (i \Pdd - m + {\cal{B}}) \: \tilde{p}_m \;=\; 0 \spc . \]
\end{Lemma}
{\Beweis}
Wir rechnen explizit
\begin{eqnarray*}
(i \Pdd - m) \: V p_m &=&
	\sum_{l=0}^\infty \: (-1)^l \!\!\! \sum_{ \scriptsize
	\begin{array}{cc} \scriptsize Q \in {\cal{P}}(l) , \\
		\scriptsize \# Q \; {\mbox{gerade}}
	\end{array} }  \frac{(\#Q-1)!!}{(\#Q/2)! \cdot 2^{\#Q/2} }
	\; (i \pi)^{\#Q} \\
&&\;\;\times \; (i \Pdd - m) \:
	C_m(Q,1) \: {\cal{B}} \: C_m(Q,2) \cdots C_m(Q,l-1) \: {\cal{B}}
	\: C_m(Q,l) \: {\cal{B}} \: p_m \\
&&\hspace*{-2.5cm}=\; \sum_{l=1}^\infty \: (-1)^l \!\!\! \sum_{ \scriptsize
	\begin{array}{cc} \scriptsize Q \in {\cal{P}}(l) , \\
		\scriptsize \# Q \; {\mbox{gerade und $1 \not \in Q$}}
	\end{array} }  \frac{(\#Q-1)!!}{(\#Q/2)! \cdot 2^{\#Q/2} }
	\; (i \pi)^{\#Q} \\
&&\;\;\times \;
	{\cal{B}} \: C_m(Q,2) \cdots C_m(Q,l-1) \: {\cal{B}}
	\: C_m(Q,l) \: {\cal{B}} \: p_m \;=\; - {\cal{B}} \: V p_m \;\;\; .
\end{eqnarray*}
\QED

Damit haben wir die St"orungsrechnung f"ur $p_m$, $k_m$ durchgef"uhrt.
Wegen der Kausalit"at bei der St"orungsrechnung f"ur die
Greensfunktionen nennen wir die Methode
{\em{kausale St"orungsentwicklung}}\index{kausale St"orungsentwicklung}.
F"ur $\tilde{k}_m$ folgt direkt aus der Konstruktion, da"s
keine nichtlokalen Linienintegrale auftreten.
F"ur $\tilde{p}_m$ ist das nach Definition \ref{2_def8} nicht unmittelbar
klar. Man mu"s dazu die Analogie der
Formeln der Lichtkegelentwicklung f"ur $\tilde{p}_m$, $\tilde{k}_m$
ausnutzen, der wir schon bei der Diskussion der Ergebnisse aus Anhang
A-D begegnet sind. Wir werden nicht allgemein
beweisen, da"s $\tilde{p}_m$ der Lokalit"atsforderung \Ref{2_fl} gen"ugt.
Bei allen expliziten Rechnungen wird sich aber zeigen, da"s
tats"achlich alle nichtlokalen Linienintegrale verschwinden.

Nach unserer Konstruktion ist klar, da"s sich die
Eindeutigkeit der St"orungsrechnung f"ur die Greensfunktionen auch auf
$\tilde{p}_m, \tilde{k}_m$ "ubertr"agt. Wir k"onnen also sagen, da"s die
kausale St"orungsentwicklung die einzige Methode der St"orungsrechnung
ist, bei der keine nichtlokalen Linienintegrale auftreten.

Um die Unterschiede zwischen der kausalen St"orungsrechnung und den
urspr"unglichen Gleichungen \Ref{2_q11} besser zu erkennen, betrachten
wir abschlie"send eine Entwicklung bis zur Ordnung ${\cal{O}}({\cal{B}}^3)$:
\begin{eqnarray}
\label{eq:2_71}
V &=& \int_{\sR \cup i \sR} dm \; \left( p_m - s_m \: {\cal{B}} \: p_m +
	s_m \: {\cal{B}} \: s_m \: {\cal{B}} \: p_m - \frac{\pi^2}{2}
	\: k_m \: {\cal{B}} \: k_m \: {\cal{B}} \: p_m \right)
	+ {\cal{O}}({\cal{B}}^3) \spc\;\: \\
\tilde{p}_m &=& p_m - s_m \: {\cal{B}} \: p_m - p_m \: {\cal{B}} \: s_m
	\;+\; p_m \: {\cal{B}} \: s_m \: {\cal{B}} \: s_m + s_m \:
	 \: {\cal{B}} \: p_m \: {\cal{B}} \: s_m + s_m \: {\cal{B}} \: s_m
	 \: {\cal{B}} \: p_m \nonumber \\
\label{eq:2_72}
&& - \frac{\pi^2}{2} \left( k_m \: {\cal{B}} \: k_m \: {\cal{B}} \: p_m
	+ p_m \: {\cal{B}} \: k_m \: {\cal{B}} \: k_m \right)
	\;+\; {\cal{O}}({\cal{B}}^3) \\
\tilde{k}_m &=& k_m - s_m \: {\cal{B}} \: k_m - k_m \: {\cal{B}} \: s_m
	\;+\; k_m \: {\cal{B}} \: s_m \: {\cal{B}} \: s_m + s_m \:
	 \: {\cal{B}} \: k_m \: {\cal{B}} \: s_m + s_m \: {\cal{B}} \: s_m
	 \: {\cal{B}} \: k_m \nonumber \\
\label{eq:2_73}
&& - \pi^2 \; k_m \: {\cal{B}} \: k_m \: {\cal{B}} \: k_m
	\;+\; {\cal{O}}({\cal{B}}^3)
\end{eqnarray}
In erster Ordnung stimmen diese Gleichungen mit \Ref{2_69} und
\Ref{2_28}, \Ref{2_29} "uberein.
In zweiter Ordnung tritt in den Formeln f"ur $\tilde{p}_m$,
$\tilde{k}_m$ gegen"uber \Ref{2_q11} zus"atzlich der Beitrag in der
zweiten Zeile von \Ref{2_72}, \Ref{2_73} auf.
Durch diese Beitr"age verschwinden die nichtlokalen Linienintegrale
in zweiter Ordnung.
Der Operator $V$ unterscheidet sich von $U$, \Ref{2_69}, durch den
zus"atzlichen Term $-\frac{\pi^2}{2}\: k_m \: {\cal{B}} \: k_m \: {\cal{B}}
\: p_m$. Wie erwartet ist $V$ dadurch in zweiter Ordnung nicht unit"ar,
genauer
\[ V \: V^* \;=\; V^* \: V \;=\; 1 - \frac{\pi^2}{2} \int_{\sR \cup i \sR} dm \;
	\left( k_m \: {\cal{B}} \: k_m \: {\cal{B}} \: p_m +
	p_m \: {\cal{B}} \: k_m \: {\cal{B}} \: k_m \right)
	\;+\; {\cal{O}}({\cal{B}}^3) \spc . \]

\subsection{Formale St"orungsentwicklung f"ur $P(x,y)$}
Die kausale St"orungsentwicklung f"ur $p_m, k_m$ l"a"st sich direkt
auf den fermionischen Projektor "ubertragen, indem man in alle
Formeln wie in den Abschnitten \ref{2_ab23}, \ref{2_ab24} die
Asymmetriematrizen $X, Y$ einf"ugt.
Zur Klarheit stellen wir die Konstruktion noch einmal in systematischer
Reihenfolge zusammen.

Wir arbeiten wieder gem"a"s \Ref{2_o0}, \Ref{2_o1} mit den Operatoren
$s, k, p$ auf $H \otimes \C^f$. Analog zu \Ref{2_f7}, \Ref{2_f8}
definieren wir die avancierte und retardierte Greensfunktion durch
\[ s^\vee_{[m]} \;=\; s_{[m]} \:+\: i \pi \: k_{[m]} \;\;\;\;,\spc
	s^\wedge_{[m]} \;=\; s_{[m]} \:-\: i \pi \: k_{[m]} \spc . \]
F"ur diese Greensfunktionen l"a"st sich die St"orungsrechnung kanonisch
durchf"uhren, wir setzen
\Equ{2_u3}
\tilde{s}^\vee \;=\; \sum_{k=0}^\infty (-s^\vee \:{\cal{B}})^k \: s^\vee
	\;\;\;\;,\spc \tilde{s}^\wedge \;=\;
		\sum_{k=0}^\infty (-s^\wedge \:{\cal{B}})^k \: s^\wedge
	\spc .
\EndEqu
Wie man direkt nachrechnet, erf"ullen die gest"orten Greensfunktionen
die Gleichungen
\[ (i \Pdd - m Y + {\cal{B}}) \: \tilde{s}^\vee \;=\;
	(i \Pdd - m Y + {\cal{B}}) \: \tilde{s}^\wedge \;=\; \1 \spc . \]
Im Gegensatz zum vorigen Abschnitt f"uhren wir nun zun"achst den
Operator $V$ ein und definieren damit die gest"orten Gr"o"sen.
Auf diese Weise k"onnen $p, k$ einheitlich behandelt werden.

\begin{Def}
Wir definieren auf $H \otimes \C^f$ den Operator
\begin{eqnarray}
V &=& \int_{\sR \cup i \sR} dm \;
	\sum_{l=0}^\infty \: (-1)^l \!\!\! \sum_{ \scriptsize
	\begin{array}{cc} \scriptsize Q \in {\cal{P}}(l) , \\
		\scriptsize \# Q \; {\mbox{gerade}}
	\end{array} }  \frac{(\#Q-1)!!}{(\#Q/2)! \cdot 2^{\#Q/2} }
	\; (i \pi)^{\#Q} \nonumber \\
\label{eq:2_u2}
&& \hspace*{1cm} \times \;
	C_{[m]}(Q,1) \: {\cal{B}} \: C_{[m]}(Q,2) \cdots
	C_{[m]}(Q,l-1) \: {\cal{B}} \: C_{[m]}(Q,l) \: {\cal{B}}
	\: p_m \;\;\;,\spc
\end{eqnarray}
dabei stehen die Faktoren $C_{[m]}(Q,n)$ f"ur die Operatoren
\[ C_{[m]}(Q, n) \;=\; \left\{ \begin{array}{ll}
	k_{[m]} & {\mbox{falls $n \in Q$}} \\
	s_{[m]} & {\mbox{falls $n \not \in Q$}} \end{array} \right.
	\;\;\;, \spc Q \subset \N \spc . \]
Die St"orungsentwicklung f"ur $p, k$ wird mit der nichtunit"aren St"ortransformation
\[ \tilde{p} \;:=\; V \:p\: V^* \spc,\spc
	\tilde{k} \;:=\; V \:k\: V^* \]
durchgef"uhrt.
\end{Def}

\begin{Satz}
Die Operatoren $\tilde{p}, \tilde{k}$ erf"ullen die gest"orte Diracgleichung
\Equ{2_u0}
(i \Pdd - m Y + {\cal{B}}) \: \tilde{p} \;=\;
	(i \Pdd - m Y + {\cal{B}}) \: \tilde{k} \;=\; 0 \spc .
\EndEqu
Die St"orungsrechnung f"ur $k$ l"a"st sich auf diejenige
f"ur die Greensfunktionen zur"uckf"uhren,
\Equ{2_u1}
\tilde{k} \;=\; \frac{1}{2 \pi i} \: (\tilde{s}^\vee - \tilde{s}^\wedge)
	\spc .
\EndEqu
\end{Satz}
{\Beweis}
Gleichung \Ref{2_u1} und die Diracgleichung \Ref{2_u0} f"ur $\tilde{k}$
folgen genau wie in Satz \ref{2_satz3}. Die Diracgleichung f"ur
$\tilde{p}$ erh"alt man analog wie in Satz \ref{2_lem0}.
\QED
Falls der fermionische Projektor nur eine Massenasymmetrie besitzt,
k"onnen wir $\tilde{P}(x,y)$ analog zu \Ref{2_n5} durch
\[ \tilde{P}(x,y) \;=\; \frac{1}{2} \: \Tr_{\cal{F}}
	(\tilde{p} - \tilde{k})(x,y) \]
definieren. F"ur lokale St"oroperatoren ${\cal{B}}$ treten bei dieser
St"orungsrechnung keine nichtlokalen Linienintegrale auf.

Im Fall mit zus"atzlicher chiraler Asymmetrie m"ussen wir die Matrix $X$
einf"ugen und setzen entsprechend zu Gleichung \Ref{2_o5}
\Equ{2_u00}
\tilde{P} \;=\; \frac{1}{2} \: \Tr_{\cal{F}} \left( V \:
	X (p-k) \: V^* \right) \spc .
\EndEqu
Als Folge der chiralen Asymmetrie ist die Lokalit"atsforderung \Ref{2_fl}
nicht mehr f"ur beliebige lokale St"oroperatoren erf"ullt.
In Verallgemeinerung unserer "Uberlegung in erster Ordnung St"orungstheorie
fallen aber f"ur die Diracoperatoren \Ref{2_p3}, \Ref{2_p6} auch in
h"oherer Ordnung alle nichtlokalen Linienintegrale weg.

\subsection{St"orungsrechnung im Ortsraum}
\label{2_ab33}
Nachdem die formale St"orungsentwicklung durchgef"uhrt ist,
k"onnen wir uns dem Studium der einzelnen St"orungsbeitr"age zuwenden.
Wir m"ussen zeigen, da"s alle Beitr"age
endlich sind. Au"serdem m"ussen f"ur die St"orungsbeitr"age explizite
Formeln im Ortsraum abgeleitet werden.
In Anhang E wurden einige Rechnungen in h"oherer Ordnung
St"orungstheorie durchgef"uhrt. Wir wollen hier
die verwendete Methode veranschaulichen und die Ergebnisse
aus Anhang E beschreiben.

Bevor wir beginnen,
k"onnen wir schon einen allgemeinen Unterschied zum Vorgehen in Abschnitt
\ref{2_ab22} festhalten:
In erster Ordnung war es ausreichend, die St"orungsrechnung
f"ur $p_m, k_m$ durchzuf"uhren. Wegen der Linearit"at lassen sich die
Ergebnisse n"amlich direkt auf den fermionischen Projektor "ubertragen,
indem man die Asymmetriematrizen $X, Y$ geeignet in die
Entwicklungsformeln einf"ugt.
In St"orungstheorie h"oherer Ordnung ist die Situation komplizierter,
weil die St"oroperatoren und Asymmetriematrizen in verschiedensten
Kombinationen, z.B.
\[ X\:Y\:{\cal{B}}\:{\cal{B}} \;\;\;,\;\;\;\; X\:{\cal{B}}\:Y\:{\cal{B}}
	\;\;\;,\;\;\;\; X \:{\cal{B}}\:{\cal{B}}\:Y \;\;\;,\;\;\;\;
	{\cal{B}}\: X \:{\cal{B}}\: Y \;\;\;,\;\;\;\;\ldots \spc , \]
auftreten k"onnen. Aus diesem Grund m"ussen wir nun die
St"orungsrechnung \Ref{2_u00} f"ur den fermionischen Projektor untersuchen.

\subsubsection*{Prinzip der Rechnung}
Da eine explizite Durchf"uhrung der St"orungsrechnung mit dem Operator $V$
wegen der kombinatorischen
Faktoren in \Ref{2_u2} un"ubersichtlich ist, wird in Anhang E
mit der St"orungsentwicklung \Ref{2_u3} f"ur die Greensfunktionen
gearbeitet.
Mit Hilfe von Gleichung \Ref{2_u1} lassen sich die Ergebnisse auf
$\tilde{k}$ und, mit einem allgemeinen Analogieargument f"ur
die Ergebnisse der Lichtkegelentwicklung, auch auf $\tilde{p}$ "ubertragen.
Schlie"slich wird in die Endformeln die chirale Asymmetriematrix $X$ eingef"ugt.

Damit die Darstellung "uberschaubar bleibt, lassen wir hier die chirale
Asymmetriematrix weg und betrachten lediglich die St"orungsentwicklung f"ur
die avancierte Greensfunktion.
Au"serdem werden wir die Methode der Rechnung nur am Beispiel eines $U(B)$-Potentials
${\cal{B}}=\Aslsh$ beschreiben. Wir konzentrieren uns also auf die formale
St"orungsreihe
\Equ{2_v9}
\tilde{s}^\vee \;=\; \sum_{n=0}^\infty (-s^\vee \:\Aslsh)^n \: s^\vee
	\spc .
\EndEqu

Zun"achst "uberlegen wir, warum die Beitr"age jeder Ordnung als
Distribution wohldefiniert sind:
F"ur den Beitrag erster Ordnung $-s^\vee \Aslsh s^\vee$ k"onnen wir mit
Hilfe von \Ref{2_u1} die Ergebnisse von Theorem \ref{a1_theorem1},
Theorem \ref{a2_theorem2} "ubertragen und erhalten
Entwicklungsformeln bis zur Ordnung ${\cal{O}}(\xi^2)$. Wir wollen
das Ergebnis intrinsisch mit der freien Greensfunktion ausdr"ucken.
Dazu entwickeln wir die avancierte Greensfunktion nach der Masse,
\begin{eqnarray*}
s^\vee_m &=& \sum_{l=0}^\infty m^l \: s^\vee_{(l)} \spc
	{\mbox{mit}} \\
s^\vee_{(0)}(x,y) &=& \frac{i}{\pi} \:\xi\slsh\: \delta^\prime(\xi^2) \:
	\Theta(\xi^0) \;\;\;\;,\spc s^\vee_{(1)}(x,y) \;=\; -\frac{1}{2\pi}
	\: l^\vee(\xi) \\
s^\vee_{(2)}(x,y) &=& -\frac{i}{4 \pi} \:\xi\slsh\: l^\vee(\xi)
	\;\;\;\;\;\;\;\;\;\;,\spc
	s^\vee_{(3)} \;=\; \frac{1}{8 \pi} \: \Theta^\vee(\xi) \spc,\;\;\;
	\ldots \;\;\; ,
\end{eqnarray*}
und schreiben den St"orungsbeitrag erster Ordnung in der Form
\Equ{2_v0}
(-s^\vee \:\Aslsh\: s^\vee)(x,y) \;=\; \sum_{l=0}^\infty F^1_{(l)}(x,y)
	\: s^\vee_{(l)}(x,y)
\EndEqu
mit glatten Funktionen $F^1_{(l)}$. Genauer sind
die Funktionen $F^1_{(l)}$ Linienintegrale "uber das Potential und
dessen partielle Ableitungen; in Theorem \ref{a1_theorem1} und
Theorem \ref{a2_theorem2}
wurden $F^1_{(0)},\ldots,F^1_{(6)}$ explizit berechnet.
Bei der St"orungsrechnung f"ur das elektromagnetische Feld haben wir
gesehen, da"s die Entwicklungsbeitr"age h"oherer Ordnung in der Masse
auf dem Lichtkegel schw"acher singul"ar sind. Daher ist einsichtig,
da"s sich die St"arke der Singularit"at auf dem Lichtkegel bei
Entwicklung nach $m$ mit der Formel
\Equ{2_v1}
(-s^\vee_{(p)} \:\Aslsh\: s^\vee_{(q)})(x,y) \;=\;
	\sum_{l=p+q}^\infty F^{1, (p,q)}_{(l)}(x,y) \: s^\vee_{(l)}(x,y)
\EndEqu
und geeigneten Funktionen $F^{1, (p,q)}_{(l)}$ beschreiben l"a"st.
Der entscheidende Schritt bei der "Ubertragung dieser Ergebnisse auf h"ohere
Ordnung St"orungstheorie ist die Tatsache, da"s die Entwicklungsformeln
\Ref{2_v0}, \Ref{2_v1} iteriert werden k"onnen.
Leider wird die Konstruktion
durch die Kombinatorik der Diracmatrizen in $F^1_{(l)}, s^\vee, {\cal{B}}$
erschwert. Zur Einfachheit werden wir diese Komplikation im folgenden
ignorieren. Dann liefert Gleichung \Ref{2_v0} bei Iteration eine
Entwicklungsformel vom gleichen Typ
\Equ{2_v2}
\left( (-s^\vee \:\Aslsh)^n \: s^\vee \right)(x,y) \;=\;
	\sum_{l=0}^\infty F^n_{(l)}(x,y) \: s^\vee_{(l)}(x,y) \spc ,
\EndEqu
nur haben die Funktionen $F^n_{(l)}$ als geschachtelte Linienintegrale
"uber $A$ und $\partial^p A$ gegen"uber $F^1_{(l)}$ eine kompliziertere
Form.
Wir sehen, da"s die St"orungsbeitr"age jeder Ordnung wohldefiniert sind.
Bei Entwicklung nach der Masse erhalten wir mit Hilfe von \Ref{2_v1}
eine Gleichung der Form
\Equ{2_v3}
\left( (-s^\vee \:\Aslsh)^n \: s^\vee \right)_{(p)}(x,y) \;=\;
	\sum_{l=p}^\infty F^{n, p}_{(l)}(x,y) \: s^\vee_{(l)}(x,y) \spc .
\EndEqu
Die Beitr"age h"oherer Ordnung in $m$ sind also
auf dem Lichtkegel schw"acher singul"ar;
damit ist auch in h"oherer Ordnung St"orungstheorie eine
Entwicklung nach der Masse sinnvoll.

Wir kommen zur Frage, wie die St"orungsbeitr"age konkret aussehen.
Bevor wir mit einer detaillierteren Untersuchung der Funktionen
$F^n_{(l)}$ in \Ref{2_v2} beginnen, beschreiben wir das weitere
Vorgehen im Prinzip:
Wir berechnen zun"achst $F^n_{(l)}$ f"ur festes $l$ und beliebiges $n$.
Die Summe "uber $n$ kann ausgef"uhrt werden, und wir erhalten explizite Ausdr"ucke
f"ur die Funktionen
\[ F_{(l)} \;:=\; \sum_{n=0}^\infty F^n_{(l)} \spc . \]
Diese Funktionen liefern nicht-perturbative
Entwicklungsformeln f"ur $\tilde{s}^\vee$, denn nach \Ref{2_v9},
\Ref{2_v2} gilt
\begin{eqnarray*}
\tilde{s}^\vee(x,y) &=& \sum_{n=0}^\infty \:\sum_{l=0}^\infty
	\: F^n_{(l)}(x,y) \: s^\vee_{(l)}(x,y) \\
&=& \sum_{l=0}^\infty \:\left( \sum_{n=0}^\infty
	\: F^n_{(l)}(x,y) \: s^\vee_{(l)}(x,y) \right) \;=\;
	\sum_{l=0}^\infty \: F_{(l)}(x,y) \: s^\vee_{(l)}(x,y) \spc .
\end{eqnarray*}
Es ist mathematisch nicht klar, da"s die Summen "uber $l, n$ vertauscht
werden k"onnen. Aus diesem Grund liefert unsere Methode, wie bereits
zu Beginn dieses Abschnittes erw"ahnt, keinen Beweis f"ur die
Konvergenz der St"orungsentwicklung. Da es uns mehr auf die explizite
Berechnung von $\tilde{s}^\vee$ ankommt, klammern wir diese eher
technischen Konvergenzfragen aus.

Wir beschreiben die Technik zur Berechnung der Funktionen $F^n_{(l)}$,
$F_{(l)}$ in mehreren Schritten und beginnen mit $l=0$, also der
f"uhrenden Singularit"at auf dem Lichtkegel:
In erster Ordnung St"orungstheorie brauchen wir nur die Eichterme
$\sim m^0$ zu ber"ucksichtigen,
\[ (-s^\vee \:\Aslsh\: s^\vee)(x,y) \;=\; -i \int_x^y A_j \xi^j \:
	s_0^\vee(x,y) \:+\: \cdots \spc . \]
Bei Iteration erhalten wir mit der Notation von \Ref{2_q3}
\begin{eqnarray*}
\left( (-s_0^\vee \:\Aslsh)^2 \: s_0^\vee \right)(x,y) &=&
	(-i)^2 \int_0^1 d\lambda_1 \int_{\lambda_1}^1 d\lambda_2 \;
	A_{j_1}(z_1) \:\xi^{j_1}\: A_{j_2}(z_2) \: \xi^{j_2} \;
	s_0^\vee(x,y) \:+\: \cdots \\
\left( (-s_0^\vee \:\Aslsh)^n \: s_0^\vee \right)(x,y) &=&
	(-i)^n \int_0^1 d\lambda_1 \int_{\lambda_1}^1 d\lambda_2 \;
	\cdots\; \int_{\lambda_{n-1}}^1 d\lambda_n \\
&&\hspace*{1.5cm} \times 
	A_{j_1}(z_1) \:\xi^{j_1}\:\cdots\: A_{j_n}(z_n) \: \xi^{j_n} \;
	s_0^\vee(x,y) \:+\: \cdots \spc .
\end{eqnarray*}
Die Beitr"age lassen sich zu einem
{\em{geordneten Exponential}}\index{geordnetes Exponential}
\Equ{2_v4}
\tilde{s}^\vee(x,y) \;=\; \Texp \left( -i \int_x^y A_j \xi^j \right)
	\: s_0^\vee(x,y) \:+\: \cdots
\EndEqu
aufsummieren, das wie "ublich durch die
absolut konvergente Dyson-Reihe gegeben ist,
\[ \Texp \left( -i \int_x^y A_j \xi^j \right) \;:=\; \sum_{n=0}^\infty
	(-i)^n \int_0^1 d\lambda_1 \int_{\lambda_1}^1 d\lambda_2 \;
	\cdots\; \int_{\lambda_{n-1}}^1 d\lambda_n \;
	A_{j_1}(z_1) \:\xi^{j_1}\:\cdots\: A_{j_n}(z_n) \: \xi^{j_n}
	\;\;\; . \]

Wir bemerken, da"s das Auftreten des geordneten Integrals in
\Ref{2_v4} nicht erstaunlich ist, sondern bereits aufgrund der nichtabelschen
Eichsymmetrie zu erwarten war: Im Fall einer $U(B)$-Eichtransformation
$\Psi(x) \rightarrow U(x) \: \Psi(x)$ haben wir
$A_j=i U (\partial_j U^{-1})$. Das geordnete Integral kann ausgef"uhrt
werden\footnote{\label{2_footexp}
Das sieht man folgenderma"sen: Wir definieren f"ur
festes $x, y$ die Funktionen
\begin{eqnarray*}
F_1(\alpha) &=& U(\alpha y + (1-\alpha) x) \; U^{-1}(y) \\
F_2(\alpha) &=& \Texp \left( \int_z^y
	U (\partial_j U^{-1}) \; (y - z)^j \right)_{|z=\alpha y +
	(1-\alpha) x} \spc .
\end{eqnarray*}
Nach einer Variablentransformation hat man
\[ F_2(\alpha) \;=\;
	\sum_{k=0}^\infty \int_\alpha^1 d\lambda_1
	\int_{\lambda_1}^1 d\lambda_2 \;
	\cdots\; \int_{\lambda_{n-1}}^1 d\lambda_n \;
	(U (\partial_{j_1} U^{-1}))_{|z_1} \:\xi^{j_1}\:\cdots\:
	(U (\partial_{j_n} U^{-1}))_{|z_n} \: \xi^{j_n} \spc . \]
Wie man direkt nachrechnet, erf"ullen $F_1, F_2$ die gew"ohnliche
Differentialgleichung
\[ \frac{d}{d\alpha} F_{1\!/\!2}(\alpha) \;=\; - (U (\partial_j U^{-1}))_{|\alpha
	y + (1-\alpha) x} \:\xi^j\: F_{1\!/\!2}(\alpha)
	\spc{\mbox{mit}}\spc F_{1\!/\!2}(1) \;=\; 1 \spc , \]
und stimmen folglich auch f"ur $\alpha=0$ "uberein.}
\[ \Texp \left(-i \int_x^y A_j \xi^j \right) \;=\; \Texp \left(
	\int_x^y U (\partial_j U^{-1}) \: \xi^j \right) \;=\;
	U(x) \: U^{-1}(y) \spc , \]
und liefert in \Ref{2_v4} die gew"unschte lokale Phasentransformation
\[ \tilde{s}^\vee(x,y) \;=\; U(x) \:s^\vee_0(x,y)\: U^{-1}(y) \:+\:
	\cdots \spc . \]

Bei der n"achstschw"acheren Singularit"at $\sim l^\vee(\xi)$
von \Ref{2_v2} gibt es zwei verschiedene Beitr"age: zum einen
k"onnen $\sim m^0$ Ableitungen des Eichpotentials auftreten, zum
anderen tragen Terme h"oherer Ordnung in der Masse bei.
Wir untersuchen diese Beitr"age nacheinander:

Den Beitrag $\sim m^0$ $n$-ter Ordnung bauen wir in Gedanken
auf, indem wir den Beitrag erster Ordnung $(n-1)$-mal von links mit
dem Operator $-s^\vee_0 \Aslsh$ multiplizieren und nach jedem Schritt
um den Lichtkegel entwickeln.
Um eine Singularit"at $\sim l^\vee(\xi)$ zu erhalten, mu"s bei den
Lichtkegelentwicklungen genau einmal der schw"acher singul"are
Beitrag \Ref{a1_112},\Ref{a1_113},\Ref{a1_114} verwendet werden,
$(n-1)$-mal jedoch die f"uhrende Singularit"at der Eichterme.
Nach Addition "uber die Ordnung der St"orungstheorie k"onnen wir also
symbolisch
\Equ{2_v6}
\tilde{s}^\vee(x,y) \;\asymp\; \left[ \sum_{k_1=0}^\infty
	(-s_0^\vee \:\Aslsh)^{k_1} \: s_0^\vee \right] \:\hat{A} \!\slsh \:
	\left[ \sum_{k_2=0}^\infty (-s_0^\vee \:\Aslsh)^{k_2} \:
		s_0^\vee \right]
\EndEqu
schreiben, dabei bezeichnet der Hut ``$\hat{\:.\:}$'' die Stelle, an der
bei Lichtkegelentwicklung die Terme \Ref{a1_112},\Ref{a1_113},\Ref{a1_114}
auftreten sollen. In den eckigen Klammern d"urfen nur die
Eichterme verwendet werden, und wir k"onnen \Ref{2_v4} einsetzen.
Nach dieser "Uberlegung ist einsichtig, da"s wir den Beitrag zu
$\tilde{s}^\vee(x,y)$ erhalten, indem wir in die Terme von Theorem
\ref{a1_theorem1} geordnete Integrale "uber $A$ einf"ugen, genauer
\begin{eqnarray}
\lefteqn{ \tilde{s}^\vee(x,y) \;\asymp\;
- \frac{i}{4 \pi} \int_x^y dz \; (\alpha^2-\alpha) \;
	\T e^{-i \int_x^z A_j \: (z-x)^j}
	\: j_k(z) \: \xi^k \:
	\T e^{-i \int_z^y A_l \: (y-z)^l}
	\; \xi\slsh \: l^\vee(\xi) } \nonumber \\
&&+ \frac{1}{4 \pi} \int_x^y dz \; (2 \alpha - 1) \:
	\T e^{-i \int_x^z A_j \: (z-x)^j} \: \xi^j \:
	\gamma^k \: F_{kj}(z) \:
	\T e^{-i \int_z^y A_l \: (y-z)^l} \; l^\vee(\xi) \nonumber \\
\label{eq:2_x9}
&&+ \frac{i}{8 \pi} \int_x^y dz \;
	\T e^{-i \int_x^z A_j \: (z-x)^j}
	\:\varepsilon^{ijkl} \; F_{ij}(z) \:
	\xi_k \; \rho \gamma_l \:
	\T e^{-i \int_z^y A_l \: (y-z)^l} \; l^\vee(\xi) \;\;\; , \spc
\end{eqnarray}
wobei wir f"ur das geordnete Integral eine Kurzschreibweise verwendet
haben. Diese "Ubertragung des Ergebnisses erster Ordnung St"orungstheorie
auf endliche St"orungen h"atten wir "ahnlich wie \Ref{2_v4} wegen der
$U(B)$-Eichsymmetrie vermuten k"onnen.

Bei den Beitr"agen $\sim l^\vee(\xi)$ h"oherer Ordnung in der Masse
m"ussen wir wegen \Ref{2_v3} nur bis zur Ordnung ${\cal{O}}(m^3)$ entwickeln.
Bei den in $m$ linearen St"orungsbeitr"agen tritt in den
Operatorprodukten genau einmal der Faktor $m Y s^\vee_{(1)}$ auf.
Nach Resummation haben wir also
\Equ{2_x0}
\tilde{s}^\vee \;\asymp\; m \left[ \sum_{k_1=0}^\infty (-s^\vee_0
	\:\Aslsh)^{k_1} \right] \:Y s^\vee_{(1)} \: \left[
	\sum_{k_2=0}^\infty (-\Aslsh\: s^\vee_0)^{k_2} \right] \spc .
\EndEqu
Bei Lichtkegelentwicklung d"urfen nur die f"uhrenden Eichterme
verwendet werden. Die eckigen Klammern liefern bei
Lichtkegelentwicklung ganz "ahnlich wie in \Ref{2_v6} geordnete Exponentiale.
Im Spezialfall $[A, Y]=0$ erh"alt man in Verallgemeinerung des
Eichterms \Ref{a2_c2} den Ausdruck
\Equ{2_x1}
\tilde{s}^\vee_{(1)}(x,y) \;=\; \Texp \left(-i \int_x^y A_j \xi^j \right)
	\; Y s^\vee_{(1)} \;+\; \cdots \spc ,
\EndEqu
der allgemeine Fall ist etwas komplizierter. Zur Ordnung $\sim m^2$
gibt es zwei Beitr"age
\begin{eqnarray*}
\tilde{s}^\vee &\asymp& m^2 \left[ \sum_{k_1=0}^\infty (-s^\vee_0
	\:\Aslsh)^{k_1} \right] \:Y^2 s^\vee_{(2)} \: \left[
	\sum_{k_2=0}^\infty (-\Aslsh\: s^\vee_0)^{k_2} \right] \\
&&-m^2 \left[ \sum_{k_1=0}^\infty (-s^\vee_0 \:\Aslsh)^{k_1} \right]
		\:Y s^\vee_{(1)} \:\Aslsh\:
	\left[ \sum_{k_1=0}^\infty (-s^\vee_0
	\:\Aslsh)^{k_1} \right] \:Y s^\vee_{(1)} \:
	\left[ \sum_{k_2=0}^\infty (-\Aslsh\: s^\vee_0)^{k_2} \right]
	\spc ,
\end{eqnarray*}
die sich wiederum unter Verwendung der Eichterme iterativ
um den Lichtkegel entwickeln lassen.

Wir betrachten noch kurz die Singularit"at $\sim \Theta^\vee(\xi)$:
Mit der Methode \Ref{2_v6} lassen sich alle Beitr"age der
St"orungsrechnung erster Ordnung "ubertragen,
beispielsweise erhalten wir aus dem Stromterm \Ref{a1_118}
den Beitrag
\[ s^\vee(x,y) \;\asymp\; \frac{1}{4 \pi} \: \int_x^y dz \;
	(\alpha^2 - \alpha) \:
	\T e^{-i \int_x^z A_j \: (z-x)^j}
	\: \gamma^k \: j_k(z) \: \T e^{-i \int_z^y A_l \: (y-z)^l}
	\; \Theta^\vee(\xi) \spc . \]
Zus"atzlich k"onnen bei der Lichtkegelentwicklung zweimal die
schw"acher singul"aren Beitr"age \Ref{a1_112},\Ref{a1_113},\Ref{a1_114}
auftreten, also symbolisch
\[ \tilde{s}^\vee(x,y) \;\asymp\; \left[ \sum_{k_1=0}^\infty
	(-s_0^\vee \:\Aslsh)^{k_1} \: s_0^\vee \right] \:\hat{A} \!\slsh\:
	\left[ \sum_{k_2=0}^\infty (-s_0^\vee \:\Aslsh)^{k_2} \:
		s_0^\vee \right] \:\hat{A} \!\slsh\:
	\left[ \sum_{k_3=0}^\infty (-s_0^\vee \:\Aslsh)^{k_3} \:
		s_0^\vee \right] \spc . \]
Auf diese Weise erh"alt man beispielsweise einen Term, der proportional zum
Energie-Impuls-Tensor des Eichfeldes $F_{ik} \: F_j^k - \frac{1}{4} \: g_{ij} \:
F_{kl} F^{kl}$ ist.
Bei den St"orungsbeitr"agen h"oherer Ordnung in der Masse m"ussen nun bei
Lichtkegelentwicklung auch die Feldst"arke- und Stromterme
ber"ucksichtigt werden.

Damit wollen wir die Diskussion der einzelnen St"orungsbeitr"age abschlie"sen.
Es ist nach unserer Beschreibung klar, da"s die Methode der
Entwicklung und Resummation der Operatorprodukte beliebig fortgesetzt
werden kann. Nat"urlich werden die Rechnungen f"ur die schw"acheren
Singularit"aten auf dem Lichtkegel immer aufwendiger, im Prinzip l"a"st
sich damit aber $\tilde{s}^\vee$ (und damit letztlich auch der
fermionische Projektor) zu beliebiger Ordnung in $\xi^2$ exakt bestimmen.
Die Methode der Rechnungen ist auch f"ur theoretische "Uberlegungen
interessant, weil damit das Verhalten der St"orungsbeitr"age
auf den Lichtkegel schon vor expliziter Lichtkegelentwicklung bestimmt
werden kann.
Insbesondere k"onnen wir genau sagen, welche Beitr"age h"oherer Ordnung f"ur
uns wichtig sind, und k"onnen diese Beitr"age dann gezielt berechnen.

Als Vorbereitung auf die Diskussion des n"achsten Unterabschnitts
betrachten wir abschlie"send, wie sich die Ergebnisse auf den Diracoperator
mit chiralen Potentialen \Ref{2_p2}, also auf die St"orungsreihe
\[ \tilde{s}^\vee \;=\; \sum_{n=0}^\infty \left(-s^\vee \:
	(\chi_L \: \Aslsh_R + \chi_R \: \Aslsh_L) \right)^n \: s^\vee \]
"ubertragen lassen. Bei den Beitr"agen $\sim m^0$ nutzen wir aus,
da"s $s^\vee_0$ ungerade ist (also mit $\rho$ antikommutiert)
und k"onnen die chiralen Projektoren durchkommutieren,
\Equ{2_x7}
\chi_{L\!/\!R} \: \tilde{s}^\vee \;\asymp\; \chi_{L\!/\!R} \:
	\sum_{n=0}^\infty (-s^\vee_0 \: \Aslsh_{L\!/\!R})^n \: s^\vee_0
	\spc .
\EndEqu
Damit l"a"st sich die St"orungsrechnung f"ur ${\cal{B}}=\Aslsh$
unmittelbar "ubertragen.
In h"oherer Ordnung in $m$ ist die Situation etwas schwieriger, weil
$s^\vee_{(l)}$ f"ur ungerades $l$ eine gerade Matrix ist.
Beispielsweise haben wir f"ur den in $m$ linearen Beitrag anstelle
von \Ref{2_x0}
\Equ{2_x8}
\chi_{L\!/\!R} \: \tilde{s}^\vee \;\asymp\;
	m \left[ \sum_{k_1=0}^\infty (-s^\vee_0
	\:\Aslsh_{L\!/\!R})^{k_1} \right] \:Y s^\vee_{(1)} \: \left[
	\sum_{k_2=0}^\infty (-\Aslsh_{R\!/\!L}\: s^\vee_0)^{k_2} \right]
\EndEqu
und erhalten folglich bei Lichtkegelentwicklung Kombinationen der Form
\Equ{2_x2}
\Texp \left(-i \int_x^z A^j_{L\!/\!R} \: (z-x)_j \right) \:Y\:
	\Texp \left(-i \int_z^y A^k_{R\!/\!L} \: (y-z)_k \right)
	\spc .
\EndEqu
Allgemein kehrt sich in zusammengesetzten Ausdr"ucken bei jedem
Faktor $Y$ der chirale Index der Potentiale $A_{L\!/\!R}$ um.

\subsubsection*{Beschreibung der Ergebnisse von Anhang E}
Die Ergebnisse der Rechnungen von Anhang E sind in den
Lichtkegelentwicklungen von Theorem \ref{a6_thm0}
auf Seite \pageref{a6_thm0} und von Theorem \ref{a6_satz26}
auf Seite \pageref{a6_satz26} zusammengestellt.
Wir wollen nun diese Formeln genauer betrachten.

Theorem \ref{a6_thm0} ist im allgemeinen Fall
mit Massenasymmetrie und chiraler Asymmetrie anwendbar und liefert einen
expliziten Ausdruck f"ur die Operatoren $V Xp V^*$,
$V Xk V^*$. Gem"a"s \Ref{2_u00} erh"alt man durch Spurbildung "uber den
Flavour-Raum unmittelbar eine Gleichung f"ur den gest"orten
fermionischen Projektor. Der gest"orte Diracoperator kann die recht
allgemeine Form
\Equ{2_x5}
\chi_L \: U_R \:(i \Pdd + \Aslsh_R)\: U_R^{-1} \:+\:
	\chi_R \: U_L \:(i \Pdd + \Aslsh_L)\: U_L^{-1} \:-\:
	m \: \Xi \:-\: i \rho \:m\: \Phi \spc .
\EndEqu
haben; er enth"alt wie \Ref{2_p3} unit"ar transformierte
chirale Potentiale und zus"atzlich eine skalare/pseudoskalare St"orung.
Die chiralen Potentiale sollen mit der chiralen Asymmetriematrix kommutieren
\Equ{2_x6}
[X_L,\:A_L] \;=\; [X_R,\:A_R] \;=\; 0 \spc .
\EndEqu

Wir m"ussen zun"achst die verwendete Notation erkl"aren: Die chirale
Asymmetriematrix wurde wieder gem"a"s \Ref{2_x3} in $X_{L\!/\!R}$ zerlegt.
Die Matrizen $Y_{L\!/\!R}$ sind Kombinationen der Massenmatrix mit
dem skalaren/pseudoskalaren Potential, genauer
\Equ{2_y6}
Y_L(x) \;:=\; Y + \Xi(x) + i \Phi(x) \;\;\;,\spc
	Y_R(x) \;:=\; Y + \Xi(x) - i \Phi(x) \spc .
\EndEqu
Wir haben folglich
\[ Y + \Xi(x) + i \rho \Phi(x) \;=\; \chi_R \: Y_L + \chi_L \: Y_R \spc . \]
Die Tensoren $F^{ij}_{L\!/\!R}$, $j^k_{L\!/\!R}$ sind der Feldst"arke-
und Stromterm der chiralen Potentiale $A_{L\!/\!R}$.
Die Menge ${\cal{O}}(\ln(|\xi^2|))$ bezeichnet alle Distributionen $f(x,y)$ mit
der Eigenschaft, da"s $|(\ln(y-x)^2)^{-1} \: f(x,y)|$ regul"ar ist.
F"ur eine kompakte Schreibweise wurde schlie"slich der Ableitungsoperator
$\hat{\Pdd}$ eingef"uhrt. Bei Anwendung auf geordnete Exponentiale
liefert er nach Definition den Exponenten als geordneten Faktor, genauer
\begin{eqnarray*}
\hat{\Pdd}_x \Texp \left( -i \int_x^y A_j \xi^j \right) &:=& (i \Aslsh(x)) \:
	\Texp \left( -i \int_x^y A_j \xi^j \right) \\
\hat{\Pdd}_y \Texp \left( -i \int_x^y A_j \xi^j \right) &:=&
	\Texp \left( -i \int_x^y A_j \xi^j \right) \: (-i \Aslsh(y)) \spc .
\end{eqnarray*}
Auf alle anderen Funktionen wirkt $\hat{\Pdd}$ wie ein gew"ohnlicher
Differentialoperator. In einem zusammengesetzten Ausdruck haben wir
beispielsweise
\begin{eqnarray*}
\lefteqn{ \hat{\Pdd}_z \: \T e^{-i \int_x^z A_j \: (z-x)^j} \:f(z)\:
	\T e^{-i \int_z^y A_j \: (y-z)^j} } \\
&=& \T e^{-i \int_x^z A_j \: (z-x)^j} \left( (-i \Aslsh) \: f \:+\: (\Pdd f)
	\:+\: f \: (i \Aslsh) \right)_{|z} \: \T e^{-i \int_z^y A_j \: (y-z)^j}
	\spc .
\end{eqnarray*}

Zur besseren "Ubersicht beginnen wir die Diskussion von Theorem
\ref{a6_thm0} mit einem Spezialfall und gehen dann schrittweise zu den
allgemeinen Voraussetzungen "uber. Zun"achst betrachten wir
den Fall $X=1$ ohne chirale Asymmetrie und nehmen mit den Voraussetzungen
$U_{L\!/\!R} \equiv 1$, $\Xi \equiv \Phi \equiv 0$ an, da"s der Diracoperator
die Form \Ref{2_p2} hat.
Es f"allt auf, da"s in \Ref{a6_405} ein nichtlokales Linienintegral auftritt.
Nach Zusammenfassen der beiden Summanden in der geschweiften Klammer
f"allt aber die Nichtlokalit"at weg.
Insgesamt vereinfacht sich die Formel von Theorem \ref{a6_thm0}
auf das Zwischenergebnis von Satz \ref{a6_satz7} (man beachte, da"s
dort die beiden Summanden \Ref{a6_400}, \Ref{a6_404}
zusammengefa"st sind).
Im f"uhrenden Summanden \Ref{a6_400} tritt genau wie in \Ref{2_v4}
ein geordnetes Exponential "uber das Potential auf;
wir wir an \Ref{2_x7} gesehen hatten, mu"s man lediglich $A$ durch
$A_L$ ersetzen. Die Summanden \Ref{a6_401},
\Ref{a6_402}, \Ref{a6_403} entsprechen den Beitr"agen \Ref{2_x9}.
Die Terme erster Ordnung in der Masse, \Ref{a6_404}, \Ref{a6_405},
erh"alt man durch Lichtkegelentwicklung von \Ref{2_x8}. Man sieht
in "Ubereinstimmung mit \Ref{2_x2}, da"s das chirale Potential
links und rechts des Faktors $Y$ umgekehrte H"andigkeit besitzt.
In erster Ordnung in $A_{L\!/\!R}$ gehen die Beitr"age \Ref{a6_404},
\Ref{a6_405} in die Eich-/Pseudoeichterme \Ref{a2_c2}, \Ref{a2_x1},
\Ref{a2_x2} "uber, wie man direkt nachrechnen kann.
Alle weiteren St"orungsbeitr"age sind wenigstens quadratisch
in der Masse oder auf dem Lichtkegel h"ochstens logarithmisch singul"ar
und wurden weggelassen.

Im n"achsten Schritt gehen wir zum Fall mit chiraler Asymmetrie "uber.
In der Lichtkegelentwicklung treten nun die Faktoren $X_L$
und $X_R$ auf, und zwar immer in Kombination mit geordneten Exponentialen
"uber $A_L$ bzw. $A_R$. Wegen Bedingung \Ref{2_x6} kommutiert $X_{L\!/\!R}$
mit diesen geordneten Exponentialen. Insbesondere k"onnen wir in der
geschweiften Klammer von \Ref{a6_405} die Faktoren $X_{L\!/\!R}$
in die Mitte kommutieren, wo sie bei Anwendung von \Ref{2_asymm}
herausfallen.
Folglich k"onnen die beiden Summanden in der geschweiften Klammer wieder
zu einem lokalen Integral zusammengefa"st werden.
In \Ref{a6_401}, \Ref{a6_402}, \Ref{a6_403} kommutiert $X_L$ jeweils
mit dem gesamten Integralausdruck.

An den auftretenden Produkten von $X_{L\!/\!R}$ mit den geordneten
Exponentialen kann man sich "uberlegen, was die Kommutatorbedingung
\Ref{2_x6} bei endlichen St"orungen bedeutet, wir diskutieren exemplarisch
den Beitrag \Ref{a6_400}:
Wir nehmen an, da"s $X_L$ nicht mit dem geordneten Exponential in \Ref{a6_400}
kommutiert. Dann ist \Ref{a6_400} nicht hermitesch und mu"s folglich
durch einen anderen Ausdruck ersetzt werden. Es zeigt sich, da"s in der
St"orungsrechnung h"oherer Ordnung ("ahnlich wie bei \Ref{2_o8})
unbeschr"ankte Linienintegrale auftreten. Man erh"alt also anstelle
des geordneten Exponentials eine unendliche Reihe geschachtelter,
nichtlokaler Linienintegrale.
Damit ist die Lokalit"atsforderung \Ref{2_fl} selbstverst"andlich verletzt.
Wenn man will, kann man sogar einen Schritt weiter gehen und die
Konvergenzprobleme dieser Reihe als mathematisches Argument
f"ur die Lokalit"atsforderung ansehen.

Wir kommen zum Fall mit zus"atzlichen chiralen Transformationen
$U_{L\!/\!R}$.
Nach der Relation\footnote{Dies kann man ganz analog wie in Fu"snote
\ref{2_footexp} auf Seite \pageref{2_footexp} verifizieren.}
\[ \Texp \left( \int_x^y (-i U A_j U^{-1} + U (\partial_j U^{-1})) \: \xi^j
	\right) \;=\; U(x) \:\Texp \left( -i \int_x^y A_j \xi^j \right)\:
	U^{-1}(y) \]
ist klar, da"s alle geordneten Exponentiale von links und rechts mit
einem Faktor $U_{L\!/\!R}$ bzw. $U_{L\!/\!R}^{-1}$ zu multiplizieren sind.
Die chiralen Asymmetriematrizen $X_{L\!/\!R}$ treten stets zwischen den
beiden zus"atzlichen Faktoren $U_{L\!/\!R}, U_{L\!/\!R}^{-1}$ auf.
Man beachte, da"s das Integral in \Ref{a6_405} nicht mehr
notwendigerweise lokal ist.
Damit die Nichtlokalit"at verschwindet, m"ussen die Produkte von
$X_{L\!/\!R}$ mit dem mittleren Faktor $U_L^{-1} Y_L U_R$ in beiden
Summanden der geschweiften Klammer "ubereinstimmen, also
\Equ{2_z1}
U_L^{-1} Y U_R \: X_R \;=\; X_L \: U_L^{-1} Y U_R\;\;\; {\mbox{und entsprechend}}
	\;\;\; U_R^{-1} Y U_L \: X_L \;=\; X_R \: U_R^{-1} Y U_L \;\;\; .
\EndEqu

Im letzten Schritt betrachten wir zus"atzlich die skalare/pseudoskalare
St"orung: Gem"a"s unserer "Uberlegung an
\Ref{2_k4a}, \Ref{2_x4} beschreiben die Potentiale $\Xi, \Phi$ f"ur die
f"uhrende Singularit"at auf dem Lichtkegel
eine skalare bzw. axiale Massenverschiebung.
Daher ist einsichtig, da"s wir zur Beschreibung der skalaren/pseudoskalaren
St"orung einfach $\Xi$, $\Phi$ gem"a"s \Ref{2_y6} mit der Massenmatrix
zusammenfassen und $Y$ durch die dynamischen Massenmatrizen
$Y_{L\!/\!R}(x)$ ersetzen m"ussen.
Der chirale Index ist dabei stets wie bei den links davon stehenden
Potentialen $A_{L\!/\!R}$ zu w"ahlen.
Damit \Ref{a6_405} ein lokales Linienintegral ist, m"ussen analog zu
\Ref{2_z1} die Gleichungen
\Equ{2_z2}
U_L^{-1} Y_L U_R \: X_R \;=\; X_L \: U_L^{-1} Y_L U_R \;\;\;{\mbox{und entsprechend}}
	\;\;\; U_R^{-1} Y_R U_L \: X_L \;=\; X_R \: U_R^{-1} Y_R U_L
\EndEqu
gelten.

Zum Abschlu"s der Diskussion von Theorem \ref{a6_thm0} stellen wir einen
Zusammenhang zum Diracoperator \Ref{2_p6} und den Bedingungen
\Ref{2_y1}, \Ref{2_p5} her genauer begr"unden:
Wir k"onnen den Diracoperator \Ref{2_x5} in der Form \Ref{2_p6}
schreiben und setzen dazu
\[ {\cal{B}}^u \;=\; \chi_L \: \Aslsh_R \:+\: \chi_R \: \Aslsh_L \;\;\;,\;\;\;\;\;
	{\cal{B}}^g \;=\; m \chi_R \: U_L^{-1} (-\Xi - i \Phi) U_L
	\:+\: m \chi_L \: U_R^{-1} (-\Xi + i \Phi) U_R \;\;\; . \]
Da die Potentiale $U_{L\!/\!R}, \Xi, \Phi$ in Theorem \ref{a6_thm0} beliebig
sein k"onnen, sind die Bedingungen \Ref{2_y1}, \Ref{2_p5} i.a. verletzt.
Die Bedingung \Ref{2_y1} folgt aus \Ref{2_z1}.
F"ur die Fermionkonfiguration des Standardmodells ist $X_L=1$, so
da"s \Ref{2_z1} und \Ref{2_y1} sogar "aquivalent sind.
Der erste Teil von Gleichung \Ref{2_p6} stimmt mit der
Kommutatorbedingung \Ref{2_x6} "uberein.
Bei Einsetzen von \Ref{2_y6} in \Ref{2_z2} erhalten wir
\[ \chi_R \:{\cal{B}}^g\: X_R \:-\: \chi_R \:X_L\: {\cal{B}}^g \;=\; 0
	\spc{\mbox{und}}\spc
   \chi_L \:{\cal{B}}^g\: X_L \:-\: \chi_L \:X_R\: {\cal{B}}^g \;=\; 0
	\spc , \]
also den zweiten Teil von Bedingung \Ref{2_p6}.
Mit dem Ergebnis von Theorem \ref{a6_thm0} l"a"st sich also
am Beispiel des Diracoperators \Ref{2_x5}
explizit "uberpr"ufen, da"s die Bedingungen \Ref{2_y1}, \Ref{2_p5}
notwendig und hinreichend sind, damit in der St"orungsrechnung keine
nichtlokalen Linienintegrale auftreten. Man sieht auch, warum der
Ansatz \Ref{2_x5} gerade in dieser Form sinnvoll ist.

Wir kommen zu Theorem \ref{a6_satz26}. Dort sind die Beitr"age
$\sim m^2$ zu $\tilde{p}$, $\tilde{k}$ aufgelistet, die ja in Theorem
\ref{a6_thm0} nicht ber"ucksichtigt wurden. Damit die Rechnung nicht
zu aufwendig wird, haben wir nur den Fall ohne chirale Asymmetrie behandelt,
au"serdem hat der Diracoperator gegen"uber \Ref{2_x5} die speziellere
Form
\[ i \Pdd \:+\: i \chi_L \: U_R (\Pdd U_R^{-1}) \:+\:
	i \chi_R \: U_L (\Pdd U_L^{-1}) \:-\: m \:\Xi
	\:-\: i \rho m \: \Phi \spc . \]
Diese Vereinfachungen sind aber unwesentlich, weil $X_{L\!/\!R}$ und
die geordneten Exponentiale "uber $A_{L\!/\!R}$ direkt in die Formel
von Theorem \ref{a6_satz26} eingef"ugt werden k"onnen.
Die Lichtkegelentwicklung wurde bis zur Ordnung ${\cal{O}}(\xi^2)$
bzw. ${\cal{O}}(\xi^0)$ durchgef"uhrt, dabei bezeichnet ${\cal{O}}(\xi^0)$
die Menge aller regul"aren Distributionen.
In erster Ordnung in den chiralen Potentialen
$U_{L\!/\!R} (\partial_j U_{L\!/\!R}^{-1})$ geht \Ref{a6_A} in den
Eich-/Pseudoeichterm \Ref{a2_c2}, \Ref{a2_x1} "uber, die Summanden
\Ref{a6_D}, \Ref{a6_E} liefern den Massenterm \Ref{a2_x6}.
In erster Ordnung in $\Xi, \Phi$ f"uhrt \Ref{a6_A} auf die
Massenverschiebung \Ref{a21_s2}, die Beitr"age \Ref{a6_B} und
\Ref{a6_D}, \Ref{a6_E} liefern die Ableitungsterme \Ref{a2_sc} bzw.
\Ref{a2_sd}, \Ref{a2_pa}.
Der Summand \Ref{a6_C} tr"agt bei St"orungsentwicklung erst ab zweiter
Ordnung bei.

Wir kommen zum Ende der Untersuchung endlicher St"orungen und fassen
die Ergebnisse noch einmal kurz zusammen:
Wir haben eine Methode beschrieben, mit welcher der fermionische Projektor
bei St"orungen des Diracoperators nicht-perturbativ um den Lichtkegel
entwickelt werden werden kann. Mit Theorem \ref{a6_thm0} und
Theorem \ref{a6_satz26} wurden f"ur alle Singularit"aten bis zur
Ordnung ${\cal{O}}(\ln(|\xi^2|))$ explizite Formeln im Ortsraum abgeleitet.
Bei den Beitr"agen $\sim m^2$ haben wir sogar die Singularit"at
$\sim \ln(|\xi^2|)$ exakt berechnet. Au"serdem lassen sich viele Ergebnisse
der St"orungsrechnung erster Ordnung unmittelbar durch Einf"ugen von
geordneten Exponentialen "uber die chiralen Potentiale auf endliche
St"orungen "ubertragen.
Schlie"slich k"onnen wir von allen nicht berechneten St"orungsbeitr"agen
h"oherer Ordnung die St"arke der Singularit"at auf dem Lichtkegel genau
angeben.
Damit haben wir gen"ugend Informationen "uber den wechselwirkenden
fermionischen Projektor zusammengetragen, um mit der Untersuchung
zusammengesetzter Ausdr"ucke in $\tilde{P}(x,y)$ beginnen zu k"onnen.

\chapter{Produkte von Distributionen}
\label{kap3}
Im vorangehenden Kapitel 2 haben wir den fermionischen Projektor
$P(x,y)$ im Kontinuum eingef"uhrt und das Verhalten dieser Distribution
bei verschiedenen St"orungen des Di\-rac\-ope\-ra\-tors untersucht
(wir lassen ab jetzt die Tilde ``$\tilde{\;\;}$'' beim gest"orten fermionischen
Projektor zur Einfachheit meist weg).
Wenn man Gleichungen der Form \Ref{1_29}, \Ref{1_30b} auf naive
Weise von der diskreten Raumzeit ins Kontinuum "ubertr"agt, treten
formale Distributionsprodukte
\Equ{3_1}
\left( P(x,y) \: P(y,x) \right)^p \;\;\;\;\;,\spc \left( P(x,y) \: P(y,x)
	\right)^p \: P(x,y)
\EndEqu
auf. In diesem Kapitel wollen wir solchen Ausdr"ucken einen mathematischen
Sinn geben.

\subsubsection*{die Klammerschreibweise $(. \:|\: .)$}
\index{Klammerschreibweise $(. \:|\: .)$}
Da sich die formalen Produkte \Ref{3_1} im Block-Index komponentenweise
untersuchen lassen, k"onnen wir uns hier auf den Fall eines Blocks, also
Spindimension 4, beschr"anken. Die Diracmatrizen lassen sich mit den
"ublichen Rechenregeln vereinfachen.
Um das Problem m"oglichst allgemein zu behandeln,
arbeiten wir anstelle der Diracmatrizen mit Ten\-sor\-in\-di\-zes, die
sp"ater durch Kontraktion miteinander verkn"upft werden.
Wir schreiben die einzelnen St"orungsbeitr"age zu $P(x,y)$ also
in der Form
\Equ{3_2}
P(x,y) \;\asymp\; f_{i_1 \cdots i_p}(x,y) \;\: \xi_{j_1} \: \cdots \: \xi_{j_q} \; D(\xi)
\EndEqu
mit glatten Funktionen $f_{i_1 \cdots i_p}(x,y)$ (z.B.
Linienintegralen "uber Str"ome oder Feldst"arken), Faktoren
$\xi_i=y_i-x_i$ und einer temperierten Distribution $D(y-x)$.
Da der fermionische Projektor gem"a"s \Ref{2_u00} aus
Diracseen $\frac{1}{2}(p-k)$ aufgebaut wird, besteht \Ref{3_2}
immer aus der Differenz entsprechender St"orungsbeitr"age zu
$p$ und $k$.
Bei Vergleich der St"orungsrechnung f"ur $p$, $k$
stellt man fest, da"s f"ur $D(\xi)$ lediglich die Kombinationen
\begin{eqnarray}
\label{eq:3_3}
\frac{1}{\xi^4} &-& i \pi \: \delta^\prime(\xi^2) \: \epsilon(\xi^0) \\
\label{eq:3_4}
\frac{1}{\xi^2} &+& i \pi \: \delta(\xi^2) \: \epsilon(\xi^0) \\
\label{eq:3_5}
\ln(|\xi^2|) &+& i \pi \: \Theta(\xi^2) \: \epsilon(\xi^0) \\
\label{eq:3_6}
\xi^2 \: \ln(|\xi^2|) &+& i \pi \: \xi^2 \: \Theta(\xi^2) \:
	\epsilon(\xi^0)
\end{eqnarray}
auftreten, wobei $\xi^{-2}$ und $\xi^{-4}$ als Hauptwert bzw.
Distributionsableitung des Hauptwertes definiert sind.
Die Beitr"age zu $P(y,x)$ erh"alt man durch komplexe Konjugation
von \Ref{3_2}, wobei sich jeweils das Vorzeichen des zweiten
Summanden in \Ref{3_3} bis \Ref{3_6} umkehrt.

Zun"achst f"uhren wir f"ur die einzelnen Beitr"age zu $P(x,y), P(y,x)$
eine einfache und zweckm"a"sige Notation ein:
Wir gehen in den Impulsraum. Bei expliziter Berechnung der
Fouriertransformierten stellt man fest, da"s der Tr"ager der
Distributionen \Ref{3_3} bis \Ref{3_6} im oberen Massenkegel,
also in der Menge $\{ k \;|\; k^2 \geq 0 {\mbox{ und }} k^0 \geq 0 \}$,
liegt. Die komplex Konjugierten haben den Tr"ager entsprechend im
unteren Massenkegel. Die jeweils ersten Summanden von \Ref{3_3}
bis \Ref{3_6},
\Equ{3_9}
\frac{1}{\xi^4} \;\;,\;\;\; \frac{1}{\xi^2} \;\;,\;\;\;
	\ln(|\xi^2|) \;\;,\;\;\; \xi^2 \: \ln(|\xi^2|) \spc ,
\EndEqu
besitzen als deren Realteil den Tr"ager sowohl im oberen als auch im
unteren Lichtkegel.
Folglich k"onnen wir die Distributionen \Ref{3_3} bis \Ref{3_6} und ihre
komplex Konjugierten durch Projektion von \Ref{3_9} auf die Zust"ande
positiver bzw. negativer Energie darstellen, also beispielsweise
\begin{eqnarray*}
\frac{1}{\xi^2} \pm i \pi \: (l^\vee(\xi)-l^\wedge(\xi)) &=&
	\int d^4 \tilde{\xi} \; \frac{1}{\tilde{\xi}^2} \; \left(
	\int \frac{d^4p}{(2 \pi)^4} \; \Theta(\pm p^0) \; e^{-ip (\xi-\tilde{\xi})}
	\right) \spc .
\end{eqnarray*}
Aus diesem Grund ist es mathematisch sinnvoll, als Kurzschreibweise
f"ur \Ref{3_3} bis \Ref{3_6} die Ausdr"ucke \Ref{3_9} in die linke Seite
einer Klammer $(.|.)$ zu schreiben
\[ \left( \xi^{-4} \;|\; 1 \right) \;\;\;,\;\;\;\;\;
	\left( \xi^{-2} \;|\; 1 \right) \;\;\;,\;\;\;\;
	\left(\ln (|\xi^2|) \;|\; 1 \right) \;\;\;,\;\;\;\;
	\left( \xi^2 \:\ln(|\xi^2|) \;|\; 1 \right) \spc . \]
Bei komplexer Konjugation vertauschen wir den ersten und den
zweiten Eintrag, also z.B. $\overline{(\ln (|\xi^2|) \:|\: 1)} =
(1 \:|\: \ln (|\xi^2|))$.
Die Faktoren $\xi_j$ schreiben wir mit in die Klammer $(.|.)$ hinein.
Die einzelnen Beitr"age zum fermionischen Projektor haben mit dieser
Klammernotation also die Form
\begin{eqnarray}
\label{eq:3_7}
P(x,y) &\asymp& f_{i_1 \cdots i_p}(x,y) \; \left( \xi_{j_1} \: \cdots \: \xi_{j_q} \; h(\xi^2)
	\;|\; 1 \right) \\
\label{eq:3_8}
P(y,x) &\asymp& f_{i_1 \cdots i_p}(x,y) \; \left( 1 \;|\; \xi_{j_1} \: \cdots \:
	\xi_{j_q} \; h(\xi^2) \right) \spc ,
\end{eqnarray}
dabei ist $h(\xi^2)$ eine der Funktionen \Ref{3_9}.

Nach Definition liegt der Tr"ager der Distributionen
$(h(\xi^2)\:|\:1)$, $(1\:|\:h(\xi^2))$ im oberen bzw. unteren Massenkegel.
Da die Multiplikation mit $\xi_j$ im Impulsraum der partiellen
Ableitung $i \partial_{p^j}$ entspricht, haben die Faktoren
\begin{eqnarray}
\label{eq:3_11}
\left( \xi_{j_1} \: \cdots \: \xi_{j_q} \; h(\xi^2) \;|\; 1 \right) \\
\label{eq:3_12}
\left( 1 \;|\; \xi_{j_1} \: \cdots \: \xi_{j_q} \; h(\xi^2) \right)
\end{eqnarray}
in \Ref{3_7}, \Ref{3_8} ebenfalls den Tr"ager im oberen bzw. unteren
Massenkegel.
Diese Tatsache haben wir empirisch aus der St"orungsrechnung erhalten.
Man kann sich auch direkt "uberlegen, warum das so sein mu"s:
Ohne Wechselwirkung ist $P(x,y)$ aus freien Diracseen,
also Zust"anden auf der unteren Massenschale, aufgebaut. 
Als Funktion von $y$ hat $P(x,y)$ also den Tr"ager im
oberen Massenkegel.
Im gest"orten Fall ist die Situation komplizierter, weil die
Zust"ande von $P$ nicht mehr nur aus negativen Frequenzen bestehen.
Im Beitrag \Ref{3_7} der St"orungsrechnung enth"alt der Faktor
$f_{i_1 \cdots i_p}$ klassische Potentiale oder Felder,
der Faktor \Ref{3_11} ist dagegen von der Dynamik der St"orung
unabh"angig. Im Grenzfall homogener, station"arer St"orungen "andern sich
Impuls und Energie der Zust"ande von $P$ beliebig wenig, so da"s dann
der Tr"ager von \Ref{3_7} und damit auch allgemein von \Ref{3_11}
im oberen Massenkegel liegt.

\subsubsection*{die Methode der variablen Regularisierung}
Bevor wir mit den mathematischen Konstruktionen beginnen, wollen wir
das grundlegende Problem herausarbeiten und die verwendete
Methode qualitativ beschreiben.
Unsere Aufgabe besteht darin, auf sinnvolle Weise Produkte der
Distributionen \Ref{3_7}, \Ref{3_8} zu definieren. Da die glatten
Funktionen $f_{i_1 \cdots i_p}$ problemlos miteinander multipliziert
werden k"onnen, lassen wir sie bei der folgenden Diskussion zur
Einfachheit weg und beschr"anken uns auf die Distributionen
\Ref{3_11}, \Ref{3_12}.

Wir betrachten zun"achst die Situation im Impulsraum: Die
Multiplikation im Ortsraum entspricht gem"a"s
\begin{eqnarray}
(\widehat{f g})(p) &=& \int d^4x \; \int \frac{d^4q_1}{(2 \pi)^4}
\int \frac{d^4q_2}{(2 \pi)^4} \; \hat{f}(q_1) \: \hat{g}(q_2) \;
	e^{-i(q_1+q_2-p) \: x} \nonumber \\
\label{eq:3_13}
&=& \int \frac{d^4q}{(2 \pi)^4} \; \hat{f}(q) \: \hat{g}(p-q) \;=\;
	\frac{1}{(2 \pi)^4} \; (\hat{f} \ast \hat{g})(p)
\end{eqnarray}
einer Faltung im Impulsraum.
Bei der Multiplikation zweier Distributionen \Ref{3_11} mit Tr"ager
im oberen Massenkegel mu"s man in \Ref{3_13} "uber den nach
oben ge"offneten Massenkegel um den Ursprung mit dem nach
unten ge"offneten Massenkegel um $p$ integrieren, also
\Equ{3_a1}
q \;\in\; \left\{ q^2 \geq 0 ,\; q^0 \geq 0 \right\} \:\cap\:
	\left\{ (q-p)^2 \geq 0 ,\; q^0 - p^0 \leq 0 \right\} \spc .
\EndEqu
Das Integrationsgebiet ist kompakt; das Integral l"a"st sich problemlos
berechnen und ist endlich.
Also k"onnen wir Distributionen vom Typ \Ref{3_11} und analog auch
vom Typ \Ref{3_12} jeweils untereinander multiplizieren und erhalten
als Ergebnis wieder eine Distribution.
Beim Produkt $\Ref{3_11} \cdot \Ref{3_12}$ zweier Distributionen mit
Tr"ager im oberen und unteren Massenkegel erh"alt man dagegen in
\Ref{3_13} das Integral "uber den Schnitt zweier nach oben ge"offneter
Massenkegel, also
\Equ{3_a0}
q \;\in\; \left\{ q^2 \geq 0 ,\; q^0 \geq 0 \right\} \:\cap\:
	\left\{ (q-p)^2 \geq 0 ,\; q^0 - p^0 \geq 0 \right\} \spc .
\EndEqu
Nun ist das Integrationsgebiet unbeschr"ankt, so da"s das
Faltungsintegral i.a. divergiert. Die Multiplikation von \Ref{3_11} mit
\Ref{3_12} f"uhrt also auf Probleme. Da die beiden Faktoren f"ur
$\xi^2 \neq 0$ regul"are Funktionen sind, k"onnen wir genauer sagen,
da"s bei der Multiplikation Divergenzen auf dem Lichtkegel auftreten.

Diese Divergenzen haben wir schon in der
Einleitung angesprochen. Wie dort beschrieben wurde, m"ussen sie
durch Regularisierung der Distributionen auf der L"angenskala
$\varepsilon$ beseitigt werden. In der Planckn"aherung f"uhrt man eine
Entwicklung nach $\varepsilon$ durch und untersucht die
einzelnen Polordnungen getrennt. Dieses Vorgehen ist nur dann
sinnvoll, wenn die Endergebnisse unabh"angig von der verwendeten
Regularisierung sind.

Die eigentliche Schwierigkeit liegt in der geforderten Unabh"angigkeit
von der Regularisierungsmethode. Auf den ersten Blick scheint dies
ein ganz prinzipielles Problem zu sein. Die Regularisierung auf der
L"angenskala $\varepsilon$ bedeutet n"amlich im Impulsraum, da"s
die Distribution f"ur Impulse der Gr"o"senordnung $2 \pi / \varepsilon$
abge"andert wird, beispielsweise durch einen Cutoff.
Da in das Faltungsintegral \Ref{3_13} im Fall \Ref{3_a0} beliebig gro"se
Impulse $q$, $q-p$ eingehen, ist zun"achst nicht klar, warum es auf
die spezielle Art der Regularisierung letztlich nicht ankommen sollte.
Gl"ucklicherweise wird die Situation besser, wenn man ber"ucksichtigt,
da"s die Divergenzen auf dem Lichtkegel auftreten:
Wie ab Seite \pageref{2_ansch} beschrieben, ist f"ur die Singularit"at
der Distribution $P(x,y)$ auf dem Lichtkegel die Flanke der
Fouriertransformierten auf dem Massenkegel verantwortlich.
Genauer kommt es im Fall $\xi^2=0$ lediglich auf die Zust"ande in einer
Umgebung der 2-Ebene
\Equ{3_a4}
e(\xi) \;:=\; \left\{ k \:|\: k^2 = 0 {\mbox{ und }} k_j \: \xi^j = 0 \right\}
\EndEqu
an. In die Divergenz des Produktes $\Ref{3_11} \cdot \Ref{3_12}$
geht dann auch nur die Form der Regularisierung
l"angs $e(\xi)$ ein; insbesondere ist das Verhalten der regularisierten
Distributionen au"serhalb einer Umgebung des Massenkegels
irrelevant. Darum k"onnen wir hoffen, da"s die Divergenzen auf dem
Lichtkegel von der Regularisierungsmethode weitgehend
unabh"angig sind.

Diese anschauliche Vorstellung ist im Moment sehr vage und
qualitativ. Um sie zu verifizieren und mathematisch zu pr"azisieren,
mu"s man eine m"oglichst allgemeine Klasse von Regularisierungen
betrachten. Erst dann l"a"st sich die Abh"angigkeit des
Dis\-tri\-bu\-tions\-pro\-duk\-tes von $\varepsilon$ und dem
Regularisierungsverfahren genau untersuchen.
Alle Aussagen, die unabh"angig von der Regularisierungsmethode sind,
k"onnen auf sinnvolle Weise in die Definition des Distributionsproduktes
"ubernommen werden. Alle Aussagen, in die das
Regularisierungsverfahren eingeht, werden wir dagegen ignorieren.
Wir nennen dieses Vorgehen {\em{Methode der variablen
Regularisierung}}\index{Methode der variablen Regularisierung}.

Es stellt sich die Frage, was wir genau unter ``m"oglichst allgemeine
Klasse von Regularisierungen'' verstehen wollen.
Auf der einen Seite mu"s die Klasse so gro"s sein, da"s sich die
Abh"angigkeit des Produktes von der Regularisierung detailliert untersuchen
l"a"st. Auf der anderen Seite soll sich der mathematische Aufwand in
Grenzen halten. Als Kompromi"s werden wir die Distributionen durch Faltung
mit einer beliebigen rationalen Funktion $\eta$ regularisieren.
Das ist technisch relativ einfach, trotzdem sollten sich damit alle wichtigen
Effekte beschreiben lassen. Unsere Konstruktionen werden auf jeden Fall
in dem Sinne kanonisch sein, da"s jede andere in sich konsistente
Methode auf die gleichen Ergebnisse f"uhrt.

Wir wollen etwas konkreter werden. Nach Regularisierung auf der L"angenskala
$\varepsilon$ k"onnen wir die Distributionsprodukte ausf"uhren und im
schwachen Sinne untersuchen. Etwas vereinfacht erh"alt man in
einem speziellen Bezugssystem $\xi = (t, \vec{x})$ ein Integral der Form
\Equ{3_a3}
\frac{1}{\varepsilon^p} \: \inti dt \int_{\sR^3} d\vec{x} \; g(t,\vec{x}) \:
f(t,\vec{x}) \; \frac{\Lambda(t,\vec{x})}{r^q} \:
\frac{\delta_\varepsilon(|t|-r)}{2r} \spc ,
\EndEqu
dabei ist $g$ eine Testfunktion, $\Lambda$ eine glatte Funktion,
$f$ ein zusammengesetzter Ausdruck in den Tensorfeldern
$f_{i_1 \cdots i_p}$ in \Ref{3_7}, \Ref{3_8} und $\delta_\varepsilon$
eine regularisierte $\delta$-Distribution.
Dieser Ausdruck ist als recht allgemeiner Ansatz f"ur ein Integral, das
f"ur $\varepsilon \rightarrow 0$ auf dem Lichtkegel divergiert, auch direkt
einsichtig. Wichtig ist, da"s $\Lambda$ wesentlich von der
Wahl der Regularisierungsfunktion $\eta$ abh"angt.

Bei der f"uhrenden Singularit"at $\sim \varepsilon^{-p}$ k"onnen wir
hoffen, da"s die Abh"angigkeit von $\eta$ letztlich keine Rolle spielt:
die Gleichung $\Ref{3_a3}=0$ liefert beispielsweise die
lorentzinvariante Bedingung
\Equ{3_a4a}
f(\pm |\vec{x}|, \vec{x}) \;=\; 0 \spc .
\EndEqu
F"ur die schw"acher singul"aren Beitr"age $\sim \varepsilon^{-p+1}$
m"ussen die Funktionen in \Ref{3_a3} um den Lichtkegel entwickelt
werden. Dabei erh"alt man zusammengesetzte Ausdr"ucke in $g,
f, \Lambda$ und deren partiellen Ableitungen, also z.B. anstelle von
\Ref{3_a4a} die Gleichung
\[ \left( \partial_j f \: \Lambda \:+\: f \: \partial_j \Lambda
	\right)_{|(\pm |\vec{x}|, \vec{x})} \;=\; 0 \spc . \]
In solchen Gleichungen kann die Funktion $\Lambda$ nicht beseitigt
werden, so da"s wir keine von der Regularisierungsmethode
unabh"angigen Bedingungen erhalten.
Allgemein kann man mit der Methode der variablen Regularisierung
h"ochstens Aussagen "uber die f"uhrende, nicht verschwindende
Divergenz auf dem Lichtkegel gewinnen.

Die Beschreibung der f"uhrenden Singularit"at mit \Ref{3_a4a}
ist leider zu einfach. Tats"achlich werden n"amlich verschiedene
Beitr"age der Form \Ref{3_a3} mit unterschiedlichen Funktionen
$\Lambda$ auftreten.
Diese Beitr"age m"ussen modulo Divergenzen der Ordnung
$\sim \varepsilon^{-p+1}$ ineinander umgeformt werden. Erst wenn
die Beitr"age genau die gleiche Form haben, l"a"st sich gem"a"s
\Ref{3_a4a} die von der Regularisierung abh"angige Funktion $\Lambda$
herausk"urzen. F"ur diese Umformungen werden wir
{\em{asymptotische Rechenregeln}}\index{asymptotische Rechenregeln}
verwenden. In die Herleitung der
asymptotischen Rechenregeln wird eine mathematisch strenge
Fassung der "Uberlegung an \Ref{3_a4} entscheidend eingehen.

Wie gerade erw"ahnt, werden wir alle vom Regularisierungsverfahren
abh"angigen Beitr"age einfach weglassen. Wir beschreiben abschlie"send,
wie dieses Vorgehen bei unserer Vorstellung der diskreten Raumzeit zu
verstehen ist:
Gem"a"s der Beschreibung in der Einleitung kann der fermionische Projektor
$P$ der diskreten Raumzeit als eine spezielle Regularisierung der
Distribution $P(x,y)$ auf der Skala der Planck-L"ange angesehen werden.
Leider k"onnen wir "uber die genaue Form von $P$
keine Aussagen machen und sind deshalb auf die Methode
der variablen Regularisierung angewiesen.
Die Abh"angigkeit gewisser Beitr"age des Distributionsproduktes von der
Regularisierungsmethode bedeutet, da"s die Euler-Lagrange-Gleichungen
auch Bedingungen an den fermionischen Projektor liefern,
die sich nicht ins Kontinuum "ubertragen lassen.
Diese zus"atzlichen Bedingungen k"onnen mit unseren Methoden nicht
genauer analysiert werden. Sie k"onnen aber, wenn man will, mit den in Abschnitt
\ref{1_ab5} angesprochenen nichtlokalen Quantenbedingungen identifiziert
und somit als Best"atigung f"ur unseren Deutungsversuch der
Feldquantisierung aufgefa"st werden.

\section{Produkte im Distributionssinn}
\label{3_ab1}
Nach diesen Vorbereitungen k"onnen wir mit der Konstruktion beginnen.
Zu\-n"achst wollen wir das Distributionsprodukt so weit wie m"oglich ohne
Regularisierung aus\-f"uh\-ren. Dazu m"ussen wir in der Klammer $(.|.)$
allgemeinere Funktionen zulassen: Wir definieren die rellen Distributionen
\Equ{3_13a}
\xi_{j_1} \cdots \xi_{j_p} \; \xi^{-2\alpha} \: \ln^\beta (|\xi^2|)
\EndEqu
mit $\alpha \in \Z, \beta \in \N_0$ analog zu $\xi^{-2}, \xi^{-4}$
als Hauptwertintegral. Die Fouriertransformierte von \Ref{3_13a} hat den
Tr"ager im Massenkegel (also in der Menge $\{k \:|\: k^2 \geq 0\}$), wie
man durch eine direkte Rechnung verifizieren kann. Daher k"onnen wir
die temperierten Distributionen
\begin{eqnarray}
\label{eq:3_14}
\left( \xi_{j_1} \cdots \xi_{j_p} \; \xi^{-2 \alpha} \; \ln^\beta(|\xi^2|)
	\;|\; 1 \right) \\
\label{eq:3_15}
\left( 1 \;|\; \xi_{j_1} \cdots \xi_{j_p} \; \xi^{-2 \alpha} \;
	\ln^\beta(|\xi^2|) \right)
\end{eqnarray}
durch Projektion von \Ref{3_13a} auf die positiven bzw. negativen
Energiezust"ande definieren. Der Realteil von \Ref{3_14},\Ref{3_15} stimmt
mit \Ref{3_13a} "uberein, auf dem Lichtkegel kommt
bei diesen Distributionen im allgemeinen ein singul"arer Beitrag hinzu.

Nach Definition liegt der Tr"ager von \Ref{3_14} im oberen
Massenkegel. Nach unserer "Uberlegung im Impulsraum
k"onnen wir die Ausdr"ucke \Ref{3_14} durch Berechnung des
Faltungsintegrals im Distributionssinne miteinander multipizieren.
Dabei gilt
\begin{eqnarray}
&&\left( \xi_{i_1} \cdots \xi_{i_p} \; \xi^{-2 \alpha} \; \ln^\beta(|\xi^2|)
	\;|\; 1 \right) \cdot
\left( \xi_{j_1} \cdots \xi_{j_q} \; \xi^{-2 \gamma} \; \ln^\delta(|\xi^2|)
	\;|\; 1 \right) \nonumber \\
\label{eq:3_16}
&&\hspace*{2cm}\;=\; \left( \xi_{i_1} \cdots \xi_{i_p} \: \xi_{j_1} \cdots \xi_{j_q}
	\; \xi^{-2 (\alpha+\gamma)} \; \ln^{(\beta+\delta)}(|\xi^2|)
	\;|\; 1 \right) \spc ,
\end{eqnarray}
wie man explizit im Impulsraum verifizieren kann.
Das Produkt der Distributionen \Ref{3_15} definieren wir analog, es gilt
\begin{eqnarray}
&&\left( 1 \;|\; \xi_{i_1} \cdots \xi_{i_p} \; \xi^{-2 \alpha} \;
	\ln^\beta(|\xi^2|) \right) \cdot
\left( 1 \;|\; \xi_{j_1} \cdots \xi_{j_q} \; \xi^{-2 \gamma} \;
	\ln^\delta(|\xi^2|) \right) \nonumber \\
\label{eq:3_17}
&&\hspace*{2cm}\;=\; \left( 1 \;|\; \xi_{i_1} \cdots \xi_{i_p} \: \xi_{j_1} \cdots
	\xi_{j_q}       \; \xi^{-2 (\alpha+\gamma)} \;
	\ln^{(\beta+\delta)}(|\xi^2|) \right) \spc .
\end{eqnarray}
Die Gleichungen \Ref{3_16}, \Ref{3_17} lassen sich auch direkt einsehen:
Au"serhalb des Lichtkegels sind die Gleichungen
f"ur den Realteil punktweise erf"ullt.
Aus der Faltungsvorschrift im Impulsraum folgt au"serdem, da"s der
Tr"ager der linken Seite, genau wie nach Definition der Tr"ager der rechten
Seite, im oberen bzw. unteren Massenkegel liegt.
Deswegen stimmen auch der Imagin"arteil und das singul"are Verhalten auf dem
Lichtkegel auf beiden Seiten "uberein.

Mit den Rechenregeln \Ref{3_16}, \Ref{3_17} k"onnen wir in dem formalen
Produkt \Ref{3_1} jeweils alle Faktoren \Ref{3_11} und \Ref{3_12}
im Distributionssinne ausmultiplizieren. Au"serdem f"uhren wir das Produkt
der glatten Funktionen $f_{i_1 \cdots i_p}$ in \Ref{3_7}, \Ref{3_8}
aus und erhalten einen formalen Ausdruck der Form
\[ f_{i_1 \cdots i_p}(x,y) \; \left( H_1(\xi) \;|\; 1 \right) \;\cdot\;
	\left( 1 \;|\; H_2(\xi) \right) \]
mit $H_j(\xi)$ gem"a"s \Ref{3_13a}.
In unseren Anwendungen wird immer nur eine der beiden Funktionen
$H_j$ den Faktor $\ln^\beta (|\xi^2|)$ enthalten.
Wir k"onnen uns auf den Fall beschr"anken, da"s dieser Faktor in
der linken Seite der Klammer $(.|.)$ steht
\Equ{3_19}
f_{i_1 \cdots i_p}(x,y) \; \left( \xi_{j_1} \cdots \xi_{j_q} \;
	\xi^{-2 \alpha} \; \ln^\beta (|\xi^2|) \;|\; 1 \right) \;\cdot\;
	\left( 1 \;|\; \xi_{k_1} \cdots \xi_{k_r} \; \xi^{-2 \gamma} \right)
	\spc ,
\EndEqu
den umgekehrten Fall erh"alt man daraus durch komplexe Konjugation.
Jetzt m"ussen wir nur noch zwei Distributionen miteinander multiplizieren, was das urspr"ungliche Problem deutlich vereinfacht.

\section{Regularisierung}
Wir f"uhren nun die Regularisierung ein. Dazu betrachten wir eine reelle,
rationale Funktion $\eta \in C^\infty(\R^4)$ mit
\Equ{3_20}
	\int_{\sR^4} \eta \;=\; 1 \;\;\;,\spc \eta(-x) \;=\; \eta(x)
\EndEqu
und definieren f"ur die Distributionen $D$ gem"a"s \Ref{3_14}, \Ref{3_15}
die $C^\infty$-Funktionen $D^\varepsilon$ durch
\begin{eqnarray}
\label{eq:3_21}
D^\varepsilon(x) &=& (D \ast \eta_\varepsilon)(x) \spc
{\mbox{mit}} \spc \eta_\varepsilon(x) \;=\; \varepsilon^{-4} \; \eta
	\left(x / \varepsilon \right) \spc .
\end{eqnarray}
Die Funktion $\eta$ soll im Unendlichen so stark abfallen, da"s
das Faltungsintegral \Ref{3_21}
existiert. Dabei ist ein geeigneter polynomialer Abfall ausreichend, weil
$D$ wegen \Ref{3_14}, \Ref{3_15} im Unendlichen nur polynomial ansteigt
(die f"ur uns interessanten Distributionen fallen sogar im Unendlichen
ab, so da"s bereits \Ref{3_20} die Existenz von \Ref{3_21} impliziert).
Die Rationalit"at von $\eta$ ist eine technische Bedingung, die es uns
sp"ater erm"oglichen wird, mit dem Residuensatz zu arbeiten.

Als Beispiel kann man f"ur $\eta$ das Produkt regularisierter
$\delta$-Distributionen
\Equ{3_22}
\eta(x) \;=\; \prod_{j=0}^3 \; \frac{1}{2 \pi i} \left(
	\frac{1}{x^j - i} - \frac{1}{x^j+i} \right)
\EndEqu
oder auch eine ``rotationssymmetrische'' Funktion
\begin{eqnarray}
\label{eq:3_23}
\eta(x) &=& \frac{1}{2 \pi^3 i} \left( \frac{1}{(\|x\|-i)^4} -
	\frac{1}{(\|x\|-i)^4} \right) \spc
{\mbox{mit }} \;\;\; \|x\|^2 \;=\; \sum_{j=0}^3 |x^j|^2
\end{eqnarray}
w"ahlen. F"ur $\eta$ sind auch Funktionen zul"assig, die aus
\Ref{3_22}, \Ref{3_23} durch Lorentztransformation hervorgehen,
au"serdem kann man allgemeinere (z.B. auch oszillierende) rationale
Funktionen f"ur $\eta$ verwenden.

Nach Regularisierung der beiden Distributionen
k"onnen wir das formale Produkt in \Ref{3_19} ausf"uhren, wir
verwenden f"ur das Ergebnis die Schreibweise
\Equ{3_24}
f_{i_1 \cdots i_p}(x,y) \; \left( \xi_{j_1} \cdots \xi_{j_q} \;
	\xi^{-2 \alpha}
	\; \ln^\beta (|\xi^2|) \;|\; \xi_{k_1} \cdots \xi_{k_r}
	\; \xi^{-2 \gamma} \right)^\varepsilon \spc .
\EndEqu

\section{Verkn"upfung der Tensorindizes}
\label{3_ab3}
Bei der bisherigen Konstruktion haben wir die Faktoren $\xi_j$ in der
linken und rechten Seite der Klammer $(.|.)$ immer sorgf"altig
voneinander getrennt. Andererseits haben wir die
ebenfalls glatten Funktionen $f_{i_1 \cdots i_p}$ einfach miteinander
multipliziert und vor die Klammer $(.|.)$ geschrieben.
Wir wollen dieses Vorgehen nachtr"aglich begr"unden:
Ganz allgemein k"onnen Multiplikation und Regularisierung nicht
miteinander vertauscht werden; f"ur $g \in C^\infty(\R^4 \times \R^4)$
und $H_j$ gem"a"s \Ref{3_13a} hat man also
\Equ{3_25}
f(x,y) \; (H_1 \;|\; H_2)^\varepsilon \;\neq\;
	(f(x,y) \: H_1 \;|\; H_2)^\varepsilon \;\neq\;
	(H_1 \;|\; f(x,y) \: H_2)^\varepsilon \spc .
\EndEqu
Wie wir an \Ref{3_a3} "uberlegt haben, k"onnen wir nur die f"uhrende
Singularit"at auf dem Lichtkegel sinnvoll beschreiben. Es ist nach
\Ref{3_a4a} einsichtig, da"s dabei lediglich der Funktionswert von $f(x,y)$
auf dem Lichtkegel eingeht (dies werden wir noch genauer an der
Konstruktion sehen). Als Folge spielt es keine Rolle, ob und wie wir
$f$ regularisieren, so da"s die Unterschiede in \Ref{3_25} verschwinden.
Darum brauchten wir mit den Funktionen $f_{i_1 \cdots i_p}$ nicht
sorgf"altig umzugehen. Bei den Faktoren $\xi_j$ mu"s man mehr aufpassen.
In Ausdr"ucken der Form
\Equ{3_26}
(\xi_j \: \xi^j \; H_1 \;|\; H_2)^\varepsilon \;\;,\;\;\;
	(\xi_j \: H_1 \;|\; \xi^j \: H_2)^\varepsilon \;\;,\;\;\;
	(H_1 \;|\; \xi_j \: \xi^j \; H_2)^\varepsilon
\EndEqu
tr"agt n"amlich die h"ochste Ordnung in $1/\varepsilon$ nicht bei,
weil der Faktor $\xi^2$ auf dem Lichtkegel verschwindet.
In diesem Fall k"onnen wir Aussagen "uber die
n"achstniedrigere Ordnung in $1/\varepsilon$ machen.
Dabei kommt es entscheidend darauf an, wie die Faktoren $\xi_j$
innerhalb der Klammer $(.|.)$ angeordnet sind.

Wir sehen an dieser "Uberlegung, da"s nur diejenigen Faktoren $\xi_j$
sauber behandelt werden m"ussen, die mit anderen $\xi_j$ kontrahiert
werden. Wir nennen diese Faktoren {\em{innere Faktoren}}\index{innerer Faktor}.
Die Kontraktion der $\xi_j$ in \Ref{3_24} mit der Funktion
$f_{i_1 \cdots i_p}(x,y)$ ist dagegen unproblematisch, wir k"onnen diese
{\em{"au"seren Faktoren}}\index{"au"serer Faktor} mit dem Vorfaktor
$f_{i_1 \cdots i_p}(x,y)$ zusammenfassen.

Bei der Kontraktion zweier innerer Faktoren in der linken Seite der Klammer
$(.|.)$ k"onnen wir \Ref{3_14} im Distributionssinne umformen
\Equ{3_90a}
\left( \xi_j \: \xi^j \: \xi_{i_1 \cdots i_p} \; \xi^{-2\alpha} \:
	\ln^\beta(|\xi^2|) \;|\; 1 \right) \;=\;
	\left( \xi_{i_1 \cdots i_p} \; \xi^{-2\alpha+2} \:
	\ln^\beta(|\xi^2|) \;|\; 1 \right)
\EndEqu
und erhalten nach Regularisierung und Multiplikation die Regel
\begin{eqnarray*}
\lefteqn{ \left( \xi_l \:\xi^l \: \xi_{j_1} \cdots \xi_{j_q} \;\xi^{-2
	\alpha} \; \ln^\beta (|\xi^2|) \;|\; \xi_{k_1} \cdots \xi_{k_r}
	\; \xi^{-2 \gamma} \right)^\varepsilon } \\
&& =\; \left( \xi_{j_1} \cdots \xi_{j_q} \; \xi^{-2 \alpha+2}
	\; \ln^\beta (|\xi^2|) \;|\; \xi_{k_1} \cdots \xi_{k_r}
	\; \xi^{-2 \gamma} \right)^\varepsilon \spc .
\end{eqnarray*}
Analog gilt
\begin{eqnarray*}
\lefteqn{ \left( \xi_{j_1} \cdots \xi_{j_q} \;\xi^{-2 \alpha}
	\; \ln^\beta (|\xi^2|) \;|\; \xi_l \: \xi^l \: \xi_{k_1} \cdots
	\xi_{k_r} \; \xi^{-2 \gamma} \right)^\varepsilon } \\
&& =\; \left( \xi_{j_1} \cdots \xi_{j_q} \; \xi^{-2 \alpha}
	\; \ln^\beta (|\xi^2|) \;|\; \xi_{k_1} \cdots \xi_{k_r}
	\; \xi^{-2 \gamma+2} \right)^\varepsilon \spc .
\end{eqnarray*}

Durch Anwendung dieser Rechenregeln und Herausnehmen der
"au"seren Faktoren k"onnen wir \Ref{3_24} in der Form
\Equ{3_27}
f_{i_1 \cdots i_s \cdots i_q}(x,y) \: \xi^{i_1} \cdots \xi^{i_s} \;
	\left( \xi_{j_1} \cdots \xi_{j_p}
	\; \xi^{-2\alpha} \; \ln^\beta(|\xi^2|) \;|\; \xi^{j_1} \cdots
	\xi^{j_p} \; \xi^{-2\gamma} \right)^\varepsilon
\EndEqu
umschreiben.
Wir werden diesen Ausdruck im schwachen Sinne untersuchen, also
f"ur eine Testfunktion $h$ und festes $x$ das Verhalten des Integrals
\[ \int d^4y \; h(y) \: f_{i_1 \cdots i_s \cdots i_q}(x,y) \: \xi^{i_1} \cdots
	\xi^{i_s} \; \left( \xi_{j_1} \cdots \xi_{j_p}
	\; \xi^{-2\alpha} \; \ln^\beta(|\xi^2|) \;|\; \xi^{j_1} \cdots
	\xi^{j_p} \; \xi^{-2\gamma} \right)^\varepsilon \]
im Limes $\varepsilon \rightarrow 0$ studieren. Um die Notation zu
vereinfachen, betrachten wir den Tensor $f_{i_1 \cdots i_q}$
komponentenweise und fassen den "au"seren Faktor mit der
Testfunktion zusammen,
\begin{eqnarray}
\label{eq:3_28}
&=&\int d^4y \; g(y) \; \left( \xi_{j_1} \cdots \xi_{j_p}
	\; \xi^{-2\alpha} \; \ln^\beta(|\xi^2|) \;|\; \xi^{j_1} \cdots
	\xi^{j_p} \; \xi^{-2\gamma} \right)^\varepsilon \\
\label{eq:3_92a}
&&{\mbox{mit }} g = h \; f_{i_1 \cdots i_s \cdots i_q} \:
	\xi^{i_1} \cdots \xi^{i_s} \spc .
\end{eqnarray}

Wir wollen unser Vorgehen kurz erl"autern: Die Unterscheidung
zwischen inneren und "au"seren Faktoren ist  nicht eindeutig;
man kann innere Faktoren nach
\Equ{3_31a}
g(y) \; \left( \xi_j \: H_1 \;|\; \xi^j \: H_2 \right) \;=\;
	g(y) \: g_{ij} \; \left( \xi^i \: H_1 \;|\; \xi^j \: H_2 \right)
\EndEqu
auch als "au"sere Faktoren auffassen, wenn man die Metrik mit dem Vorfaktor
zusammenfa"st. Dadurch geht aber Information verloren.
Wenn wir annehmen, da"s $(H_1 \:|\: H_2)$ zur f"uhrenden
Ordnung $\sim \varepsilon^{-p}$ beitr"agt, k"onnen wir n"amlich auf der rechten
Seite von \Ref{3_31a} nur aussagen, da"s der Beitrag dieser Ordnung
verschwindet; auf der linken Seite k"onnen wir zus"atzlich den Beitrag
$\sim \varepsilon^{-p+1}$ berechnen.
Nach der Methode der variablen Regularisierung m"ussen wir m"oglichst
viele der $\xi_j$ als innere Faktoren schreiben.
Auch bei einer Antisymmetrisierung der $\xi_j$ ist deren Lage innerhalb der
Klammer $(.|.)$ wichtig. Befinden sich die Faktoren auf der gleichen
Seite der Klammer, so verschwindet der regularisierte Ausdruck exakt,
also z.B.
\[ \left( \xi_i \: \xi_j \; H_1 \;|\; H_2 \right)^\varepsilon \; \sigma^{ij}
	\;=\; \varepsilon^{ijkl} \;
	\left( \xi_k \: \xi_l \; H_1 \;|\; H_2 \right)^\varepsilon
	\;=\; 0 \spc . \]
Wenn die Faktoren auf verschiedenen Seiten der Klammer angeordnet sind,
beispielsweise wie in
\Equ{3_32}
\left( \xi_i \; H_1 \;|\; \xi_j \; H_2 \right)^\varepsilon \; \sigma^{ij}
	\;,\;\;\; \varepsilon^{ijkl} \;
	\left( \xi_k \; H_1 \;|\; \xi_l \; H_2 \right)^\varepsilon
	\;,\;\;\; F_{ij} \; \left( \xi^i \; H_1 \;|\; \xi^j \;
	H_2 \right)^\varepsilon \;\;\; ,
\EndEqu
wissen wir zun"achst nur, da"s die h"ochste Ordnung in $1/\varepsilon$
verschwindet. An Rechnungen in speziellen Regularisierungen sieht
man, da"s die Beitr"age niedrigerer Ordnung nicht verschwinden,
aber wesentlich vom Regularisierungsverfahren abh"angen.
Wir werden sie gem"a"s der Methode der variablen Regularisierung
ignorieren.

Im n"achsten Schritt beseitigen wir in \Ref{3_28} alle inneren Faktoren:
Zun"achst schreiben wir die Faktoren $\xi_j$ des unregularisierten
linken Klammerfaktors als partielle Ableitungen um,
genauer\footnote{Im konkreten Fall kann man diese Umformung einfach
berechnen, wir betrachten als Beispiel die Distribution
\[ \left( \xi_i \: \xi_j \; \frac{1}{\xi^6} \;|\; 1 \right) \spc . \]
Man hat
\begin{eqnarray*}
\partial_i \left( \frac{1}{\xi^2} \;|\; 1 \right) &=& -2 \: \left( \xi_i \:
	\frac{1}{\xi^4} \;|\; 1 \right) \\
\partial_{ij} \left( \frac{1}{\xi^2} \;|\; 1 \right) &=&
	8 \ \left( \xi_i \: \xi_j \: \frac{1}{\xi^6} \;|\; 1 \right)
	- 2 \: g_{ij} \: \left( \frac{1}{\xi^4} \;|\; 1 \right)
\end{eqnarray*}
und somit
\[ \left( \xi_i \: \xi_j \: \frac{1}{\xi^6} \;|\; 1 \right) \;=\;
	\frac{1}{8} \: \partial_{ij} \left( \frac{1}{\xi^2} \;|\; 1 \right)
	+\frac{1}{4} \: g_{ij} \: \left(\frac{1}{\xi^4} \;|\; 1 \right)
	\spc .\] }
\begin{eqnarray}
\lefteqn{ \left( \xi_{j_1} \cdots \xi_{j_p} \; \xi^{-2\alpha} \; \ln(|\xi^2|)
	\;|\; 1 \right)
\;=\; \partial_{j_1 \cdots j_p} \left( K_1(\xi^2) \;|\; 1 \right) } \nonumber \\
&&+ \sum_{\sigma \in {\cal{S}}(p)} g_{j_{\sigma(1)} j_{\sigma(2)}} \;
	\partial_{j_{\sigma(3)} \cdots j_{\sigma(p)}}
	\left( K_2(\xi^2) \;|\; 1 \right) \nonumber \\
\label{eq:3_29}
&&+ \sum_{\sigma \in {\cal{S}}(p)} g_{j_{\sigma(1)} j_{\sigma(2)}} \;
	g_{j_{\sigma(3)} j_{\sigma(4)}}
	\partial_{j_{\sigma(5)} \cdots j_{\sigma(p)}}
	\left( K_3(\xi^2) \;|\; 1 \right) \;+\; \cdots
\end{eqnarray}
mit geeigneten Funktionen $K_j$ der Form $\xi^{-2\alpha} \:
\ln^\beta(|\xi^2|)$.
Da partielle Ableitungen mit Faltungen vertauschen, gilt \Ref{3_29} auch
nach der Regularisierung, wenn man die Klammern $(.|.)$ durch
$(.|.)^\varepsilon$ ersetzt.
Wir behandeln die Distribution $(1\:|\: \xi^{j_1} \cdots \xi^{j_p} \:
\xi^{-2 \gamma})$ genauso, setzen die erhaltenen Formeln in \Ref{3_28}
ein und multiplizieren die Summen aus. Die Faktoren $g_{ij}$
f"uhren dabei auf $\Box$-Operatoren, die jeweils auf eine der
regularisierten Distributionen wirken. Wir k"onnen diese Operatoren mit
der Regularisierung vertauschen und direkt ausf"uhren.
Jeder der sich ergebenden Summanden hat dann die Form
\Equ{3_30}
\int d^4y \; g(y) \; \partial_{j_1 \cdots j_p} (K_1(\xi^2)
	 \;|\; 1 )^\varepsilon \;\cdot\; \partial^{j^1 \cdots j^p}
	(1 \;|\; K_2(\xi^2) )^\varepsilon \spc .
\EndEqu
Nach iterativer Anwendung der Greenschen Formel
\begin{eqnarray}
\int d^4y \; g(y) \; \partial_j \alpha(y) \; \partial^j \beta(y) &=&
	\frac{1}{2} \int d^4y \; g \; \left( \Box(\alpha \: \beta)
	- (\Box \alpha) \: \beta - \alpha \: (\Box \beta)
	\right) \nonumber \\
\label{eq:3_97a}
&=& \frac{1}{2} \int d^4y \; \left( (\Box g) \: \alpha \: \beta
	- g \: (\Box \alpha) \: \beta - g \: \alpha \: (\Box \beta) \right)
\end{eqnarray}
verschwinden alle Tensorindizes. Wir k"onnen die $\Box$-Operatoren
bei Anwendung auf die regularisierten Distributionen wieder explizit
berechnen und erhalten schlie"slich f"ur die einzelnen Beitr"age des
Distributionsproduktes Ausdr"ucke der Form
\Equ{3_31}
{\cal{A}}^\varepsilon \;:=\; \int d^4y \; g(y) \; \left( \xi^{-2\alpha} \;
	\ln^\beta(|\xi^2|) \;|\; \xi^{-2 \gamma} \right)^\varepsilon
\EndEqu
mit geeigneten Testfunktionen $g$.

Dieses Verfahren zur Behandlung der inneren Faktoren wirkt im Moment relativ
unhandlich, wir werden damit aber in Abschnitt \ref{3_regel}
einfache asymptotische Rechenregeln ableiten.

\section{Asymptotische Entwicklung}
Das verbleibende Problem besteht darin, das Integral \Ref{3_31} im
Grenzfall $\varepsilon \rightarrow 0$ zu untersuchen. Dieses Integral
hat eine so einfache Form, da"s wir nun mit expliziten Rechnungen
beginnen k"onnen.
Da wir in diesem Abschnitt nur Funktionen einer Variablen betrachten,
verwenden wir zur einfacheren Notation $\xi$ nicht und arbeiten mit
$x,y$ als freien, unabh"angigen Variablen.

Zun"achst w"ahlen wir ein spezielles Bezugssystem $(t, \vec{x})$, $r=|\vec{x}|$
und schreiben die Distributionen $(x^{-2\alpha} \:
\ln^\beta(|x^2|) \;|\; 1)$, $(1 \;|\; x^{-2 \gamma})$ mit Konturintegralen um:
F"ur $(x^{-2} \;|\; 1)$, $(\ln(|x^2|) \;|\; 1)$ haben wir die
Relationen
\begin{eqnarray}
\label{eq:3_40}
\left(x^{-2} \;|\; 1 \right)(f) &=&\lim_{\delta \rightarrow 0} \;
	\int_{\sR^3} d\vec{x} \inti dt \; \frac{1}{(t-r-i \delta)
	(t+r-i \delta)} \; f(t,\vec{x}) \\
\left( \ln(|x^2|) \;|\; 1 \right)(f) &=& \lim_{\delta \rightarrow 0} \;
	\int_{\sR^3} d\vec{x} \inti dt \nonumber \\
\label{eq:3_41}
&& \times \; ( \ln(t-r-i\delta) + \ln(t+r-i\delta) + i \pi ) \; f(t,\vec{x}) \;\;\;,
\end{eqnarray}
wie man an \Ref{3_4}, \Ref{3_5} sowie dem Verhalten der Pole und des
Logarithmus in der komplexen $t$-Ebene direkt sieht. In \Ref{3_41} haben
wir die komplexe Ebene l"angs der beiden Strahlen $\{ {\mbox{Re }} t =
\pm r, \; {\mbox{Im }} t \geq \delta \}$ geschlitzt, damit der
Logarithmus eindeutig ist.
Die Distribution $(x^{-2\alpha} \: \ln^\beta(|x^2|) \;|\; 1)$
l"a"st sich durch Multiplikation aus \Ref{3_40}, \Ref{3_41} und der
Funktion $x^2$ aufbauen. Deswegen ist einsichtig, da"s die Gleichung
\Equ{3_42}
\left( x^{-2\alpha} \: \ln^\beta(|x^2|) \;|\; 1 \right) \;=\;
\lim_{\delta \rightarrow 0} \; \frac{
	(\ln(t-r-i\delta) + \ln(t+r-i\delta) + i \pi)^\beta }{
	(t-r-i\delta)^\alpha \: (t+r-i\delta)^\alpha }
\EndEqu
gilt\footnote{Um \Ref{3_42} sauber herzuleiten, kann man sich allgemein
"uberlegen, da"s f"ur das Distributionsprodukt auch das Produkt
der regularisierten Distributionen gebildet und anschlie"send die
Regularisierung entfernt werden kann.
Die Regularisierung zweier Distributionen $D_1, D_2$ mit Tr"ager im
oberen Massenkegel entspricht im Impulsraum der Multiplikation mit
$\widehat{\eta_\delta}$. Im Grenzfall $\delta \rightarrow 0$
konvergiert $\widehat{D_j^\delta}$ lokal gleichm"a"sig gegen $\widehat{D_j}$.
Da das Integrationsgebiet von \Ref{3_13} nach \Ref{3_a1}
kompakt ist, konvergiert das Integral \Ref{3_42} punktweise
\[ \left( \widehat{D_1^\delta} \ast \widehat{D_2^\delta} \right)
	(p) \stackrel{\delta \rightarrow 0}{\longrightarrow}
	(\widehat{D_1} \ast \widehat{D_2})(p) \spc . \]
Da diese Konvergenz lokal gleichm"a"sig in $p$ ist, konvergiert
$\widehat{D_1^\delta} \ast \widehat{D_2^\delta}$ sogar im
Distributionssinnne. Es folgt
\[ D_1 \cdot D_2 \;=\; \lim_{\delta \rightarrow 0} D_1^\delta
	D_2^\delta \spc . \] }.
Durch komplexe Konjugation erh"alt man
\Equ{3_43}
\left( 1 \;|\; x^{-2 \gamma} \right) \;=\; \lim_{\delta \rightarrow 0} \;
\frac{1}{(t-r+i \delta)^\gamma \: (t+r+i \delta)^\gamma } \spc .
\EndEqu
An der Darstellung \Ref{3_42}, \Ref{3_43} kann man direkt ablesen,
da"s der Tr"ager dieser Distributionen im oberen bzw. unteren
Massenkegel liegt: In \Ref{3_42} beispielsweise liegen die Pole
in der oberen Halbebene. Bei Fouriertransformation
\[ D(\omega,\vec{p}) \;=\; \int_{\sR^3} d\vec{x} \; e^{-i \vec{p} \: \vec{x}}
	\inti dt \; D(t,\vec{x}) \: e^{i \omega t} \]
k"onnen wir f"ur $\omega<0$ das $t$-Integral in der unteren Halbebene
schlie"sen und erhalten null. Damit ist \Ref{3_42} nur aus positiven
Frequenzen aufgebaut.

Wir betrachten nun die rationale Funktion $\eta(.,\vec{x})$ f"ur festes
$\vec{x}$: Den Grad des Nenners von $\eta$ bezeichnen wir mit $K$.
Wir k"onnen annehmen, da"s der Nenner nur einfache Nullstellen
besitzt, den allgemeinen Fall erh"alt man daraus im Grenzfall, da"s
sich mehrere Nullstellen beliebig nahe kommen.
Da $\eta$ eine glatte Funktion ist, k"onnen diese
Nullstellen nicht reell sein. Aus $\eta=\overline{\eta}$ folgt, da"s
mit $z$ auch $\overline{z}$ eine Nullstelle des Nenners ist.
Damit erhalten wir f"ur $\eta$ eine Partialbruchzerlegung der Form
\[ \eta(t,\vec{x}) \;=\; {\mbox{Re }} \left( \frac{1}{i \pi} \sum_{k=1}^K
	\; \frac{c_k(\vec{x})}{t-z_k(\vec{x})} \right) \;\;\;,\spc
	{\mbox{Im }} z_k > 0 \]
mit komplexen Koeffizienten $c_k$, die auch verschwinden k"onnen.
F"ur $\eta_\varepsilon$ ergibt sich
\Equ{3_102a}
\eta_\varepsilon(t,\vec{x}) \;=\; \frac{1}{2 \pi i}
	\frac{1}{\varepsilon^3} \sum_{k=1}^K \;
	\frac{c_k(\varepsilon^{-1} \: \vec{x})}{t-\varepsilon \:
		z_k(\varepsilon^{-1} \: \vec{x})} -
	\frac{\overline{c_k}(\varepsilon^{-1} \: \vec{x})}{t-\varepsilon \:
		\overline{z_k}(\varepsilon^{-1} \: \vec{x})} \spc .
\EndEqu
Jetzt k"onnen wir die Faltung $\Ref{3_42} \ast \eta_\varepsilon$ mit dem
Residuensatz berechnen: wir schlie"sen das $t$-Integral nach oben und
erhalten mit den Bezeichungen $y=(y^0,\vec{y})$, $\tilde{t}=y^0-t$,
$\tilde{r}=|\vec{y}-\vec{x}|$
\begin{eqnarray}
\lefteqn{ \left( y^{-2\alpha} \: \ln^\beta(|y^2|) \;|\; 1
	\right)^\varepsilon \;=\; \frac{1}{2 \pi i} \;
	\frac{1}{\varepsilon^3} \;
	\lim_{\delta \rightarrow 0} \; \int_{\sR^3} d\vec{x} \;
	\sum_{k=1}^K
	\; c_k(\varepsilon^{-1} \: \vec{x}) } \nonumber \\
&&  \times \inti dt \; \frac{1}{(t-\varepsilon \: z_k)} \;
	\frac{ (\ln(\tilde{t}-\tilde{r}-i\delta) +
		\ln(\tilde{t}+\tilde{r}-i\delta) + i \pi)^\beta }{
	(\tilde{t}-\tilde{r}-i\delta)^\alpha \:
		(\tilde{t}+\tilde{r}-i\delta)^\alpha } \nonumber \\
\label{eq:3_45}
&=& \frac{1}{\varepsilon^3} \int_{\sR^3} d\vec{x} \; \sum_{k=1}^K \;
	c_k(\varepsilon^{-1} \: \vec{x}_1) \;
	\frac{ (\ln(y^0-\tilde{r}-\varepsilon z_k)
		+ \ln(y^0+\tilde{r}-\varepsilon z_k) + i \pi)^\beta }{
	(y^0-\tilde{r}-\varepsilon z_k)^\alpha \:
	(y^0+\tilde{r}-\varepsilon z_k)^\alpha } \;\;\; , \spc
\end{eqnarray}
dabei wurde die Abh"angigkeit $z_k=z_k(\varepsilon^{-1} \: \vec{x})$
nicht ausgeschrieben. Entsprechend hat man
\Equ{3_46}
\left( 1 \;|\; y^{-2 \gamma} \right) \;=\;
\frac{1}{\varepsilon^3} \int_{\sR^3} d\vec{x} \; \sum_{l=1}^K
	\; \overline{c_l}(\varepsilon^{-1} \: \vec{x}) \;
	\frac{1}{(y^0-\tilde{r}-\varepsilon \overline{z_l})^\gamma \:
	(y^0+\tilde{r}-\varepsilon \overline{z_l})^\gamma } \spc .
\EndEqu
Wir setzen in \Ref{3_31} ein und erhalten nach Umskalierung von
$\vec{x}_1, \vec{x}_2$ die Gleichung
\begin{eqnarray}
{\cal{A}}^\varepsilon &=&
	\int_{\sR^3} d\vec{x} \int_{\sR^3} d\vec{x}_1 \int_{\sR^3}
	d\vec{x}_2 \sum_{k,l=1}^K \; c_k(\vec{x}_1) \:
	\overline{c_l}(\vec{x}_2) \inti dt \; g(t,\vec{x}) \nonumber \\
\label{eq:3_47}
&& \times  \; \frac{ (\ln(t-\tilde{r}_1-\varepsilon z_k)
	+ \ln(t+\tilde{r}_1-\varepsilon z_k) + i \pi)^\beta
	}{ (t-\tilde{r}_1-\varepsilon z_k)^\alpha \:
	(t+\tilde{r}_1-\varepsilon z_k)^\alpha \;
	(t-\tilde{r}_2-\varepsilon \overline{z_l})^\gamma \:
	(t+\tilde{r}_2-\varepsilon \overline{z_l})^\gamma }
\end{eqnarray}
mit $z_k=z_k(\vec{x}_1)$, $z_l=z_l(\vec{x}_2)$,
$\tilde{r}_1=|\vec{x}-\varepsilon \vec{x}_1|$,
$\tilde{r}_2=|\vec{x}-\varepsilon \vec{x}_2|$.
Die komplexe $t$-Ebene ist jetzt auf den Strahlen $\{ t=\pm \tilde{r}_1
+\varepsilon z_k + i \lambda \;;\; 0 \leq \lambda \in \R \}$
geschlitzt, die Pole von $(t \pm \tilde{r}_1 - \varepsilon z_k)^{-\alpha}$
und $(t \pm \tilde{r}_2 - \varepsilon \overline{z_k})^{-\gamma}$
liegen in der oberen bzw. unteren Halbebene.
Wir k"onnen das $t$-Integral in \Ref{3_47} nicht direkt mit dem
Residuensatz ausf"uhren, weil wir die Pole der Funktion $g(.,\vec{x})$
in der komplexen Ebene nicht kennen.

Im Grenzfall $\varepsilon \rightarrow 0$ treten in \Ref{3_47}
auf dem Lichtkegel Singularit"aten auf.

\subsection{Die Singularit"aten auf dem Lichtkegel}
Wir untersuchen zun"achst die Singularit"aten auf dem Lichtkegel und
nehmen dazu an, da"s $g$ in einer Umgebung des Ursprungs verschwindet.
Bei dem vereinfachten Integral
\[ \inti dt \; \frac{\ln^\beta(t-\tilde{r}_1-\varepsilon z_k)}{
	(t-\tilde{r}_1-\varepsilon z_k)^\alpha \;
	(t-\tilde{r}_2-\varepsilon \overline{z_l})^\gamma } \]
k"onnen wir den Integrationsweg nach unten schlie"sen und erhalten
\[ \;=\; - 2 \pi i \: \frac{1}{(\gamma-1)!} \: \left(\frac{d}{dt}
	\right)^{\gamma-1} \left( \frac{\ln^\beta(t-\tilde{r}_1
	-\varepsilon z_k)}{(t-\tilde{r}_1-\varepsilon z_k)^\alpha}
	\right)_{|t=\tilde{r}_2 + \varepsilon \overline{z_l}} \spc , \]
also einen Pol der Ordnung $\ln^\beta(\varepsilon) \:
\varepsilon^{-\alpha-\gamma+1}$.
In dem $t$-Integral in \Ref{3_47} treten zus"atzlich Funktionen auf, die
bei $t=r$ nicht singul"ar sind. F"ur den f"uhrenden Beitrag in $1/\varepsilon$
k"onnen wir diese Funktionen durch ihren Funktionswert bei $t=r$ ersetzen.
Wir behandeln die Singularit"aten bei $t=-r$ auf die gleiche Weise und
erhalten
\begin{eqnarray}
\lefteqn{ {\cal{A}}^\varepsilon \;=\;
	{\cal{O}}\left(\ln^\beta(\varepsilon)\:
		\varepsilon^{-\alpha-\gamma+2} \right) \;-\; 2 \pi i
	\int_{\sR^3} d\vec{x}
	\int_{\sR^3} d\vec{x}_1 \int_{\sR^3} d\vec{x}_2 \; \sum_{k,l=1}^K
	\; c_k(\vec{x}_1) \; \overline{c_l}(\vec{x}_2)
	 } \nonumber \\
&\times& \!\left\{ \frac{g(r,\vec{x})}{(2r)^{\alpha+\gamma}} \frac{1}{(\gamma-1)!}
	\left(\frac{d}{d\tau} \right)^{\gamma-1} \!\!
	\left( \frac{ (\ln(\tau)+\ln(2r) + i \pi)^\beta }{ \tau^\alpha } \right)
	_{|\tau=\tilde{r}_2 - \tilde{r}_1 + \varepsilon \overline{z_l}
	- \varepsilon z_k }
	\right. \nonumber \\
\label{eq:3_48}
&&\left.+  \frac{g(-r,\vec{x})}{(-2r)^{\alpha+\gamma}} \frac{1}{(\gamma-1)!}
	\left(\frac{d}{d\tau} \right)^{\gamma-1} \!\!
	\left( \frac{ (\ln(\tau)+\ln(2r)+i \pi)^\beta }{ \tau^\alpha } \right)
	_{|\tau=\tilde{r}_1 - \tilde{r}_2 + \varepsilon z_k -
	\varepsilon \overline{z_l} } \right\} \;\;\;. \spc
\end{eqnarray}
In diese Gleichung geht tats"achlich nur der Funktionswert von $g$
auf dem Lichtkegel ein. Wir setzen die Entwicklung
\[ \tilde{r}_2 - \tilde{r}_1 \;=\; \frac{\varepsilon}{r} \: \bra \vec{x}
	\;,\; \vec{x}_2-\vec{x}_1 \ket \;+\; {\cal{O}}(\varepsilon^2) \]
ein. Falls $\varepsilon$ klein genug ist, k"onnen wir die Integration "uber
$\vec{x}_1, \vec{x}_2$ sowie die Summe "uber $k,l$ ausf"uhren und erhalten
\begin{eqnarray}
{\cal{A}}^\varepsilon &=& \int_{\sR^3} d\vec{x} \;
	\ln^\beta(\varepsilon) \: \varepsilon^{-\alpha-\gamma+1}
\left( \frac{g(r,\vec{x})}{r^{\alpha+\gamma}} \; \Lambda_1(\vec{x})
	+ \frac{g(-r,\vec{x})}{(-r)^{\alpha+\gamma}} \; \Lambda_2(\vec{x})
	\right) \nonumber \\
\label{eq:3_50}
&&+ \left\{ \begin{array}{ll}
	{\cal{O}}\left(\ln^{\beta-1}(\varepsilon) \: \varepsilon^{-\alpha
	-\gamma+1} \right) & {\mbox{f"ur $\beta>0$}} \\
	{\cal{O}}\left( \varepsilon^{-\alpha-\gamma+2} \right) &
		{\mbox{f"ur $\beta=0$}} \end{array} \right.
\end{eqnarray}
mit geeigneten Funktionen $\Lambda_j=\Lambda_j(c_k,z_k,\vec{x})$.
Aus einem Skalierungsargument und der Relation $\eta(-x)=\eta(x)$ erh"alt
man
\begin{eqnarray}
\label{eq:3_51}
\Lambda_j(\lambda \vec{x}) &=& \Lambda_j(\vec{x}) \;\;{\mbox{f"ur}}
	\;\;\; \lambda > 0 \\
\label{eq:3_52}
\Lambda_1(-\vec{x}) &=& \Lambda_2(\vec{x}) \spc ,
\end{eqnarray}
au"serdem ist $\Lambda_j (\vec{x}) \neq 0$.
Detailliertere Aussagen k"onnen wir "uber die $\Lambda_j$ nicht machen,
weil diese Funktionen wesentlich von der Regularisierungsfunktion
$\eta$ abh"angen.

\subsection{Entwicklung nach $\varepsilon^2 z^{-1}$}
Dieser Abschnitt ist noch nicht vollst"andig ausgearbeitet.
Er wird aber auch erst in Abschnitt \ref{4_ab5} referiert.

\section{Asymptotische Rechenregeln}
\label{3_regel}
Mit den bisherigen Konstruktionen haben wir f"ur Produkte von
Distributionen der Form \Ref{3_7}, \Ref{3_8} die Singularit"at auf dem
Lichtkegel untersucht. Wir k"onnen die f"uhrende
Singularit"at mit Gleichung \Ref{3_48} beschreiben und haben die zugeh"orige
Polordnung $\sim \varepsilon^{-p} \ln^\beta \varepsilon$ bestimmt.
Leider hat der Ausdruck \Ref{3_48} f"ur verschiedene Distributionsprodukte
selbst bei gleicher Polordnung eine unterschiedliche Form, man
vergleiche z.B.
\[ (\xi^{-4} \: \ln(|\xi^2|) \:|\: \xi^{-2}) \;\;,\;\;\; (\xi^{-2} \: \ln(|\xi^2|) \:|\: \xi^{-4})
\;\;,\;\;\; (\xi_j \:\xi^{-4} \: \ln(|\xi^2|) \:|\: \xi^j \: \xi^{-2})
\;\;,\;\;\; (\xi^{-2} \:|\: \xi^{-4} \: \ln(|\xi^2|)) \;\;\; . \]
Um solche Distributionsprodukte miteinander in Beziehung setzen zu
k"onnen, m"ussen die Integralausdr"ucke \Ref{3_48} mit
{\em{asymptotischen Rechenregeln}}\index{asymptotische
Rechenregeln} umgeformt werden.

Wir verwenden oft die Abk"urzung $z=\xi^2$.
Diese Schreibweise ist g"unstig, weil in der Variablen
$z$ ``partiell integriert'' werden kann:
\begin{Lemma}
\label{3_lemma1}
F"ur $\gamma>1$ gilt
\begin{eqnarray}
\lefteqn{ \int d^4y \; g(y) \; \left(z^{-\alpha} \:
	\ln^\beta(|z|) \;|\; z^{-\gamma} \right)^\varepsilon } \nonumber \\
&=& \frac{1}{\gamma-1} \; \int d^4y \; g(y) \: \left( \frac{d}{dz}
	z^{-\alpha} \: \ln^\beta(|z|) \;|\; z^{-\gamma+1}
	\right)^\varepsilon \nonumber \\
\label{eq:3_60a}
&&+{\cal{O}}(\ln^\beta(\varepsilon) \;
	\varepsilon^{-\alpha-\gamma+2}) \;\;\; .
\end{eqnarray}
\end{Lemma}
{\Beweis}
F"ur die komplexen Vierervektoren
\[ \xi_\pm \:=\: \left( \pm(r+\tilde{r}_2-\tilde{r}_1 + \varepsilon
	\overline{z_l} - \varepsilon z_k, \: \vec{x} \right) \]
gilt
\[ \xi_\pm^2 \;=\; \pm 2r \: (\tilde{r}_2 - \tilde{r}_1 +
	\varepsilon \overline{z_l} - \varepsilon z_k)
	\;+\; {\cal{O}}(\varepsilon^2) \spc . \]
Wir k"onnen Gleichung \Ref{3_48} in der Form
\begin{eqnarray}
{\cal{A}}^\varepsilon &=&
	{\cal{O}}\left(\ln^\beta(\varepsilon)\:
		\varepsilon^{-\alpha-\gamma+2} \right) - 2 \pi i
	\int_{\sR^3} \!\! d\vec{x}
	\int_{\sR^3} \!\! d\vec{x}_1 \int_{\sR^3} \!\! d\vec{x}_2
	\sum_{k,l=1}^K c_k(\vec{x}_1) \: \overline{c_l}(\vec{x}_2) \nonumber \\
&&\times \left\{ \frac{g(r,\vec{x})}{2r} \frac{1}{(\gamma-1)!}
	\left(\frac{d}{dz} \right)^{\gamma-1} \!\!
	\left( z^{-\alpha} \; \ln^\beta(z) \right)
	_{|z=\xi_+^2} \right. \nonumber \\
\label{eq:3_63}
&&\left. \spc - \; \frac{g(-r,\vec{x})}{2r} \frac{1}{(\gamma-1)!}
	\left(\frac{d}{dz} \right)^{\gamma-1} \!\!
	\left( z^{-\alpha} \; \ln^\beta(z) \right)
	_{|z=\xi_-^2} \right\}
\end{eqnarray}
umschreiben, dabei ist $\ln(z)$ durch
$\ln(\xi^2) = \ln(\xi^0 + |\vec{\xi}|) + \ln(\xi^0 - |\vec{\xi}|) + i \pi$
definiert, die komplexe $\xi^0$-Ebene ist wieder auf zwei Strahlen
nach oben geschlitzt. F"ur die Divergenz auf dem Lichtkegel folgt
die Behauptung mit der Umformung
\[ \frac{1}{(\gamma-1)!} \: \left(\frac{d}{dz}\right)^{\gamma-1}
	\left(z^{-\alpha} \: \ln^\beta(|z|)\right) \;=\; \frac{1}{(\gamma-1)}
	\; \frac{1}{(\gamma-2)!} \: \left(\frac{d}{dz}\right)^{\gamma-2}
	\; \frac{d}{dz} \left(z^{-\alpha} \: \ln^\beta(|z|)\right) \;\;\; . \]
\QED
Wir wollen nun unsere Behandlung der inneren Faktoren mit der
Greenschen Formel, \Ref{3_97a}, in eine einfachere Form bringen und
beginnen dazu mit zwei inneren Faktoren.
\begin{Satz}
\label{3_satz1}
F"ur $\gamma>1$ gilt
\begin{eqnarray}
\lefteqn{\int d^4y \; g(y) \; \left(\xi_j \; z^{-\alpha} \:
	\ln^\beta(|z|) \;|\; \xi^j \; z^{-\gamma}
	\right)^\varepsilon } \nonumber \\
&=& \frac{1}{2} \: \int d^4y \; g(y) \: \left\{ \left(z^{-\alpha+1}
	\: \ln^\beta(|z|) \;|\; z^{-\gamma} \right)^\varepsilon
	\;+\; \left(z^{-\alpha} \: \ln^\beta(|z|) \;|\;
	z^{-\gamma+1} \right)^\varepsilon \right\} \nonumber \\
\label{eq:3_60}
&& +{\cal{O}}(\ln^\beta(\varepsilon) \;
		\varepsilon^{-\alpha-\gamma+3}) \spc .
\end{eqnarray}
\end{Satz}
{\Beweis}
Wir setzen $f(z)=z^{-\alpha} \: \ln^\beta(|z|)$ und bezeichnen die
Stammfunktion von $f$ mit $F$. Nach \Ref{3_97a} haben wir
\begin{eqnarray}
\lefteqn{ \int d^4y \; g(y) \; \left(\xi_j \; z^{-\alpha} \:
	\ln^\beta(|z|) \;|\; \xi^j \; z^{-\gamma}
	\right)^\varepsilon \;=\;
-\frac{1}{4 (\gamma-1)} \: \int d^4y \; g(y) \; \left( \partial_j
	F \;|\; \partial^j z^{-\gamma+1}
	\right)^\varepsilon } \nonumber \\
\label{eq:3_64}
&=& -\frac{1}{8 (\gamma-1)} \int d^4y \left[ (\Box g)
	\left( F \:|\: z^{-\gamma+1} \right)^\varepsilon
	- g \left( \Box F \:|\: z^{-\gamma+1}
	\right)^\varepsilon - g \left( F \:|\: \Box z^{-\gamma+1}
	\right)^\varepsilon \right] \;\: . \spc
\end{eqnarray}
Wir wenden jetzt die asymptotische Entwicklung \Ref{3_48} an. 
Der erste Summand des Integranden in \Ref{3_64} f"uhrt auf eine
Singularit"at der Ordnung $\ln^\beta(\varepsilon) \; \varepsilon^{-\alpha
-\gamma+3}$ und ist vernachl"assigbar.
Wir rechnen ab jetzt modulo Terme der in der Behauptung angegebenen
Ordnung in $1/\varepsilon$. Damit erhalten wir
\[ \left(\xi_j \; z^{-\alpha} \: \ln^\beta(|z|) \;|\; \xi^j
	\; z^{-\gamma} \right)^\varepsilon \;=\;
\frac{1}{8(\gamma-1)} \: \left(\Box F \;|\; z^{-\gamma+1} \right)^\varepsilon
	\;+\; \frac{1}{8(\gamma-1)} \: 
	\left(F \;|\; \Box z^{-\gamma+1} \right)^\varepsilon \]
und nach Einsetzen der Relationen
\begin{eqnarray*}
\Box F &=& \left(4 z \frac{d^2}{dz^2}+ 8 \frac{d}{dz} \right) F
	\;=\; \left(4 z \frac{d}{dz} + 8 \right) f \\
\Box z^{-\gamma+1} &=& 4 \: (\gamma-1)(\gamma-2) \; z^{-\gamma}
\end{eqnarray*}
sowie Anwendung von Lemma \ref{3_lemma1}
\begin{eqnarray*}
\lefteqn{ \left(\xi_j \; \xi^{-2 \alpha} \: \ln^\beta(|\xi^2|) \;|\; \xi^j
	\; \xi^{-2\gamma} \right)^\varepsilon
\;=\; \frac{1}{8(\gamma-1)} \: \left((4z \: \frac{d}{dz} + 8) f \;|\;
	z^{-\gamma+1} \right)^\varepsilon
	\;+\; \frac{\gamma-2}{2} \: 
	\left(F \;|\; z^{-\gamma} \right)^\varepsilon } \\
&=& \frac{1}{2(\gamma-1)} \: \left(\frac{d}{dz} \: z \: f \;|\;
	z^{-\gamma+1} \right)^\varepsilon + \frac{1}{2(\gamma-1)}
	\: \left(f \;|\; z^{-\gamma+1} \right)^\varepsilon
	+ \frac{\gamma-2}{2(\gamma-1)} \: \left(\frac{d}{dz} F \;|\;
	z^{-\gamma+1} \right)^\varepsilon \\
&=& \frac{1}{2} \: \left( z \: f \;|\; z^{-\gamma} \right)^\varepsilon \;+\;
	\frac{1}{2} \: \left(f \;|\; z^{-\gamma+1} \right)^\varepsilon
	\spc .
\end{eqnarray*}
\QED
Das Ergebnis dieses Satzes kann man sich leicht merken. Wir k"onnen
danach den inneren Faktor $\xi^2$ zur H"alfte auf die linke und rechte
Seite der Klammer $(.|.)$ schreiben, also formal
\Equ{3_66}
\left(\xi_j \;.\:|\: \xi^j \; .\;\right) \;=\;
	\frac{1}{2} \: \left(z \;.\:|\: .\;\right) \;+\; \frac{1}{2} \:
	\left( .\:|\: z \; .\;\right) \spc .
\EndEqu
Rechnungen in einer speziellen Regularisierung deuten darauf hin, da"s
diese Regel auch dann angewendet werden kann, wenn man in \Ref{3_66}
beliebige Funktionen der Form \Ref{3_13a} einsetzt.
Damit sollte man mit \Ref{3_66} durch Iteration unmittelbar den Fall
beliebig vieler innerer Faktoren behandeln k"onnen, genauer
\[ \left( \xi_{j_1} \cdots \xi_{j_q} \;.\;|\; \xi^{j_1} \cdots \xi^{j_q}
	\,.\;\right)
	\;=\; \frac{1}{2^q} \: \sum_{j=1}^q \;
	\left( \!\! \begin{array}{c} q \\ j \end{array} \!\! \right) \;
	\left( z^{j} \;.\;|\; z^{q-j} \;. \;\right) \spc . \]
F"ur unsere Zwecke gen"ugt es, diese Gleichung f"ur vier innere Faktoren zu
beweisen:
\begin{Satz}
\label{3_satz2}
F"ur $\gamma>2$ gilt
\begin{eqnarray}
\lefteqn{\int d^4y \; g(y) \; \left(\xi_j\:\xi_k \; z^{-\alpha} \:
	\ln^\beta(|z|) \;|\; \xi^j\:\xi^k \; z^{-\gamma}
	\right)^\varepsilon } \nonumber \\
&=& \frac{1}{4} \: \int d^4y \; g(y) \: \left\{ \left(z^{-\alpha+2}
	\: \ln^\beta(|z|) \;|\; z^{-\gamma} \right)^\varepsilon
	\;+\; 2 \: \left(z^{-\alpha+1} \: \ln^\beta(|z|) \;|\;
	z^{-\gamma+1} \right)^\varepsilon \right. \nonumber \\
&& \left. \spc\spc \;+\; \left(z^{-\alpha} \: \ln^\beta(|z|) \;|\;
	z^{-\gamma+2} \right)^\varepsilon \right\} \nonumber \\
\label{eq:3_68}
&& +{\cal{O}}(\ln^\beta(\varepsilon) \;
		\varepsilon^{-\alpha-\gamma+4}) \spc .
\end{eqnarray}
\end{Satz}
{\Beweis}
Wir verwenden die gleiche Methode wie beim Beweis von Satz \ref{3_satz1},
nur ist die Rechnung jetzt etwas aufwendiger.
Wir setzen wieder $f(z)=z^{-\alpha} \: \ln^\beta(|z|)$; $F$ sei nun
eine Funktion mit $F^{\prime \prime}=f$.

Das Umschreiben der inneren Faktoren in partielle Ableitungen
gem"a"s \Ref{3_29} liefert
\begin{eqnarray*}
\xi_j \: \xi_k \; f &=& \frac{1}{4} \: \partial_{jk} F \:-\:
	\frac{1}{2} \: g_{jk} \: F^\prime \\
\xi_j \: \xi_k \; z^{-\gamma} &=& \frac{1}{4(\gamma-1)(\gamma-2)} \:
	\partial_{jk} z^{-\gamma+2} \:+\: \frac{1}{2(\gamma-1)} \:
	g_{jk} \: z^{-\gamma+1}
\end{eqnarray*}
und unter Verwendung der Relationen
\begin{eqnarray*}
\Box F &=& 4 z \: f \:+\: 8 \: F^\prime \\
\Box z^{-\gamma+2} &=& 4 \: (\gamma-2)(\gamma-3) \; z^{-\gamma+1}
\end{eqnarray*}
schlie"slich
\begin{eqnarray}
\left(\xi_j \: \xi_k \; f \;|\; \xi^j \: \xi^k \;
	z^{-\gamma}
	\right)^\varepsilon &=& \frac{1}{16(\gamma-1)(\gamma-2)} \:
	\left(\partial_{jk} F \;|\; \partial^{jk} z^{-\gamma+2}
	\right)^\varepsilon \nonumber \\
\label{eq:3_69}
&& - \frac{\gamma-3}{2(\gamma-1)} \: \left(F^\prime \;|\;
	z^{-\gamma+1} \right)^\varepsilon + \frac{1}{2(\gamma-1)} \:
	\left(z \: f \;|\; z^{-\gamma+1} \right)^\varepsilon
	\;\;\; .
\end{eqnarray}
Wir formen jetzt die ersten beiden Summanden weiter um, dabei
lassen wir alle Terme der in
der Behauptung angegebenen Ordnung weg.

Bei dem ersten Summanden in \Ref{3_69} k"onnen wir rekursiv zweimal
gem"a"s \Ref{3_97a} partiell integrieren. Die Summanden, die $\Box g$
enthalten, sind genau wie in Satz \ref{3_satz1} von niedrigerer Ordnung und
k"onnen vernachl"assigt werden.
Damit haben wir
\begin{eqnarray*}
\lefteqn{ \left(\partial_{jk} F \;|\; \partial^{jk} z^{-\gamma+2}
	\right)^\varepsilon } \\
&=& \frac{1}{4} \: \left(\Box^2 F \;|\; z^{-\gamma+2} \right)^\varepsilon
	+\; \frac{1}{2} \: \left(\Box F \;|\; \Box z^{-\gamma+2}
	\right)^\varepsilon +\; \frac{1}{4} \: \left(F \;|\; \Box^2
	z^{-\gamma+2} \right)^\varepsilon \\
&=& 4 \: \left(\frac{d^2}{dz^2} \: z \: \frac{d^2}{dz^2} \: z \: F \;|\;
	z^{-\gamma+2} \right)^\varepsilon +\; 8 \: (\gamma-2)(\gamma-3)
	\left( \frac{d}{dz^2} \: z F \;|\; z^{-\gamma+1}
	\right)^\varepsilon \\
&&+\; 4 \: (\gamma-1)(\gamma-2)^2 (\gamma-3) \: \left(F \;|\; z^{-\gamma}
	\right)^\varepsilon \\
&=& 4 \: \left( \frac{d^2}{dz^2} \: z^2 \: f \;|\; z^{-\gamma+2}
	\right)^\varepsilon +\; 8 \: \left(\frac{d}{dz^2} \: z \: F^\prime
	\;|\; z^{-\gamma+2} \right)^\varepsilon \\
&&+\; 8 \: (\gamma-2)(\gamma-3) \: \left(z \: f \;|\; z^{-\gamma+1}
	\right)^\varepsilon +\; 16 \: (\gamma-2)(\gamma-3) \left(
	F^\prime \;|\; z^{-\gamma+1} \right)^\varepsilon \\
&&+\; 4 \: (\gamma-1)(\gamma-2)^2 (\gamma-3) \left(F \;|\; z^{-\gamma}
	\right)^\varepsilon \spc .
\end{eqnarray*}
Wir formen mit Hilfe von Lemma \ref{3_lemma1} weiter um und erhalten
\begin{eqnarray}
&=& 4 \: (\gamma-1)(\gamma-2) \left(z^2 f \;|\; z^{-\gamma}
	\right)^\varepsilon
	\;+\; 8 \: (\gamma-2)^2 \left(z \: f \;|\; z^{-\gamma+1}
	\right)^\varepsilon \nonumber \\
\label{eq:3_70}
&&+\; 4 \: (\gamma^2-\gamma+4) \left(f \;|\; z^{-\gamma+2}
	\right)^\varepsilon \spc .
\end{eqnarray}
Au"serdem haben wir
\Equ{3_71}
-\frac{\gamma-3}{2(\gamma-1)} \: \left(F^\prime \;|\; z^{-\gamma+1}
	\right)^\varepsilon \;=\; \frac{-2\gamma+6}{2(\gamma-1)(\gamma-2)}
	\: \left(f \;|\; z^{-\gamma+2} \right)^\varepsilon \spc .
\EndEqu
Bei Einsetzen von \Ref{3_70}, \Ref{3_71} in \Ref{3_69} folgt die
Behauptung.
\QED

\section{Zusammenstellung}
\label{3_sec_zus}
Zur besseren "Ubersicht wollen wir unsere Konstruktion und die
abgeleiteten Rechenregeln mit einer etwas kompakteren Schreibweise
zusammenfassen.

Wir haben zun"achst im formalen Produkt \Ref{3_1} m"oglichst viele
Faktoren im Distributionssinne ausmultipliziert und Ausdr"ucke der
Form \Ref{3_19} erhalten. Mit der Abk"urzung $z=\xi^2$ schreiben wir
f"ur das Produkt der beiden Distributionen auch einfach
\Equ{3_80}
\left(\xi_{j_1} \cdots \xi_{j_q} \; z^{-\alpha} \: \ln^\beta(|z|) \;|\;
	\xi_{k_1} \cdots \xi_{k_r} \; z^{-\gamma} \right) \spc .
\EndEqu
Falls in \Ref{3_80} der Eintrag auf der rechten oder linken Seite der Klammer
gleich $1$ ist, k"onnen wir \Ref{3_80} als eine Distribution mit
Tr"ager im oberen bzw. unteren Massenkegel definieren.
Auch im allgemeinen Fall untersuchen wir \Ref{3_80} im schwachen Sinne;
der Ausdruck erh"alt dann aber erst nach Regularisierung und
asymptotischer Entwicklung einen mathematischen Sinn.

Nach Definition der Distributionen \Ref{3_14}, \Ref{3_15} und gem"a"s
\Ref{3_16}, \Ref{3_17} haben wir die Rechenregeln
\begin{eqnarray}
\label{eq:3_81}
\overline{ \left( H_1 \:|\: H_2 \right) } &=& \left( H_2\:|\:H_1 \right) \\
\label{eq:3_82}
\left( H_1 \:|\: H_2 \right) \cdot \left( H_3 \:|\: H_4 \right)
	&=& \left( H_1 \: H_3 \;|\; H_2 \: H_4 \right) \spc .
\end{eqnarray}
Mit Hilfe von \Ref{3_81} lassen sich alle Ergebnisse unmittelbar auf den
Fall erweitern, da"s der Faktor $\ln^\beta(|z|)$ in \Ref{3_80} in der
rechten Seite der Klammer $(.|.)$ steht.
Bei Kontraktion der Tensorindizes k"onnen wir gem"a"s \Ref{3_90a} die
Umformungen
\begin{eqnarray}
\left( \xi_j \: \xi^j \;.\;|\;.\; \right) &=& \left(z \;.\;|\;.\;\right)\\
\left( \;.\;|\; \xi_j \: \xi^j\; . \; \right) &=& \left(\;.\;|\;z\;.\;\right)
\end{eqnarray}
anwenden und damit die Distributionsprodukte in die Form
\Equ{3_83}
f_{i_1 \cdots i_t \cdots i_s \cdots i_k} \; \left(\xi^{i_1} \cdots \xi^{i_t}
	\; \xi_{j_1} \cdots \xi_{j_p} \; z^{-\alpha} \: \ln^\beta(|z|)
	\;|\; \xi^{i_{t+1}} \cdots \xi^{i_s} \; \xi^{j_1} \cdots \xi^{j_p}
	\; z^{-\gamma} \right)
\EndEqu
bringen. Wir haben im Gegensatz zur Konstruktion in Abschnitt
\ref{3_ab3} die "au"seren Faktoren innerhalb der Klammer $(.|.)$ stehen
gelassen, um zu betonen, da"s das Zusammenfassen dieser $\xi_j$ mit
dem Vorfaktor erst durch die asymptotische Entwicklung
gerechtfertigt wird. Alle bisherigen Rechenregeln folgen entweder
unmittelbar aus einer Definition oder sind Umformungen im
Distributionssinne.

F"ur unsere Zwecke gen"ugt es, den Fall $p \leq 2$ zu betrachten.
Bei Regularisierung und asymptotischer Entwicklung von \Ref{3_83} treten
auf dem Lichtkegel Singularit"aten der Ordnung
\Equ{3_84}
\ln^\beta(\varepsilon) \; \varepsilon^{-\alpha-\gamma+p+1}
\EndEqu
auf. Wir k"onnen die divergenten Terme mit Hilfe von Gleichung \Ref{3_48}
bis zur Ordnung
\Equ{3_85}
{\cal{O}} (\ln^\beta(\varepsilon) \; \varepsilon^{-\alpha-\gamma+p+2})
\EndEqu
beschreiben. "Uber die genaue Form der Beitr"age in \Ref{3_48} k"onnen wir
keine Aussagen machen, weil darin die Regularisierungsfunktion
$\eta$ eingeht.
Die Bedeutung dieser Gleichung liegt darin, da"s damit die
Distributionsprodukte modulo Terme der Ordnung \Ref{3_85} umgeformt
werden k"onnen.
Mit der Schreibweise ``$\simeq$'' f"ur ``"aquivalent bis auf Terme der
Ordnung \Ref{3_85}'' haben wir die asymptotischen Rechenregeln
\begin{eqnarray}
\label{eq:3_86}
\lefteqn{ \hspace*{-2cm}
	f_{i_1 \cdots i_t \cdots i_s \cdots i_k} \; \left(\xi^{i_1}
	\cdots \xi^{i_t} \; H_1
	\;|\; \xi^{i_{t+1}} \cdots \xi^{i_s} \; H_2 \right)
\;\simeq\; f_{i_1 \cdots i_s \cdots i_k} \; \xi^{i_1} \cdots \xi^{i_s}\;
	\left( H_1 \:|\: H_2 \right) } \\
\label{eq:3_87}
\left(\;. \;|\; z^{-\gamma} \right) &\simeq& \frac{1}{\gamma-1} \:
	\left(\frac{d}{dz} \;. \;|\; z^{-\gamma+1} \right) \\
\label{eq:3_88}
\left(\xi_j \;. \;|\; \xi^j \;.\; \right) &\simeq& \frac{1}{2} \:
	\left(z \;. \;|\; .\; \right) \;+\; \frac{1}{2} \: \left(\;.
	\;|\; z \;.\; \right) \spc \spc \spc \spc \spc \;\;\;
\end{eqnarray}
hergeleitet.
Mit diesen Regeln k"onnen wir alle Tensorindizes aus der Klammer $(.|.)$
beseitigen und erhalten f"ur die Beitr"age der St"orungsrechnung
Ausdr"ucke der Form
\[ f_{i_1 \cdots i_s \cdots i_k}(x,y) \; \xi^{i_1} \cdots \xi^{i_s} \;
	\left(z^{-\alpha} \: \ln^\beta(|z|) \;|\; z^{-\gamma} \right)
	\spc . \]

Wir beschreiben abschlie"send schematisch, wie die
Euler-Lagrange-Gleichungen mit den asymptotischen Rechenregeln
in sinnvolle Bedingungen an die Tensorfelder $f_{i_1 \cdots i_k}$
umgeschrieben werden k"onnen:
Mit Gleichung \Ref{3_87} k"onnen alle
Distributionsprodukte, die das gleiche Divergenzverhalten in
$1/\varepsilon$ zeigen, in eine der beiden Normalformen
\[ \left( z^{-\alpha} \; \ln^\beta(|z|) \;|\; z^{-1} \right) \;\;\;,\spc
	\left( z^{-1} \;|\; z^{-\alpha} \; \ln^\beta(|z|) \right) \]
gebracht werden. Damit gehen die Euler-Lagrange-Gleichungen in
f"uhrender Ordnung in $1/\varepsilon$ in Gleichungen der Form
\begin{eqnarray*}
\lefteqn{ f_{i_1 \cdots i_s \cdots i_k}(x,y) \; \xi^{i_1} \cdots \xi^{i_s}
	\; \left( z^{-\alpha} \; \ln^\beta(|z|) \;|\; z^{-1} \right) } \\
&&+\; g_{j_1 \cdots j_s \cdots j_k}(x,y) \; \xi^{j_1} \cdots \xi^{j_s}
	\left( z^{-1} \;|\; z^{-\alpha} \; \ln^\beta(|z|) \right) \;=\; 0
\end{eqnarray*}
"uber. Wir werden uns im Einzelfall "uberlegen, da"s daraus die Bedingung
\[ f_{i_1 \cdots i_k}(x,y) \;=\;
	g_{j_1 \cdots j_k}(x,y) \;=\;0 \]
folgt. Auf diese Weise f"allt die Abh"angigkeit von der Regularisierung letztlich
heraus.

\chapter{Der Weg zum Modell}
\label{kap4}
In Kapitel \ref{kap2} haben wir den fermionischen Projektor $P$
eingef"uhrt und genauer untersucht. Zun"achst wurde der freie
Projektor aufgebaut. Anschlie"send haben wir durch St"orungen
des Diracoperators kollektive Anregungen der Fermionen beschrieben,
die St"orungen durch Eich- und Gravitationsfelder waren dabei ein
Spezialfall.

"Uber die Struktur der Wechselwirkung haben wir in Kapitel \ref{kap2}
noch keine Aussage gemacht. Die St"orung des Diracoperators konnte
beliebig sein; wir haben allgemein mathematisch studiert, wie
sich $P$ bei diesen St"orungen verh"alt.
Wir haben aber nicht spezifiziert, welche dieser St"orungen
tats"achlich auftreten d"urfen. Insbesondere sind die folgenden
Punkte noch unbestimmt:
\begin{itemize}
\item Durch welche bosonischen Felder kann die Dynamik des Systems
beschrieben werden?
\item Mit welchen Eichgruppen lassen sich diese bosonischen
Felder beschreiben?
\item Welchen Gleichungen gen"ugen die bosonischen Felder, wie koppeln
sie an die Fermionen an?
\end{itemize}
Zur Beschreibung der Dynamik m"ussen wir folglich
zus"atzliche Gleichungen aufstellen, die wir die {\em{Gleichungen der
diskreten Raumzeit}} nennen. Da wir nur den fermionischen
Projektor $P$ und die Projektoren $E_x$ der diskreten
Raumzeit-Punkte $x \in M$ als fundamentale physikalische Objekte
ansehen, m"ussen diese Gleichungen mit $P$, $E_x$ formuliert werden.

In diesem Kapitel wollen wir konkreter auf die Form dieser Gleichungen
eingehen. Wir werden f"ur verschiedene
Konfigurationen des fermionischen Projektors m"ogliche Gleichungen
diskutieren und auf diese Weise schrittweise Gleichungen ableiten,
die ein realistisches physikalisches Modell beschreiben k"onnten.
Diese Gleichungen bilden dann den Ausgangspunkt von Kapitel
5 (das noch nicht getippt ist) und werden dort sys\-te\-ma\-ti\-sch
untersucht.

Bei der Suche nach ``sinnvollen'' Gleichungen lassen wir uns von
der Forderung leiten, da"s die Gleichungen der diskreten Raumzeit im
Kontinuumslimes in bekannte klassische Feldgleichungen "ubergehen sollen.
Insbesondere werden wir versuchen, die Wechselwirkungen
des Standardmodells sowie das Gravitationsfeld nachzubilden.

\section{Ansatz f"ur die Gleichungen der diskreten Raumzeit}
\label{4_ab1}
In diesem Abschnitt werden wir anhand allgemeiner "Uberlegungen
einen ersten Ansatz f"ur die Gleichungen der diskreten Raumzeit
herleiten.

Da das Variationsprinzip in der klassischen Feldtheorie sehr erfolgreich
ist, wollen wir ebenfalls mit einem Variationsprinzip arbeiten.
Wir m"ussen also eine reelle Funktion $S(P)$ finden, aus der man
bei Variation des fermionischen Projektors als
``Euler-Lagrange-Gleichungen'' die Gleichungen der
diskreten Raumzeit erh"alt. In Analogie zur klassischen Feldtheorie
nennen wir $S$ die {\em{Wirkung}}\index{Wirkung} des Systems.

\subsubsection*{allgemeine Struktur der Gleichungen}
Wir untersuchen zun"achst, welche mathematischen M"oglichkeiten wir bei
der Konstruktion der Wirkung haben:
Aus $P$, $E_x$ lassen sich durch Multiplikation weitere
Operatoren bilden. Da $P$, $E_x$ Projektoren sind und die Gleichung
$E_x \: E_y \:=\: \delta_{xy} \: E_x$ erf"ullen, treten bei Produkten
die Faktoren $P$, $E_x$ immer abwechselnd auf, also in
Kombinationen der Form
\[ P \:E_{x_1} \:P\: E_{x_2} \:P\: \cdots \;\;\;{\mbox{mit}}
\spc x_j \in M \;\;\; . \]
Um aus diesen Operatoren Skalare zu bilden, kann man
Determinanten und Spuren verwenden. Determinanten sind nicht sinnvoll,
denn
\[ \det \left( P \:E_x \:P\: \cdots \right) \;=\;
	\det (P) \: \det(E_x) \:\det(P)\: \cdots \;=\; 0 \]
(man beachte, da"s $P, E_x$ als Projektoren auf echte Teilr"aume von $H$
singul"ar sind). \\
Folglich mu"s die Wirkung aus den komplexwertigen Gr"o"sen
\Equ{4_1}
\alpha_{x_1 \cdots x_p} \;:=\;
	\tr \left( P \:E_{x_1}\: P \:E_{x_2} \:\cdots\: P \: E_{x_p} \right)
	\;\;\;{\mbox{mit}} \spc x_j \in M
\EndEqu
konstruiert werden. Dazu k"onnen wir zun"achst beliebige Funktionen der
$\alpha_{x_1 \cdots x_p}$ bilden.
Um die Abh"angigkeit von den Parametern $x_1, \ldots, x_p$ auf sinnvolle
Weise zu behandeln, ben"otigen wir ein physikalisches Argument:
In der klassischen Feldtheorie ist die Wirkung invariant unter
Diffeomorphismen.
Wie in der Einleitung genauer beschrieben, mu"s diese (aktive)
Koordinateninvarianz bei Diskretisierung der Raumzeit zu einer
Per\-mu\-ta\-tions\-sym\-me\-trie in $M$ verallgemeinert werden. Folglich fordern
wir, da"s die Wirkung unter Vertauschungen der Raumzeitpunkte invariant
ist. Um die Abh"angigkeit von $x_1, \ldots, x_p$ zu beseitigen, ohne
diese Permutationssymmetrie zu zerst"oren, k"onnen wir die $x_j$
in Gruppen gleichsetzen und "uber $M$ summieren. Es w"are auch m"oglich,
Produkte "uber $M$ zu bilden, doch kann man diesen Fall durch
Logarithmieren auf Summen zur"uckf"uhren.
Wir erhalten so beispielsweise die Gr"o"sen
\begin{eqnarray*}
\sum_{x_1, x_2 \in M} f(\alpha_{x_1 x_2}, \alpha_{x_1 x_1 x_2}) \;\;\;,
	\spc \sum_{x_1, x_2, x_3 \in M} g(\alpha_{x_1 x_2 x_3})
\end{eqnarray*}
mit Funktionen $f: \C^2 \rightarrow \R$, $g: \C \rightarrow \R$.
Die Wirkung kann schlie"slich eine beliebige reelle Funktion solcher
Ausdr"ucke sein.

\subsubsection*{Punkt- und Ringbeitr"age}
Dieser Ansatz f"ur die Wirkung ist f"ur uns noch zu allgemein.
Wir verwenden qualitative Informationen "uber den Kontinuumslimes,
um die Form der Wirkung zu spezialisieren.
Nach Kapitel \ref{kap2} wissen wir, da"s der Operator
\[ P(x,y) \;\equiv\; E_x \: P \: E_y \]
als Distribution einen sinnvollen Kontinuumslimes besitzt.
Nach den "Uberlegungen zu Beginn von Kapitel \ref{kap3} kann man
den Kontinuumslimes auch durch den Grenzwert $\varepsilon \rightarrow 0$
einer regularisierten Distribution $P^\varepsilon(x,y)$ beschreiben.
Dabei darf die genaue Art der Regularisierung keine Rolle spielen.

Wir betrachten zun"achst in \Ref{4_1} einen Faktor der Form
\Equ{4_2}
	E_x \:P\: E_x \spc ,
\EndEqu
den wir als {\em{Punktbeitrag}}\index{Punktbeitrag}
bezeichnen. Bei Regularisierung
im Kontinuum tritt anstelle von \Ref{4_2} ein Faktor
$P^\varepsilon(x,x)$ auf. Wir w"ahlen neue Variablen
\[ P^\varepsilon(x,y) \;=:\; \hat{P}^\varepsilon \left( \frac{x+y}{2},
	\frac{x-y}{2} \right) \]
und bilden im zweiten Argument von $\hat{P}^\varepsilon$
die Fouriertransformierte
\Equ{4_3}
P^\varepsilon(x,x) \;=\; \hat{P}^\varepsilon(x,0) \;=\;
	\int \frac{d^4k}{(2 \pi)^4} \; \hat{P}^\varepsilon(x,k) \spc .
\EndEqu
Das erste Argument von $\hat{P}^\varepsilon$ beschreibt die
makroskopische Raumzeit-Abh"angigkeit des fermionischen Projektors,
das zweite Argument dagegen die oszillierenden Anteile der
fermionischen Wellenfunktionen
(f"ur den freien Projektor h"angt $\hat{P}^\varepsilon$ nur
vom zweiten Argument ab).
Wenn wir annehmen, da"s die makroskopischen L"angenskalen viel gr"o"ser
als die Wellenl"angen sind, was zur Beschreibung klassischer
Systeme stets ausreichend ist,
so k"onnen wir mit dem qualitativen Bild von \Ref{3_a4} arbeiten:
Wir hatten "uberlegt, da"s
wir nur dann Ergebnisse erhalten, die unabh"angig von der Regularisierung
sind, wenn die Zust"ande in der N"ahe des Massenkegels besonders eingehen,
oder, anders ausgedr"uckt, wenn es auf die Flanke von
$\hat{P}^\varepsilon(x, .)$ auf dem Massenkegel ankommt.
In \Ref{4_3} wird aber "uber alle Zust"ande gleicherma"sen integriert.
Folglich kann man f"ur Punktbeitr"age den Kontinuumslimes nicht auf
sinnvolle Weise definieren. Wir werden deshalb nur Wirkungen betrachten,
die keine Punktbeitr"age enthalten.

\label{4_ring}
Wir kommen zu dem Fall, da"s in \Ref{4_1} mehr als zwei der
$x_j$ voneinander verschieden sind, was wir als
{\em{Ringbeitrag}}\index{Ringbeitrag}
bezeichnen. Hier treten Schwierigkeiten auf, wenn wir Eichfelder
betrachten. Zur Einfachheit diskutieren wir das Problem
exemplarisch an dem Term
\[ \tr \left( P \:E_{x_1}\: P \:E_{x_2}\:
	P \:E_{x_3} \right) \spc
	{\mbox{mit $x_i \neq x_j \;\;\; \forall i \neq j$}} \]
und einem $U(1)$-Eichfeld, die "Uberlegung l"a"st sich aber unmittelbar
auf den allgemeinen Fall "ubertragen.
Nach Regularisierung im Kontinuum erh"alt man den Ausdruck
\Equ{4_4}
\Tr \left( P^\varepsilon(x_1, x_2) \: P^\varepsilon(x_2, x_3) \:
	P^\varepsilon(x_3, x_1) \right) \spc .
\EndEqu
Bei einer lokalen Eichtransformation mit Eichpotential
$A_j = \partial_j \Lambda$ wird die Phase von $P^\varepsilon(x,y)$ gem"a"s
\Equ{4_5}
P^\varepsilon(x,y) \;\longrightarrow\;
	e^{-i (\Lambda(y) - \Lambda(x))} \: P^\varepsilon(x,y)
\EndEqu 
transformiert. Wegen der Eichinvarianz bleibt dabei \Ref{4_4} unver"andert,
wie man auch explizit verifiziert.
Wir betrachten nun den Fall eines Potentials $A$,
das nicht global weggeicht werden kann:
Nach Kapitel \ref{kap2} treten in $P^\varepsilon(x,y)$
viele verschiedene St"orungsbeitr"age mit Potentialen,
Feldst"arken und Noether-Str"omen auf.
Nach Kapitel \ref{kap3} sind diejenigen Beitr"age dominant, welche auf
dem Lichtkegel am st"arksten singul"ar sind. Das sind die Eichterme,
die analog zu \Ref{4_5} eine Phasentransformation beschreiben
\Equ{4_6}
P^\varepsilon(x,y) \;\longrightarrow\;
	e^{-i \int_x^y A_j \: (y-x)^j} \: P^\varepsilon(x,y) \spc .
\EndEqu
Der Ausdruck \Ref{4_4} transformiert sich unter \Ref{4_6} gem"a"s
\begin{eqnarray*}
\lefteqn{ \Tr \left( P^\varepsilon(x_1, x_2) \: P^\varepsilon(x_2, x_3) \:
	P^\varepsilon(x_2, x_3) \right) } \\
&\longrightarrow& e^{-i \int_{\partial \Delta} A_j \: dx^j}
	\; \Tr \left( P^\varepsilon(x_1, x_2) \: P^\varepsilon(x_2, x_3) \:
	P^\varepsilon(x_2, x_3) \right) \spc ,
\end{eqnarray*}
dabei ist $\Delta$ das Dreieck mit Ecken $x_1, x_2, x_3$;
$\partial \Delta$ bezeichnet dessen Rand.
Nach dem Satz von Stokes haben wir
\[ \int_{\partial \Delta} A_j \: dx^j \;=\;
	\int_\Delta \epsilon^{ijkl} \: F_{ij} \: dx_k \: dx_l \]
mit $F_{ij}=\partial_i A_j - \partial_j A_i$.
Also bleibt der Ringbeitrag \Ref{4_4} nun (im Gegensatz zur
Eichtransformation \Ref{4_5}) nicht unver"andert; in der
Transformationsformel tritt
der Flu"s des Feldes durch das Dreieck $\Delta$ auf.

Um die Auswirkung dieses Flu"sbeitrages zu diskutieren, nehmen wir an,
da"s die Gleichungen der diskreten Raumzeit den Ringbeitrag \Ref{4_4}
enthalten. Damit diese Gleichungen sinnvoll sind, m"ussen sie im freien
Fall (also f"ur $A \equiv 0$) erf"ullt sein. F"ur $A \not \equiv 0$
tritt in \Ref{4_4} zus"atzlich der Flu"sbeitrag auf.
Damit werden die Gleichungen der diskreten
Raumzeit i.a. nur dann weiterhin erf"ullt sein, wenn der Flu"s
durch $\Delta$ verschwindet. Gilt dies f"ur beliebige Dreiecke $\Delta$,
so folgt $F_{ij} \equiv 0$. Damit haben wir zwar eine lokale $U(1)$-Eichsymmetrie;
die Potentiale k"onnen aber global weggeicht werden, so da"s die
Eichfreiheitsgrade keine Dynamik beschreiben.
Allgemeiner kommen wir zu dem Schlu"s, da"s bei Gleichungen mit
Ringbeitr"agen keine Dynamik durch Eichfelder auftritt, was physikalisch
nicht sinnvoll ist. Darum werden wir nur Wirkungen ohne Ringbeitr"age
betrachten.

Diese Argumentation ist etwas unsauber, weil nicht klar ist, wie sich
die Flu"sbeitr"age bei asymptotischer Entwicklung genau auswirken.
Wir k"onnen diesen Punkt auch nicht allgemein genauer diskutieren,
weil dabei die spezielle Form der Gleichungen eingeht.
Beispielsweise w"are es denkbar, da"s in einer geeignet konstruierten
Gleichung mit Ringtermen die Flu"sbeitr"age ganz verschwinden.
Zumindest k"onnen wir das Vermeiden von Ringbeitr"agen aber so
begr"unden:
Es ist eine allgemeine Beobachtung, da"s in die klassischen
Feldgleichungen die Str"ome der Eichpotentiale, nicht aber die
Feldst"arken eingehen. Darum scheint es nat"urlich, f"ur die Wirkung
einen Ansatz zu w"ahlen,
der diese Tatsache von Beginn ber"ucksichtigt. Daf"ur d"urfen keine
Flu"sbeitr"age auftreten.

\subsubsection*{Ansatz f"ur die Wirkung}
Wir kommen zu dem Schlu"s, da"s unsere Wirkung
keine Punkt- oder Ringbeitr"age enthalten soll.
Damit d"urfen in \Ref{4_1} nur zwei verschiedene
Parameter $x,y \in M$ vorkommen; die zugeh"origen Projektoren
$E_x, E_y$ m"ussen immer abwechselnd auftreten. Die Wirkung mu"s
folglich aus den reellen Gr"o"sen
\[ \alpha^{(q)}_{xy} \;:=\; \tr \left( (P \:E_x\: P \:E_y)^q \right)
	\;\;\; {\mbox{mit $q \in \N$; $\;x, y \in M$}} \]
aufgebaut werden, die wir {\em{Linienbeitr"age}}\index{Linienbeitrag}
nennen. Beachte, da"s $\alpha^{(q)}_{xy} = \alpha^{(q)}_{yx}$.

Zur Konstruktion der Wirkung k"onnnen wir eine beliebige Funktion
der Linienbeitr"age bilden und anschlie"send "uber $x, y$ summieren.
Das f"uhrt auf Terme der Form
\Equ{4_11}
\sum_{x, y \in M} f(\alpha^{(1)}_{xy}, \alpha^{(2)}_{xy}, \ldots)
	\spc .
\EndEqu
Die Wirkung kann eine beliebige Funktion solcher Ausdr"ucke sein.

Um die Form der Wirkung weiter zu spezialisieren, wenden wir erneut
ein Analogieargument zur klassischen Feldtheorie an:
In der klassischen Feldtheorie ist die Wirkung als Integral "uber
eine Lagrangedichte gegeben
\[ S \;=\; \int L \: d^4x \spc . \]
In der diskreten Raumzeit entspricht dem Integral eine Summe "uber $M$.
Darum sollte unsere Wirkung eine "au"sere Summe "uber $M$ enthalten.
Der Term \Ref{4_11} ist von dieser Form; diese Eigenschaft geht aber
i.a. verloren, sobald wir Funktionen von Ausdr"ucken der Form \Ref{4_11}
bilden.
Deswegen setzen wir einfach
\Equ{4_12}
S \;=\; \sum_{x, y \in M} \: L(\alpha^{(1)}_{xy}, \alpha^{(2)}_{xy}, \ldots)
\EndEqu
und nennen $L$ die {\em{Lagrangedichte}}\index{Lagrangedichte} des Systems.
Im Gegensatz zur klassischen Lagrangedichte h"angt sie von zwei
Raumzeit-Punkten $x,y$ ab\footnote{Um die Analogie zur klassischen
Feldtheorie besser zu wahren, sollten wir \Ref{4_12} in der Form
\[ S \;=\; \sum_{x \in M} \: \left[ \sum_{y \in M} \:
	 L(\alpha^{(1)}_{xy}, \alpha^{(2)}_{xy}, \ldots) \right] \]
umschreiben und den Ausdruck in eckigen Klammern als Lagrangedichte
bezeichnen. Diese Notation w"are f"ur unser weiteres Vorgehen aber nicht
zweckm"a"sig.}.

Gleichung \Ref{4_12} ist der gesuchte Ansatz f"ur die Wirkung.
Nat"urlich war unsere Ableitung nicht mathematisch streng. Sie
war auch nicht in dem Sinne zwingend, da"s wir \Ref{4_12} als
den einzig erfolgversprechenden Ansatz f"ur die Wirkung bezeichnen
k"onnten. Wir haben lediglich beschrieben, welche "Uberlegungen auf
\Ref{4_12} f"uhren. Ob dieser Ansatz physikalisch sinnvoll ist,
kann erst eine genauere mathematische Analyse zeigen.

\subsubsection*{die Euler-Lagrange-Gleichungen}
Wir leiten die
Euler-Lagrange-Gleichungen\index{Euler-Lagrange-Gleichungen}
der Wirkung \Ref{4_12} ab:
Die Variation des fermionischen Projektors wird durch eine Schar
unit"arer Transformationen beschrieben
\[ P(\tau) \;=\; U(\tau) \:P\: U^{-1}(\tau) \spc . \]
In erster Ordnung in $\tau$ haben wir
\Equ{4_13}
\delta P \;=\; i [A, P]
\EndEqu
mit dem selbstadjungierten Operator $A=-i \dot{U}(0)$ (siehe auch
Seite \pageref{unit}).
Ferner haben wir
\[ \delta \alpha^{(q)}_{xy} \;=\; q \: \tr \left(
	(P \:E_x\: P \:E_y)^{q-1} \: \delta(P \:E_x\: P \:E_y) \right) \]
und damit
\begin{eqnarray*}
\delta S &=& \sum_{x,y \in M} \: \sum_{q=1}^\infty \:
	\left( \frac{\partial}{\partial \alpha^{(q)}_{xy}}
	L(\alpha^{(1)}_{xy}, \alpha^{(2)}_{xy}, \ldots) \right) \;
	\delta \alpha^{(q)}_{xy} \\
&=&\sum_{x,y \in M} \: \sum_{q=1}^\infty \:
	\left( \frac{\partial}{\partial \alpha^{(q)}_{xy}}
	L(\alpha^{(1)}_{xy}, \alpha^{(2)}_{xy}, \ldots) \right) \\
&&\hspace*{1cm} \times \; q \: \tr \left(
	(P \:E_x\: P \:E_y)^{q-1} \: ((\delta P) \:E_x\: P \:E_y
	\:+\: P \:E_x\: (\delta P) \:E_y) \right) \spc .
\end{eqnarray*}
Wir wenden die Relation $\alpha^{(q)}_{xy} = \alpha^{(q)}_{yx}$,
Gleichung \Ref{4_13} und die zyklische Invarianz der Spur an
\begin{eqnarray}
&=& 2 \: \sum_{x,y \in M} \: \sum_{q=1}^\infty \:
	q \: \left( \frac{\partial}{\partial \alpha^{(q)}_{xy}}
	L \right) \; \tr \left(
	(P \:E_x\: P \:E_y)^{q-1} \: (\delta P) \:E_x\: P \:E_y
	\right) \nonumber \\
&=& 2i \: \sum_{x,y \in M} \: \sum_{q=1}^\infty \:
	q \: \left( \frac{\partial}{\partial \alpha^{(q)}_{xy}}
	L \right) \; \tr \left(
	(P \:E_x\: P \:E_y)^{q-1} \: [A, P] \:E_x\: P \:E_y
	\right) \nonumber \\
&=& 2i \: \sum_{x,y \in M} \: \sum_{q=1}^\infty \:
	q \: \left( \frac{\partial}{\partial \alpha^{(q)}_{xy}}
	L \right) \; \tr \left( A \: \left[ P, \:E_x\: P \:E_y \:
	(P \:E_x\: P \:E_y)^{q-1} \right] \right) \nonumber \\
\label{eq:4_14}
&=& 2i \; \tr \left( A \: [P, \: Q] \right) \spc ,
\end{eqnarray}
dabei ist $Q$ der Operator
\begin{eqnarray}
\label{eq:4_15}
Q &=& \sum_{x,y \in M} \: \sum_{q=1}^\infty \: q \:
	\left( \frac{\partial}{\partial \alpha^{(q)}_{xy}}
	L(\alpha^{(1)}_{xy}, \alpha^{(2)}_{xy}, \ldots) \right) \;
	(E_x \:P\: E_y \:P)^{q-1} \: E_x \:P\: E_y \;\;\; .
\end{eqnarray}
Damit die Euler-Lagrange-Gleichungen erf"ullt sind, mu"s die Variation
der Wirkung verschwinden, also $\delta S=0$.
Da $A$ in \Ref{4_14} ein beliebiger selbstadjungierter Operator sein
kann, folgt die Kommutatorgleichung
\Equ{4_16}
	[ P, \: Q] \;=\; 0 \spc .
\EndEqu
\Ref{4_16}, \Ref{4_15} ist unser Ansatz f"ur die Gleichungen der diskreten
Raumzeit.

\section{Analyse des Kontinuumslimes}
\label{4_ab2}
Um zu verstehen, welche klassische Dynamik das Variationsprinzip
mit Wirkung \Ref{4_12} beschreibt, m"ussen wir den Kontinuumslimes
studieren.

Wir vermeiden von nun an in allen Formeln die Projektoren $E_x$
und verwenden anstatt dessen f"ur einen Operator $A$ die
Matrixschreibweise
\[ A(x,y) \;\equiv\; E_x \:A\: E_y \spc . \]
Wir wissen nach Kapitel \ref{kap2}, da"s $P(x,y)$ im Kontinuumslimes
in eine wohldefinierte Distribution "ubergeht. Bei zusammengesetzten
Ausdr"ucken wie \Ref{4_12}, \Ref{4_15} ist zun"achst nicht klar,
ob und wie der Kontinuumslimes gebildet werden kann. Deshalb
regularisieren wir den fermionischen Projektor des Kontinuums,
setzen $P^\varepsilon(x,y)$ in \Ref{4_12}, \Ref{4_15} ein und
untersuchen den Grenzwert
$\varepsilon \rightarrow 0$. Damit dieses Vorgehen sinnvoll ist,
darf die genaue Art der Regularisierung nicht in die Endergebnisse
eingehen.

Die regularisierte Wirkung hat die Form
\begin{eqnarray}
\label{eq:4_20}
S^\varepsilon &=& \int d^4x \int d^4y \; L(\alpha^{(1, \varepsilon)}_{xy},
	\alpha^{(2, \varepsilon)}_{xy}, \ldots) \spc {\mbox{mit}} \\
\alpha^{(q,\varepsilon)}_{xy} &=& \Tr \left( (P^\varepsilon(x,y) \:
	P^\varepsilon(y,x))^q \right) \spc , \nonumber
\end{eqnarray}
die zugeh"origen Euler-Lagrange-Gleichungen lauten
\begin{eqnarray}
\label{eq:4_21}
[P^\varepsilon, \: Q^\varepsilon] &=& 0 \spc {\mbox{mit}} \\
\label{eq:4_22}
Q^\varepsilon(x,y) &=& \sum_{q=1}^\infty \: \left(
	\frac{\partial}{\partial \alpha^{(q,\varepsilon)}_{xy}}
	L(\alpha^{(1,\varepsilon)}_{xy}, \alpha^{(2,\varepsilon)}_{xy},
	\ldots) \right) \:q\:
	(P^\varepsilon(x,y) \: P^\varepsilon(y,x))^{q-1}
	\; P^\varepsilon(x,y) \; . \spc
\end{eqnarray}

Es scheint nicht m"oglich zu sein, eine direkte Beziehung zwischen
dem Kontinuumslimes von \Ref{4_20} und einer klassischen Wirkung
herzustellen. Man erh"alt zwar in diesem Grenzfall einen Ausdruck
in den Tensoren der klassischen Feldtheorie; die bei Variation der
klassischen Felder erhaltenen Euler-Lagrange-Gleichungen stimmen aber
nicht mit dem Koninuumslimes von \Ref{4_21} "uberein.
Das liegt daran, da"s bei der Variation des fermionischen Projektors
die Nebenbedingungen $P^*=P^2=P$ zu ber"ucksichtigen sind, welche
nicht unmittelbar in
Nebenbedingungen bei Variation der klassischen Felder "ubersetzt werden
k"onnen.

Aus diesem Grunde m"ussen wir den Kontinuumslimes der
Eu\-ler-La\-grange-Glei\-chun\-gen \Ref{4_21}, \Ref{4_22} untersuchen.

\subsubsection*{Ordnen nach Homogenit"aten}
Unter der Annahme, da"s $L$ eine analytische Funktion ist (was aus
physikalischer Sicht keine wesentliche Einschr"ankung darstellt),
k"onnen wir die Lagrangedichte in einer Taylorreihe entwickeln.
Man erh"alt
\begin{eqnarray}
\label{eq:4_23}
S^\varepsilon &=& \int d^4x \int d^4y \; \sum_{r=1}^\infty
	\sum_{\{p\}_r} \: c_{\{p\}} \; \left( \prod_{i=1}^r
	\alpha^{(p_i, \varepsilon)}_{xy} \right) \\
\label{eq:4_24}
Q^\varepsilon(x,y) &=& \sum_{q=1}^\infty \sum_{r=0}^\infty
	\sum_{\{p\}_r} \: c^{(q)}_{\{p\}} \; \left( \prod_{i=1}^r
	\alpha^{(p_i, \varepsilon)}_{xy} \right) \;
	(P^\varepsilon(x,y) \: P^\varepsilon(y,x))^{q-1} \;
	P^\varepsilon(x,y)
\end{eqnarray}
mit reellen Koeffizienten $c_{\{p\}}$, $c^{(q)}_{\{p\}}$; die Summe
$\sum_{\{p\}_r}$ durchl"auft alle Konfigurationen der Parameter
$p_1, \ldots, p_r$ mit $1 \leq p_1 \leq \cdots \leq p_r$.
Beachte, da"s man die Koeffizienten $c^{(q)}_{\{p\}}$ durch
$c_{\{p\}}$ ausdr"ucken kann, indem man die Taylorreihe f"ur $L$
partiell nach $\alpha^{(q,\varepsilon)}_{xy}$ ableitet.
Genauer gilt
\Equ{4_7a}
c^{(p_s)}_{\{p_1, \ldots, \widehat{p_s}, \ldots, p_r\}} \;=\;
	n(p_s) \: c_{\{p_1, \ldots, p_r\}} \spc ,
\EndEqu
dabei bedeutet $\widehat{p_s}$, da"s wir den Parameter $p_s$ aus
$\{p_1, \ldots, p_r\}$ herausnehmen; $n(p_s)$ gibt an, wie oft
$p_s$ in $p_1, \ldots, p_r$ vorkommt.
Folglich sind die Koeffizienten $c^{(q)}_{\{p\}}$ nicht voneinander
unabh"angig, sondern gen"ugen den Relationen
\Equ{4_8a}
\frac{1}{n(p_s)} \: c^{(p_s)}_{\{p_1, \ldots, \widehat{p_s}, \ldots, p_t,
	\ldots, p_r\}} \;=\;
   \frac{1}{n(p_t)} \: c^{(p_t)}_{\{p_1, \ldots, p_s, \ldots,
	\widehat{p_t}, \ldots, p_r\}} \spc .
\EndEqu
Wir verschieben die systematische Untersuchung der Beziehungen
\Ref{4_7a}, \Ref{4_8a} auf Abschnitt \ref{4_ab6}.

Der formale Kontinuumslimes von \Ref{4_23}, \Ref{4_24} enth"alt
Produkte von Distributionen vom Typ \Ref{3_1}.
Damit ist nach den Ergebnissen von Kapitel \ref{kap3} der
Limes $\varepsilon \rightarrow 0$ sinnvoll durchf"uhrbar.
Wir k"onnen in \Ref{4_23}, \Ref{4_24}
die Indizes $\varepsilon$ weglassen und meinen damit gem"a"s der Notation von
Abschnitt \ref{3_sec_zus} einen Ausdruck, der nach Regularisierung und
asymptotischer Entwicklung als Distribution wohldefiniert ist.
F"ur Umformungen k"onnen wir alle in Abschnitt \ref{3_sec_zus}
zusammengestellten Rechenregeln verwenden.

Wir m"ussen noch "uberlegen, wie in Gleichung \Ref{4_21} der Grenzwert
$\varepsilon \rightarrow 0$ durchgef"uhrt werden kann: Wir untersuchen
den Kommutator im schwachen Sinne, betrachten also den Ausdruck
\begin{eqnarray*}
\lefteqn{ \int d^4x \int d^4y \; \left[ P^\varepsilon, \:
	Q^\varepsilon \right](x,y) \; f(x) \: g(y) } \\
&=& \int d^4x \int d^4y \int d^4z \; \left( P^\varepsilon(x,z) \:
	Q^\varepsilon(z, y) \:-\: Q^\varepsilon(x,z) \:
	P^\varepsilon(z, y) \right) \; f(x) \: g(x)
\end{eqnarray*}
mit beliebigen Schwartzfunktionen $f, g$. Nach Umordnen
der Integrale
\begin{eqnarray*}
&=& \int d^4z \; \left(
	\left( \int d^4x \; f(x) \: P^\varepsilon(x,z) \right)
	\left( \int d^4y \; Q^\varepsilon(z,y) \: g(y) \right) \right. \\
&&\hspace*{3cm} \left. \:-\:
	\left( \int d^4x \; f(x) \: Q^\varepsilon(x,z) \right)
	\left( \int d^4y \; P^\varepsilon(z,y) \: g(y) \right) \right)
\end{eqnarray*}
k"onnen wir die Integration "uber $x, y$ ausf"uhren. Als Ergebnis erhalten
wir glatte Funktionen in $z$, und wir k"onnen den Limes $\varepsilon
\rightarrow 0$ bilden. Wir k"onnen sogar bei
$P^\varepsilon$ und $Q^\varepsilon$ getrennt den Grenzwert
$\varepsilon \rightarrow 0$ durchf"uhren und erhalten das gleiche
Ergebnis.
Mit anderen Worten k"onnen wir zun"achst $Q$ asymptotisch entwickeln
und anschlie"send den Kommutator $[P, Q]$ im Distributionssinne
bilden\footnote{Wir bemerken, da"s der Integralkern $[P, Q](x,y)$
(nach asymptotischer Entwicklung) sogar eine regul"are Funktion ist.
Um das zu sehen, mu"s man die Beitr"age der asymptotischen Entwicklung
genauer im Impulsraum analysieren, was f"ur unser weiteres Vorgehen
aber nicht ben"otigt wird.}.
Wir schreiben f"ur den so definierten Kontinuumslimes der 
Euler-Lagrange-Gleichungen auch einfach
\Equ{4_0c}
[P, \: Q] \;=\; 0 \spc .
\EndEqu

Nach \Ref{3_84} wird die Singularit"at von $Q$ auf dem Lichtkegel und
am Ursprung
bei steigender Potenz in $P(x,y)$ st"arker. Mit den asymptotischen
Rechenregeln k"onnen nur solche Ausdr"ucke sinnvoll (also unabh"angig von
der Regularisierung) miteinander in Beziehung gesetzt werden, welche
das gleiche Polverhalten in $\varepsilon$ zeigen.
Deshalb scheint es sinnvoll, die Distributionsprodukte nach Potenzen
in $P(x,y)$ zu ordnen.
Dazu definieren wir $|\{p\}_r|=p_1+\cdots+p_r$ und setzen
\begin{eqnarray}
\label{eq:4_25}
S &=& \sum_{g=1}^\infty S^{[g]} \;\;\;\;,\spc Q \;=\; \sum_{g=0}^\infty Q^{[g]}
	\spc {\mbox{mit}} \\
\label{eq:4_26}
S^{[g]} &=& \int d^4x \int d^4y \; \sum_{r=1}^g \;
	\sum_{\{p\}_r {\mbox{\scriptsize{ mit }}} |\{p\}_r|=g} \;
	c_{\{p\}} \; \prod_{i=1}^r \alpha^{(p_i)}_{xy} \\
\label{eq:4_27}
Q^{[g]}(x,y) &=& \sum_{q=1}^g \sum_{r=0}^g \;
	\sum_{\{p\}_r {\mbox{\scriptsize{ mit }}} |\{p\}_r|=g-q}
	\!\!\!\!\!\!
	c^{(q)}_{\{p\}} \; \left( \prod_{i=1}^r
	\alpha^{(p_i)}_{xy} \right) \;
	(P(x,y) \: P(y,x))^{q-1} \; P(x,y) \spc \\
\label{eq:4_28}
\alpha^{(q)}_{xy} &=& \Tr \left( (P(x,y) \: P(y,x))^q \right) \spc .
\end{eqnarray}
Wir nennen die Darstellungen \Ref{4_25}
{\em{Homogenit"atsreihen}}\index{Homogenit"atsreihe}.

\subsubsection*{Spektrale Analyse von $P(x,y) \: P(y,x)$}
\label{4_spk}
In Kapitel \ref{kap2} und den Anh"angen A bis E
wurde die Distribution $P(x,y)$ f"ur verschiedene St"orungen des
Diracoperators explizit berechnet.
Wir m"ussen eine Methode finden, mit der sich
die Auswirkung der einzelnen St"orbeitr"age von $P(x,y)$ in
dem Ausdruck f"ur $Q^{[g]}$, \Ref{4_27},
beschreiben l"a"st.

Im Prinzip k"onnte man dazu den gest"orten fermionischen Projektor
in \Ref{4_27}, \Ref{4_28} einsetzen und die einzelnen Beitr"age von $P(x,y)$
ausmultiplizieren. Dieses Verfahren ist aus theoretischer Sicht
v"ollig unproblematisch. Die Rechnungen werden aber wegen der
nicht-kommutierenden Dirac- und Pauli-Matrizen in unseren Formeln
f"ur $P(x,y)$ zu kompliziert und un"ubersichtlich, besonders
bei hohen Potenzen $p_i, q$.

Darum wenden wir eine andere Methode an: Wenn wir $x, y$ als feste
Parameter ansehen, kommen in \Ref{4_27}, \Ref{4_28} Polynome
in der $(4n \times 4n)$-Matrix $P(x,y) \: P(y,x)$ vor. Mit einer
Spektralzerlegung von $P(x,y) \: P(y,x)$
\Equ{4_29}
P(x,y) \: P(y,x) \;=\; \sum_j \lambda_j(x,y) \: E_j(x,y)
\EndEqu
mit Eigenwerten $\lambda_j$ und Spektralprojektoren $E_j$ lassen 
sich diese Polynome in Polynome in den Eigenwerten umschreiben
\begin{eqnarray}
\label{eq:4_30}
Q^{[g]}(x,y) &=& \sum_j \sum_{q=1}^g \: f^{(q,g)}_{xy} \: (\lambda_j(x,y))^{q-1} \;
	\left( E_j(x,y) \: P(x,y) \right) \spc {\mbox{mit}} \\
\label{eq:4_5a}
f^{(q,g)}_{xy} &=& \sum_{r=0}^g \:
	\sum_{\{p\}_r {\mbox{\scriptsize{ mit }}} |\{p\}_r|=g-q}
	c^{(q)}_{\{p\}} \; \prod_{i=1}^r \alpha^{(p_i)}_{xy} \\
\label{eq:4_5b}
\alpha^{(p)}_{xy} &=& \sum_j n_j(x,y) \: (\lambda_j(x,y))^p \spc ,
\end{eqnarray}
dabei ist $n_j = {\mbox{dim Im}} (E_j)$ die Vielfachheit der Eigenwerte.
Nun besteht $Q^{[g]}(x,y)$ aus einem Polynom in $\lambda_j$ vom Grade $g$.
Die Koeffizienten $f^{(q)}_{xy}$ sind ebenfalls polynomial aus den
Eigenwerten von $P(x,y) \: P(y,x)$ zusammengesetzt, und zwar so,
da"s $Q^{[g]}(x,y)$ in den $\lambda_j$ homogen vom Grade $g$ ist.
Der Matrixcharakter von $Q^{[g]}(x,y)$ wird durch den Faktor
$E_j(x,y) \: P(x,y)$ in \Ref{4_30} beschrieben.

Damit hat sich die Struktur der Gleichungen wesentlich vereinfacht.
Zun"achst einmal k"onnen wir mit (elementaren) algebraischen Methoden
Aussagen "uber die Koeffizienten $c^{(q)}_{\{p\}}$ gewinnen.
Wenn wir beispielsweise verlangen, da"s $Q^{[g]}$ im Fall ohne Entartung
der Eigenwerte verschwindet, mu"s das charakteristische Polynom der Matrix
$P(x,y) \: P(y,x)$ das Polynom
\Equ{4_28a}
{\cal{P}}_{xy}(\lambda) \;=\; \sum_{q=1}^g f^{(q,g)}_{xy} \: \lambda^{q-1}
\EndEqu
teilen, was sich unmittelbar in Bedingungen an die Parameter
$c^{(q)}_{\{p\}}$ umschreiben l"a"st.
Au"serdem kann man das reelle Polynom in Gleichung \Ref{4_30} 
leicht nach verschiedenen Parametern entwickeln.

Wir m"ussen pr"azisieren, wie die Spektralzerlegung \Ref{4_29}
mathematisch zu verstehen ist: Es macht sicher keinen Sinn,
$P(x,y) \: P(y,x)$ punktweise (also f"ur festes $x, y$) zu diagonalisieren,
auch wenn diese Vorstellung f"ur qualitative "Uberlegungen sehr
hilfreich ist.
Denn die Matrix $P(x,y) \: P(y,x)$ ist erst nach
asymptotischer Entwicklung als Distribution definiert.
Selbst mit Regularisierung gibt es Schwierigkeiten, weil die s.a.
Matrix $P^\varepsilon(x,y) \:P^\varepsilon(y,x)$ wegen des
indefiniten Skalarproduktes nicht diagonalisierbar zu sein braucht.
Darum werden wir die Eigenwerte und Spektralprojektoren lediglich als
formale Ausdr"ucke berechnen, denen wir keinen mathematischen Sinn
geben. Nach Einsetzen in \Ref{4_30} erh"alt man jedoch f"ur $Q^{[g]}$
einen Ausdruck, der nach der Methode von Kapitel \ref{kap3} wohldefiniert
ist. Dieses Vorgehen ist unproblematisch und f"ur unsere Zwecke v"ollig
ausreichend, weil die
Spektralzerlegung von $P(x,y) \: P(y,x)$ nur ein technisches
Hilfsmittel ist, um das Verhalten des Operators $Q^{[g]}$
bei St"orungen des fermionischen Projektors effizienter berechnen zu
k"onnen.

In Anhang F werden explizite Formeln f"ur die Eigenwerte und
Spektralprojektoren von $P(x,y) \: P(y,x)$ hergeleitet.
Insbesondere wird die Auswirkung der einzelnen
St"orbeitr"age von $P(x,y)$ auf $\lambda_j, E_j$ genau untersucht.
Dort werden auch die gerade angesprochenen mathematischen
Schwierigkeiten ausf"uhrlicher diskutiert.
In den folgenden Abschnitten werden wir einzelne Ergebnisse aus
Anhang F verwenden und gleichzeitig genauer erkl"aren.

\section{Systeme mit einer Fermionsorte}
\label{4_ab3}
In den vorangehenden Abschnitten \ref{4_ab1}, \ref{4_ab2}
haben wir mit \Ref{4_12}, \Ref{4_16}, \Ref{4_15} einen Ansatz f"ur
die Gleichungen der diskreten Raumzeit
abgeleitet und die allgemeine Methode beschrieben, mit welcher
der Kontinuumslimes dieser Gleichungen untersucht werden kann.
Die Lagrangedichte ist in \Ref{4_12} oder nach Taylorentwicklung gem"a"s
\Ref{4_23} aber noch unbestimmt, und wir haben im Moment keine Vorstellung
davon, wie eine sinnvolle Lagrangedichte aussehen sollte.
Um den Zusammenhang zwischen
der Form der Lagrangedichte und der Dynamik des Systems zu verstehen,
wollen wir nun konkrete Modelle diskutieren.

Wir beginnen mit dem einfachsten Beispiel,
n"amlich einem fermionischen Projektor, der lediglich aus einem
Diracsee aufgebaut ist. Die Spindimension ist 4. Unsere "Uberlegung
l"a"st sich direkt auf den Fall mehrerer Teilchenfamilien
(also mehreren Diracseen im gleichen $(4 \times 4)$-Block)
"ubertragen.

Nat"urlich sind erst bei h"oherer Spindimension physikalisch interessante
Wechselwirkungen zu erwarten. Als Vorbereitung auf realistischere
Modelle ist das Studium von Systemen bei Spindimension $4$ trotzdem
sinnvoll, besonders weil der freie Projektor im allgemeinen Fall eine direkte
Summe solcher $(4 \times 4)$-Bl"ocke ist.

\subsection{Massive Fermionen}
\label{4_ab31}
Im Vakuum beschreibt der fermionische Projektor einen vollst"andig
gef"ullten Diracsee. Bei Fermionen mit Ruhemasse haben wir also mit der
Bezeichnung von Definition \ref{2_def1}
\Equ{4_a}
	P(x,y) \;=\; \frac{1}{2} \: (p_m - k_m)(x,y) \spc ,
\EndEqu
dabei ist $m$ die (nackte) Masse der Fermionen. Den Fall mit Wechselwirkung
erh"alt man hieraus, indem man einzelne Fermionen hinzuf"ugt bzw.
aus dem Diracsee entfernt und anschlie"send $P$ einer unit"aren
Transformation unterwirft       .

\subsubsection*{die Bedingung $Q(x,y) \simeq 0$}
\label{4_qsim0}
Wir beginnen mit der Untersuchung des freien Projektors, was
uns bis zu Seite \pageref{4_dyn} besch"aftigen wird.

Zun"achst wollen wir begr"unden, weswegen die Lagrangedichte
so gew"ahlt werden mu"s, da"s nicht nur die Euler-Lagrange-Gleichung
\Ref{4_0c}, sondern sogar die st"arkere Bedingung
\Equ{4_41}
	Q(x,y) \;\simeq\; 0
\EndEqu
erf"ullt ist.

Einen ersten Hinweis auf diese Forderung erhalten wir durch direkte
Berechnung des Kommutators $[P,Q]$ im Vakuum. Diese Rechnung ist nicht
ganz unproblematisch, weil die Regularisierung explizit eingeht, wir k"onnen
sie aber trotzdem erkl"aren:
Wir nehmen an, da"s Bedingung \Ref{4_41} verletzt ist. Bei asymptotischer
Entwicklung von $Q^\varepsilon(x,y)$ erh"alt man dann typischerweise
Ausdr"ucke der Form
\begin{eqnarray}
\label{eq:4_42}
f(x,y) &=& \frac{1}{\varepsilon^p} \: \delta((y-x)^2) \; h(y-x) \\
\label{eq:4_43}
g(x,y) &=& \frac{1}{\varepsilon^p} \: \delta((y-x)^2) \; h(y-x) \;
	(y-x)^j \gamma_j \spc , \: p \in \N
\end{eqnarray}
mit einer auf $M \setminus \{0\}$ stetigen Funktion, die homogen vom
Grade $-q$ ist
\[ h(\lambda z) \;=\; \lambda^{-q} \: h(z) \spc . \]
Der Faktor $\varepsilon^{-p} \: \delta((y-x)^2)$ und die Funktion $h$
beschreiben die Singularit"at auf dem Lichtkegel bzw. die Singularit"at
am Ursprung.
Man beachte, da"s die Form von $h$ wesentlich von der gew"ahlten
Regularisierung abh"angt, und da"s \Ref{4_42}, \Ref{4_43}
keine lorentzinvarianten Ausdr"ucke sind. Bei Fouriertransformation
von \Ref{4_42} erh"alt man eine regul"are Funktion $\tilde{f}(k)$,
die ebenfalls die Lorentzsymmetrie verletzt.
Der zus"atzliche Faktor $(y-x)^j \gamma_j$ in \Ref{4_43} "ubersetzt
sich im Impulsraum in den Ableitungsoperator $i \Pdd_k$, also
$\tilde{g}(k) = i \Pdd_k \tilde{f}(k)$.
Mit dem Ausdruck
\[ P(k) \;=\; (k \slsh + m) \; \delta(k^2 - m^2) \: \Theta(-k^0) \]
f"ur den freien fermionischen Projektor folgt
\[ \widetilde{[P, g]}(k) \;=\; \left[ i \Pdd_k \tilde{f}(k), \:
	k \slsh \right] \; \delta(k^2 - m^2) \: \Theta(-k^0) \spc . \]
Der Kommutator auf der rechten Seite verschwindet nicht, weil der Vierervektor
$\partial_k \tilde{f}(k)$ wegen der gebrochenen Lorentzsymmetrie i.a.
nicht parallel zu $k$ ist\footnote{Bei Wahl einer speziellen
Regularisierung kann man diese Rechnung explizit durchf"uhren.
F"ur $P^\varepsilon=P * \eta^\varepsilon$ mit rein zeitabh"angigem
$\eta$ hat man beispielsweise
\[ h(z) \;=\; (z^0)^{-q} \;\;\; {\mbox{oder}} \spc h(z) \;=\; (z^0)^{-q} \:
	\epsilon(q^0) \spc . \]
Man beachte, da"s vor der Fouriertransformation die
Singularit"at am Ursprung zus"atzlich regularisiert werden mu"s.}.

Ein eleganteres Argument f"ur die Notwendigkeit von Bedingung
\Ref{4_41} erhalten wir bei der Betrachtung eines zus"atzlichen
$U(1)$-Eichfeldes. Die "Uberlegung hat "Ahnlichkeit mit der
Diskussion der Ringbeitr"age auf Seite \pageref{4_ring}; wir verwenden
auch die gleiche Notation:
Wir schreiben zun"achst die Euler-Lagrange-Gleichungen mit
Integralkernen um
\Equ{4_44}
0 \;=\; [P, Q](x,y) \;=\; \int d^4z \;
	\left( P(x,z) \: Q(z,y) \:-\:
	Q(x,z) \: P(z,y) \right) \spc .
\EndEqu
Bei einer lokalen Eichtransformation mit Potential $A_j = \partial_j \Lambda$
wird die Phase von \Ref{4_44} transformiert
\[ [P, Q](x,y) \;\longrightarrow\;
	e^{-i (\Lambda(y) - \Lambda(x))} \; [P, Q](x,y) \spc . \]
Im Fall eines allgemeinen Eichpotentials $A$ sind bei asymptotischer
Entwicklung die Eichterme dominant, unter denen sich $P, Q$ gem"a"s
\[ P(x,y) \:\longrightarrow\: e^{-i \int_x^y A_j \: (y-x)^j}
	\: P(x,y) \;\;\; , \spc
   Q(x,y) \:\longrightarrow\: e^{-i \int_x^y A_j \: (y-x)^j}
	\: Q(x,y) \]
verh"alt. Einsetzen in \Ref{4_44} liefert
\begin{eqnarray}
\lefteqn{ [P, Q](x,y) \;\longrightarrow\;
	e^{-i \int_x^y A_j \: (y-x)^j} } \nonumber \\
\label{eq:4_45}
&&\times \;
	\int d^4z \; e^{-i \int_{\partial \Delta} A_j \: dx^j} \;
	\left( P(x,z) \: Q(z,y) \:-\:
	Q(x,z) \: P(z,y) \right) \spc , \spc
\end{eqnarray}
wobei $\Delta$ das Dreieck mit Ecken $x, y, z$ bezeichnet.
Das Integral "uber $\partial \Delta$ gibt nach dem Satz von Stokes
den Flu"s durch das Dreieck $\Delta$ an. Wenn wir annehmen, da"s
\Ref{4_41} verletzt ist, sind die Integranden in \Ref{4_44}, \Ref{4_45}
nach asymptotischer Entwicklung nicht null.
Durch den zus"atzlichen Flu"sfaktor ger"at der Integrand
in \Ref{4_45} gegen"uber \Ref{4_44} au"ser Phase.
Deshalb verschwindet das Integral in \Ref{4_45} i.a. nur dann, wenn es keinen
Flu"s durch die Dreiecke $\Delta$ gibt. Folglich kann das $U(1)$-Feld
global weggeicht werden und beschreibt keine Dynamik. Das ist
physikalisch nicht sinnvoll.

Wir erw"ahnen ein weiteres Argument f"ur Bedingung \Ref{4_41}.
Es nimmt qualitativ die Methode vorweg, mit der wir
sp"ater den Zusammenhang zu klassischen Feldgleichungen
herstellen werden: \label{4_warg}
Die klassischen Feldgleichungen sind lineare Gleichungen in den
Noether- und Diracstr"omen sowie dem Energie-Impuls- und
Kr"ummungstensor. Da die Euler-Lagrange-Gleichungen
\Ref{4_0c} im Kontinuumslimes die
klassischen Feldgleichungen liefern sollen, erwarten wir, da"s
die Beitr"age der klassischen Tensoren zu $P, Q$ in linearer
St"orungstheorie
behandelt werden k"onnen\footnote{Wir werden in Abschnitt
\ref{4_ab5} zeigen, da"s diese perturbative Behandlung tats"achlich
zul"assig ist.}.
Eine Entwicklung der Euler-Lagrange-Gleichungen liefert
\Equ{4_49}
0 \;=\; [P, \Delta Q] \:+\: [\Delta P, Q] \spc ,
\EndEqu
dabei sind $P, Q$ die freien Operatoren und $\Delta P, \Delta Q$ die
Beitr"age der klassischen Tensoren.
Die St"orungsbeitr"age $\Delta P(x,y)$ zum fermionischen Projektor
wurden in den Anh"angen A-D berechnet.
Nach Anhang F sind auch die St"orungen $\Delta \lambda_j(x,y)$,
$\Delta E_j(x,y)$ der Eigenwerte und Spektralprojektoren explizit bekannt.
Damit kann $\Delta Q$ durch Entwicklung von \Ref{4_30} bestimmt
werden. Man erh"alt die beiden Beitr"age
\begin{eqnarray}
\label{eq:4_45a}
\Delta Q^{[g]} &=& \sum_j \sum_{q=1}^g \: \Delta \left( f^{(q,g)}_{xy} \:
	(\lambda_j(x,y))^{q-1} \right) \;
	\left( E_j(x,y) \: P(x,y) \right) \\
\label{eq:4_46a}
&&+\sum_j \sum_{q=1}^g \: \left( f^{(q,g)}_{xy} \: (\lambda_j(x,y))^{q-1} \;
	\right) \Delta \left( E_j(x,y) \: P(x,y) \right) \spc .
\end{eqnarray}
\Ref{4_45a} gibt die St"orung des Polynoms an und kann mit $\Delta \lambda_j$
ausgedr"uckt werden, \Ref{4_46a} h"angt dagegen von $\Delta E_j$,
$\Delta P(x,y)$ ab.
Wir f"uhren nun f"ur feste Parameter $x, y$ eine Dimensionsbetrachtung
durch. F"ur die Wahl der komplexen $(4 \times 4)$-Matrix $\Delta P(x,y)$
gibt es $2 \times 4 \times 4 = 32$ reelle Freiheitsgrade. Da $\Delta P(x,y)$
direkt in $\Delta(E_j P(x,y))$ eingeht, wird $\Delta(E_j P(x,y))$
ebenfalls durch 32 Parameter beschrieben. Bei den 4 Parametern
$\Delta \lambda_j$ gibt es dagegen nur 4 Freiheitsgrade
(beachte dazu, da"s $P(x,y) \: P(y,x)$ s.a. ist).
Folglich kann man die St"orung \Ref{4_45a} mit 4 Parametern beschreiben,
f"ur \Ref{4_46a} werden i.a. 32 Parameter ben"otigt.
Wir k"onnen nicht erwarten, da"s sich Beitr"age der beiden Summanden
in \Ref{4_49} gegenseitig kompensieren oder da"s von den Freiheitsgraden von
$\Delta P$, $\Delta Q$ bei Einsetzen in \Ref{4_49} einige wegfallen.
Folglich "ubersetzen sich in den Euler-Lagrange-Gleichungen alle Freiheitsgrade
in Bedingungen an die St"ormatrix $P(x,y)$ (und damit mittelbar in
Bedingungen an die St"orung des Diracoperators).
Es zeigt sich, da"s 32 Bedingungen f"ur sinnvolle Gleichungen zu viel
sind. (Insbesondere gibt es Probleme bei den Stromtermen, weil die
Terme der Form $\Delta P(x,y) \sim \xi^2 j_k \gamma^j,
j_k \xi^k \xi\slsh$ nicht miteinander in Beziehung gesetzt werden
k"onnen.)
Hieraus folgt zun"achst einmal, da"s die Matrix $\Delta(E_j P(x,y))$ nicht
in $\Delta Q$ eingehen darf. Dazu mu"s der Beitrag \Ref{4_46a} unabh"angig von
$\Delta(E_j P(x,y))$ verschwinden. Nach \Ref{4_25} bedeutet dies
\[ \sum_{g=1}^\infty \sum_{q=1}^g \: f^{(q, g)}_{xy} \:
	(\lambda_j(x,y))^{q-1} \;=\; 0 \spc {\mbox{f"ur alle $j$}} \spc . \]
Durch Einsetzen in \Ref{4_30} folgt Bedingung \Ref{4_41}.
Ist diese Bedingung aber erf"ullt, so verschwindet auch der
zweite Summand in \Ref{4_49}, so da"s in die
Euler-Lagrange-Gleichungen tats"achlich
nur die 4 Parameter $\Delta \lambda_j$ eingehen.

\subsubsection*{homogener Polynomansatz}
Wir kommen zu dem Schlu"s, da"s der freie fermionische
Projektor die Bedingung \Ref{4_41} erf"ullen mu"s. Wir wollen
allgemeiner untersuchen, was diese Bedingung "uber $Q$ aussagt
und dann einen konkreten Ansatz f"ur die Lagrangedichte machen.
Dazu argumentieren wir wieder qualitativ mit der Spektralzerlegung
\Ref{4_30} und halten $x, y$ fest: Im allgemeinen ist die Matrix
$P(x,y)$ invertierbar. Nach \Ref{4_25}, \Ref{4_30} impliziert damit
\Ref{4_41} die Bedingung
\[ \sum_j \; \left( \sum_{g=1}^\infty \: \sum_{q=1}^g f^{(q, g)}_{xy} \:
	\lambda_j^{q-1} \right) \: E_j \;=\; 0 \spc . \]
Nach den Eigenschaften $E_i E_j = \delta_{ij} E_i$ der Spektralprojektoren
folgt, da"s die Reihe
\Equ{4_50}
F(z) \;=\; \sum_{q=1}^\infty f^{(q)} \: z^{q-1} \;\;\;\; {\mbox{mit}} \spc
	f^{(q)} \;=\; \sum_{g=q}^\infty f^{(q, g)}_{xy}
\EndEqu
die Eigenwerte $\lambda_j$ als Nullstellen besitzt.
Au"ser diesen endlich vielen (in unserem Fall h"ochstens 4) Bedingungen
haben wir "uber die Funktion $F$ keinerlei Informationen.
Bei einfachen transzendenten Funktionen (z.B. $\exp$, $\log$,
trigonometrische oder hyperbolische Funktionen) scheint es nicht nat"urlich,
eine endliche Zahl von variablen Nullstellen zu fordern. Deswegen
machen wir f"ur $F$ einen Polynomansatz
\[ F(z) \;=\; \sum_{q=1}^h f^{(q)} \: z^{q-1} \spc
	{\mbox{mit $h \in \N$}} \spc . \]
Damit in \Ref{4_50} h"ochstens $(h-1)$-te Potenzen von $z$ auftreten, mu"s
in \Ref{4_30} und damit auch in \Ref{4_27} stets $q \leq h$ sein.
Am einfachsten kann man das erreichen, indem man die Homogenit"atsreihe
f"ur $Q$ nach dem $h$-ten Glied abbricht
\Equ{4_51}
	Q \;=\; \sum_{g=1}^h Q^{[g]} \spc .
\EndEqu
Bei asymptotischer Entwicklung sind in \Ref{4_51} die Summanden f"ur
gro"ses $g$ dominant\footnote{Man beachte, da"s dieser Schlu"s bei einer
unendlichen Reihe nicht m"oglich ist. Der Ausdruck
\[ \tanh \left( \Tr (P(x,y) \: P(y,x)) \right) \]
ist beispielsweise eine regul"are Funktion, obwohl die einzelnen
Glieder einer Potenzreihenentwicklung immer st"arkere Singularit"aten
auf dem Lichtkegel besitzen.}, und wir k"onnen gleich
\Equ{4_52}
	Q \;=\; Q^{[h]}
\EndEqu
setzen (zumindest, solange wir nur die h"ochsten Ordnungen der
Singularit"at auf dem Lichtkegel und am Ursprung untersuchen).
Da sich bei Variation der Wirkung die Potenz in $P(x,y) \: P(y,x)$
um eins erniedrigt, folgt
\Equ{4_53}
	S \;=\; S^{[h]} \spc .
\EndEqu
Wir nennen den Ansatz \Ref{4_53}, \Ref{4_52} {\em{homogenen
Polynomansatz}}\index{homogener Polynomansatz}.

Nat"urlich h"atten wir gleich in Abschnitt \ref{4_ab1} die Wirkung in
der Form \Ref{4_53} ansetzen k"onnen. Wir haben den etwas allgemeineren
Zugang gew"ahlt um herauszuarbeiten, da"s die aus Bedingung
\Ref{4_41} folgenden Nullstellenbedingungen an \Ref{4_50}
eine polynomiale Form von $L$
nahelegen, und da"s die asymptotischen Entwicklungsregeln schlie"slich
auf den homogenen Ansatz f"uhren.

\subsubsection*{die intrinsische Methode}
Mit dem homogenen Polynomansatz haben wir nur noch endlich
viele Parameter, um die Wirkung festzulegen, n"amlich den
Homogenit"atsgrad $h$ und die Koeffizienten $c_{\{p\}}$.
Allerdings gibt es (zumindest bei gro"sem $h$) sehr viele freie
Parameter. Es w"are aus theoretischer Sicht unbefriedigend oder
zumindest unsch"on, alle diese Parameter als empirische Gr"o"sen
aufzufassen und mit Informationen "uber das zu beschreibende
physikalische System zu fitten.
Darum schr"anken wir uns zur Bestimmung von $h$, $c_{\{p\}}$ mit
der folgenden Methode ein:
Den Homogenit"atsgrad $h \in \N$ geben wir empirisch vor.
Zum Berechnen der $c_{\{p\}}$ verwenden wir ausschlie"slich die Bedingung,
da"s die Euler-Lagrange-Gleichungen mathematisch sinnvoll sein sollen.
Mit ``mathematisch sinnvoll'' meinen wir, da"s die Gleichungen f"ur
den freien fermionischen Projektor erf"ullt sind und da"s man im
Kontinuumslimes partielle Differentialgleichungen in Potentialen und
Feldern erh"alt.
F"ur die Bestimmung der $c_{\{p\}}$ wollen wir aber keine physikalischen
Eigenschaften unseres Systems verwenden.
Insbesondere d"urfen die erwarteten Eichgruppen und
Wechwelwirkungen, Kopplungskonstanten und Ladungen nicht in die
Koeffizienten $c_{\{p\}}$ eingehen.
Wir nennen dieses Vorgehen {\em{in\-trin\-si\-sche
Methode}}\index{intrinsische Methode}.

Mit der intrinsischen Methode wird unser physikalisches System durch
den freien fermionischen Projektor und den Homogenit"atsgrad $h$ bereits
vollst"andig beschrieben. Die Koeffizienten $c_{\{p\}}$ k"onnen
("ahnlich wie Lagrangesche Multiplikatoren in der klassischen Physik)
als zun"achst unbestimmte Parameter angesehen werden.
Die Euler-Lagrange-Gleichungen liefern Bedingungen sowohl an
$c_{\{p\}}$ als auch an $P(x,y)$. Damit k"onnen die Koeffizienten
$c_{\{p\}}$ bestimmt werden; die Bedingungen an
$P(x,y)$ legen dann die Dynamik des Systems fest.

Durch die intrinsische Methode wird auch der Homogenit"atsgrad $h$
weitgehend fixiert: Die Anzahl der Koeffizienten $c_{\{p\}}$ nimmt mit
steigendem $h$ zu. W"ahlen wir $h$ zu klein, so gibt es nicht
gen"ugend Parameter, um mathematisch sinnvolle Gleichungen zu bilden.
Ist $h$ dagegen zu gro"s, so bleiben nach Erf"ullung aller mathematischer
Konsistenzbedingungen noch freie Parameter "ubrig, was durch die
intrinsische Methode ausgeschlossen wird.

\subsubsection*{ein Beispiel: Bestimmung von $S^{[3]}$}
\label{4_ex}
Wir wollen nun die intrinsische Methode auf den fermionischen Projektor
\Ref{4_a} anwenden und die Wirkung explizit berechnen.

Nach unserer bisherigen Diskussion haben wir als einzige Bedingung
f"ur den freien Projektor Gleichung \Ref{4_41} zu erf"ullen.
Damit m"ussen wir den kleinsten Homogenit"atsgrad $h$ und die
zugeh"origen Konstanten $c_{\{p\}}$ bestimmen, welche
\Ref{4_41} gen"ugen.
Es ist nicht zu erwarten, da"s diese Wirkung bereits auf mathematisch
sinnvolle Gleichungen (insbesondere auf  Differentialgleichungen in
Potentialen und Feldern) f"uhrt, denn dazu sind m"oglicherweise zus"atzliche,
bislang noch nicht untersuchte Bedingungen notwendig.
Wir wollen an diesem Beispiel lediglich die bisherigen Konstruktionen
und Methoden erl"autern.

F"ur die Berechnung von $h$, $c_{\{p\}}$ werden wir wieder mit den
Eigenwerten der Matrix $P(x,y) \: P(y,x)$ argumentieren.
Wie ab Seite \pageref{4_spk} beschrieben, ist dieses Verfahren aus
mathematischer Sicht nicht einwandfrei; unsere Rechnung l"a"st
sich aber mit geringem Mehraufwand auch mathematisch sauber durchf"uhren.
An der folgenden Rechnung wollen wir auch exemplarisch zeigen, wie mit Hilfe
der Spektralzerlegung \Ref{4_30} hergeleitete Ergebnisse nachtr"aglich
mathematisch gerechtfertigt werden k"onnen.

Mit der Notation \Ref{3_7} hat der freie Projektor \Ref{4_a} die Form
\Equ{4_60}
P(x,y) \;=\; (i \xi\slsh f(z) + g(z) \:|\: 1) \;\;\;, \spc z 
	\equiv \xi^2 \spc ,
\EndEqu
dabei sind $f, g$ Besselfunktionen mit Reihenentwicklung
\begin{eqnarray*}
f(z) &=& c_0 z^{-2} \:+\: c_2 m^2 z^{-1} \:+\: c_4 m^4 (\ln(|z|) + C_e)
	\:+\: \cdots \\
g(z) &=& c_1 m z^{-1} \:+\: c_3 m^3 (\ln(|z|) + C_e) \:+\: \cdots
\end{eqnarray*}
und geeigneten reellen Koeffizienten $c_k$. Durch Bildung der Adjungierten
(bzgl. des Spin\-ska\-lar\-pro\-duk\-tes) folgt
\[ P(y,x) \;=\; P(x,y)^* \;=\; (1 \:|\: -i \xi\slsh f(z) + g(z)) \]
und damit
\Equ{4_61}
P(x,y) \: P(y,x) \;=\; (i \xi\slsh f + g \:|\: -i \xi\slsh f + g) \spc .
\EndEqu
Nach den asymptotischen Rechenregeln m"ussen wir $\xi^+, \xi^-$
f"ur formale Rechnungen als verschiedene Vektoren ansehen, auch
wenn diese Vektoren au"serhalb des Lichtkegels (wo \Ref{4_61}
punktweise definiert ist) nat"urlich "ubereinstimmen.

Wir betrachten das orthogonale Komplement von $\xi^\pm$
\[ V \;:=\; \left\{ v \in M \;|\; \xi^+_j v^j = \xi^-_j v^j = 0 \right\}
	\spc . \]
$V$ ist zweidimensional. F"ur jedes $v \in V$ kommutiert die Matrix
$\rho v\slsh$ mit $\xi\slsh^\pm$
\[ \left[ \rho v\slsh, \: \xi\slsh^\pm \right] \;=\; \rho \: \left\{
	v\slsh, \: \xi\slsh^\pm \right\} \;=\; 2 \rho \: \xi_j^\pm
	v^j \;=\; 0 \]
und damit auch mit \Ref{4_61}. Folglich gibt es eine zweiparametrige
Schar unit"arer Transformationen, unter denen die Matrix $P(x,y) \: P(y,x)$
invariant ist
\[ P(x,y) \: P(y,x) \;=\; \exp \left( i \rho v\slsh \right) \;
	P(x,y) \: P(y,x) \;
	\exp \left( -i \rho v\slsh \right) \;\;\;, \spc v \in V \spc . \]
Da diese Schar au"serdem keine eindimensionalen invarianten Unterr"aume
besitzt
\[ \rho v\slsh \: \Psi \;\sim \Psi \;\; \forall v \in V \spc
{\mbox{impliziert}} \spc \Psi = 0 \;\;\; (\Psi \in \C^4) \spc , \]
m"ussen alle Eigenwerte wenigstens zweifach entartet sein.
\label{4_chent}
Wir nennen dies die {\em{chirale Entartung}}\index{chirale Entartung}
der Eigenwerte.

Aufgrund der chiralen Entartung besitzt \Ref{4_61} genau zwei
Eigenwerte $\lambda_{1\!/\!2}$, die zugeh"origen Eigenr"aume sind
zweidimensional (genau ein Eigenwert kann nicht sein, weil ansonsten
\Ref{4_61} ein Vielfaches der Einheitsmatrix w"are).
Die Funktion $F(z)$, \Ref{4_50}, vereinfacht sich mit dem homogenen
Polynomansatz zu
\Equ{4_62}
F(z) \;=\; \sum_{q=1}^h f^{(q,h)}_{xy} \: z^{q-1} \spc .
\EndEqu
Wie auf Seite \pageref{eq:4_50} allgemeiner erkl"art wurde, m"ussen
$\lambda_{1\!/\!2}$ Nullstellen dieses Polynoms sein.
Folglich mu"s der Homogenit"atsgrad $h \geq 3$ sein.
Nach der intrinsischen Methode m"ussen wir sogar $h=3$ w"ahlen.
Es folgt
\Equ{4_6b}
F(z) \;=\; (z-\lambda_1) (z-\lambda_2) \;=\; z^2 - (\lambda_1+\lambda_2)
	z + \lambda_1 \lambda_2
\EndEqu
und nach Koeffizientenvergleich mit \Ref{4_62}
\Equ{4_63}
f^{(3,3)}_{xy} = 1 \;\;\;, \spc f^{(2,3)}_{xy} = -(\lambda_1 + \lambda_2)
	\;\;\;, \spc f^{(1,3)}_{xy} = \lambda_1 \lambda_2 \spc .
\EndEqu
Die Gr"o"sen $\alpha^{(q)}_{\{p\}}$ berechnen sich mit Hilfe von
\Ref{4_5b} zu
\Equ{4_63a}
\alpha^{(0)}_{xy} = 2 \;\;\;, \spc \alpha^{(1)}_{xy} = 2 (\lambda_1
	+\lambda_2) \;\;\;, \spc \alpha^{(2)}_{xy} = 2
	(\lambda_1^2 + \lambda_2^2) \spc .
\EndEqu
Einsetzen von \Ref{4_63}, \Ref{4_63a} in \Ref{4_5a} liefert die Gleichungen
\begin{eqnarray*}
1 &=& f^{(3,3)}_{xy} \;=\; c^{(3)}_{\{\}} \: 4 \\
-(\lambda_1 + \lambda_2) &=& f^{(2,3)}_{xy} \;=\; c^{(2)}_{\{1\}} \:
	2 (\lambda_1 + \lambda_2) \\
\lambda_1 \lambda_2 &=& f^{(1,3)}_{xy} \;=\; c^{(1)}_{\{2\}} \:
	2 (\lambda_1^2 + \lambda_2^2)
	\:+\: c^{(1)}_{\{1, 1\}} \: 4 (\lambda_1 + \lambda_2)^2 \spc ,
\end{eqnarray*}
aus denen sich die Parameter $c^{(q)}_{\{p\}}$ bestimmen lassen:
\begin{eqnarray}
\label{eq:4_7c}
c^{(3)}_{\{\}} = \frac{1}{4} \spc
c^{(2)}_{\{1\}} = -\frac{1}{2} \spc
c^{(1)}_{\{2\}} = -\frac{1}{4} \spc
	c^{(1)}_{\{1,1\}} = \frac{1}{8}
\end{eqnarray}
F"ur den Operator $Q$ erhalten wir durch Einsetzen von \Ref{4_7c}
in \Ref{4_27} die explizite Formel
\begin{eqnarray}
\lefteqn{ Q(x,y) \;=\; Q^{[3]}(x,y) \;=\; \frac{1}{4} \:
	(P(x,y) \: P(y,x))^2 \: P(x,y) } \nonumber \\
\label{eq:4_64}
&&-\frac{1}{2} \: \alpha^{(2)}_{xy} \; P(x,y) \: P(y,x) \: P(x,y) \;+\;
	\frac{1}{8} \: \left( (\alpha^{(1)}_{xy})^2 - 2 \alpha^{(2)}_{xy}
	\right) \: P(x,y) \spc . \spc
\end{eqnarray}
Durch `Integration' der Konstanten $c^{(q)}_{\{p\}}$ gem"a"s \Ref{4_7a}
erh"alt man f"ur die Lagrangedichte
\begin{eqnarray}
\label{eq:4_65}
L(x,y) \;=\; L^{[3]}(x,y) &=& \frac{1}{12} \:
	\alpha^{(3)}_{xy} \:-\:\frac{1}{4} \: \alpha^{(2)}_{xy} \: \alpha^{(1)}_{xy}
	\:+\: \frac{1}{24} \: (\alpha^{(1)}_{xy})^3  \spc . \spc
\end{eqnarray}
Man kann direkt verifizieren, da"s die Wirkung
\[ S \;=\; S^{[3]} \;=\; \int d^4x \int d^4y \; L^{[3]}(x,y) \]
mit Lagrangedichte \Ref{4_65} bei Variation tats"achlich auf $Q$
gem"a"s \Ref{4_64} f"uhrt.

Damit haben wir die Wirkung vollst"andig bestimmt.
Um dieses Ergebnis mathematisch zu rechtfertigen, d"urfen wir
die Spektralzerlegung von $P(x,y) \: P(y,x)$ nicht verwenden:
Mit den asymptotischen Rechenregeln k"onnen beliebige Produkte von
$P(x,y) \: P(y,x)$ gebildet und berechnet werden, beispielsweise
\begin{eqnarray}
\lefteqn{ P(x,y) \: P(y,x) \;=\; \xi\slsh^+ \xi\slsh^- \: (f \:|\: f) \:+\:
	i \: (\xi\slsh f \:|\: g) \:-\: i \: (g \:|\: \xi\slsh f) \:+\:
	(g \:|\: g) } \\
\lefteqn{ P(x,y) \: P(y,x) \: P(x,y) } \nonumber \\
&=& i \xi\slsh^+ \xi\slsh^-
	\xi\slsh^+\: (f^2 \:|\: f) \:-\: (\xi\slsh^2 f^2 \:|\: g)
	\:+\: \xi\slsh^- \xi\slsh^+ \: (f g \:|\: f) \:+\: i \:
	(\xi\slsh f g \:|\: g) \nonumber \\
&&+ \xi\slsh^+ \xi\slsh^- \: (f g\:|\: f) \:+\:
	i \: (\xi\slsh fg \:|\: g) \:-\: i \: (g^2 \:|\: \xi\slsh f) \:+\:
	(g^2 \:|\: g) \nonumber \\
&\simeq& 2 i \: z \: (\xi\slsh f^2 \:|\: f) \:-\:
	i \: (z f^2 \:|\: \xi\slsh f) \:-\: (z f^2 \:|\: g)
	\:+\: i \: (\xi\slsh f g \:|\: g) \nonumber \\
\label{eq:4_71}
&&+ 2z \: (f g\:|\: f) \:+\: i \: (\xi\slsh fg \:|\: g)
	\:-\: i \: (g^2 \:|\: \xi\slsh f) \:+\: (g^2 \:|\: g) \spc .
\end{eqnarray}
Durch Spurbildung erh"alt man $\alpha^{(q)}_{xy}$, also z.B.
\Equ{4_72}
\alpha^{(1)}_{xy} \;=\; \Tr \left( P(x,y) \: P(y,x) \right)
	\;=\; 4 z \: (f \:|\: f) \:+\: 4 \: (g \:|\: g) \spc .
\EndEqu
Wenn man diese Formeln in \Ref{4_64} einsetzt, heben sich
alle Terme weg. Also ist Bedingung \Ref{4_41} f"ur $Q^{[3]}$
gem"a"s \Ref{4_64} tats"achlich erf"ullt.
Es bleibt zu zeigen, da"s der Homogenit"atsgrad $h=3$ minimal ist und
da"s $Q^{[3]}$ (bis auf ein Vielfaches) eindeutig bestimmt ist.
Falls $h$ nicht minimal w"are, g"abe es eine nichttriviale L"osung
der Gleichung
\[ 0 \;\simeq\; Q^{[2]}(x,y) \;=\; c^{(2)}_{\{\}} \: P(x,y) \: P(y,x)
	\: P(x,y) \:+\: c^{(1)}_{\{1\}} \: \alpha^{(1)}_{xy} \:
	P(x,y) \;\;\;, \;\;\;\;\; c^{(1)}_{\{\}}, c^{(0)}_{\{1\}} \in \R
	\;\; . \]
Nach Einsetzen von \Ref{4_60}, \Ref{4_71}, \Ref{4_72} stellt man jedoch
fest, da"s es nur die L"osung $c^{(q)}_{\{p\}} \equiv 0$ gibt.
Das ist auch direkt einsichtig, weil in dieser Rechnung Beitr"age
$\sim \1, \xi\slsh$ auftreten und es nur zwei freie Parameter gibt.
Falls $Q^{[2]}$ nicht (bis auf ein Vielfaches) eindeutig w"are,
g"abe es wegen der Linearit"at in $Q$ von \Ref{4_41} Koeffizienten
$c^{(q)}_{\{p\}} \not \equiv 0$ mit
\[ c^{(2)}_{\{1\}} \: \alpha^{(1)}_{xy} \: P(x,y) \: P(y,x) \: P(x,y)
	\:+\: \left( c^{(1)}_{\{2\}} \: \alpha^{(2)}_{xy} \:+\:
	c^{(1)}_{\{1, 1\}} \: (\alpha^{(1)}_{xy})^2 \right) \:
	P(x,y) \;\simeq\; 0 \spc . \]
Man kann wieder explizite Formeln f"ur die Distributionsprodukte
einsetzen und diese Aussage zum Widerspruch f"uhren.

Ganz allgemein kann man die formale Spektralzerlegung von
$P(x,y) \: P(y,x)$ immer umgehen und alle Ergebnisse mit Hilfe von
\Ref{4_27}, \Ref{4_28} durch eine direkte Rechnung ableiten.
Man sieht aber schon an diesem Beispiel, da"s das Arbeiten mit der
Spektralzerlegung wesentlich anschaulicher
und einfacher ist. Dieser Vorteil wird noch deutlicher, wenn
$Q^{[g]}$ sp"ater nach bestimmten St"orbeitr"agen $\Delta P(x,y)$
des fermionischen Projektors entwickelt werden mu"s.

\subsubsection*{dynamische Eichfelder}
\label{4_dyn}
Wir wollen nun das Studium des freien fermionischen Projektors
abschlie"sen und uns der urspr"unglichen Frage zuwenden, was die
Euler-Lagrange-Gleichungen "uber die Wechselwirkung der Fermionen
aussagen.
Als ersten Schritt zur Beantwortung dieser Frage betrachten wir
St"orungen durch Eichfelder und ber"ucksichtigen nur die
am st"arksten singul"aren Beitr"age zu $P(x,y)$, also die Eich- und
Pseudoeichterme.

Bei einer St"orung des Diracoperators durch ein $U(1)$-Eichpotential $A$,
\Equ{4_87a}
i \Pdd \;\longrightarrow\; i \Pdd \:+\: \Aslsh \spc ,
\EndEqu
beschreiben die Eichterme eine Phasentransformation
\[ P(x,y) \;\longrightarrow\; e^{-i \int_x^y A_j \: (y-x)^j} \;
	P(x,y) \spc . \]
Diese Phasendrehung f"allt bei der Bildung von $P(x,y) \: P(y,x)$ heraus
\[ P(x,y) \: P(y,x) \;\longrightarrow\; P(x,y) \: P(y,x) \spc . \]
Deshalb kann die Transformation von $Q$ gem"a"s \Ref{4_27} ebenfalls
durch eine einfache Phasentransformation beschrieben werden
\Equ{4_3d}
Q(x,y) \;\longrightarrow\; e^{-i \int_x^y A_j \: (y-x)^j} \;
	Q(x,y) \spc .
\EndEqu
Der Phasenfaktor in \Ref{4_3d} ist eine glatte Funktion; folglich
ist die Bedingung an den freien Projektor \Ref{4_41}
auch mit zus"atzlichen Eichtermen erf"ullt.
Die Eich-/Pseudoeichterme der St"orung \Ref{4_87a} fallen also
in den Euler-Lagrange-Gleichungen weg.

Wir nennen allgemein Eichfelder, deren Eich-/Pseudoeichterme in den
Euler-Lagrange-Gleichungen verschwinden,
{\em{dynamische Eichfelder}}\index{Eichfelder, dynamische}.
Alternativ k"onnen dynamische Eichfelder auch dadurch charakterisiert
werden, da"s ihre Potentiale nur in Form von Ableitungen
(also Feldst"arken und Str"omen) in die Euler-Lagrange-Gleichungen eingehen.
Wir beschreiben, warum dynamische Eichfelder eine erste Vorstellung
von der Dynamik des Systems vermitteln:
Die Eich-/Pseudoeichterme sind auf dem Lichtkegel st"arker singul"ar
als alle Beitr"age der klassischen Tensoren zu $P(x,y)$ (also insbesondere
als alle Strom- und Kr"ummungsterme).
Damit die Euler-Lagrange-Gleichungen mit Wechselwirkung erf"ullt
sind, m"ussen folglich die Eich-/Pseudoeichterme aller Eichfelder verschwinden.
Anders ausgedr"uckt, wird die Dynamik des Systems nur durch die
dynamischen Eichfelder beschrieben.
Umgekehrt brauchen nicht alle dynamischen Eichfelder f"ur die Dynamik
relevant zu sein, denn die noch
nicht untersuchten Beitr"age des Feldst"arketensors zu $P(x,y)$
k"onnten zus"atzliche Bedingungen an das Eichfeld liefern.
Damit ein dynamisches Eichfeld tats"achlich in L"osungen der
Euler-Lagrange-Gleichungen auftritt, m"ussen zus"atzlich die
Beitr"age des Feldst"arketensors verschwinden, und es mu"s
einen sinnvollen Zusammenhang zwischen den Beitr"agen des
Noetherstroms und geeigneten Diracstr"omen geben.

F"ur ein realistisches physikalisches Modell erwarten wir im Moment,
da"s die dynamischen Eichfelder durch die Eichgruppe\index{Eichgruppe,
dynamische}
$SU(3) \otimes SU(2) \otimes U(1)$ des Standardmodells beschrieben
werden k"onnen. In jedem Fall sollten alle Eichfelder des Standardmodells
dynamische Eichfelder sein.

\subsubsection*{chirale Eichfelder, die Wirkung $S^{[5]}$}
\label{4_chiral} \index{Eichfelder, chirale}
Wir kommen zu Eichfeldern, welche an die links- und rechtsh"andige
Komponente der Fermionen unterschiedlich ankoppeln.
Solche Felder werden in der schwachen Wechselwirkung beobachtet und
sollten deshalb auch als dynamische Eichfelder vorkommen.

Wir beschreiben chirale Felder mit der St"orung des Diracoperators
\Equ{4_74}
i \Pdd \;\longrightarrow\; i \Pdd \:+\: \chi_L \: \Aslsh_R \:+\:
	\chi_R \: \Aslsh_L
\EndEqu
und chiralen Potentialen $\Aslsh_{L\!/\!R}$.
Die Potentiale haben die Form wie bei einer lokalen $U(1)_L \otimes
U(1)_R$-Eichsymmetrie.
Es stellt sich die Frage, unter welchen Voraussetzungen an die
Lagrangedichte diesen Potentialen dynamische
Eichfelder entsprechen.

Zur Einfachheit diskutieren wir hier nur die f"uhrende
Singularit"at $\sim m^0$ des freien Projektors. Wir nehmen also
\[ P(x,y) \;=\; c_0 \: (i \xi\slsh z^{-2} \:|\: 1) \]
an, es folgt
\Equ{4_75}
P(x,y) \: P(y,x) \;=\; c_0^2 \: (\xi\slsh z^{-2} \:|\: \xi\slsh z^{-2})
	\spc .
\EndEqu
Unter \Ref{4_74} beschreiben die Eich-/Pseudoeichterme wieder
eine Phasentransformation des fermionischen Projektors
\Equ{4_54c}
\chi_{L\!/\!R} \: P(x,y) \;\longrightarrow\; \chi_{L\!/\!R} \:
	e^{-i \int_x^y A^j_{L\!/\!R} \: (y-x)_j} \: P(x,y) \spc ,
\EndEqu
bei der Bildung von $P(x,y) \: P(y,x)$ f"allt der Phasenfaktor aber
nun nicht weg
\begin{eqnarray}
\lefteqn{ \chi_{L\!/\!R} \: P(x,y) \: P(y,x) \;=\;
	\chi_{L\!/\!R} \: P(x,y) \; \chi_{R\!/\!L} \: P(y,x) }
	\nonumber \\
\label{eq:4_76}
&\longrightarrow& \chi_{L\!/\!R} \: e^{-i \int_x^y
	(A^j_{L\!/\!R} - A^j_{R\!/\!L}) \: (y-x)_j} \; P(x,y) \: P(y,x)
	\spc .
\end{eqnarray}
Um die Bedeutung der Transformationsformel \Ref{4_76} zu verstehen,
wollen wir die Eigenwerte von $P(x,y) \: P(y,x)$ bestimmen.
Dazu spalten wir zun"achst den Spinorraum $\C^4$ in die links- und
rechtsh"andige Komponente auf
\[ \C^4 \;=\; \C^4_L \oplus \C^4_R \;\;\;\; {\mbox{mit}} \spc
	\C^4_{L\!/\!R} \;=\; \chi_{L\!/\!R} \: \C^4 \spc . \]

Im freien Fall \Ref{4_75} ist die Matrix $P(x,y) \: P(y,x)$ auf
$\C^4_{L\!/\!R}$ invariant, genauer
\Equ{4_77}
P(x,y) \: P(y,x) \;=\; \left[ c_0^2 \: \chi_L \: (\xi\slsh z^{-2} \:|\:
	\xi\slsh z^{-2}) \: \chi_L \right] \;\oplus\; \left[
	c_0^2 \: \chi_R \: (\xi\slsh z^{-2} \:|\: \xi\slsh z^{-2})
	\: \chi_R \right] \;\;\; .
\EndEqu
Der erste Summand in \Ref{4_77} wirkt nur auf $\C^4_L$, der zweite Summand
nur auf $\C^4_R$. Um die Eigenwerte von \Ref{4_77} zu bestimmen, m"ussen
wir die beiden direkten Summanden diagonalisieren. Da diese Summanden aus
Symmetriegr"unden die gleichen Eigenwerte besitzen, ist jeder Eigenwert
von $P(x,y) \: P(y,x)$ wenigstens zweifach entartet. Das ist die
chirale Entartung der Eigenwerte, die wir auf Seite \pageref{4_chent}
schon unter allgemeineren Voraussetzungen durch Symmetrietransformationen
beschrieben haben.
Wir berechnen die beiden Eigenwerte von \Ref{4_75}
mit dem Funktionalkalk"ul: Gesucht ist ein quadratisches Polynom, das bei
Einsetzen von $P(x,y) \: P(y,x)$ identisch verschwindet.
Wir haben
\begin{eqnarray*}
\lefteqn{ \left( P(x,y) \: P(y,x) \right)^2 \;=\; c_0^4 \:
	\xi\slsh^+ \xi\slsh^- \xi\slsh^+ \xi\slsh^- \:
	(z^{-4} \:|\: z^{-4}) } \\
&=& 2 c_0^4 \: z \: (\xi\slsh z^{-4} \:|\: \xi\slsh z^{-4}) \:-\:
	c_0^4 \: (z^{-3} \:|\: z^{-3}) \\
&\stackrel{\Ref{3_88}}{\simeq}& c_0^2 \; ((z^{-2} \:|\: z^{-1}) \:+\:
	(z^{-1} \:|\: z^{-2})) \; P(x,y) \: P(y,x) \:-\: c_0^4 \:
	(z^{-3} \:|\: z^{-3}) \spc ,
\end{eqnarray*}
das Polynom hat also die Form
\[ \lambda^2 \:-\: c_0^2 \: ((z^{-2} \:|\: z^{-1}) \:+\: (z^{-1} \:|\: z^{-2}))
	\: \lambda \:+\: c_0^4 \: (z^{-3} \:|\: z^{-3}) \spc . \]
Die (formalen) Eigenwerte $\lambda_{1\!/\!2}$ von $P(x,y) \: P(y,x)$
sind die Nullstellen dieses Polynoms
\begin{eqnarray}
\lambda_{1\!/\!2} &\simeq& \frac{1}{2} \: c_0^2 \:
	((z^{-2} \:|\: z^{-1}) \:+\: (z^{-1} \:|\: z^{-2})) \nonumber \\
&&\hspace*{1cm} \:\pm\: \sqrt{ \frac{1}{4} \:
	c_0^4 \: ((z^{-2} \:|\: z^{-1}) \:+\:
	(z^{-1} \:|\: z^{-2}))^2 \:-\: c_0^4 \: (z^{-3} \:|\: z^{-3}) }
	\nonumber \\
&=& \frac{1}{2} \: c_0^2 \: ((z^{-2} \:|\: z^{-1}) \:+\:
	(z^{-1} \:|\: z^{-2})) \:\pm\: \frac{1}{2} \: c_0^2 \:
	\sqrt{ ((z^{-2} \:|\: z^{-1}) \:-\: (z^{-1} \:|\: z^{-2}))^2 }
	\nonumber \\
\label{eq:4_78}
&=& \left\{ \begin{array}{cc}
	c_0^2 \: (z^{-2} \:|\: z^{-1}) & {\mbox{f"ur `1'}} \\[.3em]
	c_0^2 \: (z^{-1} \:|\: z^{-2}) & {\mbox{f"ur `2'}}
	\end{array} \right. \spc .
\end{eqnarray}

Im Fall mit Eich-/Pseudoeichtermen, also $P(x,y) \: P(y,x)$ gem"a"s der
rechten Seite von \Ref{4_76}, ist die Matrix $P(x,y) \: P(y,x)$
ebenfalls auf $\C^4_{L\!/\!R}$ invariant. Mit der Abk"urzung
\Equ{4_81}
\varphi \;=\; \int_x^y (A_L^j - A_R^j) \: (y-x)_j
\EndEqu
haben wir n"amlich
\begin{eqnarray*}
P(x,y) \: P(y,x) \;=\; \left[ e^{-i \varphi} \: c_0^2 \:
	\chi_L \: (\xi\slsh z^{-2} \:|\:
	\xi\slsh z^{-2}) \: \chi_L \right] \;\oplus\; \left[
	e^{i \varphi} \:
	c_0^2 \: \chi_R \: (\xi\slsh z^{-2} \:|\: \xi\slsh z^{-2})
	\: \chi_R \right] \;\;\; .
\end{eqnarray*}
Die beiden Untermatrizen $c_0^2 \: \chi_{L\!/\!R} \: (\xi\slsh z^{-2} \:|\:
	\xi\slsh z^{-2}) \: \chi_{L\!/\!R}$ besitzen jeweils die
Eigenwerte \Ref{4_78}. Folglich hat $P(x,y) \: P(y,x)$ nun die
vier Eigenwerte
\Equ{4_79}
\lambda_{L \: 1\!/\!2} \;=\; e^{-i \varphi} \: \lambda_{1\!/\!2}
	\;\;\;, \spc \lambda_{R \: 1\!/\!2} \;=\;
	e^{i \varphi} \: \lambda_{1\!/\!2}
\EndEqu
mit $\lambda_{1\!/\!2}$ gem"a"s \Ref{4_78}\footnote{Es mag auffallen,
da"s die Eigenwerte \Ref{4_78}, \Ref{4_79} nicht reell sind, obwohl
die Matrix $P(x,y) \: P(y,x)$ s.a. ist. Wie in Anhang F
genauer beschrieben, ist dies bei indefinitem Skalarprodukt kein
Widerspruch. Allgemein liegt f"ur jeden Eigenwert $\lambda \not \in \R$
auch $\overline{\lambda}$ im Spektrum; die zugeh"origen Spektralprojektoren
gehen durch hermitesche Konjugation ineinander "uber. In \Ref{4_79}
gilt speziell
\[ \overline{\lambda_{L \: 1\!/\!2}} \;=\; \lambda_{R \: 2\!/\!1} \spc {\mbox{und}}
	\spc E^*_{L \: 1\!/\!2} \;=\; E_{R \: 2\!/\!1} \]
(dabei bezeichnet $^*$ die Adjungierte bez"uglich des Spinskalarproduktes).}.

Wir sehen also, da"s die {\em{chirale Entartung}} durch die St"orung
\Ref{4_74} des Diracoperators i.a. {\em{aufgehoben}}\index{chirale
Entartung, Aufhebung der} wird. Nach
\Ref{4_79} bleibt die chirale Entartung nur dann erhalten, wenn $\varphi$
f"ur alle $x, y$ verschwindet. Mit \Ref{4_81} folgt
\[ A_L \;\equiv\; A_R \spc , \]
so da"s \Ref{4_74} in die St"orung \Ref{4_87a} durch ein
$U(1)$-Eichpotential "ubergeht.

Die Aufhebung der chiralen Entartung hat folgende Konsequenz:
Die Wirkung $S^{[3]}$ wurde so konstruiert, da"s $Q^{[3]}$ verschwindet,
falls $P(x,y) \: P(y,x)$ zwei Eigenwerte mit jeweils zweifacher Entartung
besitzt. Hat $P(x,y) \: P(y,x)$ aber vier verschiedene Eigenwerte, so
ist die Bedingung \Ref{4_41} nicht mehr erf"ullt.
Da die Eich-/Pseudoeichterme auf dem Lichtkegel genauso stark singul"ar
wie der freie fermionische Projektor sind, k"onnen wir mit den gleichen
Argumenten wie f"ur den freien Projektor folgern, da"s auch die
Euler-Lagrange-Gleichungen \Ref{4_0c} verletzt sind.
F"ur die Wirkung $S^{[3]}$ ist $U(1)_L \otimes U(1)_R$
folglich keine dynamische Eichgruppe, als dynamisches Eichfeld tritt
lediglich das $U(1)$-Eichfeld gem"a"s \Ref{4_87a} auf.

Damit die volle $U(1)_L \otimes U(1)_R$-Gruppe zu einer dynamischen
Eichgruppe wird, m"ussen wir den Homogenit"atsgrad $h$ erh"ohen:
Das Polynom \Ref{4_62} mu"s nun die vier Nullstellen
$\lambda_{L\!/\!R \: 1\!/\!2}$ besitzen. Nach der intrinsischen
Methode ist $h=5$ zu w"ahlen, es folgt
\Equ{4_80}
F(z) \;=\; \prod_{c \in \{L, R\}, \; a \in \{1, 2\}} (z - \lambda_{ca})
	\spc .
\EndEqu
Analog wie bei der Berechnung von $Q^{[3]}$ lassen sich nun die
Koeffizienten $c^{(q)}_{\{p\}}$ bestimmen, und man erh"alt $Q^{[5]}$.
Die Parameter $c_{\{p\}}$ k"onnen wieder durch `Integration'
der $c^{(q)}_{\{p\}}$ berechnet werden, was schlie"slich die
Wirkung $S^{[5]}$ liefert.

Wir stellen fest, da"s der Homogenit"atsgrad die Dynamik wesentlich
beeinflussen kann: bei $h=3$ haben wir als dynamische Eichgruppe $U(1)$,
bei $h=5$ dagegen die gr"o"sere Gruppe $U(1)_L \otimes U(1)_R$.
Die Koeffizienten $c_{\{p\}}$ k"onnen in beiden F"allen mit der
intrinsischen Methode bestimmt werden. Man beachte, da"s die
Ankopplung der Eichfelder an die Fermionen bei unserem Vorgehen
immer eindeutig festgelegt ist.

Die Wirkung $S^{[5]}$ ist bei Spindimension $4$ aus einem anderen
Grund nicht sinnvoll:
Das Polynom \Ref{4_80} ist das charakteristische Polynom
der $(4 \times 4)$-Matrix $P(x,y) \: P(y,x)$. Folglich verschwindet
$Q$ ganz unabh"angig von der Form der Eigenwerte $\lambda_{ca}$.
Damit sind die Euler-Lagrange-Gleichungen f"ur beliebiges $P(x,y)$
erf"ullt und liefern keinerlei Bedingungen an den fermionischen
Projektor.

Bei "Ubertragung dieses Argumentes auf den Fall beliebiger Spindimension
erhalten wir f"ur sinnvolle Gleichungen die allgemeine Schranke
\Equ{4_2d}
{\mbox{Homogenit"atsgrad}} \;\leq\; {\mbox{Spindimension}} \spc .
\EndEqu
Au"serdem sehen wir, da"s chirale Eichfelder erst bei einer
Spindimension $>4$ sinnvoll auftreten k"onnen.

Abschlie"send versuchen wir,
die Ergebnisse dieses Abschnittes auf die m"ogliche Lagrangedichte
eines realistischen physikalischen Modells zu "ubertragen.
Nach Kapitel \ref{kap2} ben"otigen wir zur Beschreibung aller Fermionen
des Standardmodells die Spindimension
\[ 4 \times 2 \times (3 + 1) \;=\; 32 \]
(4 wegen Diracspinoren, 2 wegen Isospin, 3 wegen Colour, 1 wegen
Leptonen).\\
Da bei den $W$- und $Z$-Bosonen chirale Eichfelder auftreten, mu"s der
Homogenit"atsgrad $h \geq 5$ sein.
Die Absch"atzung \Ref{4_2d} ist sicher grob, wir erwarten
also als fundamentale Wirkung
\[ S \;=\; S^{[h]} \;\;\;\; {\mbox{mit}} \spc 5 \leq h \ll 32 \spc . \]
Zur Einfachheit betrachten wir einmal die Wirkung $S^{[5]}$. Der
freie fermionische Projektor ist aus einzelnen $(4 \times 4)$-Bl"ocken
aufgebaut. Damit die Ergebnisse dieses Abschnitts anwendbar sind,
nehmen wir an, da"s alle Eichfelder in diesen $(4 \times 4)$-Bl"ocken
diagonal sind (wir lassen also die $W$-Potentiale weg).
Die chiralen Potentiale k"onnen dann in jedem Block mit der St"orung
\Ref{4_74} des Diracoperators beschrieben werden, die Eigenwerte
sind durch \Ref{4_79} gegeben.
Nach Konstruktion von $S^{[5]}$ verschwindet $Q^{[5]}$
nur dann, wenn die $(32 \times 32)$-Matrix
$P(x,y) \: P(y,x)$ vier Eigenwerte besitzt.
Da die chirale Entartung in jedem $(4 \times 4)$-Block aufgehoben ist,
m"ussen die Eigenwerte von $P(x,y) \: P(y,x)$ folglich in den einzelnen
$(4 \times 4)$-Bl"ocken "ubereinstimmen. Dazu mu"s die Phase
$\varphi$, \Ref{4_81}, nach \Ref{4_79}
in allen $(4 \times 4)$-Bl"ocken bis auf ein Vorzeichen "ubereinstimmen.
Also k"onnen die axialen Potentiale $A_L - A_R$ nicht in jedem Block
beliebig sein, sondern d"urfen sich in den einzelnen Bl"ocken
nur um relative Vorzeichen unterscheiden.
Das entspricht genau der physikalischen Beobachtung: Der
axiale Anteil des $Z$-Eichfeldes hat die Form
$Y(x) \: \sigma^3_\iso$ und ist somit in jedem Block
dem Betrage nach gleich. Dies ist ein erster Hinweis, da"s der homogene
Polynomansatz physikalisch sinnvoll sein k"onnte.

Diese "Uberlegung ist vereinfacht, weil wir
nicht ber"ucksichtigt haben, da"s die Neutrinos nur in einer H"andigkeit
beobachtet werden. Bevor wir die Spindimension vergr"o"sern, wollen
wir deshalb chirale Fermionen untersuchen.

\subsection{Chirale Fermionen}
\label{4_ab32}
Zur m"oglichen Beschreibung von Neutrinos betrachten wir einen Diracsee,
der nur aus linksh"andigen Fermionen aufgebaut ist.
Damit die Lorentzkovarianz gewahrt ist, mu"s die Masse
der Fermionen verschwinden.
Der freie Projektor hat also mit der Notation \Ref{3_7}
die Form
\Equ{4_1b}
P(x,y) \;=\; \chi_L \; \frac{1}{2} \: (p_0 - k_0)(x,y) \;=\;
	\chi_L \: c_0 \: (i \xi\slsh z^{-2} \:|\: 1)
\EndEqu
mit einer reellen Konstanten $c_0=-(4 \pi^3)^{-1}$.
Wir betrachten gleich den allgemeineren Fall mit chiralen Eichfeldern:
bei der St"orung \Ref{4_74} des Diracoperators beschreiben die
Eich-/Pseudoeichterme eine Phasentransformation
\Equ{4_90}
P(x,y) \;=\; \chi_L \: e^{-i \int_x^y A^j_L \: (y-x)_j} \; c_0 \:
	(i \xi\slsh z^{-2} \:|\: 1) \;\; \left[
	+\: {\cal{O}}(\xi^{-2}) \right] \spc .
\EndEqu
Da $P(x,y)$ rein linksh"andig ist, verschwindet das Produkt
$P(x,y) \: P(y,x)$
\begin{eqnarray}
P(x,y) \: P(y,x) &=& (\chi_L + \chi_R) \; P(x,y) \: P(y,x) \nonumber \\
&=& \chi_L \:P(x,y) \; \chi_R \: P(y,x) \;+\;
	\chi_R \:P(x,y) \; \chi_L \: P(y,x) \nonumber \\
\label{eq:4_91}
&=& \chi_L \:P(x,y) \; 0 \;+\;
	0 \; \chi_L \: P(y,x) \;=\; 0 \spc .
\end{eqnarray}
Dies ist ein wesentlicher Unterschied zu den massiven Fermionen des
vorigen Abschnitts. Als Folge wird die Diskussion der
Euler-Lagrange-Gleichungen trivial: Nach \Ref{4_91} tr"agt in
\Ref{4_27} nur der Summand $q=1$ bei. Die Gr"o"sen $\alpha^{(q)}_{xy}$
verschwinden nach \Ref{4_28} ebenfalls, folglich haben wir
\[ Q^{[1]}(x,y) \;=\; c^{(1)}_{\{\}} \: P(x,y) \spc {\mbox{und}} \spc
	Q^{[h]}(x,y) \;=\; 0 \;\;\;\; {\mbox{f"ur $h>1$}} \spc , \]
so da"s die Euler-Lagrange-Gleichungen in jedem Fall erf"ullt sind.

Damit haben wir allerdings nicht die Situation behandelt, die uns
eigentlich interessiert: Wie in Kapitel \ref{kap2} beschrieben, ist
ein realistisches Modell aus
massiven und chiralen Fermionen aufgebaut; beim freien
Projektor sind diese Teilchensorten in verschiedenen
$(4 \times 4)$-Spinorbl"ocken zu finden.
In diesem Abschnitt wollen wir als Vorbereitung auf den allgemeinen Fall
den $(4 \times 4)$-Block der chiralen Fermionen
f"ur sich untersuchen.
Dieses Herausgreifen eines einzelnen Blocks ist sinnvoll, solange
keine Wechselwirkung mit anderen Bl"ocken stattfindet.
Bei dieser Sichtweise ist die Spindimension in \Ref{4_27} also gr"o"ser
als vier, wir betrachten aber $Q(x,y)$ nur auf einem vierdimensionalen
Teilraum des Spinorraumes.
F"ur den Faktor $(P(x,y) \: P(y,x))^{q-1} \: P(x,y)$ spielt das keine
Rolle, wir k"onnen weiterhin \Ref{4_90}, \Ref{4_91} anwenden.
F"ur die Gr"o"sen $\alpha^{(q)}_{xy}$ ist diese Vorstellung aber wichtig,
denn die Spur in \Ref{4_28} ist dann "uber alle Spinorkomponenten
(und nicht nur "uber den chiralen Block) zu bilden.
Dadurch verschwinden die $\alpha^{(q)}_{xy}$ i.a. nicht, sondern sind
polynomial aus den Eigenwerten von $P(x,y) \: P(y,x)$ in den
massiven Bl"ocken aufgebaut.
Das einfachste Beispiel dieser Art ist bei Spindimension 8
ein System mit einem chiralen und einem massiven Fermionblock
\Equ{4_98}
P(x,y) \;=\; \left[ \chi_L \: \frac{1}{2} \: (p_0 - k_0)(x,y) \right]
	\:\oplus\: \left[ \frac{1}{2} \; (p_m - k_m)(x,y) \right] \spc .
\EndEqu
(Der erste Summand wirkt auf die ersten vier, der zweite Summand auf
die letzten vier Spinorkomponenten.)\\
Wir betrachten die Einschr"ankung von \Ref{4_98} auf den ersten
direkten Summanden; f"ur die Koeffizienten $\alpha^{(q)}_{\{p\}}$ erh"alt
man das gleiche Ergebnis wie beim massiven Projektor \Ref{4_a}.

\subsubsection*{Nilpotenz des chiralen Blocks}
\label{4_nil}
Wir untersuchen die Gleichungen \Ref{4_41} und \Ref{4_0c} unter der
allgemeinen Annahme, da"s \Ref{4_90} die Einschr"ankung von $P$ auf
den chiralen Block ist:
Die Faktoren $\alpha^{(q)}_{xy}$ in \Ref{4_27} sind nun
nicht null zu setzen.
Wir arbeiten wieder mit der Spektralzerlegung \Ref{4_30} f"ur festes
$x, y$ und fassen $f^{(q, h)}$ als (unbestimmte) Koeffizienten eines
Polynoms in $\lambda_j$ auf. Da $P(x,y) \: P(y,x)$ nur
einen Eigenwert $\lambda=0$ besitzt, vereinfacht sich \Ref{4_30} zu
\Equ{4_94}
Q^{[h]}(x,y) \;=\; f^{(1, h)}_{xy} \: P(x,y) \spc .
\EndEqu
Damit Bedingung \Ref{4_41} erf"ullt ist, mu"s folglich der Koeffizient
$f^{(1,h)}_{xy}$ verschwinden. Anders ausgedr"uckt, mu"s $z=0$
eine Nullstelle des Polynoms $F(z)$, \Ref{4_62}, sein.
Bei einem zusammengesetzten fermionischen Projektor tritt dadurch
eine zus"atzliche Nullstellenbedingung auf, was i.a. einen h"oheren
Homogenit"atsgrad zur Folge hat. 
F"ur \Ref{4_98} mu"s beispielsweise $h=4$ gew"ahlt werden; $F(z)$ hat
im Gegensatz zu \Ref{4_6b} die Form
\Equ{4_94a}
F(z) \;=\; z \: (z-\lambda_1) (z-\lambda_2) \spc .
\EndEqu

Wir kommen zu den Euler-Lagrange-Gleichungen \Ref{4_0c}.
In Verallgemeinerung von \Ref{4_91} gilt
\Equ{4_95}
P(x,z) \: P(z,y) \;=\; 0 \spc {\mbox{f"ur alle $x, y, z \in M$}} \;\;\; .
\EndEqu
Wir nennen diese Gleichung die
{\em{Nilpotenz des chiralen Blocks}}\index{Nilpotenz des chiralen
Blocks}.
Beim Umschreiben von \Ref{4_0c} mit Integralkernen folgt mit
\Ref{4_94} und \Ref{4_95}
\begin{eqnarray}
[P, \: Q](x,y) &=& \int d^4z \; \left( P(x,z) \: Q(z,y) \:-\:
	Q(x,z) \: P(z,y) \right) \nonumber \\
\label{eq:4_96}
&=& \int d^4z \; \left( P(x,z) \: P(z,y) \right) \; (f^{(1,h)}_{zy}
	- f^{(1,h)}_{xz}) \;=\; 0 \spc .
\end{eqnarray}
Die Euler-Lagrange-Gleichungen sind also in jedem Fall erf"ullt.
Im Beispiel \Ref{4_98} k"onnen wir weiterhin mit $h=3$
und $F(z)$ in der Form \Ref{4_6b} arbeiten.

Wir kommen zu dem Schlu"s, da"s f"ur chirale Fermionen Gleichung
\Ref{4_41} eine st"arkere Bedingung als die Euler-Lagrange-Gleichung
\Ref{4_0c} ist. Dies ist ein wesentlicher Unterschied zu den
massiven Fermionen, bei denen wir ab Seite \pageref{4_qsim0}
begr"undet haben, da"s \Ref{4_0c} sogar (f"ur den freien Projektor)
\Ref{4_41} impliziert.
F"ur diesen Unterschied ist die Nilpotenz des chiralen Projektors
verantwortlich.

Gleichung \Ref{4_96} ist n"utzlich, weil sich dadurch in den
fermionischen Projektor chirale Bl"ocke einbauen lassen, ohne da"s
der Homogenit"atsgrad $h$ erh"oht werden mu"s.
Dies ist auch vom theoretischen Standpunkt befriedigend, weil dadurch
chirale Bl"ocke auf nat"urliche Weise im fermionischen Projektor
auftreten k"onnen.
In dieser Hinsicht kann \Ref{4_96} als Hinweis darauf verstanden werden,
da"s die Beschreibung von Neutrinos gem"a"s \Ref{4_1b} sinnvoll ist.
Au"serdem scheint \Ref{4_96} darauf hinzudeuten, da"s die Produktstruktur
$PQ$, $QP$ in den Euler-Lagrange-Gleichungen einen Sinn macht.
Da diese Produktstruktur eng mit den Nebenbedingungen $P^2 = P^* = P$
bei der Variation von $P$ zusammenh"angt (ohne diese Nebenbedingungen
h"atten wir anstelle von \Ref{4_0c} die Euler-Lagrange-Gleichungen
\Ref{4_41}), kann \Ref{4_96} sogar als eine erste Best"atigung f"ur
das Prinzip des fermionischen Projektors angesehen werden.

\section{Systeme bei h"oherer Spindimension}
\label{4_ab4}
Im vorangehenden Abschnitt \ref{4_ab3} haben wir die Form
der Wirkung mit dem homogenen Polynomansatz wesentlich pr"azisiert.
Nach der intrinsischen Methode haben wir nur noch den Homogenit"atsgrad
$h$ als freien Parameter, um die Dynamik des Systems festzulegen.

Damit sind wir nun in einer guten Position, um unseren Ansatz zu testen.
Wenn unser Konzept physikalisch sinnvoll sein soll, m"ussen die
Wechselwirkungen des Standardmodells mit ihren Eichgruppen und
Kopplungen aus den Euler-Lagrange-Gleichungen \Ref{4_0c} folgen.
Mit dem Begriff des dynamischen Eichfeldes steht uns eine erste Methode
zur Verf"ugung, um einen Zusammenhang zwischen den Euler-Lagrange-Gleichungen
und einer durch klassische Eichfelder beschriebenen Dynamik herzustellen.
Darum wollen wir die dynamischen Eichgruppen bei Systemen mit
mehreren Fermionsorten und der Spindimension $4n$, $n>1$
allgemeiner untersuchen.

Wir werden mit Systemen beginnen, bei denen der freie fermionische
Projektor aus zwei $(4 \times 4)$-Bl"ocken aufgebaut ist; die
Spindimension ist also $8$.
Aus physikalischer Sicht sollten diese Systeme die Isospinpartner
\[ u, s, t \;\longleftrightarrow\; d, c, b \spc {\mbox{bzw.}} \spc
	\nu_e, \nu_\mu, \nu_\tau \;\longleftrightarrow\; e, \mu, \tau \]
beschreiben. Zur Einfachheit betrachten wir nur eine Teilchenfamilie;
die Diskussion l"a"st sich aber direkt auf den allgemeinen Fall
"ubertragen, wenn man jeden $(4 \times 4)$-Block des freien
fermionischen Projektors aus mehreren Diracseen aufbaut.
Wir nennen diese Systeme {\em{vereinfachten
Quark-}}\index{Quarksektor, vereinfachter} bzw.
{\em{Leptonsektor}}\index{Leptonsektor, vereinfachter}.
Anschlie"send untersuchen wir in verschiedenen Kombinationen
direkte Summen der vereinfachten Quark- und Leptonsektoren.
Schlie"slich kommen wir zu dem System von drei Quarksektoren
und einem Leptonsektor. Dieses System ist genau aus den
Fermionen des Standardmodells aufgebaut; wir hoffen, die Eichgruppe
$U(1) \otimes SU(2) \otimes SU(3)$ wiederzufinden.

\subsection{Vereinfachter Quarksektor}
\label{4_vqu}
Als freien fermionischen Projektor w"ahlen wir bei Spindimension 8
die direkte Summe von \Ref{4_a}
\Equ{4_va}
P(x,y) \;=\; \left[ \frac{1}{2} \: (p_{m_1} - k_{m_1})(x,y) \right]
	\:\oplus\: \left[ \frac{1}{2} \: (p_{m_2} - k_{m_2})(x,y) \right]
	\spc .
\EndEqu
(Der erste Summand wirkt wieder auf die ersten vier, der zweite Summand auf
die letzten vier Spinorkomponenten.)\\
Die Parameter $m_1, m_2$ sind die (nackten) Massen der beiden
Fermionsorten.
Mit der Notation \Ref{3_7} haben wir analog zu \Ref{4_60}, \Ref{4_61}
\begin{eqnarray}
P(x,y) &=& (i \xi\slsh f_1(z) + g_1(z) \:|\: 1) \:\oplus\:
	(i \xi\slsh f_2(z) + g_2(z) \:|\: 1) \\
\label{eq:4_vb}
P(x,y) \: P(y,x) &=& (i \xi\slsh f_1 + g_1 \:|\: - i \xi\slsh f_1 + g_1)
	\:\oplus\: (i \xi\slsh f_2 + g_2 \:|\: - i \xi\slsh f_2 + g_2)
	\spc
\end{eqnarray}
mit geeigneten Besselfunktionen $f_{1\!/\!2}, g_{1\!/\!2}$.

\subsubsection*{eine Massenbedingung}
Genau wie in Abschnitt \ref{4_ab31} folgt, da"s der freie Projektor die
Bedingung \Ref{4_41} erf"ullen mu"s. Au"serdem wollen wir (mit Hinblick
auf die schwache Wechselwirkung) fordern, da"s unter den dynamischen
Eichfeldern auch chirale Eichfelder sind.
Wir begr"unden, warum dies nur sinnvoll
ist, falls die Massen aller Fermionen "ubereinstimmen:
Wir nehmen $m_1 \neq m_2$ an. Zur Berechnung der Eigenwerte
von \Ref{4_vb} kann man die beiden direkten Summanden wie in Abschnitt
\ref{4_ab31} diagonalisieren. Man erh"alt jeweils zwei
Eigenwerte $\lambda_{1\!/\!2}$ mit (zweifacher) chiraler Entartung.
Da die Massenparameter $m_1, m_2$ in $\lambda_{1\!/\!2}$ eingehen,
stimmen die Eigenwerte in den beiden Bl"ocken nicht "uberein.
Folglich besitzt \Ref{4_vb} vier Eigenwerte mit jeweils zweifacher
chiraler Entartung.
Durch axiale Eichfelder wird die chirale Entartung aufgehoben, so
da"s \Ref{4_vb} dann i.a. acht verschiedene Eigenwerte besitzt.
Wenn die axialen Eichfelder dynamische Eichfelder sind, mu"s
\Ref{4_41} also auch im Fall ohne Entartung erf"ullt sein. Dazu mu"s
$h\geq9$ sein, was Bedingung \Ref{4_2d} widerspricht.

Im allgemeineren Fall mehrerer Teilchenfamilien erh"alt man ganz
analog, da"s alle Massenparameter im oberen und unteren
$(4 \times 4)$-Block "ubereinstimmen m"ussen\footnote{In
den folgenden Abschnitten \ref{4_ab44}, \ref{4_ab45}, \ref{4_ab46}
werden wir au"serdem sehen, da"s die Bedingung $h<9$ auch bei
Spindimension $>8$ gelten mu"s. Ansonsten erh"alt man n"amlich
zu viele dynamische Eichfreiheitsgrade.}.
Damit haben wir eine erste physikalische Aussage abgeleitet:
die nackten Massen der Quarks m"ussen bei unserer Beschreibung unabh"angig
vom Isospin sein (also $m_u=m_d$, $m_c=m_s$, $m_t=m_b$).
Leider l"a"st sich diese Bedingung f"ur die schweren Quarks nur schlecht
experimentell "uberpr"ufen, weil die effektiven Massen nicht genau
aus den nackten berechnet werden k"onnen. F"ur die leichten Quarks
$u, d$ stimmt die Aussage aber sehr gut, wenn man die nackten
und effektiven Massen einfach gleichsetzt.
Wir bemerken, da"s die abgeleitete Massenbedingung nicht neu ist, sondern
auch in einigen GUT-Theorien verwendet wird.
Im Standardmodell k"onnen die Massen der Quarks jedoch unabh"angig
voneinander gew"ahlt werden.

\subsubsection*{Bestimmung der dynamischen Eichgrupen}
F"ur $m_1=m_2=m$ stimmt \Ref{4_va} im den beiden $(4 \times 4)$-Bl"ocken
"uberein. Wir spalten den Spinorraum in der Form
$\C^8=\C^4 \otimes \C^2_\iso$ auf; dabei beschreibt der erste Faktor
die Diracspinoren in jedem $(4 \times 4)$-Block und der zweite Faktor
den Index, welcher die beiden $(4 \times 4)$-Bl"ocke unterscheidet.
Wir nennen den zweiten Faktor auch den Isospinraum.
Mit dieser Notation haben wir
\[ P(x,y) \;=\; \frac{1}{2} \: (p_m - k_m)(x,y) \;\otimes\; \1_\iso
	\spc . \]
F"ur die Bestimmung der chiralen Eichgruppen gehen wir genau wie
f"ur die massiven Fermionen ab Seite \pageref{4_chiral} vor: wir
f"uhren durch eine geeignete St"orung des Diracoperators chirale
Eichfelder ein, berechnen die Eigenwerte der Matrix $P(x,y) \: P(y,x)$
und nutzen aus, da"s diese Eigenwerte f"ur St"orungen durch dynamische
Eichfelder Nullstellen des Polynoms \Ref{4_62} sein m"ussen.
Zur Einfachheit ber"ucksichtigen wir nur die f"uhrende Singularit"at
$\sim m^0$, wir nehmen also
\Equ{4_vc}
P(x,y) \;=\; c_0 \: (i \xi\slsh z^{-2} \:|\: 1) \;\otimes\; \1_{\iso}
\EndEqu
an. In der St"orung
\Equ{4_4d}
i \Pdd \;\longrightarrow\; i \Pdd \:+\: \chi_L \: \Aslsh_R \:+\:
	\chi_R \: \Aslsh_L
\EndEqu
des Diracoperators sind die chiralen Potentiale
$\Aslsh_{L\!/\!R}(x)$ nun hermitesche $(2 \times 2)$-Matrizen sind,
genauer
\Equ{4_vh}
\Aslsh_{L\!/\!R} \;=\; B \!\slsh_{L\!/\!R} \: \1_\iso\;+\;
	\sum_{a=1}^3 \; B \!\slsh^a_{L\!/\!R} \: \sigma^a_\iso
\EndEqu
mit reellen Vektorfeldern $B_{L\!/\!R}, B^a_{L\!/\!R}$.
Die Potentiale haben also die Form wie bei einer lokalen
$U(2)_L \otimes U(2)_R$-Eichsymmetrie.
$B_{L\!/\!R}$ und $B^a_{L\!/\!R}$ sind $U(1)$- bzw. $SU(2)$-Potentiale.
Wir verwenden f"ur den Index $a$ auch die Vektorschreibweise
$\vec{\;}$, also z.B.
\Equ{4_4e}
\vec{B} \!\slsh_{L\!/\!R} \: \vec{\sigma}_\iso \;\equiv\;
	\sum_{a=1}^3 \: B \!\slsh^a_{L\!/\!R} \: \sigma^a_\iso \spc .
\EndEqu
Bei der Transformation von \Ref{4_vc} durch Eich-/Pseudoeichterme
tritt in Verallgemeinerung von \Ref{4_54c} ein zeitgeordnetes
Exponential auf
\Equ{4_vz}
\chi_{L\!/\!R} \: P(x,y) \;\longrightarrow\; \chi_{L\!/\!R} \;
	\Texp \left(-i \int_x^y A^j_{L\!/\!R} \: (y-x)_j \right)
	\; P(x,y) \spc .
\EndEqu
Mit der Notation
\Equ{4_1z}
\intLR_x^y\;\;\; \;:=\; \Texp \left(-i \int_x^y A^j_{L\!/\!R} \: (y-x)_j
	\right)
\EndEqu
folgt f"ur $P(x,y)$ gem"a"s der rechten Seite von \Ref{4_vz}
\Equ{4_vd}
\chi_{L\!/\!R} \: P(x,y) \: P(y,x) \;=\; \chi_{L\!/\!R} \: c_0^2 \:
	(\xi\slsh z^{-2} \:|\: \xi\slsh z^{-2}) \;\otimes\;
	\left( \intLR_x^y \;\;\; \intRL_y^x \;\;\; \right)_\iso \spc .
\EndEqu
Die Matrix $P(x,y) \: P(y,x)$ ist auf den links- und rechtsh"andigen
Unterr"aumen des Spinorraumes invariant
\[ P(x,y) \: P(y,x) \;=\; \left[ \chi_L \: P(x,y) \: P(y,x) \: \chi_L \right]
	\;\oplus\; \left[ \chi_R \: P(x,y) \: P(y,x) \: \chi_R \right]
	\spc . \]
Folglich gen"ugt es, die Eigenwerte von \Ref{4_vd} auf
$\C^8_{L\!/\!R} := \chi_{L\!/\!R} \C^8 = \C^4_{L\!/\!R} \otimes \C^2_\iso$
zu bestimmen.
Wegen der Produktstruktur von \Ref{4_vd} m"ussen wir dazu die beiden
direkten Faktoren diagonalisieren. Der erste Faktor besitzt die
Eigenwerte \Ref{4_78}. Der zweite Faktor hat als $U(2)$-Matrix die Form
\Equ{4_vj}
\intL_x^y \intR_y^x \;=\; \exp (i \varphi) \: \exp (i \vec{v} \vec{\sigma})
	\;\;\;\; , \spc
\intR_x^y \intL_y^x \;=\; \exp (-i \varphi) \: \exp(-i \vec{v} \vec{\sigma})
\EndEqu
mit geeignetem $\varphi \in [0, 2 \pi[$, $\vec{v} \in \R^3$,
$0 \leq |\vec{v}| < 2 \pi$.
Die Parameter $\varphi$, $\vec{v}$ h"angen nur von den $U(1)$- bzw.
$SU(2)$-Potentialen in \Ref{4_vh} ab, also sehr ausf"uhrlich
\begin{eqnarray}
\label{eq:4_vk}
\exp (i \varphi) &=& \exp \left( \int_x^y (B^j_L - B^j_R) \:
	(y-x)_j \right) \\
\label{eq:4_vl}
\exp (i \vec{v} \vec{\sigma}) &=& \Texp \left( \int_x^y
	\vec{B}^j_L \: (y-x)_j \: \vec{\sigma}_\iso \right)
	\Texp \left( \int_y^x \vec{B}^j_R \: (x-y)_j \:
	\vec{\sigma}_\iso \right) \;\;\; . \spc
\end{eqnarray}
Die Matrizen \Ref{4_vj} haben die Eigenwerte
\begin{eqnarray*}
\exp(i \varphi) \: (\cos \vartheta \pm i \sin \vartheta)
	&\spc {\mbox{bzw.}} \spc& 
\exp(-i \varphi) \: (\cos \vartheta \mp i \sin \vartheta)
	\spc ,
\end{eqnarray*}
wobei $\vartheta=|\vec{v}|$ gesetzt wurde. Mit der Notation
\[ \epsilon_a \;=\; \left\{ \begin{array}{cc} 1 & {\mbox{f"ur $a=1$}} \\
	-1 & {\mbox{f"ur $a=2$}} \end{array} \right.
	\spc {\mbox{oder}} \spc
	\epsilon_c \;=\; \left\{ \begin{array}{cc} 1 & {\mbox{f"ur $c=L$}} \\
	-1 & {\mbox{f"ur $c=R$}} \end{array} \right. \]
erhalten wir f"ur $P(x,y) \: P(y,x)$ folglich die acht Eigenwerte
\Equ{4_vf}
\lambda_{cak} \;=\; c_0^2 \: \exp \left( i \epsilon_c \varphi
	+ i \epsilon_c \epsilon_k \vartheta \right)
	\;\times\; \left\{ \begin{array}{cc}
	(z^{-2} \:|\: z^{-1}) & {\mbox{f"ur $a=1$}} \\
	(z^{-1} \:|\: z^{-2}) & {\mbox{f"ur $a=2$}}
	\end{array} \right.
\EndEqu
mit $a,k=1/2, c=L/R$.

Wir untersuchen die Entartung dieser Eigenwerte in Abh"angigkeit von
$\vartheta$, $\varphi$:
Gem"a"s unserer formalen Behandlung der $\lambda_{cak}$ sind die
Eigenwerte f"ur unterschiedliches $a$ in jedem Fall als voneinander
verschieden anzusehen, also $\lambda_{c1k} \neq \lambda_{d2l}$.
F"ur $\vartheta=\varphi=0$ h"angt \Ref{4_vf} nicht von $c, k$ ab,
also besitzt $P(x,y) \: P(y,x)$ zwei Eigenwerte mit jeweils vierfacher
Entartung. F"ur $\vartheta=0$, $\varphi \neq 0$ und $\varphi=0$,
$\vartheta \neq 0$ haben wir $\lambda_{La1}=\lambda_{La2} \neq
\lambda_{Ra1} = \lambda_{Ra2}$ bzw. $\lambda_{La1}=\lambda_{Ra2} \neq
\lambda_{La2}=\lambda_{Ra1}$, so da"s es vier Eigenwerte mit jeweils
zweifacher Entartung gibt.
F"ur $\vartheta \neq 0$, $\varphi \neq 0$ ist die Entartung schlie"slich
ganz aufgehoben.

Man beachte, da"s $\varphi, \vec{v}, \vartheta$ gem"a"s \Ref{4_vk},
\Ref{4_vl} von $x, y$ abh"angen. Da wir die Singularit"aten auf
dem Lichtkegel und am Ursprung untersuchen, kommt es uns nur
auf $x, y$ auf dem Lichtkegel (also f"ur $(y-x)^2=0$) an.
Damit gehen in die folgenden "Uberlegungen genaugenommen
auch $\varphi, \vec{v}, \vartheta$ nur f"ur $x, y$ auf dem Lichtkegel ein.
Diese Einschr"ankung spielt f"ur unsere Diskussion aber letztlich
keine Rolle, zur besseren "Ubersicht lassen wir sie ganz weg.

Nach diesen Vorbereitungen k"onnen wir die dynamischen Eichfreiheitsgrade
f"ur beliebigen Homogenit"atsgrad bestimmen:

Im freien Fall hat $P(x,y) \: P(y,x)$ zwei verschiedene Eigenwerte.
Folglich mu"s \Ref{4_62} wenigstens ein quadratisches Polynom sein,
also $h \geq 3$.
Im Fall $h=3,4$ besitzt \Ref{4_62} h"ochstens 2 bzw. 3 Nullstellen.
Also mu"s $\varphi=\vartheta=0$ gelten (denn ansonsten h"atte
$P(x,y) \: P(y,x)$ wenigstens vier Eigenwerte).
Diese Bedingung ist nur dann f"ur alle $x, y$ erf"ullt, wenn
$A_L \neq A_R$ gilt. Die dynamischen Eichfelder koppeln
also an die links- und rechtsh"andige Komponente der Fermionen
gleicherma"sen an; sie beschreiben eine lokale $U(2)$-Symmetrie.

F"ur den Homogenit"atsgrad $5 \leq h < 9$ darf $P(x,y) \: P(y,x)$
vier, nicht aber acht Eigenwerte besitzen. Folglich mu"s wenigstens
einer der Parameter $\vartheta, \varphi$ verschwinden.
Da die Eichpotentiale makroskopische Gr"o"sen sind, k"onnen wir
annehmen, da"s $\vartheta, \varphi$ glatte Funktionen in $x, y$ sind.
Wenn also f"ur gegebenes $(x_0, y_0) \in M \times M$ einer der
Parameter $\vartheta, \varphi$ nicht verschwindet, so ist dies auch
noch f"ur $(x, y)$ aus einer kleinen Umgebung von $(x_0, y_0)$
der Fall. Der andere Parameter mu"s dann in dieser Umgebung
entsprechend verschwinden.
Wir diskutieren diese Situation lokal, also f"ur benachbartes $x, y$:
Wir nehmen zun"achst an, da"s $\vartheta \neq 0$, $\varphi=0$
f"ur ein festes $x$ und beliebiges $y \in U_x$ gilt, dabei ist
$U_x$ eine kleine (konvexe) Umgebung von $x$.
Nach \Ref{4_vk}, \Ref{4_vl} mu"s dann in dieser Umgebung $B_L=B_R$
und $\vec{B}_L \neq \vec{B}_R$ gelten. Es d"urfen also nur chirale
$SU(2)$-Potentiale auftreten, w"ahrend das $U(1)$-Potential
an die links- und rechtsh"andige Komponente der Fermionen auf
die gleiche Weise ankoppelt.
Falls f"ur $x$ sowie $y \in U_x$ umgekehrt $\varphi \neq 0$, $\vartheta=0$
gilt, so mu"s nach \Ref{4_vk}, \Ref{4_vl} $\vec{B}_L=\vec{B}^a_R$ und
$B_L \neq B_R$ gelten. Nun treten also nur chirale $U(1)$-Potentiale
auf.

Im Fall $h \geq 9$ darf $P(x,y) \: P(y,x)$ acht Eigenwerte besitzen.
Damit k"onnen $\vartheta, \varphi$ beliebig sein, und wir erhalten
keine Bedingungen an die chiralen Potentiale.

Unsere Ergebnisse sind in Tabelle \ref{4_tab1} zusammengestellt.
\begin{table}
\caption{Dynamische Eichgruppen im vereinfachten Quarksektor}
\label{4_tab1}
\begin{tabular}{|c|c|c|} \hline
Homogenit"atsgrad & dynamische Eichgruppe & St"orung des Diracoperators \\
\hline \hline
$h=3,4$ & $U(2)$ & $\displaystyle{B \!\slsh \:+\:
	\vec{B} \: \vec{\sigma}_\iso }$ \\[.1cm]
\hline
$h=5,\ldots, 8$ & $U(1) \otimes SU(2)_L \otimes SU(2)_R$ &
	$\displaystyle{B \!\slsh +
	(\chi_R \: \vec{B} \!\slsh_L - \chi_L \: \vec{B} \!\slsh_R)
	\: \vec{\sigma}_\iso}$ \\
& oder & \\
& $U(1)_L \otimes U(1)_R \otimes SU(2)$ & $\displaystyle{
	\chi_R \: B \!\slsh_L + \chi_L \: B \!\slsh_R +
	\vec{B} \!\slsh \: \vec{\sigma}_\iso}$ \\
\hline
$h \geq 9$ & $U(2)_L \otimes U(2)_R$
	& $\displaystyle{\chi_R (B \!\slsh_L \:+\: \vec{B} \!\slsh_L \:
	\vec{\sigma}_\iso) + \chi_L (B \!\slsh_R \:+\: \vec{B} \!\slsh_R \:
	\vec{\sigma}_\iso)}$ \\
\hline
\end{tabular}
\end{table}

Wir bemerken, da"s unsere Diskussion in zweierlei Hinsicht nicht ganz
vollst"andig ist:
Zun"achst einmal m"u"sten wir auch die Eich-/Pseudoeichterme h"oherer
Ordnung in $m$ ber"ucksichtigen.
Au"serdem haben wir die Koeffizienten $f^{(q)}_{xy}$ des Polynoms
\Ref{4_62} einfach als frei w"ahlbare Konstanten angesehen.
Es ist im Moment nicht klar, ob sich die abgeleiteten Bedingungen an
die $f^{(q)}_{xy}$ tats"achlich durch geeignete Wahl der
Parameter $c_{\{p\}}$ realisieren lassen.
Damit unsere Darstellung nicht zu technische wird, werden wir darauf
an dieser Stelle nicht genauer eingehen (siehe Abschnitt \ref{4_ab6}).
Wir nehmen vorweg, da"s das Ergebnis von Tabelle
\ref{4_tab1} auch im allgemeinen Fall g"ultig ist.

\subsubsection*{globale Bedingungen}
\label{4_global}
Wir wollen pr"azisieren, wie das Wort `oder' in Tabelle \ref{4_tab1}
zu verstehen ist: F"ur $5 \leq h \leq 8$ m"ussen die Linienintegrale
\Ref{4_vk}, \Ref{4_vl} f"ur jedes $x, y$ eine der beiden Bedingungen
$\varphi=0$ oder $\vec{v}=0$ erf"ullen. Mit einer lokalen "Uberlegung
haben wir gesehen, da"s es Gebiete in der Raumzeit gibt, wo
\Equ{4_4f}
B_L=B_R \spc {\mbox{oder}} \spc \vec{B}_L=\vec{B}_R
\EndEqu
gilt. Wir begr"unden, warum sogar in der ganzen Raumzeit die gleiche
Bedingung erf"ullt sein mu"s: Wir nehmen an, da"s die Potentiale in
zwei Gebieten $A, B$ die Form $B_L=B_R$, $\vec{B}_L \neq \vec{B}_R$
bzw. $B_L \neq B_R$, $\vec{B}_L=\vec{B}_R$ haben. Wenn wir nun $x$ im
Gebiet $A$ und $y$ in $B$ w"ahlen, tragen in den Linienintegralen
\Ref{4_vk}, \Ref{4_vl} i.a. sowohl $B_L - B_R$ als auch
$B^a_L - B^a_R$ bei. Damit folgt $\varphi \neq 0$ und $\vec{v} \neq 0$,
so da"s \Ref{4_4f} verletzt ist.

Damit gibt es genau zwei M"oglichkeiten: die lokale dynamische Eichgruppe
ist in der ganzen Raumzeit entweder $U(1)_L \otimes U(1)_R \otimes SU(2)$
oder $U(1) \otimes SU(2)_L \otimes SU(2)_R$.
Durch ihren globalen Charakter scheint die Uneindeutigkeit der dynamischen
Eichgruppe weniger problematisch zu sein. Es bleibt allerdings unbefriedigend,
da"s die Dynamik mit der intrinsischen Methode nicht eindeutig festgelegt ist.
Es  w"are also w"unschenswert, die dynamische Eichgruppe mit zus"atzlichen
mathematischen Bedingungen vollst"andig zu fixieren.

In unser Argument ging entscheidend ein, da"s in \Ref{4_vk}, \Ref{4_vl}
ausgedehnte Linienintegrale vorkommen, so da"s die Potentiale
auch an entfernten Raumzeit-Punkten miteinander in Beziehung
gesetzt werden k"onnen.
Da beim Studium des Kontinuumslimes oft Linienintegrale auftreten,
f"uhren wir folgenden n"utzlichen Begriff ein:
Wir nennen allgemein Bedingungen, die aus Relationen zwischen
ausgedehnten Linienintegralen folgen,
{\em{globale Bedingungen}}\index{globale Bedingungen}.
Das Auftreten globaler Bedingungen h"angt letztlich damit zusammen,
da"s die Wirkung \Ref{4_12} nichtlokal ist (also, da"s darin
$P(x,y)$ f"ur $x \neq y$ eingeht).

\subsection{Vereinfachter Leptonsektor}
\label{4_vlep}
Wir w"ahlen bei Spindimension 8 als freien Projektor die direkte
Summe von \Ref{4_1b} und \Ref{4_a}
\Equ{4_la}
P(x,y) \;=\; \left[ \chi_L \: \frac{1}{2} \: (p_0 - k_0)(x,y) \right]
	\:\oplus\; \left[ \frac{1}{2} \: (p_m - k_m)(x,y) \right] \spc .
\EndEqu
Dieser Projektor ist uns schon mit \Ref{4_98} begegnet, wir haben
daran die Bedeutung der Nilpotenz des chiralen Blocks erkl"art.
In diesem Abschnitt wollen wir \Ref{4_98} allgemeiner untersuchen.
Insbesondere m"ussen wir den Fall studieren, da"s die chiralen
Eichpotentiale im Isospin nicht diagonal sind.

\subsubsection*{Pinning der rechtsh"andigen Komponente}
Zur besseren physikalischen Anschauung nennen wir den ersten
und zweiten direkten Summanden in \Ref{4_la} Neutrino- bzw.
Elektronblock. Mit der St"orung \Ref{4_4d} des Diracoperators f"uhren
wir wieder chirale Eichpotentiale ein.

Wir betrachten zun"achst die St"orungsrechnung f"ur $P(x,y)$.
Mit der Notation von Kapitel \ref{kap2} besitzt \Ref{4_la}
eine chirale Asymmetrie und eine Massenasymmetrie; die Asymmetriematrizen
$X, Y$ haben (in Blockmatrixdarstellung) die Form
\Equ{4_lb}
X \;=\; \left( \begin{array}{cc} \chi_L & 0 \\ 0 & 1 \end{array} \right)
	\;\;\;\;, \spc
Y \;=\; \left( \begin{array}{cc} 0 & 0 \\ 0 & 1 \end{array} \right)
	\spc .
\EndEqu
Mit chiraler Asymmetrie treten bei der St"orungsrechnung i.a.
nichtlokale Linienintegrale auf.
Wir haben in Kapitel \ref{kap2} die Forderung aufgestellt, da"s
alle nichtlokalen Linienintegrale
verschwinden m"ussen. F"ur die St"orung \Ref{4_4d} des Diracoperators
bedeutet diese Bedingung, da"s \Ref{4_4d} in der Form
\Equ{4_lc}
i \Pdd \;\longrightarrow\; \chi_R \: U_L (i \Pdd + H \!\!\slsh_L) U_L^{-1}
	\:+\: \chi_L \: U_R (i \Pdd + H \!\!\slsh_R) U_R^{-1}
\EndEqu
mit geeigneten unit"aren $U(2)$-Matrixfelder $U_{L\!/\!R}$ und
$U(2)$-Potentialen $H_{L\!/\!R}$ darstellbar ist, welche
mit der chiralen Asymmetriematrix kommutieren
\Equ{4_ld}
[ X_{L\!/\!R}, \: H_{L\!/\!R} ] \;=\; 0 \spc .
\EndEqu
F"ur die linksh"andige Komponente ist \Ref{4_ld} wegen $X_L = \1$
trivialerweise erf"ullt, folglich kann das Potential $A_L$ in \Ref{4_4d}
beliebig sein. F"ur die rechtsh"andige Komponente liefert \Ref{4_ld}
dagegen mit \Ref{4_lb} die Bedingung
\[ [ \sigma^3_\iso, \: H_R ] \;=\; 0 \;\;\;\; , \spc {\mbox{also}} \;\;\;
	H_R \;=\; H^0_R \: \1_\iso \:+\: H^3_R \: \sigma^3_\iso \spc . \]
Damit keine nichtlokalen Linienintegrale auftreten, mu"s die
St"orung des Diracoperators also die Form
\Equ{4_le}
i \Pdd \;\longrightarrow\; \chi_R \: (i \Pdd + \Aslsh_L)
	\:+\: \chi_L \: U (i \Pdd + H \!\!\slsh \:) U^{-1}
\EndEqu
haben, dabei ist $\Aslsh_L$ ein $U(2)$-Potential, $U$ ein
unit"ares $U(2)$-Matrixfeld und $H$ ein im Isospin diagonales Potential.

Es ist g"unstig, f"ur den gest"orten Diracoperator eine andere Eichung
zu w"ahlen: Nach der verallgemeinerten Phasentransformation
\Equ{4_3e}
\Psi(x) \;\rightarrow\; U^{-1}(x) \: \Psi(x)
\EndEqu
der Wellenfunktionen verschwinden $U, U^{-1}$ im zweiten Summanden von
\Ref{4_le} und nach der Ersetzung
\[ U^{-1} \Aslsh_L U + i U^{-1} (\Pdd U) \;\rightarrow\;
	\Aslsh_L \]
auch im ersten Summanden.
Folglich gen"ugt es, anstelle von \Ref{4_le} die St"orung
des Di\-rac\-ope\-ra\-tors
\Equ{4_lf}
i \Pdd \;\longrightarrow\; i \Pdd \:+\: \chi_R \: \Aslsh_L \:+\:
	\chi_L \: H \!\!\slsh
\EndEqu
zu betrachten.

Wir k"onnen das Potential $H$ weiter vereinfachen: $H$ koppelt
an die rechtsh"andige Komponente der Fermionen an. Da $H$ diagonal
ist, findet in der rechtsh"andigen Komponente keine Mischung des
Elektron- und Neutrinoblocks statt. Wir k"onnen also sagen, da"s
die obere und untere Isospinkomponente $F_{1\!/\!2} \: H$
von $H$ ausschlie"slich an die rechtsh"andigen Neutrinos bzw.
an die rechtsh"andigen Elektronen ankoppelt (wir verwenden die
Notation $F_{1\!/\!2} = \frac{1}{2} (\1 \pm \sigma^3)_\iso$).
Da die Neutrinos rein linksh"andig sind, spielt das Potential
$F_1 \: H$ gar keine Rolle\footnote{Wir bemerken, da"s $F_1 H$ bei
der St"orungsrechnung f"ur $P(x,y)$ ganz allgemein wegf"allt.
In Anhang F sieht man dies explizit f"ur die Eich-/Pseudoeichterme
und Massenterme.}, und wir k"onnen $H$ in der Form
\[ H(x) \;=\; h(x) \: F_2 \]
w"ahlen.

Nach diesen "Uberlegungen hat die St"orung des Diracoperators die
Form
\Equ{4_lm}
i \Pdd \;\longrightarrow\; i \Pdd \:+\: \chi_R \: \Aslsh_L \:+\:
	\chi_L \: \Aslsh_R
\EndEqu
mit chiralen Potentialen
\Equ{4_lg}
\Aslsh_L \;=\; B \!\slsh_L \:+\: \vec{B} \!\slsh_L \: \vec{\sigma}_\iso
	\;\;\;\; , \spc \Aslsh_R \;=\; B \!\slsh_R \: F_2 \spc .
\EndEqu
Diese Potentiale haben die Form wie bei einer lokalen
$U(1)_L \otimes U(1)_R \otimes SU(2)_L$-Symmetrie.
Physikalisch ausgedr"uckt bedeutet die Bedingung f"ur $\Aslsh_R$ in
\Ref{4_lg}, da"s es keine Eichwechselwirkung zwischen den rechtsh"andigen
Fermionen im Elektronblock und dem Neutrinoblock geben darf.
Wir nennen diesen Effekt {\em{Pinning der rechtsh"andigen
Komponente}}\index{Pinning der rechtsh"andigen Komponente}.
Das Pinning ist in "Ubereinstimmung mit dem Standardmodell
(denn die $SU(2)$ der elektroschwachen Wechselwirkung koppelt nur an
die linksh"andige Komponente der Fermionen an).
Wir haben das Pinning aus der mathematischen Forderung
abgeleitet, da"s keine nichtlokalen Linienintegrale auftreten d"urfen.
Mit der Sprechweise von Seite \pageref{4_global} handelt es sich
bei dieser Forderung um eine globale Bedingung.

\subsubsection*{Transformation der Nilpotenz}
Nach diesen Vorbereitungen k"onnen wir mit der Untersuchung der
Eich-/Pseudoeichterme beginnen. Zur Einfachheit
ber"ucksichtigen wir wieder nur die f"uhrende Singularit"at $\sim m^0$
auf dem Lichtkegel; wir nehmen also f"ur den freien Projektor
mit der Notation \Ref{3_7}
\[ P(x,y) \;=\; \left[ \chi_L \:c_0\: (i \xi\slsh z^{-2} \:|\: 1) \right]
	\:\oplus\: \left[ c_0 \: (i \xi\slsh z^{-2} \:|\: 1) \right]
	\;=\; X_\iso \:\otimes\: c_0 \: (i \xi\slsh z^{-2} \:|\: 1) \]
an. Die Eich-/Pseudoeichterme beschreiben dann f"ur die St"orung
\Ref{4_lm}, \Ref{4_lg} des Di\-rac\-ope\-ra\-tors die Transformation
\Equ{4_ln}
\chi_{L\!/\!R} \: P(x,y) \;\longrightarrow\; c_0 \: \chi_{L\!/\!R}
	\: X_{L\!/\!R} \; \intLR \;\;\;\;\;\:
	(i \xi\slsh z^{-2} \:|\: 1) \spc ,
\EndEqu
wobei wir f"ur die zeitgeordneten Integrale wieder
die Schreibweise \Ref{4_1z} verwenden.
Als Folge von \Ref{4_lg} ist $\sintR$ im Isospin diagonal und kommutiert
mit $X_R$, also
\Equ{4_ly}
\left[ \intR\:, \: F_2 \right] \;=\; 0 \spc {\mbox{oder allgemeiner}} \spc
	\left[ \intLR\;\;\;\:, \: X_{L\!/\!R} \right] \;=\; 0 \spc .
\EndEqu
Diese Gleichung ist "ubrigens auch eine notwendige Bedingung,
damit $P$ gem"a"s \Ref{4_ln} ein hermitescher Operator ist.

Falls die chiralen Potentiale im Isospin diagonal sind, k"onnen wir
die Ergebnisse von Abschnitt \ref{4_ab3} anwenden: im Elektronblock
mu"s die Versch"arfung \Ref{4_41} der Euler-Lagrange-Gleichungen
gelten, im Neutrinoblock sind die Euler-Lagrange-Gleichungen als
Folge der Nilpotenz \Ref{4_96} trivialerweise erf"ullt. Im allgemeinen
Fall findet nach \Ref{4_ln} eine Mischung des Elektron- und Neutrinoblocks
statt. Wir m"ussen untersuchen, wie sich das in
den Euler-Lagrange-Gleichungen auswirkt.

Im ersten Schritt untersuchen wir $[P, Q](x,y)$ in Blockmatrixdarstellung:
Bei diagonalen Potentialen gilt
\Equ{4_lp}
[P, \: Q](x,y) \;=\; \left( \begin{array}{cc} 0 & 0 \\ 0 & *
	\end{array} \right) \spc ,
\EndEqu
wobei `$*$' eine beliebige $(4 \times 4)$-Untermatrix bezeichnet.
Damit m"ussen die Euler-Lagrange-Gleichungen nur auf einem
vierdimensionalen Unterraum des Spinorraums betrachtet werden.
Auch im allgemeinen Fall, also $P(x,y)$ gem"a"s der rechten Seite von
\Ref{4_ln}, ist $P(x,y)$ singul"ar, genauer
\Equ{4_lr}
F_1 \: \chi_R \: P(x,y) \;=\; \chi_R \: P(x,y) \: F_1 \;=\; 0 \spc .
\EndEqu
Au"serdem ist $P(x,y)$ ungerade
\[ \left\{ \rho, P(x,y) \right\} \;=\; 0 \spc . \]
Da der $q$-te Summand in \Ref{4_27} aus $(2q-1)$ Faktoren der Matrizen
$P(x,y), P(y,x)$ aufgebaut ist, ist $Q(x,y)$ ebenfalls ungerade.
Die Relationen \Ref{4_lr} sind auch erf"ullt, wenn wir $P$ durch $Q$
ersetzen, denn gem"a"s \Ref{4_27} gilt mit der Notation \Ref{4_5a}
\begin{eqnarray*}
F_1 \: \chi_R \: Q^{[h]}(x,y) &=& \sum_{q=1}^h f^{(q,h)}_{xy}
	\: (F_1 \: \chi_R \: P(x,y)) \; (P(y,x) \: P(x,y))^{q-1}
	\;\stackrel{\Ref{4_lr}}{=}\; 0 \\
\chi_R \: Q^{[h]}(x,y) \: F_1 &=& \sum_{q=1}^h f^{(q,h)}_{xy}
	\: (P(x,y) \: P(y,x))^{q-1} \; (\chi_R \: P(x,y) \: F_1)
	\;\stackrel{\Ref{4_lr}}{=}\; 0 \spc .
\end{eqnarray*}
Wir wenden diese Gleichungen auf den Kommutator $[P,Q](x,y)$ an
und erhalten
\begin{eqnarray}
\lefteqn{ F_1 \: \chi_R \: [P, Q](x,y) } \nonumber \\
\label{eq:4_9a}
&=& \int d^4z \; \left( (F_1 \: \chi_R \: P(x,z)) \: Q(z,y) \:-\:
	(F_1 \: \chi_R \: Q(x,z)) \: P(z,y) \right) \;=\; 0 \\
\lefteqn{\chi_L \: [P, Q](x,y) \: F_1 } \nonumber \\
\label{eq:4_9b}
&=& \int d^4z \; \left( P(x,z) \: (\chi_R \: Q(z,y) \: F_1) \:-\:
	Q(x,z) \: (\chi_R \: P(z,y) \: F_1) \right) \;=\; 0
	\;\;\; .
\end{eqnarray}
Da die Matrix $[P, Q](x,y)$ als gerade Matrix au"serdem auf
$\C^8_{L\!/\!R}$ invariant ist, hat sie die Form
\Equ{4_lz}
[P ,\:Q ](x,y) \;=\; \chi_L \: \left( \begin{array}{cc} 0 & * \\
	0 & * \end{array} \right) \: \chi_L \:\oplus\:
	\chi_R \: \left( \begin{array}{cc} 0 & 0 \\
	\ast & * \end{array} \right) \: \chi_R \spc .
\EndEqu
Folglich brauchen die Euler-Lagrange-Gleichungen wieder
nur auf einem vierdimensionalen Unterraum des Spinorraums untersucht
zu werden, nur hat dieser Unterraum gegen"uber \Ref{4_lp} eine
allgemeinere Form.

\subsubsection*{die Nilpotenz bei spektraler Zerlegung von $Q$}
Um genauer zu analysieren, welche Freiheitsgrade von $Q$ in den
Eu\-ler-La\-grange-Glei\-chun\-gen gem"a"s \Ref{4_lz} verschwinden, wollen
wir die spektrale Zerlegung von $Q$, \Ref{4_30}, in die
Euler-Lagrange-Gleichungen einsetzen.

Als Vorbereitung m"ussen wir die Eigenwerte der Matrix $P(x,y) \: P(y,x)$
bestimmen: Bei der Zerlegung $\C^8=(\C^4_L \otimes \C^2_\iso) \oplus
(\C^4_R \otimes \C^2_\iso)$ des Spinorraumes zerf"allt $P(x,y) \: P(y,x)$
in der Form
\begin{eqnarray}
P(x,y) \: P(y,x) &=& \left[ \chi_L \:c_0^2\: (\xi\slsh z^{-2} \:|\:
	\xi\slsh z^{-2}) \:\otimes\: \left( \intL_x^y \intR_y^x F_2 \right)
	\right] \nonumber \\
\label{eq:4_lpa}
&&\hspace*{2cm} \oplus\: \left[ \chi_R \:c_0^2\: (\xi\slsh z^{-2} \:|\:
	\xi\slsh z^{-2}) \:\otimes\: \left( F_2 \: \intR_x^y \intL_y^x
	\right) \right] \;\;\; . \spc
\end{eqnarray}
Die Faktoren $\chi_{L\!/\!R} \:c_0^2\: (\xi\slsh z^{-2} \:|\: \xi\slsh
z^{-2})$ besitzen jeweils die beiden Eigenwerte $\lambda_{1\!/\!2}$,
\Ref{4_78}. Damit bleiben die Isospinmatrizen zu untersuchen.
Mit der Notation
\begin{eqnarray*}
\intL_x^y &=& e^{i \varphi_L} \: \exp (i \vec{v} \vec{\sigma})
	\;=\; e^{i \varphi_L} \; (\cos \vartheta + i \vec{n} \vec{\sigma}
	\sin \vartheta) \\
F_2 \: \intR_x^y &=& \intR_x^y F_2 \;=\; e^{i \varphi_R} \: F_2
\end{eqnarray*}
und $\vartheta=|\vec{v}|$, $\vec{n}=\vec{v}/\vartheta$ (falls
$\vartheta=0$ ist, setzen wir $\vec{n}=(0,0,1)$) haben wir
\begin{eqnarray*}
\intL_x^y \intR_y^x F_2 &=& e^{i(\varphi_L - \varphi_R)} \:
	\exp(i \vec{v} \vec{\sigma}) \: F_2 \nonumber \\
\label{eq:4_l5a}
&=& e^{i(\varphi_L - \varphi_R)}
	\; \left( \begin{array}{cc} 0 & (i n_1 + n_2) \: \sin \vartheta \\
	0 & \cos \vartheta - i n_3 \sin \vartheta \end{array} \right) \\
F_2 \: \intR_x^y \intL_y^x &=& F_2 \: e^{i(\varphi_R - \varphi_L)} \:
	\exp(-i \vec{v} \vec{\sigma}) \nonumber \\
&=& e^{i(\varphi_R - \varphi_L)}
	\; \left( \begin{array}{cc} 0 & 0 \\
	(-i n_1 + n_2) \: \sin \vartheta & \cos
	\vartheta + i n_3 \sin \vartheta \end{array} \right) \spc .
	\nonumber
\end{eqnarray*}
Diese beiden Matrizen haben mit der Abk"urzung $\varphi=\varphi_L -
\varphi_R$ die Eigenwerte
\[ 0, \: e^{i \varphi} (\cos \vartheta - i n_3 \sin
	\vartheta) \spc {\mbox{bzw.}} \spc
	0, \: e^{-i \varphi_L} (\cos \vartheta + i n_3 \sin \vartheta)
	\spc . \]
F"ur die Eigenwerte $\lambda_{cak}$ ($a,k=1/2$, $c=L/R$) von \Ref{4_lpa}
folgt mit einer Notation analog zu \Ref{4_vf}
\begin{eqnarray}
\label{eq:4_l8}
\lambda_{ca1} &=& 0 \;=:\; \lambda_1 \\
\label{eq:4_l9}
\lambda_{ca2} &=& c_0^2 \: \exp (i \epsilon_c \varphi) \:
	(\cos \vartheta - i \epsilon_c n_3 \sin \vartheta)
	\;\times\; \left\{ \begin{array}{cc}
	(z^{-2} \:|\: z^{-1}) & {\mbox{f"ur $a=1$}} \\
	(z^{-1} \:|\: z^{-2}) & {\mbox{f"ur $a=2$}}
	\end{array} \;\;\; . \right. \spc
\end{eqnarray}
Die zugeh"origen Spektralprojektoren bezeichnen wir mit $E_1$, $E_{cak}$.

Nun k"onnen wir die Spektralzerlegung \Ref{4_30} in den
Euler-Lagrange-Gleichungen untersuchen.
Mit der Schreibweise \Ref{4_28a} gilt
\begin{eqnarray}
0 &=& [P, Q^{[h]}](x,y) \;=\; \int d^4z \;
	(P(x,z) \: Q^{[h]}(z,y) \:-\: Q^{[h]}(x,z) \: P(z,y)) \nonumber \\
&=& \sum_{c,a,k} \int d^4z \; \left( {\cal{P}}_{zy}(\lambda_{cak}(z,y))
	\; P(x,z) \; E_{cak}(z,y) \: P(z,y) \right. \nonumber \\
\label{eq:4_l2}
&&\hspace*{3cm} \left. - {\cal{P}}_{xz}(\lambda_{cak}(x,z)) \; E_{cak}(x,z)
	\: P(x,z) \; P(z,y) \right) \;\;\;\; . \spc
\end{eqnarray}
Gem"a"s \Ref{4_lz} erwarten wir, da"s in \Ref{4_l2} einige Beitr"age
verschwinden. In Verallgemeinerung der Situation bei diagonalen
Eichpotentialen k"onnte man vermuten, da"s alle Summanden f"ur $k=1$
als Folge der Nilpotenz wegfallen.
Nach Einsetzen von \Ref{4_l8} und Ausf"uhrung der Summen "uber $a, c$
bedeutet diese Vermutung, da"s der Ausdruck
\Equ{4_l5}
\int d^4z \; \left( {\cal{P}}_{zy}(0) \: P(x,z) \: E_1(z,y) \: P(z,y)
	\:-\: {\cal{P}}_{xz}(0) \: E_1(x,z) \: P(x,z) \: P(z,y) \right)
\EndEqu
unabh"angig von ${\cal{P}}_{zy}(0)$, ${\cal{P}}_{xz}(0)$ verschwindet.
Das folgende Lemma zeigt, da"s das tats"achlich der Fall ist:
\begin{Lemma}
F"ur alle Raumzeit-Punkte $x, y, z$ gilt
\[ P(x,z) \; E_1(z,y) \: P(z,y) \;=\; E_1(x,z) \: P(x,z) \; P(z,y)
	\;=\; 0 \spc . \]
\end{Lemma}
{\Beweis}
Da $P(x,y) \: P(y,x)$ in der Form \Ref{4_lpa} zerf"allt, ist auch
$E_1$ auf $\C^8_{L\!/\!R}$ invariant. Aus den Relationen
\Equ{4_lu}
E_1(x,y) \: P(x,y) \: P(y,x) \;=\; P(x,y) \: P(y,x) \: E_1(x,y) \;=\; 0
\EndEqu
folgt mit \Ref{4_lpa} au"serdem
\begin{eqnarray}
\label{eq:4_ls}
\chi_L \: E_1(x,y) \: \left( \intL_x^y \intR_y^x F_2 \right) &=&
	\chi_R \:E_1(x,y)\: \left( F_1 \intR_x^y \intL_y^x \right)
	\;=\; 0 \\
\label{eq:4_lt}
\chi_L \: \left( \intL_x^y \intR_y^x F_2 \right) \: E_1(x,y) &=&
	\chi_R \: \left( F_1 \intR_x^y \intL_y^x \right) \: E_1(x,y)
	\;=\; 0 \spc .
\end{eqnarray}
Mit der Schreibweise
\[ p(x,y) \;=\; c_0 \: (\xi\slsh z^{-2} \:|\: \xi\slsh z^{-2}) \]
erhalten wir schlie"slich
\begin{eqnarray*}
\lefteqn{ P(x,z) \; E_1(z,y) \: P(z,y) \;=\; P(x,z) \:(\chi_L+\chi_R)
	E_1(z,y) \: P(z,y) } \\
&=& p(x,z) \: \left[ \chi_L \: F_2 \intR_x^z \:E_1(z,y)\: \intL_z^y \:+\:
	\chi_R \: \intL_x^z \:E_1(z,y)\: \intR_z^y F_2 \right] \:
	p(z,y) \\
&\stackrel{\Ref{4_ly}}{=}&
	p(x,z) \: \left[ \intL_x^z \intR_z^y \intL_y^z 
	\chi_L \: \left( \intL_z^y \intR_y^z F_2 \right) \:E_1(z,y)\:
	\intL_z^y \right. \\
&&\hspace*{4.5cm} \left. \:+\:
	\intL_x^z \chi_R \:E_1(z,y)\: \left( F_2 \intR_z^y \intL_y^z
	\right) \intL_z^y \right] \: p(z,y) \\
&\stackrel{\Ref{4_lt}}{=}& 0 \\
\lefteqn{ E_1(x,z) \: P(x,z) \; P(z,y) \;=\; E_1(x,z) \:(\chi_L + \chi_R)
	\: P(x,z) \: P(z,y) } \\
&=& E_1(x,z) \:\left[ \chi_L \: \intL_x^z \intR_z^y F_2 \:+\:
	\chi_R \: F_2 \intR_x^z \intL_z^y \right] \; p(x,z) \: p(z,y) \\
&\stackrel{\Ref{4_ly}}{=}&
	E_1(x,z) \:\left[ \chi_L \: \left( \intL_x^z \intR_y^x F_2 \right)
	\intR_x^z \intR_z^y \:+\:
	\chi_R \: \left(F_2 \intR_x^z \intL_z^x \right)
	\intL_x^z \intL_z^y \right] \; p(x,z) \: p(z,y) \\
&\stackrel{\Ref{4_ls}}{=}& 0 \spc .
\end{eqnarray*}
\QED
Die Euler-Lagrange-Gleichungen \Ref{4_12} reduzieren sich also auf
die Bedingung
\begin{eqnarray}
0 &=& \sum_{c,a} \int d^4z \; \left( {\cal{P}}_{zy}(\lambda_{ca2}(z,y))
	\; P(x,z) \; E_{ca2}(z,y) \: P(z,y) \right. \nonumber \\
\label{eq:4_l6}
&&\hspace*{3cm} \left. - {\cal{P}}_{xz}(\lambda_{ca2}(x,z)) \; E_{ca2}(x,z)
	\: P(x,z) \; P(z,y) \right) \;\;\; . \spc
\end{eqnarray}
Wie man im Spezialfall diagonaler Eichpotentiale sieht, l"a"st sich
\Ref{4_l6} unter Ausnutzung der Nilpotenz nicht weiter vereinfachen.
Von nun an k"onnen wir genau wie f"ur massive Fermionen ab Seite
\pageref{4_qsim0} argumentieren und kommen zu dem Ergebnis, da"s
\Ref{4_l6} sogar die st"arkere Bedingung
\Equ{4_l7}
\sum_{c,a} \: {\cal{P}}_{xy}(\lambda_{ca2}(x,y) \: E_{ca2}(x,y) \:
	P(x,y) \;\simeq\; 0
\EndEqu
impliziert. Im Spezialfall diagonaler Eichpotentiale vereinigt \Ref{4_l7}
die beiden Gleichungen \Ref{4_41}, \Ref{4_96}.
Mit Hilfe von \Ref{4_l7} lassen sich unsere "Uberlegungen zur Nilpotenz
(siehe Seite \pageref{4_nil}) direkt auf den Fall au"serdiagonaler
Eichpotentiale "ubertragen. Insbesondere k"onnen wir die Nilpotenz
weiterhin f"ur eine Verkleinerung des Homogenit"atsgrades ausnutzen.

\subsubsection*{ein weiteres Argument f"ur das Pinning}
Wir haben das Pinning der rechtsh"andigen Komponente aus der
Bedingung abgeleitet, da"s alle nichtlokalen Linienintegrale
verschwinden m"ussen.
Um sehr streng zu sein, k"onnte man diese Bedingung lediglich
als eine technische Forderung ansehen, damit
die St"orungsrechnung f"ur $P(x,y)$ nicht zu kompliziert wird.
Diese Sichtweise ist zwar zu einfach (besonders, weil die St"orungsreihe
mit nichtlokalen Linienintegralen gar nicht zu konvergieren
scheint); trotzdem war unsere Begr"undung von \Ref{4_lg} nicht v"ollig
befriedigend. Es ist n"amlich nicht klar, wie sich die nichtlokalen
Linienintegrale genau in den Euler-Lagrange-Gleichungen auswirken.

Aus diesem Grund wollen wir ein weiteres Argument f"ur das
Pinning anf"uhren, das auf nichtlokale Linienintegrale keinen Bezug nimmt:

Entscheidend f"ur die Vereinfachung der Euler-Lagrange-Gleichungen durch
die Nilpotenz gem"a"s \Ref{4_lz}, \Ref{4_l6} waren die Relationen
\Ref{4_9a}, \Ref{4_9b}. Wir wollen untersuchen, wie sich diese
Gleichungen ohne Pinning verhalten. Zur Einfachheit diskutieren
wir nur einen der auftretenden Summanden
\Equ{4_l10}
0 \;=\; F_1 \: \chi_R \: (PQ)(x,y) \;=\; \int d^4z \;
	F_1 \: \chi_R \; P(x,z) \: Q(z,y) \spc ,
\EndEqu
f"ur die anderen Summanden kann man ganz analog argumentieren.

Zun"achst betrachten wir die St"orung \Ref{4_le} des Diracoperators.
Da sich dieser Fall durch die Eichtransformation \Ref{4_3e} auf
\Ref{4_lf} zur"uckf"uhren l"a"st, "ubertr"agt sich \Ref{4_l10}
unmittelbar, n"amlich
\Equ{4_9c}
0 \;=\; \int d^4z \; F_1 \: \chi_R \: U^{-1}(x) \: P(x,z) \: Q(z,y)
\EndEqu
und folglich
\Equ{4_l11}
F_1 \: U^{-1}(x) \: \chi_R \; (PQ)(x,y) \;=\; 0 \spc .
\EndEqu

Wir kommen zum Fall ohne Pinning, also der allgemeinen St"orung \Ref{4_4d}
des Diracoperators. Da nun bei der St"orungsrechnung nichtlokale
Linienintegrale auftreten, k"onnen wir "uber $P(x,y)$ keine genauen
Aussagen machen. In jedem Fall gehen in $P(x,y)$ aber die chiralen
Potentiale l"angs der Verbindungsstrecke $\overline{xy}$ (oder sogar
l"angs der Geraden $xy$) ein, also symbolisch
\[ \chi_R \: P(x,y) \;\longrightarrow\; \chi_R \: {\cal{N}}_{xy}
	\: P(x,y) \: {\cal{N}}^{-1}_{yx} \]
(siehe auch Gleichung \Ref{a5_nlok} in Anhang E).
Damit "ubertr"agt sich Gleichung \Ref{4_l10} in der Form
\Equ{4_l12}
\int d^4z \; F_1 \: \chi_R \: {\cal{N}}^{-1}_{xz} \: P(x,z) \:
	Q(z,y) \;=\; 0
\EndEqu
Im Unterschied zu \Ref{4_9c} tritt nun anstelle von $U^{-1}(x)$
der Faktor ${\cal{N}}^{-1}_{xz}$ auf.
Da dieser Faktor von $z$ abh"angt, k"onnen wir ihn nicht vor das Integral
ziehen. Damit ist es nicht m"oglich, eine Operatorgleichung
der Form \Ref{4_l11} abzuleiten.

Ohne Pinning bricht also die Nilpotenz zusammen.

\subsubsection*{Bestimmung der dynamischen Eichgruppen}
Die dynamischen Eichgruppen lassen sich mit Hilfe der Eigenwerte
\Ref{4_l8}, \Ref{4_l9} und \Ref{4_l7} ganz "ahnlich wie f"ur den
Quarksektor berechnen.

Als Folge der Nilpotenz brauchen wir gem"a"s \Ref{4_l7} nur die
Eigenwerte \Ref{4_l9} zu ber"ucksichtigen; diese m"ussen Nullstellen
des Polynoms ${\cal{P}}_{xy}(\lambda)$, \Ref{4_28a}, sein.
Damit die Euler-Lagrange-Gleichungen im freien Fall erf"ullt sind,
mu"s $h \geq 3$ gelten.

F"ur $h =3,4$ mu"s die chirale Entartung $\lambda_{La2}=\lambda_{Ra2}$
erhalten sein. Daf"ur gibt es zwei M"oglichkeiten:
\begin{enumerate}
\item $n_3=1$ und $\varphi = \vartheta$.
\item $n_3 \neq 1$ und $\varphi=0$, $n_3=0$.
\end{enumerate}
Im ersten Fall sind alle dynamischen Eichpotentiale im Isospin diagonal.
Sie beschreiben eine lokale $U(1) \otimes U(1)_L$-Symmetrie,
wobei das $U(1)$- und $U(1)_L$-Eichfeld an den Elektron- bzw.
Neutrinoblock ankoppelt. Im zweiten Fall haben die Potentiale
zun"achst die Form
\[ A_L = A + B_1 \: \sigma^1_\iso + B_2 \: \sigma^2_\iso \;\;\;\;,\spc
	A_R = A \: F_2 \spc . \]
Damit $n_3$ bei beliebigen zeitgeordneten Linienintegralen
"uber $A_L$ verschwindet, m"ussen
die Potentiale $B_{1\!/\!2}$ in einem festen Verh"altnis zueinander
stehen, also
\[ B_1(x) \;=\; \alpha \: B(x) \;\;\;,\spc B_2(x) \;=\; \beta \: B(x)
	\spc {\mbox{f"ur alle $x$}} \spc . \]
Nach einer globalen Eichtransformation k"onnen wir
$A_L=A + B \: \sigma^1_\iso$ annehmen. Wir haben also wieder eine lokale
$U(1) \otimes U(1)_L$-Symmetrie; das $U(1)_L$-Potential ist aber nun
im Isospin au"serdigonal.
Die Entscheidung zwischen den beiden M"oglichkeiten wird durch
globale Bedingungen (siehe Seite \pageref{4_global}) festgelegt.

F"ur $h>4$ k"onnen die vier Eigenwerte $\lambda_{ca2}$ beliebig sein,
so da"s man die volle $U(1)_L \otimes U(1)_R \otimes SU(2)_L$ als
dynamische Eichgruppe erh"alt.

Die Ergebnisse f"ur die dynamischen Eichgruppen sind in Tabelle
\ref{4_tab2} zusammengestellt.
\begin{table}
\caption{Dynamische Eichgruppen im vereinfachten Leptonsektor}
\label{4_tab2}
\begin{tabular}{|c|c|c|} \hline
Homogenit"atsgrad & dynamische Eichgruppe & St"orung des Diracoperators \\
\hline \hline
$h=3,4$ & $U(1)_L \otimes U(1) $ & $\displaystyle{ \chi_R \: B \!\slsh
	\: F_1 + \Aslsh \: F_2 }$ \\[-.1cm]
& oder & \\[-.1em]
& $U(1)_L \otimes U(1) $ & $\displaystyle{ \chi_R \: B \!\slsh
	\: \sigma^1_\iso + \chi_R \: \Aslsh + \chi_L \: \Aslsh \:
	F_2 }$ \\[.1cm]
\hline
$h \geq 5$ & $U(1)_L \otimes U(1)_R \otimes SU(2)_L $
	& $\displaystyle{ \chi_R \: \Aslsh_L + \chi_L \: \Aslsh_R \: F_2
	+ \chi_R \: \vec{B} \!\slsh \: \vec{\sigma}_\iso}$ \\
\hline
\end{tabular}
\end{table}

\subsection{Mehrere vereinfachte Quarksektoren}
\label{4_ab44}
Wir w"ahlen bei Spindimension $8s$, $s \geq 1$ als freien fermionischen
Projektor die direkte Summe von $s$ Quarksektoren
\Equ{4_m1}
P(x,y) \;=\; \left( \frac{1}{2} \: (p_m - k_m)(x,y) \:\otimes\: \1_\iso
	\right)^s \;=\; \left( \frac{1}{2} \: (p_m - k_m)(x,y)
	\right)^{2s} \spc .
\EndEqu
F"ur die Beschreibung der Colour-Freiheitsgrade ist der Fall
$s=3$ interessant.
Im Hinblick auf die $SU(3)$ der starken Wechselwirkung erwarten
wir allgemein, da"s die dynamische Eichgruppe die Gruppe $SU(s)$
enth"alt.
Gem"a"s der rechten Seite von \Ref{4_m1} kann $P$ nicht eindeutig in
$(8 \times 8)$-Sektoren zerlegt werden, sondern zerf"allt in $2s$
identische massive Fermionbl"ocke. Wir m"ussen eine Erkl"arung daf"ur
finden, warum sich bei Einf"uhrung einer Eichwechselwirkung
einzelne $(8 \times 8)$-Sektoren ausbilden.

\subsubsection*{Bestimmung der dynamischen Eichgruppen}
Wir f"uhren mit der St"orung
\[ i \Pdd \;\longrightarrow\; i \Pdd \:+\: \chi_R \: \Aslsh_L \:+\:
	\chi_L \: \Aslsh_R \]
des Diracoperators chirale $U(2s)$-Potentiale $\Aslsh_{L\!/\!R}$ ein.
Diese St"orung hat die Form wie bei einer lokalen
$U(2s)_L \otimes U(2s)_R$-Eichsymmetrie.
F"ur die Untersuchung der Eich-/Pseu\-do\-eich\-ter\-me ber"ucksichtigen
wir wieder nur die f"uhrende Singularit"at $\sim m^0$; wir nehmen
also bei einer Zerlegung $\C^{8s}=\C^4 \otimes \C^{2s}$ des
Spinorraumes f"ur den freien Projektor
\[ P(x,y) \;=\; c_0 \: (i \xi\slsh z^{-2} \:|\: 1) \:\otimes\: \1 \]
an. Die Eich-/Pseudoeichterme beschreiben dann die Transformation
\Equ{4_m2}
\chi_{L\!/\!R} \: P(x,y) \;\longrightarrow\; \chi_{L\!/\!R} \:c_0\:
	(i \xi\slsh z^{-2} \:|\: 1) \:\otimes\: \intLR_x^y \spc .
\EndEqu

Wir berechnen die Eigenwerte der Matrix $P(x,y) \: P(y,x)$:
F"ur $P(x,y)$ gem"a"s der rechten Seite von \Ref{4_m2} haben wir
\Equ{4_m3}
\chi_{L\!/\!R} \: P(x,y) \: P(y,x) \;=\; \chi_{L\!/\!R} \:c_0^2\:
	(\xi\slsh z^{-2} \:|\: \xi\slsh z^{-2}) \:\otimes\:
	\left( \intLR_x^y \;\;\: \intRL_y^x \;\;\: \right) \spc .
\EndEqu
Der erste direkte Faktor hat wieder die beiden Eigenwerte
$\lambda_{1\!/\!2}$, \Ref{4_78}. Wir nehmen an, da"s die unit"are
$(2s \times 2s)$-Matrix
\Equ{4_m1a}
U_{xy} \;:=\; \intL_x^y \intR_y^x
\EndEqu
die Eigenwerte $\nu_1, \ldots, \nu_{2s}$ besitzt, wobei wir die
Eigenwerte mit ihrer Vielfachheit z"ahlen. Die Matrix
\[ \intR_x^y \intL_y^x \;=\; U_{xy}^* \]
hat dann die Eigenwerte $\overline{\nu_1}, \ldots,
\overline{\nu_{2s}}$. Da die Matrix $P(x,y) \: P(y,x)$ auf
$\C^{8a}_{L\!/\!R} := \chi_{L\!/\!R} \C^{8a}$ invariant ist, erh"alt man
aus \Ref{4_m3} f"ur ihre Eigenwerte $\lambda_{cak}$
($c=L\!/\!R$, $a=1\!/\!2$, $k=1,\ldots,2s$)
\begin{eqnarray}
\label{eq:4_mq2a}
\lambda_{Lak} &=& c_0^2 \:\nu_k\:\times\:
	\left\{ \begin{array}{cc}
	(z^{-2} \:|\: z^{-1}) & {\mbox{f"ur $a=1$}} \\[.3em]
	(z^{-1} \:|\: z^{-2}) & {\mbox{f"ur $a=2$}}
	\end{array} \right. \\
\label{eq:4_mq2b}
\lambda_{Rak} &=& c_0^2 \:\overline{\nu_k}\:\times\:
	\left\{ \begin{array}{cc}
	(z^{-2} \:|\: z^{-1}) & {\mbox{f"ur $a=1$}} \\[.3em]
	(z^{-1} \:|\: z^{-2}) & {\mbox{f"ur $a=2$}}
	\end{array} \right. \spc .
\end{eqnarray}

Wir k"onnen nun die dynamischen Eichfreiheitsgrade in Abh"angigkeit des
Homogenit"atsgrades bestimmen. Damit die Euler-Lagrange-Gleichungen
erf"ullt sind, m"ussen die Eigenwerte $\lambda_{cak}$ Nullstellen des
Polynoms ${\cal{P}}_{xy}(\lambda)$, \Ref{4_28a}, sein.
F"ur den freien Projektor erhalten wir wieder die Schranke $h \geq 3$.

F"ur $h=3,4$ m"ussen die Eigenwerte in $c, k$ entartet sein, also
$\lambda_{Lak}=\lambda_{Ral}$. Es folgt $U_{xy}=\1$, so da"s lediglich
$U(2s)$-Eichpotentiale auftreten k"onnen.

Interessanter ist der Fall $h=5,6$.
Wir untersuchen die Situation zun"achst f"ur festes $x, y$. Bei den
komplexen Zahlen $\nu_k, \overline{\nu_k}$ d"urfen nun h"ochstens
zwei Werte vorkommen, also
\[ \# \left( \sigma(U_{xy}) \cup \sigma(U_{xy}^*)  \right) \;\leq\; 2 \]
($\sigma(.)$ bezeichnet das Spektrum einer $(2s \times 2s)$-Matrix).\\
Damit diese Bedingung erf"ullt ist, mu"s $U_{xy}$ in einer Untergruppe
$G$ von $U(2s)$ liegen,
\[ U_{xy} \;\in\; G \;\subset\; U(2s) \spc . \]
Zur Einfachheit nehmen wir an, da"s $U_{xy}$ durch geeignete Wahl der
chiralen Potentiale $A_{L\!/\!R}$ jeden Wert in $G$ annehmen
kann\footnote{An folgender Konstruktion sieht man, da"s dies
tats"achlich eine einschr"ankende Annahme ist: Sei $H \subset
U(2s)_L \otimes U(2s)_R$ die dynamische Eichgruppe. Wir bezeichnen
die Projektionen von $H$ auf die erste bzw. zweite Komponente mit
$\rho_{L\!/\!R}$
\[ \rho_{L\!/\!R} \;:\; H \;\longrightarrow\; U(2s)_{L\!/\!R} \spc ; \]
$\rho_{L\!/\!R}$ sind Darstellungen von $H$ auf $U(2s)$. Wir betrachten
die Abbildung
\Equ{4_m01}
U \;:\; H \:\longrightarrow\; U(2s) \;:\; h \;\longrightarrow\; \rho_L(h) \:
	\rho_R(h^{-1}) \spc .
\EndEqu
Die Matrix $U_{xy}$ liegt im Bild von $U$
\[ U_{xy} \;=\; U(h_{xy}) \spc {\mbox{mit}} \spc h_{xy} \;=\; \fintL_x^y
	\:\otimes\: \fintR_x^y \spc . \]
Bei geeigneter Wahl der dynamischen Eichpotentiale l"angs
$\overline{xy}$ durchl"auft $h_{xy}$ alle Elemente von $H$.
Das Bild von $U$ ist i.a. keine Untergruppe von $U(2s)$, insbesondere da
\[ U(g) \: U(h) \;=\; \rho_L(g) \: \rho_R(g^{-1}) \: \rho_L(h) \: \rho_R(h^{-1})
	\;\neq\; \rho_L(g) \: \rho_L(h) \: \rho_R(h^{-1}) \: \rho_R(g^{-1})
	\;=\; U(gh) \spc . \]

Um den allgemeinen Fall zu behandeln, m"u"ste man die Abbildung $U$,
\Ref{4_m01}, genauer studieren. Der Autor vermutet, da"s man dabei zu dem
gleichen Ergebnis wie unter unserer vereinfachenden Annahme
${\mbox{Im }} U = G$ kommt, konnte das aber bisher nicht allgemein beweisen.

Wir werden diese technische Unsauberkeit im folgenden ignorieren.}.

Mit dem folgenden gruppentheoretischen Lemma k"onnen wir $G$
bestimmen:
\begin{Lemma}
\label{4_lemma2}
Jede nichttriviale Untergruppe $G \subset U(2s)$, bei der alle Elemente
$g \in G$ die Bedingung
\Equ{4_m4}
\# \left( \sigma(g) \cup \sigma(g^*) \right) \;\leq\; 2
\EndEqu
erf"ullen, ist isomorph zu $U(1)$ oder $SU(2)$. Die induzierte nat"urliche
Darstellung von $U(1)$ bzw. $SU(2)$ ist unit"ar "aquivalent zu einer der
folgenden Darstellungen:
\begin{eqnarray}
\label{eq:4_m6}
U(1) &\rightarrow& U(2s) \;:\; \exp (i \varphi) \:\rightarrow\:
	\exp(i \varphi)^p \oplus \exp(-i \varphi)^q \spc
	{\mbox{mit $p+q=2s$}} \spc \\
\label{eq:4_m7}
SU(2) &\rightarrow& U(2s) \;:\; \exp (i \vec{v} \vec{\sigma})
	\:\rightarrow\: \left( \exp(i \vec{v} \vec{\sigma}) \right)^{2s}
\end{eqnarray}
\end{Lemma}
{\Beweis}
Wir fassen $G$ als abstrakte Gruppe auf und betrachten die nat"urliche
Darstellung
\[ \rho \;:\; G \:\rightarrow\: U(2s) \spc . \]
Nach Definition ist $\rho$ treu. Wir betrachten den maximalen Torus $T$
von $G$ (also eine maximale abelsche Untergruppe von $G$).
$T$ habe Dimension $m$. $\rho$ induziert eine treue Darstellung von $T$
\[ \rho \;:\; T \:\rightarrow\: U(2s) \;:\; (\varphi_1, \ldots, \varphi_m)
	\:\rightarrow\: \rho(\varphi_1, \ldots, \varphi_m) \spc . \]
Die Eigenwerte der Matrix $\rho(\varphi_1, \ldots, \varphi_m)$ enthalten
Faktoren $\exp(\pm i \varphi_j)$. Folglich kann \Ref{4_m4} nur dann
erf"ullt sein, wenn $m=1$ ist. Die einzigen Untergruppen von $U(2s)$ mit
eindimensionalen maximalen Tori sind $U(1)$, $SU(2)$.

Im Fall $G=U(1)$ gibt es lediglich die irreduziblen Darstellungen
\Equ{4_m5}
\rho_n \;:\; \exp(i \varphi) \:\rightarrow\: \exp(i n \varphi) \spc
	{\mbox{mit $n \in \Z$}} \spc .
\EndEqu
Damit die Matrix $\rho(g)$ Bedingung \Ref{4_m4} erf"ullt, mu"s sie in
irreduzible Darstellungen zerfallen, bei welchen sich die Koeffizienten
$n$ in \Ref{4_m5} nur um relative Vorzeichen unterscheiden, also
\[ \rho \;:\; \exp(i \varphi) \:\rightarrow\:
	\exp(i n \varphi)^p \oplus \exp(-i n \varphi)^q \]
mit geeigneten $p, q \geq 0$ und $p+q=2s$.
Da $\rho$ treu ist, folgt schlie"slich $n=1$.

Im Fall $G=SU(2)$ sind die irreduziblen Darstellungen die Spindarstellungen
$\rho_J \::\: SU(2) \rightarrow U(2J+1)$ mit Spin $J \in \N/2$.
Die unit"aren Matrizen $\rho_J(g)$, $g \neq 1$ besitzen
$2J+1$ verschiedene Eigenwerte\footnote{Zur Erl"auterung betrachten
wir das einfache Beispiel
\[ \rho_{\frac{1}{2}} \otimes \rho_{\frac{1}{2}} \;=\; \rho_0 \oplus
	\rho_1 \spc . \]
Die Matrix $\exp(i \vec{v} \vec{\sigma}) \otimes
\exp(i \vec{v} \vec{\sigma})$ besitzt die Eigenwerte
$\exp(\pm i |\vec{v}|) \: \exp(\pm i |\vec{v}|)$, also
\begin{eqnarray*}
1 &\spc& {\mbox{mit zweifacher Entartung}} \\
\exp(\pm 2 i |\vec{v}|) && {\mbox{ohne Entartung}} \spc .
\end{eqnarray*}
Da $\rho_0$ trivialerweise Eigenwert $1$ besitzt, hat die Matrix
$\rho_1(\exp(i \vec{v} \vec{\sigma})$ die Eigenwerte
$1, \exp(\pm 2 i |\vec{v}|)$.

F"ur die h"oheren Spindarstellungen kann man ganz analog
$(\rho_{\frac{1}{2}})^{2J}$ ausreduzieren.}.
Also kommen nur die triviale Darstellung
$\rho_0$ und die identische Darstellung $\rho_{\frac{1}{2}}$ in Frage.
Die Matrizen $\rho_0(\exp(i \vec{v} \vec{\sigma}))$,
$\rho_{\frac{1}{2}}(\exp(i \vec{v} \vec{\sigma}))$ besitzen die Eigenwerte
1 bzw. $\exp(\pm i |\vec{v}|)$. Damit \Ref{4_m4} erf"ullt ist, darf
die irreduzible Zerlegung von $\rho$ entweder nur aus direkten
Summanden $\rho_0$ oder
nur aus $\rho_{\frac{1}{2}}$ bestehen. Da $\rho$ im ersten Fall
nicht treu w"are, folgt \Ref{4_m7}.
\QED
Wir haben also $G \cong 1$, $G \cong U(1)$ oder $G \cong SU(2)$.
Nach einer geeigneten Eichtransformation hat $G$ die Form der rechten
Seite von \Ref{4_m6}, \Ref{4_m7}.

Wir wollen dieses Ergebnis etwas verallgemeinern: Wir w"ahlen einen Punkt
$z$ auf der Geraden $xy$ au"serhalb der Verbindungsstrecke $\overline{xy}$,
genauer
\[ z \;=\; \lambda y + (1-\lambda) x \spc {\mbox{mit}} \spc \lambda > 1 \spc . \]
Wenn die Potentiale $A_{L\!/\!R}$ auf $\overline{yz}$ verschwinden, haben wir
$U_{xz} = U_{xy} \in G$. Da $U_{xy}$ durch geeignete Wahl von $A_{L\!/\!R}$
ganz $G$ durchl"auft, folgt $U_{xz} \in H \supset G$. Au"ser in
trivialen Spezialf"allen ist $G$ unter Ber"ucksichtigung der Bedingung
\Ref{4_m4} maximal. Da $U_{xz}$ ebenfalls Bedingung \Ref{4_m4} erf"ullen
mu"s, haben wir also $U_{xz} \in G$. Auf den Teilstrecken $\overline{xy}$,
$\overline{yz}$ gelten somit die gleichen Bedingungen an die chiralen
Potentiale; $U_{xz}$ durchl"auft schon bei geeigneter Wahl von
$A_{L\!/\!R}$ auf $\overline{yz}$ ganz $G$.

Nun lassen sich die dynamischen Eichgruppen abstrakt konstruieren:
F"ur die drei Raumzeit-Punkte $x,y$ und $z=\lambda y +
(1-\lambda) x$, $\lambda>1$ gilt
\[ \intL_x^y \:U_{yz}\: \intL_y^x \;=\; \intL_x^y \intL_y^z
	\intR_z^y \intL_y^x \;=\; \intL_x^z \intR_z^x \intR_x^y \intL_y^x
	\;=\; U_{xz} \: U_{xy}^{-1} \;\in\; G \spc . \]
Da $U_{yz}$ durch geeignete Wahl von $\Aslsh_{L\!/\!R}$ auf $\overline{yz}$
ganz $G$ durchl"auft, folgt die Bedingung
\[ \intL_x^y \:G\: \left(\intL_x^y \right)^{-1} \;=\; G \spc 
	{\mbox{, also}} \spc \intL_x^y \;\in\; N(G) \spc , \]
wobei $N(G)$ den Normalisator von $G$ in $U(2s)$ bezeichnet
\[ N(G) \;=\; \left\{ u \in U(2s) \:|\: u G u^{-1} = G \right\} \spc . \]
Nach Definition des Normalisators ist $G$ in $N(G)$ ein Normalteiler, also
\[ N(G) \;=\; G \otimes H \]
mit $H:=N(G) / G$. Nach Wiederholung dieses Argumentes f"ur $\sintR$
anstelle von $\sintL$ erh"alt man die beiden Bedingungen
\Equ{4_m8}
\intLR_x^y \;\;\:\;\in\; G \otimes H \spc .
\EndEqu
Damit $U_{xy} \in G$ ist, mu"s der zweite Faktor in \Ref{4_m8} unabh"angig
von $L\!/\!R$ sein. Wir erhalten folglich die dynamische Eichgruppe
\[ G_L \otimes G_R \otimes H \;=\; G_L \otimes N(G) \spc . \]

F"ur die Gruppen von Lemma \ref{4_lemma2} ist der Normalisator nach
Standardergebnissen der Gruppentheorie bekannt; man erh"alt
\begin{eqnarray*}
{\mbox{f"ur die triviale Gruppe $G$}} &\;:\;& N(G) = U(2s) \\
{\mbox{f"ur $G$ gem"a"s \Ref{4_m6}}} &\;:\;& N(G) = U(p) \otimes U(q) \\
{\mbox{f"ur $G$ gem"a"s \Ref{4_m7}}} &\;:\;& H = U(s) \spc . \\
\end{eqnarray*}
Da $x, y$ beliebig sind und durch die Linienintegrale auch Potentiale an
entfernten Raumzeit-Punkten miteinander verkn"upft werden, folgt genau
wie bei der Begr"undung globaler Bedingungen auf Seite \pageref{4_global},
da"s einer dieser F"alle global erf"ullt sein mu"s.

Damit haben wir f"ur den Fall $h=5,6$ die dynamischen Eichfelder mit
ihren relativen Kopplungen vollst"andig bestimmt.

Bei einem Homogenit"atsgrad $h=7,8$ d"urfen bei $\nu_k, \overline{\nu_k}$
drei verschiedene Werte auftreten. Genau wie in Lemma \ref{4_lemma2}
mu"s der maximale Torus von $G$ eindimensional sein, es folgt
$G=U(1)$ oder $G=SU(2)$. Die m"oglichen Darstellungen von $G$ sind aber
komplizierter (insbesondere k"onnen bei $SU(2)$ auch Spin-1-Darstellungen
auftreten). Wir verzichten auf eine genaue Analyse.

Bei $h \geq 9$ kann der maximale Torus von $G$ auch zweidimensional
sein, worauf wir ebenfalls nicht n"aher eingehen.

Unsere Ergebnisse sind in Tabelle \ref{4_tab3} zusammengestellt.
Zur Deutlichkeit haben wir Matrizen, die auf $\C^{4k}$ wirken,
mit einem zus"atzlichen Index $k$ gekennzeichnet.
\begin{table}
\caption{Dynamische Eichgruppen bei $s$ Quarksektoren}
\label{4_tab3}
\begin{tabular}{|c|c|c|} \hline
Homogenit"atsgrad & dynamische Eichgruppe & St"orung des Diracoperators \\
\hline \hline
$h=3,4$ & $U(2s)$ & $\displaystyle{\Aslsh_{2s}}$ \\
\hline
$h=5,6$ & $U(2s)$ & $\displaystyle{\Aslsh_{2s}}$ \\
& oder & \\
& \parbox{5cm}{\centerline{$U(1)_L \otimes U(p) \otimes U(q)$}
	\centerline{mit $p+q=2s$}}
	& \parbox{5cm}{$\chi_R \: \Aslsh_L \: (\1_p \oplus
	(-\1)_q) \:+\:  B \!\slsh_p \oplus C \!\!\slsh_q$} \\
& oder & \\[-.3cm]
& $SU(2)_L \otimes SU(2)_R \otimes U(s)$
	& \parbox{5cm}{$(\chi_R \: \vec{B} \!\slsh_L + \chi_L \: \vec{B} \!\slsh_R) \:
	  \vec{\sigma}_\iso \otimes \1_s $ \\ \centerline{$+ \1_\iso \otimes
	\Aslsh_s$}} \\
\hline
$h=7,8$ & $U(2s)$ & $\displaystyle{\Aslsh}_{2s}$ \\
& oder & \\
& $U(1)_L \otimes U(1)_R \otimes \cdots$
	& komplizierter \\
& oder & \\
& $SU(2)_L \otimes SU(2)_R \otimes \cdots$
	&  komplizierter \\
\hline
$h \geq 9$ & komplizierter & \\
\hline
\end{tabular}
\end{table}

\subsubsection*{spontane Sektorbildung}
Wir wollen die dynamischen Eichgruppen f"ur $h=5,6$ kurz diskutieren.
Nach Tabelle \ref{4_tab3} gibt es f"ur die dynamischen Eichgruppen und
die Ankopplung der Eichfelder an die Fermionen mehrere M"oglichkeiten.
Die Entscheidung zwischen diesen M"oglichkeiten wird durch globale
Bedingungen festgelegt; man hat also in jedem Fall in der ganzen
Raumzeit die gleichen dynamischen Eichgruppen und Kopplungen.

Die globale Uneindeutigkeit der dynamischen Eichgruppen ist nicht
ganz befriedigend. Wir m"ussen nach zus"atzlichen mathematischen
Bedingungen suchen, um die Wechselwirkung mit der intrinsischen Methode
eindeutig festzulegen.

Trotzdem ist das Ergebnis schon jetzt physikalisch interessant:
Unter der allgemeinen Annahme, da"s chirale Eichfelder auftreten, welche
Teilchenumwandlungen zwischen verschiedenen Fermionsorten induzieren
k"onnen, mu"s der Fall der dynamischen Eichgruppe
$SU(2)_L \otimes SU(2)_R \otimes U(s)$ auftreten. In diesem Fall bilden
sich $s$ $(8 \times 8)$-Sektoren aus, in welchen die $SU(2)_L \otimes
SU(2)_R$-Eichfelder jeweils auf die gleiche Weise ankoppeln.
Die $U(s)$-Eichfelder beschreiben eine Wechselwirkung
der einzelnen Sektoren. Wir nennen diese Segmentierung des
fermionischen Projektors {\em spontane Sektorbildung}\index{spontane
Sektorbildung}.

Die spontane Sektorbildung ist f"ur eine Beschreibung der Wechselwirkungen
des Standardmodells unbedingt notwendig. Die dynamische Eichgruppe
ist im Moment noch etwas zu gro"s (w"unschenswert w"are
$U(1)_{\mbox{\scriptsize{em}}} \otimes SU(s)_{\mbox{\scriptsize{stark}}}
\otimes SU(2)_L$), doch scheinen wir auf dem richtigen Weg zu sein.

\subsection{Kombination des vereinfachten Quark- und Leptonsektors}
\label{4_ab45}
Wir w"ahlen bei Spindimension 16 als freien Projektor die direkte
Summe von \Ref{4_la} und \Ref{4_va}
\Equ{4_k0}
P(x,y) \;=\; \left[ \chi_L \: \frac{1}{2} \: (p_0 - k_0)(x,y) \right]
	\:\oplus\; \left[ \frac{1}{2} \: (p_m - k_m)(x,y) \right]^3 \spc .
\EndEqu
$P$ besitzt eine chirale Asymmetrie und eine Massenasymmetrie. Bei
einer Aufspaltung $\C^{16}=\C^4 \oplus \C^{12}$ des Spinorraumes
haben die Asymmetriematrizen die Form
\[ X \;=\; \chi_L \oplus \1 \;\;\;,\spc Y \;=\; 0 \oplus \1 \spc . \]
Die "Uberlegungen zum Pinning "ubertragen sich aus Abschnitt \ref{4_vlep}:
Damit keine nichtlokalen Linienintegrale auftreten, mu"s die St"orung des
Diracoperators die Form
\Equ{4_k1}
i \Pdd \;\longrightarrow\; i \Pdd + \chi_R \: \Aslsh_L + \chi_L
	(0 \oplus B \!\slsh_R)
\EndEqu
mit einem $U(4)$-Potential $A_L$ und $U(3)$-Potential $B_R$ haben.

\subsubsection*{innere und "au"sere Eichgruppen}
Wir wollen zun"achst mit einem kleinen Einschub in allgemeinem Rahmen
untersuchen, welche Freiheitsgrade der dynamischen Eichfelder die
Eigenwerte der Matrix $P(x,y) \: P(y,x)$ beeinflussen.

Dazu betrachten wir bei Spindimension $4b$, $b \geq 1$ einen fermionischen
Projektor mit chiraler Asymmetriematrix $X$. Wir bezeichnen die dynamische
Eichgruppe mit $H \subset U(b)_L \otimes U(b)_R$ und bilden die
Projektionen
\Equ{4_k10}
\rho_{L\!/\!R} \;:\; H \;\longrightarrow\; U(b)_{L\!/\!R}
\EndEqu
auf die beiden Faktoren. Die Abbildungen $\rho_{L\!/\!R}$ sind unit"are
Darstellungen von $H$. Das Pinning besagt allgemein, da"s $X$ mit
$\rho_{L\!/\!R}$ kommutiert,
\[ \left[ \rho_{L\!/\!R} , \: X_{L\!/\!R} \right] \;=\; 0 \spc . \]
Die infinitesimale Fassung von \Ref{4_k10} ist eine lineare Abbildung der
zugeh"origen Lie-Algebren
\Equ{4_k11}
d \rho_{L\!/\!R} \;:\; {\mbox{Lie }} H \;\longrightarrow\;
	SA(b)_{L\!/\!R} \spc .
\EndEqu
Die dynamischen Potentiale $A_{L\!/\!R}(x)$ liegen f"ur alle Raumzeit-Punkte
$x$ im Bild von \Ref{4_k11}.

Mit den Eich-/Pseudoeichtermen hat die Matrix $P(x,y) \: P(y,x)$ unter
Ber"ucksichtigung der f"uhrenden Singularit"at $\sim m^0$ die Form
\[ \chi_{L\!/\!R} \: P(x,y) \: P(y,x) \;=\; c_0^2 \:
	(\xi\slsh z^{-2} \:|\: \xi\slsh z^{-2}) \:\otimes\:
	\left( \intLR_x^y \;\;\;X_{L\!/\!R}\: \intRL_y^x \;\;\;X_{R\!/\!L}
	\right) \spc . \]
Die Abh"angigkeit der Matrix von den dynamischen Potentialen wird durch
den zweiten Faktor beschrieben. Wir m"ussen also die $(b \times b)$-Matrix
\Equ{4_k2z}
U_{xy} \;:=\; \intL_x^y \:X_L\: \intR_y^x \:X_R
\EndEqu
betrachten. Wir verwenden die Notation
\[ h_{xy} \;=\; \intL_x^y \otimes\: \intR_x^y \;\in\; H \]
und ordnen mit der Abbildung
\Equ{4_k9a}
U \;:\; h \;\longrightarrow\; \rho_L(h) \:X_L\: \rho_R(h^{-1}) \:X_R
\EndEqu
jedem Element von $H$ eine (nicht notwendigerweise unit"are)
$(b \times b)$-Matrix zu. Dann gilt
\[ U_{xy} \;=\; U(h_{xy}) \spc . \]
Die Abbildung $U$ ist n"utzlich, weil sich damit die Auswirkung der
dynamischen Potentiale auf die Eigenwerte von $P(x,y) \: P(y,x)$
gruppentheoretisch formulieren l"a"st.

Wir bestimmen diejenigen Freiheitsgrade der dynamischen Potentiale,
welche die Eigenwerte von $P(x,y) \: P(y,x)$ nicht beeinflussen:
Wir betrachten vier Raumzeit-Punkte $u, x, y, z$ auf einer Geraden.
Damit die Potentiale $A_{L\!/\!R}$ l"angs $\overline{xy}$ nicht in die
Eigenwerte von $P(a,z) \: P(z,a)$ eingehen, mu"s f"ur eine geeignete
unit"are $(b \times b)$-Matrix $V$ die Gleichung
\Equ{4_k12}
U(h_{ax} \: h_{xy} \: h_{yz}) \;=\; V \: U(h_{ax} \: h_{yz}) \: V^{-1}
\EndEqu
gelten. Dabei k"onnen $l:=h_{ax}$, $g:=h_{yz}$ beliebige Werte in $H$
annehmen. Nach Definition von $H$ liefert \Ref{4_k12} die Bedingung
\[ \rho_L(l) \: U(h_{xy} \: g) \: \rho_R(l^{-1}) \;=\;
	V(g,l) \: \rho_L(l) \:U(g) \: \rho_R(l^{-1}) \: V(g,l)^{-1}
	\;\;\;\; \forall g,l \in H \]
oder "aquivalent
\Equ{4_k13a}
U(h_{xy} \: g) \;=\; \left[ \rho_L(l^{-1}) \: V(g,l) \: \rho_L(l) \right]
	\;U(g) \; \left[ \rho_R(l^{-1}) \: V(g,l)^{-1} \: \rho_R(l) \right]
	\;\;\;\; \forall g,l \in H \;\;\;\; .
\EndEqu
Bei allen f"ur uns wichtigen dynamischen Eichgruppen ist diese
Bedingung nur dann erf"ullt, wenn sogar
\Equ{4_k14}
U(h_{xy} \: g) \;=\; U(g) \spc \forall g \in H
\EndEqu
(und $V=\1$) gilt. Wir bilden die Menge aller Elemente, die dieser Forderung
gen"ugen
\[ I \;:=\; \left\{ h \in H \;|\; U(hg) = U(g) \;\; \forall g \in H
	\right\} \spc . \]
$I$ ist eine Untergruppe von $H$, denn
\[ {\mbox{aus }} U(h_1 \: g) = U(h_2 \: g) = U(g) \;\; \forall g \in H
	\spc {\mbox{folgt}} \spc U(g_1 \: g_2 \: h) = U(g_2 \: h) = U(h)
	\spc . \]
$I$ ist sogar ein Normalteiler von $H$, denn wir haben f"ur $g \in I$ und
$l, h \in H$
\begin{eqnarray*}
U(l g l^{-1} h) &=& \rho_L(l) \: U(g (l^{-1} h)) \: \rho_R(l^{-1}) \\
&=& \rho_L(l) \: U(l^{-1} h) \: \rho_R(l^{-1}) \;=\; U(h)
\end{eqnarray*}
und damit $l g l^{-1} \in I$. Also faktorisiert die dynamische Eichgruppe
in der Form
\Equ{4_k13}
H = I \otimes A \spc {\mbox{mit}} \spc A := H / I \spc .
\EndEqu
Wir nennen $I$ die {\em{innere Eichgruppe}}\index{Eichgruppe, innere}
und $A$ die {\em{"au"sere Eichgruppe}}\index{Eichgruppe, "au"sere}.
Im Beispiel der Punkte $u, x, y, z$ haben wir gesehen, da"s die inneren
Eichpotentiale nicht in die Matrix $P(a,z) \: P(z,a)$ eingehen. Da
$I, A$ miteinander kommutieren, gilt sogar allgemein, da"s die inneren
Eichpotentiale bei der Bildung des Produktes $P(x,y) \; P(y,x)$ wegfallen.
Die "au"seren Eichpotentiale dagegen werden durch die Eigenwerte von
$P(x,y) \: P(y,x)$, $x,y \in M$ eindeutig festgelegt.

Zur Erl"auterung dieser Konstruktion betrachten wir einige Beispiele:
\begin{enumerate}
\item $H=U(2)$ gem"a"s Tabelle \ref{4_tab1}:
Die Potentiale beschreiben eine lokale $U(2)$-Eich\-trans\-for\-ma\-tion,
die sich in der Matrix $P(x,y) \: P(y,x)$ nicht auswirkt. Folglich haben wir
\[ I \;=\; U(2) \;\;\;,\spc A \;=\; \1 \spc . \]
\item $H=U(1)_L \otimes U(1)_R \otimes SU(2)$ gem"a"s Tabelle
	\ref{4_tab1}:
Die $SU(2)$-Eichfelder gehen in die Matrix $P(x,y) \: P(y,x)$
nicht ein. Die Gruppe $U(1)_L \otimes U(1)_R$ ist abelsch und kann in
der Form
\Equ{4_k15}
U(1)_L \otimes U(1)_R \;=\; U(1)_\vektoriell \otimes U(1)_\axial
\EndEqu
umgeschrieben werden. Die $U(1)_\vektoriell$ beschreibt lediglich
$U(1)$-Pha\-sen\-trans\-for\-ma\-tio\-nen. Es folgt
\[ I \;=\; U(1)_\vektoriell \otimes SU(2) \;\;\;, \spc
	A \;=\; U(1)_\axial \spc . \]
\item $H=U(1) \otimes SU(2)_L \otimes SU(2)_R$ gem"a"s Tabelle
	\ref{4_tab1}:
Die $U(1)$-Phasentransformationen fallen in $P(x,y) \: P(y,x)$ weg.
Die $SU(2)_L \otimes SU(2)_R$ kann im Gegensatz zu \Ref{4_k15} nicht
in eine vektorielle und axiale Gruppel zerlegt werden, folglich
\[ I \;=\; U(1) \;\;\;,\spc A \;=\; SU(2)_L \otimes SU(2)_R \spc . \]
An diesem Beispiel sieht man im Vergleich zu 1., da"s eine Vergr"o"serung
der dynamischen Eichgruppe ($U(2) \subset U(1) \otimes SU(2)_L \otimes
SU(2)_R$) auf eine kleinere innere Eichgruppe f"uhren kann. In die
Konstruktion der inneren Eichgruppe geht also die Struktur der gesamten
dynamischen Eichgruppe ein.
\item $H=SU(2)_L \otimes SU(2)_R \otimes U(2s)$ gem"a"s Tabelle
	\ref{4_tab3}:
Analog wie unter 3. folgt
\[ I \;=\; U(2s) \;\;\;,\spc A \;=\; SU(2)_L \otimes SU(2)_R \spc . \]
\item $H=U(1)_L \otimes U(1)_R \otimes SU(2)_L$ gem"a"s Tabelle
	\ref{4_tab2}:
Wir zerlegen die abelsche Gruppe $U(1)_L \otimes U(1)_R$ in der Form
\[ U(1)_L \otimes U(1)_R \;=\; U(1)_\vektoriell \otimes U(1)_R \]
und erhalten
\[ I \;=\; U(1)_\vektoriell \;\;\;,\spc A \;=\; SU(2)_L \otimes U(1)_R
	\spc . \]
\end{enumerate}
Bei allen diesen Beispielen kann man leicht "uberpr"ufen, da"s \Ref{4_k13a}
tats"achlich Bedingung \Ref{4_k14} impliziert.

\subsubsection*{Schnitt der "au"seren Eichgruppen}
Wir nennen den ersten direkten Summanden in \Ref{4_k0} Neutrinoblock.
Zwischen den Elektron- und Quarkbl"ocken k"onnen wir in \Ref{4_k0}
nicht unterscheiden. Im Hinblick auf die Wechselwirkungen des
Standardmodells erwarten wir, da"s $P$ bei Einf"uhrung von Eichfeldern
"ahnlich wie im vorigen Abschnitt \ref{4_ab44} spontan in die direkte Summe zweier
$(8 \times 8)$-Sektoren zerf"allt.

Bevor wir diese spontane Sektorbildung und die damit verbundenen
Probleme behandeln, wollen wir qualitativ diskutieren, welche
dynamischen Eichgruppen wir unter der Annahme einer spontanen Sektorbildung
erwarten. Dazu spalten wir den Spinorraum in der Form $\C^{16}=\C^8 \oplus
\C^8$ auf und betrachten nur chirale Potentiale, die auf den beiden
direkten Summanden invariant sind. Wir beschr"anken uns also im
Vergleich zu \Ref{4_k1} auf die St"orung
\Equ{4_k8}
i \Pdd \;\longrightarrow\; i \Pdd + \chi_R \: (\Aslsh_L^\lep
	\oplus \Aslsh_L^\qu) + \chi_L \: (\Aslsh_R^\lep
	\oplus \Aslsh_R^\qu)
\EndEqu
des Diracoperators mit $U(2)$-Potentialen $A_L^\qu, A_R^\qu, A_L^\lep$
und einem $U(1)$-Potential
\[ A_R^\lep \;=\; B_R \: F_2 \spc . \]
\Ref{4_k8} ist die direkte Summe der Diracoperatoren \Ref{4_lm}, \Ref{4_lg}
und \Ref{4_4d}. Folglich ist der fermionische Projektor
bei der St"orung \Ref{4_k8} die direkte Summe des fermionischen Projektors
im Lepton- und Quarksektor, und wir k"onnen die Ergebnisse der Abschnitte
\ref{4_vlep}, \ref{4_vqu} anwenden. Damit die Euler-Lagrange-Gleichungen
erf"ullt sind, mu"s im Quark- und Leptonsektor \Ref{4_41} bzw. \Ref{4_l7}
gelten. Die Eigenwerte $\lambda^\qu_{cak}$, \Ref{4_vf}, im Quarksektor
sowie die nichtverschwindenden Eigenwerte $\lambda^\lep_{ca2}$,
\Ref{4_l9}, im Leptonsektor m"ussen also Nullstellen des Polynoms
${\cal{P}}(\lambda)$, \Ref{4_28a}, sein.
Bei gegebenem Homogenit"atsgrad $h$ hat dies folgende Konsequenzen:
Zun"achst einmal d"urfen bei den Eigenwerten von $P(x,y) \: P(y,x)$ im
Quark- und Leptonsektor h"ochstens $h-1$ verschiedene Werte auftreten,
also
\[ \# \left\{ \lambda^\qu_{cak} \right\} \:\leq\: h-1 \;\;\;,\spc
	\# \left\{ \lambda^\lep_{ca2} \right\} \:\leq\: h-1 \spc . \]
Die Eichfelder m"ussen also im Quark- und Leptonsektor die Form
wie in Tabelle \ref{4_tab1} und Tabelle \ref{4_tab2} haben. Au"serdem
m"ussen die Eigenwerte in dem Sinn miteinander vertr"aglich sein, da"s
sogar
\Equ{4_k9}
\# \left( \left\{ \lambda^\qu_{cak} \right\} \:\cup\:
	\left\{ \lambda^\lep_{ca2} \right\} \right) \;\leq\; h-1
\EndEqu
gilt.

Wir wollen nun untersuchen, auf welche Weise Bedingung \Ref{4_k9} die
dynamischen Eichfreiheitsgrade einschr"ankt.
Wir bezeichnen die dynamischen Eichgruppen im Lepton- und Quarksektor
mit $H^\lep$ bzw. $H^\qu$ und faktorisieren diese Gruppen in innere und
"au"sere Eichgruppen
\Equ{4_k16}
H^\lep \;=\; I^\lep \otimes A^\lep \;\;\;,\spc
	H^\qu \;=\; I^\qu \otimes A^\qu \spc .
\EndEqu
Wir setzen f"ur beliebige Raumzeit-Punkte $x, y$
\begin{eqnarray*}
h^\lep_{xy} &=& \Texp \left(-i \int_x^y A^\lep_{L\:j} \: (y-x)^j \right)
	\otimes \Texp \left(-i \int_x^y A^\lep_{R\:j} \: (y-x)^j \right)
	\;\in\; H^\lep \\
h^\qu_{xy} &=& \Texp \left(-i \int_x^y A^\qu_{L\:j} \: (y-x)^j \right)
	\otimes \Texp \left(-i \int_x^y A^\qu_{R\:j} \: (y-x)^j \right)
	\;\in\; H^\qu \spc .
\end{eqnarray*}
Diese Gruppenelemente k"onnen gem"a"s \Ref{4_k16} in der Form
\[ h^\lep_{xy} \;=\; i^\lep_{xy} \otimes a^\lep_{xy} \;\;\;,\spc
	h^\qu_{xy} \;=\; i^\qu_{xy} \otimes a^\qu_{xy} \]
zerlegt werden.
Die Bedingung \Ref{4_k9} impliziert, da"s die Gruppenelemente
$h^\lep_{xy}, h^\qu_{xy}$
miteinander vertr"aglich sein m"ussen. Wir lassen im Moment offen, was
``miteinander vertr"aglich'' genau bedeutet. Nach Konstruktion der
inneren Eichgruppen ist aber klar, da"s \Ref{4_k9} Bedingungen an alle
Freiheitsgrade von $a^\lep_{xy}$, $a^\qu_{xy}$ liefert, w"ahrend
$i^\lep_{xy}$, $i^\qu_{xy}$ beliebig sein k"onnen.

Wir betrachten $z=\lambda y + (1-\lambda)x$ mit $\lambda>1$. Dann gilt
\Equ{4_k17}
a^\lep_{xy} \: a^\lep_{yz} \;=\; a^\lep_{xz} \;\;\;,\spc
	a^\qu_{xy} \: a^\qu_{yz} \;=\; a^\qu_{xz} \spc .
\EndEqu
Da alle Faktoren $a^\lep_.$, $a^\qu_.$ in \Ref{4_k17} miteinander vertr"aglich sein
m"ussen, ist unsere Vertr"aglichkeitsbedingung bei Gruppenoperationen erhalten.
Wir fassen nun alle Gruppen nur noch als abstrakte Gruppen auf (wir
ber"ucksichtigen also die "uber die dynamischen Potentiale gegebene Darstellung
von $H^\lep, H^\qu$ nicht und betrachten alle Operationen modulo
Gruppenoperationen). Dann impliziert \Ref{4_k17}, da"s
$a^\lep_{xy}, a^\qu_{xy}$ "ubereinstimmen
\[ a^\lep_{xy} \;=\; a^\qu_{xy} \;\in\; A^\lep \cap A^\qu \spc . \]
Als dynamische Eichgruppe der direkten Summe des Quark- und Leptonsektors
haben wir also
\Equ{4_k18}
H \;\subset\; I^\qu \otimes I^\lep \otimes (A^\lep \cap A^\qu) \spc .
\EndEqu

Um zu entscheiden, ob die dynamische Eichgruppe sogar
mit der rechten Seite von \Ref{4_k18} "ubereinstimmt, und um die
Kopplung der zugeh"origen Eichfelder zu bestimmen, mu"s die Abbildung
$U$, \Ref{4_k9a}, im Quark- und Leptonsektor detailliert untersucht werden.
Darauf werden wir im n"achsten Abschnitt zur"uckkommen.

Die Konstruktion \Ref{4_k18} l"a"st sich unmittelbar erweitern
und f"uhrt auf ein allgemeines gruppentheoretisches Konzept:
Wir nehmen an, da"s $P$ in $k$ direkte Summanden zerf"allt
\[ P \;=\; P^{(1)} \oplus \cdots \oplus P^{(k)} \spc . \]
In den einzelnen Summanden habe man die dynamischen Eichgruppen
$H^{(k)}=I^{(k)} \otimes A^{(k)}$. Dann folgt f"ur die dynamische
Eichgruppe $H$ der direkten Summe
\Equ{4_k19}
H \;\subset\; I^{(1)} \otimes \cdots \otimes I^{(k)} \otimes \left(
	\bigcap_{i=1}^k A^{(i)} \right) \spc .
\EndEqu
Wer technischen Details optimistisch gegen"ubersteht, kann sogar
erwarten, da"s in \Ref{4_k19} Gleichheit gilt.
Wir nennen die Konstruktion \Ref{4_k19} {\em{Schnitt der "au"seren
Eichgruppen}}\index{Schnitt der "au"seren Eichgruppen}.

In Tabelle \ref{4_tab4} sind die nach dem Schnitt der "au"seren
Eichgruppen erwarteten dynamischen Eichgruppen f"ur den fermionischen
Projektor \Ref{4_k0} aufgelistet. Zur Deutlichkeit haben wir bei den
inneren Eichgruppen durch einen Index `$^\lep$', `$^\qu$' gekennzeichnet,
in welchem Sektor die zugeh"origen Eichfelder wirken.
\begin{table}
\caption{Gem"a"s dem Schnitt der "au"seren Eichgruppen erwartete
	dynamische Eichgruppen bei der Kombination Quark-/Leptonsektor}
\label{4_tab4}
\begin{center}
\begin{tabular}{|c|c|} \hline
Homogenit"atsgrad & erwartete dynamische Eichgruppe \\
\hline \hline
$h=3,4$ & $U^(1)^\lep \otimes U(1)^\lep \otimes U(2)^\qu$ \\
\hline
$h = 5,\ldots,8$ & $U(1)^\lep \otimes U(1)^\qu \otimes SU(2)_L \otimes
	U(1)_R$ \\
& {\mbox{oder}} \\
& $U^(1)_\lep \otimes U(1)^\lep \otimes U(2)^\qu \otimes U(1)_\axial$ \\
\hline
$h \geq 9$ & $U(1)^\lep \otimes U(1)^\qu \otimes SU(2)_L \otimes U(1)_R$ \\
\hline
\end{tabular}
\end{center}
\end{table}

Es ist physikalisch interessant, da"s in Tabelle \ref{4_tab4} gegen"uber
Tabelle \ref{4_tab3} anstelle der Untergruppe $SU(2)_L \otimes SU(2)_R$
stets die Gruppe $SU(2)_L \otimes U(1)_R$ auftritt. Bei Hinzunahme des
Leptonsektors verkleinert sich die dynamische Eichgruppe genau
in der Weise, wie wir das im Hinblick auf die Wechselwirkungen des
Standardmodells erhoffen.
Das ist eine wichtige Best"atigung f"ur unsere bisherigen Konstruktionen.

\subsubsection*{ein Problem: Nichtvertr"aglichkeit der Eigenwerte}
Nach diesen eher abstrakten "Uberlegungen wollen wir Bedingung
\Ref{4_k9} quantitativ auswerten und versuchen, f"ur den Homogenit"atsgrad
$h=5, 6$ die dynamische Eichgruppe
\Equ{4_v1}
U(1)^\lep \otimes U(1)^\qu \otimes SU(2)_L \otimes U(1)_R
\EndEqu
zu realisieren. Diese Gruppe tritt in Tabelle \ref{4_tab4}
auf und ist dort der physikalisch interessante Fall.

Die Eigenwerte $\lambda^\qu_{cak}$, $\lambda^\lep_{ca2}$ sind durch
\Ref{4_vf}, \Ref{4_l9} gegeben. Bei einer St"orung des Diracoperators
durch Potentiale der dynamischen Eichgruppe \Ref{4_v1} haben wir i.a.
\[ \# \left\{ \lambda^\qu_{cak} \right\} \;=\; 4 \;\;\;,\spc
	\# \left\{ \lambda^\lep_{ca2} \right\} \;=\; 4 \spc . \]
Die Vertr"aglichkeitsbedingung \Ref{4_k9} besagt somit, da"s eine Entartung
zwischen Eigenwerten im Quark- und Leptonsektor vorliegen mu"s;
das bedeutet genauer
\Equ{4_v2}
\exp \left( i \epsilon_c \varphi^\qu - i \epsilon_c \vartheta^\qu \right)
\;=\; \exp(i \epsilon_c \varphi^\lep) (\cos \vartheta^\lep - i \epsilon_c
	\: n^\lep_3 \: \sin \vartheta^\lep) \spc .
\EndEqu
Nach den "Uberlegungen zum Schnitt der "au"seren Eichgruppen m"ussen die
Gruppenelemente der $SU(2)_L \otimes U(1)_R$-Eichgruppe im Quark- und
Leptonsektor "aquivalent sein, es folgt
\[ \varphi^\lep(x) \;=\; \varphi^\qu(x) \;\;\;,\spc
	v\!\slsh^\lep(x) \;=\; V \:v\!\slsh^\qu(x) \: V^{-1} \spc
	{\mbox{f"ur alle $x \in M$}} \]
mit einer geeigneten $SU(2)$-Matrix $V$.
Nach einer globalen $SU(2)$-Eichtransformation
im Quarksektor k"onnen wir sogar $\vec{v}^\lep=\vec{v}^\qu$ annehmen.
Bedingung \Ref{4_v2} ist nur erf"ullt, wenn $n_3^\lep=1$ ist.
Insgesamt haben wir also
\[ \vec{v}^\qu \;=\; \vec{v}^\lep \;=\; (0,0,\vartheta) \spc . \]
Folglich m"ussen die $SU(2)_L$-Potentiale diagonal sein, und wir
erhalten im Gegensatz zu \Ref{4_v1} lediglich die dynamische Eichgruppe
\Equ{4_k20}
U(1)^\lep \otimes U(1)^\qu \otimes U(1)_L \otimes U(1)_R \spc .
\EndEqu
Physikalisch ausgedr"uckt bedeutet dieses Ergebnis, da"s die Potentiale der
$W$-Bosonen verschwinden m"ussen, was der Beobachtung
ganz offensichtlich widerspricht. An dieser Stelle scheint unser Konzept
zum ersten Mal auf ernsthafte Schwierigkeiten zu sto"sen.

Um dieses Problem genauer zu untersuchen, wollen wir die dynamischen
Eichgruppen allgemein bestimmen. Bei Ber"ucksichtigung der f"uhrenden
Singularit"at $\sim m^0$ haben die Eigenwerte $\lambda_{cak}$ ($c=L\!/\!R$,
$a=1\!/\!2$, $k=1,\ldots,4$) der Matrix $P(x,y) \: P(y,x)$ die Form
\Ref{4_mq2a}, \Ref{4_mq2b}; dabei sind $\nu_k$ die Eigenwerte der
$(4 \times 4)$-Matrix \Ref{4_k2z},
\Equ{4_k22}
U_{xy} \;=\; \intL_x^y \intR_y^x \; (0_1 \oplus \1_3) \spc .
\EndEqu

Wir w"ahlen eine spezielle Basis in $\C^4$: Nach einem geeigneten
$SU(3)$-Basiswechsel in den unteren drei Komponenten hat $U_{xy}$
die Gestalt
\[ U_{xy} \;=\; \left( \begin{array}{cccc}
	0 & * & 0 & 0 \\ 0 & * & * & * \\ 0 & * & * & * \\ 0 & * & * & *
	\end{array} \right) \spc , \]
wobei `$*$' f"ur einen beliebigen komplexen Matrixeintrag steht.
Durch eine zus"atzliche $SU(2)$-Basistransformation in den letzten beiden
Komponenten l"a"st sich die Form von $U_{xy}$ weiter vereinfachen
\Equ{4_k21}
U_{xy} \;=\; \left( \begin{array}{cccc}
	0 & * & 0 & 0 \\ 0 & * & 0 & 0 \\ 0 & 0 & * & * \\ 0 & 0 & * & *
	\end{array} \right) \spc .
\EndEqu
Nun zerf"allt $U_{xy}$ in die direkte Summe zweier $(2 \times 2)$-Matrizen,
die wir genau wie \Ref{4_vj}, \Ref{4_l5a} diagonalisieren k"onnen.
Folglich besitzt die Matrix $P(x,y) \: P(y,x)$ ganz allgemein
die Eigenwerte \Ref{4_vf}, \Ref{4_l9}.

Im Fall $h \leq 6$ mu"s
\Equ{4_k23}
\# \left( (\sigma(U_{xy} \cup \sigma(U^*_{xy})) \setminus \{0\} \right)
	\;\leq\; 2
\EndEqu
gelten. Es folgt die Vertr"aglichkeitsbedingung \Ref{4_v2} und
$n_3^\lep=1$. Also hat $h_{xy}$ gegen"uber \Ref{4_k21} sogar die Form
\Equ{4_k24}
U_{xy} \;=\;  \left( \begin{array}{cccc}
	0 & 0 & 0 & 0 \\ 0 & * & 0 & 0 \\ 0 & 0 & * & * \\ 0 & 0 & * & *
	\end{array} \right) \spc ,
\EndEqu
oder, in der urspr"unglichen Basis von \Ref{4_k22},
\[ U_{xy} \;=\;  \left( \begin{array}{cccc}
	0 & 0 & 0 & 0 \\ 0 & * & * & * \\ 0 & * & * & * \\ 0 & * & * & *
	\end{array} \right) \spc . \]
Folglich findet keine Mischung des Neutrinoblocks mit den drei anderen
Bl"ocken statt; die St"orung des Diracoperators mu"s gegen"uber \Ref{4_k1}
die Form
\[ i \Pdd \;\longrightarrow\; i \Pdd + \chi_R \:(\Aslsh_L \oplus 0)
	+ \chi_R \:(0 \oplus B\!\slsh_L)
	+ \chi_L \:(0 \oplus B\!\slsh_R) \]
mit einem $U(1)$-Potential $A_L$ und $U(3)$-Potentialen $B_{L\!/\!R}$ haben.
Nun k"onnen wir $U_{xy}$ im unteren $(3 \times 3)$-Block "ahnlich wie
in Lemma \ref{4_lemma2} behandeln: Wir haben
$U_{xy} \in 0_1 \oplus G$ mit einer Untergruppe $G \subset U(3)$.
Wir nehmen zur Einfachheit an, da"s $U_{xy}$ durch geeignete Wahl von
$B_{L\!/\!R}$ l"angs $\overline{xy}$ alle Elemente von $0 \oplus G$
durchl"auft. Damit Bedingung \Ref{4_k23} erf"ullt ist, mu"s $G$ isomorph zu
$\1$, $U(1)$ oder $SU(2)$ sein. Im Fall $G \cong SU(2)$ darf die
nat"urliche Darstellung
\[ \rho \;:\; G \;\longrightarrow\; U(3) \]
nur aus Spin-$\frac{1}{2}$-Darstellungen aufgebaut sein. Das ist aber
bei einer Darstellung auf einem Raum ungerader Dimension unm"oglich.
Folglich haben wir $G \cong \1$ oder $G \cong U(1)$. In beiden F"allen
zerf"allt $\rho$ in die direkte Summe eindimensionaler Darstellungen.
Wir k"onnen also (nach einer geeigneten globalen Eichtransformation)
annehmen, da"s $U_{xy}$ unabh"angig von der Wahl der Potentiale
$B_{L\!/\!R}$ diagonal ist.
Die dynamischen Eichgruppen erh"alt man nun durch Berechnung
des Normalisators, was schlie"slich auf die Ergebnisse von Tabelle
\ref{4_tab5} f"uhrt.
\begin{table}
\caption{Dynamische Eichgruppen bei der Kombination Quark-/Leptonsektor}
\label{4_tab5}
\begin{tabular}{|c|c|c|} \hline
Homogenit"atsgrad & dynamische Eichgruppe & St"orung des Diracoperators \\
\hline \hline
$h=3,4$ & $U(1)_L \otimes U(3)$ & $\displaystyle{\chi_R
	(\Aslsh_L \oplus 0_3) + 0_1 \oplus B\!\slsh}$ \\
\hline
$h=5,6$ & $U(1)_L \otimes U(1)_L \otimes U(3)$ & \parbox{5.2cm}{
	\centerline{$\chi_R (\Aslsh_L \oplus 0_3) +
	\chi_R (0_1 \oplus B\!\slsh_L)$}
	\centerline{$+ 0_1 \oplus B\!\slsh$}} \\
& oder & \\
& \parbox{5.2cm}{\centerline{$U(1)_L \otimes U(1)_L \otimes U(1) \otimes U(2)$}}
	& \parbox{5.1cm}{$\chi_R (\Aslsh_L \oplus 0_3) + \chi_R \: C\!\slsh
	(0_1 \oplus \1_1 \oplus (-\1)_2) + 0_1 \oplus B\!\slsh_1 \oplus
	B\!\slsh_2$} \\
\hline
$h \geq 7$ & komplizierter & \\
\hline
\end{tabular}
\end{table}

Wir diskutieren kurz die Situation f"ur den Fall $h \geq 7$:
Die Bedingung \Ref{4_k23} schw"acht sich zu
\[ \# \left( (\sigma(U_{xy}) \cup \sigma(U_{xy}^*)) \setminus
	\{0\} \right) \;<\; \frac{h-1}{2} \]
ab. Folglich braucht die Gleichung \Ref{4_v2} nicht mehr erf"ullt zu sein.
F"ur $h=7, 8$ k"onnen wir beispielsweise bei der Zerlegung
$\C^16=\C^8 \oplus \C^8$ des Spinorraumes in Lepton- und Quarksektor
im Leptonsektor ein au"serdiagonales linksh"andiges Potential
$B \sigma^1$ (wie in Tabelle \ref{4_tab2}) und im Quarksektor
ein $SU(2)_L \otimes SU(2)_R$-Potential $B_{L\!/\!R}$ (wie in
Tabelle \ref{4_tab1}) einf"uhren.
Dieses Beispiel wird durch die St"orung
\Equ{4_k25}
i \Pdd \;\longrightarrow\; i \Pdd + \chi_R \:B\: \sigma^1 \oplus 0
	+ 0 \oplus (\chi_L \:B\!\slsh_L + \chi_R \:B\!\slsh_R)
\EndEqu
des Diracoperators beschrieben. Die Bestimmung der
dynamischen Eichgruppen wird dadurch erschwert, da"s die in \Ref{4_k24}
gew"ahlte Basis von $\C^4$ i.a. von den dynamischen Potentialen abh"angt.
Darauf wollen wir nicht n"aher eingehen.

Wir sehen, da"s das zu Beginn dieses Abschnittes aufgetretene Problem
allgemeinen Charakter hat: F"ur $h \leq 6$ findet keine spontane
Sektorbildung statt; der fermionische Projektor zerf"allt in die
direkte Summe einzelner $(4 \times 4)$-Bl"ocke, auf welchen die chiralen
Potentiale diagonal sind.
Auch f"ur $h \geq 7$ ist das Ergebnis physikalisch nicht sinnvoll,
selbst wenn wir wie im Beispiel \Ref{4_k25} eine Aufspaltung des fermionischen
Projektors in zwei $(8 \times 8)$-Sektoren annehmen. Die Potentiale in
den beiden Sektoren sind dann n"amlich voneinander unabh"angig und
k"onnen nicht sinnvoll miteinander in Beziehung gesetzt werden.

Dieses Problem h"angt letztlich damit zusammen, da"s die Eigenwerte
von $P(x,y) \: P(y,x)$ im Quark- und Leptonsektor 
selbst bei einer gleichartigen St"orung des Diracoperators nicht
"ubereinstimmen. 
Wir nennen dies die {\em{Nichtvertr"aglichkeit der
Eigenwerte}}\index{Nichtvertr"aglichkeit der Eigenwerte}
im Quark- und Leptonsektor.

\subsubsection*{der Ausweg: Massendrehung}
Um das Problem der Nichtvertr"aglichkeit der Eigenwerte zu l"osen, m"ussen
wir zu allgemeineren St"orungen des Diracoperators "ubergehen. Den
genauen Mechanismus nennen wir {\em{Massendrehung}}\index{Massendrehung}.
Wir k"onnen hier nur die Idee und die grundlegende Konstruktion
beschreiben, die detaillierten Rechnungen verschieben wir auf Abschnitt
\ref{5_abmass} (in Kapitel 5).
Daf"ur gibt es zwei Gr"unde: Zum einen m"ussen bei der Massendrehung
die Eich-/Pseudoeichterme h"oherer Ordnung in der Masse ber"ucksichtigt
werden. Au"serdem ist eine quantitative Behandlung erst bei mehreren
Teilchenfamilien sinnvoll. Es zeigt sich n"amlich, da"s die Familien
("ahnlich wie bei der CKM-Matrix im Standardmodell) miteinander gemischt
werden m"ussen.
Unsere etwas qualitative Diskussion ist deswegen ausreichend, weil es
in diesem \ref{kap4}. Kapitel noch nicht darum geht, alle Details auszuarbeiten.
Unser Ziel besteht zun"achst darin, ein anschauliches Verst"andnis zu
erhalten und gleichzeitig gen"ugend Informationen zusammenzutragen, um
die Gleichungen der diskreten Raumzeit eindeutig festlegen zu k"onnen.

Das Problem bei der dynamischen Eichgruppe \Ref{4_k20} besteht darin, da"s
alle dynamischen Potentiale in den $(4 \times 4)$-Bl"ocken des
freien fermionischen Projektors diagonal sind.
Wir wollen zun"achst an einem einfachen
Beispiel beschreiben, wie sich auch bei diagonalen Potentialen eine
Mischung der verschiedenen Fermionsorten realisieren l"a"st.
Dazu betrachten wir bei Spindimension $8$ ein System zweier
Fermionsorten $A$, $B$ unterschiedlicher Masse, also
\Equ{4_d1}
P(x,y) \;=\; \left[ \frac{1}{2} \: (p_{m_A} - k_{m_A})(x,y) \right]
	\oplus \left[ \frac{1}{2} \: (p_{m_B} - k_{m_B})(x,y) \right]
\EndEqu
mit $m_A \neq m_B$. Bei einer Zerlegung des Spinorraumes in der Form
$\C^8 = \C^4 \otimes \C^2_\iso$ nennen wir $\C^2_\iso$ den Isospinraum.
Der freie Projektor \Ref{4_d1} besitzt eine Massenasymmetrie
mit Massenmatrix
\[ Y \;=\; \frac{1}{m} \: \left( \begin{array}{cc}
	m_A & 0 \\ 0 & m_B \end{array} \right)_\iso \spc . \]
Wir betrachten zur Einfachheit nur eine $U(1)$-Untergruppe der
dynamischen Eichgruppe; die zugeh"orige St"orung des Diracoperators habe
die Form
\Equ{4_d5}
i \Pdd \;\longrightarrow\; i \Pdd \:+\: C\!\slsh \: \sigma^3_\iso \spc .
\EndEqu

Die Wellenfunktionen der Fermionen sind L"osungen der Diracgleichung
\Equ{4_d2}
(i \Pdd \:+\: C\!\slsh \: \sigma^3_\iso \:-\: m Y ) \: \Psi
	\;=\; 0 \spc .
\EndEqu
Da in dieser Gleichung alle Isospinmatrizen diagonal sind, entkoppelt
\Ref{4_d2} auf den beiden Isospinkomponenten. Wir interpretieren die
jeweiligen L"osungen als Wellenfunktionen der Teilchen $A$ bzw. $B$.
Wir wollen nun in \Ref{4_d2} eine Kopplung der beiden Fermionsorten
einf"uhren. Da die dynamische Eichgruppe $U(1)$ vorgegeben ist,
d"urfen wir dazu die Form des Potentials nicht verallgemeinern.
Wir k"onnen aber versuchen, die Massenmatrix zu ver"andern:
Es scheint nicht sinnvoll zu sein, die Parameter $m_{A\!/\!B}$
(also die Eigenwerte von $Y$) abzu"andern, weil dies in anschaulicher
physikalischer Vorstellung einer Massen- und damit Energieverschiebung
des gesamten Diracsees entsprechen w"urde (an der St"orungsrechnung
von $P(x,y)$ sieht man auch explizit, da"s eine solche Massenverschiebung
nicht auftreten darf).
Aber man kann die Massenmatrix unit"ar transformieren, also $Y$ gem"a"s
\Equ{4_d4}
Y \;\rightarrow\; U(x)^{-1} \:Y\: U(x)
\EndEqu
durch ein orts- und zeitabh"angiges Matrixfeld ersetzen, dabei ist
$U(x) \in U(2)$. Wir gehen also von \Ref{4_d2} zur Diracgleichung
\Equ{4_d00}
(i \Pdd \:+\: C\!\slsh(x) \: \sigma^3_\iso \:-\: m
	\: U(x)^{-1} \:Y\: U(x) ) \: \Psi \;=\; 0
\EndEqu
"uber. Nach Umschreiben dieser Gleichung in der Form
\[ \left( (i \Pdd \:+\: C\!\slsh(x) \: \sigma^3_\iso \:+\: m \:
	(Y - U(x)^{-1} \:Y\: U(x) ) \:-\: m Y \right) \: \Psi \;=\; 0 \]
k"onnen wir die Ersetzung \Ref{4_d4} durch eine zus"atzliche skalare
St"orung des Diracoperators beschreiben. Wir betrachten also in
Verallgemeinerung von \Ref{4_d5} die St"orung
\Equ{4_d6}
i \Pdd \;\longrightarrow\; i \Pdd \:+\: C\!\slsh \: \sigma^3_\iso
	\:+\: m \: (Y - U^{-1} \:Y\: U) \spc .
\EndEqu

Um die Diracgleichung \Ref{4_d00} besser interpretieren zu k"onnen, f"uhren
wir neue Wellenfunktionen
\Equ{4_d9}
\tilde{\Psi}(x) \;=\; U(x) \: \Psi(x)
\EndEqu
ein und erhalten f"ur $\tilde{\Psi}$ die Gleichung
\begin{eqnarray}
\label{eq:4_d10}
(i \Pdd + \tilde{A} - m Y) \Psi &=& 0 \spc {\mbox{mit}} \\
\label{eq:4_d8}
\tilde{A}^j &=& U \: C^j \: \sigma^3 \: U^{-1}
	+ i U (\partial^j U^{-1}) \spc .
\end{eqnarray}
Da die Massenmatrix in \Ref{4_d10} diagonal ist, haben wir wie in
\Ref{4_d2} die Interpretation, da"s die beiden Isospinkomponenten die
Teilchensorten $A$ bzw. $B$ beschreiben.
Im Vergleich zu \Ref{4_d2} tritt aber nun
ein allgemeineres Potential $\tilde{A}_{L\!/\!R}$ auf, das nicht mehr
notwendigerweise im Isospin diagonal ist.

An diesem Beispiel k"onnen wir bereits einige allgemeine Eigenschaften
der Massendrehung diskutieren. Wir haben die Mischung der Fermionsorten
gem"a"s \Ref{4_d6} durch eine skalare St"orung des Diracoperators
eingef"uhrt. Bei einer allgemeinen Massendrehung wird hier zus"atzlich
eine pseudoskalare St"orung auftreten.
Eine solche skalare/pseudoskalare St"orung wirkt sich bei der f"uhrenden
Singularit"at $\sim m^0$ von $P(x,y) \: P(y,x)$ nicht aus, sondern
geht erst in die Singularit"aten h"oherer Ordnung in $m$ ein.
F"ur die Diskussion der Eich-/Pseudoeichterme $\sim m^0$ spielt die
skalare/pseudoskalare St"orung in \Ref{4_d6} also keine Rolle. Da das
Potential in \Ref{4_d6} im Isospin diagonal ist, tritt das
Problem der Nichtvertr"aglichkeit der Eigenwerte nicht auf.

Nach der Transformation \Ref{4_d9} der Wellenfunktionen
erhalten wir in der Diracgleichung \Ref{4_d10} ein sogenanntes
{\em{effektives Eichpotential}}\index{Eichpotential, effektives}
$\tilde{A}$, das auch im Isospin au"serdiagonale Anteile enthalten kann.
Dieses Potential mu"s gegen"uber einem allgemeinen $U(2)$-Potentiale eine
spezielle Form \Ref{4_d8} haben. Wir nennen diese Einschr"ankung f"ur
die Wahl der effektiven Potentiale {\em{Eichbedingung}}\index{Eichbedingung}.

Die Einf"uhrung der skalaren/pseudoskalaren St"orung und die
anschlie"sende
Transformation \Ref{4_d9} der Wellenfunktionen mag zun"achst wie ein
Trick erscheinen, mit dem das Problem der Nichtvertr"aglichkeit der
Eigenwerte einfach auf die schw"acheren Singularit"aten $\sim m^k$,
$k \geq 1$, von $P(x,y) \: P(y,x)$ verlagert wird. Um zu sehen, da"s
unser Problem mit dieser Methode tats"achlich gel"ost werden kann,
mu"s man die Eich-/Pseudoeichterme h"oherer Ordnung in $m$ genau
studieren, was wir (wie gesagt) auf Abschnitt \ref{5_abmass} 
 (in Kapitel 5) verschieben.

Nach diesen Vorbereitungen k"onnen wir die allgemeine Konstruktion der
Massendrehung (bei einer Teilchenfamilie) beschreiben. Dazu betrachten wir
bei Spindimension $4b$, $b \geq 1$ einen freien fermionischen Projektor
mit chiraler Asymmetriematrix $X$ und Massenmatrix $Y$.
Die dynamische Eichgruppe sei $H \subset U(b)_L \otimes U(b)_R$; wir
f"uhren die zugeh"origen chiralen Potentiale $A_{L\!/\!R}(x) \in
{\mbox{Lie }}H$ durch die St"orung
\Equ{4_d12}
i \Pdd \;\longrightarrow\; i \Pdd + \chi_R \Aslsh_L + \chi_L \Aslsh_R
\EndEqu
des Diracoperators ein. Wir wollen die Ersetzung \Ref{4_d4} so
verallgemeinern, da"s die links- und rechtsh"andigen Komponenten
unabh"angig voneinander transformiert werden k"onnen. Dazu bilden wir
\Equ{4_d11}
Y \;\rightarrow\; \chi_L \: U_R(x)^{-1} \:Y\: U_L(x) \:+\:
	\chi_R \: U_L(x)^{-1} \:Y\: U_R(x)
\EndEqu
mit unit"aren Matrixfeldern $U_{L\!/\!R}(x) \in U(b)$.
Falls der fermionische Projektor eine chirale Asymmetrie besitzt, m"ussen
die Matrizen $U_{L\!/\!R}$ eine zus"atzliche Bedingung erf"ullen:
Gleichung \Ref{2_asymm} geht bei der Ersetzung \Ref{4_d11} in
\Equ{4_dx}
X_{L\!/\!R} \: U_{L\!/\!R}(x) \: Y \;=\; U_{L\!/\!R}(x) \: Y \spc
	{\mbox{f"ur alle $x$}}
\EndEqu
"uber.

Der Ansatz \Ref{4_d12} f"ur die Massenmatrix l"a"st sich folgenderma"sen
motivieren:
Zun"achst einmal mu"s die rechte Seite von \Ref{4_d11} eine hermitesche
Matrix sein. Eine naive Ersetzung der Art
\[ Y \;\rightarrow\; \chi_L \: U_L(x)^{-1} \:Y\: U_L(x) \:+\:
	\chi_R \: U_R(x)^{-1} \:Y\: U_R(x) \]
w"are beispielsweise nicht sinnvoll.
Au"serdem findet bei der Transformation \Ref{4_d11} "ahnlich wie bei
\Ref{4_d4} keine Massenverschiebung der Diracseen statt. Um das zu sehen,
betrachten wir die freie Diracgleichung mit einer Massenmatrix gem"a"s
der rechten Seite von \Ref{4_d11}
\Equ{4_d21}
(i \Pdd \:-\: m \:\chi_L\: U_R^{-1} \:Y\: U_L \:-\: m \:\chi_R\:
	U_L^{-1} \:Y\: U_R) \:\Psi \;=\; 0
\EndEqu
und nehmen an, da"s die Matrizen $U_{L\!/\!R}$ nicht von $x$
abh"angen. Wir setzen die Wellenfunktion
\Equ{4_d20}
\tilde{\Psi}(x) \;=\; (\chi_L \: U_L + \chi_R \: U_R) \: \Psi(x)
\EndEqu
in \Ref{4_d21} ein und multiplizieren die Gleichung au"serdem
von links mit der Matrix $\chi_L \: U_R + \chi_R U_L$.
Man erh"alt die freie Diracgleichung $(i \Pdd - m Y) \tilde{\Psi} = 0$.
Die Transformation der Massenmatrix \Ref{4_d11} wirkt sich also in der
freien Diracgleichung lediglich gem"a"s \Ref{4_d20} aus; die
Wellenzahl und Frequenz der Wellenfunktionen bleiben dabei unver"andert.
Tats"achlich ist \Ref{4_d11} die allgemeinste Transformation der
Massenmatrix mit dieser Eigenschaft.

Die Ersetzung \Ref{4_d11} kann auch durch eine zus"atzliche
ska\-la\-re/pseu\-do\-ska\-la\-re St"orung des Diracoperators
beschrieben werden, also gegen"uber \Ref{4_d12} durch
\Equ{4_d18}
i \Pdd \:\longrightarrow\: i \Pdd + \chi_R \Aslsh_L + \chi_L \Aslsh_R
	+ m \: (Y - \chi_L \: U_R(x)^{-1} \:Y\: U_L(x) -
	\chi_R \: U_L(x)^{-1} \:Y\: U_R(x)) \; . \spc
\EndEqu
Die Diracgleichung hat nun die Form
\Equ{4_d13}
(i \Pdd \:+\: \chi_R \: \Aslsh_L \:+\: \chi_L \: \Aslsh_R
	\:-\: m \:\chi_L\: U_R^{-1} \:Y\: U_L \:-\: m \:\chi_R\:
	U_L^{-1} \:Y\: U_R) \:\Psi \;=\; 0 \spc .
\EndEqu
Wir f"uhren ganz analog zu \Ref{4_d20} mit
\Equ{4_d14}
\tilde{\Psi}(x) \;=\; (\chi_L \: U_L(x) + \chi_R \: U_R(x)) \: \Psi(x)
\EndEqu
neue Wellenfunktionen ein und multiplizieren \Ref{4_d13}
von links mit $\chi_L \: U_R + \chi_R U_L$. Auf diese Weise wird
die Massenmatrix diagonal, und man erh"alt f"ur $\tilde{\Psi}$ die
Diracgleichung
\begin{eqnarray}
\label{eq:4_d15}
&&(i \Pdd + \chi_R \: \tilde{A}\!\slsh_L + \chi_L \: \tilde{A}\!\slsh_R
	- mY) \: \tilde{\Psi} \;=\; 0 \spc {\mbox{mit}} \\
\label{eq:4_d16}
&&\tilde{A}^j_{L\!/\!R} \;=\; U_{L\!/\!R} \: A_{L\!/\!R}^j \: U_{L\!/\!R}^{-1}
	\:+\: i U_{L\!/\!R} \: (\partial^j U_{L\!/\!R}^{-1})
\end{eqnarray}
In \Ref{4_d15} treten effektive Potentiale $\tilde{A}_{L\!/\!R}$ auf,
die durch die Eichbedingung \Ref{4_d16} genauer bestimmt werden.

\subsubsection*{die effektive Eichgruppe, mathematische Bedeutung
der Eichbedingung}
Wir wollen nun die Eichpotentiale $\tilde{A}_{L\!/\!R}$ in \Ref{4_d15}
und die Eichbedingung \Ref{4_d16} genauer mathematisch betrachten.
Da wir an dieser Stelle die Eich-/Pseudoeichterme h"oherer Ordnung
in der Masse noch nicht analysieren wollen, m"ussen wir dabei mit einem
allgemeinen Ansatz arbeiten: Aus der Untersuchung der Singularit"aten
$\sim m^k$, $k>0$ von $P(x,y) \: P(y,x)$ erh"alt man einschr"ankende
Bedingungen f"ur die Matrizen $U_{L\!/\!R}(x)$. Wir bezeichnen die
Menge aller Paare $(U_L, U_R)$, welche diese Bedingungen erf"ullen,
mit $T \subset U(b)_L \times U(b)_R$.
Die dynamischen Potentiale $A_{L\!/\!R} \in {\mbox{Lie }} H$ k"onnen
dabei unabh"angig von $(U_L(x), U_R(x)) \in T$ gew"ahlt werden.
Wir bezeichnen f"ur $t \in T$ die Projektion auf die
beiden Komponenten $U(b)_{L\!/\!R}$ mit $t_{L\!/\!R}$ und definieren f"ur
eine Teilmenge $A \subset U(b)_L \otimes U(b)_R$ eine Konjugation
\Equ{4_e1}
A^t \;=\; \left\{ (t_L \: a_L \: t_L^{-1}, \: t_R \: a_R \: t_R^{-1})
	{\mbox{ mit }} (a_L, a_R) \in A \right\} \spc .
\EndEqu

Wir betrachten zun"achst den Spezialfall, da"s die Matrizen $U_{L\!/\!R}$
nicht von $x$ abh"angen, was wir {\em{homogene
Massendrehung}}\index{Massendrehung, homogene} nennen.
Gleichung \Ref{4_d16} vereinfacht sich dann zu
\[ \tilde{A}^j_{L\!/\!R} \;=\; U_{L\!/\!R} \: A^j_{L\!/\!R}
	\: U_{L\!/\!R}^{-1} \spc . \]
Folglich tritt als effektive dynamische
Eichgruppe mit der Notation \Ref{4_e1} die Konjugationsgruppe
$H^{(U_L, U_R)}$ auf. Da $(U_L, U_R) \in T$ beliebig sein kann, setzen wir
\Equ{4_e5}
H^\eff \;=\; \bigcup_{t \in T} H^t
\EndEqu
und bezeichnen $H^\eff$ als {\em{effektive Eichgruppe}}\index{Eichgruppe,
effektive}.

Man beachte, da"s die effektive Eichgruppe i.a. keine Gruppe ist. F"ur
zwei Elemente $g_1 \in H^{t_1}$, $g_2 \in H^{t_2}$ aus verschiedenen
Konjugationsgruppen $t_1 \neq t_2$ kann n"amlich keine sinnvolle
Multiplikation in $H^\eff$ definiert werden. Als Folge der Eichbedingung
ist das jedoch kein mathematisches Problem:
Bei konstanter Massendrehung sind die effektiven Eichpotentiale
Elemente aus der Lie-Algebra ${\mbox{Lie }} H^{(U_L, U_R)}$ der
Konjugationsgruppe. Alle Multiplikationen k"onnen innerhalb der gleichen
Konjugationsgruppe ausgef"uhrt werden. Mit der Schreibweise
\[ \tintLR \:\;\;\;\;:=\; \Texp \left( -i \int_x^y \tilde{A}^j_{L\!/\!R} \:
	(y-x)_j \right) \]
haben wir n"amlich
\begin{eqnarray*}
\tintLR_x^y \:\;\;\; \tintLR_y^x \;\;\;&=& \left( U_{L\!/\!R} \intLR_x^y
	\;\;\; U_{L\!/\!R}^{-1} \right) \left( U_{L\!/\!R} \intLR_y^z
	\;\;\; U_{L\!/\!R}^{-1} \right) \\
&=& U_{L\!/\!R} \left( \intLR_x^y
	\;\;\; \intLR_y^z \;\;\: \right) U_{L\!/\!R}^{-1} \;\in\;
	H^{(U_L, U_R)} \spc .
\end{eqnarray*}

Falls $U_{L\!/\!R}$ von $x$ abh"angt, haben die effektiven Eichpotentiale
gem"a"s \Ref{4_d16} eine kompliziertere Struktur, denn der Summand
$i U_{L\!/\!R} (\partial^j U_{L\!/\!R}^{-1})$ mu"s zus"atzlich ber"ucksichtigt
werden. Unter Ausnutzung von \Ref{4_d16} lassen sich weiterhin
zeitgeordnete Exponentiale der effektiven Eichpotentiale bilden,
genauer
\Equ{4_e2}
\tintLR_x^y \;\;\;\;=\; U_{L\!/\!R}(x) \:\intLR_x^y\;\;\;\:
	U_{L\!/\!R}(y)^{-1} \spc .
\EndEqu
In eichinvarianten Produkten von \Ref{4_e2} heben sich die inneren Faktoren
$U_{L\!/\!R}$ weg, so da"s wir nur Multiplikationen in der dynamischen
Eichgruppe $H$ ausf"uhren m"ussen
\begin{eqnarray}
\tintLR_x^y \:\;\;\; \tintLR_y^x \:\;\;\;&=& \left( U_{L\!/\!R}(x)
	\intLR_x^y \;\;\; U_{L\!/\!R}(y)^{-1} \right) \left( U_{L\!/\!R}(y)
	\intLR_y^z \;\;\; U_{R\!/\!L}(z)^{-1} \right) \nonumber \\
\label{eq:4_e3}
&=& U_{L\!/\!R}(x) \left( \intLR_x^y
	\;\;\; \intLR_y^z \;\;\: \right) U_{L\!/\!R}(z)^{-1} \spc .
\end{eqnarray}
F"ur geschlossene Integrationswege erhalten wir auf diese Weise
Elemente aus der effektiven Eichgruppe, beispielsweise
\[ \tintLR_x^y \:\;\;\; \tintLR_y^z \:\;\;\; \tintLR_z^x \:\;\;\;\;=\;
	U_{L\!/\!R}(x) \left( \intLR_x^y \;\;\; \intLR_y^z \;\;\;
	\intLR_z^x \:\;\;\; \right) U_{L\!/\!R}(x) \;\in\;
	H^{(U_L(x), U_R(x))} \spc . \]
Wir kommen zu dem Ergebnis, da"s aufgrund der Eichbedingung
nur Multiplikationen innerhalb der gleichen Konjugationsgruppe
auftreten. Anders ausgedr"uckt, k"onnen die effektiven Eichpotentiale
mit Hilfe der Eichbedingung auf sinnvolle Weise global verkn"upft werden.

\subsubsection*{erwartete effektive Eichgruppen}
Nach diesen allgemeinen mathematischen Konstruktionen wollen wir
"uberlegen, wie die effektive Eichgruppe f"ur unser System \Ref{4_k0}
konkret aussehen sollte.
Die folgende Betrachtung ist mathematisch nicht rigoros und wegen nur
einer Teilchenfamilie auch stark vereinfacht; sie ist aber
im Wesentlichen richtig und nimmt einige Ergebnisse aus Abschnitt
\ref{5_abmass}  (in Kapitel 5) qualitativ vorweg.

Die Eich-/Pseudoeichterme $\sim m^0$ werden durch die skalare/pseudoskalare
St"orung in \Ref{4_d18} nicht beeinflu"st, so da"s wir immer noch die
dynamischen Eichgruppen von Tabelle \ref{4_tab5} erhalten.
In die Singularit"at $\sim m^2$ von $P(x,y) \: P(y,x)$
gehen jedoch die Matrizen $U_{L\!/\!R}$ ein.
Man kann die Eigenwerte von $P(x,y) \: P(y,x)$ in dieser Ordnung
"ahnlich wie f"ur die f"uhrende Singularit"at $\sim m^0$ diskutieren.
In Abh"angigkeit des Homogenit"atsgrades mu"s bei den Eigenwerten
wieder eine Entartung auftreten. Es ist in Analogie zu Lemma \ref{4_lemma2}
plausibel, da"s die Matrix $P(x,y) \: P(y,x)$ als Folge dieser Entartung
global in die direkte Summe mehrerer Untermatrizen zerfallen mu"s.
Folglich erwarten wir eine spontane Sektorbildung.
Unter dieser Annahme haben $A_{L\!/\!R}$, $U_{L\!/\!R}$ die spezielle
Form
\[ A_{L\!/\!R} \;=\; A_{L\!/\!R}^\lep \oplus A_{L\!/\!R}^\qu \spc,\spc
	U_{L\!/\!R} \;=\; U_{L\!/\!R}^\lep \oplus U_{L\!/\!R}^\qu
	\spc . \]
Die dynamischen $U(2)$-Potentiale $A^\lep_{L\!/\!R}$,
$A^\qu_{L\!/\!R}$ m"ussen gem"a"s Tabelle \ref{4_tab5} diagonal sein.
Da $U(1)$-Massendrehungen bei der Berechnung der effektiven Eichgruppe
gem"a"s \Ref{4_e5} wegfallen, k"onnen wir annehmen, da"s
$U^\lep_{L\!/\!R}$, $U^\qu_{L\!/\!R}$ unit"are $SU(2)$-Matrixfelder sind.
Die Zusatzbedingung \Ref{4_dx} impliziert, da"s die Matrix $U^\lep_R$
mit $F_2$ kommutiert, also diagonal ist.
Da wir im Quark- und Leptonsektor eine vergleichbare
Massendrehung erwarten, mu"s auch $U^\qu_R$ diagonal sein.
Die linksh"andigen Massenmatrizen $U^\lep_L = U^\qu_L \in U(2)$ k"onnen
dagegen beliebige Werte annehmen. Wir erwarten also f"ur die Menge $T$
\Equ{4_i2}
T \;=\; \left\{ (U \oplus U) \;,\;\; U \in SU(2) \right\} \;\times\;
	\left\{ \exp(i \vartheta) \: (\sigma^3 \oplus \sigma^3)
	\;,\;\;\; \vartheta \in \R \right\} \spc .
\EndEqu
Nach Diagonalisierung der Massenmatrizen erh"alt man die Diracgleichung
\Ref{4_d15} mit effektiven Eichpotentialen
\[ \tilde{A}_{L\!/\!R} \;=\; \tilde{A}^\lep_{L\!/\!R} \oplus
	\tilde{A}^\qu_{L\!/\!R} \spc , \]
welche die Eichbedingung \Ref{4_d16} erf"ullen.
Nun k"onnen wir die effektiven Eichgruppen mit Hilfe der
Definitionsgleichung \Ref{4_e5} und \Ref{4_i2} sowie den dynamischen
Eichgruppen aus Tabelle \ref{4_tab5} bestimmen.
Dabei ist zu beachten, da"s die Massendrehung \Ref{4_i2} nur dann
sinnvoll eingef"uhrt werden kann, wenn die dynamischen Eichfelder
die Zerlegung von $P$ in den Quark- und Leptonsektor respektieren.
Um dies zu erreichen, k"onnen wir auch von $H$ zu einer Untergruppe
der dynamischen Eichgruppe "ubergehen. Die Ergebnisse f"ur die
effektiven Eichgruppen sind in Tabelle \ref{4_tab6} zusammengestellt.
\begin{table}
\caption{Erwartete effektive Eichgruppen bei der Kombination des
Quark- und Leptonsektors}
\label{4_tab6}
\begin{tabular}{|c|c|c|} \hline
Homogenit"atsgrad & erwartete eff. Eichgruppe & eff. St"orung des
	Diracoperators \\
\hline \hline
$h=3,4$ & $U(1)_L \otimes U(3)$ & $\displaystyle{\chi_R
	(\Aslsh_L \oplus 0_3) + 0_1 \oplus B\!\slsh}$ \\
\hline
$h=5,6$ & $U(1)_L \otimes U(1)_L \otimes U(3)$ & \parbox{5.2cm}{
	\centerline{$\chi_R (\Aslsh_L \oplus 0_3) +
	\chi_R (0_1 \oplus B\!\slsh_L)$}
	\centerline{$+ 0_1 \oplus B\!\slsh$}} \\
& oder & \\
& \parbox{5.2cm}{\centerline{$U(1)_L \otimes U(1)_L \otimes U(1) \otimes U(2)$}}
	& \parbox{5.1cm}{$\chi_R (\Aslsh_L \oplus 0_3) + \chi_R \: C\!\slsh
	(0_1 \oplus \1_1 \oplus (-\1)_2) + 0_1 \oplus B\!\slsh_1 \oplus
	B\!\slsh_2$} \\
& oder & \\
& \parbox{5.2cm}{$SU(2)_L \otimes U(1)_R \otimes U(1)^\lep \otimes U(1)^\qu$}
	& \parbox{5.1cm}{ \centerline{$\chi_R (\tilde{A}\!\slsh_L \oplus
	 \tilde{A}\!\slsh_L) + \chi_L \: B\!\slsh_R \: (\sigma^3 \oplus \sigma^3)$}
	\centerline{ $C\!\slsh \: (\1_2 \oplus 0_2) + D\!\slsh \: (0_2 \oplus \1_2)$}} \\
\hline
$h \geq 7$ & komplizierter & \\
\hline
\end{tabular}
\end{table}
Wir haben zur Deutlichkeit nur diejenigen effektiven Potentiale mit
einer Tilde `$\:\tilde{ }\:$' versehen, in welche die Massendrehung auch
tats"achlich eingeht.

Es f"allt auf, da"s die erhaltenen effektiven Eichgruppen entgegen unserer
allgemeinen Diskussion doch Gruppen sind. Das ist zwar eine Vereinfachung, spielt
aber keine grundlegende Rolle. Interessanter ist, da"s
diese Gruppen (im Fall einer sinnvollen Zerlegung in Quark- und Leptonsektor)
mit den nach dem Schnitt der "au"seren Eichgruppen
erwarteten dynamischen Eichgruppen von Tabelle \ref{4_tab4} "ubereinstimmen.
Allerdings ist das auch nicht erstaunlich:
Tabelle \ref{4_tab4} gibt die maximale dynamische Eichgruppe an,
die nach rein gruppentheoretischen "Uberlegungen auftreten kann.
Es ist klar, da"s die effektive Eichgruppe in dieser maximalen
dynamischen Eichgruppe enthalten sein mu"s.
Umgekehrt ist es einsichtig, da"s mit den zus"atzlichen
Freiheitsgraden der Matrizen $U_{L\!/\!R}$ die Gruppen aus Tabelle
\ref{4_tab4} tats"achlich als effektive Eichgruppen realisiert werden
k"onnen.

\subsubsection*{physikalische Interpretation}
Abschlie"send wollen wir die Konstruktion der Massendrehung und der
effektiven Eichgruppe erl"autern und physikalisch diskutieren.

Durch Einf"uhrung der Massendrehung sind wir von einer reinen
Wechselwirkung durch chirale Eichfelder, \Ref{4_d12}, zu einer allgemeineren
Form der Wechselwirkung "ubergegangen. Diese allgemeinere Wechselwirkung
wird gem"a"s \Ref{4_d18} durch eine zus"atzliche skalare/pseudoskalare
St"orung beschrieben. Der Nachteil der zugeh"origen Diracgleichung
\Ref{4_d13} besteht darin, da"s die Massenmatrix nicht diagonal ist, so
da"s die Gleichung nur schwer physikalisch interpretiert werden kann.
Aus diesem Grund haben wir mit Hilfe der Transformation \Ref{4_d14} die
Massenmatrix global diagonalisiert und die Diracgleichung \Ref{4_d15}
erhalten.

Wir betonen noch einmal, da"s die Einf"uhrung der Wellenfunktion
$\tilde{\Psi}$ und die Transformation der Diracgleichung gem"a"s
\Ref{4_d15} lediglich zur besseren Anschauung dient.
Gleichung \Ref{4_d14} beschreibt keine Eichtransformation. Es ist
wichtig zu beachten, da"s der fermionische Projektor aus den negativen
Energiezust"anden $\Psi$, nicht aber aus $\tilde{\Psi}$, aufgebaut ist.

Nach Diagonalisierung der Massenmatrix treten in \Ref{4_d15} nur noch
chirale Potentiale auf, so da"s die Diracgleichung wieder die gewohnte
Form hat. Als Nachteil haben wir jedoch die Eichbedingung \Ref{4_d16}
erhalten, die nur schwierig zu handhaben ist.

Um einen ersten Eindruck von der Wechselwirkung zu erhalten, betrachten
wir den Grenzfall, da"s der zweite Summand in \Ref{4_d16} gegen"uber
dem ersten vernachl"assigbar ist,
\Equ{4_i1}
U_{L\!/\!R} \: A_{L\!/\!R}^j \: U_{L\!/\!R}^{-1}
	\;\gg\; i U_{L\!/\!R} \: (\partial^j U_{L\!/\!R}^{-1}) \spc .
\EndEqu
Diese N"aherung ist sinnvoll, wenn $U_{L\!/\!R}$ nur auf einer gro"sen
L"angenskala von $x$ abh"angt, was wir
{\em{quasihomogene Massendrehung}}\index{Massendrehung,
quasihomogene}
nennen. In diesem Fall k"onnen wir die effektiven Potentiale
aus der Lie-Algebra der effektiven Eichgruppe
\[ {\mbox{Lie }} H^\eff \;:=\; \bigcup_{t \in T} {\mbox{Lie }} H^t \]
beliebig w"ahlen; es ist lediglich darauf zu achten, da"s die
Konjugationsalgebra ${\mbox{Lie }} H^t(x) \ni (\tilde{A}_L(x),
\tilde{A}_R(x))$ nur wenig in $x$ variiert.
Im Grenzfall quasihomogener Massendrehung bereitet die Eichbedingung
also keine Schwierigkeiten. Wir erhalten einen direkten
Zusammenhang zu einer reinen Eichwechselwirkung, wobei
die effektive Eichgruppe die Rolle der gew"ohnlichen Eichgruppe
"ubernimmt.

Ohne die N"aherung \Ref{4_i1} ist es nicht mehr sinnvoll, mit der
effektiven Eichgruppe zu arbeiten, wodurch die Situation wesentlich
komplizierter wird. Wir k"onnen die Wechselwirkung nicht
auf einfache Weise physikalisch interpretieren.

Gl"ucklicherweise scheint die quasihomogene Massendrehung in den
wichtigen physikalischen Situationen eine gute N"aherung zu sein:
Die Matrix $U_L(x)$ gibt das Amplitudenverh"altnis der
$\gamma, Z-$ mit den $W^\pm$-Eichfeldern an. Falls in einem
physikalischen System einzelne Eichbosonen beobachtet werden, kann
$U_L$ in diesen Regionen der Raumzeit konstant gew"ahlt werden,
so da"s \Ref{4_i1} erf"ullt ist. Diese N"aherung verliert nur dann
ihre G"ultigkeit, wenn am gleichen Ort und gleichzeitig mehrere
verschiedene Eichbosonen auftreten, beispielsweise bei der
Bildung hochenergetischer Jets in einem Beschleunigerexperiment.
In diesem Fall ist die physikalische Situation aber sehr kompliziert
und kann meist nur mit ph"anomenologischen Modellen beschrieben werden.
Darum k"onnen wir nur schwer absch"atzen, ob und in welcher Weise
sich die Eichbedingung in Experimenten auswirkt.
Zur Einfachheit werden wir stets in der N"aherung quasihomogener
Massendrehung arbeiten.

Nach diesen "Uberlegungen k"onnen wir die effektiven Eichgruppen aus
Tabelle \ref{4_tab6}
als die Gruppen einer lokalen Eichtheorie auffassen.
Aus physikalischer Sicht ist besonders interessant, da"s in Tabelle
\ref{4_tab6} die Gruppe $SU(2)_L \otimes U(1)^\lep \otimes U(1)^\qu$
auftritt. Die $SU(2)_L$ koppelt an die Fermionen genau wie die
$SU(2)$ der elektroschwachen Wechselwirkung an. Die $U(1)$ der
GSW-Theorie ist in der $U(1)^\lep \otimes U(1)^\qu$ enthalten.
Damit ist das Problem der Nichtvertr"aglichkeit der Eigenwerte
gel"ost; wir scheinen wieder auf dem richtigen Weg zu sein.

Zusammenfassend ist zu sagen, da"s wir mit Einf"uhrung der Massendrehung
aus mathematischer Sicht eine wesentliche "Anderung gegen"uber den
Eichwechselwirkungen des Standardmodells vornehmen mu"sten.
In "ublichen physikalischen Situationen sollte sich dieser Unterschied
aber nicht auswirken.
Nat"urlich ist es ein interessantes und wichtiges Problem, ob
die Massendrehung experimentell beobachtet werden kann.
Da wir im Moment dabei sind, einen ersten Zusammenhang zwischen unserem
Konzept und dem Standardmodell herzustellen, handelt es sich dabei aber
doch um eine Detailfrage, die wir zwar erw"ahnen, aber nicht n"aher
verfolgen k"onnen.

\subsection{Kombination dreier vereinfachter Quarksektoren mit einem
vereinfachten Leptonsektor}
\label{4_ab46}
Wir w"ahlen bei Spindimension 32 als freien Projektor die direkte
Summe des Leptonsektors \Ref{4_la} mit drei Quarksektoren \Ref{4_va}
\Equ{4_d1z}
P(x,y) \;=\; \left[ \chi_L \: \frac{1}{2} \: (p_0 - k_0)(x,y) \right]
	\:\oplus\; \left[ \frac{1}{2} \: (p_m - k_m)(x,y) \right]^7 \spc .
\EndEqu
Abgesehen davon, da"s wir zur Einfachheit nur mit einer Teilchenfamilie
arbeiten, ist dieses System genau aus den Fermionsorten des
Standardmodells aufgebaut. Wir hoffen daher, bei unserer Untersuchung
die Eichgruppen und relativen Kopplungen des Standardmodells
wiederzufinden.

Die Analyse des fermionischen Projektors setzt sich aus mehreren Schritten
zusammen, die alle bereits bei der Diskussion vorheriger Systeme
aufgetreten sind. Daher k"onnen wir uns recht knapp fassen und erhalten
so eine Zusammenstellung der wichtigsten Konstruktionen dieses
Abschnitts \ref{4_ab4}.

Der freie Projektor \Ref{4_d1z} besitzt eine chirale Asymmetrie und eine
Massenasymmetrie; bei einer Zerlegung $\C^{32}=\C^4 \oplus \C^{28}$ des
Spinorraumes haben die Asymmetriematrizen die Form
\[ X \;=\; \chi_L \oplus \1 \;\;\;,\spc Y \;=\; 0 \oplus \1 \spc . \]
Den ersten direkten Summanden in \Ref{4_d1} nennen wir Neutrinoblock.
Wir f"uhren chirale Eichpotentiale ein: Nach dem Effekt des Pinning
darf in der rechtsh"andigen Komponente von $P(x,y)$ keine Mischung
des Neutrinoblocks mit den massiven Fermionbl"ocken stattfinden.
Die St"orung des Diracoperators mu"s also die Form
\Equ{4_d3}
i \Pdd \;\longrightarrow\; i \Pdd + \chi_R \: \Aslsh_L + \chi_L
	(0 \oplus B \!\slsh_R)
\EndEqu
mit einem $U(8)$-Potential $A_L$ und $U(7)$-Potential $B_R$ haben.

Um eine erste Vorstellung von der Wechselwirkung zu erhalten, berechnen
wir, welche dynamischen Eichfreiheitsgrade gem"a"s dem Schnitt der
"au"seren Eichgruppen zu erwarten sind:
Wir zerlegen den Spinorraum gem"a"s $\C^{32}=\C^8 \oplus \C^{24}$ in
einen Lepton- und drei Quarksektoren und betrachten gegen"uber
\Ref{4_d3} nur St"orungen des Diracoperators, die auf diesen beiden
Summanden invariant sind, also
\Equ{4_d4z}
i \Pdd \;\longrightarrow\; i \Pdd \:+\: \chi_R \: (\Aslsh_L^\lep \oplus
	\Aslsh_L^\qu) \:+\: \chi_L \: ((0 \oplus \Aslsh_R^\el) \oplus
	\Aslsh_R^\qu)
\EndEqu
mit $U(6)$-Potentialen $A^\qu_{L\!/\!R}$, einem $U(2)$-Potential
$A^\lep_L$ und einem $U(1)$-Potential $A^\el_R$.
Bei der St"orung \Ref{4_d4z} zerf"allt der freie Projektor in die
direkte Summe des Leptonsektors und der Quarksektoren.
Wir k"onnen beide direkten Summanden gem"a"s Abschnitt \ref{4_vlep}
und Abschnitt \ref{4_ab44}
(f"ur $s=3$) getrennt untersuchen und erhalten die dynamischen
Eichgruppen $H^\lep$, $H^\qu$ von Tabelle \ref{4_tab2} bzw. Tabelle
\ref{4_tab3}. Wir faktorisieren $H^\lep$, $H^\qu$
gem"a"s \Ref{4_k16} in innere und "au"sere
Eichgruppen. Schlie"slich berechnen wir die zu erwartenden dynamischen
Eichgruppen mit Hilfe von \Ref{4_k18}.
Die Ergebnisse sind in Tabelle \ref{4_tab7} zusammengestellt.
\begin{table}
\caption{Gem"a"s dem Schnitt der "au"seren Eichgruppen erwartete
	dynamische Eichgruppen bei der Kombination dreier Quarksektoren
	mit einem Leptonsektor}
\label{4_tab7}
\begin{center}
\begin{tabular}{|c|c|} \hline
Homogenit"atsgrad & erwartete dynamische Eichgruppe \\
\hline \hline
$h=3,4$ & $U^(1)^\lep \otimes U(1)^\lep \otimes U(6)^\qu$ \\
\hline
$h = 5,6$ & $U^(1)^\lep \otimes U(1)^\lep \otimes U(6)^\qu$ \\
& {\mbox{oder}} \\
& $U(1)^\lep \otimes U(p)^\qu \otimes U(q)^\qu \otimes
	U(1)_L$ \\
& {\mbox{mit $p+q=6$}} \\
& {\mbox{oder}} \\
& $U^(1)_\lep \otimes U(3)^\qu \otimes SU(2)_L \otimes U(1)_R$ \\
\hline
$h \geq 7$ & komplizierter \\
\hline
\end{tabular}
\end{center}
\end{table}

Um die dynamischen Eichgruppen mathematisch zu bestimmen, untersucht
man f"ur die St"orung \Ref{4_d3} die Auswirkung der Eich-/Pseudoeichterme
$\sim m^0$ auf die Eigenwerte der Matrix $P(x,y) \: P(y,x)$.
Ganz analog wie bei der Kombination des Quark- und Leptonsektors
\Ref{4_k0} tritt das Problem der Nichtvertr"aglichkeit der Eigenwerte
auf. Man erh"alt im Gegensatz zu den erwarteten Ergebnissen von
Tabelle \ref{4_tab7} die dynamischen Eichgruppen von Tabelle \ref{4_tab8}.
\begin{table}
\caption{Dynamische Eichgruppen bei der Kombination dreier Quarksektoren
	mit einem vereinfachten Leptonsektor}
\label{4_tab8}
\begin{tabular}{|c|c|c|} \hline
Homogenit"atsgrad & dynamische Eichgruppe & St"orung des Diracoperators \\
\hline \hline
$h=3,4$ & $U(1)_L \otimes U(7)$ & $\displaystyle{\chi_R
	(\Aslsh_L \oplus 0_7) + 0_1 \oplus B\!\slsh}$ \\
\hline
$h=5,6$ & \parbox{4.0cm}{$U(1)_L \otimes U(1)_L \otimes U(p) \otimes U(q)$
	\centerline{mit $p+q=7$}}
	& \parbox{6.7cm}{\centerline{$\chi_R (\Aslsh_L \oplus 0_7) +
	\chi_R \: B\!\slsh \: (0_1 \oplus \1_p \oplus (- \1_q))$}
	\centerline{$+ 0_1 \oplus C\!\slsh_p \oplus D\!\slsh_q$}} \\
\hline
$h \geq 7$ & komplizierter & \\
\hline
\end{tabular}
\end{table}
Es ist offensichtlich nicht sinnvoll, diese dynamischen Eichgruppen
als physikalische Eichgruppen zu interpretieren. Insbesondere tritt
keine spontane Sektorbildung auf.

Der Grund f"ur dieses scheinbare Problem liegt darin, da"s der Ansatz
f"ur die St"orung des Diracoperators \Ref{4_d3} zu speziell ist.
Wir m"ussen gem"a"s \Ref{4_d18} zus"atzliche skalare/pseudoskalare
St"orungen ber"ucksichtigen, was wir als Massendrehung bezeichnen.
Nach globaler Diagonalisierung der Massenmatrix erh"alt man die
gewohnte Beschreibung der Wechselwirkung durch chirale Eichfelder
\Ref{4_d15}. Die mathematischen und begrifflichen Schwierigkeiten der
auftretenden Eichbedingung \Ref{4_d16} brauchen in der physikalisch
sinnvollen N"aherung quasihomogener Massendrehung nicht ber"ucksichtigt
zu werden; die Wechselwirkung kann als lokale Eichtheorie mit effektiver
Eichgruppe $H^\eff$ beschrieben werden.

Die skalare/pseudoskalare St"orung in \Ref{4_d16} wirkt sich in der
Singularit"at $\sim m^2$ von $P(x,y) \: P(y,x)$ aus. Wie im
vorangehenden Abschnitt \ref{4_ab45} qualitativ begr"undet wurde,
erwarten wir als Ergebnis der spektralen Analyse von $P(x,y) \: P(y,x)$
eine spontane Sektorbildung. Genauer m"ussen die unit"aren
$U(8)$-Matrizen $U_{L\!/\!R}$ die Form
\[ (U_L, U_R) \;\in\; T \;=\; (SU(2))^4 \:\times\:
	\left\{ \exp (i \vartheta) \: (\sigma^3)^4 \;\;,\;\;\;
	\vartheta \in \R \right\} \]
haben. Schlie"slich bestimmen wir die effektive Eichgruppe mit Hilfe von
\Ref{4_e5} und erhalten die Ergebnisse von Tabelle \ref{4_tab9}.
\begin{table}
\caption{Erwartete effektive Eichgruppen bei der Kombination dreier
Quarksektoren mit einem Leptonsektor}
\label{4_tab9}
\begin{tabular}{|c|c|c|} \hline
Homogenit"atsgrad & erwartete eff. Eichgruppe & eff. St"orung des
	Diracoperators \\
\hline \hline
$h=3,4$ & $U(1)_L \otimes U(7)$ & $\displaystyle{\chi_R
	(\Aslsh_L \oplus 0_7) + 0_1 \oplus B\!\slsh}$ \\
\hline
$h=5,6$ & \parbox{5.0cm}{$U(1)_L \otimes U(1)_L \otimes U(p) \otimes U(q)$
	\centerline{mit $p+q=7$}}
	& \parbox{5.1cm}{$\chi_R (\Aslsh_L \oplus 0_7) +
	\chi_R \: B\!\slsh \: (0_1 \oplus \1_p \oplus (- \1_q))
	+ 0_1 \oplus C\!\slsh_p \oplus D\!\slsh_q$} \\
& oder & \\
& $SU(2)_L \otimes U(1)_R \otimes U(1)^\qu \otimes U(3)^\qu$
	& \parbox{5.1cm}{$(\chi_R \: \tilde{A}\!\slsh_L +
	\chi_L \: \tilde{B}\!\slsh_R)^4 + C\!\slsh \: (\1_2 \oplus 0_6)$
	$+ D\!\slsh \: (0_2 \oplus \1_6))$} \\
\hline
$h \geq 7$ & komplizierter & \\
\hline
\end{tabular}
\end{table}
Die effektiven Eichgruppen stimmen in den F"allen mit Sektorbildung
mit den nach dem Schnitt der "au"seren Eichgruppen erwarteten Ergebnissen
von Tabelle \ref{4_tab7} "uberein, zus"atzlich erhalten wir Aussagen
"uber die Kopplung der Eichfelder an die Fermionen.
Die genauen Rechnungen zur Massendrehung sind erst bei mehreren
Teilchenfamilien sinnvoll und wurden auf Abschnitt \ref{5_abmass}
 (in Kapitel 5) verschoben. Unsere Diskussion beschreibt die Situation aber im Wesentlichen
richtig.

Die Entscheidung zwischen den verschiedenen M"oglichkeiten f"ur die
effektiven Eichgruppen bei $h=5,6$ wird durch globale Bedingungen
festgelegt; man hat also in jedem Fall in der ganzen Raumzeit die
gleichen effektiven Eichgruppen.
Es ist zwar unbefriedigend, da"s wir uns im Moment
willk"urlich f"ur eine der m"oglichen effektiven Eichgruppen entscheiden
m"ussen, auf der anderen Seite ist die Wahl aus physikalischer
Sicht ganz eindeutig. Wir k"onnen die effektive Eichgruppe bereits
durch eine sehr allgemeine physikalische Forderung festlegen,
beispielsweise durch die Bedingung, da"s eine ganz beliebige
Wechselwirkung der Neutrinos mit den massiven Fermionen stattfindet.

In diesem Sinne sind wir zu einem interessanten Ergebnis gekommen:
Ausgehend von dem freien fermionischen Projektor \Ref{4_d1} und
dem homogenen Polynomansatz f"ur die Gleichungen der diskreten Raumzeit
erhalten wir, da"s sich die Dynamik des Systems mit
einer lokalen Eichtheorie beschreiben l"a"st; dabei ist f"ur $h=5,6$
die Eichgruppe eine Untergruppe von
\[ SU(2)_L \otimes SU(3)^\qu \otimes U(1)_R \otimes U(1)^\lep
	\otimes U(1)^\qu \spc . \]
Es findet eine spontane Sektorbildung des fermionischen Projektors
in Lepton- und Quarksektoren statt. Die effektiven $SU(2)_L$- und
$SU(3)$-Eichfelder koppeln genau wie die entsprechenden Eichfelder
des Standardmodells an die Fermionen an.
Die $U(1)$-Eichgruppe der GSW-Theorie ist in der $U(1)_R
\otimes U(1)^\lep \otimes U(1)^\qu$-Gruppe enthalten.
Da aus der Untersuchung der Feldgleichungen weitere Bedingungen an
die Eichfelder zu erwarten sind, besteht die Hoffnung, da"s sich
dann auch diese Gruppe auf nat"urliche Weise ergibt.

Dieses Ergebnis ist nach den bisher eher indirekten oder
qualitativen Hinweisen eine erste klare Best"atigung f"ur das
Prinzip des fermionischen Projektors.

\section{(Die Feldgleichungen f"ur effektive Eichstr"ome)}
\label{4_ab5}
Dieser Abschnitt ist noch nicht fertig. Es sollen dort alle f"ur
mathematisch sinnvolle Feldgleichungen notwendigen
zus"atzlichen Bedingungen hergeleitet werden.

\section{(Bestimmung des Homogenit"atsgrades)}
\label{4_ab6}
Dieser Abschnitt ist noch nicht fertig. Es wird dort mit einer
Dimensionsbetrachtung der Homogenit"atsgrad festgelegt,
man erh"alt $h=12$.

\chapter{Einige Ergebnisse aus den Anh"angen}
\section{Anhang A: St"orungsrechnung f"ur $k_0$ im Ortsraum}
\subsection{Elektromagnetisches Potential}
\begin{Thm}
\label{a1_theorem1}
   In erster Ordnung St"orungstheorie gilt
\begin{eqnarray}
\label{eq:a1_111}
 \Delta k_0(x,y) &=&  - i e \left( \int_x^y A_j \right) \: \xi^j \; k_0(x,y) \\
\label{eq:a1_112}
   && + \frac{i e}{8 \pi^2} \left( \int_x^y (\alpha^2-\alpha) \; \xi \slsh
	\: \xi^k \: j_k \right) \;\; \left( l^\vee(\xi) - l^\wedge(\xi)
	\right) \\
\label{eq:a1_113}
   && - \frac{i e}{8 \pi^2} \left( \int_x^y (2 \alpha - 1) \; \xi^j \:
	\gamma^k \: F_{kj} \right) \;\; \left( l^\vee(\xi) - \l^\wedge(\xi)
	\right) \\
\label{eq:a1_114}
   && + \frac{e}{16 \pi^2} \left( \int_x^y \varepsilon^{ijkl} \; F_{ij} \:
	\xi_k \; \rho \gamma_l \right) \;\; \left( l^\vee(\xi) -l^\wedge(\xi)
	\right) \\
\label{eq:a1_115}
   && - \frac{i e}{16 \pi^3} \left( \lint_x^y - \lint_y^x \right) dz
	\int_x^z (\alpha^4 - \alpha^3) \; \zeta \slsh \;
	\zeta_k \; \Box j^k \\
\label{eq:a1_116}
   && + \frac{i e}{16 \pi^3} \left( \lint_x^y - \lint_y^x \right) dz
	\int_x^z (4 \alpha^3 - 3 \alpha^2) \; \zeta^j \: \gamma^k
	\; \Box F_{kj} \\
\label{eq:a1_117}
   && - \frac{e}{32 \pi^3} \left( \lint_x^y - \lint_y^x \right) dz
	\int_x^z \alpha^2 \; \varepsilon^{ijkl} \; (\Box F_{ij}) \:
	\zeta_k \; \rho \gamma_l \\
\label{eq:a1_118}
   && + \frac{i e}{4 \pi^3} \left( \lint_x^y - \lint_y^x \right) dz
	\int_x^z (2 \alpha^2 - \alpha) \; \gamma^k \: j_k \spc ,
\end{eqnarray}
   wobei $F_{jk}=\partial_j A_k - \partial_k A_j$ den elektromagnetischen
   Feldst"arketensor und $j^k = \partial_l F^{kl}$ den Maxwell-Strom
   bezeichnet. Zur Abk"urzung wurde $\xi = y-x$ und $\zeta = z-x$ gesetzt.
\end{Thm}

\begin{Satz}
\label{a1_randwert}
   F"ur $y-x \in \Li$ gilt
\begin{eqnarray}
\label{eq:a1_121}
  \lim_{\I_x \ni u \rightarrow y} \tilde{k}_0(x,u) &=&
  + \frac{i e}{64 \pi^2} \: \epsilon(\xi^0) \int_x^y (\alpha^4 - 2 \alpha^3
	+ \alpha^2) \: \xi \slsh \: \xi_k \; \Box j^k \\
  && - \frac{i e}{64 \pi^2} \: \epsilon(\xi^0) \int_x^y (4 \alpha^3
	- 6 \alpha^2 + 2 \alpha) \: \xi^j \: \gamma^k \: (\Box F_{kj}) \\
  && + \frac{e}{64 \pi^2} \: \epsilon(\xi^0) \int_x^y (\alpha^2 - \alpha) \;
	\varepsilon^{ijkl} \: (\Box F_{ij}) \: \xi_k \; \rho \gamma_l \\
\label{eq:a1_122}
  && - \frac{i e}{8 \pi^2} \: \epsilon(\xi^0) \int_x^y (\alpha^2 - \alpha)
	\: \gamma^k \: j_k \spc ,
\end{eqnarray}
wobei wieder $\xi = y-x$ gesetzt wurde.
\end{Satz}

\subsection{Gravitationsfeld}
\begin{Thm}
\label{a1_theorem2}
In erster Ordnung St"orungstheorie gilt in symmetrischer Eichung
\begin{eqnarray}
\label{eq:a1_212}
\lefteqn{\Delta k_0(x,y) \;=\; - \left( \int_x^y h^k_j \right) \xi^j
	\frac{\partial}{\partial y^k} \;\; k_0(x,y) } \\
\label{eq:a1_213}
  && - \frac{1}{4 \pi^2} \left( \int_x^y (2 \alpha -1) \; \gamma^i \: \xi^j \:\xi^k
	\; (h_{jk},_i - h_{ik},_j)  \right) \;\; (m^\vee(\xi) - m^\wedge(\xi)
	) \\
\label{eq:a1_214}
  && + \frac{i}{8 \pi^2} \left( \int_x^y \varepsilon^{ijlm} \; (h_{jk},_i
	- h_{ik},_j) \: \xi^k \; \xi_l \: \rho \gamma_m \right) \;\;
	(m^\vee(\xi) - m^\wedge(\xi)) \\
\label{eq:a1_215}
  && + \frac{1}{2} \left( \int_x^y (\alpha^2 - \alpha) \; \xi^j \; \xi^k \;
	R_{jk} \right) \;\; k_0(x,y) \\
\label{eq:a1_216}
  && + \frac{1}{32 \pi^2} \left( \int_x^y (\alpha^4 - 2 \alpha^3 + \alpha^2)
	\; \xi \slsh \; \xi^j \; \xi^k \; \Box R_{jk} \right) \;\;
	(l^\vee(\xi)-l^\wedge(\xi)) \\
\label{eq:a1_217}
  && - \frac{1}{32 \pi^2} \left( \int_x^y (6 \alpha^2 - 6 \alpha + 1) \;
	\xi \slsh \; R \right) \;\; (l^\vee(\xi)-l^\wedge(\xi)) \\
\label{eq:a1_218}
  && + \frac{1}{32 \pi^2} \left( \int_x^y (4 \alpha^3 - 6 \alpha^2 + 2 \alpha)
	\; \xi^j \: \xi^k \: \gamma^l \; R_{j[k},_{l]} \right) \;\;
	(l^\vee(\xi)-l^\wedge(\xi)) \\
\label{eq:a1_219}
  && - \frac{i}{16 \pi^2} \left( \int_x^y (\alpha^2 - \alpha) \;
	\varepsilon^{ijlm} \: R_{ki},_j \: \xi^k \: \xi_l \: \rho \gamma_m
	\right) \;\; (l^\vee(\xi)-l^\wedge(\xi)) \\
\label{eq:a1_220}
  && - \frac{1}{8 \pi^2} \left( \int_x^y (\alpha^2 - \alpha) \; \xi^j \:
	\gamma^k \: G_{jk} \right) \;\; (l^\vee(\xi)-l^\wedge(\xi)) \\
\label{eq:a1_221}
  && + {\cal{O}}(\xi^0) \spc , \nonumber
\end{eqnarray}
wobei $R_{jk}$ den Ricci- und $G_{jk}=R_{jk} - \frac{1}{2} R \: g_{jk}$
den Einstein-Tensor bezeichnet.\\
($m^\vee$, $m^\wedge$ sind die Distributionen $m^\vee(y)=
\delta^\prime(y^2) \Theta(y^0)$, $m^\wedge(y)=\delta^\prime(y^2)
\Theta(-y^0)$,
ferner wurde $\xi=y-x$ gesetzt.)
\end{Thm}

\subsection{Skalare St"orung}
\begin{Thm}
\label{theorem_sk0}
In erster Ordnung St"orungstheorie gilt
\begin{eqnarray}
\label{eq:a1_s2}
\Delta k_0(x,y) &=& -\frac{1}{2} \: (\Xi(y)+\Xi(x)) \; k^{(1)}(x,y) \\
\label{eq:a1_s3}
&&+ \frac{1}{8 \pi^2} \: (l^\vee(\xi)-l^\wedge(\xi)) \;
	\int_x^y (\partial_j \Xi) \; \xi_k \; \sigma^{jk} \\
\label{a1_s4}
&&- \frac{1}{16 \pi^3} \left(\lint_x^y - \lint_y^x \right) dz \int_x^z
	\alpha^2 \: (\partial_j \Box \Xi) \; \zeta_k \; \sigma^{jk} \\
\label{a1_s5}
&&+ \frac{i}{16 \pi^3} \left(\lint_x^y - \lint_y^x \right) \Box \Xi \spc .
\end{eqnarray}
\end{Thm}

\section{Anhang B: St"orungsrechnung f"ur $k_m$ im Ortsraum}
\subsection{Elektromagnetisches Potential}
\begin{Thm}
\label{a2_theorem2}
In erster Ordnung St"orungstheorie gilt
\begin{eqnarray}
\label{eq:a2_c1}
\lefteqn{ \Delta k_m(x,y) \;=\; \Delta k_0(x,y) } \\
\label{eq:a2_c2}
&& - i e \left( \int_x^y A_j \right) \xi^j \; (k_m-k_0)(x,y) \\
\label{eq:a2_c3}
&& + \frac{ie}{16 \pi^3} \: m \; \left( \lint_x^y - \lint_y^x \right) \:
   F_{ij} \: \sigma^{ij} \\
\label{eq:a2_c4}
&& + \frac{e}{8 \pi^3} \: m \; \left( \lint_x^y - \lint_y^x \right)
   dz \: \int_x^z \alpha^2 \; j_k \: \zeta^k \\
\label{eq:a2_c5}
&& + \frac{ie}{16 \pi^3} \: m^2 \; \left( \lint_x^y - \lint_y^x \right)
   dz \; F_{ij} \; \gamma^i \: (2 z -x-y)^j \\
\label{eq:a2_c6}
&& - \frac{e}{32 \pi^3} \: m^2 \; \left( \lint_x^y - \lint_y^x \right)
   \varepsilon^{ijkl} \: F_{ij} \: \xi_k \; \rho \gamma_l \\
\label{eq:a2_c7}
&& + \frac{ie}{16 \pi^3} \: m^2 \; \left(\lint_x^y - \lint_y^x\right)
   dz \; \int_x^z \alpha^2 \; j_k \: \zeta^k \; \xi \slsh \\
\label{eq:a2_c8}
&& - \frac{i e}{128 \pi^2} \: m^3 \; \left(\int_x^y F_{ij} \: \sigma^{ij}
   \right)
   \; \left(\Theta^\vee(\xi) - \Theta^\wedge(\xi)\right) \; \xi^2 \\
\label{eq:a2_c9}
&& + \frac{e}{64 \pi^2} \: m^3 \; \left(\int_x^y (\alpha^2-\alpha) \;
   j_k \: \xi^k \right) \; \left(\Theta^\vee(\xi) -
   \Theta^\wedge(\xi)\right) \; \xi^2 \\
\label{eq:a2_c10}
&& + \frac{e}{512 \pi^2} \: m^4 \; \left(\int_x^y
   \varepsilon^{ijkl} \: F_{ij} \: \xi_k \; \rho \gamma_l \right) \;
   \left(\Theta^\vee(\xi) - \Theta^\wedge(\xi)\right) \; \xi^2 \\
\label{eq:a2_c11}
&& + \frac{ie}{256 \pi^2} \: m^4 \; \left(\int_x^y (1-2\alpha) \: F_{ij}
   \: \gamma^i \: \xi^j \right) \;
   \left(\Theta^\vee(\xi) - \Theta^\wedge(\xi)\right) \; \xi^2 \\
\label{eq:a2_c12}
&& + \frac{ie}{256 \pi^2} \: m^4 \; \left(\int_x^y (\alpha^2-\alpha) \;
   j_k \: \xi^k \; \xi \slsh \right) \;
   \left(\Theta^\vee(\xi) - \Theta^\wedge(\xi)\right) \; \xi^2 \\
&& + {\cal{O}}(\xi^4) \spc , \nonumber
\end{eqnarray}
wobei $\xi=y-x$, $\zeta=z-x$ gesetzt wurde.
\end{Thm}

\begin{Satz}
\label{a2_satzb9}
F"ur $(y-x) \in \Li$ gilt
\begin{eqnarray*}
\lefteqn{\lim_{\I_x \ni u \rightarrow y} \left( \Delta k_m(x,u) -
\Delta k_0(x,u) \right) }\\
&=& \frac{ie}{32 \pi^2} \; m \: \epsilon(\xi^0) \int_x^y F_{ij} \sigma^{ij} \\
&& - \frac{e}{16 \pi^2} \; m \: \epsilon(\xi^0) \int_x^y (\alpha^2-\alpha) \:
	j_k \: \xi^k \\
&& + \frac{ie}{32 \pi^2} \; m^2 \: \epsilon(\xi^0) \int_x^y (2 \alpha -1) \:
	\gamma^i \: F_{ij} \: \xi^j \\
&& - \frac{e}{64 \pi^2} \; m^2 \: \epsilon(\xi^0) \int_x^y \varepsilon^{ijkl}
	\; F_{ij} \:  \xi_k \; \rho \gamma_l \\
&& - \frac{ie}{32 \pi^2} \; m^2 \: \epsilon(\xi^0) \int_x^y (\alpha^2-\alpha)
	\; j_k \: \xi^k \: \xi \slsh \;+\; {\cal{O}}(m^3) \spc .
\end{eqnarray*}
\end{Satz}

\subsection{Axiales Potential}
\begin{Thm}
\label{a2_thm13}
In erster Ordnung St"orungstheorie gilt
\begin{eqnarray}
\label{eq:a2_x1}
\Delta k_m(x,y) &=& - \rho \: \Delta k_m[\Aslsh](x,y) \\
\label{eq:a2_x2}
&& + \frac{e}{4 \pi^2} \: m \: (l^\vee(\xi)-l^\wedge(\xi)) \: \int_x^y \rho
	\Aslsh \: \xi\slsh \\
\label{eq:a2_x3}
&& + \frac{e}{8 \pi^3} \: m \left( \lint_x^y - \lint_y^x \right) dz \;
	\int_x^y \alpha^2 \: j_k \: \zeta^k \; \rho \\
\label{eq:a2_x4}
&& + \frac{e}{8 \pi^3} \: m \left( \lint_x^y - \lint_y^x \right) \partial_k
	A^k \; \rho \\
\label{eq:a2_x5}
&& + \frac{ie}{4 \pi^3} \: m \left( \lint_x^y - \lint_y^x \right) h_j[A_k] \:
	\rho \sigma^{jk} \\
\label{eq:a2_x6}
&& + \frac{ie}{4 \pi^3} \: m^2 \left( \lint_x^y - \lint_y^x \right) \rho
	\Aslsh \\
\label{eq:a2_x7}
&& - \frac{e}{8 \pi^3} \: m^3 \left( \lint_x^y - \lint_y^x \right) \rho
	\Aslsh \: \xi\slsh \\
\label{eq:a2_x8}
&& - \frac{e}{16 \pi^3} \: m^3 \left( \wint_x^y - \wint_y^x \right)
	(\partial_j \Aslsh) \: \gamma^j \\
\label{eq:a2_x9}
&& - \frac{ie}{16 \pi^3} \: m^4 \left( \vint_x^y + \wint_x^y - \vint_y^x
	- \wint_y^x \right) \: \rho \Aslsh \\
&& + {\cal{O}}(m^5) \spc . \nonumber
\end{eqnarray}
\end{Thm}

\begin{Satz}
\label{a2_satz13}
F"ur $y-x \in \Li$ gilt
\begin{eqnarray*}
\lefteqn{  \lim_{\I_x \ni u \rightarrow y} (\Delta k_m(x,u) - \Delta k_0(x,u))
	} \\
&=& \frac{ie}{32 \pi^2} \: m \: \epsilon(\xi^0) \int_x^y (2\alpha -1) \:
	F_{jk} \; \rho \sigma^{jk} \\
&& + \frac{e}{16 \pi^2} \: m \: \epsilon(\xi^0) \int_x^y \partial_k A^k \;
	\rho \\
&& - \frac{ie}{16 \pi^2} \: m \: \epsilon(\xi^0) \int_x^y (\alpha^2-\alpha) \;
	\Box A_j \: \xi_k \; \rho \sigma^{jk} \\
&& + \frac{ie}{8 \pi^2} \: m^2 \: \epsilon(\xi^0) \int_x^y \rho \Aslsh \\
&& - \frac{ie}{32 \pi^2} \: m^2 \: \epsilon(\xi^0) \int_x^y (2\alpha-1) \; 
	F_{ij} \: \xi^j \; \rho \gamma^i \\
&& + \frac{e}{64 \pi^2} \: m^2 \: \epsilon(\xi^0) \int_x^y \varepsilon^{ijkl}
	\; F_{ij} \: \xi_k \; \gamma_l \\
&& + \frac{ie}{32 \pi^2} \: m^2 \: \epsilon(\xi^0) \int_x^y (\alpha^2-\alpha)
	\; j_k \: \xi^k \; \rho \xi\slsh \\
&& + {\cal{O}}(m^3) \spc .
\end{eqnarray*}
\end{Satz}

\subsection{Gravitationsfeld}
\begin{Thm}
\label{thm_kpm}
In erster Ordnung St"orungstheorie gilt
\begin{eqnarray}
\lefteqn{ \Delta k_m(x,y) \;=\; \Delta k_0(x,y) } \\
\label{eq:a2_g11}
&&- \left( \int_x^y h^k_j \right) \: \xi^j \: \frac{\partial}{\partial y^k}
	\left( k_m(x,y) - k_0(x,y) \right) \\
\label{eq:a2_g12}
&&+ \frac{i}{2} \: m \left( \int_x^y h_{ki,j} \right) \: \xi^k \; \sigma^{ij}
	\; \ke(x,y) \\
\label{eq:a2_g13}
&&+ \frac{i}{8 \pi^2} \: m \; (l^\vee(\xi) - l^\wedge(\xi)) \int_x^y
	(\alpha^2-\alpha) \; R_{jk} \: \xi^j \: \xi^k \\
\label{eq:a2_g14}
&&- \frac{i}{16 \pi^3} \: m \left( \lint_x^y - \lint_y^x \right) dz
	\int_x^z (2 \alpha^2 - \alpha) \; R \\
\label{eq:a2_g15}
&&+\frac{i}{32 \pi^3} \: m \left( \lint_x^y - \lint_y^x \right) dz \;
	\zeta^j \: \zeta^k
	\int_x^z (\alpha^4-\alpha^3) \; (R_{\:,jk} - 2 \: \Box R_{jk}) \\
\label{eq:a2_g16}
&&- \frac{1}{16 \pi^3} \: m \left( \lint_x^y - \lint_y^x \right)dz
	\; \zeta^k \int_x^z \alpha^2 \; R_{ki,j} \; \sigma^{ij} \\
\label{eq:a2_g17}
&&+ \frac{1}{16 \pi^2} \: m^2 \; (l^\vee(\xi)-l^\wedge(\xi)) \int_x^y
	(2\alpha-1) \; (h_{jk,i} - h_{ik,j}) \: \gamma^i \;
	\xi^j \: \xi^k \\
\label{eq:a2_g18}
&&+ \frac{i}{16 \pi^2} \: m^2 \; (l^\vee(\xi)-l^\wedge(\xi)) \int_x^y
	\varepsilon^{ijlm} \; h_{jk,i} \; \xi^k \: \xi_l \; \rho \gamma_m \\
\label{eq:a2_g19}
&&- \frac{1}{16 \pi^2} \: m^2 \; (l^\vee(\xi)-l^\wedge(\xi)) \int_x^y
	(\alpha^2-\alpha) \; R_{jk} \; \xi^j \: \xi^k \; \xi\slsh \\
\label{eq:a2_g20}
&&- \frac{1}{16 \pi^3} \: m^2 \left( \lint_x^y - \lint_y^x \right)dz
	\; \zeta^j \int_x^z \alpha^2 \: R_{jk} \; \gamma^k \\
\label{eq:a2_g21}
&&+ \frac{1}{32 \pi^3} \: m^2 \left( \lint_x^y - \lint_y^x \right)dz
	\int_x^z (2\alpha^2-\alpha) \; R \; \xi\slsh \\
\label{eq:a2_g22}
&&- \frac{1}{64 \pi^3} \: m^2 \left( \lint_x^y - \lint_y^x \right)dz
	\; \zeta^j \: \zeta^k \int_x^z (\alpha^4-\alpha^3) \: (
	R_{\:,jk} - 2 \: \Box R_{jk}) \; \xi\slsh \\
\label{eq:a2_g23}
&&+ \frac{1}{32 \pi^3} \: m^2 \left( \lint_x^y - \lint_y^x \right)dz
	\; \zeta^j \int_x^z \alpha^2 \; (R_{jk,i} - R_{ik,j}) \:
	(2 \alpha \zeta^k - \xi^k) \; \gamma^i \\
\label{eq:a2_g24}
&&+ \frac{i}{32 \pi^3} \: m^2 \left( \lint_x^y - \lint_y^x \right)dz
	\; \zeta^k \int_x^z \alpha^2 \; \varepsilon^{ijlm} \; R_{jk,i} \;
	\xi_l \; \rho \gamma_m \\
&&+ {\cal{O}}(m^3) \nonumber \spc .
\end{eqnarray}
\end{Thm}

\subsection{Skalare St"orung}
\begin{Thm}
\label{a2_theorem_sm}
In erster Ordnung St"orungstheorie gilt
\begin{eqnarray}
\Delta k_m(x,y) &=& \Delta k_0(x,y) \\
\label{eq:a21_s2}
&&- 2 m \left( \int_x^y \Xi \right) \; k^{(2)}(x,y) \\
\label{eq:a2_sd}
&&+ \frac{1}{8 \pi^3} \: m \left(\lint_x^y - \lint_y^x \right) dz
	\; \left( (\Pdd \Xi)(z) - 2 \int_x^z \alpha \; (\Pdd \Xi) \right) \\
\label{eq:a2_sc}
&&- \frac{1}{8 \pi^3} \: m \left(\lint_x^y - \lint_y^x \right) dz \;
	\zeta \slsh \int_x^z \alpha^2 \; (\Box \Xi) \\
\label{eq:a2_sa}
&&+ \frac{3i}{8 \pi^3} \: m^2 \left(\lint_x^y - \lint_y^x \right) \Xi \\
\label{eq:a2_449}
&&+ \frac{i}{16 \pi^3} \: m^2 \left(\lint_x^y - \lint_y^x \right) dz\;
	(\partial_j \Xi) \: (2 \zeta^j - \xi^j) \\
&&- \frac{1}{16 \pi^3} \: m^2 \left(\lint_x^y - \lint_y^x \right) dz \;
	(\partial_j \Xi) \: \xi_k \; \sigma^{jk} \\
\label{eq:a2_sb}
&&- \frac{1}{8 \pi^3} \: m^3 \left(\lint_x^y - \lint_y^x \right) \Xi \; \xi\slsh \\
&&+\frac{1}{32 \pi^3} \: m^3 \left(\vint_x^y - \wint_x^y + \vint_y^x -
	\wint_y^x \right) \Pdd \Xi \\
&&+ {\cal{O}}(m^4) \spc . \nonumber
\end{eqnarray}
\end{Thm}

\subsection{Pseudoskalare St"orung}
\begin{Thm}
\label{a2_theorem_psm}
In erster Ordnung St"orungstheorie gilt
\begin{eqnarray}
\Delta k_m(x,y) &=& -i \rho \: \Delta k_0[\Xi](x,y) \\
\label{eq:a2_pa}
&&+ \frac{i}{8 \pi^3} \: m \: \rho \left(\lint_x^y-\lint_y^x \right) \Pdd \Xi \\
\label{eq:a2_pz}
&&+ \frac{1}{8 \pi^3} \: m^2 \left(\lint_x^y - \lint_y^x \right) \Xi \; \rho \\
\label{eq:a2_p449}
&&+ \frac{1}{16 \pi^3} \: m^2 \left(\lint_x^y - \lint_y^x \right) dz\;
	(\partial_j \Xi) \: (2 \zeta^j - \xi^j) \; \rho \\
&&+ \frac{i}{16 \pi^3} \: m^2 \left(\lint_x^y - \lint_y^x \right) dz \;
	(\partial_j \Xi) \: \xi_k \; \rho \sigma^{jk} \\
&&-\frac{i}{32 \pi^3} \: m^3 \left( \vint_x^y + \wint_x^y - \vint_y^x
	- \wint_y^x \right) \rho (\Pdd \Xi) \\
&&+ {\cal{O}}(m^4) \spc . \nonumber
\end{eqnarray}
\end{Thm}

\section{Anhang C: St"orungsrechnung f"ur $p_0$ im Ortsraum}
\subsection{Elektromagnetisches Potential}
\begin{Thm}
\label{a3_theorem1}
In erster Ordnung St"orungstheorie gilt
\begin{eqnarray}
\label{eq:a3_111b}
\Delta p_0(x,y) &=& - i e \left( \int_x^y A_j \right) \: \xi^j \; p_0(x,y) \\
\label{eq:a3_112a}
   && - \frac{e}{8 \pi^3} \left( \int_x^y (\alpha^2-\alpha) \; \xi \slsh
	\: \xi^k \: j_k \right) \;\; \frac{1}{\xi^2} \\
\label{eq:a3_113a}
   && + \frac{e}{8 \pi^3} \left( \int_x^y (2 \alpha - 1) \; \xi^j \:
	\gamma^k \: F_{kj} \right) \;\; \frac{1}{\xi^2} \\
\label{eq:a3_114a}
   && + \frac{i e}{16 \pi^3} \left( \int_x^y \varepsilon^{ijkl} \; F_{ij} \:
	\xi_k \; \rho \gamma_l \right) \;\; \frac{1}{\xi^2} \\
\label{eq:a3_115a}
  && - \frac{e}{64 \pi^3} \: \int_x^y (\alpha^4 - 2 \alpha^3
	+ \alpha^2) \: \xi \slsh \: \xi_k \; \Box j^k \;\; \ln(|\xi^2|) \\
 \label{eq:a3_116a}
 && + \frac{e}{64 \pi^3} \: \int_x^y (4 \alpha^3
	- 6 \alpha^2 + 2 \alpha) \: \xi^j \: \gamma^k \: (\Box F_{kj})
	\;\; \ln(|\xi^2|) \\
  \label{eq:a3_117a}
&& + \frac{i e}{64 \pi^3} \: \int_x^y (\alpha^2 - \alpha) \;
	\varepsilon^{ijkl} \: (\Box F_{ij}) \: \xi_k \; \rho \gamma_l
	\;\; \ln(|\xi^2|) \\
  \label{eq:a3_118a}
&& + \frac{e}{8 \pi^3} \: \int_x^y (\alpha^2 - \alpha)
	\: \gamma^k \: j_k \;\; \ln(|\xi^2|) \\
  && + {\cal{O}}(\xi^0) \spc . \nonumber
\end{eqnarray}
\end{Thm}

\subsection{Gravitationsfeld}
\begin{Thm}
\label{thm_gp0}
In erster Ordnung St"orungstheorie gilt in symmetrischer Eichung
\begin{eqnarray}
\label{eq:a3_g3}
\Delta{p}_0(x,y) &=&
	- \left( \int_x^y h^k_j \right) \xi^j \frac{\partial}{\partial y^k}
	\;\; p_0(x,y) \\
\label{eq:a3_g4}
  && + \frac{i}{4 \pi^3} \; \frac{1}{\xi^4}
	\left( \int_x^y (2 \alpha -1) \; \gamma^i \: \xi^j \:\xi^k
	\; (h_{jk},_i - h_{ik},_j)  \right) \\
\label{eq:a3_g5}
  && + \frac{1}{8 \pi^3} \; \frac{1}{\xi^4}
	\left( \int_x^y \varepsilon^{ijlm} \; (h_{jk},_i
	- h_{ik},_j) \: \xi^k \; \xi_l \: \rho \gamma_m \right) \\
\label{eq:a3_g6}
  && + \frac{1}{2} \left( \int_x^y (\alpha^2 - \alpha) \; \xi^j \; \xi^k \;
	R_{jk} \right) \;\; p_0(x,y) \\
\label{eq:a3_g7}
  && + \frac{i}{32 \pi^3}  \; \frac{1}{\xi^2}
	\left( \int_x^y (\alpha^4 - 2 \alpha^3 + \alpha^2)
	\; \xi \slsh \; \xi^j \; \xi^k \; \Box R_{jk} \right) \\
\label{eq:a3_g8}
  && - \frac{i}{32 \pi^3}  \; \frac{1}{\xi^2}
	\left( \int_x^y (6 \alpha^2 - 6 \alpha + 1) \;
	\xi \slsh \; R \right) \\
\label{eq:a3_g9}
  && + \frac{i}{32 \pi^3} \; \frac{1}{\xi^2}
	\left( \int_x^y (4 \alpha^3 - 6 \alpha^2 + 2 \alpha)
	\; \xi^j \: \xi^k \: \gamma^l \; R_{j[k},_{l]} \right) \\
\label{eq:a3_g10}
  && + \frac{1}{16 \pi^3} \; \frac{1}{\xi^2}
	\left( \int_x^y (\alpha^2 - \alpha) \;
	\varepsilon^{ijlm} \: R_{ki},_j \: \xi^k \: \xi_l \: \rho \gamma_m
	\right) \\
\label{eq:a3_g11}
  && - \frac{i}{8 \pi^3} \; \frac{1}{\xi^2}
	\left( \int_x^y (\alpha^2 - \alpha) \; \xi^j \:
	\gamma^k \: G_{jk} \right) \\
\label{eq:a3_g12}
  && + {\cal{O}}(\ln(|\xi^2|)) \nonumber \spc  .
\end{eqnarray}
\end{Thm}

\subsection{Skalare St"orung}
\begin{Thm}
\label{theorem_sp0}
In erster Ordnung St"orungstheorie gilt
\begin{eqnarray}
\label{eq:a3_s2}
\Delta p_0(x,y) &=& - \frac{1}{2} \: (\Xi(y)+\Xi(x)) \; p^{(1)}(x,y) \\
&&+ \frac{i}{8 \pi^3} \frac{1}{\xi^2} \int_x^y
	\; (\partial_j \Xi) \; \xi_k \; \sigma^{jk} \\
&&+ \frac{i}{32 \pi^3} \: \ln(|\xi^2|) \int_x^y
	(\alpha^2 - \alpha) \; (\partial_j \Box \Xi) \;
	\xi_k \; \sigma^{jk} \\
&&- \frac{1}{32 \pi^3} \: \ln(|\xi^2|) \int_x^y
	\Box \Xi \\
&& +\; {\cal{O}}(\xi^0) \spc . \nonumber
\end{eqnarray}
\end{Thm}

\section{Anhang D: St"orungsrechnung f"ur $p_m$ im Ortsraum}
\subsection{Elektromagnetisches Potential}
\begin{Thm}
\label{a4_thm1}
In erster Ordnung St"orungstheorie gilt
\begin{eqnarray}
\lefteqn{\Delta p_m(x,y) \;=\; \Delta p_0(x,y) } \nonumber \\
\label{eq:a4_111f}
&& - i e \left( \int_x^y A_j \right) \xi^j \; (p_m-p_0)(x,y) \\
&& - \frac{e}{32 \pi^3} \; m \; \ln(|\xi^2|) \; \int_x^y F_{ij} \:
	\sigma^{ij} \\
\label{eq:a4_113f}
&& - \frac{ie}{16 \pi^3} \; m \: \ln(|\xi^2|) \; \int_x^y (\alpha^2-\alpha) \:
	j_k \: \xi^k \\
&& - \frac{e}{32 \pi^3} \; m^2 \: \ln(|\xi^2|) \; \int_x^y (2 \alpha -1) \:
	\gamma^i \: F_{ij} \: \xi^j \\
&& - \frac{ie}{64 \pi^3} \; m^2 \: \ln(|\xi^2|) \; \int_x^y \varepsilon^{ijkl}
	\; F_{ij} \:  \xi_k \; \rho \gamma_l \\
\label{eq:a4_116f}
&& + \frac{e}{32 \pi^3} \; m^2 \: \ln(|\xi^2|) \; \int_x^y (\alpha^2-\alpha)
	\; j_k \: \xi^k \; \xi \slsh \\
&& + {\cal{O}}(m^3) + {\cal{O}}(\xi^0) \spc . \nonumber
\end{eqnarray}
\end{Thm}

\subsection{Axiales Potential}
\begin{Thm}
\label{a4_thm13}
In erster Ordnung St"orungstheorie gilt
\begin{eqnarray}
\label{eq:a4_z1}
\Delta p_m(x,y) &=& - \rho \: \Delta p_0[\Aslsh](x,y)  \\
\label{eq:a4_z2}
&& - \frac{ie}{4 \pi^3} \: m \: \frac{1}{\xi^2} \int_x^y \rho \:
	\frac{1}{2} [\xi\slsh, \Aslsh] \\
\label{eq:a4_z3}
&& -\frac{e}{32 \pi^3} \: m \: \ln(|\xi^2|) \int_x^y (2\alpha -1) \;
	F_{jk} \; \rho \sigma^{jk} \\
\label{eq:a4_z4}
&& +\frac{ie}{16 \pi^3} \: m \:  \ln(|\xi^2|) \int_x^y \partial_j A^j \;
	\rho \\
\label{eq:a4_z5}
&& +\frac{e}{16 \pi^3} \: m \:  \ln(|\xi^2|) \int_x^y (\alpha^2-\alpha) \;
	\Box A_j \; \xi_k \; \rho \sigma^{jk} \\
\label{eq:a4_z6}
&& +\frac{e}{8 \pi^3} \: m^2 \: \frac{1}{\xi^2} \int_x^y A_j \: \xi^j \;
	\rho \xi\slsh \\
\label{eq:a4_z7}
&& -\frac{e}{8\pi^3} \: m^2 \: \ln(|\xi^2|) \int_x^y \rho \Aslsh \\
\label{eq:a4_z8}
&& +\frac{e}{32 \pi^3} \: m^2 \: \ln(|\xi^2|) \int_x^y (2\alpha-1) \;
	F_{jk} \: \xi^k \; \rho \gamma^j \\
\label{eq:a4_z9}
&& +\frac{ie}{64 \pi^3} \: m^2 \: \ln(|\xi^2|) \int_x^y \varepsilon^{ijkl}
	\; F_{ij} \: \xi_k \; \gamma_l \\
\label{eq:a4_z10}
&& -\frac{e}{32 \pi^3} \: m^2 \: \ln(|\xi^2|) \int_x^y (\alpha^2-\alpha) \;
	j_k \: \xi^k \; \rho \xi\slsh \\
\label{eq:a4_z11}
&& +\frac{ie}{16 \pi^3} \: m^3 \: \ln(|\xi^2|) \int_x^y \rho\:
	\frac{1}{2} [\xi\slsh, \Aslsh] \\
&& + {\cal{O}}(m^4) + {\cal{O}}(\xi^0) \spc . \nonumber
\end{eqnarray}
\end{Thm}

\subsection{Gravitationsfeld}
\begin{Thm}
In erster Ordnung St"orungstheorie gilt
\begin{eqnarray}
\Delta p_m(x,y) &=& \Delta p_0(x,y) \\
&&- \left( \int_x^y h^k_j \right) \: \xi^j \; \frac{\partial}{\partial y^k}
	\left( p_m(x,y)-p_0(x,y) \right) \\
&&+ \frac{i}{2} \: m \left( \int_x^y h_{ki,j} \right) \: \xi^k \;
	\sigma^{ij} \; \pe(x,y) \\
\label{eq:a4_gra}
&&+ \frac{1}{2} \: m \int_x^y (\alpha^2-\alpha) \;
	R_{jk} \; \xi^j \: \xi^k \; p^{(1)}(x,y) \\
&&+ \frac{1}{16 \pi^3} \: m \: \ln(|\xi^2|) \int_x^y
	(\alpha^2-\alpha+\frac{1}{4}) \; R \\
&&- \frac{1}{64 \pi^3} \: m \: \ln(|\xi^2|) \int_x^y (\alpha^4 - 2 \alpha^3 +
	\alpha^2) \; (\Box R_{jk}) \; \xi^j \: \xi^k \\
&&+ \frac{i}{32 \pi^3} \: m \: \ln(|\xi^2|) \int_x^y (\alpha^2-\alpha) \;
	R_{ki,j} \; \xi^k \; \sigma^{ij} \\
&&+ \frac{i}{16 \pi^3} \: m^2 \: \frac{1}{\xi^2} \int_x^y (2\alpha-1)
	\; (h_{jk,i} - h_{ik,j}) \: \gamma^i \; \xi^j \: \xi^k \\
&&- \frac{1}{16 \pi^3} \: m^2 \: \frac{1}{\xi^2} \int_x^y \varepsilon^{ijlm}
	h_{jk,i} \; \xi^k \: \xi_l \; \rho \gamma_m \\
\label{eq:a4_grb}
&&- \frac{i}{16 \pi^3} \: m^2 \: \frac{1}{\xi^2} \int_x^y (\alpha^2-\alpha)
	\; R_{jk} \; \gamma^j \: \xi^k \\
&&+ {\cal{O}}(\xi^0) + m^2 \: {\cal{O}}(\ln(|\xi^2|)) + {\cal{O}}(m^3)
	\spc . \nonumber
\end{eqnarray}
\end{Thm}

\subsection{Skalare St"orung}
\begin{Thm}
\label{a4_theorem_spm}
In erster Ordnung St"orungstheorie gilt
\begin{eqnarray}
\Delta p_m(x,y) &=& \Delta p_0(x,y) \\
\label{eq:a4_s2}
&&- 2 m \left( \int_x^y \Xi \right) \; p^{(2)}(x,y) \\
&&+ \frac{i}{16 \pi^3} \: m \: \ln(|\xi^2|) \int_x^y (2\alpha-1) \;
	(\Pdd \Xi) \\
&&+ \frac{i}{16 \pi^3} \: m \: \ln(|\xi^2|) \int_x^y (\alpha^2-\alpha) \;
	(\Box \Xi) \; \xi\slsh \\
&&- \frac{1}{32 \pi^3} \: m^2 \: \ln(|\xi^2|) \; (\Xi(y) + \Xi(x)) \\
&&- \frac{1}{8 \pi^3} \: m^2 \: \ln(|\xi^2|) \int_x^y \Xi \\
&&- \frac{i}{32 \pi^3} \: m^2 \: \ln(|\xi^2|) \int_x^y 
	(\partial_j \Xi) \: \xi_k \; \sigma^{jk} \\
&&+ {\cal{O}}(m^3) + {\cal{O}}(\xi^0) \spc . \nonumber
\end{eqnarray}
\end{Thm}

\subsection{Pseudoskalare St"orung}
\begin{Thm}
\label{a4_theorem_pspm}
In erster Ordnung St"orungstheorie gilt
\begin{eqnarray}
\Delta p_m(x,y) &=& -i \rho \: \Delta p_0[\Xi](x,y) \\
&&- \frac{1}{16 \pi^3} \: m \: \rho \: \ln(|\xi^2|) \int_x^y (\Pdd \Xi) \\
&&+ \frac{i}{32 \pi^3} \: m^2 \: \ln(|\xi^2|) \; (\Xi(y) + \Xi(x)) \; \rho \\
&&- \frac{1}{32 \pi^3} \: m^2 \: \ln(|\xi^2|) \int_x^y 
	(\partial_j \Xi) \: \xi_k \; \rho \sigma^{jk} \\
&&+ {\cal{O}}(m^3) + {\cal{O}}(\xi^0) \spc . \nonumber
\end{eqnarray}
\end{Thm}

\section{Anhang E: St"orungsrechnung h"oherer Ordnung}
\begin{Satz}
\label{a6_satz7}
Mit der symbolischen Ersetzung $C=k$ oder $C=p$ gilt
\begin{eqnarray}
\lefteqn{ \chi_L \: \tilde{C}(x,y) \;=\; \chi_L \: \Texp \left(-i \int_x^y
	A_L^j \: \xi_j \right) \; C(x,y) } \nonumber \\
&&-\frac{1}{2}\:\chi_L \int_x^y dz \; (2 \alpha-1) \;\T e^{-i \int_x^z
	A_L^a \: (z-x)_a} \; \xi_j \: \gamma_k \: F_L^{kj}
	\;\T e^{-i \int_z^y A_L^b \: (y-z)_b} \; C^{(1)}(x,y)
	\nonumber \\
&&+\frac{1}{2}\:\chi_L \int_x^y dz \;(\alpha^2-\alpha)\;\T e^{-i \int_x^z
	A_L^a \: (z-x)_a} \;\xi \slsh \: \xi_k \: j_L^k
	\;\T e^{-i \int_z^y A_L^b \: (y-z)_b} \; C^{(1)}(x,y)
	\nonumber \\
&&-\frac{i}{4}\:\chi_L \int_x^y dz \;\T e^{-i \int_x^z
	A_L^a \: (z-x)_a} \;\varepsilon_{ijkl} \: F_L^{ij} \: \xi^k
	\; \rho \gamma^l
	\;\T e^{-i \int_z^y A_L^b \: (y-z)_b} \; C^{(1)}(x,y)
	\nonumber \\
&&-\frac{m}{2}\:\chi_L \int_x^y dz \;\Texp \left(-i \int_x^z
	A_L^a \: (z-x)_a\right) \;(-i \Aslsh_L(z) \: Y + i Y \: \Aslsh_R(z))
	\:\xi\slsh \nonumber \\
&& \hspace*{4cm} \times \;\Texp \left(-i \int_z^y A_R^b \: (y-z)_b\right)
	\; C^{(1)}(x,y) \nonumber \\
\label{eq:a6_101}
&&+ {\cal{O}}(\ln(|\xi^2|)) \;+\; {\cal{O}}(m^2)
\end{eqnarray}
F"ur die rechtsh"andige Komponente hat man die analoge Gleichung, wenn
man die Indizes $L$, $R$ vertauscht.
\end{Satz}

\begin{Thm}
\label{a6_thm0}
Mit der symbolischen Ersetzung $C=p$ oder $C=k$ gilt
\begin{eqnarray}
\label{eq:a6_400}
\lefteqn{\chi_L \: (VXCV^*)(x,y) \;=\; \chi_L \: U_L(x) \:
	\Texp \left(-i \int_x^y A_L^j \: \xi_j\right) \:X_L\:
	U_L^{-1}(y) \; C_0(x,y) } \\
&&-\frac{1}{2}\:\chi_L \: U_L(x) \: X_L 
	\int_x^y dz \; (2 \alpha-1) \;\T e^{-i \int_x^z
	A_L^a \: (z-x)_a} \; \xi_j \: \gamma_k \: F_L^{kj}(z) \nonumber \\
\label{eq:a6_401}
&&\hspace*{6cm} \times \; \T e^{-i \int_z^y A_L^b \: (y-z)_b} \:U_L^{-1}(y)
	\; C^{(1)}(x,y) \\
&&+\frac{1}{2}\:\chi_L \: U_L(x) \: X_L
	\int_x^y dz \;(\alpha^2-\alpha)\;\T e^{-i \int_x^z
	A_L^a \: (z-x)_a} \;\xi \slsh \: \xi_k \: j_L^k(z) \nonumber \\
\label{eq:a6_402}
&&\hspace*{6cm} \times \;\T e^{-i \int_z^y A_L^b \: (y-z)_b} \: U_L^{-1}(y)
	\; C^{(1)}(x,y) \\
&&-\frac{i}{4}\:\chi_L \: U_L(x) \: X_L
	\int_x^y dz \;\T e^{-i \int_x^z
	A_L^a \: (z-x)_a} \;\varepsilon_{ijkl} \: F_L^{ij}(z) \: \xi^k
	\; \rho \gamma^l\nonumber \\
\label{eq:a6_403}
&&\hspace*{6cm} \times \;\T e^{-i \int_z^y A_L^b \: (y-z)_b} \: U_L^{-1}(y)
	\; C^{(1)}(x,y) \\
\label{eq:a6_404}
&&+ m \: \chi_L \: U_L(x) \: \Texp \left(-i\int_x^y A_L^j \: \xi_j \right)
	\: X_L \: U_L^{-1}(y) \:Y_L(y) \; C^{(1)}(x,y) \\
&&-\frac{m}{4} \: \chi_L \: U_L(x) \inti d\lambda \nonumber \\
&&\hspace*{.5cm} \times \; \left\{ \epsilon(\lambda) \: \hat{\Pdd}_z \left(
	\T e^{-i\int_x^z A_L^j \: (z-x)_j} \: (U_L^{-1} \:Y_L\: U_R)_{|z} \:
	\T e^{-i\int_z^y A_R^k \: (y-z)_k} \right) X_R \:\xi\slsh \right.
	\nonumber \\
&&\hspace*{1cm} \left. +\:\epsilon(1- \lambda) \: X_L \: \hat{\Pdd}_z \left(
	\T e^{-i\int_x^z A_L^j \: (z-x)_j} \: (U_L^{-1}\:Y_L\: U_R)_{|z} \:
	\T e^{-i\int_z^y A_R^k \: (y-z)_k} \right) \:\xi\slsh \right\}
	\nonumber \\
\label{eq:a6_405}
&&\hspace*{0.5cm} \times \; U_R^{-1}(y) \; C^{(1)}(x,y) \;+\;
	{\cal{O}}(\ln(|\xi^2|)) \;+\; {\cal{O}}(m^2) \spc .
\end{eqnarray}
Zur Abk"urzung wurde $z=\lambda y + (1-\lambda)x$ gesetzt.
F"ur die rechtsh"andige Komponente gilt die analoge Gleichung, wenn
man die Indizes $L$, $R$ vertauscht.
\end{Thm}

\begin{Thm}
\label{a6_satz26}
Es gilt mit der symbolischen Ersetzung $C=p$ oder $C=k$
\begin{eqnarray}
\label{eq:a6_A}
\lefteqn{ \chi_L \: \tilde{C}^{(2)}(x,y) \;=\; \chi_L \: U_L(x) \int_x^y dz \;
	(U_L^{-1} \:Y_L \: Y_R\: U_L)_{|z} \;U_L^{-1}(y) \; C^{(2)}(x,y) } \\
\label{eq:a6_B}
&&-\frac{i}{2} \chi_L \: U_L(x) \int_x^y (\alpha^2-\alpha) \; \Box
	(U_L^{-1} \:Y_L \: Y_R\: U_L)_{|z} \: U_L^{-1}(y) \:
	\xi\slsh \; C^{(3)}(x,y) \\
&&+\frac{i}{2} \chi_L \: U_L(x) \int_x^y dz \int_x^z du \;
	\Pdd (U_L^{-1} \:Y_L\: U_R)_{|u} \; (z-x)^j \gamma_j \;
	\Pdd (U_R^{-1} \:Y_R\: U_L)_{|z} \nonumber \\
\label{eq:a6_C}
&& \hspace*{5.5cm} \times U_L^{-1}(y) \; C^{(3)}(x,y) \\
&&+ i \: \chi_L \:  U_L(x) \int_x^y (1-\alpha) \;
	\Pdd (U_L^{-1} \:Y_L\: U_R)_{|z} \; (U_R^{-1} \:Y_R\: U_L)_{|z}
	\nonumber \\
\label{eq:a6_D}
&& \hspace*{5.5cm} \times U_L^{-1}(y) \; C^{(3)}(x,y) \\
\label{eq:a6_E}
&&- i \: \chi_L \:  U_L(x) \int_x^y \alpha \;
	(U_L^{-1} \:Y_L\: U_R)_{|z} \; \Pdd (U_R^{-1} \:Y_R\: U_L)_{|z} \;
	U_L^{-1}(y) \; C^{(3)}(x,y) \\
&& \;+\; \left\{ \begin{array}{cc}
	{\cal{O}}(\xi^2) & {\mbox{f"ur $C=k$}} \\
	{\cal{O}}(\xi^0) & {\mbox{f"ur $C=p$}} \end{array} \right. \spc . \nonumber
\end{eqnarray}
\end{Thm}

\section{Anhang F: Spektrale Analyse von $P(x,y) \: P(y,x)$}
Ist fertig getippt, wird aber erst ab Abschnitt \ref{4_ab6} referiert.

\section{Anhang G: (Nichtlokale St"orungen)}
Ist noch nicht ausgearbeitet.


Harvard University, Department of Mathematics, Cambridge, MA 02138, USA\\
E-mail: {\bf{felix@abel.math.harvard.edu}}


\newpage
\addcontentsline{toc}{chapter}{Referenzen}
\begin{thebibliography}{99}

\bibitem[BD1]{bjorken} Bjorken / Drell
    {\em Bibliographisches Institut}\\ Relativistische Quantenmechanik
\bibitem[BD2]{bjorken2} Bjorken / Drell
    {\em Bibliographisches Institut}\\ Relativistische Quantenfeldtheorie
\bibitem[E]{ebert} D. Ebert
    {\em Akademie-Verlag Berlin}\\ Eichtheorien
\bibitem[F1]{Physdip} F. Finster
    {\em Diplomarbeit Physik}, Universit"at Heidelberg, 1992\\ Eichfreiheiten
	bei Diracoperatoren
\bibitem[F2]{Mathdip} F. Finster
    {\em Diplomarbeit Mathematik}, Universit"at Heidelberg, 1992\\ Possible
	Explanation of Physical Gauge Freedoms
\bibitem[F]{Fl} F. G. Friedlander
    {\em{Cambridge University Press}}, 1975 \\ The Wave equation on a curved space-time
\bibitem[IZ]{IZ} C. Itzykson / J.-B. Zuber
    {\em{McGraw Hill}}, 1980 \\ Quantum Field Theory
\bibitem[RS]{reed} M. Reed / B. Simon
    {\em Academic Press}\\ Methods of Modern Mathematical Physics\\
    I. Functional Analysis
\bibitem[R]{ross} Ross,
    {\em McGraw Hill}\\ Grand Unified Theories
\end{thebibliography}
\end{document}